\documentclass[oneside,11pt]{book}

\pdfoutput=1

\usepackage{coursenotes}

\renewcommand{\documentYear}{2026}

\title{Notes on Theory of Distributed Systems}
\hypersetup{pdftitle=Notes on Theory of Distributed Systems}

\author{James Aspnes}
\hypersetup{pdfauthor=James Aspnes}

\date{\quad}

\makeindex

\begin{document}

\frontmatter

\hypersetup{pageanchor=false}
\maketitle
\hypersetup{pageanchor=true}

\pagebreak

\mbox{}

\vfill

Copyright © 2002–\documentYear{} by James Aspnes.  Distributed under a Creative
Commons Attribution-ShareAlike 4.0 International license:
\url{https://creativecommons.org/licenses/by-sa/4.0/}.

\clearpage
\phantomsection
\addcontentsline{toc}{chapter}{Table of contents}
\tableofcontents

\phantomsection
\addcontentsline{toc}{chapter}{List of figures}
\listoffigures

\phantomsection
\addcontentsline{toc}{chapter}{List of tables}
\listoftables

\phantomsection
\addcontentsline{toc}{chapter}{List of algorithms}
\listofalgorithms

\myChapter{Preface}{2026}{}

These are notes for the Yale
course CPSC 4650/5650 Theory of Distributed Systems.
This document also
incorporates the lecture schedule and assignments, as well as some
sample assignments from previous semesters.
Because this is a work in
progress, it will be updated frequently over the course of the
semester.

The most recent version of these notes will be available at
\url{https://www.cs.yale.edu/homes/aspnes/classes/4650/notes.pdf}.
More stable archival versions may be found at 
\url{https://arxiv.org/abs/2001.04235}.

Not all topics in the notes will be covered during a particular semester. Some
chapters have not been updated and are marked as possibly out of date.

Much of the structure of the course follows 
Attiya and Welch's \emph{Distributed Computing}~\cite{AttiyaW2004},
with some topics based on Lynch's \emph{Distributed
Algorithms}~\cite{Lynch1996} and additional readings from the research
literature.  In most 
cases you'll find these materials contain much more detail than what is
presented here, so it may be better to consider this document a supplement
to them than to treat it as your primary source of information.

\section*{Acknowledgments}

Many parts of these notes were improved by feedback from students
taking various versions of this course, as well as others who have
kindly pointed out errors in the notes after reading them online.
Many of these suggestions, sadly, went unrecorded, so I must 
apologize to the many students who should be thanked here but 
whose names I didn't keep track of in the past.
However,
I can thank Mike Marmar
and Hao Pan in particular for suggesting improvements to some of the
posted solutions, Guy Laden for suggesting corrections to
Figure~\ref{fig-Paxos-mixed-up-execution}, Ali Mamdouh for
pointing out an error in the original presentation of
Algorithm~\ref{alg-peterson},
and Or Sattath for pointing out several errors in the original
presentation in Section~\ref{section-Paxos-safety}.
Some other contributions are acknowledged in the text.

\mainmatter

\myChapter{Introduction}{2026}{}
\label{chapter-introduction}

\indexConcept{distributed systems}{Distributed systems} are
characterized by their structure: a typical distributed system will
consist of some large number of interacting devices that each run
their own programs but that are affected by receiving messages, or
observing shared-memory updates or the states of other devices.  Examples of distributed systems range from simple systems in which a single client talks to a single server to huge amorphous networks like the Internet as a whole.

As distributed systems get larger, it becomes harder and harder to
predict or understand their behavior.  Part of the reason for
this is that we as programmers have not yet developed a
standardized set of tools for managing complexity (like subroutines or
objects with narrow interfaces, or even simple structured programming
mechanisms like loops or if/then statements) as are found in sequential programming.  Part of the reason is that large distributed systems bring with them large amounts of inherent \concept{nondeterminism}—unpredictable events like delays in message arrivals, the sudden failure of components, or in extreme cases the nefarious actions of faulty or malicious machines opposed to the goals of the system as a whole.  Because of the unpredictability and scale of large distributed systems, it can often be difficult to test or simulate them adequately.  Thus there is a need for theoretical tools that allow us to prove properties of these systems that will let us use them with confidence.

The first task of any theory of distributed systems is modeling:
defining a mathematical structure that abstracts out all relevant
properties of a large distributed system.  There are many foundational
models in the literature for distributed systems, but for this class we will follow
\cite{AttiyaW2004} and use simple automaton-based models.  

What this means is that we model each process in the system as an
automaton that has some sort of local \concept{state}, and model local computation as
a transition rule that tells us how to update this state in response
to various \indexConcept{event}{events}.  Depending on what kinds of
system we are modeling, these events might
correspond to local computation, to delivery of a message by a
network, carrying out some operation on a shared memory, or even
something like a chemical reaction between two molecules.  The
transition rule for a system specifies how the states of all processes
involved in the event are updated, based on their previous states.  We
can think of the transition rule as an arbitrary mathematical function
(or relation if the processes are nondeterministic); this corresponds
in programming terms to implementing local computation by processes as
a gigantic table lookup.

Obviously this is not how we program systems in practice.  But what
this approach does is allow us to abstract away completely from how
individual processes work, and emphasize how all of the 
processes interact with each other.  This can lead to odd results: for
example, it's perfectly consistent with this model for some process to
be able to solve the halting problem, or carry out arbitrarily complex
calculations between receiving a message and sending its response.  A
partial justification for this assumption is that in practice, the
multi-millisecond latencies in even reasonably fast networks are eons
in terms of local computation.  And as with any assumption, we can
always modify it if it gets us into trouble.

\section{Models}

The global state consisting of all process states is called a
\concept{configuration}, and 
we think of the system as a whole as passing from one global state or
\concept{configuration} to another in response to each event.  When
this occurs the processes participating in the event update their
states, and the other processes do nothing.  This does not model
concurrency directly; instead, we interleave potentially concurrent
events in some arbitrary way.  The advantage of this interleaving
approach is that it gives us essentially the same behavior as we would
get if we modeled simultaneous events explicitly, but still allows us
to consider only one event at a time and use induction to prove
various properties of the sequence of configurations we might reach.

We will often use lowercase Greek letters for individual events or
sequences of events.  Configurations are typically written as capital
Latin letters (often $C$).  An \concept{execution} of a schedule is an
alternating sequence of configurations and events $C_0 σ_1 C_1 σ_2
C_2 \dots$, where $C_{i+1}$ is the configuration that results from
applying event $σ_i$ to configuration $C$.  A \concept{schedule} is
a sequence of events $σ_1 σ_2 \dots$ from some execution.  We say
that an event $σ$ is \concept{enabled} in $C$ if this event can be
carried out in $C$; an example would be that the event that we deliver
a particular message in a message-passing system is enabled only if
that message has been sent and not yet delivered.  When $σ$ is enabled
in $C$, it is sometime convenient to write $Cσ$ for the configuration
that results from applying $σ$ to $C$.

What events are available, and what effects they have, will depend on
what kind of model we are considering.  We may also have additional
constraints on what kinds of schedules are \concept{admissible}, which
restricts the schedules under consideration to those that have
certain desirable properties (say, every message that is sent is
eventually delivered).
There are many models in the
distributed computing literature, which can be divided into a handful
of broad categories:
\begin{itemize}
\item \indexConcept{message-passing}{Message passing} models
(which we will cover in Part~\ref{part-message-passing})
correspond to systems where processes communicate by sending messages
through a network.  In
\index{message-passing!synchronous}\concept{synchronous
message-passing}, every process sends out messages at time $t$ that are
delivered at time $t+1$, at which point more messages are sent out that
are delivered at time $t+2$, and so on: the whole system runs in
lockstep, marching forward in perfect synchrony.\footnote{In an
    interleaving model, these apparently simultaneous events are still
    recorded one at a time.  What makes the system synchronous is that
    we demand that, in any admissible schedule, all 
    $n$ events for time $t$ occur as a sequential block,
    followed by all $n$ events for time $t+1$, and so
on.} Such systems are
difficult to build when the components become too numerous or too
widely dispersed, but they are often easier to analyze than
\index{message-passing!asynchronous}\indexConcept{asynchronous message-passing}{asynchronous} systems,
where messages are only delivered eventually after some unknown delay.
Variants on these models include
\index{message-passing!semi-synchronous}\indexConcept{semi-synchronous message-passing}{semi-synchronous} 
systems, where message delays are unpredictable but bounded, and
various sorts of timed systems.  Further variations come from
restricting which processes can communicate with which others, by
allowing various sorts of failures:
\index{failure!crash}\indexConcept{crash failure}{crash failures}
that stop a process dead,
\index{failure!Byzantine}\indexConcept{Byzantine failure}{Byzantine
failures} that turn a process evil, or
\index{failure!omission}\indexConcept{omission failure}{omission failures} 
that drop messages in transit.  Or—on the helpful
side—we may supply additional tools like
\indexConcept{failure detector}{failure detectors} (Chapter~\ref{chapter-failure-detectors})
or
\concept{randomization} (Chapter~\ref{chapter-randomized-consensus}).
\item \indexConcept{shared memory}{Shared-memory} models
(Part~\ref{part-shared-memory}) correspond to systems where processes
communicate by executing operations on shared objects
        
        In the
simplest case, the objects are simple memory cells supporting read and
write operations. These are called
        \index{register!atomic}\indexConcept{atomic register}{atomic
        registers}.
But in general, the objects could be more complex hardware primitives
like \concept{compare-and-swap} (§\ref{section-compare-and-swap}), 
\concept{load-linked/store-conditional}
(§\ref{section-LLSC}), \index{queue!atomic}\indexConcept{atomic
        queue}{atomic queues}, or even more exotic objects from the seldom-visited theoretical depths.  
        
        Practical shared-memory systems may be implemented as 
\concept{distributed shared-memory}
(Chapter~\ref{chapter-distributed-shared-memory}) on top of a
message-passing system. This gives an alternative approach to
        designing message-passing systems if it turns out that shared
        memory is easier to use for a particular problem.

Like message-passing systems, shared-memory systems must also deal
with issues of asynchrony and failures, both in the processes and in
the shared objects.

Realistic shared-memory systems have additional
complications, in that modern CPUs allow out-of-order execution in the
absence of special (and expensive) operations called
\index{fence!memory}\index{memory fence}\indexConcept{fence}{fences}
or \index{barrier!memory}\indexConcept{memory barrier}{memory
barriers}.\cite{AdveG1995}
We will effectively be assuming that our shared-memory code is
liberally sprinkled with these operations so that nothing surprising
happens, but
this is not always true of real production code, and indeed there is
work in the theory of distributed computing literature on algorithms
that don't require unlimited use of memory barriers.
\item A third family of models has no communication mechanism
    independent of the processes.  Instead, the processes may
    directly observe the 
    states of other processes.
    These models are used in analyzing
\concept{self-stabilization}, 
for some \concept{biologically inspired systems}, 
and for computation by \concept{population protocols} or \concept{chemical reaction networks}.
We will discuss some of this work in
Part~\ref{part-other-models}.
\item Other specialized models emphasize particular details of
distributed systems, such as the labeled-graph models used for
analyzing routing or the topological models used to give a very
high-level picture of various distributed decision problems
(see Chapter~\ref{chapter-topological-methods}).
\end{itemize}

We'll see many of these at some point in this course, and examine which of them can simulate each other under various conditions.

\section{Properties}

Properties we might want to prove about a system include:
\begin{itemize}
\item \indexConcept{safety}{Safety} properties, of the form ``nothing
bad ever happens'' or, more precisely, ``there are no bad reachable
configurations.''  These include things like ``at most one of
the traffic lights at the intersection of Busy Road and Main Street is ever
green'' or ``every value read from a counter equals the number
        of preceding increment operations.'' Such properties are typically proved using
        an
\index{invariant}, a property of configurations
that is true initially and that is preserved by all
        transitions (this is essentially a disguised induction proof).
\item \indexConcept{liveness}{Liveness} properties, of the form
``something good eventually happens.''  An example might be ``my email
is eventually either delivered or returned to me.''  These are not
properties of particular states (I might unhappily await the eventual delivery of my email for decades without violating the liveness property just described), but of executions, where the property must hold starting at some finite time.  Liveness properties are generally proved either from other liveness properties (e.g., ``all messages in this message-passing system are eventually delivered'') or from a combination of such properties and some sort of timer argument where some progress metric improves with every transition and guarantees the desirable state when it reaches some bound (also a disguised induction proof).
\item \indexConcept{fairness}{Fairness} properties are a strong kind
of liveness property of the form ``something good eventually happens
to everybody.''  Such properties exclude \concept{starvation}, a
situation where most of the kids are happily chowing down at the
        orphanage (``some kid eventually eats something'' is a
        liveness property) but poor \index{Twist!Oliver}\index{Oliver
        Twist}Oliver Twist is dying in the corner for lack of gruel.
\item \indexConcept{simulation}{Simulations} show how to build one
kind of system from another, such as a reliable message-passing system
built on top of an unreliable system (TCP~\cite{Postel1981}), a shared-memory system
built on top of a message-passing system (distributed shared
memory—see Chapter~\ref{chapter-distributed-shared-memory}),
or a synchronous system build on top of an asynchronous system
(synchronizers—see Chapter~\ref{chapter-synchronizers}).
\item \indexConcept{impossibility}{Impossibility results} describe
things we can't do.  For example, the classic \concept{Two Generals}
impossibility result (Chapter~\ref{chapter-two-generals}) says that it's impossible to guarantee agreement between two processes across an unreliable message-passing channel if even a single message can be lost.  Other results characterize what problems can be solved if various fractions of the processes are unreliable, or if asynchrony makes timing assumptions impossible.  These results, and similar lower bounds that describe things we can't do quickly, include some of the most technically sophisticated results in distributed computing.  They stand in contrast to the situation with sequential computing, where the reliability and predictability of the underlying hardware makes proving lower bounds extremely difficult.
\end{itemize}

There are some basic proof techniques that we will see over and over
again in distributed computing.

For \index{proof!lower bound}\concept{lower bound} and
\index{proof!impossibility}\concept{impossibility} proofs,
the main tool is the \concept{indistinguishability} argument.  Here we
construct two (or more) executions in which some process has the same
input and thus behaves the same way, regardless of what algorithm it
is running.  This exploitation of process's ignorance is what makes
impossibility results possible in distributed computing despite being
notoriously difficult in most areas of computer science.\footnote{An
exception might be lower bounds for data structures, which also rely
on a process's ignorance.}

For \index{proof!safety}\indexConcept{safety property}{safety properties},
statements that some bad outcome never occurs, the main proof
technique is to construct an \index{proof!invariant}\concept{invariant}.  An invariant is
essentially an induction hypothesis on reachable configurations of the
system; an invariant proof shows that the invariant holds in all
initial configurations, and that if it holds in some configuration, it
holds in any configuration that is reachable in one step.

Induction is also useful for proving
\index{proof!termination}\concept{termination} and
\index{proof!liveness}\concept{liveness} properties, statements that
some good outcome occurs after a bounded amount of time.  Here we
typically structure the induction hypothesis as a 
\concept{progress measure}, where we argue that each time unit causes
the progress measure to advance by some predictable amount, and that
when the progress measure reaches a particular value, our desired outcome
is achieved.

\part{Message passing}
\label{part-message-passing}

\myChapter{Model}{2026}{}
\label{chapter-message-passing-basics}

Message passing models simulate networks.  Because any interaction
between physically separated processors requires transmitting
information from one place to another, all distributed systems are, at
a low enough level, message-passing systems.  We start by defining a
formal model of these systems.

\section{Basic message-passing model}
\label{section-message-passing-model}

We have a collection of $n$ \indexConcept{process}{processes}
$p_{1}\dots{}p_{2}$, each of which has a \concept{state} consisting of
a state from from state set $Q_{i}$.  
We think of these processes as nodes in a directed
\index{graph!communication}\concept{communication graph} or
\concept{network}.  The edges in this graph are a collection of
point-to-point \indexConcept{channel}{channels} or
\indexConcept{buffer}{buffers} $b_{ij}$, one for each
pair of adjacent processes $i$ and $j$, 
representing 
messages that have been sent but that have not yet been delivered.
Implicit in this definition is that messages are
point-to-point, with a single sender and recipient: if you want
broadcast, you have to build it yourself.

A \concept{configuration} of the
system consists of a vector of states, one for each process and
channel.  The
configuration of the system is updated by an \concept{event}, in
which (1) zero or more messages in channels $b_{ij}$ are delivered to
process $p_j$, removing them from $b_{ij}$; (2) $p_j$ updates its state in
response; and (3) zero or more messages are added by $p_j$ to outgoing
channels $b_{ji}$.  We generally think of these events as
\index{event!delivery}\indexConcept{delivery event}{delivery events}
when at least one message is delivered, and as
\index{event!computation}\indexConcept{computation event}{computation
events} when none are.
An
\concept{execution segment} is a sequence of alternating
configurations and events $C_{0},\phi_{1},C_{1},\phi_{2},\dots{}$,
in which each triple $C_{i}\phi_{i+1}C_{i+1}$ is consistent with the
transition rules for the event $\phi_{i+1}$, and
the last element of the sequence (if any) is a configuration.  If the
first configuration $C_{0}$ is an \index{configuration!initial}\concept{initial configuration} of the system, we have an \concept{execution}.  A \concept{schedule} is an execution with the configurations removed.

\subsection{Formal details}

Let $P$ be the set of processes, $Q$ the set of process states, and $M$ the set of possible messages.

Each process $p_i$ has a state $\State_{i}∈Q$.
Each channel $b_{ij}$ has a state $\Buffer_{ij}∈M^ℕ$, the multiset of
messages that have been sent on that channel but not yet delivered.
We assume each process has a \concept{transition function} $δ:
Q×M^ℕ → Q×(P×M)^ℕ$ that maps tuples consisting of a
state and a set of incoming messages
a new state and a set of recipients and messages to be sent.
An important feature of the transition function is
that the process's behavior can't depend on which of
its previous messages have been delivered or not.  
A delivery event $\Del(i,A)$, where $A = (j_k,m_k)^ℕ)$ removes each message $m_k$
from $b_{ji}$, updates $\State_i$ according to $δ(\State_i,A)$, and
adds the outgoing messages specified to $δ(\State_i,A)$ to the
appropriate channels.
A computation
event $\Comp(i)$ does the same thing,
except that it applies $δ(\State_j,∅)$.

Some implicit features in this definition:
\begin{itemize}
 \item A process can't tell when its outgoing messages are delivered,
 because the channel states aren't available as input to $δ$.
 \item Processes are \concept{deterministic}: The next action of each process depends only on its current state, and not on extrinsic variables like the phase of the moon, coin-flips, etc.  We may wish to relax this condition later by allowing coin-flips; to do so, we will need to extend the model to incorporate probabilities.
 \item It is possible to determine the accessible state of a process
 by looking only at events that involve that process.  Specifically,
 given a schedule $S$, define the \concept{restriction} $S|i$ to be
 the subsequence consisting of all $\Comp(i)$ and $\Del(i,A)$ events
 (ranging over all possible $A$).  Since these are the only
 events that affect the state of $i$, and only the
 state of $i$ is needed to apply the transition function,
 we can compute the state of $i$ looking only at $S|i$.  In
 particular, this means that $i$ will have the same accessible state
 after any two schedules $S$ and $S'$ where $S|i = S'|i$, and thus
 will take the same actions in both schedules.  This is the basis for
 \concept{indistinguishability proofs}
 (§\ref{section-indistinguishability-proofs}), a central technique in obtaining lower bounds and impossibility results.
\end{itemize}

Attiya and Welch~\cite{AttiyaW2004}
use a different model in which communication
channels are not modeled separately from processes, but instead are
baked into processes as \Outbuf and \Inbuf variables.  This
leads to some oddities like having to distinguish the accessible state
of a process (which excludes the \Outbuf{}s) from the full state (which
doesn't). Our approach is close to that of Lynch~\cite{Lynch1996}, in
that we have separate automata representing processes and
communication channels.
But since the resulting model produces essentially the same
executions, the exact details don't really matter.\footnote{The
late 1970s and early 1980s saw a lot of research on finding the ``right''
definition of a distributed system, and some of the disputes from that
era were hard fought.  But in the end, all the various proposed models
turned out to be more or less equivalent, which is not surprising
since the authors were ultimately trying to represent the same
intuitive understanding of these systems.  So most distributed
computing papers now just use some phrasing like ``we consider the
standard model of an asynchronous message-passing system'' and leave
it
to the reader to assume that this standard model is their favorite
one. 

An example of this trick in action is that you will never see $\Del(i,A)$ or
$\Comp(i)$ again after you finish reading this footnote.}

\subsection{Network structure}

\index{network}
It may be the case that not all processes can communicate directly; if
so, we impose a network structure in the form of a directed graph,
where $i$ can send a message to $j$ if and only if there is an edge from $i$
to $j$ in the graph.  Typically we assume that each process knows the identity of all its neighbors.

For some problems (e.g., in peer-to-peer systems or other 
\index{network!overlay}
\indexConcept{overlay network}{overlay networks}) it may be natural to assume that there is a fully-connected underlying network but that we have a dynamic network on top of it, where processes can only send to other processes that they have obtained the addresses of in some way.

\section{Asynchronous systems}
\label{section-asynchronous-message-passing}

In an 
\index{asynchronous message-passing}\index{message-passing!asynchronous}\concept{asynchronous} 
model, only minimal restrictions are placed on when
messages are delivered and when local computation occurs.  A schedule
is said to be \index{schedule!admissible}\concept{admissible} if (a)
there are infinitely many computation steps for each process, and (b)
every message is eventually delivered.  (These are \concept{fairness}
conditions.)  The first condition (a)
assumes that processes do not explicitly terminate, which is the
assumption used in \cite{AttiyaW2004}; an alternative, which we will
use when convenient, is to assume that every process either has infinitely many computation steps or reaches an explicit halting state.

\subsection{Example: client-server computing}
\label{section-client-server}

Almost every distributed system in practical use is based on
\concept{client-server} interactions.  Here one process, the
\concept{client}, sends a \concept{request} to a second process, the
\concept{server}, which in turn sends back a \concept{response}.  We
can model this interaction using our asynchronous message-passing
model by describing what the transition functions for the client and
the server look like: see Algorithms~\ref{alg-client}
and~\ref{alg-server}.

\newData{\ClientServerRequest}{request}
\newData{\ClientServerResponse}{response}

\begin{algorithm}
\Initially{
    send \ClientServerRequest to server\;
    \UponReceiving{\ClientServerResponse}{
        update state
    }
}
\caption{Client-server computation: client code}
\label{alg-client}
\end{algorithm}

\begin{algorithm}
\UponReceiving{\ClientServerRequest}{
    send \ClientServerResponse to client\;
}
\caption{Client-server computation: server code}
\label{alg-server}
\end{algorithm}

The interpretation of Algorithm~\ref{alg-client} is that the client
sends \ClientServerRequest (by adding it to its \Outbuf) 
in its very first computation event (after
which it does nothing).  The interpretation of
Algorithm~\ref{alg-server} is that in any computation event where the
server observes \ClientServerRequest in its \Inbuf, it sends \ClientServerResponse.

We want to claim that the client eventually receives
\ClientServerResponse in any admissible execution.  To prove this, observe that:
\begin{enumerate}
\item After finitely many steps, the client carries out a
computation event.  This computation event puts \ClientServerRequest
in the message buffer between the client and server.
\item After finitely many more steps, a delivery event occurs that
delivers \ClientServerRequest to the server.  This causes the server to 
send \ClientServerResponse.
\item After finitely many more steps, a delivery event
delivers \ClientServerResponse to the client,
causing it to process \ClientServerResponse.
\end{enumerate}

Each step of the proof is justified by the constraints on admissible
executions.  If we could run for infinitely many steps without a
particular process doing a computation event or a particular message
being delivered, we'd violate those constraints.

Most of the time we will not attempt to prove the correctness of a
protocol at quite this level of tedious detail.  But if you are only
interested in distributed algorithms that people actually use, you
have now seen a proof of correctness for 99.9\% of them, and do not
need to read any further.

\section{Synchronous systems}

A \index{message-passing!synchronous}\concept{synchronous
message-passing} system is exactly like an asynchronous system, except
we insist that the schedule consists of alternating phases in which
(a) every process executes a computation step (that may send messages), and (b) all messages
are delivered while none are sent.\footnote{Formally, the delivery phase consists of $n$
separate delivery events, in any order, that between them clean out all the
channels.} The combination of a computation phase and a delivery
phase is called a \concept{round}. Synchronous systems are
effectively those in which all processes execute in lock-step, and
there is no timing uncertainty.  This makes protocols much easier to
design, but makes them less resistant to real-world timing oddities.
Sometimes this can be dealt with by applying a \concept{synchronizer}
(Chapter~\ref{chapter-synchronizers}), which transforms synchronous protocols into asynchronous protocols at a small cost in complexity.

\section{Drawing message-passing executions}
\label{section-drawing-message-passing}

Though formally we can describe an execution in a message-passing
system as a long list of events, this doesn't help much with
visualizing the underlying communication pattern.  So it can sometimes
be helpful to use a more visual representation of a message-passing
execution that shows how information flows through the system.

A typical example is given in Figure~\ref{fig-world-lines}.  In this
picture, time flows from left to right, and each process is
represented by a horizontal line.  This convention reflects the fact
that processes have memory, so any information available to a process
at some time $t$ is also available at all times $t' ≥ t$.  Events are
represented by marked points on these lines, and messages are
represented by diagonal lines between events.  The resulting picture
looks like a collection of 
\indexConcept{world line}{world lines}\index{line!world}
as used in physics to illustrate the path taken by various objects
through spacetime.

\begin{figure}
    \centering
    \begin{tikzpicture}[auto]
        \foreach \x in {1,2,3}{
            \node at (0,\x) {$p_{\x}$};
            \draw (1,\x) -- (10,\x);
        }
        \node at (5.5,0) {Time $\rightarrow$};
        \draw[color=blue,fill=blue,thick,radius=0.1] (2,1) circle -- (4,2) circle -- (7.2,3) circle;
        \draw[color=blue,fill=blue,thick,radius=0.1] (4,2) -- (8,1) circle;
        \draw[color=blue,fill=blue,thick,radius=0.1] (6.5,2) circle -- (7,1) circle;
        \draw[color=blue,fill=blue,thick,radius=0.1] (2,1) -- (9,3) circle;
    \end{tikzpicture}
    \caption[Asynchronous message-passing execution]{Asynchronous
        message-passing execution.  Time flows left-to-right.
        Horizontal lines represent processes.  Nodes represent events.
    Diagonal edges between events represent messages.  In this
execution, $p_1$ executes a computation event that sends messages to
$p_2$ and $p_3$.  When $p_2$ receives this message, it sends messages to $p_1$
and $p_3$.  Later, $p_2$ executes a computation event that sends a second message to $p_1$.  Because the
system is asynchronous, there is no guarantee that messages arrive in
the same order they are sent.}
    \label{fig-world-lines}
\end{figure}

Pictures like Figure~\ref{fig-world-lines} can be helpful for
illustrating the various constraints we might put on message delivery.
In Figure~\ref{fig-world-lines}, the system is completely
asynchronous: messages can be delivered in any order, even if sent
between the same processes.  If we run the same protocol under
stronger assumptions, we will get different communication patterns.

For example, Figure~\ref{fig-world-lines-fifo} shows an execution that
is still asynchronous but that assumes FIFO (first-in first-out)
channels.  A \concept{FIFO channel}\index{channel!FIFO} from some process $p$ to another process $q$
guarantees that $q$ receives messages in the same order that $p$ sends
them (this can be simulated by a non-FIFO channel by adding a
\concept{sequence number}\index{number!sequence} to each message, and
queuing messages at the receiver until all previous messages have been
processed).

\begin{figure}
    \centering
    \begin{tikzpicture}[auto]
        \foreach \x in {1,2,3}{
            \node at (0,\x) {$p_{\x}$};
            \draw (1,\x) -- (10,\x);
        }
        \node at (5.5,0) {Time $\rightarrow$};
        \draw[color=blue,fill=blue,thick,radius=0.1] (2,1) circle -- (4,2) circle -- (7.2,3) circle;
        \draw[color=blue,fill=blue,thick,radius=0.1] (4,2) -- (7,1) circle;
        \draw[color=blue,fill=blue,thick,radius=0.1] (6.5,2) circle -- (8,1) circle;
        \draw[color=blue,fill=blue,thick,radius=0.1] (2,1) -- (9,3) circle;
    \end{tikzpicture}
    \caption[Asynchronous message-passing execution with FIFO
    channels]{Asynchronous message-passing execution with FIFO channels.  Multiple
        messages from one process to another are now guaranteed to be
    delivered in the order they are sent.}
    \label{fig-world-lines-fifo}
\end{figure}

If we go as far as to assume synchrony, we get the execution in
Figure~\ref{fig-world-lines-synchronous}.  Now all messages take
exactly one time unit to arrive, and computation events follow each
other in lockstep.

\begin{figure}
    \centering
    \begin{tikzpicture}[auto]
        \foreach \x in {1,2,3}{
            \node at (0,\x) {$p_{\x}$};
            \draw (1,\x) -- (10,\x);
        }
        \node at (5.5,0) {Time $\rightarrow$};
\foreach \y in {1,2,3}{
            \foreach \x in {2,4,6,8}{
                \draw[color=blue,fill=green,thick,radius=0.1] (\x,\y) circle;
            }
        }
        \draw[color=blue,fill=blue,thick,radius=0.1] (2,1) circle -- (4,2) circle -- (6,3) circle;
        \draw[color=blue,fill=blue,thick,radius=0.1] (4,2) -- (6,1) circle;
        \draw[color=blue,fill=blue,thick,radius=0.1] (6,2) circle -- (8,1) circle;
        \draw[color=blue,fill=blue,thick,radius=0.1] (2,1) -- (4,3) circle;
    \end{tikzpicture}
    \caption[Synchronous message-passing execution]{Synchronous
    message-passing execution. All messages are now delivered in
exactly one time unit, and computation events immediately
    follow the delivery events.}
    \label{fig-world-lines-synchronous}
\end{figure}

\section{Complexity measures}
\label{section-message-passing-complexity}

There is no explicit notion of time in the asynchronous model, but we
can define a time measure by adopting the rule that every message is
delivered and processed at most 1 time unit after it is sent.
Formally, we assign time 0 to the first event, and assign the largest
time we can to each subsequent event, subject to the constraints that
(a) no event is assigned a larger time than any later event; (b) if a
message $m$ from $i$ to $j$ is created by an event at time $t$, then the time for
the delivery of $m$ from $i$ to $j$ is
no greater than $t+1$, and (c) any computation step
is assigned a time no later than 
the previous event at the same process (or $0$ if the process has no
previous events).
This is
consistent with an assumption that message propagation takes at most 1
time unit and that local computation takes 0 time units.

Another way
to look at this is that it is a definition of a time unit in terms of
maximum message delay together with an assumption that message delays
dominate the cost of the computation.  This last assumption is pretty
much always true for real-world networks with any non-trivial physical
separation between components, thanks to speed of light
limitations.

An example of an execution annotated with times in this way is given
in Figure~\ref{fig-world-lines-times}.

\begin{figure}
    \centering
    \begin{tikzpicture}[auto]
        \foreach \x in {1,2,3}{
            \node at (0,\x) {$p_{\x}$};
            \draw (1,\x) -- (10,\x);
        }
        \node at (5.5,0) {Time $\rightarrow$};
        \draw[color=blue,fill=blue,thick,radius=0.1]
        (2,1) circle node[anchor=north] {$0$}
        -- (2.8,2) circle node[anchor=north] {$1$}
        -- (5.2,3) circle node[anchor=north] {$1$};
        \draw[color=blue,fill=blue,thick,radius=0.1] (2.8,2) -- (8,1) circle node[anchor=north] {$2$};
        \draw[color=blue,fill=blue,thick,radius=0.1] (6.5,2) circle
        node[anchor=north]{$1$} -- (7,1) circle
        node[anchor=north]{$2$};
        \draw[color=blue,fill=blue,thick,radius=0.1] (2,1) -- (6,3)
        circle node[anchor=north] {$1$};
    \end{tikzpicture}
    \caption[Asynchronous time]{Asynchronous
        message-passing execution with times.}
    \label{fig-world-lines-times}
\end{figure}

The \index{complexity!time}\concept{time complexity} of a protocol
(that terminates) is the time of the last event at any process.

Note that looking at \index{complexity!step}\concept{step complexity},
the number of computation events involving either a particular process
(\index{complexity!step!individual}\concept{individual step complexity}) 
or all processes (\index{complexity!step!total}\concept{total step
complexity}) is not useful in the asynchronous model, because a
process may be scheduled to carry out arbitrarily many computation
steps without any of its incoming or outgoing messages being
delivered, which probably means that it won't be making any progress.
These complexity measures will be more useful when we look at
shared-memory models (Part~\ref{part-shared-memory}).

For a protocol that terminates, the
\index{complexity!message}\concept{message complexity} is the total
number of messages sent.  We can also look at message length in bits,
total bits sent, and so on, if these are useful for distinguishing our new improved protocol from last year's model.

For synchronous systems, time complexity becomes just the number of
rounds until a protocol finishes.  Message complexity is still only
loosely connected to time complexity; for example, there are
synchronous \concept{leader election}
(Chapter~\ref{chapter-leader-election}) algorithms that, by virtue of grossly abusing the synchrony assumption, have unbounded time complexity but very low message complexity.

\myChapter{Broadcast and convergecast}{2026}{}
\label{chapter-broadcast-and-convergecast}

Here we'll describe protocols for propagating information throughout a
network from some central initiator and gathering information back to
that same initiator.  We do this both because the algorithms are
actually useful and because they illustrate some of the issues that
come up with keeping time complexity down in an asynchronous
message-passing system.

\section{Flooding}
\label{section-flooding}

\indexConcept{flooding}{Flooding} is about the simplest of all
distributed algorithms. It's primitive and expensive, but easy to
implement, and gives you both a broadcast mechanism and a way to build
rooted spanning trees.

We'll give a fairly simple presentation of flooding roughly
following Chapter 2 of \cite{AttiyaW2004}. For more recent work on
flooding see~\cite{HussakT2023}.

\subsection{Basic algorithm}
\label{section-flooding-basic}

\newData{\SeenMessage}{seen-message}

The basic flooding algorithm is shown in Algorithm~\ref{alg-flooding}.
The idea is that when a process receives a message $M$, it forwards it
to all of its neighbors unless it has seen it before, which it tracks
using a single bit $\SeenMessage$.

\begin{algorithm}
\Initially{
    \eIf{$\Pid = \Root$}{
        $\SeenMessage ← \True$ \;
        send $M$ to all neighbors\;
    }{
        $\SeenMessage ← \False$\;
    }
}
\UponReceiving{$M$}{
  \If{$\SeenMessage = \False$}{
   $\SeenMessage ← \True$\;
   send $M$ to all neighbors\;
   }
}
\caption{Basic flooding algorithm}
\label{alg-flooding}
\end{algorithm}

\begin{theorem}
\label{theorem-flooding}
Every process receives $M$ after at most $D$ time and at
most $\card*{E}$ messages, where $D$ is the diameter of the network
and $E$ is the set of (directed) edges in the network.
\end{theorem}
\begin{proof}
  Message complexity: Each process only sends $M$ to its
  neighbors once, so each edge carries at most one copy of $M$.

  Time complexity: By induction on $d(\Root, v)$, we'll show that
  each $v$ sets $\SeenMessage$ to $\True$ no later than time $d(\Root,
  v) ≤ D$.
  The base case is when $v = \Root$, $d(\Root, v) = 0$; here $\Root$
    does its initial computation event at time $0$.
   For the induction step,
   Let $d(\Root, v) = k > 0$.  Then $v$ has a
   neighbor $u$ such that $d(\Root, u) = k-1$.  By the induction
   hypothesis, $u$ sets $\SeenMessage$ to $\True$ no later than time
   $k-1$.  From the code, $u$ then sends $M$ to all of its neighbors,
   including $v$; $M$ arrives at $v$ no later than time $(k-1)+1 = k$.
\end{proof}

Note that the time complexity proof also demonstrates correctness:
every process receives $M$ at least once.

As written, this is a one-shot algorithm: you can't broadcast a second
message even if you wanted to.  The obvious fix is for each process to
remember which messages it has seen and only forward the new ones
(which costs memory) and/or to add a \concept{time-to-live} (TTL)
field on each message that drops by one each time it is forwarded
(which may cost extra messages and possibly prevents complete
broadcast if the initial TTL is too small).  The latter method is what
was used for searching in \index{Gnutella}\wikipedia{Gnutella}, an
early peer-to-peer system.  An interesting property of Gnutella was
that since the application of flooding was to search for huge
(multiple MiB) files using tiny ($\approx 100$ byte) query messages, the actual bit complexity of the flooding algorithm was not especially large relative to the bit complexity of sending any file that was found.

We can optimize the algorithm slightly by not sending $M$ back to the node it came from; this will slightly reduce the message complexity but makes the proof a sentence or two longer. It's all a question of what you want to optimize.

\subsection{Adding parent pointers}
\label{section-flooding-parent-pointers}

To build a spanning tree, modify Algorithm~\ref{alg-flooding} by
having each process remember who it first received $M$ from.  The
revised code is given as Algorithm~\ref{alg-flooding-parents}

\begin{algorithm}
\Initially{
    \eIf{$\Pid = \Root$}{
        $\Parent ← \Root$\;
        send $M$ to all neighbors\;
    }{
        $\Parent ← ⊥$\;
    }
}
\UponReceiving{$M$ \From $p$}{
  \If{$\Parent = ⊥$}{
   $\Parent ← p$\\\;
   send $M$ to all neighbors\;
   }
   }
\caption{Flooding with parent pointers}
\label{alg-flooding-parents}
\end{algorithm}

We can easily prove that Algorithm~\ref{alg-flooding-parents} has the same termination
properties as Algorithm~\ref{alg-flooding} by observing that if we map
\Parent to \SeenMessage by the rule $⊥
\rightarrow \False$, anything else $\rightarrow \True$, then we have
the same algorithm.  We would like one additional property, which is
that when the algorithm \indexConcept{quiesce}{quiesces} (has no outstanding messages), the set of parent pointers form a rooted spanning tree.  For this we use induction on time:
\begin{lemma}
\label{lemma-flooding-parent-invariant}
At any time during the execution of
Algorithm~\ref{alg-flooding-parents}, the following invariant holds:
\begin{enumerate}
  \item If $u.\Parent ≠ ⊥$, then
  $u.\Parent.\Parent ≠ ⊥$ and following parent
  pointers gives a path from $u$ to \Root.
  \item If there is a message $M$ in transit from $u$ to $v$, then
  $u.\Parent ≠ ⊥$.
\end{enumerate}
\end{lemma}
\begin{proof}
We have to show that the invariant is true initially, and that
    any event preserves the invariant. We'll assume
    that all events are delivery events for a single message, since we
    can have the algorithm treat a multi-message delivery event as a
    sequence of single-message delivery events.

    We'll treat the initial configuration as the result of the root
    setting its parent to itself and sending messages to all its
    neighbors. It's not hard to verify that the invariant holds in the
    resulting configuration.

    For a delivery event,
  let $v$ receive $M$ from $u$.  There are
  two cases: if $v.\Parent$ is already non-null, the only state
  change is that $M$ is no longer in transit, so we don't care about
  $u.\Parent$ any more.  If $v.\Parent$ is null, then
\begin{enumerate}
   \item $v.\Parent$ is set to $u$.  This triggers the first
   case of the invariant.  From the induction hypothesis we have that
   $u.\Parent ≠ ⊥$ and that there exists a path from $u$
   to the root.  Then $v.\Parent.\Parent = u.\Parent
   ≠ ⊥$ and the path from $v→u→\Root$
   gives the path from $v$.
   \item Message $M$ is sent to all of $v$'s neighbors.  Because $M$
   is now in transit from $v$, we need $v.\Parent ≠ ⊥$;
   but we just set it to $u$, so we are happy.
\end{enumerate}
\end{proof}

At the end of the algorithm, the invariant shows that every
process has a path to the root, i.e., that the graph represented by
the parent pointers is connected.  Since this graph has exactly
$\card*{V}-1$ edges (if we don't count the self-loop at the root), it's a tree.

Though we get a spanning tree at the end, we may not get a very good
spanning tree.  For example, suppose our friend the adversary picks
some Hamiltonian path through the network and delivers messages along
this path very quickly while delaying all other messages for the full
allowed 1 time unit.  Then the resulting spanning tree will have depth
$\card*{V}-1$, which might be much worse than $D$.  If we want the
shallowest possible spanning tree, we need to do something more
sophisticated: see the discussion of \concept{distributed
breadth-first search} in Chapter~\ref{chapter-distributed-BFS}.  However, we may be happy with the tree we get from simple flooding: if the message delay on each link is consistent, then it's not hard to prove that we in fact get a shortest-path tree.  As a special case, flooding always produces a BFS tree in the synchronous model.

Note also that while the algorithm works in a directed graph, the parent pointers may not be very useful if links aren't two-way.

\subsection{Identifying children}

By adding acknowledgment messages, it is possible for each node to
learn exactly which of its neighbors become its children. Because the
system is asynchronous, this requires each neighbor to inform the node
both whether it is a child (using an $\Ack$ message) and when it is
not (using a $\Nack$ message); only upon receiving one or the other of
these messages will the node know that it's not going to receive the
other.

The modified code is given in
Algorithm~\ref{alg-flooding-children}

\begin{algorithm}
\Initially{
    $\NonChildren ← ∅$\;
    \eIf{$\Pid = \Root$}{
        $\Parent ← \Root$\;
        $\Children ← \Set{ \Root }$\;
        send $M$ to all neighbors\;
    }{
        $\Parent ← ⊥$\;
        $\Children ← ∅$\;
    }
}
\UponReceiving{$M$ \From $p$}{
  \eIf{$\Parent = ⊥$}{
   $\Parent ← p$\\\;
   send $\Ack$ to $p$\;
   send $M$ to all neighbors\;
   }{
       send $\Nack$ to $p$\;
   }
   }
    \UponReceiving{$\Ack$ \From $p$}{
        $\Children ← \Children ∪ \Set{p}$\;
    }
    \UponReceiving{$\Nack$ \From $p$}{
        $\NonChildren ← \NonChildren ∪ \Set{p}$\;
    }
\caption{Flooding tracking children}
\label{alg-flooding-children}
\end{algorithm}

If we take an execution of
Algorithm~\ref{alg-flooding-children} and remove all the $\Ack$ and
$\Nack$ messages, we get an execution of
Algorithm~\ref{alg-flooding-parents}. So all of the properties that we
proved for Algorithm~\ref{alg-flooding-parents} continue to hold.

For the improved algorithm, we'd like to show that when the algorithm
quiesces, every node $p_i$ has a list of all the nodes $p_j$ for which
$p_j.\Parent = p_i$ in $p_i.\Children$ and a list of all the neighbors
$p_j$ for which $p_j.\Parent ≠ p_i$ in $p_i.\NonChildren$.

We can do this by showing a mix of safety and liveness properties:

\begin{enumerate}
    \item (Safety) If $p_j ∈ p_i.\Children$, then $p_j.\Parent = p_i$.
        Proof sketch: Verify the strengthening of this property that adds
        $\Ack ∈ b_{ji}$ implies $p_j.\Parent = p_i$ is an invariant.
    \item (Safety) If $p_j ∈ p_i.\NonChildren$, then $p_j.\Parent ∉
        \Set{p_i, ⊥}$.
        Proof sketch: Verify the strengthening of this property that adds
        $\Nack ∈ b_{ji}$ implies $p_j.\Parent ∉ \Set{p_i,⊥}$ is an invariant.
    \item (Liveness) Eventually, every neighbor of $p_i$ appears in
        $p_i.\Children ∪ p_i.\NonChildren$.
        Proof: We have previously shown that every node $p_i$
        eventually sets $p_i.\Parent ≠ ∅$. From the code we have that
        whenever a node does this, it sends $M$ to all neighbors. For
        each neighbor $p_j$, observe that upon receiving $M$ it 
        responds with exactly one of $\Ack$ or $\Nack$. When this
        message is eventually delivered, $p_j$ is added to
        $p_i.\Children ∪ p_i.\NonChildren$.
\end{enumerate}

Since we assume that each $p_i$ knows which nodes are its neighbors,
we can use the property that $p_i.\Children ∪ p_i.\NonChildren$
includes all neighbors as a kind of local termination test. This can
be handy if we want to use flooding as the first step
in some larger protocol.

\section{Convergecast}

A \concept{convergecast} is the inverse of broadcast:
instead of a message propagating down from a single root to all nodes, data is collected from outlying nodes to the root.  Typically some function is applied to the incoming data at each node to summarize it, with the goal being that eventually the root obtains this function of all the data in the entire system.  (Examples would be counting all the nodes or taking an average of input values at all the nodes.)

A basic convergecast algorithm is given in
Algorithm~\ref{alg-convergecast}; it propagates information up through
a previously-computed spanning tree.

\begin{algorithm}
\Initially{
    \If{I am a leaf}{
        send $\Input$ to $\Parent$\;
    }
}
\UponReceiving{$M$ \From $c$}{
    append $(c,M)$ to \Buffer\\
    \If{\Buffer contains messages from all my children}{
        $v ← f(\Buffer, \Input)$\;
        \eIf{$\Pid = \Root$}{
            \Return $v$\;
        }{
            send $v$ to $\Parent$\;
        }
    }
}
\caption{Convergecast}
\label{alg-convergecast}
\end{algorithm}

The details of what is being computed depend on the choice of $f$:

\begin{itemize}
 \item If $\Input = 1$ for all nodes and $f$ is sum, then we count the number of nodes in the system.
 \item If $\Input$ is arbitrary and $f$ is sum, then we get a total of all the input values.
 \item Combining the above lets us compute averages, by dividing the total of all the inputs by the node count.
 \item If $f$ just concatenates its arguments, the root ends up with a
 vector of all the input values.
\end{itemize}

Running time is bounded by the depth of the tree: we can prove by
induction that any node at height $h$ (height is length of the longest
path from this node to some leaf) sends a message by time $h$ at the
latest.  Message complexity is exactly $n-1$, where $n$ is the number of nodes; this is easily shown by observing that each node except the root sends exactly one message.

Proving that convergecast returns the correct value is similarly done
by induction on depth: if each child of some node computes a correct
value, then that node will compute $f$ applied to these values and its
own input.  What the result of this computation is will, of course,
depend on $f$; it generally makes the most sense when $f$ represents
some associative operation (as in the examples above).

\section{Flooding and convergecast together}
\label{section-flooding-with-convergecast}

A natural way to build the spanning tree used by convergecast is to
run flooding first.  This also provides a mechanism for letting the
leaves know that they are leaves and initiating the protocol.  The
combined algorithm is shown as
Algorithm~\ref{alg-flooding-with-convergecast}.

\newData{\FloodingInit}{init}

\begin{algorithm}
\Initially{
    $\Children ← ∅$ \;
    $\NonChildren ← ∅$ \;
    \eIf{$\Pid = \Root$}{
        $\Parent ← \Root$ \;
        send $\FloodingInit$ to all neighbors
    }{
        $\Parent ← ⊥$\;
    }
}
\UponReceiving{$\FloodingInit$ \From $p$}{
  \eIf{$\Parent = ⊥$}{
   $\Parent ← p$\;
   send $\FloodingInit$ to all neighbors\;
   }{
   send $\Nack$ to $p$\;
   }
   }
\UponReceiving{$\Nack$ \From $p$}{
    $\NonChildren ← \NonChildren ∪ \Set{p}$\;
}

\AsSoonAs{$\Children ∪ \NonChildren$ includes all
my neighbors}{
        $v ← f(\Buffer, \Input)$\;
        \eIf{$\Pid = \Root$}{
            \Return $v$\;
        }{
            send $\Ack(v)$ to $\Parent$\;
        }
}
\UponReceiving{$\Ack(v)$ \From $k$}{
    add $(k,v)$ to \Buffer\;
    add $k$ to $\Children$\;
}
\caption{Flooding and convergecast combined}
\label{alg-flooding-with-convergecast}
\end{algorithm}

However, this may lead to very bad time
complexity for the convergecast stage.  Consider a wheel-shaped network consisting
of one central node $p_0$ connected to nodes $p_1, p_2, \dots,
p_{n-1}$, where each $p_i$ is also connected to $p_{i+1}$.  By
carefully arranging for the $p_i p_{i+1}$ links to run much faster
than the $p_0 p_i$ links, the adversary can make flooding build a tree
that consists of a single path $p_0 p_1 p_2 \dots p_{n-1}$, even
though the diameter of the network is only $2$.  While it only takes 2
time units to build this tree (because every node is only one hop away
from the initiator), when we run convergecast we suddenly find that
the previously-speedy links are now running only at the guaranteed
$≤ 1$ time unit per hop rate, meaning that convergecast takes $n-1$
time.

This may be less of an issue in real networks, where the latency of
links may be more uniform over time, meaning that a deep tree of fast
links is still likely to be fast when we reach the convergecast step.
But in the worst case we will need to be more clever about building
the tree.  We show how to do this in
Chapter~\ref{chapter-distributed-BFS}.

\myChapter{Distributed breadth-first search}{2026}{}
\label{chapter-distributed-BFS}

Here we describe some algorithms for building a \concept{breadth-first
search} (\concept{BFS}) tree in a network.  All assume that there is a designated
\concept{initiator} node that starts the algorithm.  At the end of
the execution, each node except the initiator has a parent pointer and
every node has a list of children.  These are consistent and define a
BFS tree: nodes at distance $k$ from the initiator appear at level $k$ of the tree.

In a synchronous network, \concept{flooding} (§\ref{section-flooding}) solves BFS; see
\cite[Lemma 2.8, page 21]{AttiyaW2004} or \cite[\S4.2]{Lynch1996}.  So
the interesting case is when the network is asynchronous.

In an asynchronous network, the complication is that we can no longer rely on synchronous
communication to reach all nodes at distance $d$ at the same time.  So
instead we need to keep track of distances explicitly, or possibly
enforce some approximation to synchrony in the algorithm.  (A general
version of this last approach is to apply a synchronizer to one of the
synchronous algorithms using a \concept{synchronizer}; see
Chapter~\ref{chapter-synchronizers}.)

To keep things simple, we'll drop the requirement that a parent learn the IDs of its children, since this can be tacked on as a separate notification protocol, in which each child just sends one message to its parent once it figures out who its parent is.

\section{Using explicit distances}
\label{section-distributed-BFS-relaxation}

This is a translation of the AsynchBFS automaton from \cite[\S15.4]{Lynch1996}.  It's
a very simple algorithm, closely related to Dijkstra's algorithm for
shortest paths, but there is otherwise no particular reason to use it.
Not only does it not detect termination, but
it is also dominated by the $O(D)$ time and $O(DE)$ message complexity
synchronizer-based algorithm described in
§\ref{section-distributed-BFS-synchronizer}.  (Here $D$ is the
\concept{diameter} of the network, the maximum distance between any
two nodes.)

The idea is to run flooding with distances attached.  Each node sets
its distance to $1$ plus the smallest distance sent by its neighbors and its parent to the neighbor supplying that smallest distance.  A node notifies all its neighbors of its new distance whenever its distance changes.

Pseudocode is given in Algorithm~\ref{alg-asynchBFS}

\newData{\AsynchBFSdist}{distance}

\begin{algorithm}
\Initially{
    \eIf{$\Pid = \Initiator$}{
        $\AsynchBFSdist ← 0$\;
        send $\AsynchBFSdist$ to all neighbors\;
    }{
        $\AsynchBFSdist ← \infty$\;
    }
}
\UponReceiving{$d$ \From $p$}{
    \If{$d+1 < \AsynchBFSdist$}{
        $\AsynchBFSdist ← d+1$\;
        $\Parent ← p$\;
        send $\AsynchBFSdist$ to all neighbors\;
    }
}
\caption{AsynchBFS algorithm (from \cite{Lynch1996})}
\label{alg-asynchBFS}
\end{algorithm}

(See \cite{Lynch1996} for a precondition-effect description, which also includes code for buffering outgoing messages.)

The claim is that after at most $O(VE)$ messages and $O(D)$ time, all
distance values are equal to the length of the shortest path from the
initiator to the appropriate node. The proof is by showing the
following:
\begin{lemma}
\label{lemma-lynch-bfs}
The variable $\AsynchBFSdist_p$ is always the length of some path from
initiator to $p$, and any message sent by $p$ is also the length of
some path from $\Initiator$ to $p$.
\end{lemma}
\begin{proof}
The second part follows from the first; any message sent equals $p$'s
current value of distance.  For the first part, suppose $p$ updates
its distance; then it sets it to one more than the length of some path from
$\Initiator$ to $p'$, which is the length of that same path extended by
adding the $pp'$ edge.
\end{proof}

We also need a liveness argument that says that $\AsynchBFSdist_p =
d(\Initiator, p)$ no later than time
$d(\Initiator, p)$.  Note that we can't detect when
$\AsynchBFSdist$ stabilizes to the correct value without a lot of additional work.

In \cite{Lynch1996}, there's an extra $\card*{V}$ term in the time
complexity that comes from message pile-ups, since the model used
there only allows one incoming message to be processed per time
units (the model in \cite{AttiyaW2004} doesn't have this restriction).
The trick to arranging this to happen often is to build a graph where
node 1 is connected to nodes 2 and 3, node 2 to 3 and 4, node 3 to 4
and 5, etc.  This allows us to quickly generate many paths of distinct
lengths from node 1 to node $k$, which produces $k$ outgoing messages
from node $k$.  It may be that a more clever analysis can avoid this blowup, by showing that it only happens in a few places.

\section{Using layering}
\label{section-distributed-BFS-layered}

\newData{\LayeringBound}{bound}

This approach is used in the \emph{LayeredBFS} algorithm
in \cite{Lynch1996}, which is due to
Gallager~\cite{Gallager1982}.

Here we run a sequence of up to $\card*{V}$ instances of the simple algorithm with a
distance bound on each: instead of sending out just 0, the initiator
sends out $(0, \LayeringBound)$, where $\LayeringBound$ is initially 1 and increases at
each phase.  A process only sends out its improved distance if it is
less than \LayeringBound.  

Each phase of the algorithm constructs a partial BFS tree that
contains only those nodes within distance $\LayeringBound$ of the
root.  This tree is used to report back to the root when the phase is
complete.
For the following phase, notification of the increase in
$\LayeringBound$ 
increase is distributed only through the partial BFS tree constructed
so far.  With some effort, it is possible to prove that in a
bidirectional network that this approach guarantees that each edge is
only probed once with a new distance (since distance-1 nodes are
recruited before distance-2 nodes and so on), and the \LayeringBound-update and
acknowledgment messages contribute at most $\card*{V}$ messages per
phase.  So we get $O(E + VD)$ total messages.  But the time complexity
is bad: $O(D^2)$ in the worst case.

\section{Using local synchronization}
\label{section-distributed-BFS-synchronizer}

\newData{\BFSdistance}{distance}
\newData{\BFSnotDistance}{not-distance}
\newData{\BFSexactly}{exactly}
\newData{\BFSmoreThan}{more-than}

The reason the layering algorithm takes so long is that at each phase
we have to phone all the way back up the tree to the initiator to get
permission to go on to the next phase.  We need to do this to make
sure that a node is only recruited into the tree once: otherwise we
can get pile-ups on the channels as in the simple algorithm.  But we
don't necessarily need to do this globally.  Instead, we'll require
each node at distance $d$ to delay sending out a recruiting message
until it has confirmed that none of its neighbors will be sending it a
smaller distance.  We do this by having two classes of
messages:\footnote{In an earlier version of these notes, these
messages where called $\BFSdistance(d)$ and
$\BFSnotDistance(d)$; the more self-explanatory $\BFSexactly$ and
$\BFSmoreThan$ terminology is taken from \cite{Boulinier2008}.}

\begin{itemize}
 \item $\BFSexactly(d)$: ``I know that my distance is $d$.''
 \item $\BFSmoreThan(d)$: ``I know that my distance is $>d$.''
\end{itemize}

The rules for sending these messages for a non-initiator are:

\begin{enumerate}
 \item  I can send $\BFSexactly(d)$ as soon as I have received
 $\BFSexactly(d-1)$ from at least one neighbor and $\BFSmoreThan(d-2)$ from all neighbors.
 \item  I can send $\BFSmoreThan(d)$ if $d = 0$ or as soon as I have
 received $\BFSmoreThan(d-1)$ from all neighbors.
\end{enumerate}

The initiator sends $\BFSexactly(0)$ to all neighbors at the start of the protocol (these are the only messages the initiator sends).

My distance will be the unique distance that I am allowed to send in
an $\BFSexactly(d)$ messages.  Note that this algorithm terminates in the sense that every node learns its distance at some finite time.  

If you read the discussion of synchronizers in
Chapter~\ref{chapter-synchronizers}, this algorithm essentially corresponds to building the
\index{synchronizer!alpha}\concept{alpha synchronizer} into the synchronous BFS algorithm, just
as the layered model builds in the \index{synchronizer!beta}\concept{beta synchronizer}.  See
\cite[\S11.3.2]{AttiyaW2004} for a discussion of BFS using
synchronizers.  The original approach of applying synchronizers to get
BFS is due to Awerbuch~\cite{Awerbuch1985}.

We now show correctness.  Under the assumption that local computation
takes zero time and message delivery takes at most 1 time unit, we'll
show that if $d(\Initiator, p) = d$, (a) $p$ sends
$\BFSmoreThan(d')$ for any $d' < d$ by time $d'$, (b) $p$ sends
$\BFSexactly(d)$ by time $d$, (c) $p$ never sends $\BFSmoreThan(d')$
for any $d' ≥ d$, and (d) $p$ never sends $\BFSexactly(d')$ for any
$d' ≠ d$.  For parts (c) and (d) we use induction on $d'$; for (a) and (b), induction on time.  This is not terribly surprising: (c) and (d) are safety properties, so we don't need to talk about time.  But (a) and (b) are liveness properties so time comes in.

Let's start with (c) and (d).  The base case is that the initiator 
never sends any $\BFSmoreThan$ messages at all, and so never sends
$\BFSmoreThan(0)$, and any
non-initiator never sends $\BFSexactly(0)$.
For larger $d'$, observe that if a non-initiator $p$ sends $\BFSmoreThan(d')$ for
$d'≥d$, it must first have received $\BFSmoreThan(d'-1)$ from all
neighbors, including some neighbor $p'$ at distance $d-1$.  But the
induction hypothesis tells us that $p'$ can't send
$\BFSmoreThan(d'-1)$ for $d'-1 ≥ d-1$.  Similarly, to send
$\BFSexactly(d')$ for $d' < d$, $p$ must first have received
$\BFSexactly(d'-1)$ from some neighbor $p'$, but again $p'$ must be at
distance at least $d-1$ from the initiator and so can't send this
message either.  In the other direction, to send $\BFSexactly(d')$ for
$d' > d$, $p$ 
must first receive
$\BFSmoreThan(d'-2)$ from this closer neighbor $p'$, 
but then $d'-2 > d-2 ≥ d-1$ so
$\BFSmoreThan(d'-2)$ is not sent by $p'$.

Now for (a) and (b).  The base case is that the initiator sends
$\BFSexactly(0)$ to all nodes at time 0, giving (a), and there is no
$\BFSmoreThan(d')$ with $d' < 0$ for it to send, giving (b) vacuously;
and any non-initiator sends $\BFSmoreThan(0)$ immediately.  At time
$t+1$, we have that (a) $\BFSmoreThan(t)$ was sent by any node at
distance $t+1$ or greater by time $t$ and (b) $\BFSexactly(t)$ was
sent by any node at distance $t$ by time $t$; so for any node at
distance $t+2$ we send $\BFSmoreThan(t+1)$ no later than time $t+1$
(because we already received $\BFSmoreThan(t)$ from all our neighbors)
and for any node at distance $t+1$ we send $\BFSexactly(t+1)$ no later
than time $t+1$ (because we received all the preconditions for doing so by this time).

Message complexity: A node at distance $d$ sends $\BFSmoreThan(d')$
for all $0 < d' < d$ and $\BFSexactly(d)$ and no other messages.  So we have
message complexity bounded by $\card*{E} \cdot D$ in the worst case.
Note that this is gives a bound of $O(DE)$, which is slightly worse
than the $O(E+DV)$ bound for the layered algorithm.

Time complexity: It's immediate from (a) and (b) that all messages
that are sent are sent by time $D$, and indeed that any node $p$ learns its
distance at time $d(\Initiator, p)$.  So we have optimal time
complexity, at the cost of higher message complexity.  I don't
know if this trade-off is necessary, or if a more sophisticated
algorithm could optimize both.

Our time proof assumes that messages don't pile up on edges, or that
such pile-ups don't affect delivery time (this is the default
assumption used in \cite{AttiyaW2004}).  A more sophisticated proof could remove this assumption.

One downside of this algorithm is that it has to be started
simultaneously at all nodes.  Alternatively, we could trigger ``time
0'' at each node by a broadcast from the initiator, using the usual
asynchronous broadcast algorithm; this would give us a BFS tree in
$O(\card*{E}\cdot D)$ messages (since the $O(\card*{E})$ messages of the
broadcast disappear into the constant) and $2D$ time.  The analysis of
time goes through as before, except that the starting time $0$ becomes
the time at which the last node in the system is woken up by the
broadcast.  Further optimizations are possible; see, for example, the
paper of Boulinier~\etal~\cite{Boulinier2008}, which shows how to
run the same algorithm with constant-size messages.

\myChapter{Leader election}{2026}{}
\label{chapter-leader-election}

(See also \cite[Chapter 3]{AttiyaW2004} or \cite[Chapter
3]{Lynch1996}.)

The idea of leader election is that we want a single process to
declare itself leader and the others to declare themselves
non-leaders. The non-leaders may or may not learn the identity of the
leader as part of the protocol; if not, we can usually add an extra
phase where the leader broadcasts its identity to the others.  The
leader should be unique in the sense that there is exactly one process
that ever decides it is the leader. This excludes protocols that might
accidentally elect two or more leaders even if we eventually remove
the extras.

Traditionally, leader election has been used as a way to study the
effects of symmetry, and many leader election algorithms are designed
for networks in the form of a \concept{ring}. These networks consist of a
sequence of processes $p_0, p_1, \dots, p_{n-1}$, with each process
$p_i$ able to send messages only to its immediate neighbors $p_{i-1}$
and $p_{i+1}$ (mod $n$). Some algorithms work in the weaker model of a
\index{ring!unidirectional}\concept{unidirectional ring} where
$p_i$ can only send messages to $p_{i+1}$.

A classic result of Angluin~\cite{Angluin1980} shows that leader
election in a ring is impossible if the processes do not start with
distinct identities. The proof is that if the processes run
synchronously, they all receive and send the same messages in each
round, update their state identically, and 
in the end all put on the crown at the same time.  We
discuss this result in §\ref{section-symmetry}.

With ordered identities, a simple algorithm due to Le
 Lann~\cite{LeLann1977} and
 Chang and Roberts~\cite{ChangR1979} solves the problem in $O(n)$ time with
 $O(n^{2})$ messages: I send out my own ID clockwise and forward any
 ID bigger than mine.  If I get my ID back, I win.  This works with a
 unidirectional ring, doesn't require synchrony, and never produces
 multiple leaders.  See §\ref{section-LCR}.

On a bidirectional ring we can get $O(n \log n)$ messages and
 $O(n)$ time with power-of-2 probing, using an algorithm of Hirschberg
 and Sinclair~\cite{HirschbergS1980}.  
 See §\ref{section-hirschberg-sinclair}.
 
A sneaky trick: if we have synchronized starting and known $n$, and 
IDs that are natural numbers (or that can be converted to natural
numbers), we can have process with ID $i$ wait until round $i\cdot n$
to start sending its ID around, and have everybody else drop out when
they receive it; this way only one process (the one with smallest ID)
ever starts a message and only $n$ messages are sent. But the running
time can be pretty bad. If we are willing to do a bit more tinkering,
we can follow~\cite[Lemma 1]{FredericksonL1987} and have ID $i$
be forwarded by each process only after $2^i$ steps; this also gets
$O(n)$ message complexity, at the cost of even worse time complexity,
but it does not require knowing $n$.

For general networks, we can apply the same basic strategy as in
Le Lann-Chang-Roberts by having each process initiate a
broadcast/convergecast algorithm that succeeds only if the initiator
has the smallest ID. See
§\ref{section-leader-election-in-general-networks}.

Some additional algorithms for the asynchronous ring are given in
§§\ref{section-Peterson-leader-election}
and~\ref{section-randomized-leader-election}.  Lower bounds
are shown in §\ref{section-leader-election-lower-bounds}.

\section{Symmetry}
\label{section-symmetry}

A system exhibits \concept{symmetry} if we can permute the nodes
without changing the behavior of the system.  More formally, we can
define a symmetry as an \concept{equivalence relation} on processes,
where we have the additional properties that all processes in the same
equivalence class run the same code; and whenever $p$ is equivalent
to $p'$, each neighbor $q$ of $p$ is equivalent to a corresponding
neighbor $q'$ of $p'$.

An example of a network with
a lot of symmetries would be an \concept{anonymous} \concept{ring},
which is a network in the form of a cycle (the ring part) in which
every process runs the same code (the anonymous part).  In this case
all nodes are equivalent.  If we have a line, then we might or might
not have any non-trivial symmetries: if each node has a
\concept{sense of direction} that tells it which neighbor is to the
left and which is to the right, then we can identify each node
uniquely by its distance from the left edge.  But if the nodes don't
have a sense of direction,
we can flip the line over and pair up nodes that map to each other.\footnote{Typically, this does not mean that
    the nodes can't tell their neighbors apart.  But it does mean that
    if we swap the labels for all the neighbors (corresponding to
    flipping the entire line from left to right), we get the
same executions.}

Symmetries are convenient for proving impossibility results, as
observed by Angluin~\cite{Angluin1980}.  The underlying theme is that
without some mechanism for \concept{symmetry breaking}, a
message-passing system escape from a symmetric initial configuration.
The following lemma holds for \concept{deterministic} systems,
basically those in which processes can't flip coins:
\begin{lemma}
\label{lemma-symmetry}
A symmetric deterministic message-passing system that starts in an initial
configuration in which equivalent processes have the same state has a
synchronous execution in which equivalent processes continue to have the same
state.
\end{lemma}
\begin{proof}
Easy induction on rounds: if in some round $p$ and $p'$ are equivalent and have the
same state, and all their neighbors are equivalent and have the same
state, then $p$ and $p'$ receive the same messages from their
neighbors and can proceed to the same state (including outgoing messages)
in the next round.
\end{proof}

An immediate corollary is that you can't do 
leader election in an anonymous system with a symmetry that puts each
node in a non-trivial equivalence class, because as soon as I stick my hand up
to declare I'm the leader, so do all my equivalence-class buddies.

With \concept{randomization}, Lemma~\ref{lemma-symmetry} doesn't
directly apply, since we can break symmetry by having my coin-flips
come up differently from yours.  It does show that we can't guarantee
convergence to a single leader in any fixed amount of time (because
otherwise we could just fix all the coin flips to get a deterministic
algorithm).  Depending on what the processes know about the size of
the system, it may still be possible to show that a randomized
algorithm necessarily fails in some cases.\footnote{Specifically, if
    the processes don't know the size of the ring, we can imagine a
    ring of size $2n$ in which the first $n$ processes happen to get
    exactly the same coin-flips as the second $n$ processes for long
    enough that two matching processes, one in each region,
    both think they have won the fight in a ring of size $n$ and 
    declare themself to be the leader.}

A more direct way to break symmetry is to assume that all processes
have \indexConcept{identity}{identities}; now processes can break
symmetry by just declaring that the one with the smaller or larger
identity wins.  This approach is taken in the algorithms in the
following sections.

\section{Leader election in rings}
\label{section-leader-election-in-rings}

Here we'll describe some basic leader election algorithms for rings.
Historically, rings were the first networks in which leader election
was studied, because they are the simplest networks whose symmetry
makes the problem difficult, and because of the connection to
token-ring networks, a method for congestion control in local-area
networks that is no longer used much.

\subsection{The Le Lann-Chang-Roberts algorithm}
\label{section-LCR}

This is about the simplest leader election algorithm there is.  It
works in a \concept{unidirectional ring}, where messages can only
travel clockwise.\footnote{We'll see later in
§\ref{section-Peterson-leader-election} that the distinction between
unidirectional rings and bidirectional rings is not a big deal, but
for now let's imagine that having a unidirectional ring is a serious
hardship.} The algorithm does not require synchrony.

Formally, we'll let the state space for each process $i$ consist of
two variables: \Leader, initially $0$, which is set to $1$ if $i$
decides it's a leader; and \MaxId, the largest ID seen so far.  We
assume that $i$ denotes $i$'s position rather than its ID, which we'll
write as $\Id_i$.  We will also treat all positions as values mod $n$, to simplify the arithmetic.

The initial version of this algorithm was proposed by Le
Lann~\cite{LeLann1977}; it involved sending every ID all the way
around the ring, and having a node decide it was a leader if it had
the largest ID. Chang and Roberts~\cite{ChangR1979} improved on this
by having nodes refuse to forward any ID smaller than the maximum ID
seen so far. This means that only the largest ID makes it all the way
around the ring, so a node can declare itself leader the moment it
sees its own ID. Depending on the writer, 
the resulting algorithm is known as either
Chang-Roberts or Le Lann-Chang-Roberts (LCR). We'll go with the latter
because it is always polite to be generous with credit.

Code for the LCR algorithm is given in Algorithm~\ref{alg-LCR}.

\begin{algorithm}
\Initially{
    $\Leader ← 0$\;
    $\MaxId ← \Id_i$\;
    send $\Id_i$ to clockwise neighbor
}
\UponReceiving{$j$}{
    \If{$j = \Id_i$}{
        $\Leader ← 1$\;
    }
    \If{$j > \MaxId$}{
        $\MaxId ← j$\;
        send $j$ to clockwise neighbor\;
    }
}
\caption{LCR leader election}
\label{alg-LCR}
\end{algorithm}

Intuitively, this protocol works because whichever process $p_{max}$ holds
the maximum ID $\Id_{max}$ will (a) refuse to forward any smaller ID,
and (b) eventually have its value forwarded through all of the other
processes, causing it to eventually set its $\Leader$ bit to $1$.

Looking closely at this intuition we see that (a) is a safety property
and (b) a liveness property. So we obtain a proof of
correctness by converting (a) into an invariant that for each $p_i ≠
p_{max}$, $\Id_i$ is never sent by any process in the range $p_{max}
\dots p_{i-1}$; and converting (b) into an induction argument that
each process $p_{\max+j}$ sends $\Id_{max}$ to $p_{\max+j+1}$ no later
than time $j$. Because the code only has a process $p_i$ set $\Leader$
to $1$ if it receives $\Id_i$ from $p_{i-1}$, the invariant tells us
that no $p_i ≠ p_{max}$ becomes the leader, while the induction
argument tells use that eventually $p_{max}$ does.

\subsubsection{Performance}

It's immediate from the correctness proof that the protocol elects a
leader within at most $n$ time in the asynchronous model or exactly
$n$ rounds in a synchronous model.

To bound message traffic, observe that each process sends at most one
copy of each of the $n$ process IDs, for a total of $O(n^{2})$
messages. This is a tight bound since if the IDs are in decreasing
order $n, n-1, n-2, \dots{} 1$, then no messages get eaten until they
hit $n$.

There is a subtlety with the termination guarantee: at the moment the
unique leader $p_{max}$ sets its leader bit, the other processes all
have $\MaxId = \Id_{max}$, but they don't actually \emph{know} that
they have the correct leader ID, since there is no information
available locally at a non-leader process that allows it to detect
that there can't be some larger ID out there that just hasn't reached
it yet. As with all leader election algorithms, we can have the leader
confirm its election with an additional broadcast protocol, which in
this case raises the time complexity from $n$ to $2n$ (still $O(n)$) and adds an
extra $n$ messages (still $O(n^2)$ in total).

\subsection{The Hirschberg-Sinclair algorithm}
\label{section-hirschberg-sinclair}

This algorithm improves on Le Lann-Chang-Roberts by reducing the
message complexity.  The idea is that instead of having each process
send a message all the way around a ring, each process will first probe
locally to see if it has the largest ID within a short distance.  If
it wins among its immediate neighbors, it doubles the size of the
neighborhood it checks, and continues as long as it has a winning ID.
This means that most nodes drop out quickly, giving a total message
complexity of $O(n \log n)$.  The running time is a constant factor
worse than LCR, but still $O(n)$. The algorithm assumes a
bidirectional ring, since the reverse edges are needed to send back
responses to probes.

To specify the protocol, it may help to think of
messages as mobile agents and the state of each process as being of
the form $(\text{local-state}, \{ \text{agents I'm carrying} \})$.
Then the sending rule for a process becomes \emph{ship any agents in
whatever direction they want to go} and the transition rule is
\emph{accept any incoming agents and update their state in terms of
their own internal transition rules}.  An agent state for LCR will be
something like (original-sender, direction, hop-count, max-seen) where
direction is $R$ or $L$ depending on which way the agent is going,
hop-count in phase $k$ is initially $2^{k}$ when the agent is sent and drops by $1$ each time the agent moves, and max-seen is the biggest ID of any node the agent has visited.  An agent turns around (switches direction) when hop-count reaches 0.

To prove this works, we can mostly ignore the early phases (though we have to show that the max-id node doesn't drop out early, which is not too hard).  The last phase involves any surviving node probing all the way around the ring, so it will declare itself leader only when it receives its own agent from the left.  That exactly one node does so is immediate from the same argument for LCR.

Complexity analysis is mildly painful but basically
comes down to the fact that any node that sends a message $2^{k}$
hops had to be a winner in phase $2k-1$, which means that it is
the largest of some group of $2^{k-1}$ IDs.  Thus the $2^{k}$-hop
senders are spaced at least $2^{k-1}$ away from each other and there
are at most $n/2^{k-1}$ of them.  Summing up over all $\ceil{\lg n}$
phases, we get $\sum_{k=0}^{\ceil{\lg n}} 2^{k} n / 2^{k-1} = O(n \log
n)$ messages and  $\sum_{k=0}^{\ceil{\lg n}} 2^{k} = O(n)$ time.

\subsection{Peterson's algorithm for the unidirectional ring}
\label{section-Peterson-leader-election}

This algorithm is due to Peterson~\cite{Peterson1982} and assumes an
asynchronous, unidirectional ring.
It gets $O(n \log n)$ message complexity in all executions.

Let's start by describing a version with two-way communication. 
Start with $n$ candidate
leaders.  In each of at most $\lg n$ asynchronous phases, each
candidate probes its nearest surviving neighbors to the left and right; if its
ID is larger than the IDs of both neighbors, it survives to the next
phase. Non-candidates act as relays passing messages between
candidates. As in Hirschberg and Sinclair
(§\ref{section-hirschberg-sinclair}), the probing operations in each
phase take $O(n)$ messages, and at least half of the candidates drop
out in each phase.  The last surviving candidate wins when it finds
that it's its own surviving neighbor.

To make this work in a 1-way ring, we have to simulate 2-way
communication by moving the candidates clockwise around the ring to
catch up with their unsendable counterclockwise messages.  Peterson's
algorithm does this with a two-hop approach that is inspired by the
2-way case above; in each phase $k$, a candidate effectively moves two
positions to the right, allowing it to look at the IDs of three
phase-$k$ candidates before deciding to continue in phase $k+1$ or
not.  Here is a very high-level description; it assumes that we can
buffer and ignore incoming messages from the later phases until we get
to the right phase, and that we can execute sends immediately upon
receiving messages.  Doing this formally in terms of 
the model of §\ref{section-message-passing-model} means that we have
to build explicit internal buffers into our processes, which we can
easily do but won't do here (see \cite[pp.~483--484]{Lynch1996} for
the right way to do this).

We can use a similar trick to transform any bidirectional-ring
algorithm into a unidirectional-ring algorithm: alternate between
phases where we send a message right, then send a virtual process
right to pick up any left-going messages deposited for us.  The
problem with this trick is that it requires two messages per process
per phase, which gives us a total message complexity of $O(n^2)$ if we
start with an $O(n)$-time algorithm. Peterson's algorithm avoids this
by propagating only the surviving candidates.

Pseudocode for Peterson's algorithm is given in Algorithm~\ref{alg-peterson}.

\begin{algorithm}
\newData{\PetersonPhase}{phase}
\newData{\PetersonCurrent}{current}
\newFunc{\PetersonProbe}{probe}
\newFunc{\PetersonCandidate}{candidate}
\newFunc{\PetersonRelay}{relay}

\Procedure{$\PetersonCandidate()$}{
    $\PetersonPhase ← 0$ \;
    $\PetersonCurrent ← \Pid$ \;

    \While{\True}{
        send $\PetersonProbe(\PetersonPhase, \PetersonCurrent)$ \;
        wait for $\PetersonProbe(\PetersonPhase, x)$ \;
        $\Id_2 ← x$ \;
        send $\PetersonProbe(\PetersonPhase + 1/2, \Id_2)$ \;
        wait for $\PetersonProbe(\PetersonPhase + 1/2, x)$ \;
        $\Id_3 ← x$ \;
        \uIf{$\Id_2 = \PetersonCurrent$}{
            I am the leader!\;
            \Return\;
        }
        \uElseIf{$\Id_2 > \PetersonCurrent$ \KwAnd $\Id_2 > \Id_3$}{
            $\PetersonCurrent ← \Id_2$\;
            $\PetersonPhase ← \PetersonPhase + 1$\;
        }
        \Else{
            switch to $\PetersonRelay()$\;
        }
    }
}
\Procedure{$\PetersonRelay()$}{
    \UponReceiving{$\PetersonProbe(p, i)$}{
        send $\PetersonProbe(p, i)$\;
    }
}
\caption{Peterson's leader-election algorithm}
\label{alg-peterson}
\end{algorithm}

Note: The phase arguments in the probe messages are useless if one has
FIFO channels, which is why \cite{Lynch1996} doesn't use them. 

Proof of correctness is essentially the same as for the 2-way
algorithm.  For any pair of adjacent candidates, at most one of their
current IDs survives to the next phase.  So we get a sole survivor
after $\ceil{\lg n}$ phases.  Each process sends or relays at most 2 messages
per phase, so we get at most $2 n \ceil{\lg n}$ total messages.

Curiously, the time complexity of Peterson's algorithm may be worse
than $O(n)$. It's not hard to construct an identity assignment in
which all nodes in half the ring drop out, leaving $n/4$ candidates on
the other side of the ring. Each subsequent phase may then require as
much as $n/2$ time to transmit a message across the missing half.
If it takes $Θ(\log n)$ phases to reduce these $n/4$ candidates to
one, this gives $Θ(n \log n)$ total time.

\subsection{A simple randomized \texorpdfstring{$O(n \log n)$}{O(n log
n)}-message algorithm}
\label{section-randomized-leader-election}

An alternative to running a more sophisticated algorithm is to reduce
the average cost of LCR 
using randomization.  The presentation here follows the average-case analysis done
by Chang and Roberts~\cite{ChangR1979}.

Run LCR where each ID is constructed by prepending a long random
bit-string to the real ID. This gives uniqueness (since the real IDs
act as tie-breakers) and something very close to a random permutation
on the constructed IDs. When we have unique random IDs, a simple
argument shows that the $i$-th largest ID only propagates an expected
$n/i$ hops, giving a total of $O(n H_{n}) = O(n \log n)$
hops.\footnote{Alternatively, we could consider the
\concept{average-case complexity} of the algorithm when we assume all
$n!$ orderings of the IDs are equally likely; this also gives $O(n
\log n)$ expected message complexity~\cite{ChangR1979}.} Unique
random IDs occur with high probability provided the range of the
random sequence is $\gg n^{2}$.

The downside of this algorithm compared to Peterson's is that
knowledge of $n$ is required to pick random IDs from a large enough
range.
It also has higher bit complexity, since Peterson's algorithm
is sending only IDs (in the FIFO-channel version) without any random
padding.  An possible upside is that if the range of random IDs is
large enough, we can run it without any initial IDs at all, as long as
we are willing to accept a small probability of accidentally electing
two leaders.

\section{Leader election in general networks}
\label{section-leader-election-in-general-networks}

For general networks, a simple approach is to have each node initiate
a breadth-first-search and convergecast, with nodes refusing to
participate in the protocol for any initiator with a lower ID.
It follows that only the node with the maximum ID can finish its
protocol; this node becomes the leader.  If messages from parallel
broadcasts are combined, it's possible to keep the message complexity
of this algorithm down to $O(DE)$.

More sophisticated algorithms reduce the message complexity by
coalescing local neighborhoods similar to what happens in the
Hirschberg-Sinclair and Peterson algorithms.  A noteworthy example is
an $O(n \log n)$ message-complexity algorithm of Afek and
Gafni~\cite{AfekG1991}, who also show an $Ω(n \log n)$ lower
bound on message complexity for any synchronous algorithm in a
complete network.

\section{Lower bounds}
\label{section-leader-election-lower-bounds}

Here we present two classic $\Omega(\log n)$ lower bounds on message
complexity for leader election in the ring.  The first, due to
Burns~\cite{Burns1980},
assumes that the system is asynchronous and that 
the algorithm is \concept{uniform}: it does not depend on the size of
the ring.  The second, due to Frederickson and
Lynch~\cite{FredericksonL1987}, allows a synchronous system and relaxes the
uniformity assumption, but requires that the algorithm can't do
anything to IDs but copy and compare them.

\subsection{Lower bound on asynchronous message complexity}
\label{section-asynchronous-leader-election-lower-bound}

Here we describe a lower bound for uniform asynchronous leader election
in the ring.
The description here is based on~\cite[\S3.3.3]{AttiyaW2004}; a
slightly different presentation can also be found
in~\cite[\S15.1.4]{Lynch1996}.  The original result is due to
Burns~\cite{Burns1980}.  We assume the system is
deterministic.

The proof constructs a bad execution in which $n$ processes send lots of
messages recursively, by first constructing two bad $(n/2)$-process
executions and pasting them together in a way that generates many
extra messages.  If the pasting step produces
$Θ(n)$ additional messages, we get a recurrence $T(n) ≥ 2T(n/2) +
Θ(n)$ for the total message traffic, which has solution $T(n) =
\Omega(n \log n)$.

We'll assume that
all processes are trying to learn the identity of the process with the
smallest ID.  This is a slightly stronger problem that mere leader
election, but it can be solved with at most an additional $2n$ messages
once we actually elect a leader.  So if we get a lower bound of $f(n)$
messages on this problem, we immediately get a lower bound of
$f(n)-2n$ on leader election.

To construct the bad execution, we consider ``open executions'' on
rings of size $n$ where no message is delivered across some edge
(these will be partial executions, because otherwise the guarantee of
eventual delivery kicks in).
Because no message is delivered across this edge, the processes can't
tell if there is really a single edge there or some enormous
unexplored fragment of a much larger ring.  Our induction hypothesis
will show that a line of $n/2$ processes can be made to send at least
$T(n/2)$
messages in an open execution (before seeing any messages across the
open edge); we'll then show that a linear number of additional
messages can be generated by pasting two such executions together
end-to-end, while still getting an open execution with $n$ processes.

In the base case, we let $n=1$.  Somebody has to send a message
eventually, giving $T(2) ≥ 1$.

For larger $n$, suppose that we have two open executions on $n/2$
processes that each send at least $T(n/2)$ messages.  Break the open edges in
both executions and replace them with new edges to create a ring of
of size $n$; similarly paste the schedules $σ_1$ and $σ_2$
of the two executions together to get a combined schedule
$σ_1σ_2$ with at least $2T(n/2)$ messages.  Note that in
the combined schedule no messages are passed between the two sides, so
the processes continue to behave as they did in their separate
executions.

Let $e$ and $e'$ be the edges we used to past together the two rings.
Extend $σ_1σ_2$ by the longest possible suffix $σ_3$ in
which no messages are delivered across $e$ and $e'$.  Since $σ_3$
is as long as possible, after $σ_1σ_2σ_3$, there are no
messages waiting to be delivered across any edge except $e$ and $e'$
and all processes are \concept{quiescent}—they will send no
additional messages until they receive one.

We now consider some suffix $σ_4$ that causes the protocol to finish
when appended to $σ_1σ_2σ_3$.
While executing $σ_4$, construct two sets of processes $S$ and $S'$ by
the following rules:
\begin{enumerate}
    \item If a process is not yet in $S$ or $S'$ and receives a
        message delivered across $e$, put it in $S$; similarly if it
        receives a message delivered across $e'$, put it in $S'$.
    \item If a process is not yet in $S$ or $S'$ and receives a
        message that was sent by a process in $S$, put it in $S$;
        similarly for $S'$.
\end{enumerate}

Observe that this process must eventually make $S$ and $S'$ adjacent,
because if there is some node in the half to the ring with the larger
minimum id that receives no messages in $σ_4$ (and thus is never added
to $S$ or $S'$?), that node doesn't learn the global minimum.

So now imagine stopping the process after the shortest prefix $σ'_4$ of $σ_4$
that makes $S$ and $S'$ adjacent. This gives $\card{S∪S'} ≥ n/2$,
because we include all nodes between $e$ and $e'$ on one side or the
other. It follows that at least one of $S$ and $S'$ contains at least
$n/4$ nodes after $σ'_4$.

Assume without loss of generality that it is $\card{S}$ that is at
least $n/4$.  Except for the two processes incident to $e$, every
process that is added to $S$ is added in response to a message sent in
$σ'_4$.  So there are at least $n/4-2$ such messages.  We can also
argue that all of these messages are sent in the subschedule $τ$ of
$σ'_4$ that contains only messages that do not depend on messages
delivered across $e'$.  It follows that $σ_1σ_2σ_3τ$ is an open
execution on $n$ processes with at least $2T(n/2) + n/4 - 2$ sent
messages. This gives $T(n) ≥ 2T(n/2) + n/4 - 2 = 2T(n/2) + Ω(n)$ as
claimed.

\subsection{Lower bound for comparison-based protocols}

Here we give an $\Omega(n \log n)$ lower bound on
messages for synchronous-start comparison-based protocols in
bidirectional synchronous rings. For full details see
\cite[\S3.6]{Lynch1996}, \cite[\S3.4.2]{AttiyaW2004}, or
the original JACM paper by Frederickson and
Lynch~\cite{FredericksonL1987}.

The argument proceeds as follows:
\begin{itemize}
 \item Two fragments $i \dots i+k$ and $j\dots j+k$ of a ring are
 \concept{order-equivalent} provided $\Id_{i+a} > \Id_{i+b}$ if and
 only if $\Id_{j+a} > \Id_{j+b}$ for $b = 0\dots k$.
 \item A protocol is \index{protocol!comparison-based}\indexConcept{comparison-based
 protocol}{comparison-based} if it can't do anything to IDs but copy
 them and test for $<$.  The state of such an protocol is modeled by
 some non-ID state together with a big bag of IDs, messages have a
 pile of IDs attached to them, etc.  Two states/messages are
 equivalent under some mapping of IDs if you can translate the first
 to the second by running all IDs through the mapping.
 
 An equivalent version uses an explicit equivalence relation between
 processes.
 Let executions of $p_{1}$ and $p_{2}$ be \concept{similar} if
 both processes send messages in the same direction(s) in the same
 rounds and both processes 
 declare themselves leader (or not) at the same round.
 Then an protocol is
 comparison-based based if order-equivalent rings yield similar
 executions for corresponding processes.  This can be turned into the
 explicit-copying-ids model by replacing the original protocol with a
 \concept{full-information protocol} in which each message is replaced by the ID and a complete history of the sending process (including all messages it has every received).
 \item Define an \concept{active round} as a round in which at least
 one message is sent.  Claim: Actions of $i$ after $k$ active rounds
 depends, up to an order-equivalent mapping of IDs, only on the
 order-equivalence class of IDs in $i-k\dots i+k$, the
 \concept{$k$-neighborhood} of $i$.  Proof: by induction on $k$.
 Suppose $i$ and $j$ have order-equivalent $(k-1)$-neighborhoods; then
 after $k-1$ active rounds they have equivalent states by the
 induction hypothesis.  In inactive rounds, $i$ and $j$ both receive
 no messages and update their states in the same way.  In active
 rounds, $i$ and $j$ receive order-equivalent messages and update their states in an order-equivalent way.
 \item If we have an order of IDs with a lot of order-equivalent
 $k$-neighborhoods, then after $k$ active rounds if one process sends a message, so do a lot of other ones.
\end{itemize}

\begin{figure}
\centering
\includegraphics[scale=0.8]{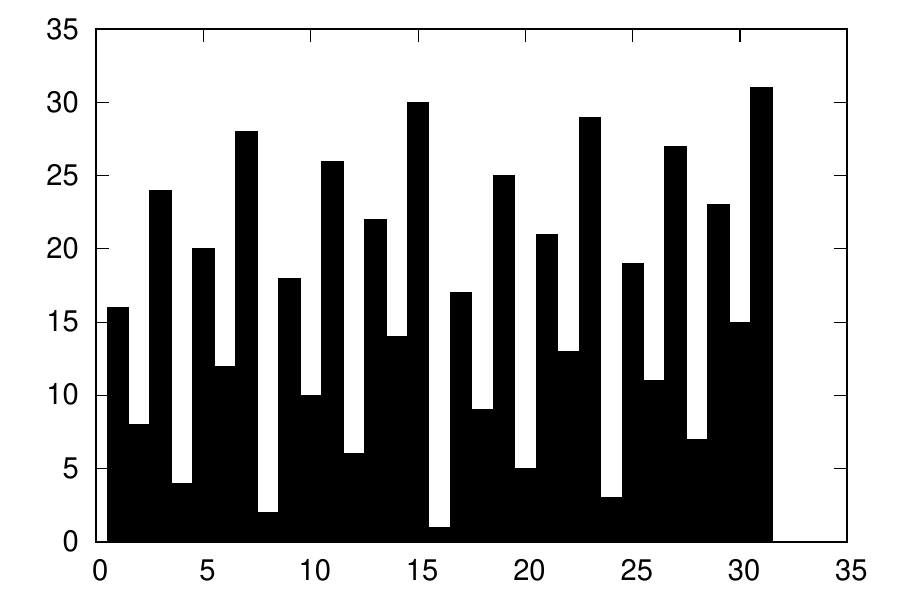}
\caption{Labels in the bit-reversal ring with $n=32$}
\label{figure-bit-reversal-ring}
\end{figure}

Now we just need to build a ring with a lot of order-equivalent
neighborhoods.  For $n$ a power of $2$ we can use the bit-reversal
ring, e.g., ID sequence $000, 100, 010, 110, 001, 101, 011, 111$ (in
binary) when $n=8$.  Figure~\ref{figure-bit-reversal-ring} gives a picture
of what this looks like for $n=32$.

For $n$ not a power of $2$ we look up Frederickson and
Lynch~\cite{FredericksonL1987} or
Attiya~\etal~\cite{AttiyaSW1988}.
In either case we get
$\Omega(n/k)$ order-equivalent members of each equivalence class after
$k$ active rounds, giving $\Omega(n/k)$ messages per active round,
which sums to $\Omega(n \log n)$.

For non-comparison-based protocols we can still prove $\Omega(n \log
n)$ messages for time-bounded protocols, but it requires
techniques from \concept{Ramsey theory}, the branch of combinatorics
that studies when large enough structures inevitably contain
substructures with certain properties.\footnote{The classic example is
\concept{Ramsey's Theorem}, which says that if you color the edges of
a complete graph red or blue, while trying to avoid having any subsets
of $k$ vertices with all edges between them the same color, 
you will no longer be able to once the graph is large enough (for any
fixed $k$).  See~\cite{GrahamRS1990} for much more on the subject of
Ramsey theory.}
Here
``time-bounded'' means that the running time can't depend on the size
of the ID space.
See
\cite[\S3.4.2]{AttiyaW2004} or \cite[\S3.7]{Lynch1996} for the
textbook version, or
\cite[\S7]{FredericksonL1987} for the original result.  

The intuition is that for any fixed protocol, if the
ID space is large enough, then there exists a
subset of the ID space where the protocol acts like a comparison-based
protocol.  So the existence of an $O(f(n))$-message time-bounded
protocol implies the existence of an $O(f(n))$-message
comparison-based protocol, and from the previous lower bound we know
$f(n)$ is $\Omega(n \log n)$.  Note that time-boundedness is
necessary: we can't prove the lower bound for non-time-bounded
algorithms because of the $i\cdot n$ trick.

\myChapter{Causal ordering and logical clocks}{2026}{}
\label{chapter-logical-clocks}

\index{clock!logical}\indexConcept{logical clock}{Logical clocks}
assign a timestamp to all events in an asynchronous message-passing
system that simulates real time, thereby allowing timing-based
algorithms to run despite asynchrony. In general, they don't have
anything to do with clock synchronization or wall-clock time; instead,
they provide numerical values that increase over time and are
consistent with the observable behavior of the system. This means
that local events on a single process have increasing times, and
messages are never delivered before they are sent, when time is
measured using the logical clock.

Because the processes in a system don't necessarily know the relative
order of distant events, a totally-ordered logical clock may impose
an ordering on events that is not observable by the processes. We can
capture the observable (partial) ordering using a \concept{causal
ordering}, defined in §\ref{section-causal-ordering}.
A totally-ordered logical clock is correct if it gives an ordering
that is a refinement of the causal ordering; some examples are given
in §\ref{section-logical-clock-implementations}.
Alternatively, by using
partially-ordered set for the values of our logical clock, it may be
possible to capture the causal ordering precisely
(§\ref{section-vector-clocks}).

One application of logical clocks is to implement a
\concept{snapshot}, as described in
§\ref{section-chandy-lamport}. The simplest version of this
is to have each process record its state at some particular logical
clock time. This is not quite an description of the global
configuration of the system at some real-time instant in the execution, because
asynchronous processes can't guarantee that they all take a snapshot
at the same real time. Instead, it's a description of a global
configuration that is consistent with the observations of the
processes, in the sense that there exists an execution
indistinguishable from the real one that contains this configuration.
Causal ordering is the tool that lets us argue that this hypothetical
execution exists.

\section{Causal ordering}
\label{section-causal-ordering}

Here we define the
\index{ordering!causal}\concept{causal ordering}, a partial order on
events that describes when one event $e$ can shown to occur before some
other event $e'$ based only on the sequences of events observed by
each process.

For the purpose of defining the causal ordering and logical clocks, we
will assume that a schedule consists of
\index{event!send}\indexConcept{send event}{send events} and
\index{event!receive}\indexConcept{receive event}{receive events},
which correspond to some process sending a single message or receiving
a single message, respectively. This is not quite the same as our
usual model that allows many messages to be received and sent as part
of the same delivery event, but for asynchronous systems we can treat
the definitions as equivalent by splitting a multi-message delivery
event into a sequence of events, one for each message.

Given two schedules $S$ and $S'$, call $S$ and $S'$ \concept{similar}
if $S|p = S'|p$ for all processes $p$; in other words, $S$ and $S'$
are similar if they are indistinguishable by all participants.  We can define a causal
ordering on the events of some schedule $S$ implicitly by considering
all schedules $S'$ similar to $S$, and declare that $e < e'$ if $e$
precedes $e'$ in all such $S$.  But it is usually more useful to make this ordering explicit.

Following~\cite[\S6.1.1]{AttiyaW2004} (and
ultimately~\cite{Lamport1978}), define the
\concept{happens-before} relation $\happensBefore{S}$ on a schedule $S$
to consist of:
\begin{enumerate}
 \item All pairs $(e, e')$ where $e$ precedes $e'$ in $S$ and $e$ and
 $e'$ are events of the same process.
 \item All pairs $(e, e')$ where $e$ is a send event and $e'$ is the receive event for the same message.
 \item All pairs $(e, e')$ where there exists a third event $e''$ such
 that $e \happensBefore{S} e''$ and $e'' \happensBefore{S} e'$.  (In
 other words, we take the \concept{transitive closure} of the relation
 defined by the previous two cases.)
\end{enumerate}

It is not terribly hard to show that this gives a partial order; the
main observation is that if $e \happensBefore{S} e'$, then $e$
precedes $e'$ in $S$.  So $\happensBefore{S}$ is a subset of the total
order $<_{S}$ given by the order of events in $S$.

A \concept{causal shuffle} $S'$ of a schedule $S$ is a permutation of
$S$ that is consistent with the happens-before relation on $S$; that
is, if $e$ happens-before $e'$ in $S$, then $e$ precedes $e'$ in $S'$.
The importance of the happens-before relation follows from the
following lemma,
which says that the causal shuffles of $S$ are precisely the schedules
$S'$ that are similar to $S$.
\begin{lemma}
\label{lemma-happens-before}
 Let $S'$ be a permutation of the events in $S$.  Then the following two statements are equivalent:
\begin{enumerate}
  \item $S'$ is a causal shuffle of $S$.
  \item $S'$ is the schedule of an execution fragment of a
  message-passing system with $S|p = S'|p$ for all $S'$.
\end{enumerate}
\end{lemma}
\begin{proof}
  $(1 ⇒ 2)$.  We need to show both similarity and
  that $S'$ corresponds to some execution fragment.  We'll show
  similarity first.  Pick some $p$; then every event at $p$ in $S$
  also occurs in $S'$, and they must occur in the same order by the
  first case of the definition of the happens-before relation.  This
  gets us halfway to showing $S'$ is the schedule of some execution
  fragment, since it says that any events initiated by $p$ are
  consistent with $p$'s programming.  To get the rest of the way,
  observe that any other events are receive events.  For each receive
  event $e'$ in $S$, there must be some matching send event $e$ also
  in $S$; thus $e$ and $e'$ are both in $S'$ and occur in the right order by the second case of the definition of happens-before.

  $(2 ⇒ 1)$.  First observe that since every event
  $e$ in $S'$ occurs at some process $p$, if $S'|p = S|p$ for all $p$,
  then there is a one-to-one correspondence between events in $S'$ and
  $S$, and thus $S'$ is a permutation of $S$.  Now we need to show
  that $S'$ is consistent with $\happensBefore{S}$.  Let $e
  \happensBefore{S} e'$.  There are three cases.
\begin{enumerate}
   \item $e$ and $e'$ are events of the same process $p$ and $e <_{S}
   e'$.  But then $e <_{S'} e'$ because $S|p = S'|p$.
   \item $e$ is a send event and $e'$ is the corresponding receive
   event.  Then $e <_{S'} e'$ because $S'$ is the schedule of an execution fragment.
   \item $e \happensBefore{S} e'$ by transitivity.  Then each step in
   the chain connecting $e$ to $e'$ uses one of the previous cases,
   and $e <_{S'} e'$ by transitivity of $<_{S'}$.
\end{enumerate}
\end{proof}

There are two main applications for causal shuffles:
\begin{enumerate}
    \item We can prove upper bounds by using a causal shuffle to turn
        some arbitrary $S$ into a nice $S'$, and argue that the
        niceness of $S'$ means that even if $S$ might not be nice, 
        it looks nice to the
        processes.  An example of this can be found in
        Lemma~\ref{lemma-local-synchronizer}.
    \item We can prove lower bounds by using a causal shuffle to turn
        some specific $S$ into a nasty $S'$, and argue that the existence
        of $S'$ tells us that there exist nasty schedules for some
        particular problem.  An example of this can be found in
        §\ref{section-session-problem}.  This works particularly well
        because $\happensBefore{S}$ includes enough information to
        determine the latest possible time of any event in either $S$
        or $S'$, so
        rearranging schedules like this doesn't change the worst-case time.
\end{enumerate}

In both cases, we are using the fact that if I tell you $\happensBefore{S}$, then you know
everything there is to know about the order of events in $S$ that you
can deduce from reports from each process together with the fact that
messages don't travel back in time.  

In the case that we want to use this information \emph{inside} an
algorithm, we run into the issue that
$\happensBefore{S}$ is a
pretty big relation ($Θ(\card*{S}^2)$ bits with a naive encoding), and
seems to require global knowledge of $<_{S}$ to compute.  So we can
ask if there is some simpler, easily computable description that works
almost as well.  This is where logical clocks come in.

\section{Logical clocks}
\label{section-logical-clock-implementations}

The idea of a logical clock is to compute a \concept{timestamp}
for each event, so that comparing timestamps gives information about
$\happensBefore{S}$.  Note that these timestamps need not be totally ordered.  In
general, we will have a relation $<_L$ between timestamps such that 
$e \happensBefore{S} e'$ implies $e <_L e'$, but it may be that there are some pairs
of events that are ordered by the logical clock despite being
incomparable in the happens-before relation.

Examples of logical clocks that use small timestamps but add extra
ordering are Lamport clocks~\cite{Lamport1978}, discussed in
§\ref{section-Lamport-clock}; and Neiger-Toueg-Welch
clocks~\cite{NeigerT1987,Welch1987}, discussed in
§\ref{section-Neiger-Toueg-Welch-clock}.  These both assign integer
timestamps to events and may order events that are not causally
related.  
The main difference between them is that Lamport clocks do not alter
the underlying execution, but may allow arbitrarily large jumps in the
logical clock values; while Neiger-Toueg-Welch clocks guarantee small
increments at the cost of possibly delaying parts of the
system.\footnote{This makes them similar to
    \indexConcept{synchronizer}{synchronizers}, which we will
discuss in Chapter~\ref{chapter-synchronizers}.}

More informative are
\indexConcept{vector clock}{vector clocks}~\cite{Fidge1991,Mattern1993}, discussed in
§\ref{section-vector-clocks}.  These use
$n$-dimensional vectors of integers to capture $\happensBefore{S}$ exactly, at the
cost of much higher overhead.

\subsection{Lamport clock}
\label{section-Lamport-clock}

\newData{\Clock}{clock}

\index{Lamport clock}
\index{logical clock!Lamport}
\index{clock!logical!Lamport}
Lamport's \concept{logical clock}~\cite{Lamport1978} runs on top of
any other
message-passing protocol, adding additional state at each process and
additional content to the messages (which is invisible to the
underlying protocol).  Every process maintains a local variable
\Clock.  When a process sends a message or executes an internal step,
it sets $\Clock ← \Clock + 1$ and assigns the resulting
value as the clock value of the event.  If it sends a message, it
piggybacks the resulting clock value on the message.  When a process
receives a message with timestamp $t$, it sets $\Clock ←
\max(\Clock, t)+1$; the resulting clock value is taken as the time of receipt of the message.  (To make life easier, we assume messages are received one at a time.)

\begin{theorem}
\label{theorem-lamport-clock}
If we order all events by clock value, we get an execution of the underlying protocol that is locally indistinguishable from the original execution.
\end{theorem}
\begin{proof}
Let $e <_{L} e'$ if $e$ has a lower clock value than $e'$.  If $e$ and
$e'$ are two events of the same process, then $e <_{L} e'$.  If $e$
and $e'$ are send and receive events of the same message, then again
$e <_{L} e'$.  So for \emph{any} events $e$, $e'$, if $e
\happensBefore{S} e'$, then $e <_{L} e'$.  Now apply
Lemma~\ref{lemma-happens-before}.
\end{proof}

\subsection{Neiger-Toueg-Welch clock}
\label{section-Neiger-Toueg-Welch-clock}

\index{Neiger-Toueg-Welch clock}
\index{logical clock!Neiger-Toueg-Welch}
\index{clock!logical!Neiger-Toueg-Welch}
Lamport's clock has the advantage of requiring no changes in the
behavior of the underlying protocol, but has the disadvantage that
clocks are entirely under the control of the logical-clock protocol
and may as a result make huge jumps when a message is received.  If
this is unacceptable—perhaps the protocol needs to do some
unskippable maintenance task every 1000 clock ticks—then an
alternative approach due to Neiger and
Toueg~\cite{NeigerT1987} and Welch~\cite{Welch1987} can be used.

Method: Each process maintains its own variable \Clock, which it
increments whenever it feels like it.  To break ties, the process
extends the clock value to 
$\langle\Clock, \Id, \DataSty{eventCount}\rangle$ where
\DataSty{eventCount} is a count of send and receive events
(and possibly local computation steps).  As in Lamport's clock, each
message in the underlying protocol is timestamped with the current
extended clock value.  Because the protocol can't change the clock
values on its own, when a message is received with a timestamp later
than the current extended clock value, its delivery is delayed until \Clock exceeds the message timestamp, at which point the receive event is assigned the extended clock value of the time of delivery.

\begin{theorem}
\label{theorem-welch-clock}
If we order all events by clock value, we get an execution of the underlying protocol that is locally indistinguishable from the original execution.
\end{theorem}
\begin{proof}
Again, we have that (a) all events at the same process occur in
increasing order (since the event count rises even if the clock value
doesn't, and we assume that the clock value doesn't drop) and (b) all
receive events occur later than the corresponding send event (since we
force them to).  So Lemma~\ref{lemma-happens-before} applies.
\end{proof}

The advantage of the Neiger-Toueg-Welch clock is that it doesn't impose any
assumptions on the clock values, so it is possible to make \Clock be a
real-time clock at each process and nonetheless have a
causally-consistent ordering of timestamps even if the local clocks
are not perfectly synchronized.  If some process's clock is too far
off, it will have trouble getting its messages delivered quickly (if
its clock is ahead) or receiving messages (if its clock is
behind)—the net effect is to add a round-trip delay to that process
equal to the difference between its clock and the clock of its
correspondent.  But the protocol works well when the processes' clocks
are closely synchronized, which is a reasonable assumption in many
systems thanks to the Network Time Protocol, cheap GPS
receivers, and clock synchronization mechanisms built into most
cellular phone networks.\footnote{As I write this, my computer reports
that its clock is an estimated 289 microseconds off from the
timeserver it is synchronized to, which is less than a tenth of the
round-trip delay to machines on the same local-area network and a tiny
fraction of the round-trip delay to machines elsewhere, including
the timeserver machine.}

\subsection{Vector clocks}
\label{section-vector-clocks}

Logical clocks give a \emph{superset} of the happens-before relation:
if $e \happensBefore{S} e'$, then $e <_{L} e'$ (or conversely, if $e
\not<_{L} e'$, then it is not the case that $e \happensBefore{S} e'$).
This is good enough for most applications, but what if we want to
compute $\happensBefore{S}$ exactly?

\newData{\VC}{VC}

Here we can use a \concept{vector clock}, invented independently by
Fidge~\cite{Fidge1991} and Mattern~\cite{Mattern1993}. Instead of a single clock
value, each event is stamped with a vector of values, one for each
process.  

A process $p$ starts with a vector $t^p = 0$ (all components $0$).
When a process executes a local event or a send event, it
increments only its own component $t^p_{p}$ of the vector, and includes
the updated vector clock value with its message.  When it
receives a message, it increments $t^p_{p}$ and sets $t^p_q$ for each
$q$ to the max
max of its previous value and the value of $t_{q}$ piggybacked on the
message.  We define $\VC(e)$ were $e$ is an event $p$ to be the value of $t^p$ at the end of
event $e$. We define $\VC(e) ≤ \VC(e')$, where $\VC(e)$ is the
value of the vector clock for $e$, if $\VC(e)_{i} ≤ \VC(e')_{i}$
for all $i$.

\begin{theorem}
\label{theorem-vector-clock}
 Fix a schedule $S$; then for any $e$, $e'$, $VC(e) < VC(e')$ if and
 only if $e \happensBefore{S} e'$.
\end{theorem}
\begin{proof}
    We'll start by showing that for any event $e$ at a process $p$,
    the value of $\VC(e)_q$ for any $q≠p$ is equal to the max $\VC(e')_q$ for any
    event $e'$ of $q$ such that $e' \happensBefore{S} e$, or $0$ if there is no such
    $e'$.

    The proof is by induction on the schedule so far.

    If $e$ is a local event or a send event, then there is either no
    preceding event at the same process (and thus no event $e'$ of $q$
    with $e'\happensBefore{S} e$) and $\VC(e)_q = 0$ as required; or there is some
    preceding event $e''$ of $p$. Since $e''$ is the only immediate
    predecessor of $e'$ in $\happensBefore{S}$, if there is an event
    $e'$ of $q$ maximizing $\VC_(e')_q$ such that $e' \happensBefore{S} e$, $e'
    \happensBefore{S} e''$ and so $\VC(e)_q = \VC(e'')_q = \VC(e')_q$
    as required.

    Alternatively, if $e$ is a receive event, then there is at most
    one immediately preceding event $e_1$ of the same process and a send event $e_2$
    of the same message such that $\VC(e)_q = \max(\VC(e_1)_q,
    \VC(e_2),q)$. Since any event $e'$ of $q$ with $e'
    \happensBefore{S} e$ has either $e' \happensBefore{S} e_1$ or $e
    \happensBefore{S} e_2$, we can apply the induction hypothesis to
    both $e_1$ and $e_2$ and then observe that $\VC(e)_q =
    \max(\VC(e_1)_q, \VC(e_2)_q)$ satisfies the requirements of the
    induction hypothesis.

    Given this characterization of $\VC(e)_q$,
    the if part follows immediately from the update rules for the vector
    clock. For events $e \happensBefore{S} e'$ of the same process, observe that both
    update rules strictly increase that process's clock, so $\VC(e) <
    \VC(e')$. Similarly the update rule for receiving a message
    implies that $\VC(e) < \VC(e')$ when $e$ and $e'$ are matching
    send and receive events, with the minor issue that we do need to
    use the observation above to verify that $e_p < e'_p$ for the
    receiver $p$.

    For the only if part, suppose $e$ does not happen-before $e'$.
Then $e$ and $e'$ are events of distinct processes $p$ and $p'$.  For
$\VC(e) < \VC(e')$ to hold, we must have $\VC(e)_{p} ≤ \VC(e')_{p}$;
but as shown above, this can occur only if 
$e \happensBefore{S} e'$.
\end{proof}

\section{Consistent snapshots}
\label{section-chandy-lamport}

A \concept{consistent snapshot} of a message-passing computation is a
description of the states of the processes (and possibly messages in
transit, but we can reduce this down to just states by keeping logs of
messages sent and received) that gives the global configuration at
some instant of a schedule that is a consistent reordering of the real
schedule (a \concept{consistent cut} in the terminology of
\cite[\S6.1.2]{AttiyaW2004}.  Without shutting down the protocol before taking a snapshot this is the about the best we can hope for in a message-passing system.

Logical clocks can be used to obtain consistent snapshots: pick some
logical clock time and have each process record its state at this time
(i.e.,
immediately after its last step before the time or immediately before
its first step after the time).  We have already argued that the logical
clock gives a consistent reordering of the original schedule, so the
set of values recorded is just the configuration at the end of an
appropriate prefix of this reordering.  In other words, it's a consistent snapshot.

\newData{\CLsnap}{snap}
\newData{\CLmarker}{marker}

If we aren't building logical clocks anyway, there is a simpler
consistent snapshot algorithm due to Chandy and
Lamport~\cite{ChandyL1985}.  Here some central initiator broadcasts a
\CLsnap message, and each process records its state and immediately
forwards the \CLsnap message to all neighbors when it first receives a
snap message.  To show that the resulting configuration is a
configuration of some consistent reordering, observe that (with FIFO
channels) no process receives a message before receiving \CLsnap that
was sent after the sender sent \CLsnap: thus causality is not violated
by lining up all the pre-snap operations before all the post-snap
ones.\footnote{If FIFO channels are not available, they can be
simulated in the absence of failures by adding a sequence number to each outgoing
message on a given channel, and processing messages at the recipient
only when all previous messages have been processed.}

The full Chandy-Lamport algorithm adds a second \CLmarker message that
is used to sweep messages in transit out of the communications
channels, which avoids the need to keep logs if we want to reconstruct
what messages are in transit (this can also be done with the logical
clock version).  The idea is that when a process records its state
after receiving the \CLsnap message, it issues a \CLmarker message on
each outgoing channel.  For incoming channels, each process records all messages received between the snapshot and receiving a marker message on that channel (or nothing if it receives \CLmarker before receiving \CLsnap).  A process only reports its value when it has received a marker on each channel.  The \CLmarker and \CLsnap messages can also be combined if the broadcast algorithm for \CLsnap resends it on all channels anyway, and a further optimization is often to piggyback both on messages of the underlying protocol if the underlying protocol is chatty enough.

Note that Chandy-Lamport is equivalent to the logical-time snapshot using Lamport clocks, if the snap message is treated as a message with a very large timestamp.  For Neiger-Toueg-Welch clocks, we get an algorithm where processes spontaneously decide to take snapshots (since Neiger-Toueg-Welch clocks aren't under the control of the snapshot algorithm) and delay post-snapshot messages until the local snapshot has been taken.  This can be implemented as in Chandy-Lamport by separating pre-snapshot messages from post-snapshot messages with a marker message, and essentially turns into Chandy-Lamport if we insist that a process advance its clock to the snapshot time when it receives a marker.

\subsection{Property testing}

Consistent snapshots are in principle useful for debugging (since one
can gather a consistent state of the system without being able to talk
to every process simultaneously), and in practice are mostly used for
detecting \index{property!stable}\indexConcept{stable property}{stable
properties} of the system.  Here a stable property is some predicate
on global configurations that remains true in any successor to a
configuration in which it is true, or (bending the notion of
properties a bit) functions on configurations whose values don't
change as the protocol runs.  Typical examples are quiescence and its
evil twin, deadlock.  More exotic examples include total money supply
in a banking system that cannot create or destroy money, or the fact
that every process has cast an irrevocable vote in favor of some
proposal or advanced its Neiger-Toueg-Welch-style clock past some threshold.

The reason we can test such properties using consistent snapshot is
that when the snapshot terminates with value $C$ in some configuration
$C'$, even though $C$ may never have occurred during the actual
execution of the protocol, there \emph{is} an execution which leads
from $C$ to $C'$.  So if $P$ holds in $C$, stability means that it
holds in $C'$.  

Naturally, if $P$ doesn't hold in $C$, we can't say much.  So in this
case we re-run the snapshot protocol and hope we win next time.  If
$P$ eventually holds, we will eventually start the snapshot protocol
after it holds and obtain a configuration (which again may not
correspond to any global configuration that actually occurs) in which
$P$ holds.

\myChapter{Synchronizers}{2026}{}
\label{chapter-synchronizers}

\indexConcept{synchronizer}{Synchronizers} simulate an execution of a
failure-free synchronous system in a failure-free asynchronous system.
See \cite[Chapter 11]{AttiyaW2004} or \cite[Chapter 16]{Lynch1996} for a detailed (and rigorous) presentation.

\section{Definitions}

Formally, a synchronizer sits between the underlying network and the processes and does one of two things:
\begin{itemize}
 \item A \index{synchronizer!global}\concept{global synchronizer} guarantees that no process
 receives a message from round $r$ until \emph{all processes} have sent
 their messages for round $r$.
 \item A \index{synchronizer!local}\concept{local synchronizer} guarantees that no process
 receives a message from round $r$ until \emph{all of that process's
 neighbors} have sent their messages for round $r$.
\end{itemize}

In both cases, the synchronizer packages all the incoming round $r$
messages $m$ for a single process together and delivers them as a single
action $\Recv(p, m, r)$.  Similarly, a
process is required to hand over all of its outgoing round-$r$
messages to the synchronizer as a single action $\Send(p, m, r)$—this prevents a process from changing its mind and sending an
extra round-$r$ message or two.  It is easy to see that the global
synchronizer produces executions that are effectively
indistinguishable from synchronous executions, assuming that a
synchronous execution is allowed to have some variability in exactly
when within a given round each process does its thing.  The local
synchronizer only guarantees an execution that is locally
indistinguishable from an execution of the global synchronizer: an
individual process can't tell the difference, but comparing actions at
different (especially widely separated) processes may reveal some
process finishing round $r+1$ while others are still stuck in round
$r$ or earlier.  Whether this is good enough depends on what you want:
it's bad for coordinating simultaneous missile launches, but may be
just fine for adapting a synchronous message-passing algorithm (as
with distributed breadth-first search as described in
§\ref{section-distributed-BFS-synchronizer}) to an asynchronous system, if we only care about the final states of the processes and not when precisely those states are reached.

Formally, the relation between global and local synchronization is
described by the following lemma:
\begin{lemma}
\label{lemma-local-synchronizer}
For any schedule $S$ of a locally synchronous execution, there is a
schedule $S'$ of a globally synchronous execution such that $S|p =
S'|p$ for all processes $p$.
\end{lemma}
\begin{proof}
Essentially, we use the same \concept{happens-before} relation as in
Chapter~\ref{chapter-logical-clocks}, and the fact that if a schedule
$S'$ is a causal shuffle of another schedule $S$ (i.e., a permutation
of $T$ that preserves causality), then $S'|p = S|p$ for all $p$
(Lemma~\ref{lemma-happens-before}).

Given a schedule $S$, consider a schedule $S'$ in which the events are
ordered first by increasing round and then by putting all sends before
receives.  This ordering is consistent with $\happensBefore{S}$, so it's
a causal shuffle of $S$ and $S'|p = S|p$.  But it is globally
synchronized, because no round $r$ operation ever happens before a
round $(r-1)$ operation.
\end{proof}

\section{Implementations}
\label{section-synchronizer-implementations}

Here we describe several implementations of synchronizers.
All of them give at least local synchrony.
One of them, the beta
synchronizer (§\ref{section-beta-synchronizer}), also gives
global synchrony.  

The names were chosen by their inventor,
Baruch Awerbuch~\cite{Awerbuch1985}.
The main difference between them is the mechanism used to determine
when round-$r$ messages have been delivered.

In the \index{synchronizer!alpha}\concept{alpha synchronizer}, every node sends a message to every
neighbor in every round (possibly a dummy message if the underlying
protocol doesn't send a message); this allows the receiver to detect
when it's gotten all its round-$r$ messages (because it expects to get a message from every neighbor) but may produce huge blow-ups in message complexity in a dense graph.

In the \index{synchronizer!beta}\concept{beta synchronizer}, messages are acknowledged by their receivers (doubling the message complexity), so the senders can detect when all of their messages are delivered.  But now we need a centralized mechanism to collect this information from the senders and distribute it to the receivers, since any particular receiver doesn't know which potential senders to wait for.  This blows up time complexity, as we essentially end up building a global synchronizer with a central leader.

The \index{synchronizer!gamma}\concept{gamma synchronizer} combines the two approaches at different levels to obtain a trade-off between messages and time that depends on the structure of the graph and how the protocol is organized.

Details of each synchronizer are given below.

\subsection{The alpha synchronizer}
\label{section-alpha-synchronizer}

The alpha synchronizer uses local information to construct a local
synchronizer.
In round $r$, the synchronizer at $p$ sends $p$'s message (tagged with
the round number) to each neighbor $p'$ or $\NoMsg(r)$ if it has no messages.  When it collects a
message or \NoMsg from each neighbor for round $r$, it delivers all the messages.  It's easy to see that this satisfies the local synchronization specification.

This produces no change in time but may drastically increase message
complexity because of all the extra \NoMsg messages flying around.
For a synchronous protocol that runs in $T$ rounds with $M$ messages, the same
protocol running with the alpha synchronizer will still run in $T$ time
units, but the message complexity will go up to $T\cdot\card*{E}$
messages, or worse if the original algorithm doesn't detect
termination.

\subsection{The beta synchronizer}
\label{section-beta-synchronizer}

\newData{\betaAck}{ack}
\newData{\betaOK}{OK}
\newData{\betaGo}{go}

The beta synchronizer centralizes detection of message delivery using a rooted directed
spanning tree (previously constructed).  When $p'$ receives a
round-$r$ message from $p$, it responds with $\betaAck(r)$.  When $p$
collects an \betaAck for all the messages it sent plus an \betaOK from
all of its children, it sends \betaOK to its parent.  When the root
has all the \betaAck and \betaOK messages it is expecting, it
broadcasts \betaGo.  Receiving \betaGo makes $p$ deliver the queued
round-$r$ messages.

This works because in order for the root to issue \betaGo, every
round-$r$ message has to have gotten an acknowledgment, which means
that all round-$r$ messages are waiting in the receivers' buffers to
be delivered.  For the beta synchronizer, message complexity for one
round increases
slightly from $M$ to $2M+2(n-1)$, but time complexity goes up by a factor proportional to the depth of the tree.

\subsection{The gamma synchronizer}

The gamma synchronizer combines the alpha and beta synchronizers to try to get low blowups on both time complexity and message complexity.  The essential idea is to cover the graph with a spanning forest and run beta within each tree and alpha between trees.  Specifically:

\newData{\GammaReady}{ready}
\newData{\GammaGo}{go}

\begin{itemize}
 \item Every message in the underlying protocol gets $\Ack$ed (including messages that pass between trees).
 \item When a process has collected all of its outstanding round-$r$
 $\Ack$s, it sends \OK up its tree.
 \item When the root of a tree gets all {\Ack}s and \OK, it sends \GammaReady to the roots of all adjacent trees (and itself).  Two trees are adjacent if any of their members are adjacent.
 \item When the root collects \GammaReady from itself and all adjacent roots, it broadcasts \GammaGo through its own tree.
\end{itemize}

As in the alpha synchronizer, we can show that no root issues \GammaGo unless it and all its neighbors issue \GammaReady, which happens only after both all nodes in the root's tree and all their neighbors (some of whom might be in adjacent trees) have received acks for all messages.  This means that when a node receives \GammaGo it can safely deliver its bucket of messages.

Message complexity is comparable to the beta synchronizer assuming
there aren't too many adjacent trees: $2M$ messages for sends and
acks, plus $O(n)$ messages for in-tree communication, plus
$O(E_{\text{roots}})$ messages for root-to-root communication.  Time complexity per synchronous round is proportional to the depth of the trees: this includes both the time for in-tree communication, and the time for root-to-root communication, which might need to be routed through leaves.

In a particularly nice graph, the gamma synchronizer can give costs
comparable to the costs of the original synchronous algorithm.  An
example in \cite{Lynch1996} is a ring of $k$-cliques, where we build a
tree in each clique and get $O(1)$ time blowup and $O(n)$ added
messages.  This is compared to $O(n/k)$ time blowup for the beta
synchronizer and
$O(k)$ message blowup (or worse) for the alpha synchronizer.  Other graphs may favor
tuning the size of the trees in the forest toward the alpha or beta
ends of the spectrum, e.g., if the whole graph is a clique (and we
didn't worry about contention issues), we might as well just use beta and get
$O(1)$ time blowup and $O(n)$ added messages.

\section{Applications}

See \cite[\S11.3.2]{AttiyaW2004} or \cite[\S16.5]{Lynch1996}.  The one
we have seen is distributed breadth-first search, where the two
asynchronous algorithms we described in
Chapter~\ref{chapter-distributed-BFS} were essentially the synchronous
algorithms with the beta and alpha synchronizers embedded in them.
But what synchronizers give us in general is the ability to forget about problems resulting from asynchrony provided we can assume no failures (which may be a very strong assumption) and are willing to accept a bit of overhead.

\section{Limitations of synchronizers}

Here we show some lower bounds on synchronizers, justifying our
previous claim that failures are trouble and showing that global
synchronizers are necessarily slow in a high-diameter network.

\subsection{Impossibility with crash failures}

These synchronizers all fail badly if some process crashes. In the
$α$ synchronizer, the system slowly shuts down as a wave of
waiting propagates out from the dead process.  In the $β$
synchronizer, the root never gives the green light for the next round.
The $\gamma$ synchronizer, true to its hybrid nature, fails in a way
that is a hybrid of these two disasters.

This is unavoidable in the basic asynchronous model, although we don't
have all the results we need to prove this yet.  The idea is
that if we are in a synchronous system with crash failures, it's
possible to solve \concept{agreement}, the problem of getting all the
processes to agree on a bit (see
Chapter~\ref{chapter-synchronous-agreement}).  But it's not
possible to solve this problem in an asynchronous system with even one
crash failure (see Chapter~\ref{chapter-FLP}).  Since a
synchronous-with-crash-failure agreement protocol on top of a
fault-tolerant synchronizer would give a solution to an unsolvable
problem, the element of this stack that we don't know an algorithm for
must be the one we can't do.  Hence there are no fault-tolerant synchronizers.

We'll see more examples of this trick of showing that a
particular simulation is impossible because it would allow us to
violate impossibility results later, especially when we start looking
at the strength of shared-memory objects
in Chapter~\ref{chapter-wait-free-hierarchy}.

\subsection{Unavoidable slowdown with global synchronization}
\label{section-session-problem}

The \concept{session problem}~\cite{ArjomandiFL1983} gives a lower bound on the speed
of a global synchronizer, or more generally on
any protocol that tries to approximate synchrony in a certain sense.
Recall that in a global synchronizer, our
goal is to produce a simulation that looks synchronous \emph{from the
outside}; that is, that looks synchronous to an observer that can see
the entire schedule.  In contrast, a local synchronizer produces a
simulation that looks synchronous \emph{from the inside}—the resulting execution is indistinguishable from a synchronous execution to any of the processes, but an outside observer can see that different processes execute different rounds at different times.  
The global synchronizer we've seen takes more time than a local
synchronizer; the session problem shows that this is necessary.

In our description, we will mostly follow \cite[\S6.2.2]{AttiyaW2004}.

A solution to the session problem is an asynchronous protocol in which
each process repeatedly executes some \concept{special
action}\index{action!special}.  Our
goal is to guarantee that these special actions group into $s$
\indexConcept{session}{sessions}, where a session is an interval of time in which every process executes at least one special action.  We also want the protocol to terminate: this means that in every execution, every process executes a finite number of special actions.

A synchronous system can solve this problem trivially in $s$ rounds:
each process executes one special action per round.  For an
asynchronous system, a lower bound of Attiya and
Mavronicolas~\cite{AttiyaM1994} (based on an earlier bound of
Arjomandi, Fischer, and Lynch~\cite{ArjomandiFL1983}, who defined the problem in a slightly different
communication model), shows that if the diameter of the network is
$D$, any solution to the $s$-session problem takes
$(s-1)D$ time or more in the worst case.  The argument is based on
reordering events in any synchronous execution that takes less time
to produce fewer than $s$
sessions, using the happens-before relation described in
Chapter~\ref{chapter-logical-clocks}.

We now give an outline of the proof that this is expensive.
(See \cite[\S6.2.2]{AttiyaW2004} for the real proof.)

Fix some algorithm $A$ for solving the $s$-session problem, and
suppose that its worst-case time complexity is $(s-1)D$ or less.
Consider some synchronous execution of $A$ (that is, one where the
adversary scheduler happens to arrange the schedule to be synchronous)
that takes $(s-1)D$ rounds or less.  Divide this execution into two
segments: an initial segment $γ$ that includes all rounds with
special actions, and a suffix $δ$ that includes any extra rounds
where the algorithm is still floundering around.  We will mostly
ignore $δ$, but we have to leave it in to allow for the possibility
that whatever is happening there is important for the algorithm to
work (say, to detect termination).

We now want to perform a causal shuffle on $γ$ that leaves it with
only $s-1$ sessions.  Because causal shuffles don't affect time
complexity, this will give us a new bad execution $γ'δ$ that has only
$s-1$ sessions despite taking $(s-1)D$ time.

The first step is to chop $γ$ into 
$s-1$ segments $γ_{1},γ_{2}, \dots γ_{s-1}$ of at most $D$ rounds
each.  Because a message sent in round $i$ is not delivered until
round $i+1$, if we have a chain of $k$ messages, each of which
triggers the next, then if the first message is sent in round $i$, the
last message is
not delivered until round $i+k$.  If the chain has length $D$, its
events (including the initial send and the final delivery) span $D+1$
rounds $i, i+1, \dots, i+D$.  In this case the initial send and final
delivery are necessarily in different segments $γ_i$ and $γ_{i+1}$.

Now pick
processes $p$ and $q$ at distance $D$ from each other.
Then any chain of messages starting
at $p$ within some segment reaches $q$ after the end of the
segment.  It follows that for any events $e_{p}$ of $p$ and
$e_{q}$ of $q$ in the \emph{same} segment $γ_{i}$, 
$e_{p}\not\happensBefore{γδ}e_{q}$.  So there exists a causal
shuffle of $γ_{i}$ that puts all events of $p$ after all
events of $q$.\footnote{Proof: Because $e_p\not\happensBefore{γδ}e_q$, we can
    add $e_q < e_p$ for all events $e_q$ and $e_p$ in $γ_i$ and still
    have a partial order consistent with $\happensBefore{γδ}$.  Now apply topological sort to get the
shuffle.}
By a symmetrical argument, we can similarly put
all events of $q$ in a segment after all events of $p$ in the same
segment. In both cases the resulting schedule is indistinguishable by all processes from the original.

So now we apply these shuffles to each of the segments $γ_{i}$ in
alternating order: $p$ goes first in the odd-numbered segments
and $q$ goes first in the odd-numbered segments.
Let's write the shuffled version of $γ_i$ as $α_iβ_i$ for odd $i$ and
$β_i α_i$ for even $i$; in each case, $α_i$ contains only events of
$p$ and other processes that aren't $q$ and $β_i$ contains only events of
$q$ and other processes that aren't $p$.

When we put these alternating shuffles together, we get an execution
that looks like this example with $s-1 = 4$:
\begin{align*}
α_1 β_1 β_2 α_2 α_3 β_3 β_4 α_4 δ
\end{align*}

Now let's count sessions. Since a session includes special actions by
both $p$ and $q$, it can't lie entirely within $α$ intervals or $β$
intervals.
So any session has to span one of the points in the
schedule marked by slashes below:
\begin{align*}
α_1/ β_1 β_2/ α_2 α_3/ β_3 β_4/ α_4 δ
\end{align*}

There is one such point for each of our original $s-1$ intervals, so
we get at most $s-1$ sessions.

This means that any algorithm that runs
in time $(s-1)D$ in the worst case (here, the original synchronous
execution) can't guarantee to give $s$ sessions in all cases (it fails
in the shuffled asynchronous execution). Note that this is not quite the same as
saying that any execution with at least $s$ sessions must take
$(s-1)D$ time. Instead, we've shown that algorithm that guarantees we get at least
$s$ sessions sometimes takes more than $(s-1)D$ time, even though it
might sometimes use less time if it gets lucky.

\myChapter{Coordinated attack}{2026}{}
\label{chapter-coordinated-attack}
\label{chapter-two-generals}

(See also \cite[\S5.1]{Lynch1996}.)

The \concept{Two Generals} problem was the first widely-known
distributed consensus problem, described in 1978 by Jim
Gray~\cite[§5.8.3.3.1]{Gray1978}, although the same problem
previously appeared under a different name~\cite{AkkoyunluEH1975}.

The setup of the problem is that we have two generals on opposite sides of an enemy army, who must choose whether to attack the army or retreat.  If only one general attacks, his troops will be slaughtered.  So the generals need to reach agreement on their strategy.

To complicate matters, the generals can only communicate by sending
messages by (unreliable) carrier pigeon.  We also suppose that at some
point each general must make an irrevocable decision to attack or
retreat.  The interesting property of the problem is that if carrier
pigeons can become lost, there is no protocol that guarantees
agreement in all cases unless the outcome is predetermined (e.g., the
generals always attack no matter what happens).  The essential idea of
the proof is that any protocol that does guarantee agreement can be
shortened by deleting the last message; iterating this process
eventually leaves a protocol with no messages.

Adding more generals turns this into the \concept{coordinated attack}
problem, a variant of \concept{consensus}, but it doesn't make things
any easier.

\section{Formal description}
\label{section-two-generals-formalized}

To formalize this intuition, 
suppose that we have $n ≥ 2$ generals in a synchronous
system with unreliable channels—the set of messages received in
round $i+1$ is always a subset of the set sent in round $i$, but it
may be a proper subset (even the empty set).  Each general starts with an input 0 (retreat) or 1 (attack) and must output 0 or 1 after some bounded number of rounds.  The requirements for the protocol are that, in all executions:

\begin{description}
 \item[Agreement]\index{agreement} All processes output the same decision (0 or 1).
 \item[Validity]\index{validity} If all processes have the same input
 $x$, and no messages are lost, all processes produce output $x$.  (If
 processes start with different inputs or one or more messages are
 lost, processes can output 0 or 1 as long as they all agree.)
 \item[Termination]\index{termination} All processes terminate in a bounded number of
 rounds.\footnote{\indexConcept{bounded}{Bounded} means that there is a fixed
 upper bound on the length of any execution.  We could also demand
 merely that all processes terminate in a \emph{finite} number of rounds.
 In general, finite is a weaker requirement than bounded, but if the
 number of possible outcomes at each step is finite (as they are in
 this case), they're
 equivalent.  The reason is that if we build a tree of all
 configurations, each configuration has only finitely many
 successors, and the length of each path is finite, then 
 \concept{König's lemma} (see \wikipedia{Konig's_lemma})
 says that there are only finitely many paths.  So we can take the
 length of the longest of these paths as our fixed
 bound.~\cite[Lemma 3.1]{BorowskyG1997}
 }
\end{description}

Sadly, there is not protocol that satisfies all three conditions.  We
show this in the next section.

\section{Impossibility proof}
\label{section-two-generals-impossible}
\label{section-indistinguishability-proofs}

To show coordinated attack is impossible,\footnote{Without making
additional assumptions, always a caveat when discussing
impossibility.} we use an \concept{indistinguishability
proof}\index{proof!indistinguishability}.

The key steps of an indistinguishability proof usually look like this:
\begin{itemize}
 \item Show that execution $A$ is \concept{indistinguishable} from execution $B$
 for some process $p$, meaning that $p$ sees the same things (messages or
 operation results) in both executions.
 \item Observe that if $A$ is indistinguishable from $B$ for $p$, then because $p$
 can't tell which of these two possible worlds it is in, it returns
 the same output in both.
\end{itemize}

So far, pretty dull.  But now let's consider a chain of hypothetical executions $A
= A_{0} A_{1} \dots{} A_{k} = B$, where each $A_{i}$ is indistinguishable
from $A_{i+1}$ for some process $p_{i}$.
Suppose also that we are
trying to solve an agreement task, where every process must output the
same value.  Then since $p_{i}$ outputs the same value in $A_{i}$ and
$A_{i+1}$, every process outputs the same value in $A_{i}$ and
$A_{i+1}$.  By induction on $k$, every process outputs the same value
in $A$ and $B$, even though $A$ and $B$ may be very different executions.  

This gives us a tool for proving impossibility results for agreement: show that there is a path of indistinguishable executions between two executions that are supposed to produce different output.  Another way to picture this: consider a graph whose nodes are all possible executions with an edge between any two indistinguishable executions; then the set of output-0 executions can't be adjacent to the set of output-1 executions.  If we prove the graph is connected, we prove the output is the same for all executions.

For coordinated attack, we will show
that no protocol satisfies all of agreement, validity, and termination
using an indistinguishability argument.  The key idea is to construct
a path between the all-0-input and all-1-input executions with no message loss via intermediate executions that are indistinguishable to at least one process.

Let's start with $A = A_{0}$ being an execution in which all inputs
are 1 and all messages are delivered.  We'll build executions $A_{1},
A_{2},$ etc., by pruning messages.  Consider $A_{i}$ and let $m$ be
some message that is delivered in the last round in which any message
is delivered.  Construct $A_{i+1}$ by not delivering $m$.  Observe
that while $A_{i}$ is distinguishable from $A_{i+1}$ by the recipient
of $m$, on the assumption that $n ≥ 2$ there is some other process
that can't tell whether $m$ was delivered or not (the recipient can't
let that other process know, because no subsequent message it sends
are delivered in either execution).  Continue until we reach an
execution $A_{k}$ in which all inputs are 1 and no messages are sent.
Next, let $A_{k+1}$ through $A_{k+n}$ be obtained by
changing one input at a time from 1 to 0; each such execution is
indistinguishable from its predecessor by any process whose input
didn't change. 
Finally, construct $A_{k+n}$ through $A_{k+n+k'}$ by adding back
messages in the reverse process used for $A_0$ through $A_k$; note
that this might not result in exactly $k$ new messages, because the
number of messages might depend on the inputs.  This
gets us to an execution $A_{k+n+k'}$ in which all processes have input
$0$ and no messages are lost.
If agreement holds, then the indistinguishability of adjacent
executions to some process means that the common
output in $A_{0}$ is the same as in $A_{k+n+k'}$.  But validity requires
that $A_{0}$ outputs 1 and $A_{k+n+k'}$ outputs 0: so either agreement
or validity is violated in some execution.

\section{Randomized coordinated attack}
\label{section-randomized-coordinated-attack}

So we now know that we can't solve the coordinated attack problem.  But maybe we want to solve it anyway.  The solution is to change the problem.

\index{coordinated attack!randomized}
\indexConcept{randomized coordinated attack}{Randomized coordinated attack} is like
standard coordinated attack, but with less coordination.
Specifically, we'll allow the processes to flip coins to decide what
to do, and assume that the communication pattern (which messages get
delivered in each round) is fixed and independent of the coin-flips.
This corresponds to assuming an 
\index{adversary!oblivious}
\concept{oblivious adversary}
that can't see what is going on at all
or perhaps 
a 
\index{adversary!content-oblivious}
\concept{content-oblivious adversary} that
can only see where messages are being sent but not the contents of the
messages.  
We'll also relax the agreement property to only hold with some high probability:

\begin{description}
 \item[Randomized agreement]\index{agreement!randomized}\index{randomized agreement}
 For any adversary $A$, the probability that some process decides $0$ and some
 other process decides $1$ given $A$ is at most $ε$.
\end{description}

Validity and termination are as before.

\subsection{An algorithm}

Here's an algorithm that gives $ε = 1/r$.  (See
\cite[\S5.2.2]{Lynch1996} for details or~\cite{VargheseL1992} for the
original version.)  A simplifying assumption is
that network is complete, although a strongly-connected network with
$r$ greater than or equal to the diameter also works.

\newData{\RandAttackLevel}{level}
\newData{\RandAttackKey}{key}

\begin{itemize}
 \item First part: tracking information levels
\begin{itemize}
  \item Each process tracks its ``information level,'' initially $0$.
  The state of a process consists of a vector of (input,
  information-level) pairs for all processes in the system.  Initially
  this is (my-input, 0) for itself and $(⊥, -1)$ for everybody else.
  \item Every process sends its entire state to every other process in every round.
  \item Upon receiving a message $m$, process $i$ stores any inputs
  carried in $m$ and, for each process $j$, sets
  $\RandAttackLevel_{i}[j]$ to
  $\max(\RandAttackLevel_{i}[j], \RandAttackLevel_{m}[j])$.  It then
  sets its own information level to
  $\min_{j}(\RandAttackLevel_{i}[j])+1$.
\end{itemize}
 \item Second part: deciding the output
\begin{itemize}
  \item Process 1 chooses a random key value uniformly in the range
  $[1,r]$.
  \item This key is distributed along with 
  $\RandAttackLevel_i[1]$, 
  so that every process with $\RandAttackLevel_i[1] ≥ 0$
  knows the key.
  \item A process decides $1$ at round $r$ if and only if it knows the
  key, its information level is greater than or equal to the key, and all inputs are 1.
\end{itemize}
\end{itemize}

\subsection{Why it works}
\begin{description}
 \item[Termination] Immediate from the algorithm.
 \item[Validity] 
\begin{itemize}
  \item If all inputs are 0, no process sees all 1 inputs (technically requires an invariant that processes' non-null views are consistent with the inputs, but that's not hard to prove.)
  \item If all inputs are 1 and no messages are lost, then the information
  level of each process after $k$ rounds is $k$ (prove by induction)
  and all processes learn the key and all inputs (immediate from first
  round).  So all processes decide $1$.
\end{itemize}
 \item[Randomized Agreement] 
\begin{itemize}
  \item First prove a lemma: Define $\RandAttackLevel_{i}^{t}[k]$ to
  be the value of $\RandAttackLevel_{i}[k]$ after $t$ rounds.  Then
  for all $i, j, k, t$, (1) $\RandAttackLevel_{i}[j]^{t} ≤
  \RandAttackLevel_{j}[j]^{t-1}$ and (2)
  $\abs*{\RandAttackLevel_{i}[k]^{t} - \RandAttackLevel_{j}[k]^{t} }
  ≤ 1$.  As always, the proof is by induction on rounds.  Part (1) is easy and boring so we'll skip it.  For part (2), we have:
\begin{itemize}
   \item After $0$ rounds, $\RandAttackLevel_{i}^{0}[k] =
   \RandAttackLevel_{j}^{0}[k] = -1$ if neither $i$ nor $j$ equals
   $k$; if one of them is $k$, we have $\RandAttackLevel_{k}^{0}[k] =
   0$, which is still close enough.
   \item After $t$ rounds, consider $\RandAttackLevel_{i}^{t}[k] -
   \RandAttackLevel_{i}^{t-1}[k]$ and similarly
   $\RandAttackLevel_{j}^{t}[k] - \RandAttackLevel_{j}^{t-1}[k]$.
   It's not hard to show that each can jump by at most 1.  If both
   deltas are $+1$ or both are 0, there's no change in the difference
   in views and we win from the induction hypothesis.  So the
   interesting case is when $\RandAttackLevel_{i}[k]$ stays the same
   and $\RandAttackLevel_{j}[k]$ increases or vice versa.
   \item There are two ways for $\RandAttackLevel_{j}[k]$ to increase:
\begin{itemize}
    \item If $j \ne k$, then $j$ received a message from some $j'$
    with $\RandAttackLevel_{j'}^{t-1}[k] >
    \RandAttackLevel_{j}^{t-1}[k]$.  From the induction hypothesis,
    $\RandAttackLevel_{j'}^{t-1}[k] ≤ \RandAttackLevel_{i}^{t-1}[k]
    + 1 = \RandAttackLevel_{i}^{t}[k]$.  So we are happy.
    \item If $j = k$, then $j$ has $\RandAttackLevel_{j}^{t}[j] = 1 +
    \min_{k \ne j} \RandAttackLevel_{j}^{t}[k] ≤ 1 +
    \RandAttackLevel_{j}^{t}[i] ≤ 1 + \RandAttackLevel_{i}^{t}[i]$.  Again we are happy.
\end{itemize}
\end{itemize}
  \item Note that in the preceding, the key value didn't figure in; so
  everybody's \RandAttackLevel at round $r$ is independent of the key.
  \item So now we have that $\RandAttackLevel_{i}^{r}[i]$ is in $\{ \ell,
  \ell+1 \}$, where $\ell$ is some fixed value uncorrelated with the key.
  The only way to get some process to decide 1 while others decide 0
  is if $\ell+1 ≥ \RandAttackKey$ but $\ell < \RandAttackKey$.  (If
  $\ell = 0$, a process at this level doesn't know $\RandAttackKey$,
  but it can still reason that $0 < \RandAttackKey$ since
  $\RandAttackKey$ is in $[1,r]$.)  This can only occur if
  $\RandAttackKey = \ell+1$, which occurs with probability at most
  $1/r$ since $\RandAttackKey$ was chosen uniformly.
\end{itemize}
\end{description}

\subsection{Almost-matching lower bound}
\label{section-randomized-coordinated-attack-lower-bound}

The bound on the probability of disagreement in the previous algorithm
is almost tight. Varghese and Lynch~\cite{VargheseL1992} show that no synchronous
algorithm can get a probability of disagreement less than
$\frac{1}{r+1}$, using a stronger validity condition
that requires that the processes
output $0$ if any input is $0$.  This is a natural assumption for
database commit, where we don't want to commit if any process wants to
abort.  We restate their result below:
\begin{theorem}
\label{theorem-randomized-coordinated-attack-lower-bound}
For any synchronous algorithm for randomized coordinated attack that runs in $r$
rounds that satisfies the additional condition that all non-faulty processes
decide $0$ if any input is $0$,
$\Pr[\text{disagreement}] ≥ 1/(r+1)$.
\end{theorem}
\begin{proof}
Let $ε$ be the bound on the probability of disagreement.
Define $\RandAttackLevel_{i}^{t}[k]$ as in the previous algorithm (whatever the real algorithm is doing).
We'll show $\Pr[\text{$i$ decides $1$}] ≤ ε
\cdot (\RandAttackLevel_{i}^{r}[i] + 1)$, by induction on
$\RandAttackLevel_{i}^{r}[i]$.
\begin{itemize}
   \item If $\RandAttackLevel_{i}^{r}[i] = 0$, the real execution is
   indistinguishable (to $i$) from an execution in which some other
   process $j$ starts with 0 and receives no messages at all.  In that
   execution, $j$ must decide 0 or risk violating the strong validity
   assumption. 
   So $i$
   decides $1$ with probability at most $ε$ (from the disagreement bound).
   \item If $\RandAttackLevel_{i}^{r}[i] = k > 0$, the real execution
   is indistinguishable (to $i$) from an execution in which some other
   process $j$ only reaches level $k-1$ and thereafter receives no
   messages.  From the induction hypothesis, $\Pr[\text{$j$ decides
   $1$}] ≤ ε k$ in that pruned execution, and so
   $\Pr[\text{$i$ decides $1$}] ≤ ε(k+1)$ in the pruned
   execution.  But by indistinguishability, we also have
   $\Pr[\text{$i$ decides
   $1$}] ≤ ε(k+1)$ in the original execution.
\end{itemize}

Now observe that in the all-1 input execution with no messages lost,
$\RandAttackLevel_{i}^{r}[i] = r$ and $\Pr[\text{$i$ decides $1$}] =
1$ (by validity).  So $1 ≤ ε(r+1)$, which implies $ε
≥ 1/(r+1)$.
\end{proof}

\myChapter{Synchronous agreement}{2026}{}
\label{chapter-synchronous-agreement}

\index{synchronous agreement}
\index{agreement!synchronous}
\index{consensus!synchronous}
Here we'll consider synchronous agreement algorithm with stopping
failures, where a process stops dead at some point, sending and receiving no further
messages.  
We'll also consider Byzantine failures,
where a process deviates from its programming by sending arbitrary
messages, but mostly just to see how crash-failure algorithms hold up;
for algorithms designed specifically for a Byzantine model, see
Chapter~\ref{chapter-Byzantine-agreement}.

If the model has communication failures instead,
we have the coordinated attack problem from
Chapter~\ref{chapter-coordinated-attack}.  

\section{Problem definition}
\label{section-synchronous-agreement-problem}

We use the usual synchronous model with $n$ processes with binary inputs and binary
outputs.  Up to $f$ processes may fail at some point; when a process
fails, one or
one or more of its outgoing messages are lost in the round of failure and all outgoing
messages are lost thereafter.  

There are two variants on the problem, depending on whether we want a
useful algorithm (and so want strong conditions to make our algorithm
more useful) or a lower bound (and so want weak conditions to make our
lower bound more general).
For algorithms, we will ask for these 
conditions to hold:
\begin{description}
 \item[Agreement] \index{agreement}All non-faulty processes decide the same value.
 \item[Validity] \index{validity}If all processes start with the same
     input, all non-faulty processes decide it.
 \item[Termination] \index{termination}All non-faulty processes eventually decide.
\end{description}

For lower bounds, we'll replace validity with \concept{non-triviality}
(often called validity in the literature):
\begin{description}
 \item[Non-triviality] There exist failure-free executions $A$ and $B$ that produce different outputs.
\end{description}

Non-triviality follows from validity but doesn't imply validity; for
example, a non-trivial algorithm might have the property that if all
non-faulty processes start with the same input, they all decide
something else.

In 
§\ref{section-synchronous-agreement-flooding}, we'll show that a simple algorithm gives
agreement, termination, and validity with $f$ failures using $f+1$
rounds.  We'll then show in §\ref{section-synchronous-agreement-lower-bound}
that non-triviality, agreement, and
termination imply that $f+1$ rounds is the best possible.
In Chapter~\ref{chapter-Byzantine-agreement}, we'll show that the
agreement is still possible in $f+1$ rounds even if faulty processes
can send arbitrary messages instead of just crashing, but only if the
number of faulty processes is strictly less than $n/3$.

\section{Solution using flooding}
\label{section-synchronous-agreement-flooding}

The flooding algorithm, due to Dolev and Strong~\cite{DolevS1983}
gives a straightforward solution to synchronous agreement for the
crash failure case.  It runs in $f+1$ rounds assuming $f$ crash
failures. The algorithm given here is a gross simplification of Dolev
and Strong's original algorithm, which solves the harder problem of
authenticated Byzantine agreement. (This algorithm is also described in more detail in
\cite[\S5.1.3]{AttiyaW2004} or \cite[\S6.2.1]{Lynch1996}.)

Each process keeps a set of
$\Tuple{\DataSty{id}, \DataSty{input}}$ pairs, initially just $\{(\DataSty{myId},
\DataSty{myInput})\}$.  At
round $r$, I broadcast my set to everybody and take the union of my set
and all sets I receive.  At round $f+1$, I decide on $f(S)$, where $f$
is some fixed function from sets of process-input pairs to outputs
that picks some input in $S$:
for example, 
$f$ might take the input with the smallest process-id attached to it, take the max of all known input values, or take the majority of all known input values.
\begin{lemma}
\label{lemma-synchronous-agreement-flooding-gives-same-set}
After $f+1$ rounds, all non-faulty processes have the same set.
\end{lemma}
\begin{proof}
Let $S_{i}^{r}$ be the set stored by process $i$ after $r$ rounds.  What
we'll really show is that if there are no failures in round $k$, then
$S_{i}^{r} = S_{j}^{r} = S_{i}^{k+1}$ for all $i$, $j$, and $r > k$.
To show this, observe that no faults in round $k$ means that all
processes that are still alive at the start of round $k$ send their
message to all other processes.  Let $L$ be the set of live processes
in round $k$.  At the end of round $k$, for $i$ in $L$ we have
$S_{i}^{k+1} = \bigcup_{j\in{}L} S_{j}^{k} = S$.  Now we'll consider
some round $r = k+1+m$ and show by induction on $m$ that $S_{i}^{k+m}
= S$; we already did $m = 0$, so for larger $m$ notice that all
messages are equal to $S$ and so $S_{i}^{k+1+m}$ is the union of a
whole bunch of $S$'s.  So in particular we have $S_{i}^{f+1} = S$
(since some failure-free round occurred in the preceding $f+1$ rounds)
and everybody decides the same value $f(S)$.
\end{proof}

\subsection{Authenticated version}

Flooding depends on being able to trust second-hand descriptions of
values; it may be that process 1 fails in round 0 so that only process
2 learns its input.  If process 2 can suddenly tell 3 (but nobody
else) about the input in round $f+1$—or worse, tell a different
value to 3 and 4—then we may get disagreement.

Usually we assume that we don't have access to cryptography, but if we
include an authentication mechanism that allows processes to attach
unforgeable signatures to messages, then the full version of the
Dolev-Strong algorithm solves agreement in $f+1$ even with $f$
\concept{Byzantine} faults, where a process can send any messages it
likes regardless of the protocol.  The idea is that instead of sending
around unauthenticated input values, I send around input values that
are authenticated by a sequence of signatures, one for each process
that forwarded it. So a value $v_1$ that started as the input to
process $p_1$ and reached me via processes $p_2$ and $p_3$ might
arrive in a message as $\Tuple{v_1, 123, S_3(S_2(S_1(v_1)))}$, giving
the value, the path it reached me by, and a nested sequence of
signatures allowing me to verify that it did in fact travel this path.

To avoid mischief, a process will accept in round $r$ only a message
that appears to have traveled a path involving $f+1$ processes, and
will only resend values it accepts. We can limit message complexity by
having each process resend only the first copy of each value it
accepts, and only to processes that are not already listed in the
history.

We now have the property that any value a non-faulty process accepts
in round $f+1$ passed through $f+1$ processes, including at least one
non-faulty process. That non-faulty process will have forwarded it to
all non-faulty processes. If a process accepts a value earlier than
round $f+1$, then it forwards it itself. In either case, if you and I
are both non-faulty, then I know that my eventual set $S$ is a subset
of yours. Since this holds in reverse as well, my $S$ equals your $S'$
and so we decide the same value $f(S)=f(S')$.

The intuition here is that if a Byzantine process can be forced to
show its work, Byzantine failures essentially reduce to omission
failures, since a non-faulty process can discard any incoming messages
that are obviously bogus. For the most part we will not assume that we
have the tools to do this, and that catching Byzantine processes will
require more careful protocols.

\section{Lower bound on rounds}
\label{section-synchronous-agreement-lower-bound}

Here we show that synchronous agreement requires at least $f+1$ rounds
if $f$ processes can fail.
This proof is modeled on
the one in \cite[§{}6.7]{Lynch1996} and works backwards from the
final state; for a proof of the same result that works in the opposite
direction, see \cite[§{}5.1.4]{AttiyaW2004}.
The original result (stated for Byzantine failures)
is due to Dolev and Strong~\cite{DolevS1983}, based on
a more complicated proof due to 
Fischer and Lynch~\cite{FischerL1982}; see the chapter
notes for Chapter 5 of~\cite{AttiyaW2004} for more discussion of the
history.

Note that unlike the algorithms in the preceding and following
sections, which provide validity, the lower bound applies even if we
only demand non-triviality.

Like the similar proof for coordinated attack
(§\ref{section-two-generals-impossible}), the proof uses
an \index{indistinguishability}indistinguishability argument.
But we have to construct a more complicated chain of intermediate
executions.

A \index{failure!crash}\concept{crash failure} at process $i$ means
that (a) in some round $r$, some or all of the messages sent by $i$
are not delivered, and (b) in subsequent rounds, no messages sent by
$i$ are delivered.  The intuition is that $i$ keels over dead in the
middle of generating its outgoing messages for a round.  Otherwise $i$
behaves perfectly correctly.  
A process that crashes at some point during an execution is called \concept{faulty}

We will show that if up to $f$ processes
can crash, and there are at least $f+2$ processes,\footnote{With only
    $f+1$ processes, we can solve agreement in $f$ rounds using
    flooding.  The idea is that either (a) at most $f-1$ processes
    crash, in which case the flooding algorithm guarantees agreement;
    or (b) exactly $f$ processes crash, in which case the one
    remaining non-faulty process agrees with itself.  So $f+2$
processes are needed for the lower bound to work, and we should be
suspicious of any lower bound proof that does not use this assumption.} then at least
$f+1$ rounds are needed (in some execution) for any algorithm that
satisfies agreement, termination, and non-triviality.
In particular, we will show that
if all executions run in $f$ or fewer rounds, then the
indistinguishability graph is connected; this implies non-triviality doesn't
hold, because (as in §\ref{section-two-generals-impossible}), two
adjacent states must decide the same value because of the agreement
property.\footnote{The same argument works with even a weaker version of
non-triviality that omits
the requirement that $A$ and $B$ are failure-free, but we'll
keep things simple.}

Now for the proof.  To simplify the argument, let's assume that all
executions terminate in exactly $f$ rounds (we can always have
processes send pointless chitchat to pad out short executions) and
that every processes sends a message to every other process in every
round where it has not crashed (more pointless chitchat).  Formally,
this means we have a sequence of rounds $0, 1, 2, \dots{}, f-1$ where
each process sends a message to every other process (assuming no
crashes), and a final round $f$ where all processes decide on a value (without sending any additional messages).

We now want to take any two executions $A$ and $B$ and show that both
produce the same output.  To do this, we'll transform $A$'s inputs into
$B$'s inputs one process at a time, crashing processes to hide the changes.  The problem is that just crashing the process whose input changed might change the decision value—so we have to crash later witnesses carefully to maintain indistinguishability all the way across the chain.

Let's say that a process $p$ \concept{crashes fully} in round $r$ if
it crashes in round $r$ and no round-$r$ messages from $p$ are delivered.  The \concept{communication pattern} of an execution describes which messages are delivered between processes without considering their contents—in particular, it tells us which processes crash and what other processes they manage to talk to in the round in which they crash.

With these definitions, we can state and prove a rather complicated
induction hypothesis:
\begin{lemma}
    \label{lemma-synchronous-agreement-lower-bound}
For any $f$-round protocol with $n≥f+2$ processes permitting up to $f$ crash failures; any
process $p$; and any execution $A$ in which at most one process crashes
per round in rounds $0\dots r-1$, $p$ crashes fully in round $r+1$,
and no other processes crash; there is a sequence of executions $A =
A_{0} A_{1} \dots{} A_{k}$ such that each $A_{i}$ is indistinguishable
from $A_{i+1}$ by some process, each $A_{i}$ has at most one crash per
round, and the communication pattern in $A_{k}$ is identical to $A$
except that $p$ crashes fully in round $r$.
\end{lemma}
\begin{proof}
By induction on $f-r$.
If $r = f$, we just crash $p$ in round $r$ and
nobody else notices.  For $r < f$, first crash $p$ in round $r$
instead of $r+1$, but deliver all of its round-$r$ messages anyway
(this is needed to make space for some other process to crash in round
$r+1$).  Then choose some message $m$ sent by $p$ in round $r$, and let
$p'$ be the recipient of $m$.  We will show that we can produce a
chain of indistinguishable executions between any execution in which
$m$ is delivered and the corresponding execution in which it is not.

If $r=f-1$, this is easy; only $p'$ knows whether $m$ has been
delivered, and since $n ≥ f+2$, there exists another non-faulty
$p''$ that can't distinguish between these two executions, since $p'$
sends no messages in round $f$ or later.  If $r < f-1$, we have to 
make sure $p'$ doesn't tell anybody about the missing message.

By the induction hypothesis, there is a
sequence of executions starting with $A$ and ending with $p'$ crashing
fully in round $r+1$, such that each execution is indistinguishable
from its predecessor.  Now construct the sequence 
\begin{align*}
A &\rightarrow{}
(\text{$A$ with $p'$ crashing fully in $r+1$}) 
\\& \rightarrow{}
(\text{$A$ with $p'$ crashing fully in $r+1$ and $m$ lost})
\\& \rightarrow{} (\text{$A$ with $m$ lost and $p'$ not
crashing}).
\end{align*}
The
first and last step apply the induction hypothesis; the middle one
yields indistinguishable executions since only $p'$ can tell the
difference between $m$ arriving or not and its lips are sealed.

We've shown that we can remove one message through a sequence of
executions where each pair of adjacent executions is indistinguishable
to some process.
Now
paste together $n-1$ such sequences (one per message) to prove the lemma.
\end{proof}

The rest of the proof: Crash some process fully in round 0 and then change its input.  Repeat until all inputs are changed.

\section{Variants}

So far we have described \index{consensus!binary}\concept{binary
consensus}, since all inputs are $0$ or $1$.  We can also allow larger
input sets.  With crash failures, this allows a stronger validity
condition: the output must be equal to some non-faulty process's input. 
It's not hard to see that Dolev-Strong
(§\ref{section-synchronous-agreement-flooding}) gives this stronger
condition.

\myChapter{Byzantine agreement}{2026}{}
\label{chapter-Byzantine-agreement}

\index{agreement!Byzantine}\index{Byzantine agreement}
Like synchronous agreement (as in
Chapter~\ref{chapter-synchronous-agreement}) except that we replace
crash failures with \index{failure!Byzantine}\indexConcept{Byzantine failure}{Byzantine failures}, where a faulty process can ignore its programming and send any messages it likes.  Since we are operating under a universal quantifier, this includes the case where the Byzantine processes appear to be colluding with each other under the control of a centralized adversary.

\section{Lower bounds}
\label{section-Byzantine-lower-bounds}

We'll start by looking at lower bounds.

\subsection{Minimum number of rounds}

We've already seen an $f+1$ lower bound on rounds for crash failures
(see §\ref{section-synchronous-agreement-lower-bound}).  This lower
bound applies \emph{a fortiori} to Byzantine failures, since Byzantine failures can simulate crash failures.

\subsection{Minimum number of processes}
\label{section-Byzantine-minimum-processes}

We can also show that we need $n > 3f$ processes. For $n = 3$ and $f
= 1$ the intuition is that Byzantine $B$ can play non-faulty $A$ and
$C$ off against each other, telling $A$ that $C$ is Byzantine and $C$
that $A$ is Byzantine. Since $A$ is telling $C$ the same thing about
$B$ that $B$ is saying about $A$, $C$ can't tell the difference and
doesn't know who to believe. Unfortunately, this tragic soap opera is
not a real proof, since we haven't actually shown that $B$ can say
exactly the right thing to keep $A$ and $C$ from guessing that $B$ is
evil.

Here is a real proof, which works by explicitly showing how to
construct a bad execution for any given algorithm.\footnote{The presentation here is based on
\cite[\S5.2.3]{AttiyaW2004}. The original
impossibility result is
due to Pease, Shostak, and Lamport~\cite{PeaseSL1980}.  This 
particular proof is due to Fischer, Lynch, and
Merritt~\cite{FischerLM1986}.} Consider an
artificial execution where (non-Byzantine) $A$, $B$, and $C$ are
duplicated and then placed in a ring 
$A_{0} B_{0} C_{0} A_{1} B_{1} C_{1}$,
where the digits indicate inputs.  We'll still keep
the same code for $n=3$ on each process, but when $A_{0}$
tries to send a message to what it thinks of as just $C$ we'll send it
to $C_{1}$ while messages from $B_{0}$ will instead go to $C_{0}$.
For any adjacent pair of processes (e.g. $A_{0}$ and $B_{0}$), the
behavior of the rest of the ring could be simulated by a single
Byzantine process ($\evil{C}$), so each process in the 6-process ring
behaves just as it does in some 3-process execution with 1 Byzantine
process.  It follows that all of the processes terminate and decide in
this unholy 6-process
\index{Frankenexecution}Frankenexecution\footnote{Not a real word.} the same value that they would in the corresponding 3-process Byzantine execution.  So what do they decide?

\begin{figure}
    \centering
    \begin{tikzpicture}[auto,node distance=4cm]
    \node (3) {
\begin{tikzpicture}
    \node (A) at (150:1) {$A_0$};
    \node (B) at (30:1) {$B_0$};
    \node[color=red] (C) at (270:1) {$\evil{C}$};
    \path
        (A) edge (B)
        (B) edge (C)
        (C) edge (A)
        ;
\end{tikzpicture}
};
\node (6) [right of=3] {
\begin{tikzpicture}
    \node (A0) at (120:1) {$A_0$};
    \node (B0) at (60:1) {$B_0$};
    \node (C0) [color=red] at (0:1) {$C_0$};
    \node (A1) [color=red] at (300:1) {$A_1$};
    \node (B1) [color=red] at (240:1) {$B_1$};
    \node (C1) [color=red] at (180:1) {$C_1$};

    \path
        (A0) edge (B0)
        (B0) edge (C0)
        (C0) edge [color=red] (A1)
        (A1) edge [color=red] (B1)
        (B1) edge [color=red] (C1)
        (C1) edge (A0)
    ;
\end{tikzpicture}
};
\end{tikzpicture}
\caption[Synthetic execution for Byzantine
agreement lower bound]{Three-process vs.~six-process execution in Byzantine
agreement lower bound.  Processes $A_0$ and $B_0$ in right-hand
execution receive same messages as in left-hand three-process
execution with Byzantine $\evil{C}$ simulation $C_0$ through $C_1$.
So validity forces them to decide $0$.  A similar argument using
Byzantine $\evil{A}$ shows the same for $C_0$.}
\label{fig-Byzantine-agreement-six-processes}
\end{figure}

Given two processes with the same input, say, $A_{0}$ and $B_{0}$, the
giant execution is indistinguishable from an $A_{0} B_{0} \evil{C}$ execution
where $\evil{C}$ is Byzantine (see
Figure~\ref{fig-Byzantine-agreement-six-processes}.  Validity says $A_{0}$ and $B_{0}$ must both
decide 0.  Since this works for any pair of processes with the same
input, we have each process deciding its input.  But now consider the
execution of $C_{0} A_{1} \evil{B}$, where $\evil{B}$ is Byzantine.  In the big
execution, we just proved that $C_{0}$ decides 0 and $A_{1}$ decides
1, but since the $C_{0} A_{1} \evil{B}$ execution is indistinguishable from the big execution to $C_{0}$ and $A_{1}$, they do the same thing here and violate agreement.

This shows that with $n=3$ and $f=1$, we can't win.  We can generalize
this to $n = 3f$.  Suppose that there were an algorithm that solved
Byzantine agreement with $n=3f$ processes.  Group the processes into
groups of size $f$, and let each of the $n=3$ processes simulate one
group, with everybody in the group getting the same input, which can only
make things easier.  Then we get a protocol for
$n=3$ and $f=1$, an impossibility.

\subsection{Minimum connectivity}
\label{section-Byzantine-minimum-connectivity}

So far, we've been assuming a complete communication graph.  If the
graph is not complete, we may not be able to tolerate as many
failures.  In particular, we need the connectivity of the graph
(minimum number of nodes that must be removed to split it into two
components) to be at least $2f+1$.  See \cite[\S6.5]{Lynch1996} for
the full proof.  The essential idea is that if we have an arbitrary
graph with a vertex cut of size $k < 2f+1$, we can simulate it on a
4-process graph where $A$ is connected to $B$ and $C$ (but not $D$),
$B$ and $C$ are connected to each other, and $D$ is connected only to
$B$ and $C$.  Here $B$ and $C$ each simulate half the processes in the
size-$k$ cut, $A$ simulates all the processes on one side of the cut
and $D$ all the processes on the other side.  We then construct an
8-process artificial execution with two non-faulty copies of each of
$A$, $B$, $C$, and $D$ and argue that if one of $B$ or $C$ can be
Byzantine then the 8-process execution is indistinguishable to the
remaining processes from a normal 4-process execution.  (See
Figure~\ref{fig-Byzantine-agreement-six-processes}.)

\begin{figure}
    \centering
    \begin{tikzpicture}[auto,node distance=4cm]
    \node (4) {
\begin{tikzpicture}[auto,node distance=1.5cm]
    \node (A) {$A_0$};
    \node (B) [above right of=A] {$B_0$};
    \node (C) [color=red,below right of=A] {$\evil{C}$};
    \node (D) [below right of=B] {$D_0$};
    \path
        (A) edge (B) edge (C)
        (B) edge (C) edge (D)
        (C) edge (D)
        ;
\end{tikzpicture}
};
\node (8) [right of=4] {
\begin{tikzpicture}[auto,node distance=1.5cm]
    \node (A0) {$A_0$};
    \node (B0) [above right of=A0] {$B_0$};
    \node (D0) [below right of=B0] {$D_0$};
    \node (C0) [color=red,below left of=D0] {$C_0$};
    \node (A1) [color=red,below right of=C0] {$A_1$};
    \node (B1) [color=red,below of=C0] {$B_1$};
    \node (C1) [color=red,below of=A0] {$C_1$};
    \node (D1) [color=red,below of=C1] {$D_1$};
    \path
        (A0) edge (B0) edge (C1)
        (B0) edge (C0) edge (D0)
        (C0) edge (D0) edge [color=red] (A1)
        (A1) edge [color=red] (B1)
        (B1) edge [color=red] (C1) edge [color=red] (D1)
        (C1) edge [color=red] (D1)
    ;
\end{tikzpicture}
};
\end{tikzpicture}
\caption[Synthetic execution for Byzantine
agreement connectivity]{Four-process vs.~eight-process execution in Byzantine
agreement connectivity lower bound.  Because Byzantine $\evil{C}$ can
simulate $C_0,D_1,B_1,A_1,$ and $C_1$, good processes $A_0$, $B_0$
and $D_0$ must all decide $0$ or risk violating validity.}
\label{fig-Byzantine-agreement-connectivity}
\end{figure}

An argument
similar to the $n > 3f$ proof then shows we violate one of validity or
agreement: if we replacing $C_0$, $C_1$, and all the nodes on one side
of the $C_0+C_1$ cut with a single Byzantine $\evil{C}$, we force the
remaining non-faulty nodes to decide their inputs or violate validity.
But then doing the same thing with $B_0$ and $B_1$ yields an
execution that violates agreement.

Conversely, if we have connectivity $2f+1$, then the processes can
simulate a general graph by sending each other messages along $2f+1$
predetermined vertex-disjoint paths and taking the majority value as
the correct message.  Since the $f$ Byzantine processes can only
corrupt one path each (assuming the non-faulty processes are careful
about who they forward messages from), we get at least $f+1$ good
copies overwhelming the $f$ bad copies.  This reduces the problem on a general graph with sufficiently high connectivity to the problem on a complete graph, allowing Byzantine agreement to be solved if the other lower bounds are met.

\subsection{Weak Byzantine agreement}

(Here we are following \cite[\S6.6]{Lynch1996}.  The original result
is due to Lamport~\cite{Lamport1983}.)

\index{Byzantine agreement!weak}\indexConcept{weak Byzantine
agreement}{Weak Byzantine agreement} is like 
regular Byzantine agreement, but validity is only required to
hold if there are no faulty processes at all.
If there is a single
faulty process, the non-faulty processes can output any value
regardless of their inputs (as long as they agree on it).  Sadly, this
weakening doesn't improve things much: even weak Byzantine agreement
can be solved only if $n ≥ 3f+1$.

Proof: As in the strong Byzantine agreement case, we'll construct a many-process Frankenexecution to figure out a strategy for a single Byzantine process in a 3-process execution.  The difference is that now the number of processes in our synthetic execution is much larger, since we want to build an execution where at least some of our test subjects think they are in a non-Byzantine environment.  The trick is to build a very big, highly-symmetric ring so that at least some of the processes are so far away from the few points of asymmetry that might clue them in to their odd condition that the protocol terminates before they notice.

Fix some protocol that allegedly solves weak Byzantine agreement, and
let $r$ be the number of rounds for the protocol.  Construct a ring of
$6r$ processes $A_{01} B_{01} C_{01} A_{02} B_{02} C_{02} \dots{}
A_{0r} B_{0r} C_{0r} A_{10} B_{10} C_{10} \dots{} A_{1r} B_{1r}
C_{1r}$, where each $X_{ij}$ runs the code for process $X$ in the
3-process protocol with input $i$.  For each adjacent pair of
processes, there is a 3-process Byzantine execution which is
indistinguishable from the $6r$-process execution for that pair: since
agreement holds in all Byzantine executions, each adjacent pair
decides the same value in the big execution and so either everybody
decides $0$ or everybody decides $1$ in the big execution.

Now we'll show that means that validity is violated in some
no-failures 3-process execution.  We'll extract this execution by
looking at the execution of processes $A_{0,r/2} B_{0,r/2} C_{0,r/2}$.
The argument is that up to round $r$, any input-0 process that is at
least $r$ steps in the ring away from the nearest 1-input process acts
like the corresponding process in the all-0 no-failures 3-process
execution.  Since $A_{0,r/2}$ is $3r/2 > r$ hops away from $A_{1r}$
and similarly for $C_{0,r/2}$, our 3 stooges all decide 0 by validity.
But now repeat the same argument for $A_{1,r/2} B_{1,r/2} C_{1,r/2}$ and get 3 new stooges that all decide 1.  This means that somewhere in between we have two adjacent processes where one decides 0 and one decides 1, violating agreement in the corresponding 3-process execution where the rest of the ring is replaced by a single Byzantine process.
This concludes the proof.

This result is a little surprising: we might expect that weak Byzantine
agreement could be solved by allowing a process to return a default
value if it notices anything that might hint at a fault somewhere.
But this would allow a Byzantine process to create disagreement
revealing its bad behavior to just one other process in the very last
round of an execution otherwise headed for agreement on the
non-default value.  The chosen victim decides the default value, but
since it's the last round, nobody else finds out.  Even if the
algorithm is doing something more sophisticated, examining the
$6r$-process execution will tell the Byzantine process exactly when 
and how to start acting badly.

\section{Upper bounds}
\label{section-Byzantine-upper-bounds}
\label{section-Byzantine-algorithms}

Here we describe two upper bounds for Byzantine agreement, one of
which gets an optimal number of rounds at the cost of many large messages,
and the other of which gets smaller messages at the cost of more
rounds.  (We are following §\S5.2.4--5.2.5 of \cite{AttiyaW2004} in
choosing these algorithms.)  Neither of these algorithms is
state-of-the-art, but they demonstrate some of the issues in solving 
Byzantine agreement without the sometimes-complicated optimizations
needed to get all the parameters of the algorithm down simultaneously.

\subsection{Exponential information gathering gets \texorpdfstring{$n
= 3f+1$}{n = 3f+1}}
\label{section-Byzantine-exponential-information-gathering}

\newData{\EIGval}{val}
\newData{\EIGpath}{path}
\newData{\EIGvalue}{value}
\newData{\EIGround}{round}

The idea of \concept{exponential information gathering}
is that each process will do a lot of gossiping, but now its 
state is no longer just a flat set of inputs, but a tree describing
who it heard what from.  We build this tree out of pairs of the form
$\Tuple{\EIGpath, \Input}$ where $\EIGpath$ is a sequence of
intermediaries with no repetitions and $\Input$ is some input.  A
process $i$'s state at each round is just a set of such pairs, represented
by the variables $\EIGval{\EIGpath, i} = \Input$.  At the end
of $f+1$ rounds of communication (necessary because of the lower bound
for crash failures), each non-faulty process $i$ attempts to untangle the complex web
of hearsay and second-hand lies to compute the same decision value as
the other processes, by computing reconstructed values
$\EIGval^*(\EIGpath, i)$ that, we hope, will eventually converge to
the same values for all processes.

This technique was used
by Pease, Shostak, and Lamport~\cite{PeaseSL1980} to show that
their impossibility result is tight: there exists an algorithm for
Byzantine agreement that runs in $f+1$ synchronous rounds and
guarantees agreement and validity as long as $n ≥ 3f+1$.

\begin{algorithm}
    \tcp{Set my value to my input}
    $\EIGval(\Tuple{}, i) ← \Input$\;
    \For{$\EIGround ← 0 \dots f$}{
        \tcp{send step for this round}
        \Foreach{non-repeating $w$, $\card{w} = \EIGround$, $i∉w$}{
            Send $\Tuple{wi,\EIGval(w,i})$ to all processes\;
        }
        \tcp{receive step for this round}
        \Foreach{non-repeating $w$, $\card{w} = \EIGround$}{
            \eIf{$j$ sent $\Tuple{wj,v}$}{
                \tcp{Record reported value}
                $\EIGval(wj,i) ← v$\;
            }{
                \tcp{Record default value}
                $\EIGval(wj,i) ← 0$\;
            }
        }
    }
    \tcp{Compute decision value}
    \Foreach{path $w$ of length $f+1$ with no repeats}{
        $\EIGval^*(w,i) ← \EIGval(w,i)$\;
    }
    \For{$\ell ← f$ \DownTo $0$}{
        \Foreach{non-repeating $w$, $\card{w} = \ell$}{
            $\EIGval^*(w,i) ← \maj_{j∉w} \EIGval^*(wj,i)$\;
        }
    }
    Decide $\EIGval^*(\Tuple{},i)$\;
        \caption[Exponential information gathering]{Exponential
        information gathering.  Code for process $i$.}
    \label{alg-exponential-information-gathering}
\end{algorithm}

The algorithm is given in
Algorithm~\ref{alg-exponential-information-gathering}.
The communication phase is just gossiping, where each process starts with its
only its input and forwards any values it hears about along with their
provenance to all of the other processes.  At the end of this phase,
each process $i$ has set $\EIGval(\EIGpath, i)$ to some value
$\EIGvalue$,
where \EIGpath spans all sequences of 0 to $f+1$ distinct
IDs and \EIGvalue is the input value forwarded along that path.

Because we can't trust
these $\EIGval(w,i)$ values to be an accurate description of any
process's input if there is a Byzantine process in $w$, each process
computes for itself reconstructed values $\EIGval^*(w,i)$ that use
majority voting to try to get a more trustworthy picture of the
original inputs.

Formally, we
think of the set of paths as a tree where $w$ is the parent of $wj$
for each path $w$ and each ID $j$ not in $w$.  To apply EIG in the
Byzantine model, ill-formed or missing messages from $j$ are 
replaced by default values, but otherwise the data-collecting part of EIG
proceeds as in the crash failure model.  However, we compute the
decision value from the last-round values
recursively as follows.  First, set $\EIGval^*(w,i)$ for 
any path $w$ with $\card*{w} = f+1$ to $\EIGval(w,i)$. Then for each
path $w$ with $\card{w} < f+1$,
define $\EIGval^*(w, i)$ to be the majority value among
$\EIGval^*(wj, i)$ for all $j$.
Finally, have process $i$ decide $\EIGval^*(\Tuple{},
i)$. Note that this entire reconstruction process can be computed
locally by each process, although we haven't yet shown that $i$'s
decision value
$\EIGval^*(\Tuple{}, i)$ will necessarily be the same as
$j$'s decision value $\EIGval^*(\Tuple{},j)$.

The majority rule for $w = \Tuple{}$ makes the decision value
$\EIGval^*(\Tuple{}, i)$ a majority of reconstructed inputs $\EIGval^*(j, i)$.
One way to
think about this is that I never trust $j$ to give me the correct value
for $wj$—even when $w = \Tuple{}$ and $j$ is claiming to report
its own input—so instead I take a majority of values of $wj$ that
$j$ allegedly reported to other people. But since I don't trust those
other people either, I use the same process recursively to construct
those reports, and hope that all the lies are eventually overcome by
the truth.

\subsubsection{Proof of correctness}

This is just a sketch of the proof from \cite[§{}6.3.2]{Lynch1996};
essentially the same argument appears in \cite[§{}5.2.4]{AttiyaW2004}.

We start with a basic observation that good processes send and record
values correctly. Throughout the proof, we use $\EIGval(w,i)$ for the
final value of $\EIGval(w,i)$ recorded by $i$.

\begin{lemma}
\label{lemma-EIG-trivial}
If $i$ and $j$ are both non-faulty, then for all $w$, $\EIGval(wj,
i) = \EIGval(w, j)$.
\end{lemma}
\begin{proof}
    Trivial: $j$ sends $\Tuple{wj,\EIGval(w, i)}$ to $i$, and $i$ records it in
    $\EIGval(wj,i)$.
\end{proof}

More involved is this lemma, which says that when we reconstruct a
value for a trustworthy process at some level, we get the same value
that it sent us.  In particular this will be used to show that the
reconstructed inputs $\EIGval^*(j,i)$ are all equal to the real inputs
for good processes.
\begin{lemma}
\label{lemma-EIG-common}
If $i$ and $j$ are non-faulty, then for all $w$, $\EIGval^*(wj, i) =
    \EIGval(w, j)$.
\end{lemma}
\begin{proof}
By induction on $f+1-\card*{wj}$.  If $\card*{wj} = f+1$, then
$\EIGval^*(wj, i) = \EIGval(wj, i) = \EIGval(w, j)$.
If $\card*{wj} < f+1$, then 
    then $\EIGval^*(wj,i) = \maj_{k∉wj} \EIGval^*(wjk,i)$.
    The induction hypothesis says $\EIGval^*(wjk,i) = \EIGval(wj,k)$,
    which equals $\EIGval(w,j)$ by Lemma~\ref{lemma-EIG-trivial}.
    Now observe that there are at least $3f+1-\card{wj} ≥ 2f+1$
    possible $k$, of which at most $f$ are faulty, leaving a
    non-faulty majority all of which have $\EIGval^*(wjk,i) =
    \EIGval(w,j)$.
\end{proof}

We call a node $w$ \index{node!common}\indexConcept{common node}{common}
if $\EIGval^*(w, i) = \EIGval^*(w, j)$ for all non-faulty $i, j$.  
Lemma~\ref{lemma-EIG-common} implies that $wk$ is common if $k$ is
non-faulty.  We can also show that any node whose children are all
common is also common, whether or not the last process in its label is
faulty.

\begin{lemma}
    \label{lemma-EIG-common-children}
    Let $wk$ be common for all $k$.  Then $w$ is common.
\end{lemma}
\begin{proof}
    Recall that, for $\card*{w} < f+1$, $\EIGval^*(w,i)$ is the majority
    value among all $\EIGval^*(wk,i)$.  If all $wk$ are common, then
    $\EIGval^*(wk,i) = \EIGval^*(wk,j)$ for all non-faulty $i$ and $j$.
    so $i$ and $j$ compute the same majority values and get
    $\EIGval^*(w,i) = \EIGval^*(w,j)$.
\end{proof}

We can now prove the full result.

\begin{theorem}
\label{theorem-Byzantine-EIG}
Exponential information gathering using $f+1$ rounds
in a synchronous Byzantine system
with at most $f$ faulty processes satisfies validity and agreement,
provided $n ≥ 3f+1$.
\end{theorem}
\begin{proof}
    Termination: Protocol finishes after $f+1$ rounds.

Validity: Immediate application of Lemmas~\ref{lemma-EIG-trivial} and~\ref{lemma-EIG-common} when
    $w = \Tuple{}$.  We have $\EIGval^*(j, i) = \EIGval(j, i) =
    \EIGval(\Tuple{}, j)$ for all non-faulty $j$ and $i$, which
means that a majority of the $\EIGval^*(j, i)$ values equal the common
    input and thus so does $\EIGval^*(\Tuple{}, i)$.

Agreement: 
Observe that every path has a common node on it, since a path travels
through $f+1$ nodes and one of them is good.  If we then suppose that the root is
not common: by Lemma~\ref{lemma-EIG-common-children},
it must have a not-common child, that node must have a
not-common child, etc.  But this constructs a path from the root to a
leaf with no not-common nodes, which we just proved can't happen.
\end{proof}

\subsection{Phase king gets constant-size messages}
\label{section-Byzantine-phase-king}

The following algorithm, based on work of Berman, Garay, and
Perry~\cite{BermanGP1989},
achieves Byzantine agreement in $2(f+1)$ rounds using constant-size
messages, provided $n ≥ 4f+1$.  The description here is drawn from
\cite[\S5.2.5]{AttiyaW2004}.  The original Berman-Garay-Perry paper
gives somewhat better bounds, but the algorithm and its analysis are more complicated.

\subsubsection{The algorithm}

The main idea of the algorithm is that we avoid the recursive
majority voting of EIG by running a vote in each of $f+1$
\emph{phases} through a \concept{phase king}, some process chosen in advance to run the phase.  Since the number of phases exceeds the number of faults, we eventually get a non-faulty phase king.  The algorithm is structured so that one non-faulty phase king is enough to generate agreement and subsequent faulty phase kings can't undo the agreement.

\newData{\PKpref}{pref}
\newData{\PKmajority}{majority}
\newData{\PKmultiplicity}{multiplicity}
\newData{\PKkingMajority}{kingMajority}

Pseudocode appears in Algorithm~\ref{alg-phase-king}.
Each processes $i$ maintains an array $\PKpref_{i}[j]$, where
$j$ ranges over all process IDs.
There are also utility values \PKmajority, \PKkingMajority and
\PKmultiplicity for each process that are used to keep track of what
it hears from the other processes.
Initially, $\PKpref_{i}[i]$ is just $i$'s input and $\PKpref_{i}[j] = 0$ for $j \ne i$.

\begin{algorithm}
$\PKpref_{i}[i] = \DataSty{input}$\;
\lFor{$j \ne i$}{$\PKpref_{i}[j] = 0$}
\For{$k ← 1$ \KwTo $f+1$}{
    \tcp{First round of phase $k$}
    send $\PKpref_{i}[i]$ to all processes (including myself) \;
    $\PKpref_{i}[j] ← v_{j}$, where $v_{j}$ is the value received
    from process $j$ \;
    $\PKmajority ←$ majority value in $\PKpref_{i}$ \;
    $\PKmultiplicity ←$ number of times $\PKmajority$ appears in
        $\PKpref_{i}$ \;
    \tcp{Second round of phase $k$}
    \If{$i=k$}{
        \tcp{I am the phase king}
        send $\PKmajority$ to all processes \;
    }
    \eIf{received $m$ from phase king}{
        $\PKkingMajority \gets m$\;
    }{
        $\PKkingMajority \gets 0$\;
    }
    \eIf{$\PKmultiplicity > n/2 + f$}{
        $\PKpref_{i}[i] = \PKmajority$\;
    }{
        $\PKpref_{i}[i] = \PKkingMajority$\;
    }
}
\Return $\PKpref_{i}[i]$
\caption{Byzantine agreement: phase king}
\label{alg-phase-king}
\end{algorithm}

The idea of the algorithm is that in each phase, everybody announces
their current preference (initially the inputs).  If the majority of
these preferences is large enough (e.g., all inputs are the same),
everybody adopts the majority preference.  Otherwise everybody adopts
the preference of the phase king.  The majority rule means that once
the processes agree, they continue to agree despite bad phase kings.
The phase king rule allows a good phase king to end disagreement.  By
choosing a different king in each phase, after $f+1$ phases, some king
must be good.  This intuitive description is justified below.

\subsubsection{Proof of correctness}

Termination is immediate from the algorithm.

For validity, suppose all inputs are $v$.  We'll show that all
non-faulty $i$ have $\PKpref_{i}[i] = v$ after every phase.  In the
first round of each phase, process $i$ receives at least $n-f$
messages containing $v$; since $n ≥ 4f + 1$, we have $n-f ≥ 3f+1$
and $n/2 + f ≤ (4f+1)/2 + f = 3f+1/2$, and thus these $n-f$ messages
exceed the $n/2+f$ threshold for adopting them as the new preference.
So all non-faulty processes ignore the phase king and stick with $v$,
eventually deciding $v$ after round $2(f+1)$.

For agreement, we'll ignore all phases up to the first phase with a
non-faulty phase king.  Let $k$ be the first such phase, and assume
that the \PKpref values are set arbitrarily at the start of this phase.  We want to argue that at the end of the phase, all non-faulty processes have the same preference.  There are two ways that a process can set its new preference in the second round of the phase:

\begin{enumerate}
 \item The process $i$ observes a majority of more than $n/2+f$
 identical values $v$ and ignores the phase king.  Of these values,
 more than $n/2$ of them were sent by non-faulty processes.  So the
 phase king also receives these values (even if the faulty processes
 change their stories) and chooses $v$ as its majority value.
 Similarly, if any other process $j$ observes a majority of $n/2+f$
 identical values, the two $>n/2$ non-faulty parts of the majorities
 overlap, and so $j$ also chooses $v$.
 \item The process $i$ takes its value from the phase king.  We've
 already shown that $i$ then agrees with any $j$ that sees a big
 majority; but since the phase king is non-faulty, process $i$ will
 agree with any process $j$ that also takes its new preference from the phase king.
\end{enumerate}

This shows that after any phase with a non-faulty king, all processes agree.  The proof that the non-faulty processes continue to agree is the same as for validity.

\subsubsection{Performance of phase king}
It's not hard to see that this algorithm sends exactly $(f+1)(n^2+n)$
messages of $1$ bit each (assuming $1$-bit inputs).  The cost is
doubling the minimum number of rounds and reducing the tolerance for
Byzantine processes.  As mentioned earlier, a variant of phase-king
with 3-round phases gets optimal fault-tolerance with $3(f+1)$ rounds
(but 2-bit messages).  Still better is a rather complicated descendant
of the EIG algorithm due to Garay and Moses~\cite{GarayM1998}, which gets
$f+1$ rounds with $n ≥ 3f+1$ while still having polynomial message traffic.

\myChapter{Impossibility of asynchronous agreement}{2026}{}
\label{chapter-FLP}

There's an easy argument that says that you can't do most things in an
asynchronous message-passing system with $n/2$ crash failures:
partition the processes into two subsets $S$ and $T$ of size $n/2$
each, and allow no messages between the two sides of the partition for
some long period of time. Since the processes
in each side can't distinguish between the other side being slow and
being dead, eventually each has to take action on their own.
For many problems, we can show that this leads
to a bad configuration. For example, for agreement, we can supply each
side of the partition with a different common input value, forcing
disagreement because of validity.  We can then satisfy the fairness
condition that says all messages are eventually delivered by
delivering the delayed messages across the partition, but it's too
late for the protocol.

The Fischer-Lynch-Paterson (FLP) result~\cite{FischerLP1985} says
something much stronger: you can't do
agreement in an asynchronous message-passing system if even \emph{one} crash
failure is allowed.\footnote{Unless you augment the basic model in some way,
say by adding randomization
(Chapter~\ref{chapter-randomized-consensus}) or failure detectors
(Chapter~\ref{chapter-failure-detectors}).}  After its initial
publication, it was quickly generalized to other models including
asynchronous shared memory~\cite{LouiA1987}, and indeed the presentation of the result
in~\cite[\S12.2]{Lynch1996} is given for shared-memory first, with the original
result appearing in \cite[\S17.2.3]{Lynch1996} as a corollary of the ability of
message passing to simulate shared memory.  In these notes, I'll
present the original result; the dependence on the model is
surprisingly limited, and so most of the proof is the same for both
shared memory (even strong versions of shared memory that support
operations like
atomic snapshots\footnote{Chapter~\ref{chapter-atomic-snapshots}.}) and message passing.

Section 5.3 of \cite{AttiyaW2004} gives a very different version of
the proof, where it is shown first for two processes in shared memory,
then generalized to $n$ processes in shared memory by adapting the
classic Borowsky-Gafni simulation~\cite{BorowskyG1993} to show that two
processes with one failure can simulate $n$ processes with one
failure. This is worth looking at (it's an excellent example of the
power of simulation arguments, and BG simulation is useful in many
other contexts) but we will stick with the original argument, which is simpler.
We will look at this again when we consider BG simulation in
Chapter~\ref{chapter-BG-simulation}.

\section{Agreement}

Usual rules: \concept{agreement} (all non-faulty processes decide the same value), \concept{termination} (all non-faulty processes eventually decide some value), \concept{validity} (for each possible decision value, there an execution in which that value is chosen).  Validity can be tinkered with without affecting the proof much.

To keep things simple, we assume the only two decision values are 0 and 1.

\section{Failures}

A failure is an internal action after which all send operations are
disabled. The adversary is allowed one failure per execution.
Effectively, this means that any group of $n-1$ processes must
eventually decide without waiting for the $n$-th, because it might have failed.

With asynchronous scheduling and required termination, this is
equivalent to a limited version of fairness in which one process is
labeled as faulty and the adversary is not required to deliver
messages from that process.  Having an active failure step (as opposed
to the adversary just choosing internally not to deliver some
process's messages) mostly just lets us more easily describe which
process the adversary is doing this to.

\section{Steps}

The FLP paper uses a notion of \emph{steps} that is slightly different
from the send and receive actions of the asynchronous message-passing
model we've been using.  Essentially a step consists of receiving zero
or more messages followed by doing a finite number of sends.  To fit
it into the model we've been using, we'll define a step as either a
pair $(p,m)$, where $p$ receives message $m$ and performs zero or more
sends in response, or $(p,⊥)$, where $p$ receives nothing and
performs zero or more sends.  We assume that the processes are
deterministic, so the messages sent (if any) are determined by $p$'s
previous state and the message received.  Note that these steps do not
correspond precisely to delivery and send events or even pairs of
delivery and send events, because what message gets sent in response to a particular delivery may change as the result of delivering some other message; but this won't affect the proof.

The fairness condition essentially says that if $(p,m)$ or
$(p,⊥)$ is continuously enabled it eventually happens.  Since
messages are not lost, once $(p,m)$ is enabled in some configuration
$C$, it is enabled in all successor configurations until it occurs;
similarly $(p,⊥)$ is always enabled.  So to ensure fairness, we have to ensure that any non-faulty process eventually performs any enabled step.

Comment on notation: I like writing the new configuration reached by
applying a step $e$ to $C$ like this: $Ce$.  The FLP paper uses
$e(C)$.

\section{Bivalence and univalence}

The core of the FLP argument is a strategy allowing the adversary (who
controls scheduling) to steer the execution away from any
configuration in which the processes reach agreement.  The guidepost
for this strategy is the notion of \concept{bivalence}, where a
configuration $C$ is \concept{bivalent} if there exist traces $T_{0}$
and $T_{1}$ starting from $C$ that lead to configurations $CT_{0}$ and
$CT_{1}$ where all processes decide 0 and 1 respectively.  A
configuration that is not bivalent is \concept{univalent}, or more
specifically \concept{0-valent} or \concept{1-valent} depending on
whether all executions starting in the configuration produce 0 or 1 as
the decision value.  (Note that bivalence or univalence are the only
possibilities because of termination.)  The important fact we will use
about univalent configurations is that any successor to an $x$-valent
configuration is also $x$-valent.

It's clear that any configuration where some process has decided is
not bivalent, so if the adversary can keep the protocol in a bivalent
configuration forever, it can prevent the processes from ever
deciding.  The adversary's strategy is to start in an initial bivalent
configuration $C_{0}$ (which we must prove exists) and then choose
only bivalent successor configurations (which we must prove is
possible).  A complication is that if the adversary is only allowed
one failure, it must eventually allow any message in transit to a non-faulty process to be received and any non-faulty process to send its outgoing messages, so we have to show that the policy of avoiding univalent configurations doesn't cause problems here.

\section{Existence of an initial bivalent configuration}

We can specify an initial configuration by specifying the inputs to
all processes.  If one of these initial configurations is bivalent, we
are done.  Otherwise, let $C$ and $C'$ be two initial configurations
that differ only in the input of one process $p$; by assumption, both
$C$ and $C'$ are univalent.  Consider two executions starting with $C$
and $C'$ in which process $p$ is faulty; we can arrange for these
executions to be indistinguishable to all the other processes, so both
decide the same value $x$.  It follows that both $C$ and $C'$ are
$x$-valent.  But since any two initial configurations can be connected
by some chain of such indistinguishable configurations, we have that
all initial configurations are $x$-valent, which violations validity.

\section{Staying in a bivalent configuration}
\label{section-FLP-staying-bivalent}

Now start in a failure-free bivalent configuration $C$ with some step
$e = (p,m)$ or $e = (p,⊥)$ enabled in $C$.  Let $S$ be the set of
configurations reachable from $C$ without doing $e$ or failing any
processes, and let $e(S)$ be the set of configurations of the form
$C'e$ where $C'$ is in $S$.  (Note that $e$ is always enabled in $S$,
since once enabled the only way to get rid of it is to deliver the
message.)  We want to show that $e(S)$ contains a failure-free bivalent configuration.

The proof is by contradiction: suppose that $C'e$ is univalent for all
$C'$ in $S$.  We will show first that there are $C_{0}$ and $C_{1}$ in
$S$ such that each $C_{i}e$ is $i$-valent.  To do so, consider any
pair of $i$-valent $A_{i}$ reachable from $C$; if $A_{i}$ is in $S$,
let $C_{i} = A_{i}$.  If $A_{i}$ is not in $S$, let $C_{i}$ be the
last configuration before executing $e$ on the path from $C$ to
$A_{i}$ ($C_{i}e$ is univalent in this case by assumption).

So now we have $C_{0}e$ and $C_{1}e$ with $C_{i}e$ $i$-valent in each
case.  We'll now go hunting for some configuration $D$ in $S$ and step
$e'$ such that $De$ is 0-valent but $De'e$ is 1-valent (or vice
versa); such a pair exists because $S$ is connected and so some step
$e'$ crosses the boundary between the $C'e =$ 0-valent and the $C'e =$
1-valent regions.

By a case analysis on $e$ and $e'$ we derive a contradiction:
\begin{enumerate}
 \item Suppose $e$ and $e'$ are steps of different processes $p$ and
 $p'$.  Let both steps go through in either order.  Then $Dee' =
 De'e$, since in an asynchronous system we can't tell which process
 received its message first.  But $De$ is 0-valent, which implies
 $Dee'$ is also 0-valent, which contradicts $De'e$ being 1-valent.
 \item Now suppose $e$ and $e'$ are steps of the same process $p$.
 Again we let both go through in either order.  It is not the case now
 that $Dee' = De'e$, since $p$ knows which step happened first (and
 may have sent messages telling the other processes).  But now we
 consider some finite sequence of steps $e_{1}e_{2}\dots{}e_{k}$ in
 which no message sent by $p$ is delivered and some process decides in
 $Dee_{1}\dots{}e_{k}$ (this occurs since the other processes can't
 distinguish $Dee'$ from the configuration in which $p$ died in $D$,
 and so have to decide without waiting for messages from $p)$.  This
 execution fragment is indistinguishable to all processes except $p$
 from $De'ee_{1}\dots{}e_{k}$, so the deciding process decides the
 same value $i$ in both executions.  But $Dee'$ is $0$-valent and
 $De'e$ is 1-valent, giving a contradiction.
\end{enumerate}

It follows that our assumption was false, and there is some reachable
bivalent configuration $C'e$.

Now to construct a fair execution that never decides, we start with a bivalent configuration, choose the oldest enabled action and use the above to make it happen while staying in a bivalent configuration, and repeat.

\section{Generalization to other models}

The FLP results extends to any asynchronous model where it is
impossible to tell which of two events happened first. 
The main idea is to replace
the definition of a step to whatever is available in the new model,
and adapt the resulting case analysis of 0-valent
$De'e$ vs 1-valent $Dee'$ as appropriate.
For example, in asynchronous shared memory, if $e$ and $e'$
are operations on different memory locations, they commute (just like
steps of different processes), and if they are operations on the same
location, either they commute (two reads) or only one process can
tell whether both happened (with a write and a read, only the
reader knows, and with two writes, only the first writer knows).
Killing the witness yields two indistinguishable configurations with
different valencies, a contradiction.

Loui and Abu-Amara~\cite{LouiA1987} first proved this
generalization to shared memory using standard read-write registers.
Herlihy~\cite{Herlihy1991waitfree}
later provided similar arguments for a wide variety of shared-memory
primitives that may provide additional operations beyond reads and
writes.
We will see many of these latter arguments in
Chapter~\ref{chapter-wait-free-hierarchy}.

\myChapter{Paxos}{2026}{}
\label{chapter-Paxos}

The \concept{Paxos} algorithm for consensus in a message-passing system  was
first described by Lamport in 1990 in a tech report that was widely
considered to be a joke (see
\url{http://research.microsoft.com/users/lamport/pubs/pubs.html#lamport-paxos}
for Lamport's description of the history).  The algorithm was finally
published in 1998~\cite{Lamport1998}, and after the algorithm continued to be
ignored, Lamport finally gave up and translated the results into
readable English~\cite{Lamport2001}.  It is now understood to be one of the most efficient practical algorithms for achieving consensus in a message-passing system with failure detectors, mechanisms that allow processes to give up on other stalled processes after some amount of time (which can't be done in a normal asynchronous system because giving up can be made to happen immediately by the adversary).

We will describe the basic Paxos algorithm in
§\ref{section-Paxos-algorithm}.  This is a one-shot version of Paxos
that solves a single agreement problem.  The version that is more
typically used, called \concept{multi-Paxos}, uses repeated executions
of the basic Paxos algorithm to implement a replicated state machine;
we'll describe this in §\ref{section-multi-Paxos}.

There are many more variants of Paxos in use.
The WikiPedia article on Paxos (\wikipedia{Paxos_(computer_science)})
gives a reasonably good survey of subsequent developments and applications.

\section{The Paxos algorithm}
\label{section-Paxos-algorithm}

The algorithm runs in a message-passing model with asynchrony and
fewer than $n/2$ crash failures (but not Byzantine failures, at least
in the original algorithm).
As always, we want to get agreement,
validity, and termination.

The Paxos algorithm itself is mostly
concerned with guaranteeing agreement and validity, while allowing for the possibility of termination if there is a long enough interval in which no process restarts the protocol.
A noteworthy feature of Paxos is that it is robust even to omission
failures, in the sense that lost messages can prevent termination,
but if new messages start being delivered again, the protocol can
recover.

Processes are classified as \indexConcept{proposer}{proposers},
\indexConcept{accepter}{accepters}, and \indexConcept{learner}{learners} (a single process may have all three roles).  The idea is that a proposer attempts to ratify a proposed decision value (from an arbitrary input set) by collecting acceptances from a majority of the accepters, and this ratification is observed by the learners.  Agreement is enforced by guaranteeing that only one proposal can get the votes of a majority of accepters, and validity follows from only allowing input values to be proposed.  The tricky part is ensuring that we don't get deadlock when there are more than two proposals or when some of the processes fail.  The intuition behind how this works is that any proposer can effectively restart the protocol by issuing a new proposal (thus dealing with lockups), and there is a procedure to release accepters from their old votes if we can prove that the old votes were for a value that won't be getting a majority any time soon.

To organize this vote-release process, we attach a distinct proposal
number to each proposal.  The safety properties of the algorithm don't
depend on anything but the proposal numbers being distinct, but since
higher numbers override lower numbers, to make progress we'll need
them to increase over time.  The simplest way to do this in practice
is to make the proposal number be a timestamp with the proposer's ID
appended to break ties.  We could also have the proposer poll the other processes for the most recent proposal number they've seen and add 1 to it.

The revoting mechanism now works like this: before taking a vote, a
proposer tests the waters by sending a $\PaxosPrepare(r)$ message to
all accepters, where $r$ is the proposal number.  An accepter responds
to this with a promise never to accept any proposal with a number less
than $r$ (so that old proposals don't suddenly get ratified) together
with the highest-numbered proposal that the accepter has accepted (so
that the proposer can substitute this value for its own, in case the
previous value was in fact ratified).  If the proposer receives a
response from a majority of the accepters, the proposer then does a
second phase of voting where it sends $\PaxosAccept(r, v)$ to all accepters and wins if receives a majority of votes.
(The exclamation point on $\PaxosAccept$ is not in the original paper,
but has become a common convention to emphasize that it's a command,
not a response.)

So for each proposal, the algorithm proceeds as follows:
\begin{enumerate}
 \item The proposer sends a message $\PaxosPrepare(r)$ to all accepters.  (Sending to only a majority of the accepters is enough, assuming they will all respond.)
 \item Each accepter compares $r$ to the highest-numbered proposal for
 which it has responded to a $\PaxosPrepare$ message and the
        highest-numbered proposal it has accepted.  If $r$ is
 greater than both, it responds with $\Ack(r, v, r_{v})$, where $v$ is the
 highest-numbered proposal it has accepted and $r_{v}$ is the number
 of that proposal (or $⊥$ and $-∞$ if there is no such proposal).

 An optimization at this point is to allow the accepter to send back
 $\Nack(r,r')$ where $r'$ is some higher number to let the proposer
 know that it's doomed and should back off and try again with a higher
 proposal number.  (This keeps a confused proposer who thinks it's the future from locking up the protocol until 2087.)
 \item The proposer waits (possibly forever) to receive \Ack from
 a majority of accepters.  If any \Ack contained a value, it sets
 $v$ to the most recent (in proposal number ordering) value that it
 received.  It then sends $\PaxosAccept(r, v)$ to all accepters (or
 just a majority).  You should think of \PaxosAccept as a demand (``Accept!'') rather than acquiescence (``I accept'')—the accepters still need to choose whether to accept or not.
 \item Upon receiving $\PaxosAccept(r, v)$, an accepter accepts $v$
 unless it has already received $\PaxosPrepare(r')$ for some $r' > r$.  If a majority of accepters accept the value of a given proposal, that value becomes the decision value of the protocol.
\end{enumerate}

Implementing these rules require only that each accepter track $r_{\Ack}$,
the highest number of any proposal for which it sent an $\Ack$, and
$\Tuple{v,r_v}$, the last proposal that
it accepted. Pseudocode showing the behavior of proposer and accepters in
the core Paxos protocol is given in Algorithm~\ref{alg-Paxos}.

\begin{algorithm}
    \Procedure{$\FuncSty{Propose}(r,v)$}{
        \tcp{Issue proposal number $r$ with value $v$}
        \tcp{Assumes $r$ is unique}
        send $\PaxosPrepare(r)$ to all accepters\;
        wait to receive $\Ack(r,v',r_{v'})$ from a majority of accepters\;
        \If{some $v'$ is not $⊥$}{
            $v \gets v'$ with maximum $r_{v'}$\;
        }
        send $\PaxosAccept(r,v)$ to all accepters\;
    }
    \Procedure{$\FuncSty{accepter}()$}{
        \Initially{
            $r_{\Ack} \gets -∞$\;
            $v \gets ⊥$\;
            $r_v \gets -∞$\;
        }
        \UponReceiving{$\PaxosPrepare(r)$ from $p$}{
            \If{$r > \max(r_{\Ack},r_v)$}{
                \tcp{Respond to proposal}
                send $\Ack(r,v,r_v)$ to $p$\;
                $r_{\Ack} \gets r$\;
            }
        }
        \UponReceiving{$\PaxosAccept(r,v')$}{
            \If{$r ≥ \max(r_{\Ack},r_v)$}{
                \tcp{Accept proposal}
                send $\PaxosAccepted(r,v')$ to all learners\;
                \If{$r > r_v$}{
                    \tcp{Update highest accepted proposal}
                    $\Tuple{r_v,v} \gets \Tuple{r,v'}$\;
                }
            }
        }
    }
    \caption{Paxos}
    \label{alg-Paxos}
\end{algorithm}

Note that acceptance is a purely local phenomenon; additional messages
are needed to detect which if any proposals have been accepted by a
majority of accepters.  Typically this involves a fourth round, where
accepters send $\PaxosAccepted(r, v)$ to all learners.

There is no requirement that only a single proposal is sent out (indeed, if proposers can fail we will need to send out more to jump-start the protocol).  The protocol guarantees agreement and validity no matter how many proposers there are and no matter how often they start.

\section{Informal analysis: how information flows between rounds}
Call a \concept{round} the collection of all messages labeled with
some particular proposal $r$.  The structure of the algorithm simulates a sequential execution in which higher-numbered rounds follow lower-numbered ones, even though there is no guarantee that this is actually the case in a real execution.

When an accepter sends $\Ack(r, v, r_{v})$, it is telling the
round-$r$ proposer the last value preceding round $r$ that it
accepted.  The rule that an accepter only acknowledges a proposal
higher than any proposal it has previously accepted prevents it
from sending information ``back in time''—the round $r_{v}$ in an
acknowledgment is always less than $r$.  The rule that an accepter
doesn't accept any proposal earlier than a round it has acknowledged
means that the value $v$ in an $\Ack(r, v, r_{v})$ message never
goes out of date—there is no possibility that an accepter might
retroactively accept some later value in round $r'$ with $r_{v} < r' <
r$.  So the \Ack message values tell a consistent story about the history of the protocol, even if the rounds execute out of order.

The second trick is to use overlapping majorities to make sure that
any value that is accepted is not lost.
If the only way to
decide on a value in round $r$ is to get a majority of accepters to
accept it, and the only way to make progress in round $r'$ is to get
acknowledgments from a majority of accepters, these two majorities
overlap.  So in particular the overlapping process reports the
round-$r$ proposal value to the proposer in round $r'$, and we can
show by induction on $r'$ that this round-$r$ proposal value becomes
the proposal value in all subsequent rounds that proceed past the
acknowledgment stage.  So even though it may not be possible to detect
that a decision has been reached in round $r$ (say, because some of the accepters in the accepting majority die without telling anybody what they did), no later round will be able to choose a different value.  This ultimately guarantees agreement.

\section{Example execution}

For Paxos to work well, proposal numbers should increase over time.
But there is no requirement that proposal numbers are increasing or
even that proposals with different proposal numbers don't overlap.
When thinking about Paxos, it is easy to make the mistake of ignore
cases where proposals are processed concurrently or out of order.  In
Figure~\ref{fig-Paxos-mixed-up-execution}, we give an example of an
execution with three proposals running concurrently.

\begin{figure}
    \begin{displaymath}
        \begin{array}{cccccc}
p_1 &p_2 &p_3 &a_1 &a_2 &a_3 \\
    & &\PaxosPrepare(3) & & &\\
    &\PaxosPrepare(2) & & & &\\
        \PaxosPrepare(1) & & & & &\\
& & & & &\Ack(3,⊥,0)\\
& & &\Ack(1,⊥,0)& &\\
& & &&\Ack(1,⊥,0) &\\
            \PaxosAccept(1,1)&&&&&\\
&&&\PaxosAccepted(1,1)&&\\
&&&&&\Nack(1,3)\\
            & & &\Ack(2,1,1) & &\\
            & & & &\Ack(2,⊥,0) &\\
&\PaxosAccept(2,1)&&&&\\
& & & & &\Nack(2,3)\\
            &&&&\PaxosAccepted(2,1)&\\
            & && &\Ack(3,1,2) &\\
&&\PaxosAccept(3,1)&&&\\
& & &\PaxosAccepted(3,1)& &\\
            && & & & \PaxosAccepted(3,1)\\
    \end{array}
\end{displaymath}
    \caption[Example execution of Paxos]{Example execution of Paxos.
    Time increases downward. Each column records messages sent by one
    of
        three proposers $p_1, p_2$, and $p_3$ and three
    accepters $a_1, a_2,$ and $a_3$.  Proposer $p_1$'s proposed value
$1$ is not accepted by a majority of processes in round $1$, but it is
picked up by proposer $p_2$ in round $2$, and is eventually accepted
by a majority in round $3$.}
    \label{fig-Paxos-mixed-up-execution}
\end{figure}

\section{Safety properties}
\label{section-Paxos-safety}

We now present a more formal analysis of the Paxos protocol.  We
consider only the safety properties of the protocol, corresponding to
validity and agreement.  Without additional assumptions, Paxos does \emph{not} guarantee termination.

Call a value \emph{chosen} with proposal number $r$ if it is accepted by a
majority of accepters with that proposal number. The safety properties of Paxos are:
\begin{itemize}
 \item No value is chosen with any proposal number unless it is first proposed. (This gives validity.)
 \item No two distinct values are ever both chosen, even with
     different proposal numbers. (This gives agreement.)
\end{itemize}

The first property is immediate from examination of the algorithm:
every value propagated through the algorithm is ultimately a
copy of some proposer's original input. We can formalize this
observation by checking that, for any set of values $S$,
the property that all values contained in messages or processes'
internal state are in $S$ is an invariant.

For the second property, let $r$ be the smallest proposal number which
is accepted by a majority of accepters. We'll show by induction on
$r'≥r$ that the value $v$ chosen with proposal number $r$ is the value proposed by any
proposer $p_{r'}$ with proposal number $r'$. There are two claims that
make this true:
\begin{enumerate}
    \item Any $\Ack(r',v',r_{v'})$ message received by $p_{r'}$ has
        $r_{v'} < r'$.

        Proof: Immediate from the code.
    \item If a majority of accepters accept a proposal with number
        $r$ at some point during the execution,
        and $p_{r'}$ receives $\Ack(r',-,-)$ messages from a
        majority of accepters, then $p_{r'}$ receives 
        at least one $\Ack(r',v',r_{v'})$ message with $r_{v'} ≥ r$.

        Proof: Let $S$ be the set of processes
        that issue $\PaxosAccepted(r,v)$ and let $T$ be the set of processes
        that send $\Ack(r',-,-)$ to $p'$.
        Because $S$ and $T$ are both majorities, there is at least one
        accepter $a$ in $S∩T$.  Suppose $p_{r'}$ receives $\Ack(r,v'',r'')$
        from $a$.  If $r'' < r$, then at the time $a$ sends its
        $\Ack(r,v'',r'')$ message, it has not yet accepted a proposal
        with number $r$.  But then when it does receive
        $\PaxosAccept(r,v)$, it rejects it.  This contradicts $a∈S$.
\end{enumerate}

Suppose now that $p_{r'}$ proposes $v'$ with proposal number $r'$.
Then $p_{r'}$ received $\Ack(r',-,-)$ messages from a majority of
accepters, so it receives at least one $\Ack(r',v'',r'')$ message with
$r < r''$ (2) and no $\Ack(r',v'',r'')$ messages with $r''>r'$ (1).
So the maximum proposal number it sees is some $r''$ with
$r≤r''<r'$, which has corresponding value $v$ by the induction
hypothesis. So $p_{r'}$ proposes $v'=v$.

\section{Learning the results}

Somebody has to find out that a majority accepted a proposal in order
to get a decision value out.  The usual way to do this is to have a
fourth round of messages where the accepters send $\PaxosAccepted(v,
r)$ to some designated learners. These are often the processes that
need to implement whatever decision was made by the agreement
protocol, but in principle could be any processes that care about the
outcome.

\section{Liveness properties}

We'd like the protocol to terminate eventually.  Suppose there is a
single proposer, and that it survives long enough to collect a
majority of \Ack{}s and to send out \PaxosAccept{}s to a majority of
the accepters.  If everybody else cooperates, we get termination in 4
message delays, including the time for the learners to detect acceptance.

If there are multiple proposers, then they can step on each other.
For example, it's enough to have two carefully-synchronized proposers
alternate sending out \PaxosPrepare messages to prevent any accepter from
every accepting (since an accepter promises not to accept
$\PaxosAccept(r,v)$ once it has responded to $\PaxosPrepare(r+1)$).  The solution is
to ensure that there is eventually some interval during which there is
exactly one proposer who doesn't fail.  One way to do this is to use
exponential random backoff (as popularized by Ethernet): when a
proposer decides it's not going to win a round (e.g., by receiving a
\Nack or by waiting long enough to realize it won't be getting any
more acks soon), it picks some increasingly large random delay before
starting a new round.  Unless something strange is going on, 
new rounds will eventually start far enough apart in time that one will get done without interference.  

A more abstract solution is to assume some sort of weak leader election mechanism, which tells each accepter who the ``legitimate'' proposer is at each time.  The accepters then discard messages from illegitimate proposers, which prevents conflict at the cost of possibly preventing progress.  Progress is however obtained if the mechanism eventually reaches a state where a majority of the accepters bow to the same non-faulty proposer long enough for the proposal to go through.

Such a weak leader election method is an example of a more general
class of mechanisms known as \indexConcept{failure detector}{failure
detectors}, in which each process gets hints about what other
processes are faulty that eventually converge to reality.  The
weak-leader-election failure detector needed for Paxos is called the
$Ω$ failure detector~\cite{ChandraHT1996}, and there is a sense in
which it is the weakest possible failure detector that can be used to
solve consensus for $f < n/2$ using any algorithm.
We will discuss
failure detectors in detail in Chapter~\ref{chapter-failure-detectors}.

Since implementing this kind of leader election allows us to solve
consensus, the FLP result (Chapter~\ref{chapter-FLP}) implies that we
can't build it using only the tools available in the asynchronous
message-passing model. In practice, detecting failures and electing a
non-faulty leader involves using lots of timeouts. An example of a
Paxos-like protocol that does this is the Raft protocol of Ongaro and
Osterhout~\cite{OngaroO2014}, which may be the most commonly implemented
protocol in this family.

\section{Replicated state machines and multi-Paxos}
\label{section-multi-Paxos}
\label{section-replicated-state-machines}

The most common practical use of Paxos is to implement a
\concept{replicated state machine}~\cite{Lamport1978}.
The idea is to maintain many copies of some data structure, each on a
separate machine, and guarantee that each copy (or \concept{replica})
stays in sync with all the others as
new operations are applied to them.  
This requires some mechanism to
ensure that all the different replicas apply the same sequence of
operations, or in other words that the machines that hold the replicas
solve a sequence of agreement problems to agree on these operations.
The payoff is that the state of the data structure survives the failure of some of
the machines, without having to copy the entire structure every time
it changes.

Making all copies consistent requires solving a new version of
agreement every time we want to add another operation. Paxos works
well for this because we can have the proposer simply issue a new
proposal without taking into account any lower-numbered values,
assuming that it has verified that lower-numbered values have in fact
been accepted. The round-number mechanism means that all of the
accepters will switch to working on the new proposal without any
modifications to their code.

Typically for this application, we'll have a single active proposer
that is responsible for serializing any incoming operations to the
replicated state machine.
If the proposer doesn't change very often, a further
optimization allows skipping the \PaxosPrepare and \Ack messages in
between agreement protocols for consecutive operations.  This reduces
the time to certify each operation to a single round-trip for the
\PaxosAccept and \PaxosAccepted messages, which is about the best we
can reasonably hope for.

One detail is that to make this work, we need to distinguish between consecutive 
proposals by the same proposer, and ``new'' proposals that change the
proposer in addition to reaching agreement on some value.  This is
done by splitting the proposal number into a major and minor number,
with proposals ordered lexicographically.  A proposer that wins
$\Tuple{x,0}$ is allowed to make further proposals numbered
$\Tuple{x,1}, \Tuple{x,2},$ etc.  But a different proposer will need
to increment $x$.

Lamport calls this optimization Paxos in~\cite{Lamport2001}; other
authors have called it \concept{multi-Paxos} to distinguish it from
the basic Paxos algorithm.

\myChapter{Failure detectors}{2026}{}
\label{chapter-failure-detectors}

\index{detector!failure}\indexConcept{failure detector}{Failure
detectors} were proposed by Chandra and Toueg~\cite{ChandraT1996} as a
mechanism for solving consensus in an asynchronous message-passing
system with crash failures by distinguishing between slow processes
and dead processes. This involves extending the model by giving each process 
a \concept{failure detector}\index{detector!failure} module that continuously outputs an estimate
of which processes in the system have failed.  The output does not
need to be
correct; indeed, the main contribution of Chandra and Toueg's paper
(and a companion paper by Chandra, Hadzilacos, and
Toueg~\cite{ChandraHT1996}) is characterizing just how bogus the output of a failure detector can be and still be useful.

We will mostly follow Chandra and Toueg in these notes; see the paper for the full technical details.

To emphasize that the output of a failure detector is merely a hint at
the actual state of the world, a failure detector (or the process it's
attached to) is said to \concept{suspect} a process at time $t$ if it
outputs \DataSty{failed} at that time.  Failure detectors can then be classified based on when their suspicions are correct.

We use the usual asynchronous message-passing model, and in particular assume that non-faulty processes execute infinitely often, get all their messages delivered, etc.  From time to time we will need to talk about time, and unless we are clearly talking about real time this just means any steadily increasing count (e.g., of total events), and will be used only to describe the ordering of events.

\section{How to build a failure detector}

Failure detectors are only interesting if you can actually build them.
In a fully asynchronous system, you can't (this follows from the FLP
result and the existence of failure-detector-based consensus
protocols).  But with timeouts, it's not hard: have each process ping
each other process from time to time, and suspect the other process if
it doesn't respond to the ping within twice the maximum round-trip
time for any previous ping.  Assuming that ping packets are never lost
and there is an (unknown) upper bound on message delay, this gives
what is known as an \index{failure detector!eventually
perfect}\concept{eventually perfect failure detector}: once the max round-trip times rise enough and enough time has elapsed for the live processes to give up on the dead ones, all and only dead processes are suspected.

\section{Classification of failure detectors}

Chandra and Toueg define eight classes of failure detectors, based on when they suspect faulty processes and non-faulty processes.  Suspicion of faulty processes comes under the heading of \concept{completeness}; of non-faulty processes, \concept{accuracy}.

\subsection{Degrees of completeness}

\begin{description}
 \item[Strong completeness] Every faulty process is eventually permanently suspected by every non-faulty process.
 \item[Weak completeness] Every faulty process is eventually permanently suspected by some non-faulty process.
\end{description}

There are two temporal logic operators embedded in these statements:
``eventually permanently'' means that there is some time $t_{0}$ such
that for all times $t ≥ t_{0}$, the process is suspected.  Note that completeness says nothing about suspecting non-faulty processes: a paranoid failure detector that permanently suspects everybody has strong completeness.

\subsection{Degrees of accuracy}
These describe what happens with non-faulty processes, and with faulty processes that haven't crashed yet.

\begin{description}
 \item[Strong accuracy] No process is suspected (by anybody) before it crashes.
 \item[Weak accuracy] Some non-faulty process is never suspected.
 \item[Eventual strong accuracy] After some initial period of confusion, no process is suspected before it crashes.  This can be simplified to say that no non-faulty process is suspected after some time, since we can take end of the initial period of chaos as the time at which the last crash occurs.
 \item[Eventual weak accuracy] After some initial period of confusion, some non-faulty process is never suspected.
\end{description}

Note that ``strong'' and ``weak'' mean different things for accuracy vs completeness: for accuracy, we are quantifying over suspects, and for completeness, we are quantifying over suspectors.  Even a weakly-accurate failure detector guarantees that all processes trust the one visibly good process.

\subsection{Boosting completeness}
\label{section-failure-detector-boosting-completeness}

It turns out that any weakly-complete failure detector can be boosted
to give strong completeness.
Recall that the difference between weak completeness and strong completeness is that with weak completeness, somebody suspects a dead process, while with strong completeness, everybody suspects it.  So to boost completeness we need to spread the suspicion around a bit.  On the other hand, we don't want to break accuracy in the process, so there needs to be some way to undo a premature rumor of somebody's death.  The simplest way to do this is to let the alleged corpse speak for itself: I will suspect you from the moment somebody else reports you dead until the moment you tell me otherwise.

Pseudocode is given in Algorithm~\ref{alg-boosting-completeness}.
\begin{algorithm}
\newData{\BCsuspects}{suspects}
\Initially{
    $\BCsuspects ← \emptyset$\;
}

\While{\True}{
    Let $S$ be the set of all processes my weak detector suspects.\;
    Send $S$ to all processes.\;
}

\UponReceiving{$S$ from $q$}{
    $\BCsuspects ← (\BCsuspects ∪ S) ∖ \Set{q}$\;
    }
\caption{Boosting completeness}
\label{alg-boosting-completeness}
\end{algorithm}

It's not hard to see that this boosts completeness: if $p$ crashes,
somebody's weak detector eventually suspects it, this process tells
everybody else, and $p$ never contradicts it.  So eventually everybody
suspects $p$.

What is slightly trickier is showing that it preserves accuracy.  The
essential idea is this: if there is some good-guy process $p$ that
everybody trusts forever (as in weak accuracy), then nobody ever
reports $p$ as suspect—this also covers strong accuracy since the
only difference is that now every non-faulty process falls into this
category.  For eventual weak accuracy, wait for everybody to stop
suspecting $p$, wait for every message ratting out $p$ to be
delivered, and then wait for $p$ to send a message to everybody.  Now
everybody trusts $p$, and nobody every suspects $p$ again.  Eventual strong accuracy is again similar.

This will justify ignoring the weakly-complete classes.

\subsection{Failure detector classes}

Two degrees of completeness times four degrees of accuracy gives eight
classes of failure detectors, each of which gets its own name.
But since we can boost weak completeness to strong completeness, we
can use this as an excuse to consider only the strongly-complete
classes.

\begin{description}
 \item[$P$ (perfect)]\index{failure detector!perfect}
 \index{perfect failure detector} Strongly complete and strongly accurate: non-faulty processes are never suspected; faulty processes are eventually suspected by everybody.  Easily achieved in synchronous systems.
 \item[$S$ (strong)] \index{failure detector!strong}
 \index{strong failure detector}
 Strongly complete and weakly accurate.  The name is misleading if
 we've already forgotten about weak completeness, but the
 corresponding $W$ (weak) class is only weakly complete and weakly
 accurate, so it's the strong completeness that the $S$ is referring to.
 \item[$◇P$ (eventually perfect)]
 \index{failure detector!eventually perfect}
 \index{eventually perfect failure detector}
 Strongly complete and eventually strongly accurate.
 \item[$◇S$ (eventually strong)]
 \index{failure detector!eventually strong}
 \index{eventually strong failure detector}
 Strongly complete and eventually weakly accurate.
\end{description}

Jumping to the punch line: $P$ can simulate any of the others, $S$ and
$◇P$ can both simulate $◇S$ but can't simulate $P$
or each other, and $◇S$ can't simulate any of the others
(See Figure~\ref{fig-failure-detectors}—we'll prove all of this
later.)
Thus $◇S$ is the weakest class of failure detectors in this
list.  However, $◇S$ \emph{is} strong enough to solve
consensus, and in fact any failure detector (whatever its properties)
that can solve consensus is strong enough to simulate $◇S$
(this is the result in the Chandra-Hadzilacos-Toueg
paper~\cite{ChandraHT1996})—this makes $◇S$ the ``weakest
failure detector for solving consensus'' as advertised.
Continuing
our tour through Chandra and Toueg~\cite{ChandraT1996}, we'll show the simulation results and that $◇S$ can solve consensus, but we'll skip the rather involved proof of $◇S$'s special role from Chandra-Hadzilacos-Toueg.

\begin{figure}
    \centering
    \begin{tikzpicture}[auto,node distance=2cm]
        \node (P) {P};
        \node (S) [below left of=P] {$S$};
        \node (dP) [below right of=P] {$◇P$};
        \node (dS) [below right of=S] {$◇S$};
        \path
            (P) edge (S) edge (dP)
            (S) edge (dS)
            (dP) edge (dS)
            ;
    \end{tikzpicture}
    \caption[Failure detector classes]{Partial order of failure detector classes.  
    Higher classes can simulate lower classes but not vice versa.}
    \label{fig-failure-detectors}
\end{figure}

\section{Consensus with \texorpdfstring{$S$}{S}}
\label{section-consensus-with-strong}

With the strong failure detector $S$, we can solve consensus for any
number of failures.

In this model, the failure detectors as applied to most processes are completely
useless.  However, there is some non-faulty process $c$ that nobody
every suspects, and this is enough to solve consensus with as many as
$n-1$ failures.

The protocol is carried out in three phases.  In the first
phase, the processes gossip about input values for $n-1$ asynchronous
rounds.  In the second, they exchange all the values they've seen and
prune out any that are not universally known.  In the third, each
process decides on the lowest-id input that hasn't been pruned
(minimum input also works since at this point everybody has the same view of the inputs).

\begin{algorithm}
    $V_p \gets \Set{\Tuple{p,v_p}}$ \tcp{All values known to $p$}
    $δ_p \gets \Set{\Tuple{p,v_p}}$ \tcp{New values $p$ learned last round}
    \tcp{Phase 1: add values} 
    \For{$i \gets 1$ \KwTo $n-1$}{
        Send $\Tuple{i, δ_p}$ to all processes.\;
        Wait to receive $\Tuple{i, δ_q}$ from all $q$ I do not
        suspect.\;
        $δ_p \gets \left(\bigcup_{q} δ_q\right) ∖ V_p$\;
        $V_p \gets \left(\bigcup_{q} δ_q\right) ∪ V_p$\;
    }
    \tcp{Phase 2: subtract values}
    Send $\Tuple{n, V_p}$ to all processes.\;
    Wait to receive $\Tuple{n, V_q}$ from all $q$ I do not
    suspect.\;
    $V_p \gets \left(\bigcap_{q} V_q\right) ∩ V_p$\;
    \tcp{Phase 3: decide on something}
    \Return some input from $V_p$ chosen via a consistent rule.\;
    \caption{Consensus with a strong failure detector}
    \label{alg-strong-failure-detector-consensus}
\end{algorithm}

Pseudocode is given in
Algorithm~\ref{alg-strong-failure-detector-consensus}

In Phase 1, each process $p$ maintains two partial
functions $V_{p}$ and $δ_{p}$, where $V_{p}$ lists all the
input values $\Tuple{q,v_{q}}$ that $p$ has ever seen and $δ_{p}$
lists only those input values seen in the most recent of $n-1$
asynchronous rounds.  Both $V_{p}$ and $δ_{p}$ are initialized
to $\Set{\Tuple{p,v_{p}}}$.  In round $i$, $p$ sends $(i,δ_{p})$ to
all processes.  It then collects $\Tuple{i,δ_{q}}$ from each $q$ that
it doesn't suspect and sets $δ_{p}$ to
$\bigcup_{q} δ_{q} ∖ V_{p}$ (where $q$ ranges over the
processes from which $p$ received a message in round $i$) and sets
$V_{p}$ to $V_{p} \cup δ_{p}$.  In the next round, it repeats the
process.  Note that each pair $\Tuple{q,v_{q}}$ is only sent by a particular
process $p$ the first round after $p$ learns it: so any value that is
still kicking around in round $n-1$ had to go through $n-1$ processes.

In Phase 2, each process $p$ sends $\Tuple{n,V_{p}}$, waits to receive
$\Tuple{n,V_{q}}$ from every process it does not suspect, and sets $V_{p}$
to the intersection of $V_{p}$ and all received $V_{q}$.  At the end
of this phase all $V_{p}$ values will in fact be equal, as we will show.

In Phase 3, everybody picks some input from their $V_{p}$ vector according to a consistent rule.

\subsection{Proof of correctness}
Let $c$ be a non-faulty process that nobody every suspects.

The first observation is that the protocol satisfies validity, since
every $V_{p}$ contains $v_{c}$ after round 1 and each $V_{p}$ can only
contain input values by examination of the protocol.   Whatever it may
do to the other values, taking intersections in Phase 2 still leaves
$v_{c}$, so all processes pick some input value from a nonempty list in Phase 3.

To get termination we have to prove that nobody ever waits forever for
a message it wants; this basically comes down to showing that the
first non-faulty process that gets stuck eventually is informed by the
$S$-detector that the process it is waiting for is dead.

For agreement, we must show that in Phase 3, every $V_{p}$ is equal;
in particular, we'll show that every $V_{p} = V_{c}$.  First it is
necessary to show that at the end of Phase 1, $V_{c} \subseteq V_{p}$
for all $p$.  This is done by considering two cases:
\begin{enumerate}
 \item If $\Tuple{q,v_{q}} \in V_{c}$ and $c$ learns $\Tuple{q,v_{q}}$ before
 round $n-1,$ then $c$ sends $\Tuple{q,v_{q}}$ to $p$ no later than round
 $n-1$, $p$ waits for it (since nobody ever suspects $c$), and adds it
 to $V_{p}$.
 \item If $\Tuple{q,v_{q}} \in V_{c}$ and $c$ learns $\Tuple{q,v_{q}}$ only in
 round $n-1,$ then $\Tuple{q,v_{q}}$ was previously sent through $n-1$ other
 processes, i.e., all of them.  Each process $p \ne c$ thus added
 $\Tuple{q,v_{q}}$ to $V_{p}$ before sending it and again $\Tuple{q,v_{q}}$ is in
 $V_{p}$.
\end{enumerate}

(The missing case where $\Tuple{q,v_{q}}$ isn't in $V_{c}$ we don't care about.)

But now Phase 2 knocks out any extra elements in $V_{p}$, since
$V_{p}$ gets set to $V_{p}\cap V_{c}\cap (\text{some other $V_{q}$'s
that are supersets of $V_{c}$)}$.  It follows that, at the end of Phase
2, $V_{p} = V_{c}$ for all $p$.  Finally, in Phase 3, everybody applies the same selection rule to these identical sets and we get agreement.

\section{Consensus with \texorpdfstring{$◇S$}{<>S} and
\texorpdfstring{$f < n/2$}{f < n/2}}
\label{section-consensus-eventually-strong}
\label{section-Chandra-Toueg}

The consensus protocol for $S$ depends on some process $c$ never being
suspected; if $c$ is suspected during the entire (finite) execution of
the protocol—as can happen with $◇S$—then it is possible
that no process will wait to hear from $c$ (or anybody else) and the
processes will all decide their own inputs.  So to solve consensus
with $◇S$ we will need to assume fewer than $n/2$ failures, allowing any process to wait to hear from a majority no matter what lies its failure detector is telling it.

The resulting protocol, known as the 
\index{consensus!Chandra-Toeug}\concept{Chandra-Toueg consensus
protocol}, is structurally similar to the consensus protocol in
Paxos.\footnote{See Chapter~\ref{chapter-Paxos}.}  The difference is
that instead of proposers blindly showing up, the protocol is divided
into rounds with a rotating \concept{coordinator} $p_{i}$ in each
round $r$ with $r = i \pmod{n}$.  The termination proof is based on showing that in any round where the coordinator is not faulty and nobody suspects it, the protocol finishes.

The consensus protocol uses as a subroutine a
protocol for \index{broadcast!reliable}
\concept{reliable broadcast}, which guarantees that any message that
is sent is either received by no non-faulty processes or exactly once by all non-faulty processes.
Pseudocode for reliable broadcast is given as
Algorithm~\ref{alg-reliable-broadcast}. 
It's easy to see that if a process $p$ is non-faulty and receives $m$,
then the fact that $p$ is non-faulty means that is successfully sends
$m$ to everybody else, and that the other non-faulty processes also receive the message at least once and deliver it.

\newFunc{\CTbroadcast}{broadcast}
\newFunc{\CTbroadcasrReceive}{receiveBroadcast}

\begin{algorithm}
\caption{Reliable broadcast}
\label{alg-reliable-broadcast}
\Procedure{$\CTbroadcast(m)$}{
    send $m$ to all processes.\;
}
\UponReceiving{$m$}{
    \If{I haven't seen $m$ before}{
        send $m$ to all processes\;
        deliver $m$ to myself\;
    }
}
\end{algorithm}

Here's a sketch of the actual consensus protocol:
\begin{itemize}
 \item Each process keeps track of a preference (initially its own input) and a timestamp, the round number in which it last updated its preference.
 \item The processes go through a sequence of asynchronous rounds, each divided into four phases:
\begin{enumerate}
  \item All processes send (round, preference, timestamp) to the coordinator for the round.
  \item The coordinator waits to hear from a majority of the processes (possibly including itself).  The coordinator sets its own preference to some preference with the largest timestamp of those it receives and sends (round, preference) to all processes.
  \item Each process waits for the new proposal from the coordinator
  \emph{or} for the failure detector to suspect the coordinator.  If
  it receives a new preference, it adopts it as its own, sets timestamp
  to the current round, and sends (round, ack) to the coordinator.  Otherwise, it sends (round, nack) to the coordinator.
  \item The coordinator waits to receive ack or nack from a majority
  of processes.  If it receives ack from a majority, it announces the
  current preference as the protocol decision value using reliable
  broadcast.
\end{enumerate}
 \item Any process that receives a value in a reliable broadcast decides on it immediately.
\end{itemize}

Pseudocode is in Algorithm~\ref{alg-Chandra-Toueg}.

\begin{algorithm}
    \newData{\CTpreference}{preference}
    \newData{\CTtimestamp}{timestamp}
    \newData{\CTround}{round}
    $\CTpreference \gets \Input$\;
    $\CTtimestamp \gets 0$\;
    \For{$\CTround \gets 1 \dots \infty$}{
        Send $\Tuple{\CTround, \CTpreference, \CTtimestamp}$ to
        coordinator\;
        \If{I am the coordinator}{
            Wait to receive $\Tuple{\CTround, \CTpreference, \CTtimestamp}$ from majority of processes.\;
            Set $\CTpreference$ to value with largest
            $\CTtimestamp$.\;
            Send $\Tuple{\CTround, \CTpreference}$ to all processes.\;
        }
        Wait to receive $\Tuple{\CTround, \CTpreference'}$ from
        coordinator or to suspect coordinator.\;
        \eIf{I received $\Tuple{\CTround, \CTpreference'}$} {
            $\CTpreference \gets \CTpreference'$\;
            $\CTtimestamp \gets \CTround$\;
            Send $\Ack(\CTround)$ to coordinator.\;
        }{
            Send $\Nack(\CTround)$ to coordinator.\;
        }
        \If{I am the coordinator}{
            Wait to receive $\Ack(\CTround)$ or $\Nack(\CTround)$ from a majority of
            processes.\;
            \If{I received no $\Nack(\CTround)$ messages}{
                Broadcast $\CTpreference$ using reliable broadcast.\;
            }
        }
    }
    \caption{Consensus with an eventually-strong failure detector}
    \label{alg-Chandra-Toueg}
\end{algorithm}

\subsection{Proof of correctness}
For validity, observe that the decision value is an estimate and all estimates start out as inputs.

For termination, observe that no process gets stuck in Phase 1, 2, or 4, because either it isn't waiting or it is waiting for a majority of non-faulty processes who all sent messages unless they have already decided (this is why we need the nacks in Phase 3).  The loophole here is that processes that decide stop participating in the protocol; but because any non-faulty process retransmits the decision value in the reliable broadcast, if a process is waiting for a response from a non-faulty process that already terminated, eventually it will get the reliable broadcast instead and terminate itself.  In Phase 3, a process might get stuck waiting for a dead coordinator, but the strong completeness of $◇S$ means that it suspects the dead coordinator eventually and escapes.  So at worst we do finitely many rounds.

Now suppose that after some time $t$ there is a process $c$ that is
never suspected by any process.  Then in the next round in which $c$
is the coordinator, in Phase 3 all surviving processes wait for $c$
and respond with ack, $c$ decides on the current estimate, and triggers the reliable broadcast protocol to ensure everybody else decides on the same value.  Since reliable broadcast guarantees that everybody receives the message, everybody decides this value \emph{or some value previously broadcast}—but in either case everybody decides.

Agreement is the tricky part.  It's possible that two coordinators
both initiate a reliable broadcast and some processes choose the value
from the first and some the value from the second.  But in this case
the first coordinator collected acks from a majority of processes in
some round $r$, and all subsequent coordinators collected estimates
from an overlapping majority of processes in some round $r' > r$.  By
applying the same induction argument as for Paxos, we get that all subsequent coordinators choose the same estimate as the first coordinator, and so we get agreement.

\section{\texorpdfstring{$f < n/2$}{f < n/2} is still required even
with \texorpdfstring{$◇P$}{<>P}}
\label{section-failure-detector-eventually-perfect-requires-majority}

We can show that with a majority of failures, we're in trouble with just $◇P$ (and thus with $◇S$, which is trivially simulated by $◇P$).  The reason is that $◇P$ can lie to us for some long initial interval of the protocol, and consensus is required to terminate eventually despite these lies.  So the usual partition argument works: start half of the processes with input 0, half with 1, and run both halves independently with $◇P$ suspecting the other half until the processes in both halves decide on their common inputs.  We can now make $◇P$ happy by letting it stop suspecting the processes, but it's too late.

\section{Relationships among the classes}
\label{section-failure-detector-relationships}

It's easy to see that $P$ simulates $S$ and $◇P$ simulates
$◇S$ without modification.  It's also immediate that $P$
simulates $◇P$ and $S$ simulates $◇S$ (make
``eventually'' be ``now''), which gives a diamond-shaped lattice
structure between the classes.  What is trickier is to show that this
structure doesn't collapse: 
$◇P$
can't simulate $S$, $S$ can't simulate $◇P$, and
$◇S$ can't simulate any of the
other classes.

First let's observe that $◇P$ can't simulate $S$:
if it could, we would get a consensus protocol for $f ≥ n/2$
failures, which we can't do.  It follows that $◇P$ also can't
simulate $P$ (because $P$ can simulate $S$).

To show that $S$ can't simulate $◇P$, choose some non-faulty
victim process $v$ and consider an execution in which $S$ periodically
suspects $v$ (which it is allowed to do as long as there is some other
non-faulty process it never suspects).  If the $◇P$-simulator
ever responds to this by refusing to suspect $v$, there is an
execution in which $v$ really is dead, and the simulator violates
strong completeness.  But if not, we violate eventual strong accuracy.
Note that this also implies $S$ can't simulate $P$, since $P$ can
simulate $◇P$.  It also shows that $◇S$ can't
simulate either of $◇P$ or $P$.

We are left with showing $◇S$ can't simulate $S$.  Consider a
system where $p$'s $◇S$ detector suspects $q$ but not $p$
from the start of the execution. Run $p$ 
until $p$'s $S$-simulator gives up and suspects $q$, which it
must do eventually by strong completeness, since this run is
indistinguishable from one in which $q$ is faulty. Then wake up $q$
and crash $p$.  Since $q$ is the only non-faulty process, and the
alleged $S$-simulator suspected it, we've violated weak accuracy.

\section{Terminating reliable broadcast with \texorpdfstring{$P$}{P}}

If we look carefully at the arguments so far, we haven't actually
shown anything that $P$ is good for: we only know that $S$ and $◇P$
can't simulate $P$ because neither can simulate the other. This raises
the obvious question of whether there is something we might actually
want to do that requires $P$.

Chandra and Toueg~\cite{ChandraT1996} give as an example of a natural
problem that can be solved only with $P$ the problem of
\index{broadcast!terminating reliable} \index{reliable
broadcast!terminating} \concept{terminating reliable broadcast}.  In
this problem, a leader process $\ell$ sends a message $m$, and all
processes eventually decide on $m$ or a no-message value $⊥$.
Validity in this case says that if $\ell$ is non-faulty, every
non-faulty process decides $m$. Agreement says that all non-faulty
processes must decide the same value (which will be one of $m$ or $⊥$)
whether $\ell$ is faulty or not. Terminating is the usual condition
that all processes eventually decide on some value.

This problem is equivalent to having the processes reach consensus on
a value that defaults to $⊥$ if no message is received from $\ell$.
Since $P$ implements $S$, we can do this using our already-known
algorithm for solving consensus for any number of failures using $S$.
The resulting algorithm runs in two phases:
\begin{enumerate}
    \item In the first phase, $\ell$ transmits $m$ to all processes,
        and each process waits to either receive $m$ (and use $m$ as
        the input to the next phase) or suspect $\ell$ (and use $⊥$ as
        the input to the next phase).
    \item In the second phase, use
        Algorithm~\ref{alg-strong-failure-detector-consensus} to reach
        agreement on $m$ or $⊥$. (We can do this because $P$ is also
        an instance of $S$.)
\end{enumerate}

If $\ell$ is non-faulty, all non-faulty processes start the consensus
phase with $m$ and end with $m$. Whether $\ell$ is faulty or not, all
non-faulty processes end the consensus phase with the same value. So
validity and agreement are satisfied.

It's not hard to see that we can't solve terminating reliable
broadcast with either $S$ or $◇P$. 
If we try to solve it using $S$,
the weak accuracy of $S$ means that some non-faulty $p$ is never
suspected, but $p$ doesn't have to be $\ell$. So if all the processes
start off suspecting $\ell$, either they wait forever for a faulty
$\ell$ to wake up (violating termination), or they finish the protocol
and decide on the wrong value before a non-faulty $\ell$ wakes up (violating
validity). 
The same argument works for $◇P$: during the initial period of
confusion, a non-faulty $\ell$ might be suspected by all processes,
and if we wait to decide until $\ell$ starts sending messages or
becomes non-suspect, we violate termination in the case where $\ell$
really is faulty.

\myChapter{Quorum systems}{2014}{}
\label{chapter-quorum-systems}

\section{Basics}

In the past few chapters,
we've seen many protocols that depend on the fact that if I talk to more than
$n/2$ processes and you talk to more than $n/2$ processes, the two
groups overlap.  This is a special case of a \concept{quorum system}, a family of subsets of the set of processes with the property that any two subsets in the family overlap.  By choosing an appropriate family, we may be able to achieve lower load on each system member, higher availability, defense against Byzantine faults, etc.

The exciting thing from a theoretical perspective is that these turn a
systems problem into a combinatorial problem: this means we can ask combinatorialists how to solve it.

\section{Simple quorum systems}
\begin{itemize}
 \item Majority and weighted majorities
 \item Specialized read/write systems where write quorum is a column and read quorum a row of some grid.
 \item Dynamic quorum systems: get more than half of the most recent copy.
 \item Crumbling walls~\cite{PelegW1997dc,PelegW1997dam}: optimal
 small-quorum system for good choice of wall sizes.
\end{itemize}

\section{Goals}
\begin{itemize}
 \item Minimize \concept{quorum size}.
 \item Minimize \concept{load}, defined as the minimum over all access
 strategies (probability distributions on quorums) of the maximum over all servers of probability it gets hit.
 \item Maximize \concept{capacity}, defined as the maximum number of quorum accesses per time unit in the limit if each quorum access ties up a quorum member for 1 time unit (but we are allowed to stagger a quorum access over multiple time units).
 \item Maximize \indexConcept{fault-tolerance!quorum system}{fault-tolerance}: minimum number of server failures that blocks all quorums.  Note that for standard quorum systems this is directly opposed to minimizing quorum size, since killing the smallest quorum stops us dead.
 \item Minimize \indexConcept{failure probability!quorum system}{failure probability} = probability that every quorum contains at least one bad server, assuming each server fails with independent probability.
\end{itemize}

Naor and Wool~\cite{NaorW1998} describe trade-offs between these goals (some of these were previously known, see the paper for citations):
\begin{itemize}
 \item $\text{capacity} = 1/\text{load}$; this is obtained by selecting the quorums independently at random according to the load-minimizing distribution.  In particular this means we can forget about capacity and just concentrate on minimizing load.
 \item $\text{load} ≥ \max(c/n, 1/c)$ where $c$ is the minimum
 quorum size.
 The first case is obvious: if every access hits $c$ nodes, spreading
 them out as evenly as possible still hits each node $c/n$ of the
 time.  The second is trickier: Naor and Wool prove it using LP
 duality, but the argument essentially says that if we have some
 quorum $Q$ of size $c$, then since every other quorum $Q'$ intersects $Q$ in at
 least one place, we can show that every $Q'$ adds at least $1$
 unit of load in total to the $c$ members of $Q$.  So if we pick a random
 quorum $Q'$, the average load added to all of $Q$ is at least $1$, so
 the average load added to some particular element of $Q$ is at least $1/\card*{Q}
 = 1/c$.  Combining the two cases, we can't hope to get load better
 than $1/\sqrt{n}$, and to get this load we need quorums of size at
 least $\sqrt{n}$.
 \item failure probability is at least $p$ when $p > 1/2$ (and optimal
 system is to just pick a single leader in this case), failure
 probability can be made exponentially small in size of smallest
 quorum when $p < 1/2$ (with many quorums).  These results are due to
 Peleg and Wool~\cite{PelegW1995}.
\end{itemize}

\section{Paths system}

This is an optimal-load system from Naor and Wool~\cite{NaorW1998} with exponentially low failure probability, based on percolation theory.

For this system, we build a $d\times{}d$ mesh-like graph where a
quorum consists of the union of a top-to-bottom path (TB path) and a
left-to-right path (LR path); this gives quorum size $O(\sqrt{n})$ and
load $O(1/\sqrt{n})$.  Note that the TB and LR paths are not necessarily
direct: they may wander around for a while in order to get where they
are going, especially if there are a lot of failures to avoid.
But the smallest quorums will have size $2d+1 = O(\sqrt{n})$.

\begin{figure}
\centering
\includegraphics[scale=0.5]{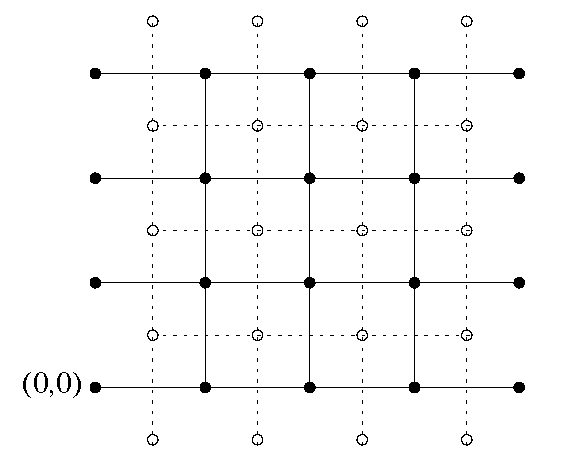}
\caption[Figure 2 from~\cite{NaorW1998}]{Figure 2
    from~\cite{NaorW1998}.  Solid lines are $G(3)$;
dashed lines are $G^*(3)$.}
\label{figure-naor-wool-grid}
\end{figure}

The actual mesh is a little more complicated.
Figure~\ref{figure-naor-wool-grid} reproduces the picture of
the $d=3$ case from the Naor and Wool paper.

Each server corresponds to a \emph{pair} of intersecting edges, one
from the $G(d)$ grid and one from the $G^{*}(d)$ grid (the star
indicates that $G^{*}(d)$ is the \concept{dual graph}\footnote{The dual of a graph $G$
embedded in the plane has a vertex for each region of $G$, and an edge
connecting each pair of vertices corresponding to adjacent regions,
where a region is a subset of the plane that is bounded by
edges of $G$.} of $G(d)$.  A quorum consists of a set of servers that
produce an LR path in $G(d)$ and a TB path in $G^{*}(d)$.  Quorums
intersect, because any LR path in $G(d)$ must cross some TB path in
$G^{*}(d)$ at some server (in fact, each pair of quorums intersects in
at least two places).  The total number of elements $n$ is $(d+1)^{2}$
and the minimum size of a quorum is $2d+1 = Θ(\sqrt{n})$.

The symmetry of the mesh gives that there exists a LR path in the mesh
if and only if there does not exist a TB path in its
\concept{complement}, the graph that has an edge only if the mesh
doesn't.  For a mesh with failure probability $p < 1/2$, the
complement is a mesh with failure probability $q = 1-p > 1/2$.  Using
results in percolation theory, it can be shown that for failure
probability $q > 1/2$, the probability that there exists a
left-to-right path is exponentially small in $d$ (formally, for each
$p$ there is a constant $\phi(p)$ such that $\Pr[\exists \text{LR
path}] ≤ \exp(-\phi(p)d)$).  We then have 
\begin{align*}
\Pr[\exists \text{(live quorum)}] 
&
= \Pr[\exists \text{(TB path)} ∧ \exists \text{(LR path)}]
\\&
= \Pr[¬\exists \text{(LR path in complement)} ∨ ¬\exists
\text{(TB path in complement)}] 
\\&
≤ \Pr[¬\exists \text{(LR path in complement)}] + \Pr[¬\exists \text{(TB path in complement)}]
\\&
≤ 2 \exp(-\phi(1-p)d)
\\&
= 2 \exp(-Θ(\sqrt{n})).
\end{align*}
So the failure probability of this system is exponentially small for
any fixed $p < 1/2$.

See the paper~\cite{NaorW1998} for more details.

\section{Byzantine quorum systems}
Standard quorum systems are great when you only have crash failures,
but with Byzantine failures you have to worry about finding a quorum
that includes a Byzantine serve who lies about the data.  For this
purpose you need something stronger.  Following
Malkhi and Reiter~\cite{MalkhiR1998}
and
Malkhi~\etal~\cite{MalkhiRWW2001},
one can define:

\begin{itemize}
 \item A 
 \index{quorum system!$b$-disseminating}
 \concept{$b$-disseminating quorum system} guarantees
 $\card*{Q_{1}\cap{}Q_{2}} ≥ b+1$ for all quorums $Q_{1}$ and
 $Q_{2}$.  This guarantees that if I update a quorum $Q_{1}$ and you
 update a quorum Q$_{2}$, and there are at most $b$ Byzantine
 processes, then there is some non-Byzantine process in both our
 quorums.  Mostly useful if data is ``self-verifying,'' that is, signed with digital signatures that the Byzantine processes can't forge.  Otherwise, I can't tell which of the allegedly most recent data values is the right one since the Byzantine processes lie.
 \item A 
 \index{quorum system!$b$-masking}
 \concept{$b$-masking quorum system} guarantees 
 $\card*{Q_{1}\cap{}Q_{2}} ≥ 2b+1$ for all quorums $Q_{1}$ and
 $Q_{2}$.  (In other words, it's the same as a $2b$-disseminating
 quorum system.)  This allows me to defeat the Byzantine processes
 through voting: given $2b+1$ overlapping servers, if I want the most
 recent value of the data I take the one with the most recent
 timestamp that appears on at least $b+1$ servers, which the Byzantine guys can't fake.
\end{itemize}

An additional requirement in both cases is that for any set of servers
$B$ with $\card*{B} ≤ b$, there is some quorum $Q$ such that
$Q\cap{}B = \emptyset$.  This prevents the Byzantine processes from stopping the system by simply refusing to participate.

Note: these definitions are based on the assumption that there is some
fixed bound on the number of Byzantine processes.  Malkhi and
Reiter~\cite{MalkhiR1998} give more complicated definitions for the
case where one has an arbitrary family $\{ \mathcal{B} \}$ of
potential Byzantine sets.  The definitions above are actually
simplified versions from~\cite{MalkhiRWW2001}.

The simplest way to build a $b$-disseminating quorum system is to use
supermajorities of size at least $(n+b+1)/2$; the overlap between any
two such supermajorities is at least $(n+b+1)-n = b+1$.  This gives a
load of substantially more than $\frac{1}{2}$.  There are better
constructions that knock the load down to $Θ(\sqrt{b/n})$;
see~\cite{MalkhiRWW2001}.

For more on this topic in general, see the survey by
by Merideth and Reiter~\cite{MeridethR2010}.

\section{Probabilistic quorum systems}
The problem with all standard (or 
\index{quorum system!strict}
\indexConcept{strict quorum system}{strict}) quorum systems is that we need big quorums to get high fault tolerance, since the adversary can always stop us by knocking out our smallest quorum.  A 
\index{quorum system!probabilistic}
\concept{probabilistic quorum system} or more specifically an 
\index{quorum system!$ε$-intersecting}
\concept{$ε$-intersecting quorum system}~\cite{MalkhiRWW2001} 
improves the fault-tolerance by relaxing the
requirements.  For such a system we have not only a set system $Q$,
but also a probability distribution $w$ supplied by the quorum system
designer, with the property that $\Pr[Q_{1}\cap Q_{2} = \emptyset] ≤
ε$ when $Q_{1}$ and $Q_{2}$ are chosen independently according to their weights.

\subsection{Example}
Let a quorum be any set of size $k\sqrt{n}$ for some $k$ and let all
quorums be chosen uniformly at random.  Pick some quorum $Q_{1}$; what
is the probability that a random $Q_{2}$ does not intersect $Q_{1}$?
Imagine we choose the elements of $Q_{2}$ one at a time.  The chance
that the first element $x_{1}$ of $Q_{2}$ misses $Q_{1}$ is exactly
$(n-k\sqrt{n})/n = 1 - k/\sqrt{n}$, and conditioning on $x_{1}$
through $x_{i-1}$ missing $Q_{1}$ the probability that $x_{i}$ also
misses it is $(n-k\sqrt{n}-i+1)/(n-i+1) ≤ (n-k\sqrt{n})/n = 1 -
k/\sqrt{n}$.  So taking the product over all $i$ gives $\Pr[\text{all
miss $Q_{1}$}] ≤ (1-k/\sqrt{n})^{k\sqrt{n}} ≤
\exp(-k\sqrt{n})^{k/\sqrt{n})} = \exp(-k^{2})$.  So by setting $k =
Θ(\ln 1/ε)$, we can get our desired $ε$-intersecting system.

\subsection{Performance}
Failure probabilities, if naively defined, can be made arbitrarily
small: add low-probability singleton quorums that are hardly ever
picked unless massive failures occur.  But the resulting system is
still $ε$-intersecting.

One way to look at this is that it points out a flaw in the
$ε$-intersecting definition: $ε$-intersecting quorums
may cease to be $ε$-intersecting conditioned on a particular
failure pattern (e.g., when all the non-singleton quorums are knocked
out by massive failures).  But Malkhi~\etal~\cite{MalkhiRWW2001} address the
problem in a different way, by considering only survival of 
\index{quorum!high quality}\indexConcept{high quality quorum}{high
quality} quorums, where a particular quorum $Q$ is 
\index{quorum!$δ$-high-quality}
\indexConcept{$δ$-high-quality quorum}{$δ$-high-quality} if 
$\Pr[Q_{1}\cap Q_{2} = \emptyset | Q_{1} = Q] ≤ δ$ and high
quality if it's $\sqrt{ε}$-high-quality. It's not hard to show
that a random quorum is $δ$-high-quality with probability at
least $ε/δ$, so a high quality quorum is one that fails to
intersect a random quorum with probability at most $\sqrt{ε}$
and a high quality quorum is picked with probability at least
$1-\sqrt{ε}$.

We can also consider load; Malkhi~\etal~\cite{MalkhiRWW2001} show that
essentially the same bounds on load for strict quorum systems also
hold for $ε$-intersecting quorum systems: 
$\QuorumLoad(S) ≥ \max((\E(\card*{Q})/n,
(1-\sqrt{ε})^{2}/\E(\card*{Q}))$, where $\E(\card*{Q})$ is the
expected size of a quorum.  The left-hand branch of the max is just
the average load applied to a uniformly-chosen server.  For the
right-hand side, pick some high quality quorum $Q'$ with size less
than or equal to $(1-\sqrt{ε})\E(\card*{Q})$ and consider the
load applied to its most loaded member by its nonempty intersection
(which occurs with probability at least $1-\sqrt{ε})$ with a random quorum.

\section{Signed quorum systems}
A further generalization of probabilistic quorum systems gives
\index{quorum system!signed}
\indexConcept{signed quorum system}{signed quorum
systems}~\cite{Yu2006}.
In these systems, a quorum consists of
some set of positive members (servers you reached) and negative
members (servers you tried to reach but couldn't).  These allow
$O(1)$-sized quorums while tolerating $n-O(1)$ failures, under certain natural probabilistic assumptions.  Because the quorums are small, the load on some servers may be very high: so these are most useful for fault-tolerance rather than load-balancing.  See the paper for more details.

\myChapter{Permissionless systems}{2026}{}
\label{chapter-permissionless-systems}

All of the results we have considered so far for message-passing
systems have made a critical assumption: the number of processes $n$
is known and fixed, so we can sensibly talk about things like
majorities of processes, fewer than $n/3$ Byzantine faults, and so on.
This assumption is not unreasonable for systems operated by a single
organization, but it may not make sense for large distributed systems
that can in principle be joined by anybody. In this case, to solve a
problem like agreement, we need some mechanism other than simply
counting machines to produce overlapping quorums or to outvote
Byzantine coalitions.

This is particularly tricky because it is possible for a single
machine on the Internet to masquerade as many, by using routing
trickery to simulate many distinct IP addresses. This is not something
we can practically remove from the IP protocol stack, since it's used
for positive ends like allowing a single machine to simulate multiple
low-use servers or, in the other direction, allowing a warehouse full
of machines to simulate a single high-use server. But this possibility
allows for a \concept{Sybil attack}~\cite{Douceur2002}, where an algorithm naively
implemented on the assumption that faulty processes form a small
minority is suddenly overwhelmed by a single faulty machine backed by
an army of virtual clones. This requires re-examining how (or if) we
can achieve consensus in systems that allow arbitrary participants.

The current dominant strategy for doing so is to use cryptographic
mechanisms to substitute majorities expressed in terms of unforgeable resources
like computing power, storage, or simulated currencies for majorities
expressed in mere counts of IP addresses. This is often coupled with a
certified replicated-state-machine approach that replaces agreement
with weaker various \concept{eventual
consistency}\index{consistency!eventual} guarantees, all of which 
is encompassed by the notion of a \concept{blockchain}, which has no
universally-accepted formal definition, but which we can
think of roughly as a global-scale replicated-state-machine algorithm 
that allows arbitrary participants
and enforces consistency in the long run using a combination of
cryptographic tools and social engineering.

The blockchain world is a bit of a moving target, and constructing a
practical blockchain that will see wide adoption involves solving a number of
economic and social issues that go beyond simply putting together the
right technology. But from a distributed-systems perspective, we can
look at the systems that people have actually built and try to learn
something from them. In this chapter, we'll start by looking at the
problem of defending against arbitrary participants in a distributed
system, and then look at how the Bitcoin system~\cite{Nakamoto2008}
appears to do so successfully even though it arguably shouldn't.

\section{Sybil attacks}

The idea of the \concept{Sybil attack}
is that one bad machine can masquerade as many different machines
using routing tricks. This defeats any distributed algorithm based on
assuming a fixed fraction of the processes are bad. This is
particularly difficult to defend against in the current Internet as
most machines are now buried behind Network Address Translation (NAT)
mechanisms to allow a single IP to be shared between multiple
machines, making it trivial for an army of bogus clones to masquerade
as separate machines behind a NAT.

Whatever the source of the bogus clones, they are a problem for any
system with open admissions, where any machine on the network can join
it. Examples of such systems are the SMTP-based mail system, the
HTTP-based World-Wide Web, and many multiplayer games. The openness of
these systems makes them inherently vulnerable to malicious actors
(spammers for SMTP, various kinds of undesirable users for HTTP,
cheaters in games), especially if new identities can be manufactured
for free.

The name ''Sybil attack'' was popularized by a paper by John
Douceur~\cite{Douceur2002}, in a paper that analyzes several methods
for attempting to defeat them.  The term itself is
credited to Brian Zill in Douceur's paper, and is based on the book
(and later movie) \emph{Sybil}~\cite{Schreiber1973} about a
psychiatric patient diagnosed with multiple-personality disorder.

Note that Sybil attacks do not include attacks using botnets, where
the problem is that we really do have 10,000,000 bad nodes
overwhelming our system; instead, we are worrying about the case where
a bad router can claim to have 10,000,000 bad nodes behind it but
these nodes are simulated by only a small number of machines.

\subsection{Resource-based defenses}

Douceur considers an abstract model involving interactions between
\indexConcept{entity}{entities}, which may or may not correspond to
actual machines. (For consistency with the rest of these notes, we
will just call them ``processes.'') The communication model is a generic broadcast
channel (called a ``cloud'' in Douceur's terminology) that, unlike our
usual model, does not record the source of messages. It is assumed
that processes are computationally bounded, allowing the use of
public-key cryptography. In particular, an process can establish an
\concept{identity} by creating a public/secret key pair and signing
all of its messages using the secret key.

Non-faulty processes will do this once, establishing a single
\concept{legitimate} identity. Faulty processes will attempt to
construct as many extra \concept{counterfeit} identities as they can
get away with. 

Assuming that there is no external agent (like a
centralized certificate authority) that prevents them from doing this,
we need a mechanism to constrain how many identities a faulty process
can construct. 
One solution is to assume that all processes have
limited access to some resource needed to construct identities.
Typically this is computational power, allowing for a
\concept{proof-of-work} strategy where any new identity is validated
by demonstrating that the process using it has burned some minimal
amount of computational time.

This approach was first proposed by Dwork and Naor~\cite{DworkN1992}
as a tool for combating email spam, and is frequently reinvented. The
usual approach is to pick a cryptographically-secure hash function $h$
that produces $n$ bits of output, a puzzle input $x$ that should be
unique to this instance of the problem, and demand that the process
find some value $y$ such that $h(xy) = 0$; if we assume that $h$ acts
like a random function, it should require $2^n$ computations of $h$ on
average to find such a $y$.

Proof-of-work allows for direct validation of identities: if you
present me with an identity that incorporates $xy$ with $h(xy) = 0$, I
can be reasonably confident that you spent computed approximately
$2^n$ hashes since you learned $x$. The cost of checking a valid
identity is relatively cheap, since I only have to compute one hash
(although the cost of checking a bogus identity might be more
expensive than generating the bogus identity). Assuming that the
each faulty process spends at most $ρ$ times as much processing
power than any non-faulty process, Douceur observes that the expected
number of counterfeit identities for each faulty process will average
around $ρ$.

A key assumption here is that the proof-of-work tasks are carried out
over a bounded time period. If the faulty processes can prepare
identities well in advance, Douceur observes that a faulty process can
spend as much time as it likes to construct as many identities as it
likes.

There is a third main result in this paper, which shows that using an
initial assignment of identities to validate later identities using
some sort of vouching processes just leads to an initial army of
counterfeits validating more counterfeits. This is mostly interesting
because it was still used at the time as a way to try to validate 
identities in PGP~\cite{rfc1991}, a popular open-source public-key
encryption system for email messages.

\subsection{Limitations of resource-based defenses}

Douceur's paper was interpreted by many researchers as a sign that
proof-of-work is fundamentally useless for defending against Sybil
attacks, at least in the context of problems like consensus where a
constant fraction of faulty agents can disrupt the protocol.
The usual argument goes like this:\footnote{As far as I can tell, this
argument initially arose as a
folk theorem. But see~\cite{BadertscherGMTZ2018} for references to
more serious game-theoretic analyses that are similarly pessimistic
and some reasons to be less pessimistic.}
\begin{enumerate}
    \item For any instance of a problem to be solved using
        proof-of-work, non-faulty processes need to burn resources
        that are a constant fraction of the resources burned by faulty
        processes.
    \item This resource burning needs to exceed the value of whatever
        target is being defended, or the faulty processes can obtain a net
        profit by burning enough resources to overwhelm the non-faulty
        processes.
    \item The resource burning by the non-faulty processes
        needs to be repeated every time the
        target is defended, because the non-faulty processes only need
        to get lucky once. In contrast, the faulty processes can wait
        and burn their resources for only one instance of the
        protocol.
\end{enumerate}

It follows that the non-faulty processes quickly expend more resources
defending the target than the target is worth: proof-of-work can't
work.\footnote{This did not stop some of us from trying anyway. One of
the earliest written examples of attempting to use proof-of-work to
solve Byzantine agreement despite Sybil attacks is a Yale CS tech
report derived from Collin Jackson's CPSC 490 senior
project~\cite{AspnesJK2005}. Sadly the two co-authors who advised him
on this project didn't think it was worth trying to publish this
idea anywhere more visible.}

The description above is a little vague about what it means to protect
a target. As a concrete example, suppose I am purchasing some
real-world good from you using a transaction that is recorded in a
\concept{distributed ledger}\index{ledger!distributed}, a replicated
state machine that records payments. If I can subvert the consensus
protocol used to update the distributed ledger, I might be able to
show you a ledger that includes a payment from me to you (causing you
to turn over the valuable good), but then convince all other
participants to adopt a different update in which this transaction
never happened, and explain that you are simply a Byzantine process
trying to steal my money.\footnote{A reasonable objection is that if
you demand that I sign my transactions using a private key, I won't be
able to repudiate my payment even if you only have a private copy. In
this case what I do is show you a version of the ledger where I have
plenty of funds to pay you, and then show everybody else a version
where my payment to you sadly does not go through because I already
gave all my money to my suspiciously identical twin.}

\subsection{Alternative defenses}

\indexConcept{CAPTCHA}{CAPTCHAs}~\cite{vonAhnBHL2003} have been used
in the context of web sites interacting with human users, by requiring
the users to complete tasks that are hard to automate. This raises the
cost of a fake identity by a bit, since a human being needs to be
involved in the process somewhere, but it's still fundamentally a
resource-burning technique, just now the resource is human time
instead of computer time. As with proof-of work, the problem is that
non-faulty users are required to spend the same effort as faulty
users, and this adds up fast via the folk theorem.  This becomes
particularly annoying when attackers can arbitrage low wage rates in
some countries or even apply man-in-the-middle attacks that get
would-be visitors to one site to solve CAPTCHAs for another site.

Some other approaches that have been proposed use physical locations
or social networks to attempt to detect counterfeit identities
generated by the same process. Bazzi and Konjevod~\cite{BazziK2007}
proposed that a process that wants to authenticate itself could a
\concept{geometric certificate}\index{certificate!geometric}
consisting of verified ping times to a collection of standardized
beacon nodes. Multiple virtual machines located at the same physical
location will end up with essentially the same certificate, and can be
treated as one (possibly corrupted) node. Unfortunateley, so will
multiple users at large institutions with a single outgoing pipe to
the rest of the network. The idea does avoid resource-burning, but it
never really caught on on practice, and if tried now could probably be
easily defeated by geographically-distributed botnets.

\concept{SybilGuard}~\cite{YuKGF2006} was proposed by Yu~\etal{}
as a defense against Sybil attacks based on the structure of social
networks. The idea is that a social network graph with many Sybil
nodes is likely to decompose into a subnetwork consisting mostly of
legitimate nodes and a subnetwork consisting mostly of counterfeit
nodes, with the majority of links between nodes within each subnetwork
and few links between legitimate nodes and counterfeit nodes.
This approach is pretty clever, and subsequent work explored in depth
efficient algorithms for separating these two subnetworks, but it
causes trouble for users that wish to disconnect their activities from
their social-network identity, and more practically is trivially
defeated if the faulty processes can amass enough bogus social network
accounts that they are not longer an obvious disconnected minority.

\section{Bitcoin}

Since proof-of-work is too expensive, and other approaches are easily
defeated, what do we do if we really want to solve consensus in an
open system? It turns out we bite the bullet and accept the huge cost
of proof-of-work. This was the approach taken in
Bitcoin~\cite{Nakamoto2008}, developed by the pseudonymous
person or persons going by Satoshi Nakamoto.
This system evades some of the issues in the folk theorem by (a)
convincing lots of non-faulty processes to join by including a lottery
awarding tokens to participants and (b) relying on the would-be faulty
processes not to be coordinated enough or have enough available
processing power relative to the huge horde of non-faulty
lottery-ticket buyers to target a specific round of the protocol.
All of this requires Bitcoin to be able to dispense rewards that are
comparable to the costs incurred by the non-faulty participants.

Bitcoin can do this because it implements a \concept{cryptocurrency}, a
mechanism for exchanging cryptographic tokens between users that can
be used analogously to standard currencies. To make all transfers
visible thus prevent \concept{double-spending}, it constructs what is
now usually called a \concept{distributed
ledger}\index{ledger!distributed} consisting of a \concept{chain}
(sequence) of \indexConcept{block}{blocks}, each of which contains a set of
transactions that record transfers of tokens between participants.
Participants are identified by cryptographic keys, and a transaction
must be signed by the sender of the tokens to be valid.  

A cryptographic hash of the entire ledger is updated with the addition
of each block, to prevent tampering and to construct the key for the
proof-of-work puzzle used to select the next block to be added.  This
technique, which gave rise to the name \concept{blockchain} for
systems of this kind, was originally developed by Haber and
Stornetta~\cite{HaberS1990}, without the proof-of-work consensus
algorithm, as a tool for making it difficult to backdate digital
documents by storing their hashes in a centrally-maintained sequence
of signed blocks of this type whose full hash is published from time
to time in a difficult-to-corrupt location. (Haber and Stornetta's
company Surety used a weekly classified advertisement in the New York
Times.)

Bitcoin takes this idea and adds a proof-of-work based
consensus protocol on top, while including side payments to reward
participation in the protocol. The rule for the consensus protocol is
that every interested process tries to extend the current chain as best it can,
but only a process that provably solves a cryptographic puzzle can do
so. So the first process to solve the puzzle wins, and if the majority
of the computation power belongs to non-faulty processes, this process
is likely to be non-faulty. In the case of a tie (possibly created by
faulty processes that refuse to admit defeat), longer chain wins. In
this way the computationally-strong majority eventually
overcomes the computationally-weak minority, since even if the
minority gets lucky a few times they are unlikely to win the race 
against the more powerful faction.

To analyze this, let's assume
a synchronous message-passing system
where messages are distributed through an anonymous broadcast channel.
Synchrony is obtained by assuming roughly-synchronized clocks and
setting a very long timeout of 10 minutes for each round.
Because the identities of processes are not relevant to the protocol,
there is no need to identify the sender of a message, although the
proof-of-work mechanism used to select blocks also has the useful side
effect of limiting propagation of spam updates.

In distributed computing terms, Bitcoin implements a replicated
state machine, using a probabilistic version of consensus to choose
between possible extensions. Using randomization evades the
Dolev-Strong~\cite{DolevS1983} and FLP~\cite{FischerLP1985} lower
bounds, because the bad executions constructed in these bounds are
either (a) highly improbable or (b) require the adversary to predict
the future (we'll come back to this idea in
Chapter~\ref{chapter-randomized-consensus}). The Nakamoto paper does
not reference the distributed computing literature, and its definition
of consensus deviates substantially from the traditional
termination-validity-agreement framework of
Pease~\etal~\cite{PeaseSL1980}. Instead of guaranteeing termination
and validity, the protocol attempts to provide an \concept{eventual
consistency}\index{consistency!eventual} where over time, the copies
of the state machine continuously converge to agreeing on an initial
prefix of the operation history that includes all but a few
recently-added blocks.

\subsection{Obtaining eventual consistency}

In this section, we'll describe the operation of the Bitcoin consensus
protocol, often called \concept{Nakamoto
consensus}\index{consensus!Nakamoto}. There is a somewhat heuristic
analysis of this protocol in the original Bitcoin white paper. We'll
give a slightly less heuristic analysis that is still pretty sloppy.
For a more serious analysis, see~\cite{GarayKL2015}, which influenced
some of the less suspicious parts of the discussion below.

Our model is already strong enough to trivially guarantee agreement in each
round: since every non-faulty process sees the same chains in the
broadcast channel, it's enough to discard any invalid chains (which we
will define soon), and apply some consistent tie-breaking rule to
choose among the remaining valid chains. So the goal of the consensus
step will be to guarantee \concept{eventual consistency} between
rounds, which we will take to mean that any block buried deep enough
in the chain $C_r$ for round $r$ also appears in any chain $C_{r'}$
for $r' > r$.

The mechanism for doing this is to generate each $C_{r+1}$ as an
extension of $C_r$. To construct an extension, a process $i$ that
wishes to add block $x_i$ must first solve a hash puzzle by finding
some $y$ such that $h(C_r, x_i, y) ≤ D$, where $h$ is a hash function
that is sufficiently cryptographically secure that we can pretend it's
a random function, and $D$ is a difficulty
parameter that can be tuned to adjust the likelihood of finding a
solution within the time bounds associated with the round. If
successful, the process can propose an extension $C_r \Tuple{x_i, y}$
that is valid if it satisfies both application-specific requirements
like $x_i$ doesn't include transactions that spend money the spender
doesn't have after $C_r$, and protocol-specific requirements like
$C_r$ is valid and $h(C_r, x_i, y) ≤ D$. These conditions are easily
checked by any process.

For the tie-breaking rule, we will favor longer chains over shorter
ones, and otherwise break ties consistently. As noted previously,
consistent tie-breaking means all non-faulty processes adopt the same
value $C_r$ for each $r$. To replace a buried block, the faulty processes
will need to supply an alternative chain that wins the tie-breaking
rule by being the same length or longer as the chain built by the
non-faulty processes.

The resulting protocol is given in Algorithm~\ref{alg-Nakamoto-consensus}.

\begin{algorithm}
    $C \gets$ some initial chain.\;
    \tcp{Do infinitely many synchronous rounds}
    \For{$r ← 0\dots \infty$}{
        Let $x$ be the block I want to add to $C$\;
        \tcp{Attempt to extend $C$}
        \For{$i \gets 1 \dots q$}{
            Choose a random value $y$\;
            \If{$h(C, x, y) ≤ D$}{
                $C \gets C \Tuple{x,y}$\;
                Broadcast $C$\;
                \Break\;
            }
        }
        \tcp{Take best valid $C'$}
        \Foreach{$C'$ received this round}{
            \If{$C'$ is valid and tie-breaker favors $C'$ over $C$}{
                $C \gets C'$\;
            }
        }
    }
    \caption{Nakamoto consensus}
    \label{alg-Nakamoto-consensus}
\end{algorithm}

The main issue with this protocol is that if the faulty processes get
lucky, they can construct a chain that is longer than the chain of the
non-faulty processes, and use this to hijack the protocol. We'd like
to show that when this happens, the bad chain shares all but a small
suffix with the good chain it displaces. If we are willing to cut a
few corners in the argument, this comes down to demonstrating that the
faulty processes can't win the race to extend their evil chain past
the non-faulty processes' preferred chain over long sequences of
rounds. We will consider the specific case where the non-faulty and
faulty processes both start off with some common $C_r = \evil{C}_r$,
and over the next $m$ rounds the non-faulty processes extend $C_r$ as
best they can using Algorithm~\ref{alg-Nakamoto-consensus} while the
faulty processes extend $\evil{C}_r$ in secret.  The faulty processes
win if the resulting $\evil{C}_{r+m}$ is longer than the non-faulty
processes' $C_{r+m}$. (There is a lot of unjustified simplification
sneaking in here. For a much more sophisticated argument that doesn't
cheat, see~\cite{GarayKL2015}.)

For each process $i$, let $p_i$ be the expected number of puzzle
solutions it finds in a single round. If $i$ is non-faulty, this
is just the probability that it finds a solution, since non-faulty
processes stop after finding one solution. If $i$ is faulty,
$i$ can generate more than one solution, which might make
$p_i$ a bit larger than it would be for a non-faulty process
with the same computational power. If $p_i$ is very small,
the difference will be slight.

To simplify things, we'll assume that the set of processes and their
$p_i$ values are fixed over time. Let $α$ be the sum of $p_i$ over all
the non-faulty processes, and $β$ the sum of $p_i$ over all the faulty
processes. These give the expected number of solutions obtained in one
round by all non-faulty or faulty processes respectively.

Inclusion-exclusion says that the probability that the non-faulty
processes solve at least one puzzle in a given round is at least $∑_i
p_i - ∑_{i≠j} p_i p_j ≥ α - α^2$. Letting $X_i$ be the indicator
for the event that the non-faulty processes add a new block
in round $r+i$, they add at least an expected $∑ \Exp{X_i} ≥ m(α-α^2)$ blocks in $m$
rounds. We can similarly argue that the faulty processes add at most
an expected $mβ = ∑ \Exp{Y_i}$ blocks in $m$ rounds, where $Y_i$
is the indicator variable for success of the $i$-th puzzle attempt by
a non-faulty process. In both cases we are looking at a sum of 0–1
random variable with known mean, so Chernoff bounds apply and we get,
for any $δ$,
\begin{align*}
    \Prob{∑ X_i ≤ (1-δ)m(α-α^2)} &≤ e^{-δ^2 m(α-α^2)/2}\\
    \Prob{∑ Y_i ≥ (1+δ)mβ} &≤ e^{-δ^2 mβ/2}\\
\end{align*}

Let's suppose $β$ is less than half of $α-α^2$, corresponding to the
faulty processes having a bit less than a third of the total
computational power. Writing $k=m(α-α^2)=\Exp{∑X_i}$ we get
$\Exp{∑Y_i} = mβ ≤ k/2$. Set $δ=1/3$ to get
\begin{align*}
    \Prob{∑ X_i ≤ (1-δ)k = (2/3)k} &≤ e^{-k/18}\\
    \Prob{∑ Y_i ≥ (1+δ)(k/2) = (2/3)k} &≤ e^{-k/36}.
\end{align*}

This gives a total probability of at most $e^{-k/18} + e^{-k/36} =
e^{-Θ(k)}$ that either the bad chain gets extended by 
$(2/3)k$ or more blocks the good chain gets extended by $(2/3)k$ or
fewer blocks. If neither of these events happen, the good chain wins.
This means that as we consider longer and longer suffixes, it becomes
exponentially more improbable that the suffix in the chain the
non-faulty processes currently agree on will suddenly be replaced by
an alternative chain prepared in secret.

This is not as good a guarantee as we get from iterating traditional
Byzantine agreement, where the output of the protocol at each step
will never be retracted, but it seems to be good enough in practice
that users are willing to tolerate it. 

\subsection{Does Bitcoin disprove the folk theorem?}

The short answer is no, and a proof can be found in a paper by
Leshno~\etal~\cite{LeshnoPS2023} (which also gives an alternative
open distributed ledger construction that is less vulnerable).
In fact, Leshno~\etal~show that the situation is even worse: the
reward mechanism built into Bitcoin means that an adversary who can
rent enough resources can substitute its own ledger for the real one
at close to zero net cost.

And yet Bitcoin is still in use.

I'm not really qualified to answer
why Bitcoin seems to work anyway, but I suspect that some of its
survival is a result of it being uniquely huge.
This has consequences that don't apply to a smaller system:
\begin{enumerate}
    \item The amount of work needed for a sustained attack on Bitcoin
        is enormous. Given that most of the proof-of-work puzzles
        in the Bitcoin as currently implemented are solved using
        custom parallel hardware running off of low-cost power
        sources, the likelihood of any attacker (other than a few large
        state actors) amassing comparable hardware in secret is low.
    \item While the volume of transactions on the Bitcoin blockchain
        increases the potential rewards of a successful attack, their
        diversity makes the chances of collecting that reward low. It's much easier
        to imagine convincing a single participant of a low-volume
        blockchain to trade their valuable cartoon ape for a handful
        of virtual fairy gold that turns into virtual dirt by dawn.
        It is probably much harder to do this to every participant in
        a high-volume chain over a long enough interval to make a
        sufficient profit.
    \item Though Bitcoin was designed to be decentralized, in practice
        economies of scale mean that most of the protocol is run by a
        small number of organizations. A profitable attack on Bitcoin
        might lead these organization to simply fork the
        chain, erasing the attacker's gains. (A similar fork 
        happened after a 2016 attack on Etherium; see
        \url{https://blog.ethereum.org/2016/07/20/hard-fork-completed}.)
        So the political and
        social factors surrounding successful blockchains aren't taken
        into account in the abstract model underlying the folk
        theorem.
\end{enumerate}

At the same time, Bitcoin is still absurdly costly, and the guarantees
it provides are not as strong as can be obtained by running iterated
Byzantine agreement on a small number of semi-trusted parties. This
may be why more recent systems have been moving away from
proof-of-work, and suggests that Bitcoin's unusual status as the
first widely-used blockchain may, in the long run, not save it from
being outcompeted by other systems.

Perhaps the way to think about the enormous cost of proof-of-work
based systems is that they are paying a \concept{price of
anarchy}\index{anarchy!price of}~\cite{KoutsoupiasP2009} for avoiding any
kind of centralized management in the form of a privileged set of
servers. Unfortunately, much of this cost appears to be unavoidable
without such management~\cite{PassS2018}. 

\part{Shared memory}
\label{part-shared-memory}

\myChapter{Model}{2026}{}
\label{chapter-shared-memory-model}

Shared memory models describe the behavior of processes in a
multiprocessing system.  These processes might correspond to actual
physical processors or processor cores, or might use time-sharing on a
single physical processor.  In either case the assumption is that
communication is through some sort of shared data structures.

Here we describe the basic shared-memory model.  See also \cite[§{}4.1]{AttiyaW2004}.

Where shared memory differs from message passing is that processes
can't communicate with each other directly, but instead
communicate through a pool of shared
\indexConcept{object}{objects}.  These are typically
\indexConcept{register}{registers} supporting read and write
operations, but fancier objects corresponding to more sophisticated
data structures or synchronization primitives may also be included in
the model.  

It is usually assumed that the shared objects do not
experience faults.  This means that the shared memory can be used as a
tool to prevent partitions and other problems that can arise in
message passing if the number of faults get too high. As a result,
for large numbers of processor failures, shared memory is a more
powerful model than message passing, although we will see in
Chapter~\ref{chapter-distributed-shared-memory} that both models can
simulate each other provided a majority of processes are non-faulty.

\section{Atomic registers}
\label{section-atomic-registers}

An \index{register!atomic}\concept{atomic register} supports read and
write operations.
We think of these as happening instantaneously, and think of operations of different processes as interleaved in some sequence.
Each read operation on a particular register returns the value written by the last previous write operation.  Write operations return nothing.

A process is defined by giving, for each state, the operation that it
would like to do next, together with a transition function that
specifies how the state will be updated in response to the return
value of that operation.  A configuration of the system consists of a
vector of states for the processes and a vector of value for the
registers.  A sequential execution consists of a sequence of alternating
configurations and operations $C_{0}, π_{1}, C_{1}, π_{2},
C_{2} \dots{}$, where in each triple $C_{i}, π_{i+1}, C_{i+1}$,
the configuration $C_{i+1}$ is the result of applying $π_{i+1}$ to
configuration $C_{i}$.  For read operations, this means that the state of the reading process is updated according to its transition function.  For write operations, the state of the writing process is updated, and the state of the written register is also updated.

\newData{\SMdone}{done}
\newData{\SMleftDone}{leftDone}
\newData{\SMrightDone}{rightDone}
\newData{\SMleftIsDone}{leftIsDone}
\newData{\SMrightIsDone}{rightIsDone}

Pseudocode for shared-memory protocols is usually written using standard pseudocode conventions, with the register operations appearing either as explicit subroutine calls or implicitly as references to shared variables.  Sometimes this can lead to ambiguity; for example, in the code fragment
\begin{align*}
\SMdone &← \SMleftDone ∧ \SMrightDone,
\end{align*}
it is clear that the operation $\Write(\Done, -)$ happens after
$\Read(\SMleftDone)$ and $\Read(\SMrightDone)$, but it is not clear
which of $\Read(\SMleftDone$ and $\Read(\SMrightDone)$ happens first.  When the order is important, we'll write the sequence out explicitly:
\begin{algorithm}[h]
    $\SMleftIsDone ← \Read(\SMleftDone)$ \;
    $\SMrightIsDone ← \Read(\SMrightDone)$ \;
    $\Write(\Done, \SMleftIsDone ∧ \SMrightIsDone)$ \;
\end{algorithm}

Here \SMleftIsDone and \SMrightIsDone are internal variables of the process, so using them does not require read or write operations to the shared memory.

\section{Single-writer versus multi-writer registers}
One variation that does come up even with atomic registers is what
processes are allowed to read or write a particular register.  A
typical assumption is that registers are 
\index{register!single-writer}
\index{single-writer register}
\indexConcept{single-writer multi-reader register}{single-writer multi-reader}—there is only one process that can write to the register (which simplifies implementation since we don't have to arbitrate which of two near-simultaneous writes gets in last and thus leaves the long-term value), although it's also common to assume 
\index{register!multi-writer}
\index{multi-writer register}
\indexConcept{multi-writer multi-reader register}{multi-writer
multi-reader} registers, which if not otherwise available can be built
from single-writer multi-reader registers using atomic snapshot (see
Chapter~\ref{chapter-atomic-snapshots}).  Less common are
\index{register!single-reader}
\indexConcept{single-writer single-reader register}{single-writer
single-reader} registers, which act much like message-passing channels except that the receiver has to make an explicit effort to pick up its mail.

\section{Fairness and crashes}

From the perspective of a schedule, the fairness condition says that
every processes gets to perform an operation infinitely often, unless
it enters either a crashed or halting state where it invokes no
further operations.  (Note that unlike in asynchronous
message-passing, there is no way to wake up a process once it stops
doing operations, since the only way to detect that any activity is
happening is to read a register and notice it changed.)  Because the
registers (at least in in multi-reader models) provide a permanent
fault-free record of past history, shared-memory systems are much less
vulnerable to crash failures than message-passing systems (though
a version FLP\footnote{See Chapter~\ref{chapter-FLP}.} still
applies~\cite{LouiA1987}); so in
extreme cases, we may assume as many as $n-1$ crash failures, which
makes the fairness condition very weak.  The $n-1$ crash failures case
is called the \concept{wait-free} case—since no process can wait for any other process to do anything—and has been extensively studied in the literature.

For historical reasons, work on shared-memory systems has tended to assume crash failures rather than Byzantine failures—possibly because Byzantine failures are easier to prevent when you have several processes sitting in the same machine than when they are spread across the network, or possibly because in multi-writer situations a Byzantine process can do much more damage.  But the model by itself doesn't put any constraints on the kinds of process failures that might occur.

\section{Concurrent executions}

Often, the operations on our shared objects will be implemented using
lower-level operations.  When this happens, it no longer makes sense
to assume that the high-level operations occur one at a
time—although an implementation may try to give that impression to
its users.  To model to possibility of concurrency between operations,
we split an operation into an \concept{invocation} and
\concept{response}, corresponding roughly to a procedure call and its
return.  The user is responsible for invoking the object; the object's
implementation (or the shared memory system, if the object is taken as
a primitive) is responsible for responding.  Typically we will imagine
that an operation is invoked at the moment it becomes pending, but
there may be executions in which that does not occur.  The time
between the invocation and the response for an operation is the
\concept{interval} of the operation.

A \index{execution!concurrent}\concept{concurrent execution} is a
sequence of invocations and responses, where after any prefix of the
execution, every response corresponds to some preceding invocation,
and there is at most one invocation for each process—always the
last—that does not have a corresponding response.  A concurrent
execution is \indexConcept{complete
execution}{complete}\index{execution!complete} if every invocation has
a matching response, and it is \indexConcept{sequential
execution}{sequential} if the operations don't overlap, meaning that
there is at most one invocation without a corresponding response in
any prefix of the execution.

Sequential executions correspond to executions of a sequential object,
which doesn't allow (or at least doesn't experience) concurrent
operations. How a given concurrent execution may or may not relate to a
sequential execution depends on the consistency properties of the
implementation, as described below.

\section{Consistency properties}
\label{section-consistency-properties}

Different shared-memory systems may provide various \index{consistency
property} \conceptFormat{consistency properties}, which describe how
views of an object by different processes mesh with each other.  The
strongest consistency property generally used is
\concept{linearizability}~\cite{HerlihyW1990}, which says roughly that
an implementation of an object is \concept{linearizable} if, for any
complete concurrent execution of the object, there is a sequential
execution of the object with the same operations and return values,
where the (total) order of operations in the sequential execution is a
linearization of the (partial) order of operations in the concurrent
execution. The order in each case is defined as $a <_H b$ if the
response event for operation $a$ in execution $H$ precedes the invoke
event for operation $b$ in the same execution.

The actual definition is a little bit more technical, since it has to
deal with the issue of concurrent executions that may include
incomplete operations for which there is an invoke event but no
response. We'd like to give the implementation the flexibility of
deciding whether these operations have taken effect or not, so given
an incomplete concurrent execution $H$, a \concept{linearization} of $H$
involves three steps:
\begin{enumerate}
    \item Extend $H$ by adding zero or more response events, obtaining
        a new execution $H'$.
    \item Remove any invoke events in $H'$ that don't have a matching
        response event, obtaining a new execution $H''$.
    \item Construct a sequential $S$ such that $S$ meets the
        sequential specification of the object,  $H''|p = S|p$ for all
        $p$, and $<_{H''} ⊆ <_S$.\footnote{There is a subtle issue here:
        the original definition of linearizability by Herlihy and Wing
        only required $<_H ⊆ <_S$. In some rare cases, this allows
        objects with strange behavior, as observed by
        Sela~\etal~\cite{SelaHP2021}, who proposed the fix of
        requiring the completion $H''$ to be linearizable with respect to
        its own observable operation ordering. The
        revised definition behaves better in corner cases and
        is more consistent with
        the usual approach of proving linearizability only for
        executions in which all operations are complete.}
\end{enumerate}

An execution is now linearizable if it has a linearization as defined
above.

Most of the complexity of the above definition is needed only to be
able to decide if incomplete executions are linearizable. If we
consider only complete executions, we can skip the $H'$ and $H''$
steps, since neither changes $H$. Even better, if we are asking if an
implementation of an object is linearizable—meaning that all
executions of the object are linearizable—then we can usually prove
this by proving it only for complete executions, since if the
implementation has the property that any operation in progress can
eventually finish, we can extend any incomplete $H$ to a complete
$H'=H''$ by simply running any pending operations to completion. (If
our implementation does not have this property, we will need to use
the more general definition, but this may be the least of our
problems.)

Linearization is usually proved for complete $H$
by constructing the total order $<_S$ explicitly, which gives $S$ as the
unique sequential execution equivalent to $H$ that assigns this order
to operations.
An alternative method is to assign each operation a
\concept{linearization point} somewhere between when its invocation
and response, and obtain $S$ by assuming that
all operations occur atomically at their linearization points is
consistent with the specification of the object; this is equivalent to
constructing $<_S ⊇ <_H$ since given $<_S$ we can always find
consistent linearization points. I personally find constructing a
linearization ordering easier for most implementations, but
linearization points are useful because they emphasize that to the
user, it really does look like a linearizable implementation executes
all operations atomically.
Using either definition, we are given a fair bit of flexibility in how to order
overlapping operations, which can sometimes be exploited by clever
implementations (or lower bounds).

A weaker condition is \concept{sequential consistency}~\cite{Lamport1979}.
This says
that for any concurrent execution of the object, there exists some
sequential execution that is indistinguishable to all processes;
however, this sequential execution might include operations that occur
out of order from a global perspective.  
(Essentially we are dropping the requirement $<_H ⊆ <_S$ from the
linearizability definition.)
For example, we could have an execution of an atomic register where
you write to it, then I read from it, but I get the initial value that
precedes your write. This is sequentially consistent but not
linearizable.

Linearizability has the useful property of being composable, in the
sense that if $H|A$ is linearizable for any particular object $A$,
then $H$ is linearizable. Sequential consistency does not generally
have this property.  For this reason, we will usually ask any
implementations we consider to be linearizable.  However, both
linearizability and sequential consistency are much stronger than the
consistency conditions provided by real multiprocessors.  For some
examples of weaker memory consistency rules, a good place to start
might be the dissertation of Kawash~\cite{Kawash2000}.

\section{Complexity measures}

There are several complexity measures for shared-memory systems.

\begin{description}
 \item[Time] Assume that no process takes more than 1 time unit
 between operations (but some fast processes may take less).  Assign
 the first operation in the schedule time 1 and each subsequent
 operation the largest time consistent with the bound.  The time of
 the last operation is the 
 \index{complexity!time}
 \concept{time complexity}.
 This is also known as the
 \concept{big-step} or \concept{round} measure because the time increases by 1 precisely when every non-faulty process has taken at least one step, and a minimum interval during which this occurs counts as a big step or a round.
 \item[Total work] The \index{work!total}\concept{total work} or 
 \index{complexity!step!total}
 \index{step complexity!total}
 \concept{total step complexity} is just the length of the
 schedule, i.e., the number of operations.  This doesn't consider how
 the work is divided among the processes, e.g., an $O(n^{2})$ total
 work protocol might dump all $O(n^{2})$ operations on a single
 process and leave the rest with almost nothing to do.  There is
 usually not much of a direct correspondence between total work and
 time.  For example, any algorithm that involves
 \concept{busy-waiting}—where a process repeatedly reads a register
 until it changes—may have unbounded total work (because the
 busy-waiter might spin very fast) even though it runs in bounded time
 (because the register gets written to as soon as some slower process
 gets around to it).  However, it is trivially the case that the time
 complexity is never greater than the total work.
 \item[Per-process work] The 
 \index{work!per-process}
 \concept{per-process work},
 \index{work!individual}
 \concept{individual work},
 \index{complexity!step!per-process}
 \index{step complexity!per-process}
 \concept{per-process step complexity},
 or
 \index{complexity!step!individual}
 \index{step complexity!individual}
 \concept{individual step complexity}
 measures the maximum number of operations performed by any single
 process.  Optimizing for per-process work produces more equitably
        distributed workloads (or reveals inequitably distributed
        workloads).  Like total work, per-process work gives an upper
        bound on time, since each time unit includes at least one
        operation from the longest-running process, but time
        complexity might be much less than per-process work (e.g., in the busy-waiting case above).
 \item[Remote memory references] As we've seen, step complexity doesn't make much
     sense for processes that busy-wait.  An alternative measure is
     \concept{remote memory reference} complexity or \concept{RMR}
     complexity.  This measure charges one unit for write operations
     and the first read operation by each process following a write,
     but charges nothing for subsequent read operations if there are
     no intervening writes (see §\ref{section-RMR-complexity} for
     details).  In this measure, a busy-waiting operation is only
     charged one unit.  RMR complexity can be justified to a certain
     extent by the cost structure of multi-processor
     caching~\cite{Mellor-CrummeyS91,Anderson1990}.
 \item[Contention] In multi-writer or multi-reader situations, it may
 be bad to have too many processes pounding on the same register at
 once.  The \concept{contention} measures the maximum number of
 pending operations on any single register during the schedule (this
 is the simplest of several definitions out there).  A single-reader
 single-writer algorithm always has contention at most 2, but
 achieving such low contention may be harder for multi-reader
 multi-writer algorithms.  Of course, the contention is never worse
 that $n$, since we assume each process has at most one pending operation at a time.
 \item[Space] Just how big are those registers anyway?  Much of the
     work in this area assumes they are \emph{very} big.\footnote{A
     typical justification for this assumption is that an
     arbitrarily-large register can be simulated by a smaller register
     that holds pointers to single-use collections of registers
     holding the actual values. But even using this technique there
     are problems for which individual registers of unbounded size are
     necessary~\cite{DelporteFFRT2023}.}  But we can ask
 for the maximum number of bits in any one register (\concept{width})
 or the total size (\index{complexity!bit}\concept{bit complexity}) or
 number (\index{complexity!space}\concept{space complexity}) of all registers, and will try to minimize these quantities when possible.  We can also look at the size of the internal states of the processes for another measure of space complexity.
\end{description}

\section{Fancier registers}

In addition to stock read-write registers, one can also imagine more
tricked-out registers that provide additional operations.  These
usually go by the name of \concept{read-modify-write} (\concept{RMW}) registers, since the additional operations consist of reading the state, applying some function to it, and writing the state back, all as a single atomic action.  Examples of RMW registers that have appeared in real machines at various times in the past include:
\begin{description}
 \item[Test-and-set bits] A \concept{test-and-set} operation sets the bit to 1 and returns the old value.
 \item[Fetch-and-add registers] A \concept{fetch-and-add} operation adds some increment (typically -1 or 1) to the register and returns the old value.
 \item[Compare-and-swap registers] A \concept{compare-and-swap}
 operation writes a new value only if the previous value is equal to a supplied test value.
\end{description}

These are all designed to solve various forms of 
\concept{mutual exclusion} or locking,
where we want at most one process at a time to work on some shared
data structure.

Some more exotic read-modify-write registers that have appeared in the literature are

\begin{description}
 \item[Fetch-and-cons] Here the contents of the register is a linked
 list; a \concept{fetch-and-cons} adds a new head and returns the old list.
 \item[Sticky bits (or sticky registers)] With a \concept{sticky bit}
 or \concept{sticky register}~\cite{Plotkin1989}, once the initial
 empty value is overwritten, all further writes fail.  The writer is
 not notified that the write fails, but may be able to detect this
 fact by reading the register in a subsequent operation.
 \item[Bank accounts] Replace the write operation with
     \FuncSty{deposit}, which adds a non-negative amount to the state,
     and \FuncSty{withdraw}, which subtracts a non-negative amount from the state provided the result would not go below 0; otherwise, it has no effect.
\end{description}

These solve problems that are hard for ordinary read/write registers
under bad conditions.  Note that they all
have to return something in response to an invocation. 

There are also blocking objects like locks or semaphores, but these don't fit into the RMW framework.

We can also consider generic read-modify-write registers that can
compute arbitrary functions (passed as an argument to the
read-modify-write operation) in the modify step.  Here we typically
assume that the read-modify-write operation returns the old value of
the register.  Generic read-modify-write registers are not commonly
found in hardware but can be easily simulated (in the absence of
failures) using mutual exclusion.\footnote{See
Chapter~\ref{chapter-mutex}.}

\myChapter{Distributed shared memory}{2026}{}
\label{chapter-distributed-shared-memory}

In 
\index{shared memory!distributed}
\concept{distributed shared memory},
our goal is to simulate a collection of memory locations or
\indexConcept{register}{registers}, each of which supports a \concept{read}
operation that returns the current state of the register and a
\concept{write} operation that updates the state.  Our implementation
should be \concept{linearizable}~\cite{HerlihyW1990}, meaning that
read and write operations appear to occur instantaneously
(\indexConcept{atomic}{atomically}) at some point in between when the
operation starts and the operation finishes; equivalently, there
should be some way to order all the operations on the registers to
obtain a \concept{sequential execution} consistent with the behavior
of a real register (each read returns the value of the most recent
write) while preserving the observable partial order on operations
(where $π_1$ precedes $π_2$ if $π_1$ finishes before $π_2$
starts).  Implicit in this definition is the assumption that
implemented operations take place over some interval, between an
\concept{invocation} that starts the operation and a
\concept{response} that ends the operation and returns its
value.\footnote{More details on the shared-memory model are given in
Chapter~\ref{chapter-shared-memory-model}.}

In the absence of process failures, we can just assign each register
to some process, and implement both read and write operations by
remote procedure calls to the process (in fact, this works for
arbitrary shared-memory objects).  With process failures, we need to
make enough copies of the register that failures can't destroy all of
them.  This creates an asymmetry between simulations of
message-passing from shared-memory and vice versa; in the former case
(discussed briefly in
§\ref{section-message-passing-from-shared-memory} below), a process
that fails in the underlying shared-memory system only means that the
same process fails in the simulated message-passing system.  But in
the other direction, not only does the failure of a process in the
underlying message-passing system mean that the same process fails in
the simulated shared-memory system, but the simulation collapses
completely
if a majority of processes fail.

\section{Message passing from shared memory}
\label{section-message-passing-from-shared-memory}

We'll start with the easy direction.  We can build a reliable FIFO channel from
single-writer single-reader registers using polling.  The naive
approach is that for each edge $uv$ in the message-passing system, we
create a (very big) register $r_{uv}$, and $u$ writes the entire
sequence of every message it has ever sent to $v$ to $r_{uv}$ every
time it wants to do a new send.  To receive messages, $v$ polls all of
its incoming registers periodically and delivers any messages in the
histories that it hasn't processed yet.\footnote{If we are really
cheap about using registers, and are willing to accept even more
absurdity in the register size, we can just have $u$ write every
message it ever sends to $r_{u}$, and have each $v$ poll all the $r_u$
and filter out any messages intended for other processes.}

The ludicrous register width can be reduced
by adding in an acknowledgment mechanism in a
separate register $\Ack_{vu}$; the idea is that $u$ will only write
one message at a time to $r_{uv}$, and will queue subsequent messages
until $v$ writes in $\Ack_{vu}$ that the message in $r_{uv}$ has been
received.  With some tinkering, it is possible to knock $r_{uv}$ down
to only three possible states (sending 0, sending 1, and reset) and
$\Ack_{vu}$ down to a single bit (value-received, reset-received), but that's probably overkill for most applications.

Process failures don't affect any of these protocols, except that a dead process stops sending and receiving.

\section{Shared memory from message passing: the Attiya-Bar-Noy-Dolev algorithm}
\label{section-ABD}

\newData{\ABDvalue}{value}
\newData{\ABDtimestamp}{timestamp}

Here we show how to implement shared memory from message passing.
We'll assume that our system is asynchronous, that the network
is complete, and that we are only dealing with $f < n/2$ crash
failures.  We'll also assume we only want to build single-writer
registers, just to keep things simple; we can extend to multi-writer
registers later.

Here's the algorithm, which is due to Attiya, Bar-Noy, and
Dolev~\cite{AttiyaBD1995}; see also \cite[\S17.1.3]{Lynch1996}.
(Section~9.3 of
\cite{AttiyaW2004} gives an equivalent algorithm, but the details are
buried in an implementation of totally-ordered broadcast).  We'll make
$n$ copies of the register, one on each process.  Each process's copy
will hold a pair $(\ABDvalue, \ABDtimestamp)$ where timestamps are
(unbounded) integer values.  Initially, everybody starts with
$(⊥, 0)$.  A process updates its copy with new values $(v,t)$ upon
receiving $\Write(v,t)$ from any other process $p$, provided $t$ is
greater than the process's current timestamp.  It then responds to $p$
with $\Ack(v,t)$, whether or not it updated its local copy.  A process
will also respond to a message $\Read(u)$ with a response $\Ack(\ABDvalue,
\ABDtimestamp, u)$; here $u$ is a \concept{nonce}\footnote{A
\conceptFormat{nonce} is any value that is guaranteed to be used at most
once (the term originally comes from cryptography, which in turn got
it from linguistics).  In practice, a
reader will most likely generate a nonce by combining its process ID
with a local timestamp.} used to distinguish between different read operations so that a process can't be confused by out-of-date acknowledgments.

To write a value, the writer increments its timestamp, updates its
value and sends $\Write(\ABDvalue, \ABDtimestamp)$ to all other processes.  The write operation terminates when the writer has received acknowledgments containing the new timestamp value from a majority of processes.

To read a value, a reader does two steps:
\begin{enumerate}
 \item It sends $\Read(u)$ to all processes (where $u$ is any value it
 hasn't used before) and waits to receive acknowledgments from a
 majority of the processes.  It takes the value $v$ associated with
 the maximum timestamp $t$ as its return value (no matter how many processes sent it).
 \item It then sends $\Write(v,t)$ to all processes, and waits for
 response $\Ack(v,t)$ from a majority of the processes.  Only then does it return.
\end{enumerate}

(Any extra messages, messages with the wrong nonce, etc., are discarded.)

Both reads and writes cost $Θ(n)$ messages ($Θ(1)$ per process).

Intuition: Nobody can return from a write or a read until they are
sure that subsequent reads will return the same (or a later) value.  A
process can only be sure of this if it knows that the values collected
by a read will include at least one copy of the value written or read.
But since majorities overlap, if a majority of the processes have a
current copy of $v$, then the majority read quorum will include it.
Sending $\Write(v,t)$ to all processes and waiting for acknowledgments
from a majority is just a way of ensuring that a majority do in fact
have timestamps that are at least $t$.

If we omit the $\Write$ stage of a $\Read$ operation, we may violate
linearizability.  An example would be a situation where two values
($1$ and $2$, say), have been written to exactly one process each,
with the rest still holding the initial value $⊥$.  A reader that
observes $1$ and $(n-1)/2$ copies of $⊥$ will return $1$, while a
reader that observes $2$ and $(n-1)/2$ copies of $⊥$ will return
$2$.  In the absence of the $\Write$ stage, we could have an
arbitrarily long sequence of readers return $1$, $2$, $1$, $2$, \dots,
all with no concurrency.  This would not be consistent with any
sequential execution in which $1$ and $2$ are only written once.

\section{Proof of linearizability}
\label{section-ABD-linearizable}

Our intuition may be strong, but we still need a proof the algorithm
works.  In particular, we want to show that
for any trace $T$ of the ABD protocol, there is an trace of an atomic
register object that gives the same sequence of invoke and response
events.  The usual way to do this is to find a
\index{linearizability}\concept{linearization} of the read and write
operations: a total order that extends the observed order in $T$ where
$π_1 < π_2$ in $T$ if and only if $π_1$ ends before $π_2$
starts.  Sometimes it's hard to construct such an order, but in this
case it's easy: we can just use the timestamps associated with the
values written or read in each operation.  Specifically, we define the
timestamp of a write or read operation as the timestamp used in the
$\Write(v,t)$ messages sent out during the implementation of that
operation, and we put $π_1$ before $π_2$ if:
\begin{enumerate}
 \item $π_1$ has a lower timestamp than $π_2$, or
 \item $π_1$ has the same timestamp as $π_2$, $π_1$ is a write,
 and $π_2$ is a read, or
 \item $π_1$ has the same timestamp as $π_2$ and $π_1 <_{T}
 π_2$, or
 \item none of the other cases applies, and we feel like putting
 $π_1$ first.
\end{enumerate}

The intent is that we pick some total ordering that is consistent with
both $<_{T}$ and the timestamp ordering (with writes before reads when timestamps are equal).  To make this work we have to show (a) that these two orderings are in fact consistent, and (b) that the resulting ordering produces values consistent with an atomic register: in particular, that each read returns the value of the last preceding write.

Part (b) is easy: since timestamps only increase in response to
writes, each write is followed by precisely those reads with the same
timestamp, which are precisely those that returned the value written.

For part (a), suppose that $π_1 <_{T} π_2$.  The first case is
when $π_2$ is a read.  Then before the end of $π_1$, a set $S$ of
more than $n/2$ processes send the $π_1$ process an $\Ack(v1,t_1)$
message.  Since local timestamps only increase, from this point on any
$\Ack(v_2,t_2,u)$ message sent by a process in $S$ has $t_2 ≥ t_1$.
Let $S'$ be the set of processes sending $\Ack(v_2,t_2,u)$ messages
processed by $π_2$.  Since $\card*{S} > n/2$ and $\card*{S'} > n/2$,
we have $S\cap S'$ is nonempty and so $S'$ includes a process that
sent $\Ack(v_2,t_2)$ with $t_2 ≥ t_1$.  So $π_2$ is serialized
after $π_1$.  The second case is when $π_2$ is a write; but then
$π_1$ returns a timestamp that precedes the writer's increment in
$π_2,$ and so again is serialized first.

\section{Proof that \texorpdfstring{$f < n/2$}{f < n/2} is necessary}
\label{section-distributed-shared-memory-requires-majority}

This is pretty much the standard partition argument that $f < n/2$ is
necessary to do anything useful in a message-passing system.  Split
the processes into two sets $S$ and $S'$ of size $n/2$ each.  Suppose
the writer is in $S$.  Consider an execution where the writer does a
write operation, but all messages between $S$ and $S'$ are delayed.
Since the writer can't tell if the $S'$ processes are slow or dead, it
eventually returns.  Now let some reader in $S'$ attempt to read the
simulated register, again delaying all messages between $S$ and $S'$;
now the reader is forced to return some value without knowing whether
the $S$ processes are slow or dead.  If the reader doesn't return the
value written, we lose.  If by some miracle it does, then we lose in
the execution where the write didn't happen and all the processes in
$S$ really were dead.

\section{Multiple writers}
\label{section-ABD-multi-writer}

\newData{\ABDcount}{count}

So far we have assumed a single writer.  The main advantage of this
approach is that we don't have to do much to manage timestamps: the
single writer can just keep track of its own.  With multiple writers
we can use essentially the same algorithm, but each write needs to
perform an initial round of gathering timestamps so that it can pick a
new timestamp bigger than those that have come before.  We also extend
the timestamps to be of the form $\Tuple{\ABDcount, \Id}$, lexicographically ordered, so that two timestamps with the same count field are ordered by process ID.  The modified write algorithm is:
\begin{enumerate}
 \item Send $\Read(u)$ to all processes and wait to receive acknowledgments from a majority of the processes.
 \item Set my timestamp to  $t = (\max_q \ABDcount_q + 1, \Id)$ where
 the max is taken over all processes $q$ that sent me an
 acknowledgment.  Note that this is a two-field timestamp that is
 compared lexicographically, with the \Id field used only to prevent duplicate timestamps.
 \item Send $\Write(v,t)$ to all processes, and wait for a response
 $\Ack(v,t)$ from a majority of processes.
\end{enumerate}

This increases the cost of a write by a constant factor, but in the
end we still have only a linear number of messages.  The proof of
linearizability is essentially the same as for the single-writer
algorithm, except now we must consider the case of two write
operations by different processes.  Here we have that if $π_1 <_{T}
π_2$, then $π_1$ gets acknowledgments of its write with timestamp
$t_1$ from a majority of processes before $π_2$ starts its initial 
phase to compute \ABDcount.
Since $π_2$ waits for acknowledgments from a majority of
processes as well, these majorities overlap, so $π_2$'s timestamp
$t_2$ must exceed $t_1$.  So the linearization ordering previously defined still works.

\section{Other operations}

The basic ABD framework can be extended to support other operations.

One such operation is a \concept{collect}~\cite{SaksSW1991},
where we read $n$ registers
in parallel with no guarantee that they are read at the same time.
This can trivially be implemented by running $n$ copies of ABD in
parallel, and can be implemented with the same time and message complexity as
ABD for a single register by combining the messages from the parallel
executions into single (possibly very large) messages.

The ABD algorithm can also be used to implement a max register, which
is a register that returns the largest value ever written to it
instead of the most recent value (see
Chapter~\ref{chapter-restricted-use}). The idea is that the
multi-writer version of ABD already implements a max register for
timestamps. So we can discard the value field entirely and just set
each timestamp to a writer's input, and have each reader return
the largest timestamp it sees.

\section{Byzantine failures}

With effort, it is possible to adapt the ABD algorithm~\cite{AttiyaBD1995} to handle
Byzantine failures. Because a Byzantine writer can overwrite a
simulated register with garbage, this mostly makes sense for SWMR
registers, where we can limit the damage done by a Byzantine process
to the contents of its own simulated register.

Mostéfaoui~\etal~\cite{MostefaouiPRJ2017} give an ABD-like algorithm
that simulates a SWMR register in an asynchronous message-passing
system with $t < n/3$ Byzantine faults, without resorting to
cryptography. The main change is to replace the broadcast done by the
writer with a Byzantine reliable broadcast due to
Bracha~\cite{Bracha1987}. This has the unfortunate side-effect of
increasing the message complexity of a write operation to $O(n^2)$.
Fortunately, the authors are able to show that read operations can
skip the reliable broadcast and still run in $O(n)$ messages. The
details are messy enough that we will not attempt to reproduce them
here; see the cited paper if you are interested.

\myChapter{Mutual exclusion}{2026}{}
\label{chapter-mutex}

For more details see \cite[Chapter 4]{AttiyaW2004} or \cite[Chapter
10]{Lynch1996}.

\section{The problem}
\label{section-mutex-definition}

The goal is to share some critical resource between processes without
more than one using it at a time—this is \emph{the} fundamental
problem in time-sharing systems.  

The solution is to only allow access while in a specially-marked block
of code called a \concept{critical section}, and only allow one process at a time to be in a critical section.

A \concept{mutual exclusion protocol} guarantees this, usually in an
asynchronous shared-memory model.

Formally: We want a process to cycle between states \concept{trying} (trying to get into critical section), \concept{critical} (in critical section), \concept{exiting} (cleaning up so that other processes can enter their critical sections), and \concept{remainder} (everything else—essentially just going about its non-critical business).  Only in the trying and exiting states does the process run the mutual exclusion protocol to decide when to switch to the next state; in the critical or remainder states it switches to the next state on its own.

The ultimate payoff is that mutual exclusion solves for
systems without failures what consensus solves for systems with
failures: if the only way to update a data structure is to hold a lock
on it, we are guaranteed to get a nice clean sequence of
atomic-looking updates.  Of course, once we allow failures back in,
mutex becomes less useful, as our faulty processes start crashing
without releasing their locks, and with the data structure in some
broken, half-updated state.\footnote{In principle, if we can detect
    that a process has failed, we can work around this problem by
    allowing some other process to bypass the lock and clean up.  This
    may require that the original process leaves behind notes about
    what it was trying to do, or perhaps copies the data it is going
to modify somewhere else before modifying it.  But even this doesn't
work if some zombie process can suddenly lurch to life and scribble its ancient
out-of-date values all over our shiny modern data structure.}

\section{Goals}
\label{section-mutex-goals}

(See also \cite[\S4.2]{AttiyaW2004}, \cite[\S10.2]{Lynch1996}.)

Core mutual exclusion requirements:
\begin{description}
 \item[Mutual exclusion] \index{mutual exclusion} At most one process is in the critical state at a time.
 \item[Deadlock freedom (progress)]
     \index{deadlock}\index{progress}\index{deadlock-freedom}
 If there is at least one process in a trying state, then eventually some process enters a critical state; similarly for exiting and remainder states.
\end{description}

Note that the protocol is not required to guarantee that processes leave the critical or remainder state, but we generally have to insist that the processes at least leave the critical state on their own to make progress.

An additional useful property (that is not necessarily satisfied by
all mutual exclusion protocols):

\begin{description}
    \item[Starvation freedom (lockout freedom)]
        \index{lockout}\index{lockout-freedom}
        \index{starvation}\index{starvation-freedom}\index{starvation-freedom} If there is a particular process in a trying or exiting state, that process eventually leaves that state.  This means that I don't starve because somebody else keeps jumping past me and seizing the critical resource before I can.
\end{description}

Stronger starvation guarantees include explicit time bounds (how
many rounds can go by before I get in) or \concept{bounded bypass}
(nobody gets in more than $k$ times before I do). Each of these imply
starvation-freedom assuming no deadlock.

\section{Mutual exclusion using strong primitives}
\label{section-mutex-strong-primitives}

See \cite[\S4.3]{AttiyaW2004} or \cite[10.9]{Lynch1996}.  The idea is
that we will use some sort of \concept{read-modify-write} register, where the RMW operation computes a new value based on the old value of the register and writes it back as a single atomic operation, usually returning the old value to the caller as well.

\subsection{Test and set}
\label{section-mutex-test-and-set}

A \concept{test-and-set} operation does the following sequence of actions atomically:
\begin{algorithm}[h]
\newData{\TASoldValue}{oldValue}
\newData{\TASbit}{bit}
    $\TASoldValue ← \Read(\TASbit)$\;
    $\Write(\TASbit, 1)$\;
    \Return \TASoldValue\;
\end{algorithm}

Typically there is also a second \concept{reset} operation for setting
the bit back to zero.  For some implementations, this reset operation
may only be used safely by the last process to get $0$ from the
test-and-set bit.

Because a test-and-set operation is atomic, if two processes both try to perform test-and-set on the same bit, only one of them will see a return value of 0.  This is not true if each process simply executes the above code on a stock atomic register: there is an execution in which both processes read 0, then both write 1, then both return 0 to whatever called the non-atomic test-and-set subroutine.

Test-and-set provides a trivial implementation of mutual exclusion,
shown in Algorithm~\ref{alg-mutex-TAS}.

\newFunc{\MTreset}{reset}
\newData{\MTlock}{lock}

\begin{algorithm}
\While{\True}{
        \tcp{trying}
        \lWhile{$\TAS(\MTlock) = 1$}{nothing}
\tcp{critical}
        (do critical section stuff) \;
        \tcp{exiting}
        $\MTreset(\MTlock)$\;
        \tcp{remainder}
        (do remainder stuff) \;
}
\caption{Mutual exclusion using test-and-set}
\label{alg-mutex-TAS}
\end{algorithm}

It is easy to see that this code provides mutual exclusion, as once
one process gets a 0 out of \MTlock, no other can escape the inner
while loop until that process calls the \MTreset operation in its
exiting state.  It also provides progress (assuming the lock is
initially set to 0); the only part of the code that is not
straight-line code (which gets executed eventually by the fairness
condition) is the inner loop, and if \MTlock is 0, some process
escapes it, while if \MTlock is 1, some process is in the region
between the \TAS call and the \MTreset call, and so it eventually
gets to \MTreset and lets the next process in (or itself, if it is very fast).

The algorithm does \emph{not} provide starvation-freedom: nothing
prevents a single fast process from scooping up the lock bit every
time it goes through the outer loop, while the other processes
ineffectually grab at it just after it is taken away.
Starvation-freedom requires a more sophisticated turn-taking strategy.

\subsection{A starvation-free algorithm using an atomic queue}
\label{section-mutex-queue}

Basic idea: In the trying phase, each process enqueues itself on the
end of a shared queue (assumed to be an atomic operation).  When a
process comes to the head of the queue, it enters the critical
section, and when exiting it dequeues itself.  So the code would look
something like Algorithm~\ref{alg-mutex-queue}.

Note that this requires a queue that supports a \Peek operation that
returns the head of the queue.
Not all implementations of queues have this property.

\begin{algorithm}
\While{\True}{
        \tcp{trying}
        $\Enq(q, \MyId)$\;
        \lWhile{$\Peek(q) \ne \MyId$}{nothing}
        \tcp{critical}
        (do critical section stuff) \;
        \tcp{exiting}
        $\Deq(q)$\;
        \tcp{remainder}
        (do remainder stuff) \;
}
\caption{Mutual exclusion using a queue}
\label{alg-mutex-queue}
\end{algorithm}

Here the proof of mutual exclusion is that only the process whose ID
is at the head of the queue can enter its critical section.  Formally,
we maintain an invariant that any process whose program counter is
between the inner while loop and the call to $\Deq(q)$ must be
at the head of the queue; this invariant is easy to show because a
process can't leave the while loop unless the test fails (i.e., it is
already at the head of the queue), no \Enq operation changes the
head value (if the queue is nonempty), and the \Deq operation (which does change the head value) can only be executed by a process already at the head (from the invariant).

Deadlock-freedom follows from proving a similar invariant that every element of the queue is the ID of some process in the trying, critical, or exiting states, so eventually the process at the head of the queue passes the inner loop, executes its critical section, and dequeues its ID.

Starvation-freedom follows from the fact that once a process is at
position $k$ in the queue, every execution of a critical section
reduces its position by $1$; when it reaches the front of the queue (after some finite number of critical sections), it gets the critical section itself.
Alternatively, we can argue starvation-freedom by showing bounded bypass: once I am in the queue, no process
can execute two critical sections before I do, because once it leaves
its first critical section, it enqueues behind me.

\subsubsection{Replacing the queue with RMW}
\label{section-mutex-RMW}

Following \cite[§4.3.2]{AttiyaW2004}, we can give an implementation of this algorithm using a single read-modify-write (RMW) register instead of a queue; this drastically reduces the (shared) space needed by the algorithm.  The reason this works is because we don't really need to keep track of the position of each process in the queue itself; instead, we can hand out numerical tickets to each process and have the process take responsibility for remembering where its place in line is.

\newData{\MTfirst}{first}
\newData{\MTlast}{last}
\newFunc{\MTrmw}{RMW}

The RMW register has two fields, \MTfirst and \MTlast, both initially
0.  Incrementing \MTlast simulates an enqueue, while incrementing
\MTfirst simulates a dequeue.  The trick is that instead of testing if
it is at the head of the queue, a process simply remembers the value
of the \MTlast field when it ``enqueued'' itself, and waits for the
\MTfirst field to equal it.

Algorithm~\ref{alg-mutex-RMW} shows the code from
Algorithm~\ref{alg-mutex-queue} rewritten to use this technique.  The way to
read the \MTrmw operations is that the \MTfirst argument specifies the
variable to update and the second specifies an expression for
computing the new value.  Each \MTrmw operation returns the old state of the object, before the update.

\newData{\MTposition}{position}

\begin{algorithm}
\While{\True}{
        \tcp{trying}
        $\MTposition ←
            \MTrmw(V, \langle V.\MTfirst, V.\MTlast+1 \rangle)$\;
            \tcp{enqueue}
        \While{
            $\MTrmw(V,V).\MTfirst \ne \MTposition.\MTlast$
           }{nothing}
        \tcp{critical}
        (do critical section stuff) \;
        \tcp{exiting}
        $\MTrmw(V, \langle V.\MTfirst+1, V.\MTlast \rangle)$\;
            \tcp{dequeue}
        \tcp{remainder}
        (do remainder stuff) \;
}
\caption{Mutual exclusion using read-modify-write} 
\label{alg-mutex-RMW}
\end{algorithm}

It's also possible to implement this algorithm using two simpler objects,
one of which implements a \concept{fetch-and-increment} operation that
increments a register and returns the value before the increment, and
one of which is an ordinary atomic register. As in
Algorithm~\ref{alg-mutex-RMW}, a process takes a position in line by
calling the fetch-and-increment, and the head of the line is marked by
the second register, which can only be incremented by a process in the
exiting section. This implementation has the same properties of mutual
exclusion and starvation-freedom as the single-RMW version.

\section{Mutual exclusion and linearizability}

Beyond controlling access to shared resources, mutual exclusion
can instantly give us a linearizable implementation of any object for
which we have a sequential implementation. The reason is that we can
use a mutex to guard access to the shared data structure implementing
the object.

Formally, we imagine that we have a read-modify-write object of some
sort and an implementation from atomic registers that works for
sequential executions. The simplest way to model this is to imagine
that we have a single register $r$ that contains the entire state of
the object. A read-modify-write operation reads an old state $q$ from
$r$, computes a new state $f(q)$ and writes it back to $r$, and
finally returns the old value $q$. This works as long as we don't have
two or more processes executing operations concurrently. But we can
enforce this with a mutex, as in
Algorithm~\ref{alg-linearizability-from-mutex}.

\begin{algorithm}
    \Procedure{$\FuncSty{RMW}(f)$}{
        Enter critical section.\;
        $q ← r$\;
        $r ← f(q)$\;
        Leave critical section.\;
        \Return $q$\;
    }
    \caption{Building a concurrent RMW object using mutex}
    \label{alg-linearizability-from-mutex}
\end{algorithm}

To show that this implementation is linearizable, observe that for any
concurrent history $H$ we can construct a sequential history $S$ by
assigning the invoke/respond times for each operation to when that
operation enters and leaves the critical section. This gives a total
order $<_S$ since no process can enter the critical section until the
previous one leaves.  Since the processes carry out the same
operations on $r$ in both $H$ and $S$, both produce identical views.
Given two operations $a <_H b$, $a$ leaves its critical section before
$b$ enters its critical section, so $<_H ⊆ <_S$. We thus have a
linearization of any given $H$.

\section{Mutual exclusion using only atomic registers}
\label{section-mutex-registers}

While mutual exclusion is easier using powerful primitives, we can
also solve the problem using only registers.

\subsection{Peterson's algorithm}
\label{section-mutex-Peterson}

Algorithm~\ref{alg-mutex-Peterson} shows Peterson's starvation-free
mutual exclusion protocol for two
processes $p_{0}$ and $p_{1}$~\cite{Peterson1981} (see also
\cite[\S4.4.2]{AttiyaW2004} or \cite[\S10.5.1]{Lynch1996}).  It uses only atomic registers.

\begin{algorithm}
\SharedData\\
\Waiting, initially arbitrary\\
$\Present[i]$ for $i\in\{0,1\}$, initially $0$\\

Code for process $i$:\\
\While{\True}{
        \tcp{trying}
        $\Present[i] ← 1$
        \nllabel{line-alg-Peterson-present}\;
        $\Waiting ← i$
        \nllabel{line-alg-Peterson-waiting}\;

        \While{\True}{
            \If{$\Present[¬i] = 0$}{
                \nllabel{line-alg-Peterson-break-1}
                \Break\;
            }
\If{$\Waiting \ne i$}{
                \nllabel{line-alg-Peterson-break-2}
                \Break\;
            }
        }

        \tcp{critical}
        (do critical section stuff)
        \nllabel{line-alg-Peterson-critical}\;
        \tcp{exiting}
        $\Present[i] = 0$
        \nllabel{line-alg-Peterson-reset-present}\;
        \tcp{remainder}
        (do remainder stuff) 
        \nllabel{line-alg-Peterson-remainder}\;
}
\caption{Peterson's mutual exclusion algorithm for two processes}
\label{alg-mutex-Peterson}
\end{algorithm}

This uses three bits to communicate: $\Present[0]$ and
$\Present[1]$ indicate which of $p_0$ and $p_1$ are participating,
and $\Waiting$ enforces turn-taking.
The protocol requires that $\Waiting$ be multi-writer, but it's
OK for $\Present[0]$ and $\Present[1]$ to be single-writer.  

In the description of the protocol, we write Lines~\ref{line-alg-Peterson-break-1}
and~\ref{line-alg-Peterson-break-2}
as two separate lines because they
include two separate read operations, and the order of these reads is
important.

\subsubsection{Correctness of Peterson's protocol}

Intuitively, let's consider all the different ways that the entry code
of the two processes could interact.  There are basically two things
that each process does: it sets its own $\Present$ variable in
Line~\ref{line-alg-Peterson-present} and grabs the $\Waiting$ variable in
Line~\ref{line-alg-Peterson-waiting}.  Here's a typical case where one process gets in first:

\begin{enumerate}
 \item $p_0$ sets $\Present[0] ← 1$
 \item $p_0$ sets $\Waiting ← 0$
 \item $p_0$ reads $\Present[1] = 0$ and enters critical section
 \item $p_1$ sets $\Present[1] ← 1$
 \item $p_1$ sets $\Waiting ← 1$
 \item $p_1$ reads $\Present[0] = 1$ and $\Waiting = 1$ and loops
 \item $p_0$ sets $\Present[0] ← 0$
 \item $p_1$ reads $\Present[0] = 0$ and enters critical section
\end{enumerate}

The idea is that if I see a $0$ in your $\Present$ variable, I know that you aren't playing, and can just go in.

Here's a more interleaved execution where the waiting variable decides the winner:

\begin{enumerate}
 \item $p_0$ sets $\Present[0] ← 1$
 \item $p_0$ sets $\Waiting ← 0$
 \item $p_1$ sets $\Present[1] ← 1$
 \item $p_1$ sets $\Waiting ← 1$
 \item $p_0$ reads $\Present[1] = 1$
 \item $p_1$ reads $\Present[0] = 1$
 \item $p_0$ reads $\Waiting = 1$ and enters critical section
 \item $p_1$ reads $\Present[0] = 1$ and $\Waiting = 1$ and loops
 \item $p_0$ sets $\Present[0] ← 0$
 \item $p_1$ reads $\Present[0] = 0$ and enters critical section
\end{enumerate}

Note that it's the process that set the \Waiting variable last (and thus sees its own value) that stalls.  This is necessary because the earlier process might long since have entered the critical section.

Sadly, examples are not proofs, so to show that this works in general,
we need to formally verify each of mutual exclusion and
starvation-freedom.  Mutual exclusion is a safety property, so we expect
to prove it using invariants.  The proof
in~\cite{Lynch1996} is based on translating the pseudocode directly
into automata (including explicit program counter variables); we'll do
essentially the same proof but without doing the full translation to
automata.  Below, we write that $p_{i}$ is at line $k$ if it the
operation in line $k$ is enabled but has not occurred yet.

\begin{lemma}
\label{lemma-alg-Peterson-invariant-1}
If $\Present[i] = 0$, then $p_{i}$ is at
Line~\ref{line-alg-Peterson-present}
or~\ref{line-alg-Peterson-remainder}.
\end{lemma}
\begin{proof}
Immediate from the code.
\end{proof}

\begin{lemma}
\label{lemma-alg-Peterson-invariant-2}
If $p_{i}$ is at Line~\ref{line-alg-Peterson-critical},
and
$p_{¬{}i}$ is at Line~\ref{line-alg-Peterson-break-1},
\ref{line-alg-Peterson-break-2},
or~\ref{line-alg-Peterson-critical},
then $\Waiting = ¬{}i$.
\end{lemma}
\begin{proof}
We'll do the case $i = 0$; the other case is symmetric.  The proof is
by induction on the schedule.  We need to check that any event that
makes the left-hand side of the invariant true or the right-hand side false also makes the whole invariant true.  The relevant events are:
\begin{itemize}
  \item Transitions by $p_{0}$ from
  Line~\ref{line-alg-Peterson-break-1} to
  Line~\ref{line-alg-Peterson-critical}.  These occur only if
  $\Present[1] = 0$, implying $p_{1}$ is at
Line~\ref{line-alg-Peterson-present}
or~\ref{line-alg-Peterson-remainder} by
Lemma~\ref{lemma-alg-Peterson-invariant-1}.
In this case the second part of the left-hand side is false.
  \item Transitions by $p_{0}$ from
  Line~\ref{line-alg-Peterson-break-2} to
  Line~\ref{line-alg-Peterson-critical}.  These occur only if
  $\Waiting \ne 0$, so the right-hand side is true.
  \item Transitions by $p_{1}$ from
  Line~\ref{line-alg-Peterson-waiting} to
  Line~\ref{line-alg-Peterson-break-1}.  These
  set $\Waiting$ to $1$, making the right-hand side true.
  \item Transitions that set $\Waiting$ to $0$.  These are
  transitions by $p_{0}$ from Line~\ref{line-alg-Peterson-waiting} to
  Line~\ref{line-alg-Peterson-break-2}, making the left-hand side false.
\end{itemize}
\end{proof}

We can now read mutual exclusion directly off of
Lemma~\ref{lemma-alg-Peterson-invariant-2}: if both $p_{0}$ and
$p_{1}$ are at Line~\ref{line-alg-Peterson-critical},
then we get $\Waiting = 1$ and $\Waiting = 0$, a contradiction.

To show progress, observe that the only place where both processes can
get stuck forever is in the loop at
Lines~\ref{line-alg-Peterson-break-1}
and~\ref{line-alg-Peterson-break-2}.  But then $\Waiting$ isn't
changing, and so some process $i$ reads $\Waiting = ¬{}i$ and
leaves.  To show starvation-freedom, observe that if $p_0$ is stuck in
the loop while $p_1$ enters the critical section, then after $p_1$
leaves it sets $\Present[1]$ to $0$ in
Line~\ref{line-alg-Peterson-reset-present} (which lets $p_0$ in if
$p_0$ reads $\Present[1]$ in time), but even if it then sets
$\Present[1]$ back to $1$ in Line~\ref{line-alg-Peterson-present},
it still sets \Waiting to $1$ in
Line~\ref{line-alg-Peterson-waiting}, which lets $p_0$ into the
critical section.  With some more tinkering this argument shows that
$p_1$ enters the critical section at most twice while $p_0$ is in the
trying state, giving $2$-bounded bypass; see
\cite[Lemma~10.12]{Lynch1996}.  With even more tinkering we get a
constant time bound on the waiting time for process $i$ to enter the
critical section, assuming the other process never spends more than
$O(1)$ time inside the critical section.

\subsubsection{Generalization to \texorpdfstring{$n$}{n} processes}
\label{section-mutex-tournament}

(See also \cite[\S4.4.3]{AttiyaW2004}.)

The easiest way to generalize Peterson's two-process algorithm to $n$
processes is to organize a tournament in the form of log-depth binary
tree; this method was invented by Peterson and
Fischer~\cite{PetersonF1977}.  At each node of the tree, the roles of
the two processes are taken by the winners of the subtrees, i.e., the
processes who have entered their critical sections in the two-process
algorithms corresponding to the child nodes.  The winner of the
tournament as a whole enters the real critical section, and afterwards
walks back down the tree unlocking all the nodes it won in reverse
order.  It's easy to see that this satisfies mutual exclusion, and not
much harder to show that it satisfies starvation-freedom. For
starvation-freedom,
the essential idea is that if a winner at some node reaches the
root infinitely often, then starvation-freedom at that node means that a
winner of each child node reaches the root infinitely often.

The most natural way to implement the nodes is to have $\Present[0]$
and $\Present[1]$ at each node be multi-writer variables that can be
written to by any process in the appropriate subtree.  Because the
\Present variables don't do much, we can also implement them as the
OR of many single-writer variables (this is what is done in
\cite[\S10.5.3]{Lynch1996}), but there is no immediate payoff to doing this since the waiting variables are still multi-writer.

Nice properties of this algorithm are that it uses only bits and that
it's very fast: $O(\log n)$ time in the absence of contention.

\subsection{Fast mutual exclusion}
\label{section-mutex-fast}

With a bit of extra work, we can reduce the no-contention cost of
mutual exclusion to $O(1)$, while keeping whatever performance we
previously had in the high-contention case.  The trick (due to
Lamport~\cite{Lamport1987}) is to put an object at the entrance to the
protocol that diverts a solo process onto a ``fast path'' that lets it
bypass the $n$-process mutex that everybody else ends up on.

Our presentation mostly follows \cite{AttiyaW2004}[\S4.4.5], which uses the
\concept{splitter} abstraction of 
Moir and Anderson~\cite{MoirA1995} to separate out the mechanism for
diverting a lone process.\footnote{Moir and Anderson call these things
\concept{one-time building blocks}, but the name \concept{splitter} has become standard in subsequent work.}  Code for a splitter is given in
Algorithm~\ref{alg-splitter}.

\newcommand{\AlgSplitterBody}[1]{
\SharedData\\
atomic register $\SplitterRace$, big enough to hold an ID, initially $⊥$\\
atomic register $\SplitterDoor$, big enough to hold a bit, initially
$\SplitterOpen$\\
\Procedure{$\Splitter(\Id)$}{
    $\SplitterRace ← \Id$\;
    \If{$\SplitterDoor = \SplitterClosed$}{
        \Return \SplitterRight\;
    }
    $\SplitterDoor ← \SplitterClosed$\;
    \eIf{$\SplitterRace = \Id$}{
        \Return \SplitterStop
        \nllabel{line-#1-stop}\;
    }{
        \Return \SplitterDown\;
    }
}
}

\begin{algorithm}
\AlgSplitterBody{alg-splitter}
\caption{Implementation of a splitter}
\label{alg-splitter}
\end{algorithm}

A splitter assigns to each processes that arrives at it the value
\SplitterRight, \SplitterDown, or \SplitterStop.
The useful properties of
splitters are that if at least one process arrives at a splitter, then
(a) at least one process returns \SplitterRight or \SplitterStop; and (b) at
least one process returns \SplitterDown or \SplitterStop; (c) at most
one process returns \SplitterStop; and (d) any process that runs by
itself returns \SplitterStop.  
The first two properties will be useful when we consider
the problem of \concept{renaming} in Chapter~\ref{chapter-renaming};
we will prove them there.
The last two properties are what we want for mutual exclusion.

The names of the variables $\SplitterRace$ and $\SplitterDoor$ follow
the presentation in \cite[\S4.4.5]{AttiyaW2004}; 
Moir and Anderson~\cite{MoirA1995}, following Lamport~\cite{Lamport1987}, call these $X$ and $Y$.  As in~\cite{MoirA1995}, 
we separate out the \SplitterRight and
\SplitterDown outcomes—even though they are equivalent for
mutex—because we will need them later for other applications.

The intuition behind Algorithm~\ref{alg-splitter} is that setting
\SplitterDoor to
$\SplitterClosed$ closes the door to new entrants, and the last entrant to write
its ID to
\SplitterRace wins (it's a slow race), 
assuming nobody else writes \SplitterRace and messes things up.  The added cost
of the splitter is always $O(1)$, since there are no loops.

To reset the splitter, write \SplitterOpen to \SplitterDoor.  This
allows new processes to enter the splitter and possibly return
\SplitterStop.

\begin{lemma}
\label{lemma-splitter-mutex}
After each time that \SplitterDoor is set to \SplitterOpen,
at most one process running Algorithm~\ref{alg-splitter} returns $\SplitterStop$.
\end{lemma}
\begin{proof}
To simplify the argument, we assume that each process calls
$\Splitter$ at most once.

Let $t$ be some time at which $\SplitterDoor$ is set to
$\SplitterOpen$ ($-\infty$ in the case of the initial value).  Let
$S_t$ be the set of processes that read $\SplitterOpen$ from
$\SplitterDoor$ after time $t$ and before the next time at which some
process writes $\SplitterClosed$ to $\SplitterDoor$, and that later
return $\SplitterStop$ by reaching Line~\ref{line-alg-splitter-stop}.

Then every process in $S_t$ reads $\SplitterDoor$ before any process
in $S_t$ writes $\SplitterDoor$.  It follows that every process in
$S_t$ writes $\SplitterRace$ before any process in $S_t$ reads
$\SplitterRace$.  If some process $p$ is not the \emph{last} process
in $S_t$
to write $\SplitterRace$, it will not see its own ID, and will not
return \SplitterStop.  But only one process can be the last process in
$S_t$ to
write $\SplitterRace$.\footnote{It's worth noting that this last process still might not
return \SplitterStop, because some later process—not in
$S_t$—might overwrite $\SplitterRace$.  This can happen even if nobody
ever resets the splitter.}
\end{proof}

\begin{lemma}
\label{lemma-splitter-solo-wins}
If a process runs Algorithm~\ref{alg-splitter} by itself starting from
a configuration in which $\SplitterDoor = \SplitterOpen$, it returns
\SplitterStop.
\end{lemma}
\begin{proof}
Follows from examining a solo execution: the process sets $\SplitterRace$ to
$\Id$, reads $\SplitterOpen$ from $\SplitterDoor$, then reads $\Id$
from $\SplitterRace$.  This causes
it to return \SplitterStop as claimed.
\end{proof}

To turn this into an $n$-process mutex algorithm, we use the splitter
to separate out at most one process (the one that gets \SplitterStop)
onto a \concept{fast path} that bypasses the \concept{slow path} taken
by the rest of the processes.  The slow-path process first fight among
themselves to get through an $n$-process mutex; the winner then fights
in a $2$-process mutex with the process (if any) on the fast path.

Releasing the mutex is the reverse of acquiring it.  If I followed the
fast path, I release the $2$-process mutex first then reset the
splitter.  If I followed the slow path, I release the $2$-process
mutex first then the $n$-process mutex.  This gives mutual exclusion
with $O(1)$ cost for any process that arrives before there is any
contention ($O(1)$ for the splitter plus $O(1)$ for the $2$-process
mutex).

A complication is that if nobody wins the splitter, there is no
fast-path process to reset it.  If we don't want to accept that the
fast path just breaks forever in this case, we have to include a
mechanism for a slow-path process to reset the splitter if it can be
assured that there is no fast-path process left in the system.  The
simplest way to do this is to have each process mark a bit in an array
to show it is present, and have each slow-path process, while still
holding all the mutexes, check on its way out if the $\SplitterDoor$
bit is set and no processes claim to be present.  If it sees all
zeros (except for itself) after seeing $\SplitterDoor =
\SplitterClosed$, it can safely conclude that there is no fast-path
process and reset the splitter itself.  The argument then is that the
last slow-path process to leave will do this, re-enabling the fast
path once there is no contention again.  This approach is taken
implicitly in Lamport's original algorithm, which combines the
splitter and the mutex algorithms into a single miraculous blob.

\subsection{Lamport's Bakery algorithm}
\label{section-Lamport-bakery-algorithm}

See \cite[§{}4.4.1]{AttiyaW2004} or \cite[§{}10.7]{Lynch1996} for some
textbook presentations; the original algorithm is found
in~\cite{Lamport1974}.

This is a starvation-free mutual exclusion algorithm that uses only
single-writer registers (although some of the registers may end up
holding arbitrarily large values).  Code for the Bakery algorithm is
given as Algorithm~\ref{alg-bakery}.

\newData{\MTchoosing}{choosing}
\newData{\MTnumber}{number}

\begin{algorithm}
\SharedData\\
    $\MTchoosing[i]$, an atomic bit for each $i$, initially 0\\
    $\MTnumber[i]$, an \emph{unbounded} atomic register, initially 0\\

Code for process $i$:\\
\While{\True}{
    \tcp{trying}
    $\MTchoosing[i] ← 1$\;
    $\MTnumber[i] ← 1 + \max_{j\ne i} \MTnumber[j]$
    \nllabel{line-alg-bakery-get-number} \;
    $\MTchoosing[i] ← 0$\;

    \For{$j \ne i$}{
        loop until $\MTchoosing[j] = 0$
        \nllabel{line-alg-bakery-loop-choosing}\;
        loop until $\MTnumber[j] = 0$ or 
           $\langle \MTnumber[i], i \rangle
           <\langle \MTnumber[j], j \rangle$
        \nllabel{line-alg-bakery-loop-number} \;
    }

    \tcp{critical}
    (do critical section stuff)\;
    \tcp{exiting}
    $\MTnumber[i] ← 0$\;
    \tcp{remainder}
    (do remainder stuff) \;
}
\caption{Lamport's Bakery algorithm}
\label{alg-bakery}
\end{algorithm}

Note that several of these lines are actually loops; this is obvious
for Lines~\ref{line-alg-bakery-loop-choosing}
and~\ref{line-alg-bakery-loop-number}, but is also true for
Line~\ref{line-alg-bakery-get-number}, which includes an
implicit loop to read all $n-1$ values of $\MTnumber[j]$.

Intuition for mutual exclusion is that if you have a lower number than
I do, then I block waiting for you; for starvation-freedom, eventually I
have the smallest number.  
(There are some additional complications involving the \MTchoosing
bits that we are sweeping under the rug here.)
For a real proof 
see~\cite[\S4.4.1]{AttiyaW2004} or~\cite[\S10.7]{Lynch1996}.

Selling point is a strong near-FIFO guarantee and the use of only single-writer registers (which need not even be atomic—it's enough that they return correct values when no write is in progress).  Weak point is unbounded registers.

\section{RMR complexity}
\label{section-RMR-complexity}

It's not hard to see that we can't build a shared-memory mutex without
busy-waiting: any process that is waiting can't detect that the
critical section is safe to enter without reading a register, but if
that register tells it that it should keep waiting, it is back where
it started and has to read it again.  This makes our standard
step-counting complexity measures useless for describe the worst-case
complexity of a mutual exclusion algorithm.

However, the same argument that suggests we can ignore local
computation in a message-passing model suggests that we can ignore
local operations on registers in a shared-memory model.  Real
multiprocessors have memory hierarchies where memory that is close to
the CPU (or one of the CPUs) is generally much faster than memory that
is more distant.  This suggests charging only for
\index{RMR}
\indexConcept{remote memory reference}{remote memory references},
or RMRs, where each register is local to one of the processes and only
operations on non-local registers are expensive.  This has the advantage of more
accurately modeling real costs~\cite{Mellor-CrummeyS91,Anderson1990},
and allowing us to build busy-waiting mutual exclusion algorithms with
costs we can actually analyze.

As usual, there is a bit of a divergence here between theory and
practice.  Practically, we are interested in algorithms with good
real-time performance, and RMR complexity becomes a heuristic for
choosing how to assign memory locations.  This gives rise to very
efficient mutual exclusion algorithms for real machines, of which the
most widely used is the beautiful MCS algorithm of Mellor-Crummey and
Scott~\cite{Mellor-CrummeyS91}.  Theoretically, we are interested in
the question of how efficiently we can solve mutual exclusion in 
our formal model, and RMR complexity becomes just another complexity
measure, one that happens to allow busy-waiting on local variables.

\subsection{Cache-coherence vs.\ distributed shared memory}

The basic idea of RMR complexity is that a process doesn't pay for
operations on local registers.  But what determines which operations
are local?

In the \concept{cache-coherent} model (CC for short), once a process
reads a register it retains a local copy as long as nobody updates it.
So if I do a sequence
of read operations with no intervening operations by other processes,
I may pay an RMR for the first one (if my cache is out of date), but the rest are free.  
The assumption is that each process can cache registers, and there is some
cache-coherence protocol that guarantees that all the caches stay up
to date.  We may or may not pay RMRs for write operations or other
read operations, depending on the details of the cache-coherence
protocol, but for upper bounds it is safest to assume that we do.

In the
\concept{distributed shared memory} model (DSM), each register is
assigned permanently to a single process.  Other processes can read or
write the register, but only the owner gets to do so without paying an
RMR.  Here memory locations are nailed down to specific processes.

In general, we expect the cache-coherent model to be cheaper than the
distributed shared-memory model, if we ignore constant factors.  The
reason is that if we run a DSM algorithm in a CC model, then the
process $p$ to which a register $r$ is assigned incurs an RMR only if
some other process $q$ accesses $p$ since $p$'s last access.
But then we can amortize $p$'s RMR by charging $q$ double.  Since $q$
incurs an RMR in the CC model, this tells us that we pay at most twice
as many RMRs in DSM as in CC for any algorithm.

The converse is not true: there are (mildly exotic) problems for which
it is known that CC algorithms are asymptotically more efficient than
DSM algorithms~\cite{Golab2011,DanekH2004}.

\subsection{RMR complexity of Peterson's algorithm}
\label{section-mutex-RMR-of-Peterson}

As a warm-up,
let's look at the RMR complexity of
Peterson's two-process mutual exclusion algorithm
(Algorithm~\ref{alg-mutex-Peterson}).
Acquiring the mutex requires going through mostly straight-line code,
except for the loop that tests $\Present[¬i]$ and $\Waiting$.

In the DSM model, spinning on $\Present[¬i]$ is not a problem
(we can make it a local variable of process $i$).  But $\Waiting$ is
trouble.  Whichever process we don't assign it to will pay an RMR
every time it looks at it.  So Peterson's algorithm behaves badly by
the RMR measure in this model.

Things are better in the CC model.  Now process $i$ may pay RMRs for
its first reads of $\Present[¬i]$ and $\Waiting$, but any
subsequent reads are free unless process $¬i$ changes one of them.
But any change to either of the variables causes process $i$ to leave
the loop.  It follows that process $i$ pays at most 3 RMRs to get
through the busy-waiting loop, giving an RMR complexity of $O(1)$.

RMR complexities for parts of a protocol that access different
registers add just like step complexities, so the Peterson-Fischer
tree construction described in §\ref{section-mutex-tournament}
works here too.  The result is $O(\log n)$ RMRs per critical section
access, but only in the CC model.

\subsection{Mutual exclusion in the DSM model}
\label{section-Yang-Anderson}

Yang and Anderson~\cite{YangA1995} give a mutual exclusion algorithm
for the DSM model that requires $Θ(\log n)$ RMRs to reach the
critical section.
This is now known to be optimal for deterministic
algorithms~\cite{AttiyaHW2008}.  The core of the algorithm is a
$2$-process mutex similar to Peterson's, with some tweaks so that each
process spins only on its own registers.  Pseudocode is given in
Algorithm~\ref{alg-yang-anderson}; this is adapted from
\cite[Figure 1]{YangA1995}.

\newFunc{\YAside}{side}
\newData{\YArival}{rival}

\begin{algorithm}
$C[\YAside(i)] ← i$ \;
$T ← i$\;
$P[i] ← 0$\;
$\YArival ← C[¬\YAside(i)]$\;
\If{$\YArival \ne ⊥$ \KwAnd $T = i$}{
    \If{$P[\YArival] = 0$}{
        $P[\YArival] = 1$\;
    }
    \lWhile{$P[i] = 0$}{spin}
    \If{$T = i$} {
        \lWhile{$P[i] ≤ 1$}{spin}
    }
}
\tcp{critical section goes here}
$C[\YAside(i)] ← ⊥$\;
$\YArival ← T$\;
\If{$\YArival \ne i$}{
    $P[\YArival] ← 2$\;
}
\caption{Yang-Anderson mutex for two processes}
\label{alg-yang-anderson}
\end{algorithm}

The algorithm is designed to be used in a tree construction where a
process with ID in the range $\{1\dots n/2\}$ first fights with all
other processes in this range, and similarly for processes in the
range $\{n/2+1 \dots n\}$.  The function $\YAside(i)$ is $0$ for the
first group of processes and $1$ for the second.  The variables $C[0]$
and $C[1]$ are used to record which process is the winner for each
side, and also take the place of the $\Present$ variables in
Peterson's algorithm.  Each process has its own variable $P[i]$ that it spins on when
blocked; this variable is initially $0$ and ranges over $\{0,1,2\}$;
this is used to signal a process that it is safe to proceed, and
tests on $P$ substitute for tests on the non-local variables in Peterson's algorithm.  
Finally, the variable $T$ is used (like $\Waiting$ in
Peterson's algorithm) to break ties: when $T = i$, it's $i$'s turn to
wait.

Initially, $C[0] = C[1] = ⊥$ and $P[i] = 0$ for all $i$.

When I want to enter my critical section, I first set 
$C[\YAside(i)]$ so you can find me; this also has the same effect as
setting $\Present[\YAside(i)]$ in Peterson's algorithm.
I then point $T$ to
myself and look for you.
I'll block if I see $C[¬\YAside(i)] ≠ ⊥$ and $T=i$.  This can
occur in two ways: one is that I really write $T$ after you did, but
the other is that you only wrote $C[¬\YAside(i)]$ but haven't
written $T$ yet.  In the latter case, you will signal to me that $T$
may have changed by setting $P[i]$ to $1$.  I have to check $T$ again
(because maybe I really did write $T$ later), and if it is still $i$,
then I know that you are ahead of me and will succeed in entering your
critical section.  In this case I can safely spin on $P[i]$ waiting
for it to become $2$, which signals that you have left.

There is a proof that this actually works in~\cite{YangA1995}, but it's 27
pages of very meticulously-demonstrated invariants (in fairness, this
includes the entire algorithm, including the tree parts that we
omitted here).  For intuition, this is not much more helpful than
having a program mechanically check all the transitions, since the
algorithm for two processes is effectively finite-state if we ignore
the issue with different processes $i$ jumping into the role of
$\YAside(i)$.

A slightly less rigorous but more human-accessible proof
would be analogous to the proof of Peterson's algorithm.  
We need to show two things: first, that no two processes ever both
enter the critical section, and second, that no process gets stuck.

For the first part, consider two processes $i$ and $j$, where
$\YAside(i) = 0$ and $\YAside(j) = 1$.  We can't have both $i$ and $j$
skip the loops, because whichever one writes $T$ last sees itself in
$T$.  Suppose that this is process $i$ and that $j$ skips the loops.
Then $T=i$ and $P[i] = 0$ as long as $j$ is in the critical section,
so $i$ blocks.  Alternatively, suppose $i$ writes $T$ last but does so
after $j$ first reads $T$.  Now $i$ and $j$ both enter the loops.  But
again $i$ sees $T=i$ on its second test and blocks on the second loop
until $j$ sets $P[i]$ to $2$, which doesn't happen until after $j$
finishes its critical section.

Now let us show that $i$ doesn't get stuck.  Again we'll assume that
$i$ wrote $T$ second.

If $j$ skips the loops, then 
$j$ sets $P[i] = 2$ on its way out as long as $T=i$; this falsifies both loop tests.
If this happens after $i$ first sets $P[i]$ to 0,
only $i$ can set $P[i]$ back to $0$, so $i$ escapes its first loop,
and
any $j'$ that enters from the $1$ side will see $P[i]=2$ before
attempting to set $P[i]$ to $1$, so $P[i]$ remains at $2$ until $i$
comes back around again.  If $j$ sets $P[i]$ to 2 before $i$ sets $P[i]$ to
$0$ (or doesn't set it at all because $T=j$, 
then $C[\YAside(j)]$ is set to $⊥$ before $i$ reads it, so $i$
skips the loops.

If $j$ doesn't skip the loops, then $P[i]$ and $P[j]$ are both set to
$1$ after $i$ and $j$ enter the loopy part.  Because $j$ waits for
$P[j] \ne 0$, when it looks at $T$ the second time it will see $T=i
\ne j$ and will skip the second loop.  This causes it to eventually
set $P[i]$ to 2 or set $C[\YAside(j)]$ to $⊥$ before $i$ reads it
as in the previous case, so again $i$ eventually
reaches its critical section.

Since the only operations inside a loop are on local variables, the
algorithm has $O(1)$ RMR complexity.  For the full tree this becomes
$O(\log n)$.

\subsection{Lower bounds}
\label{section-mutex-lower-bounds}

For deterministic algorithms, there is a lower bound due to Attiya,
Hendler, and Woelfel~\cite{AttiyaHW2008} that shows that any one-shot
mutual exclusion algorithm for $n$ processes incurs $\Omega(n \log n)$
total RMRs in either the CC or DSM models (which implies that some
single process incurs $\Omega(\log n)$ RMRs).  This is based on an earlier
breakthrough lower bound of Fan and Lynch~\cite{FanL2006} that proved
the same lower bound for the number of times a register changes state.
Both bounds are information-theoretic:
a family of $n!$
executions is constructed containing all possible orders in which the processes enter
the critical section, and it is shown that each RMR or state change only
contributes $O(1)$ bits to choosing between them.

For randomized algorithms, Hendler and Woelfel~\cite{HendlerW2011}
have an algorithm that uses $O(\log n/\log \log n)$ expected RMRs
against an adaptive adversary, beating the deterministic lower bound.
This is the best possible for an adaptive adversary, due to a matching lower bound of
Giakkoupis and Woelfel~\cite{GiakkoupisW2012RMR} that holds even for
systems that provide compare-and-swap objects.

For an oblivious adversary, 
an algorithm of Giakkoupis and Woelfel~\cite{GiakkoupisW2014} achieves
$O(1$) expected RMRs using compare-and-swap in the DSM model.
A more recent algorithm of Giakkoupis and
Woelfel~\cite{GiakkoupisW2017} gives the same $O(1)$ expected RMRs in the CC
model; this also uses compare-and-swap.
Curiously, there also exist linearizable $O(1)$-RMR implementations of CAS
from registers in this model~\cite{GolabHHW2012}; however, it is not clear that these
implementations can be combined with the Giakkoupis-Woelfel algorithm
to give $O(1)$ expected RMRs using registers, because variations in
scheduling of randomized implementations may produce subtle
conditioning that gives different behavior from actual atomic objects
in the context of a randomized algorithm~\cite{GolabHW2011}.

\section{Space complexity}
\label{section-Burns-Lynch}

There is a famous result due to Burns and Lynch~\cite{BurnsL1993} that
any mutual exclusion protocol using only read/write registers requires
at least $n$ of them.  Details are in \cite[\S10.8]{Lynch1996}.
A slightly different version of the
argument is given in~\cite[§4.4.4]{AttiyaW2004}.  
The proof is another nice example of an indistinguishability proof, where we use the fact that if a group of processes can't tell the difference between two executions, they behave the same in both.

Assumptions: We have a protocol that guarantees mutual exclusion and
progress. Our base objects are all atomic registers.

Key idea: In order for some process $p$ to enter the critical section,
it has to do at least one write to let the other processes know it is
doing so.  If not, they can't tell if $p$ ever showed up at all, so
eventually either some $p'$ will enter the critical section and
violate mutual exclusion or (in the no-$p$ execution) nobody enters
the critical section and we violate progress.  Now suppose we can park
a process $p_{i}$ on each register $r_{i}$ with a pending write to
$i$; in this case we say that $p_{i}$ \indexConcept{cover}{covers}
$r_{i}$.  If every register is so covered, we can let $p$ go ahead and
do whatever writes it likes and then deliver all the covering writes
at once, wiping out anything $p$ did.  Now the other processes again
don't know if $p$ exists or not.  So we can say something stronger:
before some process $p$ can enter a critical section, it has to write to an uncovered register.

The hard part is showing that we can cover all the registers without
letting $p$ know that there are other processes waiting—if $p$ can
see that other processes are waiting, it can just sit back and wait
for them to go through the critical section and make progress that
way.  So our goal is to produce states in which (a) processes
$p_{1}\dots{},p_{k}$ (for some $k$) between them cover $k$ registers,
and (b) the resulting configuration is indistinguishable from an
\index{configuration!idle}\concept{idle
configuration} to $p_{k+1}\dots{}p_{n}$, where an idle configuration
is one in which every process is in its remainder section.
\begin{lemma}
Starting from any idle configuration $C$, there exists an execution in
which only processes $p_{1}\dots{}p_{k}$ take steps that leads to a
configuration $C'$ such that (a) $C'$ is indistinguishable by any of
$p_{k+1}\dots{}p_{n}$ from some idle configuration $C''$ and (b)
$k$ distinct registers are covered by $p_{1}\dots{}p_{k}$ in $C'$.
\end{lemma}
\begin{proof}
The proof is by induction on $k$.  For $k=0$, let $C'' = C' = C$.

For larger $k$, the essential idea is that starting from $C$, we first
run to a configuration $C_{1}$ where $p_{1}\dots{}p_{k-1}$ cover $k-1$
registers and $C_1$ is indistinguishable from an idle configuration by
the remaining processes, and then run $p_{k}$ until it covers one more
register.  If we let $p_{1}\dots{}p_{k-1}$ go, they overwrite anything
$p_{k}$ wrote.  Unfortunately, they may not come back to covering the
same registers as before if we rerun the induction hypothesis (and in
particular might cover the same register that $p_{k}$ does).  So we
have to look for a particular configuration $C_{1}$ that not only
covers $k-1$ registers but also has an extension that covers the same
$k-1$ registers.

Here's how we find it: Start in $C$.  Run the induction hypothesis to
get $C_{1}$; here there is a set $W_{1}$ of $k-1$ registers covered in
$C_{1}$.  Now let processes $p_{1}$ through $p_{k-1}$ do their pending
writes, then each enter the critical section, leave it, and finish,
and rerun the induction hypothesis to get to a state $C_{2}$,
indistinguishable from an idle configuration by $p_{k}$ and up, in
which $k-1$ registers in $W_{2}$ are covered.  Repeat to get sets
$W_{3}$, $W_{4}$, etc.  Since this sequence is unbounded, and there
are only $\binom{r}{k-1}$ distinct sets of registers to cover (where
$r$ is the number of registers), eventually we have $W_{i} = W_{j}$
for some $i≠j$.  The configurations $C_{i}$ and $C_{j}$ are now
our desired configurations covering the same $k-1$ registers.

Now that we have $C_{i}$ and $C_{j}$, we run until we get to $C_{i}$.
We now run $p_{k}$ until it is about to write some register not
covered by $C_{i}$ (it must do so, or otherwise we can wipe out all of
its writes while it's in the critical section and then go on to
violate mutual exclusion).  Then we let the rest of $p_{1}$ through
$p_{k-1}$ do all their writes (which immediately destroys any evidence
that $p_{k}$ ran at all) and run the execution that gets them to
$C_{j}$.  We now have $k-1$ registers covered by $p_{1}$ through
$p_{k-1}$ and a $k$-th register covered by $p_{k}$, in a configuration that is indistinguishable from idle: this proves the induction step.
\end{proof}

The final result follows by the fact that when $k=n$ we cover $n$
registers; this implies that there are $n$ registers to cover.

It's worth noting that the execution constructed in this proof might
be \emph{very, very long}. It's not clear what happens if we consider
executions in which, say, the critical section is only entered a
polynomial number of times.  If we are willing to accept a small
probability of failure over polynomially-many entries, there is a
randomized mutual exclusion protocol that uses $O(\log n)$
space~\cite{AspnesHTW2018}, at the cost of $O(n)$ amortized RMR
complexity in the cache-coherent model. It is still open whether it
is possible to reduce the space complexity below $O(n)$ for
polynomial-length executions without allowing for a small probability
of failure or without having such high RMR complexity.

\myChapter{The wait-free hierarchy}{2026}{}
\label{chapter-wait-free-hierarchy}
   
In a shared memory model, it may be possible to solve some problems
using \concept{wait-free} protocols, in which any process can finish
the protocol in a bounded number of steps, no matter what the other
processes are doing (see Chapter~\ref{chapter-obstruction-freedom} for more on this and some variants).

The \index{hierarchy!wait-free}\concept{wait-free hierarchy}
$h^{r}_{m}$ classifies asynchronous shared-memory object types $T$ by
\concept{consensus number}, where a type $T$ has consensus number $n$
if with objects of type $T$ and atomic registers (all initialized to
appropriate values\footnote{The justification for assuming that the
objects can be initialized to an arbitrary state is a little tricky.
The idea is that if we are trying to implement consensus from objects
of type $T$ that are themselves implemented in terms of objects of
type $S$, then it's natural to assume that we initialize our simulated
type-$T$ objects to whatever states are convenient. Conversely, if we
are using the ability of type-$T$ objects to solve $n$-process
consensus to show that they can't be implemented from type-$S$ objects
(which can't solve $n$-process consensus), then for both the type-$T$
and type-$S$ objects we want these claims to hold no matter how they
are initialized.

If we don't like the convenient initialization assumption, we can also
use the algorithm of Borowsky~\etal~\cite{BorowskyGA1994} to enforce
initialization to any reachable state. See
§\ref{section-consensus-initialization} for a discussion of how
this works.}) it is possible to solve wait-free consensus (i.e.,
agreement, validity, wait-free termination) for $n$ processes but not
for $n+1$ processes.  The consensus number of any type is at least
$1$, since $1$-process consensus requires no interaction, and may
range up to $\infty$ for particularly powerful objects.

The general idea is that a type $T$ with consensus number $m$ can't
simulate a type $T'$ with a higher consensus number $m'$ for $n>m$
processes, because then we could use the simulation to convert a
$m'$-process consensus protocol using $T'$ into a $m'$-process
consensus protocol using $T$. The converse claim, that objects with
the same or higher consensus numbers can simulate those with lower
ones, is not necessarily true: even though a type $T$ with consensus
number $m$ can implement any object for $n≤m$ processes (see
§\ref{section-universal-construction}), in a system with $n>m$
processes there may be objects with consensus number $m$ that $T$
can't implement. In particular, a result of Afek, Ellen, and
Gafni~\cite{AfekEG2016} gives an infinitely family of objects
$O_{m,k}$ for all $m,k≥2$ with the property that every object
$O_{m,k}$ has consensus number $m$, but $O_{m,k}$ can implement
$O_{m,k'}$ in a system with enough processes if and only if $k'≤k$. So
consensus numbers give only a partial picture of the relative
strengths of different objects. But this partial picture may still be
useful, particularly for proving impossibility results.

The wait-free hierarchy originated in a paper by
Herlihy~\cite{Herlihy1991waitfree}
that classified many common (and some uncommon) shared-memory objects
by consensus number, and showed that an unbounded collection of
objects with consensus number $n$ together with atomic registers gives
a wait-free implementation of any object in an $n$-process system.
We'll describe many of the consensus number proofs from Herlihy's paper in
§\ref{section-wait-free-hierarchy-examples}.
The procedure in each case will be to show an upper bound
on the consensus number using a variant of Fischer-Lynch-Paterson
(made easier because we are wait-free and don't have to worry about
fairness) and then show a matching lower bound (for non-trivial upper
bounds) by exhibiting an $n$-process consensus protocol for some $n$.
This will give a reasonable characterization of the relative strength of
various objects when combined with atomic registers.

Unfortunately, this high-level picture of consensus number hides many
tricky details, which we'll discuss below in
§\ref{section-consensus-numbers-formal-version} before jumping into actually
proving consensus numbers.

\section{Formal version}
\label{section-consensus-numbers-formal-version}

Various subsequent authors noticed that Herlihy's slightly informal
definition of consensus number did not give a
\index{hierarchy!robust}\concept{robust hierarchy} in the sense that combining two types of objects
with consensus number $n$ could solve wait-free consensus for larger
$n$,
and the hierarchy $h^{r}_{m}$ was proposed by
Jayanti~\cite{Jayanti1997}
as a way of
classifying objects that might be robust: an object is at level $n$ of
the $h^{r}_{m}$ hierarchy if having unboundedly many objects plus
unboundedly many registers solves $n$-process wait-free consensus but
not $(n+1)$-process wait-free consensus.\footnote{The $r$ in $h^r_m$
stands for the registers, the $m$ for having many objects of the given
type.  Jayanti~\cite{Jayanti1997} also defines a hierarchy $h^r_1$
where you only get finitely many objects.  
The $h$ stands for ``hierarchy,'' or, more specifically, $h(T)$ stands
for the level of the hierarchy at which $T$ appears~\cite{Jayanti2011}.}

There is some flexibility in what assumptions we make about
initialization and what version of consensus we solve. This is
discussed below in §§\ref{section-consensus-initialization}
and~\ref{section-consensus-output}.

\subsection{Robustness}

Whether or not the resulting
hierarchy is in fact robust for arbitrary deterministic objects is
still open, but Ruppert~\cite{Ruppert2000}
subsequently showed that it is robust for RMW registers and objects
with a read operation that returns the current state, and there is a
paper by Borowsky, Gafni, and Afek~\cite{BorowskyGA1994}
that
sketches a proof based on a topological characterization of
computability\footnote{See Chapter~\ref{chapter-topological-methods}.} that $h^{r}_{m}$ is robust for deterministic objects that
don't discriminate between processes (unlike, say, single-writer
registers).  So for well-behaved shared-memory objects (deterministic, symmetrically accessible, with read operations, etc.),
consensus number appears to give a real classification that allows us
to say for example that any collection of read-write registers
(consensus number $1$), fetch-and-increments ($2$), test-and-set bits
($2$), and queues ($2$) is not enough to build a compare-and-swap
($\infty$).

Ruppert's result is particularly handy because it is based on an
algorithm for computing the consensus number of the objects it
considers, which we'll describe in §\ref{section-n-discerning}.
However, for infinite-state objects, this requires solving the halting
problem (as previously shown by Jayanti and
Toueg~\cite{JayantiT1992}).
We won't attempt to do the robustness proofs of
Borowsky~\etal~\cite{BorowskyGA1994} or Ruppert~\cite{Ruppert2000}.

\subsection{Initialization}
\label{section-consensus-initialization}

Another useful result from the Borowsky~\etal
paper~\cite{BorowskyGA1994} mentioned above is that the
consensus number is not generally dependent on what assumptions we
make about the initial state of the objects. Specifically, \cite[Lemma
3.2]{BorowskyGA1994} states that as long as there is some sequence of
operations that takes an object from a fixed initial state to a
desirable initial state for consensus, then we can safely assume that
the object is in the desirable state. The core idea of the proof is
that each process can initialize its own copy of the object and then
announce that it is ready; each process will then participate in a
sequence of consensus protocols using the objects that they observe are
ready, with the output of each protocol used as the input to the next.
Because the first object $S_i$ to be announced as initialized will be
visible to all processes, they will all do consensus using $S_i$. Any
subsequent protocols that may be used by only a subset of the
processes will not change the common agreed output from the $S_i$
protocol.\footnote{The result in the paper is stated for a consensus
protocol that uses a single copy of the
object, but it generalizes in the obvious way to those that use multiple copies of
the object.} This justifies our assumption that objects can be
initialized to any desired value.

\subsection{Output value of the consensus protocol}
\label{section-consensus-output}

Depending on what we are interested in, we can imagine several
different conventions for the output of a consensus protocol.
These correspond to different choices for the validity condition:
\begin{enumerate}
    \item \indexConcept{binary consensus}{Binary
        consensus}\index{consensus!binary} outputs a value $0$ or
        $1$ that is equal to the input of some participating
        process.
    \item \indexConcept{id consensus}{Id
        consensus}\index{consensus!id} outputs the id of some
        participating process.
    \item \indexConcept{multivalued consensus}{Multivalued
        consensus}\index{consensus!multivalued} outputs a value that is
        equal to the input of some participating process. Unlike
        binary consensus, the range of inputs and outputs is
        arbitrary.
\end{enumerate}

It is trivial to show that multivalued consensus can implement
both binary consensus and id consensus.

In the other direction, if we have id consensus, we can implement
multivalued consensus using a standard trick: have each process $i$ write
its input to a register $r_i$ not used by the id-consensus protocol.
Then each process that learns a winner $j$ from the id-consensus
protocol can read $r_j$ to obtain $j$'s value.

The tricky case is going from binary consensus to id-consensus. Here
the idea is to perform a tournament similar to
Peterson-Fischer~\cite{PetersonF1977}. Build a binary tree whose
internal nodes are binary-consensus protocols $C_b$, each indexed by a
binary string of length equal to its depth. Each process starts at a
leaf determined by the binary expansion of its id and fights its way
to the top. Unlike mutual exclusion, a process continues to fight on
behalf of its subtree even if it loses. Once the outcome at the root
$C_{\Tuple{}}$ is determined, we can work backwards to figure out
which leaf is the actual winner. (See
Algorithm~\ref{alg-id-consensus-from-binary-consensus}.)

\begin{algorithm}
    \tcp{Returns the id of a participating process}
    \Procedure{$\FuncSty{idConsensus}()$}{
        Let $x_1\dots x_\ell = $ binary expansion of my id\;
        \For{$i ← \ell-1$ \DownTo $0$}{
            \tcp{$C_{x_1\dots x_{i-1}}$ is a binary consensus object}
            $C_{x_1 \dots x_{i-1}}(x_i)$\;
        }
        \tcp{Reconstruct winning sequence}
        \For{$i ← 0$ \KwTo $\ell - 1$}{
            \tcp{Get previously decided output}
            $y_{i+1} ← C_{y_1 \dots y_i}(0)$
        }
        \Return $y_1 \dots y_\ell$
    }
    \caption{Id consensus from binary consensus}
    \label{alg-id-consensus-from-binary-consensus}
\end{algorithm}

A complication here is that
this may require processes that didn't participate in a particular
subtree on the way up to be able to detect the outcome of the
consensus protocol for that subtree on the way down. Fortunately,
since we only do this after the winner of the subtree is determined,
it's safe for a curious process to just join the subtree's consensus
protocol with a default input value, since this default input won't
change the outcome. We'll leave the actual proof of correctness as an
exercise.

\subsection{Multiple objects vs multiple operations}
\label{section-consensus-multiple-objects}

When considering multiple objects, the usual assumption is that
objects are combined by putting them next to each other.  If we
can combine two objects by constructing a single object with
operations of both—which is essentially what happens when we apply
different machine language instructions to the same memory
location—then the object with both operations may have a higher consensus number
than the object with either operation individually.  
This was observed by Ellen~\etal~\cite{EllenGSZ2020}. A simple
example would be a register than supports increment ($+1$) and
doubling ($×2$) operations. A register with only one of these
operations is equivalent to a counter and has consensus number $1$.
But a register with both operations has consensus number at least $2$,
since if it is initialized to $2$, we can tell which of the two
operations went first by looking at the final value: $3 = 2+1, 4 =
2×2, 5=(2×2)+1, 6 = (2+1)×2$.  

\section{Classification by consensus number}
\label{section-wait-free-hierarchy-examples}

Here we show the position of various types in the wait-free hierarchy.
The quick description is shown in
Table~\ref{table-wait-free-hierarchy}; more details (mostly adapted
from~\cite{Herlihy1991waitfree}) are given below. A general approach
to computing consensus numbers from~\cite{Ruppert2000}
is deferred to the following section
(§\ref{section-n-discerning}).

\begin{table}
\begin{tabular}{p{0.14\textwidth}  p{0.2\textwidth}  p{0.5\textwidth} }
\toprule
Consensus number&Defining \mbox{characteristic}&Examples\\
\midrule
1&Read with \mbox{interfering} no-return RMW.&\mbox{Registers, counters,}
\mbox{generalized~counters}, \mbox{max~registers}, atomic~snapshots. \\
2&Interfering RMW; queue-like structures.&Test-and-set,
\mbox{fetch-and-add}, queues, \mbox{process-to-memory swap}.\\
$m$& First of $≤m$ write-like operations wins &$m$-process consensus
    objects, $m$-sliding window registers, $m$-bit logical shift
    registers, $m$-discerning objects.\\
$2m-2$& &Atomic $m$-register write.\\
$\infty$&First write-like operation wins.&Queue with peek,
fetch-and-cons, sticky bits, compare-and-swap, memory-to-memory swap,
memory-to-memory copy, arithmetic shift registers.\\
\bottomrule
\end{tabular}
\caption{Position of various types in the wait-free hierarchy}
\label{table-wait-free-hierarchy}
\end{table}

\subsection[Level 1: registers etc.]{Level 1: atomic registers, counters, other interfering RMW registers that don't return the old value}
\label{section-wait-free-level-1}

First observe that any type has consensus number at least 1, since 1-process consensus is trivial.

We'll argue that a large class of particularly weak objects has
consensus number exactly 1, by running FLP with 2 processes.  Recall
from Chapter~\ref{chapter-FLP} that in the 
Fischer-Lynch-Paterson~\cite{FischerLP1985}
proof we classify states as bivalent or univalent depending on whether
both decision values are still possible, and that with at least one
failure we can always start in a bivalent state (this doesn't depend
on what objects we are using, since it depends only on having
invisible inputs).  Since the system is wait-free there is no
constraint on adversary scheduling, and so if any bivalent state has a
bivalent successor we can just do it.  So to solve consensus we have
to reach a bivalent configuration $C$ that has only univalent
successors, and in particular has a 0-valent and a 1-valent successor
produced by applying operations $x$ and $y$ of processes $p_{x}$ and
$p_{y}$.

Assuming objects don't interact with each other behind the scenes, $x$
and $y$ must be operations of the same object.  Otherwise $Cxy = Cyx$ and we get a contradiction.

Now let's suppose we are looking at atomic registers, and consider cases:

\begin{itemize}
 \item $x$ and $y$ are both reads,   Then $x$ and $y$ commute: $Cxy =
 Cyx$, and we get a contradiction.
 \item $x$ is a read and $y$ is a write.  Then $p_{y}$ can't tell the
 difference between $Cyx$ and $Cxy$, so running $p_{y}$ to completion
 gives the same decision value from both $Cyx$ and $Cxy$, another contradiction.
 \item $x$ and $y$ are both writes.  Now $p_{y}$ can't tell the
 difference between $Cxy$ and $Cy$, so we get the same decision value
 for both, again contradicting that $Cx$ is 0-valent and $Cy$ is 1-valent.
\end{itemize}

There's a pattern to these cases that generalizes to other objects.
Suppose that an object has a read operation that returns its state and
one or more read-modify-write operations that don't return anything
(perhaps we could call them ``modify-write'' operations).  We'll say
that the MW operations are 
\index{operations!interfering}
\indexConcept{interfering operations}{interfering} 
if, for any two operations $x$ and $y$, either:
\begin{itemize}
 \item $x$ and $y$
 \index{operations!commuting}
 \indexConcept{commuting operations}{commute}: $Cxy = Cyx$.
 \item One of $x$ and $y$ 
 \index{operations!overwriting}
 \indexConcept{overwriting operations}{overwrites} the other: $Cxy =
 Cy$ or $Cyx = Cx$.
\end{itemize}

Then no pair of read or modify-write operations can get us out of a
bivalent state, because (a) reads commute; (b) for a read and MW, the
non-reader can't tell which operation happened first; (c) and for any
two MW operations, either they commute or the overwriter can't detect
that the first operation happened.  So any MW object with
uninformative, interfering MW operations has consensus number 1.

For
example, consider a counter that supports operations read, increment,
decrement, and write: a write overwrites any other operation, and
increments and decrements commute with each other, so the counter has
consensus number 1.  The same applies to a generalized counter that
supports an atomic $x ← x+a$ operation; as long as this operation
doesn't return the old value, it still commutes with other atomic
increments.  

Max registers~\cite{AspnesAC2012}, which have read operations that return the largest value
previously written, also have commutative updates, so they also have
consensus number 1.  This gives an example of an object not invented
at the time of Herlihy's paper that is still covered by Herlihy's
argument.

\subsection[Level 2: interfering RMW objects etc.]{Level 2: interfering RMW objects that return the old value, queues (without peek)}
\label{section-wait-free-level-2}

Suppose now that we have a RMW object that returns the old value, and
suppose that it is \emph{non-trivial} in the sense that it has at
least one RMW operation where the embedded function $f$ that
determines the new value is not the identity (otherwise RMW is just
read).  Then there is some value $v$ such that $f(v) \ne v$.  To solve
two-process consensus, have each process $p_{i}$ first write its
preferred value to a register $r_{i}$, then execute the non-trivial
RMW operation on the RMW object initialized to $v$.  The first process
to execute its operation sees $v$ and decides its own value.  The
second process sees $f(v)$ and decides the first process's value
(which it reads from the register).\footnote{The extra registers are
just implementing the standard construction of multivalued consensus
from id-consensus; see §\ref{section-consensus-output}.}
It follows that a non-trivial RMW object has consensus number \emph{at least} 2.

In many cases, this is all we get.  Suppose that the operations of
some RMW type $T$ are non-interfering in a way analogous to the previous
definition, where now we say that $x$ and $y$ commute if they leave
the object in the same state (regardless of what values are returned)
and that $y$ overwrites $x$ if the object is always in the same state
after both $x$ and $xy$ (again regardless of what is returned).  The
two processes $p_x$ and $p_y$ that carry out $x$ and $y$ know what happened, but a
third process $p_z$ doesn't.  So if we run $p_z$ to completion we get the
same decision value after both $Cx$ and $Cy$, which means that $Cx$
and $Cy$ can't be 0-valent and 1-valent.  It follows that no collection of RMW registers with interfering operations can solve 3-process consensus, and thus all such objects have consensus number 2.
Examples of these objects include \concept{test-and-set} bits,
\concept{fetch-and-add} registers, and \concept{swap} registers that
support an operation \Swap that writes a new value and 
returns the previous value.

There are some other objects with consensus number 2 that don't fit this pattern.  Define a 
\index{queue!wait-free}\concept{wait-free queue} as an object with
enqueue and dequeue operations (like normal queues), where dequeue
returns $⊥$ if the queue is empty (instead of blocking).  To solve
2-process consensus with a wait-free queue, initialize the queue with
a single value (it doesn't matter what the value is).  We can then
treat the queue as a non-trivial RMW register where a process wins if
it successfully dequeues the initial value and loses if it gets
empty.\footnote{But wait!  What if the queue starts empty?

This turns out to be a surprisingly annoying problem, and was one of
the motivating examples for $h^r_m$ as opposed to Herlihy's vaguer
initial definition.

With one empty queue and nothing else, Jayanti and 
Toueg~\cite[Theorem~7]{JayantiT1992} show that there is no solution to
consensus for two processes.  This is also true for stacks (Theorem 8
from the same paper).  But adding a register (Theorem 9) lets you do
it.  A second empty queue also works.}

However, enqueue operations are non-interfering: if $p_{x}$ enqueues
$v_{x}$ and $p_{y}$ enqueues $v_{y}$, then any third process can
detect which happened first; similarly we can distinguish
$\Enq(x) \Deq()$ from $\Deq() \Enq(x)$.  So to show we can't do three
process consensus we do something sneakier: given a bivalent state $C$
with allegedly 0- and 1-valent successors $C \Enq(x)$ and $C \Enq(y)$,
consider both $C \Enq(x) \Enq(y)$ and $C \Enq(y) \Enq(x)$ and run $p_x$ until it
does a $\Deq()$ (which it must, because otherwise it can't tell what to
decide) and then stop it.  Now run $p_y$ until it also does a $\Deq()$ and
then stop it.  We've now destroyed the evidence of the split and poor
hapless $p_z$ is stuck.  
In the case of $C \Deq() \Enq(x)$ and $C \Enq(x) \Deq()$
on a non-empty queue we can kill the initial dequeuer immediately and
then kill whoever dequeues $x$ or the value it replaced, and if the
queue is empty only the dequeuer knows.  In either case we reach
indistinguishable states after killing only 2 witnesses, and the queue
has consensus number at most 2.

Similar arguments work on stacks, deques, and so forth—these all have consensus number exactly 2.

\subsection{Level \texorpdfstring{$\infty$}{infinity}: objects where the first write wins}
\label{section-wait-free-level-infinity}
\label{section-compare-and-swap}
\label{section-LLSC}

\newData{\SwapVictory}{victory}
\newData{\SwapPrize}{prize}

These are objects that can solve consensus for any number of
processes.
Here are a bunch of level-$\infty$ objects:
\begin{description}
 \item[Queue with peek] \index{queue!with peek} Has operations
 $\Enq(x)$ and $\Peek()$, which
 returns the first value enqueued.  (Maybe also $\Deq()$, but we don't
 need it for consensus).  Protocol is to enqueue my input and then
 peek and return the first value in the queue.
 \item[Fetch-and-cons] \index{fetch-and-cons} Returns old \DataSty{cdr} and adds new
 \DataSty{car} on to the head of a list.  Use preceding protocol where
 $\Peek() = \FuncSty{tail}(\DataSty{car}::\DataSty{cdr})$.
 \item[Sticky bit] \index{sticky bit} Has a $\Write$ operation that
     has no effect unless register
 is in the initial $⊥$ state.  Whether the $\Write$ succeeds or
 fails, it returns nothing.
 The consensus protocol is to write my input and then return result of a read.
 \item[Compare-and-swap] \index{compare-and-swap}\index{CAS} Has $\FuncSty{CAS}(\DataSty{old},
 \DataSty{new})$ operation that writes \DataSty{new} only if previous
 value is \DataSty{old}.  Use it to build a sticky bit.
 \item[Load-linked/store-conditional]
 \index{load-linked/stored-conditional}\index{LL/SC} Like
 compare-and-swap split into
 two operations.  The \indexConcept{load-linked} operation reads a
 memory location and marks it.  The \indexConcept{store-conditional}
 operation succeeds only if the location has not been changed since
 the preceding load-linked by the same process.  Can be used to build
 a sticky bit.
 \item[Memory-to-memory swap]
 \index{swap!memory-to-memory}\index{memory-to-memory swap}
 Has $\Swap(r_{i}, r_{j})$
 operation that atomically swaps contents of $r_{i}$ with $r_{j}$, as
 well as the usual read and write operations for all registers.  Use
 to implement fetch-and-cons.  Alternatively, use two registers
 $\Input[i]$ and $\SwapVictory[i]$ for each process $i$, where
 $\SwapVictory[i]$ is initialized to 0, and a single central register
 \SwapPrize, initialized to 1.  To execute consensus, write your
 input to $\Input[i]$, then swap $\SwapVictory[i]$ with
 \SwapPrize.  The winning value is obtained by scanning all the
 \SwapVictory registers for the one that contains a 1, then returning
 the corresponding $\Input$ value.)
 \item[Memory-to-memory copy] 
 \index{copy!memory-to-memory}
 \index{memory-to-memory copy}
 Has a $\FuncSty{copy}(r_{i}, r_{j})$
 operation that copies $r_{i}$ to $r_{j}$ atomically.  
 Use the same trick as for memory-to-memory swap, where a process
 copies \SwapPrize to $\SwapVictory[i]$.  But now we have a process
 follow up by writing $0$ to \SwapPrize.  As soon as this happens, the
 $\SwapVictory$ values are now fixed; take the leftmost $1$ as the
 winner.\footnote{Or use any other rule that all processes apply
 consistently.}

 Herlihy~\cite{Herlihy1991waitfree} gives a slightly more complicated version
 of this procedure, where there is a separate $\SwapPrize[i]$
 register for each $i$, and after doing its copy a process writes $0$
 to all of the \SwapPrize registers.  This shows that memory-to-memory
 copy solves consensus for arbitrarily many processes even if we
 insist that copy operations can never overlap.  
 The same trick also works for memory-to-memory swap, since we can
 treat a memory-to-memory swap as a memory-to-memory copy given that we don't
 care what value it puts in the $\SwapPrize[i]$ register.
 \item[Bank accounts] A \concept{bank account} object stores a
     non-negative integer, and supports a \Read operation that returns
     the current value and a $\FuncSty{withdraw}(k)$ operation that
     reduces the value by $k$, unless this would reduce the value
     below $0$, in which case it has no effect.

     To solve (binary) consensus with a bank account, start it with $3$, and
     have each process with input $b$ attempt to withdraw $3-b$ from
     the account.  After the first withdrawal, the object will hold
     either $0$ or $1$, and no further withdrawals will have any
     effect.  So the bank account acts exactly like a sticky bit where
     $3$ represents $⊥$.\footnote{If you have more money,
     you can extend this construction to any fixed set of
     values.  For example, to choose among values $v$ in $0\dots m-1$, start with
     $2m$ and have a process with input $v$ subtract $2m-v$.}

     For many years, I assumed that this example demonstrated why
     cryptocurrencies all seem to use embedded consensus protocols of
     some sort. However, it turns out that there is a critical
     assumption needed for this proof, which is that more than one
     process can spend from the same account. 
     Without this assumption, it has been shown by
     Guerraoui~\etal~\cite{GuerraouiKMPS2019} that the consensus number of
     a single-spender bank account is $1$, and more generally that the
     consensus number of a $k$-spender bank account is exactly $k$.
\end{description}

\subsection{Level \texorpdfstring{$2m-2$}{2m-2}: simultaneous
\texorpdfstring{$m$}{m}-register write}
\label{section-wait-free-multi-register-writes}

\newData{\MultiWriteShared}{shared}

Here we have a (large) collection of atomic registers augmented by an
$m$-register write operation that performs all the writes
simultaneously.  The intuition for why this is helpful is that if
$p_{1}$ writes $r_{1}$ and $r_{\MultiWriteShared}$ while $p_{2}$ writes
$r_{2}$ and $r_{\MultiWriteShared}$ then any process can look at the
state of $r_{1}$, $r_{2}$ and $r_{\MultiWriteShared}$ and tell which
write happened first.  Code for this procedure is given in
Algorithm~\ref{alg-multiwrite-race}; note that up to 4 reads may be
necessary to determine the winner because of timing
issues.\footnote{The main issue is that processes can only read the
    registers one at a time.  An alternative to running
    Algorithm~\ref{alg-multiwrite-race} is to use a double-collect
    snapshot (see §\ref{section-double-collect}) to simulate reading
    all three registers at once.  However, this might require as many
    as twelve read operations, since a process doing a snapshot has to re-read
all three registers if any of them change.}

\begin{algorithm}
$v_1 ← r_1$ \;
$v_2 ← r_2$ \;
\If{$v_1 = v_2 = ⊥$}{
    \Return no winner\;
}
\If{$v_1 = 1$ \KwAnd $v_2 = ⊥$}{
    \tcp{$p_1$ went first}
    \Return 1\;
}
\tcp{read $r_1$ again}
$v_1' ← r_1$\;
\If{$v_2 = 2$ \KwAnd $v_1' = ⊥$}{
    \tcp{$p_2$ went first}
    \Return 2\;
}
\tcp{both $p_1$ and $p_2$ wrote}
\eIf{$r_{\MultiWriteShared} = 1$}{
    \Return 2\;
}{
    \Return 1\;
}
\caption[Determining the winner of a race between 2-register
writes]{Determining the winner of a race between 2-register writes.
    The assumption is that $p_1$ and $p_2$ each wrote their own IDs to
$r_i$ and $r_\MultiWriteShared$ simultaneously.  This code can be
executed by any process (including but not limited to $p_1$ or $p_2$)
to determine which of these 2-register writes happened first.}
\label{alg-multiwrite-race}
\end{algorithm}

The workings of Algorithm~\ref{alg-multiwrite-race} are straightforward:
\begin{itemize}
 \item If the process reads $r_{1} = r_{2} = ⊥$, then we don't care which went first, because the reader (or somebody else) already won.
 \item If the process reads $r_{1} = 1$ and then $r_{2} = ⊥$, then
 $p_{1}$ went first.
 \item If the process reads $r_{2} = 2$ and then $r_{1} = ⊥$, then
 $p_{2}$ went first.  (This requires at least one more read after checking the first case.)
 \item Otherwise the process saw $r_{1} = 1$ and $r_{2} = 2$.  Now
 read $r_{\MultiWriteShared}$: if it's 1, $p_{2}$ went first; if it's
 $2$, $p_1$ went first.
\end{itemize}

Algorithm~\ref{alg-multiwrite-race}
requires 2-register writes, and will give us a protocol for 2
processes (since the reader above has to participate somewhere to make
the first case work).  For $m$ processes, we can do the same thing
with $m$-register writes.  We have a register $r_{pq} = r_{qp}$ for
each pair of distinct processes $p$ and $q$, plus a register $r_{pp}$
for each $p$; this gives a total of $\binom{m}{2} + m = O(m^{2})$
registers.  All registers are initialized to $⊥$.  Process $p$ then
writes its initial preference to some single-writer register
$\DataSty{pref}_{p}$ and then simultaneously writes $p$ to $r_{pq}$
for all $q$ (including $r_{pp}$).  It then attempts to figure out the
first writer by applying the above test for each $q$ to $r_{pq}$
(standing in for $r_{\MultiWriteShared}$), $r_{pp}$ ($r_{1}$) and
$r_{qq}$ ($r_{2}$).  If it won against all the other processes, it
decides its own value.  If not, it repeats the test recursively for
some $p'$ that beat it until it finds a process that beat everybody,
and returns its value.  So $m$-register writes solve $m$-process wait-free consensus.

A further tweak gets $2m-2$: run two copies of an $(m-1)$-process
protocol using separate arrays of registers to decide a winner for
each group.  Then add a second phase where processes contend across
the groups.  This
involves each process $p$ from group $1$ writing the
winning ID for its group simultaneously into $s_{p}$ and $s_{pq}$ for
each $q$ in the other group. The first process to do this will be the
only process that wins against every process in the other group, so we
can pick a winning group by looking for some such process. We can then
return the input value for whichever process won within the
winning group.

One thing to note about the second phase is that, unlike mutex, we
can't just have the winners of the two groups fight each other, since
this would not give the wait-free property for non-winners.  Instead,
we have to allow a non-winner $p$ to pick up the slack for a slow winner
and fight on behalf of the entire group.  This requires an $m$-process
write operation to write $s_p$ and all $s_{pq}$ at once.

\subsubsection{Matching impossibility result}

It might seem that the technique used to boost from $m$-process
consensus to $(2m-2)$-process consensus could be repeated to get up to
at least $Θ(m^{2})$, but this turns out not to be the case.  The
essential idea is to show that in order to escape bivalence, we have
to get to a configuration $C$ where \emph{every} process is about to
do an $m$-register write leading to a univalent configuration (since
reads don't help for the usual reasons, and normal writes can be
simulated by $m$-register writes with an extra $m-1$ dummy registers),
and then argue that these writes can't overlap too much.  So suppose
we are in such a configuration, and suppose that $Cx$ is 0-valent and
$Cy$ is 1-valent, and we also have many other operations
$z_{1}\dots{}z_{k}$ that lead to univalent states.  Following
Herlihy~\cite{Herlihy1991waitfree}, we argue in two steps:

\begin{enumerate}
 \item There is some register that is written to by $x$ alone out of
 all the pending operations.  Proof: Suppose not.  Then the 0-valent
 configuration $Cxyz_{1}\dots{}z_{k}$ is indistinguishable from the
 1-valent configuration $Cyz_{1}\dots{}z_{k}$ by any process except
 $p_{x}$, and we're in trouble.
 \item There is some register that is written to by $x$ and $y$ but
 not by any of the $z_{i}$.  Proof:: Suppose not.
 The each register is written by at most one of $x$ and $y$, making it
 useless for telling which went first; or it is overwritten by some
 $z_i$, hiding the value that tells which went first.  So
 $Cxyz_{1}\dots{}z_{k}$ is indistinguishable from
 $Cyxz_{1}\dots{}z_{k}$ for any process other than $p_{x}$ and
 $p_{y}$, and we're still in trouble.
\end{enumerate}

Now suppose we have $2m-1$ processes.  The first part says that each
of the pending operations ($x$, $y$, all of the $z_{i}$) writes to 1
single-writer register and at least $k$ two-writer registers where $k$
is the number of processes leading to a different univalent value.
This gives $k+1$ total registers simultaneously written by this
operation.  Now observe that with $2m-1$ process, there is some set of
$m$ processes whose operations all lead to a $b$-valent state; so for
any process to get to a ($¬{}b$)-valent state, it must write $m+1$
registers simultaneously. It follows that with only $m$ simultaneous
writes we can only do $(2m-2)$-consensus.

Curiously, we can see the last bivalent configuration in the algorithm
given earlier: as long as we have not had any process contend with the
processes in the other group, it is still possible for the winner of
either group to win the overall protocol. If we run each process until
it is about to do its final $m$-register write, we get exactly the
situation where the processes in one group give exactly $m-1$ pending
writes that lead to $0$-valent configurations and the processes in the
other group give exactly $m-1$ pending writes that lead to $1$-valent
configurations, with all of these pending writes overlapping in
exactly the way required by the impossibility argument. In principle
this happens for any consensus implementation that is subject to this
kind of bivalence argument, but it is nice to see the structure of the
upper bound and lower bound matching up so directly in this case.

\subsection[Level \texorpdfstring{$m$}{m}:
various \texorpdfstring{$m$}{m}-bounded objects]{Level
\texorpdfstring{$m$}{m}: \texorpdfstring{$m$}{m}-process consensus
objects, $m$-sliding window registers}
\label{section-m-consensus}

An $m$-process \concept{consensus object} has a single
\FuncSty{consensus} operation that, the first $m$ times it is called,
returns the input value in the first operation, and thereafter returns
only $⊥$.  Clearly this solves $m$-process consensus.  To show that
it doesn't solve $(m+1)$-process consensus even when augmented with
registers, run a bivalent initial configuration to a configuration $C$
where any further operation yields a univalent state.  By an argument
similar to the $m$-register write case, we can show that the pending
operations in $C$ must all be consensus operations on the same
consensus object (anything else commutes or overwrites).  Now run
$Cxyz_{1}\dots{}z_{m-1}$ and $Cyxz_{1}\dots{}z_{m-1}$, where $x$ and $y$
lead to 0-valent and 1-valent states, and observe that the process
that did $z_{m-1}$ can't
distinguish the resulting configurations because all it got was
$⊥$.  (Note: this works even if the consensus object isn't in its
initial state, since we know that before $x$ or $y$ the configuration is still bivalent.)  

So the $m$-process consensus object has consensus number $m$.  This
shows that $h^{r}_{m}$ is nonempty at each level.

A natural question at this point is whether the inability of
$m$-process consensus objects to solve $(m+1)$-process consensus
implies robustness of the hierarchy.  One might consider the following
argument: given any object at level $m$, we can simulate it with an
$m$-process consensus object, and since we can't combine $m$-process
consensus objects to boost the consensus number, we can't combine any
objects they can simulate either.  The problem here is that while
$m$-process consensus objects can simulate any object in a system with
$m$ processes (see below), it may be that some objects can do more in
a system with $m+1$ objects while still not solving $(m+1)$-process
consensus.  A simple way to see this would be to imagine a variant of
the $m$-process consensus object that doesn't fail completely after
$m$ operations; for example, it might return one of the first two
inputs given to it instead of $⊥$.  This doesn't help with solving
consensus, but it might (or might not) make it too powerful to
implement using standard $m$-process consensus objects.

An $m$-process consensus object is arguably a very artificial way to
populate all levels of the consensus hierarchy.
Mostefaoui~\etal~\cite{MostefaouiPR2018} proposed $m$-\conceptFormat{sliding
window registers}\index{sliding window
register}\index{register!sliding window} as a
more natural class of objects that has this property.
An $m$-sliding window register $\text{RW}_m$ possesses a write
operation and a read operation
that returns the last $m$ values written to the register in the order
they were written.\footnote{This particular class of objects has
been independently invented on at least three occasions.
Ellen~\etal~\cite{EllenGSZ2020} define a
\concept{$b$-buffer object}\index{object!$b$-buffer}
that is essentially equivalent, as is the \concept{ring buffer
object}\index{object!ring buffer}
that once appeared on a final exam in this course (see
§\ref{section-MMVIII-exam-ring-buffer}).}

It's easy to solve $m$-process consensus using this object.
We assume that the initial state of the register does not contain any
process IDs, and have each contending process write its ID to the
register. The first writer wins.

The proof that an $m$-sliding window register can't solve consensus
for $m+1$ processes is similar to that for $m$-process consensus
objects. Given a system consisting of read-write registers and
$\text{RW}_m$ objects, choosing the bivalent successor of any
configuration either works forever or eventually reaches a
configuration $C$ with only univalent successors. By the usual
argument, the $m+1$ pending operations in $C$ must all be operations
on the same $m$-sliding window register.

We can easily show that none of these operations can be read
operations. Suppose $x$ is a read operation such that $Cx$ is
$b$-valent, and let $y$ be any operation such that $Cy$ is
$¬b$-valent. Then $Cxy$ and $Cy$ are indistinguishable to the $n-1$
processes that do not execute $x$, giving a contradiction.

Now let $x$ and $y$ be write operations where $Cx$ is $0$-valent and
$Cy$ is $1$-valent. Let $z_1, \dots, z_{m-1}$ be the remaining
operations enabled in $C$. Then $Cxyz_1\dots z_{m-1}$ and $Cyz_1 \dots
z_{m-1}$ apply the same last $m$ writes to the sliding window
register, leaving the resulting configurations indistinguishable to
all processes if the process carrying out $x$ takes no more steps.

Mostefaoui~\etal observe that taking this argument to the limit shows
that a unbounded distributed ledger has infinite consensus number,
which is not entirely surprising given that such an object is
equivalent to fetch-and-cons (§\ref{section-wait-free-level-infinity}).

For more practical objects, a similar result shows that $m$-bit
logical shift registers have consensus number $m$~\cite{Aspnes2025shift}. 
Curiously, arithmetic shift registers (which duplicate the leftmost
bit for right shifts instead of just shifting in extra zeros) have
infinite consensus number.

\section{Consensus numbers of readable objects}
\label{section-n-discerning}

Ruppert~\cite{Ruppert2000} gives a general rule for determining the
consensus number of a \concept{readable object}\index{object!readable}.
In its simplest form, which we will
discuss here, this is an object that supports a $\Read$ operation that returns its entire
state, although Ruppert also considers objects where obtaining the
entire state requires multiple operations, as well as various other
RMW objects that are not necessarily readable even in this sense.
A readable object may also have update operations that may or may not
return a value in addition to changing the state of the object.

A readable object type $T$ is
\conceptFormat{$n$-discerning}\index{discerning}
if it has a reachable state $q_0$ in which it is possible to split processes 
$p_1,\dots,p_n$ into two nonempty
teams $A_0$ and $A_1$ and assign each process $p_i$ an operation $π_i$,
so that after any sequence of operations $π_{i_1},\dots,π_{i_k}$, each
process can tell if the first operation $π_{i_1}$ was performed by a
process in team $A_0$ or team
$A_1$.\footnote{The two teams are called $A$ and $B$ in Ruppert's
original definition, but using subscripts simplifies the presentation
a bit.}

To formalize this, let us define $R_{A_b,j}$ for each $b∈\Set{0,1}$
and each process id $j$ as
the set of all pairs $\Tuple{r,q}$ for which there is an execution 
$q_0 π_{i_1},\dots,π_{i_k}$ where (a) all the $i_\ell$ are distinct; (b)
one of them is $p_j$, (c) $p_{i_1} ∈ A_b$, (d) $p_j$'s operation returns $r$, and (e) the
final state of the object is $q$. Then $T$ is $n$-discerning if, there exist a
state $q_0$, teams $A_0$ and $A_1$, and operations $π_1,\dots,π_n$, such that
$R_{A_0,j} ∩ R_{A_1,j} = ∅$ for all $j$.

The idea here is that this captures what happens in an FLP-style
bivalence proof when we fail to escape bivalence: each team $A_b$ will be all
the processes whose operations lead to a $b$-valent configuration,
and $R_{A_0,j} ∩ R_{A_1,j} ≠ ∅$ means that there
exists a pair of executions with different valencies that are
indistinguishable to $p_j$ running alone,
giving the usual contradiction. The
trickier part is showing that having an $n$-discerning type is enough
to implement wait-free consensus for $n$-processes.

\subsection{Example}

Define an \concept{$m$-bounded counter with overflow bit}\index{counter!$m$-bounded with overflow bit}
as an object
whose state is a pair $\Tuple{c,e}$ where $c$ is a counter value in
$\Set{0,\dots,m-1}$ and $e$ is the overflow bit, initially $0$. Let
$\Inc$ and $\Dec$ increment or decrement $c$, instead setting $e$ permanently
to $1$ if applied when $c=m-1$ or $c=0$ respectively. Let $\Read$
return the object's state.

\begin{lemma}
    \label{lemma-bounded-counter-with-overflow}
    An $m$-bounded counter with overflow bit is $m$-discerning but not
    $(m+1)$-discerning.
\end{lemma}
\begin{proof}
    To show it is $m$-discerning: Let $A_0 = \Set{p_1}$, $A_1 =
    \Set{p_2,\dots,p_m}$, $q_0 = 0$, $π_1 = \Dec$, $π_2,\dots,π_m =
    \Inc$.

    Any sequence of operations that starts with $π_1 = \Dec$ will set
    the overflow bit. Any sequence of operations that starts with an
    $\Inc$ will increment the counter at most $m-1$ time (not setting
    the overflow bit) and if $π_1$ appears in the sequence, it will do
    so only when $c>0$ (again not setting the overflow bit). So
    $R_{A_0,j}$ contains only final states with the overflow bit set
    and $R_{A_1,j}$ contains only final states with the overflow bit
    clear, giving $R_{A_0,j} ∩ R_{A_1,j} = ∅$ for all $j$.

    To show it is not $(m+1)$-discerning: Assume without loss of
    generality that $A_0 = \Set{p_1,\dots,p_k}$ and $A_1 =
    \Set{p_{k+1},\dots,p_{m+1}}$ for some $k$ with $1≤k≤m+1$. Consider the
    sets of operations $Π_b = \SetWhere{π_i}{p_i∈A_b}$. If $Π_0 ∩ Π_1
    ≠ ∅$, then there are operations $π_{i_0} ∈ Π_0$ and $π_{i_1} ∈
    Π_1$ that are both $\Dec$ or both $\Inc$. In either case, for any
    choice of $q_0$, the executions $q_0 π_{i_0} π_{i+1}$ and $q_0
    π_{i_1} π_{i_0}$ give $R_{A_0,j} ∩ R_{A_1,j} ≠ ∅$ for any process
    $p_j$. We can thus assume without loss of generality that
    $π_1,\dots,π_k = \Dec$ and $π_{k+1},\dots,π_{m+1} = \Inc$.

    We now want to show that there are no good choices for $q_0$, by
    doing a somewhat nasty case analysis:
    \begin{enumerate}
        \item If $q_0 = \Tuple{c,e}$ where $c∉\Set{0,m}$, then $q_0
            π_1 π_{m+1}$ and $q_0 π_{m+1} π_1$ both produce the view $\Tuple{r,q} =
            \Tuple{⊥, q_0}$ for $p_1$, giving $R_{A_0,1}∩R_{A_1,1}≠∅$.
        \item If $q_0 = \Tuple{0,e}$, then
            \begin{enumerate}
                \item If $k=1$, then the object 
                    reaches state $\Tuple{m-1,1}$ after either 
                    $q_0 π_1 π_2 \dots π_{m+1}$ ($\Dec$ followed by
                    $m$ $\Inc$s) or $q_0 π_2 \dots
                    π_{m+1}$ ($m$ $\Inc$s). This gives $R_{A_0,2} ∩ R_{A_1,2} ≠ ∅$.
                \item If $k>1$, then $q_0 π_1$ ($\Dec$) and $q_0 π_{m+1} π_1
                    π_2$ ($\Inc$ followed by two $\Dec$s) both leave
                    the object in state $(0,1)$, giving  $R_{A_0,1} ∩ R_{A_1,1} ≠ ∅$.
            \end{enumerate}
        \item If $q_0 = \Tuple{m-1,e}$, an argument symmetric to the
            $q = \Tuple{0,e}$ case gives the same bad result.
    \end{enumerate}
\end{proof}

The reader may be noticing at this point that the 
proof of Lemma~\ref{lemma-bounded-counter-with-overflow}
looks suspiciously like our usual strategy of giving a consensus
algorithm for $m$ processes and an FLP-style impossibility proof for
$m+1$ processes, and may reasonably wonder why we'd bother with the whole notion
of $n$-discerning objects when we could just write these proofs out
directly. The selling point for the $n$-discerning definition is that
(a) we can in principle determine if an object type $T$ is
$n$-discerning entirely mechanically by enumerating all possible
choices of $A_0, A_1, q_0,$ and $π_1,\dots,π_n$, and computing
$R_{A_b,j}$ for each $b$ and $j$; and (b) much of the setup for both
an $n$-process consensus algorithm and an $(n+1)$-process
impossibility result can be factored out into a theorem that applies
to all $n$-discerning (but not $(n+1)$-discerning) objects, which will
we see in Theorem~\ref{theorem-n-discerning} in the next section. So
this approach lets us omit the boring parts we don't need to prove,
and in principle may even let us turn the boring parts we do need to
prove over to a machine, at least in the case of an object with
finitely many states and operations.

\subsection{Consensus number of an $n$-discerning object}
\label{section-n-discerning-implies-consensus}

In this section, we will show that knowing the maximum $n$ for which an object
type is $n$-discerning gives us its consensus number.

First let's show the impossibility result.
\begin{lemma}
    \label{lemma-n-discerning-impossiblity}
    If $T$ is not $n$-discerning, then $h^r_m(T) < n$.
\end{lemma}
\begin{proof}
    Start in a bivalent configuration $C$ as usual and run until all successor
    configurations are not bivalent. By the usual argument, all pending
    operations $π_i$ must be operations on the same object $O$ of type $T$.
    Let team $A_b$ for each $b∈\Set{0,1}$ consist of all processes $i$ for which $C π_i$ is
    $b$-valent. Since $T$ is not $n$-discerning, whatever
    state $q_0$ the object is in, there is some process $p_j$ such
    that $R_{A_0,j} ∩ R_{A_1,j} ≠ ∅$. This means that for each
    $b∈\Set{0,1}$ there is a sequence of pending operations on $O$
    that leads to a $b$-valent configuration $C_b$ where $p_j$ observes the same
    return value from its operation and sees the same state in $O$
    (and every other object, since the pending operations don't change
    the state of those objects). So in
    a solo execution, $p_j$ can't distinguish $C_0$ from $C_1$,
    and so can't decide on either value.
\end{proof}

For the possibility result, we need to construct an algorithm. 
At first glance, the fact that we don't have control over the sizes of
the two teams seems like a problem, since we can't just have all the
processes with input $b$ join team $A_b$. Instead, we'll solve
consensus using a tree of $T$-objects, and add some registers to
propagate the input values.
\begin{lemma}
    \label{lemma-n-discerning-possiblity}
    If $T$ is $n$-discerning, then $h^r_m(T) ≥ n$.
\end{lemma}
\begin{proof}
    We'll show by induction on $k$ that there is a consensus
    protocol for arbitrary input values
    using only $T$  and registers for $k$ processes when $k≤n$. The trivial
    base case is when $k=1$.

    For larger $k$, let $q_0$, $A_0$, and $A_1$ have the property that
    $R_{A_0,j} ∩ R_{A_1,j} ≠ ∅$ for all $p_j$. 
    Let $O$ be an object of type $T$ initialized to $q_0$.
    Have the members of
    each team $A_b$ solve consensus among themselves recursively and write
    the result to a register $r_b$. Each process $p_i$ then applies
    operation $π_i$ to $O$. Since all processes can tell which team
    $A_b$ did the first operation to $O$, they can each read and return the
    value of $r_b$.
\end{proof}

Combining the lemmas gives:
\begin{theorem}
    \label{theorem-n-discerning}
    For any readable object $T$, $h^r_m(T) ≥ n$ if and only if $T$ is
    $n$-discerning.
\end{theorem}

The same argument works for readable objects in the more general sense
where multiple non-updating operations are needed to extract the
entire object state. This in particular applies to collections of
$n$-discerning objects, which can be thought of as a single object as
described in §\ref{section-consensus-multiple-objects}. A consequence
is that the consensus hierarchy is robust for this class of objects. For
more details and some further extensions, see the original
paper~\cite{Ruppert2000}.

\section{Universality of consensus}
\label{section-universal-construction}

\index{consensus!universality of}
\indexConcept{universality of consensus}{Universality of consensus} says that any type that can implement $n$-process consensus can,
together with atomic registers, give a wait-free implementation of any
object in a system with $n$ processes.  That consensus is universal
was shown by Herlihy~\cite{Herlihy1991waitfree} and
Plotkin~\cite{Plotkin1989}.  Both of these papers spend a lot of
effort on making sure that both the cost of each operation and the
amount of space used is bounded.  But if we ignore these constraints,
the same result can be shown using a mechanism similar to the 
replicated state machines of
§\ref{section-replicated-state-machines}.

Here the processes repeatedly use consensus to decide between
candidate histories of the simulated object, and a process
successfully completes an operation when its operation (tagged to
distinguish it from other similar operations) appears in a winning
history.  A round structure avoids too much confusion.

Details are given in Algorithm~\ref{alg-universal-construction}.

\newFunc{\UniversalApply}{apply}
\newFunc{\UniversalConsensus}{consensus}
\newData{\UniversalOp}{op}
\newData{\UniversalRound}{round}

\begin{algorithm}
\Procedure{$\UniversalApply(π)$}{
    \tcp{announce my intended operation}
    $\UniversalOp[i] ← π$\;
    \While{\True}{
        \tcp{find a recent round}
        $r ← \max_j \UniversalRound[j]$\;
        \tcp{obtain the history as of that round}
        \If{$h_r = ⊥$}{
            $h_r ← \UniversalConsensus(c[r], ⊥)$\;
        }
        \If{$π \in h_r$}{
            \Return value $π$ returns in $h_r$\;
        }
        \tcp{else attempt to advance}
        $h' ← h_r$\;
        \Foreach{$j$}{
            \If{$\UniversalOp[j] \not\in h'$}{
                append $\UniversalOp[j]$ to $h'$\;
            }
        }
        $h_{r+1} ← \UniversalConsensus(c[r+1], h')$ \;
        $\UniversalRound[i] ← r+1$\;
    }
}
\caption{A universal construction based on consensus}
\label{alg-universal-construction}
\end{algorithm}

There are some subtleties to this algorithm.  The first time that a
process calls consensus (on $c[r]$), it may supply a dummy input; the
idea is that it is only using the consensus object to obtain the
agreed-upon history from a round it missed.  It's safe to do this,
because no process writes $r$ to its \UniversalRound register until
$c[r]$ is complete, so the dummy input can't be accidentally chosen as
the correct value.

It's not hard to see that whatever $h_{r+1}$ is chosen in $c[r+1]$ is an
extension of $h_r$ (it is constructed by appending operations to
$h_r$), and that all processes agree on it (by the agreement property
of the consensus object $c[r+1]$.  So this gives us an increasing
sequence of consistent histories.
We also need
to show that these histories are linearizable.  
The obvious linearization is just the most
recent version of $h_r$.  Suppose some call to
$\UniversalApply(π_1)$ finishes before a call to
$\UniversalApply(π_2)$ starts.  Then $π_1$ is contained in 
some $h_r$ when $\UniversalApply(π_1)$ finishes, and since $π_2$
can only enter $h$ by being appended at the end, we get $π_1$
linearized before $π_2$.

Finally, we need to show termination.  The algorithm is written with
a loop, so in principle it could run forever.  But we can argue that
no process after executes the loop more than twice.
The reason is that a process $p$ puts its operation in $\UniversalOp[p]$ before it
calculates $r$; so any process that writes $r' > r$ to \UniversalRound
sees $p$'s operation before the next round.  It follows that $p$'s
value gets included in the history no later than round $r+2$.
(We'll see this sort of thing again when we do atomic snapshots in
Chapter~\ref{chapter-atomic-snapshots}.)

A minor complication with this construction is that it assumes
consensus over arbitrary inputs, while some objects directly implement
only binary consensus. Fortunately there is a straightforward
reduction of general consensus to a tree of binary consensus
protocols. Assign a register to the root of each subtree (including
leaves representing the individual processes). To do consensus, I
first write my input to my leaf. I then fight my way up through the
tree solving binary consensus at each node, with input equal to the
side (left or right) I am coming from. Whichever value wins a node,
each process participating in the node will copy the winning value
from the appropriate subtree to the register for that node. Eventually
a single value prevails at the root.

Building a consistent shared history is easier with some particular
objects that solve consensus. For example, a \concept{fetch-and-cons}
object that supplies an operation that pushes a new head onto a linked
list and returns the old head trivially implements the common history
above without the need for helping.  One way to implement
fetch-and-cons is with memory-to-memory swap; to add a new element to the
list, create a cell with its \DataSty{next} pointer pointing to
itself, then swap the \DataSty{next} field with the \DataSty{head}
pointer for the entire list.

The solutions we've described here have a number of deficiencies that
make them impractical in a real system (even more so than many of the
algorithms we've described).  If we store entire histories in a
register, the register will need to be very, very wide.  If we store
entire histories as a linked list, it will take an unbounded amount of
time to read the list.  For solutions to these problems, see
\cite[15.3]{AttiyaW2004} or the papers of Herlihy~\cite{Herlihy1991waitfree} and
Plotkin~\cite{Plotkin1989}.

\myChapter{Atomic snapshots}{2026}{}
\label{chapter-atomic-snapshots}

We've seen in the previous chapter that there are a lot of things we
can't make wait-free with just registers.  But there are a lot of
things we can.  Atomic snapshots are a tool that let us do a lot of
these things easily.

An
\index{object!snapshot}
\concept{atomic snapshot object} acts like a collection of $n$
single-writer multi-reader atomic registers with a special
\concept{snapshot} operation that returns (what appears to be) the
state of all $n$ registers at the same time.  This is easy without failures: we simply lock the whole register file, read them all, and unlock them to let all the starving writers in.  But it gets harder if we want a protocol that is wait-free, where any process can finish its own snapshot or write even if all the others lock up.

We'll give the usual sketchy description of a couple of snapshot
algorithms.  More details on early snapshot results can be found in~\cite[\S10.3]{AttiyaW2004}
or~\cite[\S13.3]{Lynch1996}. A summary of some subsequent results
can be found in a survey by Fich on upper and lower
bounds for the problem~\cite{Fich2005}.

\section{The basic trick: two identical collects equals a snapshot}
\label{section-double-collect}

\newData{\SeqNo}{seqno}

Let's tag any value written with a sequence number, so that each value
written has a \SeqNo field attached to it that increases over time.
We can now detect if a new write has occurred between two reads of the
same variable.  Suppose now that we repeatedly perform
\indexConcept{collect}{collects}—reads of all $n$ registers—until
two successive collects return exactly the same vector of values and
sequence numbers.  We can then conclude that precisely these values
were present in the registers at some time in between the two
collects.  This gives us a very simple algorithm for
snapshot.
Unfortunately, it doesn't terminate if there are a lot of writers
around.\footnote{This isn't always a problem, since there may be
external factors that keep the writers from showing up too much.
Maurice Herlihy and I got away with using exactly this snapshot
algorithm in an ancient, pre-snapshot paper on randomized
consensus~\cite{AspnesH1990consensus}.  The reread-until-no-change idea
was used as early as 1977 by Lamport~\cite{Lamport1977}.}
So we need some way to slow the writers down, or at least get them to do snapshots for us.

\section{Snapshots using double collects with helping}
\label{section-AADGMS}

This is the approach taken by Afek and his five illustrious
co-authors~\cite{AfekADGMS1993} (see also
\cite[§10.3]{AttiyaW2004} or
\cite[§13.3.2]{Lynch1996}): before a process can write to its
register, it first has to complete a snapshot and leave the results
behind with its write.\footnote{The algorithm is usually called the
AADGMS algorithm by people who can remember all the names—or at
least the initials—of the
team of superheroes who came up with it (Afek, Attiya, Dolev, Gafni,
Merritt, and Shavit).
Historically, this was one of three
independent solutions to the problem that appeared at about the same
time.  A
similar algorithm for 
\index{register!composite}
\indexConcept{composite register}{composite registers} 
was given by James Anderson~\cite{Anderson1994} 
and a somewhat different algorithm for \concept{consistent scan} was
given by Aspnes and Herlihy~\cite{AspnesH1990waitfree}.  The
Afek~\etal{} algorithm had the advantage of using bounded registers (in
its full version), and so it and its name for atomic snapshot
prevailed over its competitors.} 
This means that if some slow process (including a slow writer, since now writers need to do snapshots too) is prevented from doing the two-collect snapshot because too much writing is going on, eventually it can just grab and return some pre-packaged snapshot gathered by one of the many successful writers.

Specifically, if a process executing a single snapshot operation $σ$ sees
values written by a single process $i$ with three different sequence
numbers $s_{1}$, $s_{2}$ and $s_3$, then it can be assured that the
snapshot $σ_3$ gathered with sequence number $s_{3}$ started no
earlier than $s_{2}$ was written (and thus no earlier than 
$σ$ started, since $σ$ read $s_{1}$ after it started)
and ended no later than $σ$ ended (because $σ$ saw it).
It follows that the snapshot can safely return
$σ_3$, since that
represents the value of the registers at
some time inside $σ_3$'s interval, which is contained
completely within $σ$'s interval.

So a snapshot repeatedly does collects until either (a) it gets two
identical collects, in which case it can return the results (a
\index{scan!direct}
\concept{direct scan}, or (b) it
sees three different values from the same process, in which case it
can take the snapshot collected by the second write (an
\index{scan!indirect}
\concept{indirect scan}).  See pseudocode in
Algorithm~\ref{alg-simplified-AADGMS}.

Amazingly,
despite the fact that updates keep coming and 
everybody is trying to do snapshots all the
time, a snapshot operation of a single process is guaranteed to
terminate after at most $n+1$ collects.  The reason is that in order
to prevent case (a) from holding, the adversary has to supply at least
one new value in each collect after the first.  But it can only supply
one new value for each of the $n-1$ processes that aren't doing
collects before case (b) is triggered (it's triggered by the first
process that shows up with a second new value).  Adding up all the collects
gives $1 + (n-1) + 1 = n+1$ collects before one of the cases holds.
Since each collect takes $n-1$ read operations (assuming the process
is smart enough not to read its own register), a snapshot operation
terminates after at most $n^{2}-1$ reads.

\newData{\AADGMScount}{count}
\newData{\AADGMSvalue}{value}
\newData{\AADGMSsnapshot}{snapshot}
\newData{\AADGMSinitial}{initial}
\newData{\AADGMSprevious}{previous}

\begin{algorithm}
    \Procedure{$\Update_i(A,v)$}{
        $s ← \Scan(A)$ \;
        $A[i] ← \Tuple{A[i].\AADGMScount + 1, v, s}$\;
    }

    \Procedure{$\Scan(A)$}{
        $\AADGMSinitial ← \Collect(A)$ \;
        $\AADGMSprevious ← \AADGMSinitial$
        \While{\True}{
            $s ← \Collect(A)$ \;
            \uIf{$s = \AADGMSprevious$}{
                \tcp{Two identical collects}
                \Return $s$\;
            }
            \ElseIf{$∃j: s[j].\AADGMScount ≥
            \AADGMSinitial[j].\AADGMScount + 2$}{
                \tcp{Three different counts from $j$}
                \Return $s[j].\AADGMSsnapshot$\;
            }
            \Else{
                \tcp{Nothing useful, try again}
                $\AADGMSprevious ← s$\;
            }
        }
    }
    \caption{Snapshot of~\cite{AfekADGMS1993} using unbounded registers}
    \label{alg-simplified-AADGMS}
\end{algorithm}

For a write operation, a process first performs a snapshot, then
writes the new value, the new sequence number, and the result of the
snapshot to its register (these are very wide registers).  The total
cost is $n^{2}-1$ read operations for the snapshot plus $1$ write
operation.

\subsection{Linearizability}
\label{section-AADGMS-linearizability}

We now need to argue that the snapshot vectors returned by the
Afek~\etal{} algorithm really work, that is, that between each matching
\FuncSty{invoke-snapshot} and \FuncSty{respond-snapshot} there was some
actual time where the registers in the array contained precisely the
values returned in the respond-snapshot action.  We do so by assigning
a \concept{linearization point} to each snapshot vector, a time at
which it appears in the registers (which for correctness of the
protocol had better lie within the interval between the snapshot
invocation and response).  For snapshots obtained through case (a),
take any time between the two collects.  For snapshots obtained
through case (b), take the linearization point already assigned to the
snapshot vector provided by the third write.  In the latter case we
argue by induction on termination times that the linearization point lies inside the snapshot's interval.

Note that this means that all snapshots were ultimately collected by
two successive collects returning identical values, since any case-(b)
snapshot sits on top of a finite regression of case-(b) snapshots that
must end with a case-(a) snapshot. This means that any snapshot
corresponds to an actual global state of the registers at some point
in the execution, which is not true of all snapshot algorithms. It
also means that we can replace the registers in the snapshot array
with other objects that allow us to detect updates (say, counters or
max registers) and still get snapshots.

In an actual execution, the fact that we are waiting for double
collects with no intervening updates means that if there are many
writers, eventually all of them will stall waiting for a case-(a)
snapshot to complete. So that snapshot will complete because all the
writers are stuck. In a sense, requiring writers to do
snapshots first almost gives us a form of locking, but without the
vulnerability to failures of a real lock.

\subsection{Using bounded registers}

The simple version of the Afek~\etal{} algorithm requires unbounded
registers (since sequence numbers may grow forever).  One of the
reasons why this algorithm required so many smart people was to get
rid of this assumption: the paper describes a (rather elaborate)
mechanism for recycling sequence numbers that prevents unbounded
growth (see also \cite[13.3.3]{Lynch1996}).  In practice, unbounded registers are probably not really an issue once one has accepted very large registers, but getting rid of them is an interesting theoretical problem.

It turns out that with a little cleverness we can drop the sequence
numbers entirely.  The idea is that we just need a mechanism to detect
when somebody has done a lot of writes while a snapshot is in
progress.  A naive approach would be to have sequence numbers wrap
around mod $m$ for some small constant modulus $m$; this fails because
if enough snapshots happen between two of my collects, I may notice
none of them because all the sequence numbers wrapped around all the
way.  But we can augment mod-$m$ sequence numbers with a second handshaking mechanism that detects when a large enough number of snapshots have occurred; this acts like the guard bit on an automobile odometer, than signals when the odometer has overflowed to prevent odometer fraud by just running the odometer forward an extra million miles or so.

\newFunc{\TryHandshake}{tryHandshake}
\newFunc{\CheckHandshake}{checkHandshake}

The result is the full version of Afek~\etal~\cite{AfekADGMS1993}.
(Our presentation here
follows \cite[10.3]{AttiyaW2004}.)  The key mechanism for detecting
odometer fraud is a \concept{handshake}, a pair of single-writer bits
used by two processes to signal each other that they have done
something.  Call the processes $S$ (for \emph{same}) and $D$ (for
\emph{different}), and supposed we have handshake bits $h_{S}$ and
$h_{D}$.  We then provide operations
\TryHandshake (signal that something is happening) and \CheckHandshake (check if something happened) for each process; these operations are asymmetric.  The code is:

\begin{description}
 \item[$\TryHandshake(S)$:] $h_{S} ← h_{D}$ (make the two bits the same)
 \item[$\TryHandshake(D)$:] $h_{D} ← ¬{}h_{S}$ (make the two bits different)
 \item[$\CheckHandshake(S)$:] return $h_{S} ≠ h_{D}$ (return true if D changed its bit)
 \item[$\CheckHandshake(D)$:] return $h_{S} = h_{D}$ (return true if S changed its bit)
\end{description}

The intent is that \CheckHandshake returns true if the other process
called \TryHandshake after I did.  The situation is a bit messy,
however, since \TryHandshake involves two register operations (reading
the other bit and then writing my own).  So in fact we have to look at
the ordering of these read and write events.  Let's assume that
\CheckHandshake is called by $S$ (so it returns true if and only if it sees different values).  Then we have two cases:
\begin{enumerate}
 \item $\CheckHandshake(S)$ returns true.  Then $S$ reads a different
 value in $h_{D}$ from the value it read during its previous call to
 $\TryHandshake(S)$.  It follows that $D$ executed a write as part of a
 $\TryHandshake(D)$ operation in between $S$'s previous read and its current read.
 \item $\CheckHandshake(S)$ returns false.  Then $S$ reads the same
 value in $h_{D}$ as it read previously.  This does not necessarily
 mean that $D$ didn't write $h_{D}$ during this interval—it is
 possible that $D$ is just very out of date, and did a write that
 didn't change the register value—but it does mean that $D$ didn't
 perform both a read and a write since $S$'s previous read.
\end{enumerate}

How do we use this in a snapshot algorithm?  The idea is that before
performing my two collects, I will execute \TryHandshake on my end of
a pair of handshake bits for every other process.  After performing my
two collects, I'll execute \CheckHandshake.  I will also assume each
update (after performing a snapshot) toggles a mod-2 sequence number
bit on the value stored in its segment of the snapshot array.  The
hope is that between the toggle and the handshake, I detect any
changes.  (See \cite[Algorithm 30]{AttiyaW2004} for the actual code.)

Does this work?  Let's look at cases:
\begin{enumerate}
 \item The toggle bit for some process $q$ is unchanged between the
 two snapshots taken by $p$.  Since the bit is toggled with each
 update, this means that an even number of updates to $q's$ segment
 occurred during the interval between $p$'s writes.  If this even
 number is 0, we are happy: no updates means no call to \TryHandshake
 by $q$, which means we don't see any change in $q$'s segment, which
 is good, because there wasn't any.  If this even number is 2 or more,
 then we observe that each of these events precedes the following one:
 \begin{itemize}
 \item $p$'s call to \TryHandshake.
 \item $p$'s first read.
 \item $q$'s first write.
 \item $q$'s call to \TryHandshake at the start of its second scan.
 \item $q$'s second write.
 \item $p$'s second read.
 \item $p$'s call to \CheckHandshake.  
 \end{itemize}
 It follows that $q$ both reads and writes the handshake bits in
 between $p$'s calls to \TryHandshake and \CheckHandshake, so $p$
 correctly sees that $q$ has updated its segment.
 \item The toggle bit for $q$ has changed.  Then $q$ did an odd number
 of updates (i.e., at least one), and $p$ correctly detects this fact.
\end{enumerate}

What does $p$ do with this information?  Each time it sees that $q$
has done a scan, it updates a count for $q$.  If the count reaches 3,
then $p$ can determine that $q$'s last scanned value is from a scan
that is contained completely within the time interval of $p$'s scan.
Either this is a \index{scan!direct}\concept{direct scan}, where $q$
actually performs two collects with no changes between them, or it's
an \index{scan!indirect}\concept{indirect scan}, where $q$ got its
value from some other scan completely contained within $q$'s scan.  In
the first case $p$ is immediately happy; in the second, we observe
that this other scan is also contained within the interval of $p$'s
scan, and so (after chasing down a chain of at most $n-1$ indirect
scans) we eventually reach a direct scan contained within it that
provided the actual value.  In either case $p$ returns the value of pair of adjacent collects with no changes between them that occurred during the execution of its scan operation, which gives us linearizability.

\section{Faster snapshots using lattice agreement}
\label{section-snapshots-lattice-agreement}

The Afek~\etal{} algorithm and its contemporaries all require $O(n^2)$
operations for each snapshot.  It is possible to get this bound down
to $O(n)$ using a more clever algorithm,~\cite{InoueMCT1994} which is the best we can
reasonably hope for in the worst case given that (a) even a collect (which doesn't guarantee
anything about linearizability) requires $Θ(n)$
operations when implemented in the obvious way, and (b) there is a linear lower
bound, due to Jayanti, Tan, and Toueg~\cite{JayantiTT2000}, on a large
class of wait-free objects that includes snapshot.\footnote{But see
    §\ref{section-max-register-snapshots} for a faster alternative if we
    allow either randomization or limits on the number of times the
array is updated.}

The first step, due to Attiya, Herlihy, and
Rachman~\cite{AttiyaHR1995}, is a reduction to a related problem called 
\concept{lattice agreement}.

\subsection{Lattice agreement}
\label{section-lattice-agreement-definition}

A \concept{lattice} is a partial order in which every pair of elements
$x$, $y$ has a least upper bound $x∨{}y$ called the \concept{join}
of $x$ and $y$
and a greatest lower bound $x∧{}y$ called the \concept{meet} of
$x$ and $y$.  For example, we can make a lattice out of sets by
letting join be union and meet be intersection; or we can make a
lattice out of integers by making join be $\max$ and meet be $\min$.

In the lattice agreement problem, each process starts with an input
$x_{i}$ and produces an output $y_{i}$, where both are elements of some lattice.  The requirements of the problem are:
\begin{description}
 \item[Comparability]\index{comparability} For all $i$, $j$, $y_{i} ≤ y_{j}$ or
 $y_{j} ≤ y_{i}$.
 \item[Downward validity]\index{validity!downward}
 \index{downward validity} For all $i$, $x_{i} ≤ y_{i}$.
 \item[Upward validity]
 \index{validity!upward}
 \index{upward validity} For all $i$, $y_{i} ≤
 x_{1}∨{}x_{2}∨{}x_{3}∨{}\dots{}∨{}x_{n}$.
\end{description}

These requirements are analogous to the requirements for consensus.
Comparability acts like agreement: the views 
returned by the lattice-agreement protocol are totally ordered.
Downward validity says that each process will include its own input in
its output.  Upward validity acts like validity: an output can't
include anything that didn't show up in some input.

For the snapshot algorithm, we also demand \concept{wait-freedom}: each process terminates after a bounded number of its own steps, even if other processes fail.

Note that if we are really picky, we can observe that we don't
actually need meets; a \concept{semi-lattice} that provides only joins
is enough.  In practice we almost always end up with a full-blown
lattice, because (a) we are working with finite sets, and (b) we
generally want to include a bottom element $⊥$ that is less than
all the other elements, to represent the ``empty'' state of our data
structure.  But any finite join-semi-lattice with a bottom element
turns out to be a lattice, since we can define $x ∧ y$ as the
join of all elements $z$ such that $z ≤ x$ and $z ≤ y$.  We don't
\emph{use} the fact that we are in a lattice anywhere, but it does
save us two syllables not to have to say ``semi-lattice agreement.''

\subsection{Connection to vector clocks}
\label{section-lattice-agreement-and-vector-clocks}

The first step in reducing snapshot to lattice agreement
is to have each writer
generate a sequence of increasing timestamps $r_{1}, r_{2}, \dots{},$
and a snapshot corresponds to some vector of timestamps $\Tuple{t_{1}, t_{2}
\dots{} t_{n}}$, where $t_{i}$ indicates the most recent write by
$p_{i}$ that is included in the snapshot (in other words, we are using
vector clocks again; see §\ref{section-vector-clocks}).  
Now define $v≤v'$ if $v_{i}≤v'_{i}$ for all $i$; the resulting
partial order is a lattice, and in particular we can compute
$x∨{}y$ by the rule $(x∨{}y)_{i} = x_{i}∨{}y_{i}$.

Suppose now that we have a bunch of snapshots that satisfy the
comparability requirement. This means they are totally ordered.  Then we
can construct a sequential execution by ordering the snapshots in
increasing order with each update operation placed before the first
snapshot that includes it.  This sequential execution is not
necessarily a linearization of the original execution, and a single
lattice agreement object won't support more than one operation for
each process, but the idea is that we can nonetheless use lattice
agreement objects to enforce comparability between concurrent
executions of snapshot, while doing some other tricks (exploiting,
among other things, the validity properties of the lattice agreement
objects) to get linearizability over the full execution.

\subsection{The full reduction}
\label{section-lattice-agreement-reduction}

\newData{\LACollect}{collect}
\newData{\LAView}{view}
\newData{\LALA}{LA}

The Attiya-Herlihy-Rachman algorithm is given as
Algorithm~\ref{alg-lattice-agreement-snapshot}.  It uses an array of registers
$R_{i}$ to hold round numbers (timestamps); an array $S_{i}$ to hold
values to scan; an unboundedly humongous array $V_{ir}$ to hold views
obtained by each process in some round; and a collection of
lattice-agreement objects $\LALA_r$, one for each round.

\begin{algorithm}
\Procedure{$\Scan()$}{
    \For{$\DataSty{attempt} ← 1$ \KwTo $2$}{
       $R_{i} ← r ← \max(R_{1}\dots{}R_{n}; R_{i}+1)$ \;
       $\LACollect ← \Read(S_{1}\dots{}S_{n})$ \;
       $\LAView ← \LALA_r(\LACollect)$ \;
       \tcp{max computation requires a collect}
       \If{$\max(R_{1}\dots{}R_{n}) ≤ R_{i}$\nllabel{line-LA-check}}{ 
        $V_{ir} ← \LAView$\;
        \Return $V_{ir}$\;
       }
   }
    \tcp{finding nonempty $V_{jr}$ also requires a collect}
   $V_{ir} ←$ some nonempty $V_{jr}$
   \nllabel{line-LA-take-indirect}\;
   \Return $V_{ir}$\;
}
\caption{Lattice agreement snapshot}
\label{alg-lattice-agreement-snapshot}
\end{algorithm}

The algorithm makes two attempts to obtain a snapshot.  In both cases,
the algorithm advances to the most recent round it sees (or its
previous round plus one, if nobody else has reached this round yet),
attempts a collect, and then runs lattice-agreement to try to get a
consistent view.  If after getting its first view it finds that some
other process has already advanced to a later round, it makes a second
attempt at a new, higher round $r'$ and uses some view that it obtains in this second round, either directly from lattice agreement, or (if it discovers that it has again fallen behind), it uses an indirect view from some speedier process.

The reason why I throw away my view if I find out you have advanced to
a later round is not because the view is bad for me but because it's
bad for you: I might have included some late values in my view that you
didn't see, breaking consistency between rounds.  But I don't have to
do this more than once; if the same thing happens on my second
attempt, I can use an indirect view as in~\cite{AfekADGMS1993},
knowing that it is safe to do so because any collect that went into
this indirect view started after I did.

The update operation is the usual update-and-scan procedure; for
completeness this is given as
Algorithm~\ref{alg-lattice-agreement-update}.  To make it easier to
reason about the algorithm, we assume that an update returns the
result of the embedded scan.

\begin{algorithm}
    \Procedure{$\Update_i(v)$}{
$S_{i} ← (S_{i}.\SeqNo + 1, v)$ \;
\Return $\Scan()$\;
}
\caption{Update for lattice agreement snapshot}
\label{alg-lattice-agreement-update}
\end{algorithm}

\subsection{Why this works}
\label{section-lattice-agreement-reduction-proof}

We need to show three facts:

\begin{enumerate}
 \item All views returned by the scan operation are comparable; that is, there exists a total order on the set of views (which can be extended to a total order on scan operations by breaking ties using the execution order).
 \item The view returned by an update operation includes the update (this implies that future views will also include the update, giving the correct behavior for snapshot).
 \item The total order on views respects the execution order: if
 $π_1$ and $π_2$ are scan operations that return $v_1$ and $v_2$,
 then
 $π_{1} <_{S} π_{2}$ implies $v_{1} ≤ v_{2}$.  (This gives us linearization.)
\end{enumerate}

Let's start with comparability.  First observe that any view returned
is either a direct view (obtained from $\LALA_r)$ or an indirect view
(obtained from $V_{jr}$ for some other process $j$).  In the latter
case, following the chain of indirect views eventually reaches some
direct view.  So all views returned for a given round are ultimately
outputs of $\LALA_r$ and thus satisfy comparability.

But what happens with views from different rounds?  The
lattice-agreement objects only operate within each round, so we need
to ensure that any view returned in round $r$ is included in any
subsequent rounds.  This is where checking round numbers after calling
$\LALA_r$ comes in.

Suppose some process $i$ returns a direct view; that is, it sees no
higher round number in either its first attempt or its
second attempt.  Then at the time it starts checking the round number
in Line~\ref{line-LA-check}, no process
has yet written a round number higher than the round number of $i$'s
view (otherwise $i$ would have seen it).  So no process with a higher
round number has yet executed the corresponding collect operation.
When such a process does so, it obtains values that are at least as
large as
those fed into $\LALA_r$, and $i$'s round-$r$ view is less than or equal to
the vector of these values by upward validity of $\LALA_r$, and thus less
than or equal to the vector of values returned by $LA_{r'}$ for $r' >
r$, by
downward validity of $\LALA_{r'}$.  So we have comparability of all direct views, which
implies comparability of all indirect views as well.

To show that each view returned by a scan includes any preceding update, we observe that either a process returns its first-try scan (which includes the update by downward validity) or it returns the results of a scan in the second-try round (which includes the update by downward validity in the later round, since any collect in the second-try round starts after the update occurs).  So no updates are missed.

Now let's consider two scan operations $π_1$ and $π_2$ where $π_{1}$ precedes
$π_{2}$ in the execution.  We want to show that, for the
views $v_1$ and $v_2$ that these scans return, $v_{1} ≤
v_{2}$. Pick some time between when $π_1$ finishes and $π_2$ starts,
and let $s$ be the contents of the registers at this time. Then $v_1
≤ s$ by upward validity, since any input fed to a lattice agreement
object before $π_1$ finishes was collected from a register whose value
was no greater than it is in $s$. Similarly, $s ≤ v_2$ by downward
validity, because $v_2$ is at least as large as the $\LACollect$ value
read by $π_2$, and this is at least as large as $s$. So $v_1 ≤ s ≤
v_2$.

\subsection{Implementing lattice agreement}
\label{section-lattice-agreement-implementation}

There are several known algorithms for implementing lattice
agreement, including the original algorithm of Attiya, Herlihy, and
Rachman~\cite{AttiyaHR1995} and an adaptive algorithm of Attiya and
Fouren~\cite{AttiyaF2001}.
The best of them (assuming multi-writer registers) is Inoue~\etal's
linear-time lattice agreement protocol~\cite{InoueMCT1994}.

The intuition behind this protocol is to implement lattice agreement using divide-and-conquer.  The processes are organized into a tree, with each leaf in the tree corresponding to some process's input.  Internal nodes of the tree hold data structures that will report increasingly large subsets of the inputs under them as they become available.  At each internal node, a double-collect snapshot is used to ensure that the value stored at that node is always the union of two values that appear in its children at the same time.  This is used to guarantee that, so long as each child stores an increasing sequence of sets of inputs, the parent does so also.  

Each process ascends the tree updating nodes as it goes to ensure that
its value is included in the final result.  A clever data
structure is used to ensure that out-of-date smaller sets don't
overwrite larger ones at any node, and the cost of using this data
structure and carrying out the double-collect snapshot at a node with
$m$ leaves below it is shown to be $O(m)$.  So the total cost of a
snapshot is $O(n + n/2 + n/4 + \dots{} 1) = O(n)$, giving the linear time bound.

\newFunc{\InoueUnion}{Union}
\newFunc{\InoueReadSet}{ReadSet}
\newFunc{\InoueWriteSet}{WriteSet}

Let's now look at the details of this protocol.  There are two main
components: the \InoueUnion algorithm used to compute a new value for each node of the tree, and the \InoueReadSet and \InoueWriteSet operations used to store the data in the node.  These are both rather specialized algorithms and depend on the details of the other, so it is not trivial to describe them in isolation from each other; but with a little effort we can describe exactly what each component demands from the other, and show that it gets it.

The \InoueUnion algorithm does the usual two-collects-without change
trick to get the values of the children and then stores the result.
In slightly more detail:
\begin{enumerate}
 \item Perform \InoueReadSet on both children.  This returns a set of leaf values.
 \item Perform \InoueReadSet on both children again.
 \item If the values obtained are the same in both collects, call \InoueWriteSet on the current node to store the union of the two sets and proceed to the parent node.  Otherwise repeat the preceding step.
\end{enumerate}

The requirement of the \InoueUnion algorithm is that calling \InoueReadSet on a
given node returns a non-decreasing sequence of sets of values; that
is, if \InoueReadSet returns some set $S$ at a particular time and later
returns $S'$, then $S\subseteq{}S'$.  We also require that the set
returned by \InoueReadSet is a superset of any set written by a \InoueWriteSet that precedes it, and that it is equal to some such set.  This last property only works if we guarantee that the values stored by \InoueWriteSet are all comparable (which is shown by induction on the behavior of \InoueUnion at lower levels of the tree).

Suppose that all these conditions hold; we want to show that the
values written by successive calls to \InoueUnion are all comparable, that
is, for any values $S$, $S'$ written by union we have $S\subseteq{}S'$
or $S'\subseteq{}S$.  Observe that $S = L\cup{}R$ and $S' =
L'\cup{}R'$ where $L$, $R$ and $L'$, $R'$ are sets read from the
children.  Suppose that the \InoueUnion operation producing $S$ completes
its snapshot before the operation producing $S'$.  Then
$L\subseteq{}L'$ (by the induction hypothesis) and $R\subseteq{}R'$,
giving $S\subseteq{}S'$.

We now show how to implement the \InoueReadSet and \InoueWriteSet operations.  The
main thing we want to avoid is the possibility that some large set
gets overwritten by a smaller, older one.
The solution is to have $m$
registers $a[1\dots m]$, and write a set of size $s$ to every register
in $a[1\dots s]$ (each register gets a copy of the entire set).
Because register $a[s]$ gets only sets of size $s$ or larger, there is no possibility that our set is overwritten by a smaller one.
If we are clever about how we organize this, we can guarantee that the
total cost of all calls to \InoueReadSet by a particular process is
$O(m)$, as is the cost of the single call to \InoueWriteSet in
\InoueUnion.

Pseudocode for both is given as Algorithm~\ref{alg-LA-set}.  This is a
simplified version of the original algorithm from~\cite{InoueMCT1994},
which does the writes in increasing order and thus forces readers
to finish incomplete writes that they observe, as in
Attiya-Bar-Noy-Dolev~\cite{AttiyaBD1995} (see also
Chapter~\ref{chapter-distributed-shared-memory}).

\begin{algorithm}
\SharedData{array $a[1\dots m]$ of sets, initially $\emptyset$}
\LocalData{index $p$, initially $0$}
\bigskip
\Procedure{$\InoueWriteSet(S)$}{
    \For{$i ← \card*{S}$ \DownTo $1$}{
        $a[i] ← S$\;
    }
}
\bigskip
\Procedure{$\InoueReadSet()$}{
    \tcp{update $p$ to last nonempty position}
    \While{$\True$}{
        $s ← a[p]$ \;
        \eIf{$p = m$ \KwOr $a[p+1] = \emptyset$}{
            \Break\;
        }{
            $p ← p+1$\;
        }
    }
    \Return $s$\;
}
\caption{Increasing set data structure}
\label{alg-LA-set}
\end{algorithm}

Naively, one might think that we could just write directly to
$a[\card*{S}]$
and skip the previous ones, but this makes it harder for a reader to
detect that $a[\card*{S}]$ is occupied.  By writing all the previous
registers, we make it easy to tell if there is a set of size $\card*{S}$ or
bigger in the sequence, and so a reader can start at the beginning and
scan forward until it reaches an empty register, secure in the
knowledge that no larger value has been written.\footnote{This trick
    of reading in one direction and writing in another dates back to a
paper by Lamport from 1977~\cite{Lamport1977}.}  Since we want to
guarantee that no reader every spends more that $O(m)$ operations on
an array of $m$ registers (even if it does multiple calls to
\InoueReadSet), we also have it remember the last location read in
each call to \InoueReadSet and start there again on its next call.
For \InoueWriteSet, because we only call it once, we don't have to be
so clever, and can just have it write all $\card*{S} ≤ m$ registers.

We need to show linearizability.
We'll do so by assigning a specific linearization point to each
high-level operation.  Linearize each call to \InoueReadSet at
the last time that it reads $a[p]$.  Linearize each call to
$\InoueWriteSet(S)$ at the first time at which $a[\card*{S}] = S$ and
$a[i] \ne \emptyset$ for every $i < \card*{S}$ (in other words, at the
first time that some reader might be able to find and return $S$); if
there is no such time, linearize the call at the time at which it
returns.  Since every linearization point is inside its call's 
interval, this gives a linearization that is consistent with the
actual execution.  But we have to argue that it is also consistent
with a sequential execution, which means that we need to show that
every \InoueReadSet operation returns the largest set among those whose
corresponding \InoueWriteSet operations are linearized earlier.

Let $R$ be a call to \InoueReadSet and $W$ a call to
$\InoueWriteSet(S)$.
If $R$ returns $S$, then at the time that $R$ reads $S$ from
$a[\card*{S}]$, we have that (a) every register $a[i]$ with $i <
\card*{S}$ is non-empty (otherwise $R$ would have stopped earlier), and
(b) $\card*{S} = m$ or $a[\card*{S}+1] = \emptyset$ (as otherwise $R$
would have kept going after later reading $a[\card*{S}+1]$.  
From the rule for when \InoueWriteSet calls are linearized, we see
that the linearization point of $W$ precedes this time and that the
linearization point of any call to \InoueWriteSet with a larger set
follows it.  So the return value of $R$ is consistent.

The payoff: unless we do more updates than snapshots, don't want to
assume multi-writer registers, are worried about unbounded space,
have a beef with huge registers, or care about constant factors, it
costs no more time to do a snapshot than a collect.  
So in theory we can get away with assuming snapshots pretty much wherever we need them.

\section{Practical snapshots using LL/SC}
\label{section-snapshots-LL/SC}

\newData{\RSTcurrSeq}{currSeq}
\newData{\RSTmem}{memory}
\newData{\RSThigh}{high}
\newData{\RSTlow}{low}
\newFunc{\RSTscan}{scan}
\newData{\RSTview}{view}
\newFunc{\RSTupdate}{update}
\newData{\RSTvalue}{value}
\newData{\RSTseq}{seq}

Though atomic registers are enough for snapshots, it is possible to
get a much more efficient snapshot algorithm using stronger
synchronization primitives.  An algorithm of Riany, Shavit, and
Touitou~\cite{RianyST2001} uses
\concept{load-linked/store-conditional} objects to build an atomic snapshot protocol with linear-time snapshots and constant-time updates using small registers.  We'll give a sketch of this algorithm here.

The RST algorithm involves two basic ideas: the first is a snapshot
algorithm for a single scanner (i.e., only one process can do
snapshots) in which each updater maintains two copies of its segment,
a \RSThigh copy (that may be more recent than the current scan) and a
\RSTlow copy (that is guaranteed to be no more recent than the current scan).  The idea is that when a scan is in progress, updaters ensure that the values in memory at the start of the scan are not overwritten before the scan is completed, by copying them to the low registers, while the high registers allow new values to be written without waiting for the scan to complete.  Unbounded sequence numbers, generated by the scanner, are used to tell which values are recent or not.

As long as there is only one scanner, nothing needs to be done to
ensure that all scans are consistent, and indeed the single-scanner
algorithm can be implemented using only atomic registers.  But extending the algorithm to
multiple scanners is tricky.  A simple approach would be to keep a
separate low register for each concurrent scan—however, this would
require up to $n$ low registers and greatly increase the cost of an
update.  Instead, the authors devise a mechanism, called a
\index{collect!coordinated}\concept{coordinated collect}, that allows
the scanners collectively to implement a sequence of \emph{virtual
scans} that do not overlap.  Each virtual scan is implemented using
the single-scanner algorithm, with its output written to a common
\emph{view} array that is protected from inconsistent updates using
LL/SC operations (CAS also works).  A scanner participates in virtual scans until it obtains a virtual scan that is useful to it (this means that the virtual scan has to take place entirely within the interval of the process's actual scan operation); the simplest way to arrange this is to have each scanner perform two virtual scans and return the value obtained by the second one.

The paper puts a fair bit of work into ensuring that only $O(n)$ view arrays are needed, which requires handling some extra special cases where particularly slow processes don't manage to grab a view before it is reallocated for a later virtual scan.  We avoid this complication by simply assuming an unbounded collection of view arrays; see the paper for how to do this right.

A  more recent
paper by Fatourou and Kallimanis~\cite{FatourouK2007} gives improved time and space complexity using the same basic technique.

\subsection{Details of the single-scanner snapshot}

The single-scanner snapshot is implemented using a shared \RSTcurrSeq
variable (incremented by the scanner but used by all processes) and an
array \RSTmem of n snapshot segments, each of which is divided into a
\RSThigh and \RSTlow component consisting of a value and a timestamp.
Initially, \RSTcurrSeq is 0, and all memory locations are initialized to
$(⊥, 0)$.  This part of the algorithm does not require LL/SC.

A call to \RSTscan\ copies the first of $\RSTmem[j].\RSThigh$ or
$\RSTmem[j].\RSTlow$ that has a sequence number less than the current
sequence number.  Pseudocode is given as Algorithm~\ref{alg-RST-scan}.

\begin{algorithm}
\Procedure{$\RSTscan()$}{
    $\RSTcurrSeq ← \RSTcurrSeq + 1$\;
    \For{$j ← 0$ \KwTo $n-1$}{
        $h ← \RSTmem[j].\RSThigh$ \;
        \eIf{$h.\RSTseq < \RSTcurrSeq$}{
            $\RSTview[j] ← h.\RSTvalue$\;
        }{
            $\RSTview[j] ← \RSTmem[j].\RSTlow.\RSTvalue$\;
        }
    }
}
\caption{Single-scanner snapshot: \FuncSty{scan}}
\label{alg-RST-scan}
\end{algorithm}

The \RSTupdate operation for process $i$ cooperates by copying
$\RSTmem[i].\RSThigh$ to $\RSTmem[i].\RSTlow$ if it's old.

The \RSTupdate\ operation always writes its value to
$\RSTmem[i].\RSThigh$, but preserves the previous value in
$\RSTmem[i].\RSTlow$ if its sequence number indicates that it may have
been present at the start of the most recent call to \RSTscan.
This means that $\RSTscan$ can get the old value if the new value is
too recent.
Pseudocode
is given in Algorithm~\ref{alg-RST-update}.

\begin{algorithm}
\Procedure{$\RSTupdate()$}{
    $\RSTseq ← \RSTcurrSeq$ \;
    $h ← \RSTmem[i].\RSThigh$ \;
    \If{$h.\RSTseq \ne \RSTseq$}{
        $\RSTmem[i].\RSTlow ← h$\;
    }
    $\RSTmem[i].\RSThigh ← (\RSTvalue, \RSTseq)$\;
}
\caption{Single-scanner snapshot: \FuncSty{update}}
\label{alg-RST-update}
\end{algorithm}

To show this actually works, we need to show that there is a
linearization of the scans and updates that has each scan return
precisely those values whose corresponding updates are linearized
before it.  The ordering is based on when each \RSTscan operation $S$
increments \RSTcurrSeq and when each \RSTupdate\ operation $U$ reads it; specifically:
\begin{itemize}
 \item If $U$ reads \RSTcurrSeq after $S$ increments it, then $S < U$.
 \item If $U$ reads \RSTcurrSeq before $S$ increments it and $S$ reads
 $\RSTmem[i].\RSThigh$ (where $i$ is the process carrying out $U$)
 before $U$ writes it, then $S < U$.
 \item If $U$ reads \RSTcurrSeq before $S$ increments it, but $S$ reads
 $\RSTmem[i].\RSThigh$ after $U$ writes it, then $U < S$.
\end{itemize}

Updates are ordered based on intervening scans (i.e., $U_1 < U_2$ if
$U_1<S$ and $S<U_2$ by the above rules), or by the order in which they read \RSTcurrSeq if there is no intervening scan.

To show this is a linearization, we need first to show that it extends
the ordering between operations in the original schedule.  Each of the
above rules has $π_1 < π_2$ only if some low-level operation of
$π_1$ precedes some low-level operation of $π_2$, with the
exception of the transitive ordering of two update events with an
intervening scan.  But in this last case we observe that if $U_1<S$,
then $U_1$ writes $\RSTmem[i].\RSThigh$ before $S$ reads it, so if
$U_1$ precedes $U_2$ in the actual execution, $U_2$ must write
$\RSTmem[i].\RSThigh$ after $S$ reads it, implying $S<U_2$.

Now we show that the values returned by \RSTscan are consistent with
the linearization ordering; that, is, for each $i$, \RSTscan copies to
$\RSTview[i]$ the value in the last \RSTupdate by process $i$ in the
linearization.   Examining the code for \RSTscan, we see that a
\RSTscan operation $S$ takes $\RSTmem[i].\RSThigh$ if its sequence
number is less than $\RSTcurrSeq$, i.e., if the \RSTupdate operation $U$
that wrote it read \RSTcurrSeq before $S$ incremented it and wrote
$\RSTmem[i].\RSThigh$ before $S$ read it; this gives $U<S$.
Alternatively, if \RSTscan takes $\RSTmem[i].\RSTlow$, then
$\RSTmem[i].\RSTlow$ was copied by some update operation $U'$ from the
value written to $\RSTmem[i].\RSThigh$ by some update $U$ that read
\RSTcurrSeq before $S$ incremented it.  Here $U'$ must have written
$\RSTmem[i].\RSThigh$ before $S$ read it (otherwise $S$ would have
taken the old value left by $U$) and since $U$ precedes $U'$ (being an
operation of the same process) it must therefor also have written
$\RSTmem[i].\RSThigh$ before $S$ read it.  So again we get the first
case of the linearization ordering and $U<S$.

So far we have shown only that $S$ obtains values that were linearized
before it, but not that it ignores values that were linearized after
it.  So now let's consider some $U$ with $S < U$.  Then one of two cases holds:
\begin{itemize}
 \item $U$ reads \RSTcurrSeq after $S$ increments it.  Then $U$ writes a
 sequence number in $\RSTmem[i].\RSThigh$ that is greater than or equal to
 the \RSTcurrSeq value used by $S$; so $S$ returns $\RSTmem[i].\RSTlow$
 instead, which can't have a sequence number equal to \RSTcurrSeq and thus
 can't be $U$'s value either.
 \item $U$ reads \RSTcurrSeq before $S$ increments it but writes
 $\RSTmem[i].\RSThigh$ after $S$ reads it.  Now $S$ won't return $U$'s
 value from $\RSTmem[i].\RSThigh$ (it didn't read it), and won't get
 it from $\RSTmem[i].\RSTlow$ either (because the value that \emph{is}
 in $\RSTmem[i].\RSThigh$ will have $\DataSty{seq} < \RSTcurrSeq$, and so
 $S$ will take that instead).
\end{itemize}

So in either case, if $S<U$, then $S$ doesn't return $U$'s value.  This concludes the proof of correctness.

\subsection{Extension to multiple scanners}
See the paper for details.

The essential idea: \RSTview now represents a \emph{virtual scan}
$\RSTview_{r}$ generated cooperatively by all the scanners working
together in some asynchronous round $r$.  To avoid conflicts, we
update $\RSTview_{r}$ using LL/SC or compare-and-swap (so that only
the first scanner to write wins), and pretend that reads of
$\RSTmem[i]$ by losers didn't happen.  When $\RSTview_{r}$ is full,
start a new virtual scan and advance to the next round (and thus the
next $\RSTview_{r+1}$).

\section{Applications}
\label{section-snapshot-applications}

Here we describe a few things we can do with snapshots.

\subsection{Multi-writer registers from single-writer registers}
One application of atomic snapshot is building multi-writer registers
from single-writer registers.  The idea is straightforward: to perform
a write, a process does a snapshot to obtain the maximum sequence
number, tags its own value with this sequence number plus one, and
then writes it.  A read consists of a snapshot followed by returning
the value associated with the largest sequence number (breaking ties
by process ID).  (See~\cite[\S13.5]{Lynch1996} for a proof that this
actually works.)  This requires using a snapshot that doesn't use
multi-writer registers, and turns out to be overkill in practice; there are
simpler algorithms that give $O(n)$ cost for reads and writes based on
timestamps (see~\cite[10.2.3]{AttiyaW2004}).  

With additional work, it is even possible to eliminate the requirement
of multi-reader registers, and get a simulation of multi-writer
multi-reader registers that goes all the way down to single-writer
single-read registers, or even single-writer single-reader bits.  See
\cite[§\S10.2.1--10.2.2]{AttiyaW2004} or \cite[\S13.4]{Lynch1996} for details.

\subsection{Counters}

Given atomic snapshots, it's easy to build a counter (supporting
increment, decrement, and read operations); or, in more generality, a
generalized counter (supporting increments by arbitrary amounts); or, in even
more generality, an object supporting any collection of commutative
and associative
update operations (as long as these operations don't return anything).  The idea is that each process stores in its segment the total of all operations it has performed so far, and a read operation is implemented using a snapshot followed by summing the results.  This is a case where it is reasonable to consider multi-writer registers in building the snapshot implementation, because there is not necessarily any circularity in doing so.

\subsection{Resilient snapshot objects}

The previous examples can be generalized to objects with operations
that either read the current state of the object but don't update it
or update the state but return nothing, provided the update operations
either overwrite each other (so that $Cxy = Cy$ or $Cyx = Cx$) or
commute (so that $Cxy = Cyx$).

This was shown by Aspnes and
Herlihy~\cite{AspnesH1990waitfree} and improved on by Anderson and
Moir~\cite{AndersonM1993} by eliminating unbounded space usage.
Anderson and Moir
also defined the terms 
\index{object!snapshot}\indexConcept{snapshot object}{snapshot
objects} for those with separate read and update operations and
\index{object!resilient}
\index{resilient object}
\concept{resilience} for the property that all operations commute or
overwrite.
The basic idea underneath both of these papers is to use the
multi-writer register construction given above, but break ties among
operations with the same sequence numbers by first placing overwritten
operations before overwriting operations and only then using process
IDs.

This \emph{almost} shows that snapshots can implement any object with
consensus number 1 where update operations return nothing, because an
object that is not resilient violates the commute-or-overwrite condition in some
configuration has consensus number at least 2 (see
§\ref{section-wait-free-level-2})—in Herlihy's terminology,
non-resilient objects have interfering operations.  It doesn't quite work (as
observed in the Anderson-Moir paper), because the tie-breaking
procedure assumes a static ordering on which operations overwrite each
other, so that given
operations $x$ and $y$ where $y$ overwrites $x$, $y$ overwrites $x$ in
any configuration.  But there may be objects with a \emph{dynamic}
ordering to how operations interfere, where $y$ overwrites $x$ in some configuration, $x$
overwrites $y$ in another, and perhaps even the two operations commute
in yet another.  This prevents us from achieving consensus, but also
breaks the tie-breaking technique.  So it may be possible that there
are objects with consensus number 1 and no-return updates 
that we still can't implement using only
registers.

\myChapter{Lower bounds on perturbable objects}{2026}{}
\label{chapter-JTT}

Being able to do snapshots in linear time means that we can build
linearizable counters, generalized counters, max registers, and so on, in
linear time, by having each reader take a snapshot and combine the
contributions of each updater using the appropriate commutative and
associative operation.  A natural question is whether we can do better
by exploiting the particular features of these objects.

Unfortunately, the Jayanti-Tan-Toueg~\cite{JayantiTT2000} lower bound for
\concept{perturbable} objects says each of these objects requires
$n-1$ space and $n-1$ steps for a read operation in the worst case,
for any solo-terminating deterministic implementation from historyless
objects.  Like Burns-Lynch, this is a worst-case bound based on a
covering argument, so it may be possible to evade it in some cases 
using either randomization or a restriction on the length of an
execution (see Chapter~\ref{chapter-restricted-use}).

\concept{Perturbable} means that the object has a particular property that makes the proof work, essentially that the outcome of certain special executions can be changed by stuffing lots of extra update operations in the middle (see below for details).  

\index{termination!solo}
\index{solo termination}
\indexConcept{solo-terminating}{Solo-terminating} means that a process finishes its current operation in a finite number of steps if no other process takes steps in between; it is a much weaker condition, for example, than wait-freedom.  
\index{object!historyless}

\indexConcept{historyless object}{Historyless objects} are those for
which any operation either never changes the state (like a read, but
it could be weaker) or always sets the state to a value that depends
only on the operation and not the previous value (like a write, but it
may also return some information about the old state).  The point of
historyless objects is that covering arguments work for them: if there
is a process with a pending update operations on some object, the
adversary can use it at any time to wipe out the state of the object
and hide any previous operations from any process except the updater
(who, in a typical covering argument, is quietly killed to keep it
from telling anybody what it saw).

Atomic
registers are a common example of a historyless object: the read never
changes the state, and the write always replaces it.
\index{object!swap}
\indexConcept{swap object}{Swap objects} (with a swap operation that
writes a new state while returning the old state) are the canonical
example, since they can implement any other historyless object (and
even have consensus number 2, showing that even extra consensus power doesn't necessarily help here).
Test-and-sets (which are basically one-bit swap objects where you can
only swap in $1$) are also historyless.  In contrast, anything that
looks like a counter or similar object where the new state is a
combination of the old state and the operation is \emph{not}
historyless.   This is important because many of these objects turn out
to be perturbable, and if they were also historyless, we'd get a
contradiction.

Below is a sketch of the proof.  See the original
paper~\cite{JayantiTT2000} for more details.

The basic idea is to build a sequence of executions of the form $Λ_{k}Σ_{k}π$,
where $Λ_{k}$ is a preamble consisting of various complete
update operations and $k$ incomplete update operations by processes
$p_1$ through $p_{n-1}$,
$Σ_{k}$ delivers $k$ delayed writes from the incomplete
operations in $Λ_{k},$ and $π$ is a operation by $p_n$
that returns
some information about the object that is affected by previous
operations.
To make our life easier, we'll assume that $π$ performs only read
steps.\footnote{The idea is that if $π$ does anything else, then the
return values of other steps can be simulated by doing a $\Read$
in place of the first step and using the property of being historyless
to compute the return values of subsequent steps.  There is still a
possible objection 
that we might have some historyless objects that don't even provide
$\Read$ steps.
The easiest way to work around this is to assume that our objects do
in fact provide a $\Read$ step, because taking the $\Read$
step away isn't going to make implementing the candidate
perturbable object any easier.}

We'll expand $Λ_k Σ_k$ to $Λ_{k+1} Σ_{k+1}$ by inserting new operations in between $Λ_k$ and
$Σ_k$, and argue that because those operations can change the value returned
by $π$, one of them must write an object
not covered in $Σ_k$, which will (after some more work) allow us to cover yet another object.

In order for these covered objects to keep accumulating, the reader
has to keep looking at them.
To a first approximation, this means that we want the first $k$ reads done by $π$ to
be from objects written in $Σ_k$: since the values seen by the reader
for these objects never change, the (deterministic) reader will
continue to read them even as we add more operations before $Σ_k$.  Unfortunately, this does not quite
match all possible cases, because it may be that $π$ performs useless reads of objects that aren't
covered in $Σ_k$ but that aren't written to by anybody anyway.  So we
have the more technical condition that $π$ has an initial prefix that
only includes covered reads and useless reads: formally, there is a
prefix $π'$ of $π$ that includes at least one read operation of every
object covered by $Σ_k$, such that any other read operation in $π'$
reads an object whose state cannot be 
changed by any step that can be performed by any sequence of
operations by processes $p_1$ through $p_{n-1}$ that
can be inserted between $Λ_k$ and $Σ_k π$.

The induction hypothesis is that an execution $Λ_k Σ_k$ with these
properties exists for each $k ≤ n-1$.

For the base case, $Λ_{0}Σ_{0} = \Tuple{}$.  This
covers $0$ reads by $π$.

For the induction step, we start with $Λ_k Σ_k$, and look for a
partial execution $γ$ that we can insert in between $Λ_k$ and
$Σ_k$ that changes what $π$ returns in $Λ_k γ Σ_k π$ from what it
returned in $Λ_k γ Σ_k$.

This is where perturbability comes in: an
object is defined to be \concept{perturbable} if such a partial
execution $γ$ always exists.

Some examples of $γ$:
\begin{itemize}
   \item For a snapshot object, let $γ$ write to a component that is
       not written to by any of the operations in $Σ_k$.
   \item For a max register, let $γ$ include a bigger write than all the others.
   \item For a counter, let $γ$ include at least $n$
       increments.  We need $n$ increments, because with fewer increments, we can make $π$ return
    the same value by being sneaky about when the partial increments
    represented in $Σ_{k}$ are linearized.
    The same choice works for a mod-$m$ counter if $m$ is at
        least $2n$, and similarly we can argue that a
        fetch-and-increment or fetch-and-add is perturbable by a $γ$
        that includes at least $n$ fetch-and-increments.
\end{itemize}

In contrast, historyless objects (including atomic registers) are not perturbable: if $Σ_{k}$ includes a write that sets the value of the object, no set of operations inserted before it will change this value.  This is good, because we know that it only takes one atomic register to implement an atomic register.

Assuming that our object is perturbable, now we want to use the
existence of $γ$ to generate our bigger execution $Λ_{k+1} Σ_{k+1}$.
As in the Burns-Lynch mutex bound~\cite{BurnsL1993}, we will be
arguing that $γ$ must include a write to an object that is not
covered by the $k$ delayed writes.  Also as in Burns-Lynch, it turns
out that it is not enough just to delay this particular write, because
it might not cover the specific object we want.

Instead, we look for an alternative $γ'$ that changes the value of the
earliest object read by $π$ that can be changed.  We know that some
such $γ'$ exists, because $γ$ writes to some such object, so there
must be a first place in the execution of $π$ where the output of an
object can change, and there must be some $γ'$ that makes that change.
Note however that $γ'$ that hits that earliest object need not be the
same as the $γ$ used to demonstrate perturbability, and indeed it may
be that $γ'$ is very different from $γ$—in particular, it may be much
longer.

So now we expand $γ' = αβδ$, where $β$ is the magic write to the
uncovered object, and let $Λ_{k+1} = Λ_{k}αδ'$ and $Σ_{k+1} =
β Σ_{k}$, where $δ'$ consists of running all incomplete operations in
$α$ except the one that includes $β$ to completion.  We've now covered
$k+1$ distinct objects in $Σ_k$ and have no incomplete operations in
$Λ_{k+1}$ except the $k+1$ operations that cover these objects.
It remains only to show that the
technical condition that any uncovered object that $π$ reads before
reading all the covered objects can't have its value changed by
inserting additional operations.

Suppose that there is a sequence of operations $κ$ such that $Λ_{k+1}
κ$ changes one of these forbidden uncovered objects.  But $Λ_{k+1} κ =
Λ_k α κ$, and so $γ'' = ακ$ changes an object that either (a) can't be
changed because of the technical condition in the induction hypothesis
for $k$, or (b) changes an object that $π$ reads before the object
covered by $β$.  In the second case, this $γ''$ changes an earlier
object that $γ'$, contradicting the choice of $γ'$.

It follows that we do in fact manage to cover $k+1$ objects while satisfying
the technical condition, and the induction hypothesis holds for $k+1$.

We can repeat this step until we've covered $n-1$ objects.  This
implies that there \emph{are} at least $n-1$ objects (the space lower bound), and in the
worst case some reader reads all of them (the step complexity lower bound).

\myChapter{Restricted-use objects}{2026}{}
\label{chapter-restricted-use}

The Jayanti-Tan-Toueg bound puts a hard floor under the worst-case
complexity of almost anything interesting we'd like to implement with
solo termination in a system that provides only historyless objects as primitives. As with
the consensus hierarchy lower bounds, we could interpret
this as a reason to demand stronger primitives. Or we could look for
ways to bypass the JTT bound.

One approach is to modify our target objects so that they are no
longer perturbable. This can be done by limiting their use: a counter
or max register that can only change its value a limited number of
times is not perturbable, because once we hit the limit, there is no
perturbing sequence of operations that we can insert between $Λ_k$ and
$Σ_k$ in the JTT execution that changes the value returned by the
eventual reader. This observation motivated a line of work on
restricted-use max registers~\cite{AspnesAC2012} and restricted-use
snapshots~\cite{AspnesACHE2015} that have polylogarithmic worst-case
individual step complexity assuming a polynomial limit on updates. While
restricted-use objects might not be all that exciting on their own,
they in turn have served as building blocks for implementations of
snapshots with polylogarithmic polylogarithmic amortized individual
step complexity~\cite{AhadBaigHMT2020}.

In this chapter, we will concentrate on the original restricted-use
max register construction of
Aspnes, Attiya, and
Censor-Hillel~\cite{AspnesAC2012}, and its extension to give
restricted-use snapshots by
Aspnes~\etal~\cite{AspnesACHE2015}.

\section{Max registers}
\label{section-max-register-definition}

We will start by implementing a restricted-use
\index{register!max}\concept{max register}~\cite{AspnesAC2012}, 
for which read operation returns the largest value previously written, as opposed to the last value previously written.  So after writes of 0, 3, 5, 2, 6, 11, 7, 1, 9, a read operation will return 11.

In general, max registers are perturbable objects in the sense of the Jayanti-Tan-Toueg
bound, so in the worst case a
max-register read will have to read at least $n-1$ distinct atomic
registers, giving an $n-1$ lower bound on both step complexity and
space.  But we can get around this by considering bounded max
registers, which only hold values in some range $0\dots m-1$.  These
are not perturbable because once we hit the upper bound we can no
longer insert new operations to change the value returned by a read.
This allows for a much more efficient implementation (at least in
terms of step complexity) when $m$ is not too big.

\section{Implementing bounded max registers}
\label{section-max-register-implementation}

This implementation is from a paper by Aspnes, Attiya, and
Censor-Hillel~\cite{AspnesAC2012}.  The same paper shows that it is in
a certain sense the only possible implementation of a wait-free
restricted max register (see §\ref{section-max-register-lower-bound}).

For $m=1$, the implementation is trivial: write does nothing and read
always returns $0$.

For larger $m$, we'll show how to paste together two max registers
\MRleft and \MRright with $m_{0}$ and $m_{1}$ values
together to get a max register $r$ with $m_{0}+m_{1}$ values.  We'll
think of each value stored in the max register as a bit-vector, with
bit-vectors ordered lexicographically.  In addition to \MRleft and
\MRright, we will need a 1-bit atomic register \MRswitch used to
choose between them.  The read procedure is straightforward and is
shown in Algorithm~\ref{alg-max-register-read}; essentially we just
look at \MRswitch, read the appropriate register, and prepend the
value of \MRswitch to what we get.

\begin{algorithm}
\Procedure{$\Read(r)$}{
  \eIf{$\MRswitch = 0$}{
   \Return $0:\Read(\MRleft)$\;
  }{
   \Return $1:\Read(\MRright)$\;
}
}
\caption{Max register read operation}
\label{alg-max-register-read}
\end{algorithm}

For write operations, we have two somewhat asymmetrical cases
depending on whether the value we are writing starts with a 0 bit or a
1 bit.  These are shown in Algorithm~\ref{alg-max-register-write}.

\begin{algorithm}
\Procedure{$\Write(r, 0x)$}{
    \If{$\MRswitch = 0$}{
       $\Write(\MRleft, x)$\;
    }
}

\Procedure{$\Write(r, 1x)$}{
    $\Write(\MRright, x)$ \;
    $\MRswitch ← 1$\;
}
\caption{Max register write operations}
\label{alg-max-register-write}
\end{algorithm}

The intuition is that the max register is really a big tree of switch
variables, and we store a particular bit-vector in the max register by
setting to 1 the switches needed to make $\Read$ follow the path
corresponding to that bit-vector.  The procedure for writing $0x$
tests \MRswitch first, because once \MRswitch gets set to $1$, any
$0x$ values are smaller than the largest value, and we don't want them
getting written to \MRleft where they might confuse particularly slow
readers into returning a value we can't linearize.  The procedure for
writing $1x$ sets \MRswitch second, because (a) it doesn't need to
test \MRswitch, since $1x$ always beats $0x$, and (b) it's not safe to
send a reader down into \MRright until some value has actually been written there.

It's easy to see that $\Read$ and $\Write$ operations both require exactly one operation per bit of the value read or written.  To show that we get linearizability, we give an explicit linearization ordering (see the paper for a full proof that this works):

\begin{enumerate}
 \item All operations that read $0$ from \MRswitch go in the first pile.
\begin{enumerate}
  \item Within this pile, we sort operations using the linearization
  ordering for \MRleft.
\end{enumerate}
 \item All operations that read $1$ from \MRswitch or write $1$ to \MRswitch go in the second pile, which is ordered after the first pile.
\begin{enumerate}
  \item Within this pile, operations that touch \MRright are ordered
  using the linearization ordering for \MRright.  Operations that
  don't (which are the ``do nothing'' writes for $0x$ values) are placed consistently with the actual execution order.
\end{enumerate}
\end{enumerate}

To show that this gives a valid linearization, we have to argue first
that any $\Read$ operation returns the largest earlier $\Write$ argument and that we don't put any non-concurrent operations out of order.

For the first part, any $\Read$ in the $0$ pile returns $0 : \Read(\MRleft)$, and $\Read(\MRleft)$ returns (assuming \MRleft is a
linearizable max register) the largest value previously written to
left, which will be the largest value linearized before the \Read, or
the all-0 vector if there is no such value.  In either case we are
happy.  Any \Read in the 1 pile returns $1 : \Read(\MRright)$.  Here
we have to guard against the possibility of getting an all-0 vector
from $\Read(\MRright)$ if
no $\Write$ operations linearize before the $\Read$.  But any $\Write$
operation that writes $1x$ doesn't set \MRswitch to 1 until after it
writes to \MRright, so no \Read operation ever starts
$\Read(\MRright)$ until after at least one $\Write$ to \MRright has
completed, implying that that $\Write$ to \MRright linearizes before
the \Read from \MRright.  So in all the second-pile operations
linearize as well.

\section{Encoding the set of values}
\label{section-max-register-encoding}

If we structure our max register as a balanced tree of depth $k$, we
are essentially encoding the values $0\dots 2^{k}-1$ in binary, and
the cost of performing a read or write operation on an $m$-valued
register is exactly $k = \left\lceil\lg m\right\rceil$.  But if we are
willing to build an unbalanced tree, any \concept{prefix code} will work.

The paper describes a method of building a max register where the cost
of each operation that writes or reads a value $v$ is $O(\log v)$.
The essential idea is to build a tree consisting of a rightward path
with increasingly large left subtrees hanging off of it, where each of
these left subtrees is twice as big as the previous.  This means that
after following a path encoded as $1^{k}0$, we hit a $2^{k}$-valued
max register.  The value returned after reading some $v'$ from this
max register is $v' + (2^{k}-1)$, where the $2^{k}-1$ term takes into
account all the values represented by earlier max registers in the
chain.  Formally, this is equivalent to encoding values using an
\concept{Elias gamma code}~\cite{Elias1975}, tweaked slightly by changing the prefixes
from $0^{k}1$ to $1^{k}0$ to get the ordering right.

\section{Unbounded max registers}
\label{section-max-register-unbounded}

While the unbalanced-tree construction could be used to get an
unbounded max register, it is possible that read operations might not
terminate (if enough writes keep setting 1 bits on the right path
before the read gets to them) and for very large values the cost even
of terminating reads becomes higher than what we can get out of a
snapshot.

Here is the snapshot-based method: if each process writes its own
contribution to the max register to a single-writer register, then we
can read the max register by taking a snapshot and returning the
maximum value.  (It is not hard to show that this is
linearizable.)  This gives an unbounded max register with read and
write cost $O(n)$.
So by choosing this in preference to the
balanced tree when $m$ is large, the cost of either operation on a max register is
$\min\left(\left\lceil\lg m\right\rceil, O(n)\right)$.

We can combine this with the unbalanced tree by terminating the right
path with a snapshot-based max register.  This gives a cost for reads and
writes of values $v$ of $O(\min(\log v, n))$.

\section{Lower bound}
\label{section-max-register-lower-bound}

The $\min(\ceil{\lg m}, O(n))$ cost of a max register
read turns out to be exactly optimal, at least for the $\ceil{\lg m}$
part; there is a lower bound~\cite{AspnesAC2012} of $\min(\ceil{\lg
m}, n-1)$. Intuitively, we can show by a
covering argument that once some process attempts to write to a
particular atomic register, then any subsequent writes convey no
additional information (because they can be overwritten by the first
delayed write). So in effect, no algorithm can use get more than one
bit of information out of each atomic register, and any max register
read ends up looking like chasing a path through a tree of switches. But as always,
turning this intuition into an actual proof requires a bit more work.

We will consider solo-terminating executions in
which $n-1$ writers do any number of max-register writes in some
initial prefix $Λ$, followed by a single max-register read
$π$ by process $p_n$.  Let $T(m,n)$ be the optimal reader cost for
executions with this structure with $m$ values, and let $r$ be the
first register read by process $p_n$, assuming it is running an algorithm optimized for this class of executions (we do not even require it to be correct for other executions).

We are now going split up our set of values based on which will cause
a write operation to write to $r$.  Let $S_{k}$ be the set of all sequences of
writes that only write values $≤ k$.  Let $t$ be the smallest
value such that some execution in $S_{t}$ writes to $r$ (there must be
some such $t$, or our reader can omit reading $r$, which contradicts
the assumption that it is optimal).

\begin{description}
 \item[Case 1] Since $t$ is smallest, no execution in $S_{t-1}$ writes
 to $r$.  If we restrict writes to values $≤ t-1$, we can omit
 reading $r$, giving $T(t,n) ≤ T(m,n) - 1$, from which $T(m,n) ≥ T(t,n)
 + 1$.
 \item[Case 2] Let $α$ be some execution in $S_{t}$ that writes
 to $r$.
\begin{itemize}
  \item Split $α$ as $α'δβ$ where $δ$ is the
  first write to $r$ by some process $p_{i}$.
  \item Construct a new execution $α'\eta$ by letting
  all the max-register writes except the one performing $δ$ finish.
  \item Now consider any execution $α'\eta\gammaδ$, where
  $\gamma$ is any sequence of max-register writes with values $≥ t$
  that excludes $p_{i}$ and $p_{n}$.  Then $p_{n}$ always sees the
  same value in $r$ following these executions, but otherwise
  (starting after $α'\eta$) we have an $(n-1)$-process
  max-register with values $t$ through $m-1$.
  \item Omit the read of $r$ again to get $T(m,n) ≥ T(m-t, n-1) +
  1$.
\end{itemize}
\end{description}

We've shown the recurrence $T(m,n) ≥ \min_{t}(\max(T(t,n),
T(m-t,n))) + 1$, with base cases $T(1,n) = 0$ and $T(m,1) = 0$.  The
solution to this recurrence is exactly $\min(\left\lceil\lg
m\right\rceil, n-1)$, which is the same, except for a constant factor on 
$n$, as the upper bound we got by choosing
between a balanced tree for small $m$ and a snapshot for $m ≥
2^{n-1}$.  For small $m$, the recursive split we get is also the same
as in the tree-based algorithm: call the $r$ register \MRswitch and
you can extract a tree from whatever algorithm somebody gives you.  So
this says that the tree-based algorithm is (up to choice of the tree)
essentially the unique optimal bounded max register implementation for
$m ≤ 2^{n-1}$.

It is also possible to show lower bounds on randomized implementations
of max registers and other restricted-use objects.
See~\cite{AspnesAC2012,AspnesCAH2016,HendlerK2014} for examples.

\section{Max-register snapshots}
\label{section-max-register-snapshots}

With some tinkering, it's possible to extend the max-register
construction to get an array of max registers
that supports snapshots.  The description in this section 
follows~\cite{AspnesACHE2015}, with some updates to fix a bug noted in
the original paper in an erratum published by the
authors~\cite{AspnesACHE2015erratum}.

Formally, a \index{array!max}\concept{max array} is an object $a$ that supports
an operation $\Write(a[i], v)$ that sets $a[i] ← \max(v,
a[i])$, and an
operation $\Read(a)$ that returns a snapshot of all components of the
array.
The first step in building this beast is to do it for two components.
The resulting 
\index{array!max!2-component}
\index{max array!2-component}
\concept{$2$-component max array} can then
be used as a building block for larger max arrays and for fast
restricted-used snapshots in general.

A $k \times \ell$ max array $a$ is one that permits values in the range
$0 \dots k-1$ in $a[0]$ and $0 \dots \ell-1$ in $a[1]$.
We think of $a[0]$ as the \MRhead of the max array and $a[1]$ as the
\MRtail.
We'll show how to
construct such an object recursively from smaller objects of the same
type, analogous to the construction of an $m$-valued max register (which we can
think of as a $m \times 1$ max array).  The idea is to split \MRhead
into two pieces \MRleft and \MRright as before, while representing
\MRtail as a master copy stored in a max register 
at the top of the tree plus cached copies at
every internal node.  These cached copies are updated by readers 
at times carefully chosen to ensure linearizability.

The base of the construction is an $\ell$-valued max register
$r$, used directly as a $1 \times \ell$ max array; this is the case where
the \MRhead component is trivial and we only need to store $a.\MRtail
= r$.  Here calling $\Write(a[0], v)$ does nothing, while $\Write(a[1], v)$ maps to $\Write(r, v)$,
and $\Read(a)$ returns
$\Tuple{0, \Read(r)}$.

For larger values of $k$, paste a $k_{\MRleft}\times \ell$ max array
\MRleft and a $k_\MRright \times
\ell$ max array \MRright together
to get a $(k_{\MRleft}+k_\MRright)\times \ell$ max array.
This construction
uses a \MRswitch variable as in the basic construction,
along with an $\ell$-valued max register $\MRtail$ that is used to store the
value of $a[1]$.  

Calls to $\Write(a[0], v)$ and $\Read(a)$
follow the structure of the corresponding operations for a
simple max register, with some extra work in $\Read$ to make sure that the
value in $\MRtail$ propagates into \MRleft and \MRright
as needed to ensure the correct value is returned.

A call to $\Write(a[1], v)$
operation writes $\MRtail$ directly, and then calls $\Read(a)$ to
propagate the new value as well.\footnote{This call to $\Read(a)$ was
omitted in the original published version of the
algorithm~\cite{AspnesACHE2015}, but was added in an erratum by the
authors~\cite{AspnesACHE2015erratum}.  Without it, the implementation
can violate linearizability in some executions.}

Pseudocode is given in Algorithm~\ref{alg-max-array}.

\begin{algorithm}
    \Procedure{$\Write(a[i], v)$}{
    \eIf{$i = 0$} {
        \eIf{$v < k_{\MRleft}$}{
            \If{$a.\MRswitch = 0$}{
                $\Write(a.\MRleft[0], v)$\;
            }
        }{
            $\Write(a.\MRright[0], v-k_{\MRleft})$ \;
            $a.\MRswitch ← 1$\;
        }
    }{
        $\Write(a.\MRtail, v)$\;
        $\Read(a)$\;
    }
}
\bigskip
\Procedure{$\Read(a)$}{
    $x ← \Read(a.\MRtail)$ \;
    \eIf{$a.\MRswitch = 0$}{
        $\Write(a.\MRleft[1], x)$ \;
        \Return $\Read(a.\MRleft)$\;
    }{
        $x ← \Read(a.\MRtail)$ \;
        $\Write(a.\MRright[1], x)$ \;
        \Return $\Tuple{k_{\MRleft}, 0} + \Read(a.\MRright)$\;
    }
}
\caption{Recursive construction of a $2$-component max array}
\label{alg-max-array}
\end{algorithm}

The individual step complexity of each operation is easily computed.
Assuming a balanced tree, $\Write(a[0], v)$ takes exactly $\ceil{\lg
k}$
steps, while $\Write(a[1], v)$ costs exactly $\ceil{\lg \ell}$
steps plus the cost of $\Read(a)$.  Read operations are more complicated.  In the worst case, we 
have two reads of $a.\MRtail$ and a write to $a.\MRright[1]$ at each
level, plus up to two operations on $a.\MRswitch$, for a total cost of at most
$(3 \ceil{\lg k} - 1) (\ceil{\lg \ell} + 2) = O(\log k \log \ell)$ steps.  This
dominates other costs in $\Write(a[1],v)$, so the asymptotic cost of
both $\Write$ and $\Read$ operations is $O(\log k \log \ell)$.

In the
special case where $k = \ell$, both writes and reads have their step
complexities squared compared 
to a single-component $k$-valued max register.

\subsection{Linearizability}

In broad outline, the proof of linearizability follows the proof for a
simple max register.  But as with snapshots, we have to show that the
ordering of the head and tail components are consistent.

The key observation is the following lemma.

\begin{lemma}
\label{lemma-max-array-consistency}
Fix some execution of a max array $a$ implemented as in
Algorithm~\ref{alg-max-array}.
Suppose this execution contains a $\Read(a)$ operation $π_{\MRleft}$ that returns
$v_{\MRleft}$ from $a.\MRleft$ and a $\Read(a)$ operation $π_\MRright$ that returns
$v_\MRright$ from $a.\MRright$.
Then $v_{\MRleft}[1] ≤ v_\MRright[1]$.
\end{lemma}
\begin{proof}
Both $v_{\MRleft}[1]$ and $v_\MRright[1]$ are values that were
previously written to their respective max arrays by $\Read(a)$
operations (such writes necessarily exist because any process that
reads $a.\MRleft$ or $a.\MRright$ writes $a.\MRleft[1]$ or
$a.\MRright[1]$ first).
From examining the code, we have that any value written
to $a.\MRleft[1]$ was read from $a.\MRtail$ before $a.\MRswitch$ was
set to $1$, while any value written to $a.\MRright[1]$ was read from
$a.\MRtail$ after $a.\MRswitch$ was set to $1$.  Since 
max-register reads are non-decreasing, we have than any value written
to $a.\MRleft[1]$ is less than or equal to any value written to
$a.\MRright[1]$, proving the claim.
\end{proof}

The rest of the proof is tedious but straightforward: we linearize the
$\Read(a)$ and $\Write(a[0])$ operations as in the max-register proof,
then fit the $\Write(a[1])$ operations in based on the \MRtail values of
the reads.  The full result is:

\begin{theorem}
\label{theorem-max-array}
If $a.\MRleft$ and $a.\MRright$ are linearizable max arrays, and
$a.\MRtail$ is a linearizable max register, then
Algorithm~\ref{alg-max-array} implements a linearizable max array.
\end{theorem}

It's worth noting that the same unbalanced-tree construction used in
§§\ref{section-max-register-encoding}
and~\ref{section-max-register-unbounded} can be used here as well.
This makes the step complexity for $\Read(a)$ scale as $O(\log v[0]
\log v[1])$, where $v$ is the value returned.  For writes the step
complexity may depend in a complicated way on what values are being
written and to which side, but in the worst case, it is $O(\log v[0]
\log v[1])$, where $v$ is the value in the register when the write
finishes. (This is a consequence of the embedded $\Read(a)$ in
$\Write(a,1,v)$.)

\section{Restricted-use snapshots}

To build an ordinary snapshot object from $2$-component max arrays, we construct a
balanced binary tree in which each leaves holds a pointer to an individual snapshot
element and each internal node holds a pointer to a partial snapshot
containing all of the elements in the subtree of which it is the root.
The pointers themselves are non-decreasing indices into arrays of
values that consist of ordinary (although possibly very wide)
atomic registers.

When a process writes a new value to its component
of the snapshot object, it increases the pointer value in its leaf and
then propagates the new value up the tree by combining together
partial snapshots at each step, using $2$-component max arrays to
ensure linearizability.  The resulting algorithm is similar in many
ways to the lattice agreement procedure of
Inoue~\etal~\cite{InoueMCT1994} (see 
§\ref{section-lattice-agreement-implementation}), except that it uses a
more contention-tolerant snapshot algorithm than double collects and
we allow processes to update their values more than once.  It is also
similar to the \concept{$f$-array}
construction of Jayanti~\cite{Jayanti2002} for
efficient computation of array aggregates (sum, min, max, etc.) using
LL/SC, the main difference being that because the index values are
non-decreasing, max arrays can substitute for LL/SC.

Each node in the tree except the root is represented by one component
of a $2$-component max array that we can think of as being owned by
its parent, with the other component being the
node's sibling in the tree.  To propagate a value up the tree, at
each level the process takes a snapshot of the two children of the node
and writes
the sum of the indices to the node's component in its parent's max array (or to an
ordinary max register if we are at the root).  Before doing this last
write, a process will combine the partial snapshots from the two child
nodes and write the result into a separate array indexed by the sum.
In this way any process that reads the node's component can obtain the
corresponding partial snapshot in a single register operation.  At the
root this means that the cost of obtaining a complete snapshot is
dominated by the cost of the max-register read, at $O(\log v)$, where
$v$ is the number of updates ever performed.

\begin{figure}
\centering
\includegraphics[scale=0.5]{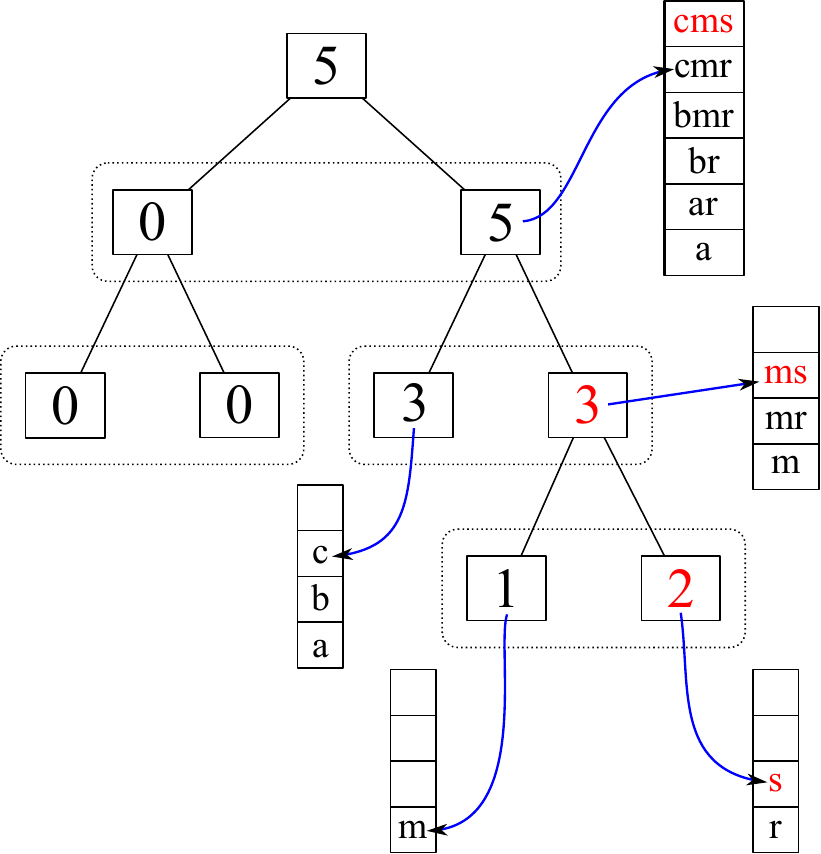}
    \caption{Snapshot from max arrays; taken from~\protect{\cite[Fig.~2]{AspnesACHE2015}}}
\label{figure-snapshot-tree}
\end{figure}

A picture of this structure, adapted from the proceedings version of~\cite{AspnesACHE2015},
appears in Figure~\ref{figure-snapshot-tree}.  The figure depicts an
update in progress, with red values being the new values written
as part of the update.  Only some of the tables associated with the
nodes are shown.

The cost of an update is dominated by the $O(\log n)$ max-array
operations needed to propagate the new value to the root.  This takes
$O(\log^2 v \log n)$ steps. Here $v$ can be taken to be the number of
update operations, which controls the maximum value on either side of
the 2-component max arrays.

The linearizability proof is trivial: linearize each update by the
time at which a snapshot containing its value is written to the root
(which necessarily occurs within the interval of the update, since we
don't let an update finish until it has propagated its value to the
top), and linearize reads by when they read the root.  This
immediately gives us an $O(\log^3 n)$ implementation—as long as we
only want to use it polynomially many times—of anything we
can build from snapshot, including counters, generalized counters, and
(by~\cite{AspnesH1990waitfree,AndersonM1993}) any other object whose
operations all commute with or overwrite each other in a
static pattern.

\subsection{Randomized and amortized snapshots}

Aspnes and Censor-Hillel~\cite{AspnesC2013} claimed to give an
unrestricted, randomized snapshot with $O(\log^3 n)$
This claimed result is somewhat suspect because (a) it is based on the original,
uncorrected version of the max array from~\cite{AspnesACHE2015}, (b)
the paper incorrectly computes the running time of the algorithm, and
(c) the claim is supported by a rather rococo proof of linearizability
that is dubious in various additional
ways. So it is not clear that this algorithm actually works.

Fortunately, this result is largely dominated by a much less
questionable result by Ahad Baig~\etal~\cite{AhadBaigHMT2020} that gives a
deterministic snapshot implementation with $O(\log^3 n)$ amortized
individual step complexity.

As in the restricted-use case, the Ahad Baig~\etal~snapshot assumes
arbitrarily-wide registers. An alternative suggested by Bashari and
Woelfel~\cite{BashariW2021} is to implement an \concept{adaptive partial
snapshot}\index{snapshot!partial}\index{snapshot!adaptive partial} where a scan effectively returns a
sensibly-sized index from which individual values can be extracted 
using a separate $\FuncSty{observe}$ operation. Bashari and Woelfel
show that such snapshots can be implemented in $O(\log n)$ steps using
fetch-and-add and compare-and-swap primitives. Whether it is possible
to improve on the $O(\log^3 n)$ bound of Ahad Baig~\etal without using
stronger primitives is still open.

Neither of these algorithms contradict the JTT lower bound: in
the worst case, each will have operations that take $Ω(n)$ steps. But
the hope is that these operations are rare, and in the amortized case,
paid for by many cheap operations.
Also, even though we may beat JTT most of the time, 
other lower bounds may still apply; see for
example~\cite{AspnesCAH2016,HendlerK2014}.

\myChapter{Common2}{2019}{}
\label{chapter-common2}

The \concept{common2} class, defined by Afek, Weisberger, and
Weisman~\cite{AfekWW1993} consists of all read-modify-write objects
where the modify functions either (a) all commute with each other or
(b) all overwrite each other.  We can think of it as the union of two
simpler classes, the set of read-modify-write objects 
where all update operations commute,
called 
\index{object!commuting}
\indexConcept{commuting object}{commuting objects}~\cite{AfekW1999};
and the set of read-modify-write objects where all updates
produce a value that doesn't depend on the previous state,
called \index{object!historyless}
\indexConcept{historyless object}{historyless
objects}~\cite{FichHS1998}).

From §\ref{section-wait-free-level-2}, we know that both
commuting objects and historyless objects have consensus number at
most 2, and that these objects have consensus number exactly 2
provided they supply at least one non-trivial update operation.
The main result of Afek~\etal~\cite{AfekWW1993} is that commuting and
historyless
objects can all be implemented from any object with consensus number
2, even in systems with more than 2 processes.  This gives a
\concept{completeness} result analogous to completeness results in
complexity theory: any non-trivial common2 object can be used to
implement any other common2 object.

The \concept{common2 conjecture} was that common2 objects could also
implement any object with consensus number $2$,  This is now known to
be false~\cite{AfekEG2016}.

The main result in the paper has two parts, reflecting the two parts 
of the common2 class: a proof that 2-process consensus plus registers
is enough to implement all commuting objects (which essentially comes
down to building a generalized fetch-and-add that 
returns an unordered list of all preceding operations); and a
proof that 2-process consensus plus registers is enough to implement
all overwriting objects (which is equivalent to showing that we can
implement swap objects).  The construction of the generalized
fetch-and-add is pretty nasty, so we'll concentrate on the
implementation of swap objects.  We will also skip the swap
implementation in~\cite{AfekWW1993}, and instead describe, in
§§\ref{section-obstruction-free-swap}
and~\ref{section-wait-free-swap}, a simpler
(though possibly less efficient) algorithm from a later paper by Afek,
Morrison, and Wertheim~\cite{AfekMW2011}.  Before we do this, we'll
start with some easier results from the older paper, including an
implementation of $n$-process test-and-set from $2$-process consensus.
This will show that anything we can do with test-and-set we can do
with any common2 object.

\section{Test-and-set and swap for two processes}
\label{section-TAS-2-from-consensus-2}

\newcommand{\TwoTAS}{\FuncSty{TAS2}}
\newcommand{\TwoConsensus}{\FuncSty{Consensus2}}

The first step is to get test-and-set.

Algorithm~\ref{alg-2TAS-from-2cons} shows how to turn $2$-process
consensus into $2$-process test-and-set.  The idea is that whoever
wins the consensus protocol wins the test-and-set.  This is
linearizable, because if I run $\TwoTAS$ before you do, I win the
consensus protocol by validity.

\begin{algorithm}
\Procedure{$\TwoTAS()$}{
    \eIf{$\TwoConsensus(\MyId) = \MyId$}{
        \Return $0$\;
    }{
        \Return $1$\;
    }
}
\caption{Building $2$-process TAS from $2$-process consensus}
\label{alg-2TAS-from-2cons}
\end{algorithm}

Once we have test-and-set for two processes, we can easily get
one-shot swap for two processes.  The trick is that a one-shot swap
object always returns $⊥$ to the first process to access it and
returns the other process's value to the second process.  We can
distinguish these two roles using test-and-set and add a register to
send the value across.  Pseudocode is in
Algorithm~\ref{alg-2swap-from-2TAS}.

\begin{algorithm}
\Procedure{$\Swap(v)$}{
    $a[\MyId] = v$ \;
    \eIf{$\TwoTAS() = 0$}{
        \Return $⊥$\;
    }{
        \Return $a[¬\MyId]$\;
    }
}
\caption{Two-process one-shot swap from TAS}
\label{alg-2swap-from-2TAS}
\end{algorithm}

\section{Building \texorpdfstring{$n$}{n}-process TAS from 2-process TAS}
\label{section-test-and-set-from-two-process-consensus}

\index{test-and-set}
To turn the $\TwoTAS$ into full-blown $n$-process $\TAS$, start by
staging a
tournament along the lines of~\cite{PetersonF1977}
(§\ref{section-mutex-tournament}).  Each process walks up a tree of
nodes, and at each node it attempts to beat every process from the
other subtree using a $\TAS_2$ object (we can't just have it fight one
process, because we don't know which one process will have won the
other subtree, and our $\TAS_2$ objects may only work for two
specific processes).  A process drops out if it ever sees a $1$.  We can
easily show that at most one process leaves each subtree with all
zeros, including the whole tree itself.

\newData{\AWWgate}{gate}
\newFunc{\AWWcompete}{compete}
\newData{\AWWnode}{node}
\newFunc{\AWWblock}{block}
\newFunc{\AWWpass}{pass}
\newFunc{\AWWpassAll}{passAll}
\newFunc{\AWWfindValue}{findValue}

Unfortunately, this process does not give a \emph{linearizable}
test-and-set object.  It is possible that $p_1$ loses early to $p_2$,
but then $p_3$ starts (elsewhere in the tree) after $p_1$ finishes,
and races to the top, beating out $p_2$.  To avoid this, we can
follow~\cite{AfekWW1993} and add a \AWWgate bit that locks out
latecomers.\footnote{The original version of this trick is from an
earlier paper~\cite{AfekGTV1992}, where the \AWWgate bit is
implemented as an array of single-writer registers.}  

The resulting
construction looks something like Algorithm~\ref{alg-n-TAS-from-2TAS}.
This gives a slightly different interface from straight \TAS; instead
of returning $0$ for winning and $1$ for losing, the algorithm returns
$⊥$ if you win and the \Id of some process that beats you  
if you lose.\footnote{Note that this process may also be a loser, just
    one that made it further up the tree than you did.  We can't
    expect to learn the ID of the ultimate winner, because that would
solve $n$-process consensus.} It's not hard to see that this gives a linearizable
test-and-set after translating the values back to $0$ and $1$ (the
trick for linearizability is that any process that wins saw an empty
gate, and so started before any other process finished).  It
also sorts the processes into a rooted tree, with each
process linearizing after its parent (this latter claim is a
little trickier, but basically comes down to a loser linearizing after
the process that defeated it either on \AWWgate or on one of the
$\TwoTAS$ objects).

\begin{algorithm}
\Procedure{$\AWWcompete(i)$}{
    \tcp{check the gate}
    \If{$\AWWgate \ne ⊥$}{
        \Return \AWWgate\;
    }
    $\AWWgate ← i$ \;
    \tcp{Do tournament, returning $\Id$ of whoever I lose to}
    $\AWWnode ← $ leaf for $i$ \;
    \While{$\AWWnode \ne \DataSty{root}$}{
        \Foreach{$j$ whose leaf is below sibling of $\AWWnode$}{
            \If{$\TwoTAS(t[i,j]) = 1$}{
                \Return $j$\;
            }
        }
        $\AWWnode ← \AWWnode.\DataSty{parent}$\;
    }
    \tcp{I win!}
    \Return $⊥$\;
}
\caption{Tournament algorithm with gate}
\label{alg-n-TAS-from-2TAS}
\end{algorithm}

This algorithm is kind of expensive: the losers that drop out early
are relatively lucky, but the winning process has to win a $\TwoTAS$
against everybody, for a total of $Θ(n)$ $\TAS$ operations.
We can reduce the cost to $O(\log n)$ if our $\TwoTAS$ objects allow
arbitrary processes to execute them.  This is done, for example, in
the RatRace test-and-set implementation of
Alistarh~\etal~\cite{AlistarhAGGG2010}, using a randomized
implementation of $\TwoTAS$ due to Tromp and Vitányi~\cite{TrompV2002}
(see §\ref{section-ratrace}).

\section{Obstruction-free swap from test-and-set}
\label{section-obstruction-free-swap}

We'll start by describing the ``strawman algorithm'' from the AMW
paper.  This is presented by the authors as a stepping-stone to their
real algorithm, which we will describe below in
§\ref{section-wait-free-swap}.  

The code is given in Algorithm~\ref{alg-obstruction-free-swap}.  This
implements a swap object that is linearizable but not wait-free.

\begin{algorithm}
    \Procedure{$\Swap(v)$}{
        $i ← 0$\;
        \While{\True}{
            \tcp{Look for a starting point}
            \While{$\TAS(s_i) = 1$}{
                \label{line-obstruction-free-swap-fetch-and-increment}
                $i ← i+1$\;
            }
            $v_i ← v$\;
            \tcp{Check if we've been blocked}
            \eIf{$\TAS(t_i) = 0$}{
                \label{line-obstruction-free-swap-grab-t}
                \tcp{We win, find our predecessor}
                \For{$j ← i-1$ \DownTo $0$}{
                    \If{$\TAS(t_j) = 1$}{
                        \tcp{Use this value}
                        \Return $v_j$\;
                    }
                }
                \tcp{Didn't find anybody, we are first}
                \Return $⊥$\;
            }{
                \tcp{Pick a new start and try again}
            }
        }
        
    }

    \caption{Obstruction-free swap from test-and-set}
    \label{alg-obstruction-free-swap}
\end{algorithm}

This algorithm uses two infinite arrays $s$ and $t$ of test-and-set
objects and an infinite array $r$ of atomic registers.  The $s_i$
objects are essentially being used to implement a fetch-and-increment,
and if we have a fetch-and-increment lying around we can
replace the loop at
Line~\ref{line-obstruction-free-swap-fetch-and-increment} with an
operation on that object instead.  The $r_i$ registers record values
to return.  The $t_i$ registers implement a block/pass mechanism where
a later process can force an earlier process to try again if it didn't
record its value in time.  This solves the problem of a process going
to sleep after acquiring a particular slot $i$ from the
fetch-and-increment but before writing down a value that somebody else
can use.

The algorithm is obstruction-free, because in any reachable
configuration, only finitely many test-and-sets have been accessed, so
there is some value $i$ with $s_j = t_j = 0$ for all $j ≥ i$.  A process running in
isolation will eventually hit one of these slots, win both
test-and-sets, and return.

For linearizability, the value of $i$ when each operation returns
gives an obvious linearization ordering.  This ordering is consistent
with the observed history, because if I finish with value $i_1$ before
you start, then at the time that I finish all $s_j$ for $j ≤ i_1$ have
$s_j = 1$.  So you can't win any of them, and get a slot $i_2 > i_1$.
But we still have to show that the return values make sense.

Consider some swap operation $π$.

Suppose that $π$ starts at position
$i$ and wins every $t_j$ down to position $k$, where it loses.  Then
no other operation wins any $t_j$ with $k < j < i$, so there is no
process that leaves with any slot between $k$ and $i$.  In addition,
the operation $π'$that did win $t_k$ must have taken slot $k$ in
Line~\ref{line-obstruction-free-swap-grab-t}, because any
other process would have needed to win $t_{k+1}$ before attempting to
win $t_k$.  So $π'$ linearizes immediately before $π$, which is good,
because $π$ returns the value $v_k$ that $π'$ wrote before it won
$t_k$.

Alternatively, suppose that $π$ never loses $t_j$ for any $j ≤ i$.
Then no other operation takes a slot less than $i$, and $π$ linearizes
first.  In this case, it must return $⊥$, which it does.

\section{Wait-free swap from test-and-set}
\label{section-wait-free-swap}

Now we want to make the strawman algorithm wait-free.  The basic idea
is similar: we will have an ordered collection of test-and-set
objects, and a process will move right until it can capture one that
determines its place in the linearization ordering, and then it will
move left to block any other processes from taking an earlier place
unless they have already written out their values.  To avoid
starvation, we assign a disjoint collection of test-and-set objects to
each operation, so that every operation eventually wins one of its own
test-and-sets.  Unfortunately this only works if we make the ordering
dense, so that between each pair of test-and-sets there are infinitely
many other test-and-sets.

\newData{\AMWmaxDepth}{maxDepth}
\newData{\AMWaccessed}{accessed}
\newData{\AMWdepth}{depth}
\newData{\AMWwin}{win}
\newData{\AMWcap}{cap}
\newData{\AMWreg}{reg}
\newData{\AMWtst}{tst}
\newData{\AMWmaxPreviousDepth}{maxPreviousDepth}
\newData{\AMWret}{ret}

AMW do this in terms of a binary tree, but I find it easier to think
of the test-and-sets as being indexed by dyadic
rationals strictly between $0$ and $1$.\footnote{The two representations are
    isomorphic: make each value $k/2^q$ be the parent of $k/2^q ±
1/2^{q+1}$.} The idea
is that the $i$-th operation to start executing the swap object will
use test-and-sets $t_q$ where $q = k/2^i$ for all odd $k$ in the range
$1\dots 2^i-1$.  In order to avoid having to check the infinitely many
possible values smaller than $q$, we will use two auxiliary objects: a
readable fetch-and-increment $\AMWmaxDepth$ that hands out denominators
and tracks the largest denominator used so far, and a max register
$\AMWaccessed$ that keeps track of the largest position accessed so
far.

AMW implement $\AMWaccessed$ using a snapshot, which we will do as
well to avoid complications from trying to build a max register out of
an infinitely deep tree.\footnote{The issue is not so much that we
    can't store arbitrary dyadics, since we can encode them using an
    order-preserving prefix-free code, but that, without some sort of
    helping mechanism, a read running concurrently with endlessly
    increasing writes (e.g. $1/2, 3/4, 7/8, \dots$) might not be
wait-free.  Plus as soon as the denominator exceeds
$2^n$, which happens after only $n$ calls to $\Swap$, $O(n)$-step snapshots are cheaper anyway.}
Note that AMW don't call this data structure a max register, but we
will, because we like max registers.

Code for the swap procedure is given in
Algorithm~\ref{alg-wait-free-swap}.

\begin{algorithm}
    \Procedure{$\Swap(v)$}{
        \tcp{Pick a new row just for me}
        $\AMWdepth ← \FuncSty{fetchAndIncrement}(\AMWmaxDepth)$\;
        \tcp{Capture phase}
        \Repeat{\AMWwin}{
            \tcp{Pick leftmost node in my row greater than \AMWaccessed}
            $\AMWcap ← \min \SetWhere{x}{\text{$x = k/2^\AMWdepth$ for odd $k$}, x > \AMWaccessed}$\;
            \tcp{Post my value}
            $\AMWreg[\AMWcap] ← v$\;
            \tcp{Try to capture the test-and-set}
            $\AMWwin ← \TAS(\AMWtst[\AMWcap]) = 0$\;
            $\FuncSty{writeMax}(\AMWaccessed, \AMWcap)$\;
        }
        \tcp{Return phase}
        \tcp{Max depth reached by anybody left of $\AMWcap$}
        $\AMWmaxPreviousDepth ← \Read(\AMWmaxDepth)$\;
        $\AMWret ← \AMWcap$\;
        \tcp{Block previous nodes until we find one we can take}
        \Repeat{$\TAS(\AMWtst[\AMWret]) = 1$}{
            $\AMWret ← \max \SetWhere{x = k/2^q}{q ≤
            \AMWmaxPreviousDepth, \text{$k$ odd}, x < \AMWret}$\;
            \If{$\AMWret < 0$}{
                \Return $⊥$\;
            }
        }
        \Return $\AMWreg[\AMWret]$\;
    }
    \caption{Wait-free swap from test-and-set~\cite{AfekMW2011}}
    \label{alg-wait-free-swap}
\end{algorithm}

To show Algorithm~\ref{alg-wait-free-swap} works, we need the
following technical lemma, which, among other things, implies that
node $1-2^{\AMWdepth}$ is
always available to be captured by the process at depth $\AMWdepth$.
This is essentially just a restatement of Lemma 1
from~\cite{AfekMW2011}.

\begin{lemma}
    \label{lemma-alg-wait-free-swap-subtree}
    For any $x = k/2^q$, where $k$ is odd, no process attempts to capture
    any $y ∈ [x,x+1/2^q)$ before some process writes $x$ to $\AMWaccessed$.
\end{lemma}
\begin{proof}
    Suppose that the lemma fails, let $y = \ell/2^r$ be the first node captured
    in violation of the lemma, and let $x = k/2^q$ be such that $y ∈
    [x,x+1/2^q)$ but $x$ has not been written to $\AMWaccessed$ when
    $y$ is captured.  Let $p$ be the process that captures $y$.

    Now consider $y' = x-1/2^r$, the last node to the left of
    $x$ at the same depth as $y$.  Why didn't $p$ capture $y'$?  

    One possibility is that some other process $p'$ blocked $y'$ during
    its return phase.  This $p'$ must have captured a node $z > y'$.
    If $z > y$, then $p'$ would have blocked $y$ first, preventing $p$
    from capturing it.  So $y' < z < y$.  

    The other possibility is that $p$ never tried to capture $y'$,
    because some other process $p'$ wrote some value $z > y'$ to
    $\AMWaccessed$ first.  This value $z$ must also be less than $y$
    (or else $p$ would not have tried to capture $y$).
    
    In both
    cases, there is a process $p'$ that captures a value $z$ with $y' <
    z < y$, before $p$ captures $y$ and thus before anybody writes $x$
    to $\AMWaccessed$.

    Since $y' < x$ and $y' < z$, either $y' < z < x$ or $y' < x < z$.  In the
    first case, $z ∈ [y',y'+1/2^r)$ is captured before $y'$ is written
        to $\AMWaccessed$.
        In the second case $z ∈ [x,x+1/2^q)$ is captured before $x$ is
            written to $\AMWaccessed$.  Either way, $y$ is not the
            first capture to violate the lemma, contradicting our
            initial assumption.
\end{proof}

Using Lemma~\ref{lemma-alg-wait-free-swap-subtree}, it is
straightforward to show that Algorithm~\ref{alg-wait-free-swap} is
wait-free.  If I get $q$ for my value of $\AMWdepth$, then no process
will attempt to capture any $y$ in $[1-2^q,1)$ before I write $1-2^q$
to $\AMWaccessed$.  But this means that nobody can block me from
capturing $1-2^q$, because processes can only block values smaller than
the one they already captured.  I also can't get stuck in the return
phase, because there are only finitely many values with denominator
less than $2^{\AMWmaxPreviousDepth}$.

It remains to show that the implementation is linearizable.  The obvious
linearization ordering is given by sorting each operation $i$ by its
captured node $\AMWcap$.  Linearizability requires then that if we
imagine a directed graph containing an edge $ij$ for each pair of
operations $i$ and $j$ such
that $i$ captures $\AMWcap_i$ and returns $\AMWreg[\AMWcap_j]$, then this graph
forms a path that corresponds to this linearization ordering.

Since each process only returns one value, it trivially holds
that each node in the graph has out-degree at most $1$.  
For the in-degree, suppose that we have operations $i$, $j$, and $k$
with $\AMWcap_i < \AMWcap_j < \AMWcap_k$ such that $j$ and $k$ both
return $\AMWreg[\AMWcap_i]$.  Before $k$ reaches $\AMWtst[\AMWcap_i]$,
it must first capture all the test-and-sets between $\AMWcap_i$ and
$\AMWcap_k$ that have depth less than or equal to
$\AMWmaxPreviousDepth_k$.  This will include $\AMWtst[\AMWcap_j]$,
because $j$ must write to $\AMWmaxDepth$ before doing anything, and
this must occur before $k$ starts the return phase if $j$ sees a value
of $\AMWaccessed$ that is less that $\AMWcap_k$.

A similar argument show that there is at most one process that returns
$⊥$; this implies that there is at most one process with out-degree
$0$.

So now we have a directed graph where every process has in-degree and
out-degree at most one, which implies that each weakly-connected
component will be a path.  But each component will also have exactly
one terminal node with out-degree $0$.  Since there is only one such
node, there is only one component, and the entire graph is a single
path.  This concludes the proof of linearizability.

\section{Implementations using stronger base objects}
\label{section-implementations-using-stronger-base-objects}

The terrible step complexity of known wait-free implementations of
Common2 objects like $\Swap$ or $\FetchAndIncrement$ from $2$-process
consensus objects and registers has led to work on finding better
implementations assuming stronger base objects.  Using
load-linked/store-conditional, Ellen and Woelfel~\cite{EllenW2013}
provide implementations of several
Common2 objects, including $\FetchAndIncrement$, $\FetchAndAdd$, and
$\Swap$ that all have $O(\log n)$ individual step
complexity.\footnote{What they actually implement is the ability to do
fetch-and-$f$, where $f$ is any binary associative function,
using an object they call an \concept{aggregator}.  
Each of these objects is obtained by choosing an appropriate $f$.}
This is
know to be optimal due to a lower bound of
Jayanti~\cite{Jayanti1998}.

The lower bound applies \emph{a fortiori} to the case where we don't
have LL/SC or CAS and have to rely on 2-process consensus objects.
But it's not out of the question that there is a matching upper bound
in this case.

\myChapter{Randomized consensus and test-and-set}{2026}{}
\label{chapter-randomized-consensus}

We've seen that we can't solve \concept{consensus} in an asynchronous
system message-passing or shared-memory system with one crash failure~\cite{FischerLP1985,LouiA1987}, but that
the problem becomes solvable using failure
detectors~\cite{ChandraT1996}.  An alternative that also allows us to
solve consensus is to allow the processes to use randomization, by
providing each process with a \concept{local coin} that can generate
random values that are immediately visible only to that process.
The resulting 
\index{consensus!randomized}
\concept{randomized consensus} problem replaces
the \concept{termination} requirement with
\concept{probabilistic termination}: all processes terminate with
probability $1$.  The agreement and validity requirements remain the
same.

In this chapter, we will describe how randomization interacts with the
adversary, give a bit of history of randomized consensus, and then
concentrate on recent algorithms for randomized consensus and the
closely-related problem of randomized test-and-set.
Much of the material in this chapter is adapted from notes
for a previous course on randomized algorithms~\cite{Aspnes2011randomizedNotes} and 
a few of my own papers~\cite{Aspnes2012modular,AspnesE2011,Aspnes2012}.

\section{Role of the adversary in randomized algorithms}

Because randomized processes are unpredictable, we need to become a
little more sophisticated in our handling of the adversary.  As in
previous asynchronous protocols, we assume that the adversary has
control over timing, which we model by allowing the adversary to
choose at each step which process performs the next operation.  But
now the adversary may do so based on knowledge of the state of the
protocol and its past evolution.  How much knowledge we give the
adversary affects its power.  Several classes of adversaries have been
considered in the literature; ranging from strongest to weakest, we
have:
\begin{enumerate}
\item An \index{adversary!adaptive}\concept{adaptive adversary}.  This
adversary is a function from the state of the system to the set of
processes; it can see everything that has happened so far (including
coin-flips internal to processes that have not yet been revealed to
anybody else), but can't predict the future.  It's known that an
adaptive adversary can force any randomized consensus protocol to take
$Θ(n^2)$ total steps~\cite{AttiyaC2008jacm}.  The adaptive
adversary is also called a
\index{adversary!strong}
\concept{strong adversary} following a foundational paper of
Abrahamson~\cite{Abrahamson1988}.
\item An 
\index{adversary!intermediate}
\concept{intermediate adversary}
or
\index{adversary!weak}
\concept{weak adversary}~\cite{Abrahamson1988}
is one that limits the adversary's ability to observe or control the system in
some way, without completely eliminating it.
For example, a
\index{adversary!content-oblivious}
\concept{content-oblivious adversary}~\cite{Chandra1996}
or
\index{adversary!value-oblivious}
\concept{value-oblivious adversary}~\cite{Aumann1997}
is restricted from seeing the values contained in registers or pending
write operations and from observing the internal states of processes
directly.  A
\index{adversary!location-oblivious}
\concept{location-oblivious adversary}~\cite{Aspnes2012modular}
can distinguish between values and the types of pending operations,
but can't discriminate between pending operations based one which
register they are operating on.
These classes of adversaries are modeled by imposing an equivalence
relation on partial executions and insisting that the adversary make
the same choice of processes to go next in equivalent situations.
Typically they arise because somebody invented a consensus protocol
        for the oblivious adversary (described below) and then looked for the next most
powerful adversary that still let the protocol work.

Weak adversaries often allow much faster consensus protocols than
adaptive adversaries.  Each of the above adversaries permits consensus
to be achieved in $O(\log n)$ expected individual work using an
appropriate algorithm.  But from a mathematical standpoint, weak
adversaries are a bit messy, and once you start combining algorithms
designed for different weak adversaries, it's natural to move all the
way down to the weakest reasonable adversary, the oblivious adversary.

\item A \index{adversary!oblivious}\concept{oblivious adversary}
has no ability to observe the system at all; instead, it fixes a
sequence of process IDs in advance, and at each step the next process
in the sequence runs.

We will
describe below a protocol that guarantees 
$O(\log \log n)$ expected individual work for an oblivious
adversary.
It is not
known whether this is optimal; in fact, is is consistent with the best
known lower bound (due to Attiya and Censor~\cite{AttiyaC2008jacm})
that consensus can be solved in $O(1)$ expected individual steps
against an oblivious adversary.
\end{enumerate}

Each of these adversaries is defined based on choosing steps of
particular objects, with particular constraints on knowledge based on
the states of those objects. This interacts badly with
abstractions like linearizability: an adversary might be able to play
games with the internals of an implementation of an object that allows
it more power than it would have with an actual sequential version of the
object. So even though linearizable implementations are
indistinguishable from sequential objects for deterministic protocols,
for randomized protocols they can give very different results for both
adaptive and oblivious adversaries~\cite{GolabHW2011}; and in the
specific case of consensus, it can be shown that there are randomized
consensus protocols that terminate with probability $1$ against an
adaptive adversary when implemented with atomic registers, but fail
to terminate with nonzero probability when implemented using an
arbitrary linearizable implementation~\cite{HadzilacosHT2020}.

These results don't necessarily imply the failure of any specific consensus
protocol implemented using a specific atomic register simulation, 
but they do justify suspicion.  The easiest way to
deal with this suspicion is to assume that our atomic registers are,
in fact, atomic, so that's what we will do here.

\section{History}

The use of randomization to solve consensus in an asynchronous system
with crash failures
was proposed by
Ben-Or~\cite{Ben-Or1983} for a message-passing model.  Chor, Israeli,
and Li~\cite{ChorIL1994} gave the first wait-free consensus protocol for a
shared-memory system, which assumed a particular kind of 
weak adversary.  Abrahamson~\cite{Abrahamson1988} defined strong and weak
adversaries and gave the first wait-free consensus protocol for a
strong adversary; its expected step complexity was
$Θ\left(2^{n^2}\right)$.  After failing to show that exponential
time was necessary, Aspnes and Herlihy~\cite{AspnesH1990consensus}
showed how to do consensus in $O(n^4)$ total step complexity, a value that was
soon reduced to $O(n^2 \log n)$ by Bracha and
Rachman~\cite{BrachaR1991}.  This remained the best known bound for
the strong-adversary model until Attiya and
Censor~\cite{AttiyaC2008jacm} showed matching $Θ(n^2)$ upper and
lower bounds on total step complexity.  A later paper by Aspnes and Censor~\cite{AspnesC2009}
showed that it was also possible to get an $O(n)$ bound on individual
step complexity.

For weak adversaries, the best known upper bound on individual step
complexity was $O(\log
n)$ for a long time~\cite{Chandra1996,Aumann1997,Aspnes2012modular},
with an $O(n)$ bound on total step complexity for some
models~\cite{Aspnes2012modular}.  
More recent work has lowered the individual step complexity 
bound to $O(\log \log n)$, under the
assumption of an oblivious adversary~\cite{Aspnes2012}.
No non-trivial lower bound on expected
individual step complexity is known, although there is a known lower
bound on the distribution of the individual step
complexity~\cite{AttiyaC2010}.

In the following sections, we will concentrate on the more recent
weak-adversary algorithms.  These have the advantage of being fast
enough that one might reasonably consider using them in practice,
assuming that the weak-adversary assumption does not create trouble,
and they are also require less probabilistic machinery to analyze than
the strong-adversary algorithms.

\section{Reduction to simpler primitives}

To show how to solve consensus using randomization, it helps to split
the problem in two: we will first see how to detect \emph{when} we've
achieved agreement, and then look at \emph{how} to achieve agreement.

\subsection{Adopt-commit objects}
\label{section-adopt-commit-objects}

Most known randomized consensus protocols have a round-based structure
that alternates between generating and detecting agreement.
Gafni~\cite{Gafni1998} proposed
\indexConcept{adopt-commit protocol}{adopt-commit protocols} as a tool
for detecting agreement, and these protocols were later abstracted as
\indexConcept{adopt-commit object}{adopt-commit
objects}~\cite{MostefaouiRRT2008,AlistarhGGT2009}.
The version described here is largely taken from~\cite{AspnesE2011}, which
shows bounds on the complexity of adopt-commit objects.

An adopt-commit object 
supports a single operation, $\AdoptCommit(u)$,
where $u$ is an input from a set of $m$ \concept{values}.
The result of this operation is an output of the form $(\ACcommit,v)$
or $(\ACadopt,v)$,
where the second component is a value from this set
and the first component
is a \concept{decision bit} that indicates whether the
process should decide value $v$ immediately or adopt it as its
preferred value in later rounds of the protocol.

The requirements for an adopt-commit object are the usual requirements
of validity and termination, plus:
\begin{enumerate}
\item \indexConcept{coherence}{Coherence.}  If the output of some
operation is $(\ACcommit,v)$,
then every output is either $(\ACadopt,v)$ or $(\ACcommit,v)$.
\item \indexConcept{convergence}{Convergence.}  If all inputs are $v$, all 
outputs are $(\ACcommit,v)$.
\end{enumerate}
These last two requirement replace the agreement property of
consensus.  They are also strictly weaker than consensus, which means
that a consensus object (with all its output labeled \ACcommit) is
also an adopt-commit object.

The reason we like adopt-commit objects is that they allow the simple
consensus protocol shown in
Algorithm~\ref{alg-consensus-from-adopt-commit}.
\begin{algorithm}
\newcommand{\ACpref}{\DataSty{preference}\xspace}
$\ACpref ← \DataSty{input}$ \\
\For{$r ← 1 \dots \infty$}{
    $(b, \ACpref) ← \AdoptCommit(AC[r], \ACpref)$\;
    \eIf{$b = \ACcommit$}{
        \Return $\ACpref$\;
    }{
        do something to generate a new \ACpref\;
    }
}
\caption{Consensus using adopt-commit}
\label{alg-consensus-from-adopt-commit}
\end{algorithm}

The idea is that the adopt-commit takes care of ensuring that once
somebody returns a value (after receiving \ACcommit), everybody else
who doesn't return adopts the same value (follows from coherence).
Conversely, if everybody already has the same value, everybody returns
it (follows from convergence).  The only missing piece is the part
where we try to shake all the processes into agreement.  For this we
need a separate object called a \emph{conciliator}.

\subsection{Conciliators}
\label{section-conciliators}

A \concept{conciliator}~\cite{Aspnes2012modular} is a
randomized weakening of consensus that replaces agreement
with \index{agreement!probabilistic}\concept{probabilistic agreement}:
the processes can disagree sometimes, but must agree with
constant probability no matter what the adversary does.
Formally, a conciliator is any protocol that satisfies
termination, validity, and probabilistic agreement.

The important feature of conciliators is that if we plug a conciliator
that guarantees agreement with probability at least $δ$ into
Algorithm~\ref{alg-consensus-from-adopt-commit}, then on average we
only have to execute the loop $1/δ$ times before every process
agrees.  This gives an expected cost equal to $1/δ$ times the
total cost of \AdoptCommit and the conciliator.  Typically we will aim
for constant $δ$.

\section{Implementing an adopt-commit object}

What's nice about adopt-commit objects is that they can be implemented
deterministically.  Here we'll give a simple adopt-commit object for
two values, $0$ and $1$.  Optimal (under certain assumptions)
constructions of $m$-valued adopt-commits can be found
in~\cite{AspnesE2011}.

Pseudocode is given in Algorithm~\ref{alg-adopt-commit}.

\newData{\ACproposal}{proposal}

\begin{algorithm}
\SharedData{$a[0]$, $a[1]$, initially $0$; \ACproposal, initially $⊥$}
\Procedure{$\AdoptCommit(v)$}{
    $a[v] ← 1$\;
    \eIf{$\ACproposal = ⊥$}{
        $\ACproposal ← v$\;
    }{
        $v ← \ACproposal$\;
    }
    \eIf{$a[¬v] = 0$}{
        \Return $(\ACcommit, v)$\;
    }{
        \Return $(\ACadopt, v)$\;
    }
}
\caption{A 2-valued adopt-commit object}
\label{alg-adopt-commit}
\end{algorithm}

Structurally, this is pretty similar to a splitter (see
§\ref{section-mutex-fast}), except that we use values instead of
process IDs.  

We now show correctness.
Termination and validity are trivial.  For coherence,
observe that if I return $(\ACcommit, v)$ I must have read 
$a[¬v] = 0$ before any process with $¬v$ writes $a[¬
v]$; it follows that all such processes will see $\ACproposal \ne
⊥$ and return $(\ACadopt, v)$.  For convergence, observe that if all
processes have the same input $v$, they all write it to $\ACproposal$
and all observe $a[¬v] = 0$, causing them all to return
$(\ACcommit, v)$.

\section{Conciliators and shared coins}
\label{section-conciliator-shared-coin}

For an adaptive adversary, the usual way to implement a conciliator is
from a \index{coin!weak shared}\index{shared coin!weak}\concept{weak
shared coin}~\cite{AspnesH1990consensus}, which is basically a
non-cryptographic version of the \index{coin!common}\concept{common
coin}~\cite{Rabin1983} found in many cryptographic Byzantine agreement protocols.  
Formally, a weak shared coin is
an object that has no inputs and returns either $0$ or $1$ to all processes with some
minimum probability $δ$.  By itself this does not give validity, so
converting a weak shared coin into a conciliator requires extra
machinery to bypass the coin if the processes that have accessed the
conciliator so far are all in agreement; see
Algorithm~\ref{alg-conciliator-shared-coin}.  The intuition is that having
some processes (who all agree with each other) skip the shared coin is not a problem, because with
probability $δ$ the remaining processes will agree with them as well.

\newFunc{\CoinConciliator}{coinCoinciliator}
\newFunc{\SharedCoin}{sharedCoin}

\begin{algorithm}
\SharedData{\\
\quad binary registers $r_0$ and $r_1$, initially 0;\\
\quad weak shared coin $\SharedCoin$}
\Procedure{$\CoinConciliator()$}{
$r_v ← 1$\;
\eIf{$r_{¬v} = 1$}{
    \Return $\SharedCoin()$ \;
}{
    \Return $v$ \;
}
}
\caption{Shared coin conciliator from~\cite{Aspnes2012modular}}
    \label{alg-conciliator-shared-coin}
\end{algorithm}

This still leaves the problem of how to build a shared coin.  In the
message-passing literature, the usual approach is to use
cryptography,\footnote{For example, Canetti and
Rabin~\cite{CanettiR1993} solved Byzantine agreement in $O(1)$ time
by
building a shared coin on top of secret sharing.}
but because we are assuming an arbitrarily powerful adversary, we
can't use cryptography.

If we
don't care how small $δ$ gets, we could just have each process flip
its own local coin and hope that they all come up the same.  
(This is more or less what was done by
Abrahamson~\cite{Abrahamson1988}.)
But that might take a while.  If we aren't willing to wait
exponentially long, a better approach is to combine many individual
local coins using some sort of voting.

A version of this approach, based on a random walk, was used by
Aspnes and Herlihy~\cite{AspnesH1990consensus} to get consensus in (bad)
polynomial expected time against an adaptive adversary.  A better
version was developed by Bracha and Rachman~\cite{BrachaR1991}.  In
their version, each process repeatedly generates a random $±1$ vote
and adds it to a common pool (which just means writing the sum and
count of all
its votes so far out to a single-writer register).  Every $Θ(n / \log
n)$ votes, the process does a collect (giving an overhead of $Θ(\log
n)$ operations per vote) and checks to see if the total number of
votes is greater than a $Θ(n^2)$ threshold.  If it is, the process
returns the sign of the total vote.

Bracha and Rachman showed that despite processes seeing different
combinations of votes (due to the collects running at possibly very
different speeds), the difference between what each process sees and
the actual sum of all votes ever generated is at most $O(n)$ with high
probability.  This means that if the total vote is more than $cn$ from
$0$ for some $c$, which occurs with constant probability, then every
processes is likely to return the same value.  This gives a weak
shared coin with constant bias, and thus also a consensus protocol,
that runs in $O(n^2 \log n)$ expected total steps.

This remained the best known protocol for many years, leaving an
annoying gap between the upper bound and the best known lower bound of
$Ω(n^2/\log^2 n)$~\cite{Aspnes1998coin}.  Eventually,
Attiya and Censor~\cite{AttiyaC2008jacm} produced an entirely new
argument to bring the lower bound up to $Ω(n^2)$ and at the same time
gave a simple tweak to the Bracha-Rachman protocol to bring the upper
bound down to $O(n^2)$, completely settling (up to constant factors)
the asymptotic expected total step complexity of strong-adversary consensus.
But the question of how quickly one could solve weak-adversary
adversary consensus remained (and still remains) open.

\section{A one-register conciliator for an oblivious adversary}
\label{section-conciliator-one-register}

\begin{algorithm}
\caption{Impatient first-mover conciliator
from~\cite{Aspnes2012modular}}
\label{algorithm-impatient-first-mover-conciliator}
\SharedData{register $r$, initially $⊥$}
$k \gets 0$\;
\While{$r = ⊥$}{
    \eWithProbability{$\frac{2^k}{2n}$}{
        write $v$ to $r$\;
    }{
        do a dummy operation\;
    }
    $k \gets k+1$\;
}
\Return $r$\;
\end{algorithm}

Algorithm~\ref{algorithm-impatient-first-mover-conciliator} implements
a conciliator for an oblivious adversary\footnote{Or any adversary
weak enough not to be able to
block the write based on how the coin-flip turned out.}
using a single register.
This particular construction is taken from~\cite{Aspnes2012modular},
and is based on an earlier algorithm of Chor, Israeli, and
Li~\cite{ChorIL1994}.  The cost of this algorithm is expected $O(n)$ total work
and $O(\log n)$ individual work.
Later (§\ref{section-sifter-consensus}), we will see a different
algorithm~\cite{Aspnes2012} that reduces the individual work to $O(\log \log n)$,
although the total work for that algorithm may be $O(n \log \log n)$.

The basic idea is that processes alternate between reading
a register $r$ and (maybe) writing to the register; if a process reads
a non-null value from the register, it returns it.  Any other process
that reads the same non-null value will agree with the first process;
the only way that this can't happen is if some process writes a
different value to the register before it notices the first write.

The random choice of whether to write the register or not avoids this
problem.  The idea is that even though the adversary can schedule a
write at a particular time, because it's oblivious, it won't be able
to tell if the process wrote (or was about to write) or did a no-op
instead.  

The basic version of this algorithm, due to Chor, Israeli,
and Li~\cite{ChorIL1994}, uses a fixed $\frac{1}{2n}$ probability of
writing to the register.  So once some process writes to the register,
the chance that any of the remaining $n-1$ processes write to it
before noticing that it's non-null is at most $\frac{n-1}{2n} < 1/2$.
It's also not hard to see that this algorithm uses $O(n)$ total
operations, although it may be that one single process running by
itself has to go through the loop $2n$ times before it finally writes
the register and escapes.

Using increasing probabilities avoids this problem, because any
process that executes the main loop $\ceil{\lg n} + 1$ times will write the
register.  This establishes the $O(\log n)$ per-process bound on
operations.  At the same time, an $O(n)$ bound on total operations
still holds, since each write has at least a $\frac{1}{2n}$ chance of
succeeding.
The price we pay for the improvement is that 
we increase the chance that an initial value written to the
register gets overwritten by some high-probability write.  But the
intuition is that the probabilities can't grow too much, because the
probability that I write on my next write is close to the sum of the
probabilities that I wrote on my previous writes—suggesting that if
I have a high probability of writing next time, I should have done a
write already.

Formalizing this intuition requires a little bit of work.
Fix the schedule, and let $p_i$ be the probability that the $i$-th
write operation in this schedule succeeds.
Let $t$ be the least value for which $\sum_{i=1}^{t} p_i ≥ 1/4$.
We're going to argue that with constant probability 
one of the first $t$ writes succeeds, and that the
next $n-1$ writes by different processes all fail.

The probability that none of the first $t$ writes succeed is
\begin{align*}
\prod_{i=1}^t (1-p_i)
&≤ \prod_{i=1}^t e^{-p_i}
\\&= \exp\left(\sum_{i=1}^t p_i\right)
\\&≤ e^{-1/4}.
\end{align*}

Now observe that if some process $p$ writes at or before the $t$-th write,
then any process $q$ with a pending write either did no writes previously,
or its last write was among the first $t-1$ writes, whose
probabilities sum to less than $1/4$.  
In either case,
$q$ has a $\sum_{i\in S_q} p_i + \frac{1}{2n}$ chance of
writing on its pending attempt, where $S_q$ is the set of indices in
$1\dots t-1$ where $q$ previously attempted to write.

Summing up these probabilities over all processes gives
a total of
$\frac{n-1}{2n} + \sum_q \sum_{i \in S_q} p_i ≤ 1/2 + 1/4 = 3/4$.
So with probability at least $e^{-1/4}(1-3/4) = e^{-1/4}/4$, we get
agreement.

\section{Sifters}
\label{section-sifters}

A faster conciliator can be obtained using a \concept{sifter}, which
is a mechanism for rapidly discarding processes using
randomization~\cite{AlistarhA2011} while keeping at least one process
around.
The simplest sifter has each process either write a register (with
low probability) or read it (with high probability); all writers and
all readers that see $⊥$ continue to the next stage of the protocol,
while all readers who see a non-null value drop out.  
If the probability of writing is tuned carefully, this
will reduce $n$ processes to at most $2\sqrt{n}$ processes on average;
by iterating this mechanism, the expected number of remaining
processes can be reduced to $1+ε$ after $O(\log \log n + \log (1/ε))$
phases.

As with previous implementations of test-and-set (see
Algorithm~\ref{alg-n-TAS-from-2TAS}),
it's often helpful to have a sifter return
not only that a process lost but which process it lost to.  This gives
the implementation shown in Algorithm~\ref{alg-sifter}.

\newFunc{\Sifter}{sifter}

\begin{algorithm}
    \Procedure{$\Sifter(p,r)$}{
    \eWithProbability{$p$}{
        $r ← \Id$ \;
        \Return $⊥$\;
    }{
        \Return $r$\;
    }
}
    \caption{A sifter}
    \label{alg-sifter}
\end{algorithm}

To use a sifter effectively, $p$ should be tuned to match the number
of processes that are likely to use it.  This is because of the
following lemma:
\begin{lemma}
    \label{lemma-sifter}
    Fix $p$, and let $X$ processes executed a sifter with parameter
    $p$.  Let $Y$ be the number of processes for which the sifter
    returns $⊥$.  Then 
    \begin{equation}
\ExpCond{X}{Y} ≤ pX + \frac{1}{p}.
        \label{eq-lemma-sifter}
    \end{equation}
\end{lemma}
\begin{proof}
In order to return $⊥$, a process must either (a) write to $r$, which
occurs with probability $p$, or (b) read $r$ before any other process
writes to it.  The expected number of writers, conditioned on $X$, is
exactly $pX$.  The expected number of readers before the first write
has a geometric distribution truncated by $X$.  Removing the truncation
gives exactly $\frac{1}{p}$ expected readers, which is an upper bound
on the correct value.
\end{proof}

For $n$ initial processes, the choice of $p$ that minimizes the bound
in \eqref{eq-lemma-sifter} is $\frac{1}{\sqrt{n}}$, giving at most
$2\sqrt{n}$ expected survivors.  Iterating this process with optimal
$p$ at each step gives a
sequence of at most $n$, $2\sqrt{n}$, $2\sqrt{2\sqrt{n}}$, etc., expected survivors
after each sifter.  The twos are a little annoying, but a
straightforward induction bounds the expected survivors after $i$
rounds by $4⋅n^{2^{-i}}$.  In particular, 
we get at most $8$ expected survivors after $\ceil{\lg \lg n}$ rounds.

At this point it makes sense to switch to a fixed $p$ and a different
analysis.  For $p=1/2$, the first process to access $r$ always
survives, and each subsequent process survives with probability at
most $3/4$ (because it leaves if the first process writes and it
reads).  So the number of ``excess'' processes drops as $(3/4)^i$, and
an additional $\ceil{\log_{4/3} (7/ε)}$ rounds are enough to reduce the
expected number of survivors from $1+7$ to $1+ε$ for any fixed
$ε$.\footnote{This argument essentially follows the
    proof of~\cite[Theorem 2]{Aspnes2012}, which, because of
    neglecting to subtract off a $1$ at one point, ends up with $8/ε$
instead of $7/ε$.}

It follows that
\begin{theorem}
    \label{theorem-sifter}
    An initial set of $n$ processes can be reduced to $1$ with
    probability at least $1-ε$
    using $O(\log \log n + \log
    (1/ε))$ rounds of sifters.
\end{theorem}
\begin{proof}
    Let $X$ be the number of survivors after $\ceil{\lg \lg n} +
    \ceil{\log_{4/3}(7/ε)}$ rounds of sifters, with probabilities tuned
    as described above.  We've shown that $\Exp{X} ≤ 1+ε$, so
    $\Exp{X-1} ≤ ε$.  Since $X-1 ≥ 0$, from Markov's inequality we
    have $\Prob{X ≥ 2} = \Prob{X-1 ≥ 1} ≤ \Exp{X-1}/1 ≤ ε$.
\end{proof}

\subsection{Test-and-set using sifters}
\label{section-sifter-TAS}

Sifters were initially designed to be used for test-and-set.  For this
purpose, we treat a return value of $⊥$ as ``keep going'' and anything
else as ``leave with value $1$.''  Using $O(\log \log n)$ rounds of
sifters, we can get down to one process that hasn't left with
probability at least $1-\log^{-c} n$ for any fixed constant $c$.
We then need a fall-back TAS to
handle the $\log^{-c} n$ chance that we get more than one such
survivor.

Alistarh and Aspnes~\cite{AlistarhA2011} used the \FuncSty{RatRace}
algorithm of Alistarh~\etal~\cite{AlistarhAGGG2010} for this purpose.
This is an adaptive randomized test-and-set built from splitters and
two-process consensus objects that runs in $O(\log k)$ expected
time, where $k$ is the number of processes that access the
test-and-set; a sketch of this algorithm is given in
§\ref{section-ratrace-and-reshuffle}.  If we want to avoid appealing to this algorithm, a
somewhat simpler approach is to use an approach similar to the
Lamport's fast-path mutual exclusion algorithm (described in
§\ref{section-mutex-fast}): any process that survives the sifters
tries to rush to a two-process TAS at the top of a tree of
two-processes TASes by winning a splitter, and if it doesn't win the
splitter, it enters at a leaf and pays $O(\log n)$ expected steps.
By setting $\epsilon = 1/\log n$,
the overall expected cost of this final stage is $O(1)$.

This algorithm does not guarantee linearizability.  I might lose a
sifter early on only to have a later process win all the sifters (say,
by writing to each one) and return $0$.  A $\AWWgate$ bit as in
Algorithm~\ref{alg-n-TAS-from-2TAS} solves this problem.
The full code is given in Algorithm~\ref{alg-log-log-TAS}.

\begin{algorithm}
    \eIf{$\AWWgate ≠ ⊥$}{
        \Return $1$ \;
    }{
        $\AWWgate ← \MyId$ \;
        \For{$i ← 1 \dots \ceil{\log \log n} + \ceil{\log_{4/3} (7 \log n)}$}{
            \eWithProbability{$\min\left(1/2, 2^{1-2^{-i+1}}\right)$}{
                $r_i ← \MyId$\;
            }{
                $w ← r_i$ \;
                \If{$w ≠ ⊥$}{
                    \Return $1$\;
                }
            }
        }
    }

    \eIf{$\Splitter() = \SplitterStop$}{
        \Return $0$\;
    }{
        \Return $\FuncSty{AWWTAS()}$
    }
    \caption{Test-and-set in $O(\log \log n)$ expected time}
    \label{alg-log-log-TAS}
\end{algorithm}

\subsection{Consensus using sifters}
\label{section-sifter-consensus}

\newData{\ChooseWrite}{chooseWrite}
\newFunc{\Conciliator}{conciliator}
\newData{\Persona}{persona}

With some trickery, the sifter mechanism can be adapted to solve
consensus, still in $O(\log \log n)$ expected individual
work~\cite{Aspnes2012}.  The main difficulty is that a process
can no longer drop out as soon as it knows that it lost: it still
needs to figure out who won, and possible help that winner over the
finish line.

The basic idea is that when a process $p$ loses a sifter to some
other process $q$, $p$ will act like a clone of $q$ from that point
on.  In order to make this work, each process writes down at the start
of the protocol all of the coin-flips it intends to use to decide
whether to read or write at each round of sifting.  Together with its
input, these coin-flips make up the process's \concept{persona}.  In
analyzing the progress of the sifter, we count surviving personae
(with multiple copies of the same persona counting as one) instead of
surviving processes.

Pseudocode for this algorithm is given in
Algorithm~\ref{alg-conciliator-sifter}.  Note that the loop body is
essentially the same as the code in Algorithm~\ref{alg-sifter}, except
that the random choice is replaced by a lookup in
$\Persona.\ChooseWrite$.
\begin{algorithm}
\Procedure{$\Conciliator(\Input)$}{
    Let $R = \ceil{\log \log n} + \ceil{\log_{4/3} (7/\epsilon)}$\;
    Let $\ChooseWrite$ be a vector of 
        $R$
    independent
    random Boolean variables with $\Pr[\ChooseWrite[i] = 1] = p_i$,
    where $p_i = 2^{1-2^{-i+1}} (n)^{-2^{-i}}$
    for $i \le \ceil{\log \log n}$ and $p_i = 1/2$ for larger $i$.   \;
    $\Persona ← \Tuple{\Input, \ChooseWrite, \MyId}$
    \label{line-initial-persona} \;
    \For{$i ← 1 \dots R$
    }{
        \eIf{$\Persona.\ChooseWrite[i] = 1$}{
            \label{line-conciliator-sifter-if-start}
            $r_i ← \Persona$\;
        }{
            $v ← r_i$\;
            \If{$v \ne \bot$}{
                $\Persona ← v$
                 \label{line-conciliator-sifter-if-end}\;
            }
        }
    }
    \Return $\Persona.\Input$\;
}
\caption{Sifting conciliator (from~\protect{\cite{Aspnes2012}})}
\label{alg-conciliator-sifter}
\end{algorithm}

To show that this works, we need to argue that having multiple copies
of a persona around doesn't change the behavior of the sifter.
In each round, we will call the first process with a given persona
$p$ to access $r_i$ the \concept{representative} of $p$, and argue
that a persona survives round $i$ in this algorithm precisely when its
representative would survive round $i$ in a corresponding test-and-set
sifter with the schedule restricted only to the representatives.

There are three cases:
\begin{enumerate}
    \item The representative of $p$ writes.  Then at least one copy of
        $p$ survives.
    \item The representative of $p$ reads a null value.  Again at
        least one copy of $p$ survives.
    \item The representative of $p$ reads a non-null value.  Then no
        copy of $p$ survives: all subsequent reads by processes
        carrying $p$ also read a non-null value and discard $p$, and since no process with
        $p$ writes, no other process adopts $p$.
\end{enumerate}

From the preceding analysis for test-and-set, we have that after 
$O(\log \log n + \log 1/ε)$ rounds with appropriate probabilities of
writing,
at most $1+ε$ values survive on average.  This gives a probability of
at most $ε$ of disagreement.  By alternating these conciliators with
adopt-commit objects, we get agreement in $O(\log \log n + \log m /
\log \log m)$ expected time, where $m$ is the number of possible input
values.

I don't think the $O(\log \log n)$ part of this expression is optimal,
but I don't know how to do better.

\subsection{A better sifter for test-and-set}

A more sophisticated sifter due to Giakkoupis and
Woelfel~\cite{GiakkoupisW2012} removes all
but $O(\log n)$ processes, on average, using two operations for each
process.  Iterating this
sifter reduces the expected survivors to $O(1)$ in $O(\log^* n)$
rounds.  A particularly nice feature of the Giakkoupis-Woelfel
algorithm is that (if you don't care about space) it doesn't have any
parameters that require tuning to $n$: this means that exactly the same structure can
be used in each round.  An unfortunate feature is that it's not
possible to guarantee that every process that leaves learns the
identity of a process that stays: this means that it can't adapted
into a consensus protocol using the persona trick described in
§\ref{section-sifter-consensus}.

Pseudocode is given in Algorithm~\ref{alg-log-star-sifter}.  In this
simplified version, we assume an infinitely long array $A[1\dots]$, so
that we don't need to worry about $n$.  Truncating the array at $\log
n$ also works, but the analysis requires handling the last position as
a special case, which I am too lazy to do here.
\begin{algorithm}
    Choose $r∈ℤ^{+}$ such that $\Prob{r=i} = 2^{-i}$\;
    $A[r] ← 1$\;
    \eIf{$A[r+1] = 0$}{
        stay\;
    }{
        leave\;
    }
    \caption{Giakkoupis-Woelfel sifter~\protect{\cite{GiakkoupisW2012}}}
    \label{alg-log-star-sifter}
\end{algorithm}

\begin{lemma}
    \label{lemma-GW}
    In any execution of Algorithm~\ref{alg-log-star-sifter} with an
    oblivious adversary and $n$ processes, at least one process stays,
    and the expected number of
    processes that stay is $O(\log n)$.
\end{lemma}
\begin{proof}
    For the first part, observe that any process that picks the
    largest value of $r$ among all processes will survive; since the
    number of processes is finite, there is at least one such
    survivor.

    For the second part, let $X_i$ be the number of survivors with
    $r=i$.  Then $\Exp{X_i}$ is bounded by $n⋅2^{-i}$, since no process
    survives with $r=i$ without first choosing $r=i$.  But we can also
    argue that $\Exp{X_i} ≤ 3$ for any value of $n$, by considering
    the sequence of write operations in the execution.

    Because the adversary is oblivious, the location of these writes
    is uncorrelated with their ordering.  If we assume that the
    adversary is trying to maximize the number of survivors, its best
    strategy is to allow each process to read immediately after
    writing, as delaying this read can only increase the probability
    that $A[r+1]$ is nonzero.  So in computing $X_i$, we are counting
    the number of writes to $A[i]$ before the first write to $A[i+1]$.
    Let's ignore all writes to other registers; then the $j$-th write
    to either of $A[i]$ or $A[i+1]$ has a conditional probability of
    $2/3$ of landing on $A[i]$ and $1/3$ on $A[i+1]$.  We are thus
    looking at a geometric distribution with parameter $1/3$, which
    has expectation $3$.

    Combining these two bounds gives $\Exp{X_i} ≤ \min(3, 2^{-i})$.
    So then
    \begin{align*}
        \Exp{\mbox{survivors}}
        &≤ ∑_{i=1}^{∞} \min(3, n⋅2^{-i}) \\
        &= 3 \lg n + O(1),
    \end{align*}
    because once $n⋅2^{-i}$ drops below $3$, the remaining terms form
    a geometric series.
\end{proof}

Like square root, logarithm is concave, so Jensen's inequality applies
here as well.  So $O(\log^* n)$ rounds of
Algorithm~\ref{alg-log-star-sifter} reduces us to an expected constant
number of survivors, which can then be fed to RatRace.

With an adaptive adversary, all of the sifter-based test-and-sets fail
badly: in this particular case, an adaptive adversary can sort the
processes in order of increasing write location so that every process
survives.  The best known $n$-process test-and-set for an adaptive
adversary is still a tree of $2$-process randomized test-and-sets, as
in the Afek~\etal~\cite{AfekWW1993} algorithm described in
§\ref{section-test-and-set-from-two-process-consensus}.  Whether
$O(\log n)$ expected steps is in fact necessary is still open (as is
the exact complexity of test-and-set with an oblivious adversary).

\section{Space bounds}

A classic result of Fich, Herlihy, and Shavit~\cite{FichHS1998} showed
that $Ω(\sqrt{n})$ registers are needed to solve consensus even under
the very weak requirement of \concept{nondeterministic solo
termination}, which says that for every reachable configuration 
and every process $p$, there exists some continuation of the execution
in which the protocol terminates with only $p$ running.  The best
known upper bound is the trivial bound of $n$—one single-writer register per
process—since any algorithm that uses multi-writer registers can be translated
into one that uses only single-writer registers, and (assuming wide enough registers)
multiple registers of a single process can be combined into one.

For many years,
there was very little progress in closing the gap between these
two bounds.  In 2013, we got a hint that FHS might be tight when
Giakkoupis~\etal~\cite{GiakkoupisHHW2013}
gave a surprising $O(\sqrt{n})$-space algorithm for the closely
related problem of obstruction-free one-shot test-and-set.

But then 
Gelashvili~\cite{Gelashvili2015} showed an $n/20$ space lower
bound for consensus for anonymous processes, and Zhu quickly
followed this with a lower bound for non-anonymous
processes~\cite{Zhu2016}, showing that at least $n-1$ registers are
required, using
a clever combination of bivalence and covering arguments.
Around the same time, Giakkoupis~\etal~\cite{GiakkoupisHHW2015}
further improved the space complexity of obstruction-free test-and-set
to $O(\log n)$, using a deterministic obstruction-free implementation
of a sifter.  So the brief coincidence of the $Ω(\sqrt{n})$ lower bound
on consensus and the $O(\sqrt{n})$ upper bound on test-and-set turned
out to be an accident.

For consensus, there is still a gap, but it's a very small gap.  Whether the
actual space needed is $n-1$ or $n$ remains open.

\myChapter{Renaming}{2026}{}
\label{chapter-renaming}

We will start by following the presentation in
\cite[§{}16.3]{AttiyaW2004}.  This mostly describes results of the
original paper of Attiya~\etal~\cite{AttiyaBDPR1990} that defined the
renaming problem and gave a solution for message-passing; however, 
it's now more common to treat renaming in the context of
shared-memory, so we will follow Attiya and Welch's translation of
these results to a shared-memory setting.

\section{Renaming}
\label{section-renaming-definition}

In the \concept{renaming} problem, we have $n$ processes, each starts
with a name from some huge namespace, and we'd like to assign them
each unique names from a much smaller namespace.  The main application
is allowing us to run algorithms that assume that the processes are
given contiguous numbers, e.g., the various collect or atomic snapshot
algorithms in which each process is assigned a unique register and we
have to read all of the registers.  With renaming, instead of reading a huge pile of registers in order to find the few that are actually used, we can map the processes down to a much smaller set.

Formally, we have a decision problem where each process has input
$x_{i}$ (its original name) and output $y_{i}$, with the requirements:
\begin{description}
 \item[Termination] Every non-faulty process eventually decides.
 \item[Uniqueness] If $p_{i} \ne p_{j}$, then $y_{i} \ne y_{j}$.
 \item[Anonymity] The code executed by any process depends only on its
 input $x_{i}$: for any execution of processes $p_{1}\dots p_{n}$ with
 inputs $x_{1}\dots x_{n}$, and any permutation $π$ of $[1\dots n]$,
 there is a corresponding execution of processes $p_{π(1)}\dots
 p_{π(n)}$ with inputs $x_{1}\dots x_{n}$ in which $p_{π(i)}$
 performs exactly the same operations as $p_{i}$ and obtains the same
 output $y_{i}$.
\end{description}

The last condition is like non-triviality for consensus: it excludes
algorithms where $p_{i}$ just returns $i$ in all executions.  Typically we do not have to do much to prove anonymity other than observing that all processes are running the same code.

We will be considering renaming in a shared-memory system, where we only have atomic registers to work with.

\section{Performance}
\label{section-renaming-performance}

Conventions on counting processes:

\begin{itemize}
 \item $N$ = number of possible original names.
 \item $n$ = maximum number of processes.
 \item $k$ = number of processes that actually execute the algorithm.
\end{itemize}

Ideally, we'd like any performance measures we get to depend on $k$
alone if possible (giving an \concept{adaptive} algorithm).  Next best
would be something polynomial in $n$ and $k$.  Anything involving $N$ is bad.

We'd also like to minimize the size of the output namespace.  How well
we can do this depends on what assumptions we make.  For deterministic
algorithms using only read-write registers, a lower bound due to
Herlihy and Shavit~\cite{HerlihyS1999} shows that we can't get fewer
than $2n-1$ names for general $n$.\footnote{This lower bound was
    further refined by
Castañeda and Rajsbaum~\cite{CastanedaR2008}, who show that $2n-2$
(but no less!) is possible for certain special values of $n$; all of these lower
bounds make extensive use of combinatorial topology, so we won't try
to present them here.}  Our target thus
will be exactly $2n-1$ output names if possible, or $2k-1$ if we are
trying to be adaptive.  For randomized algorithms, it is possible to
solve 
\index{renaming!strong}
\indexConcept{strong renaming}{strong} or 
\index{renaming!tight}
\indexConcept{tight renaming}{tight}
renaming, where the size of the namespace is exactly $k$; we'll see
how to do this in §\ref{section-randomized-renaming}.

A small note on bounds: There is a lot of variation in the literature
on how bounds on the size of the output namespace are stated.  The
original Herlihy-Shavit lower bound~\cite{HerlihyS1999} says that there is no general
renaming algorithm that uses $2n$ names for $n+1$
processes; in other
words, any $n$-process algorithm uses at least $2n-1$ names.  Many
subsequent papers discussing lower bounds on the namespace follow the
approach of Herlihy and Shavit and quote lower bounds that are
generally $2$ higher than the minimum number of names needed for $n$
processes.  This requires a certain amount of translation when
comparing these lower bounds with upper bounds, which use the more
natural convention.

\section{Order-preserving renaming}

Before we jump into upper bounds, let's do an easy lower bound from
the Attiya~\etal{} paper~\cite{AttiyaBDPR1990}.  This bound works on a
variant of renaming called 
\index{renaming!order-preserving}
\concept{order-preserving renaming}, where we require that $y_{i} < y_{j}$
whenever $x_{i} < x_{j}$.  Unfortunately, this requires a very large
output namespace: with $t$ failures, any asynchronous algorithm for
order-preserving renaming requires $2^t(n-t+1)-1$ possible output names.  This lower bound applies regardless of the model, as long as some processes may start after other processes have already been assigned names.

For the wait-free case, we have $t = n-1$, and the bound becomes just
$2^{n}-1$.  This is a simpler case than the general $t$-failure case, but the essential idea is the same: if I've only seen a few of the processes, I need to leave room for the others.

\begin{theorem}
\label{theorem-order-preserving-renaming-lower-bound}
There is no order-preserving renaming algorithm for $n$ processes
using fewer than $2^{n}-1$ names.
\end{theorem}
\begin{proof}
By induction on $n$.  For $n=1$, we use $2^{1}-1=1$ names; this is the
base case.  For larger $n$, suppose we use $m$ names, and consider an
execution in which one process $p_{n}$ runs to completion first.  This
consumes one name $y_{n}$ and leaves $k$ names less than $y_{n}$ and
$m-k-1$ names greater than $y_{n}$.  By setting all the inputs $x_{i}$
for $i < n$ either less than $x_{n}$ or greater than $x_{n}$, we can
force the remaining processes to choose from the remaining $k$ or
$m-k-1$ names.  Applying the induction hypothesis, this gives $k ≥
2^{n-1}-1$ and $m-k-1 ≥ 2^{n-1}-1$, so $m = k+(m-k-1)+1 ≥
2(2^{n-1}-1)+1 = 2^{n}-1$.
\end{proof}

\section{Deterministic renaming}
\label{section-deterministic-renaming}

In
\index{renaming!deterministic}
\concept{deterministic renaming}, we can't use randomization, and may
or may not have any primitives stronger than atomic registers.  With
just atomic registers, we can only solve loose renaming; with
test-and-set, we can solve tight renaming.  In this section, we
describe some basic algorithms for deterministic renaming.

\subsection{Wait-free renaming with \texorpdfstring{$2n-1$}{2n-1} names}
\label{section-renaming-snapshots}

Here we use Algorithm 55 from~\cite{AttiyaW2004}, which is an
adaptation to shared memory of the message-passing renaming algorithm
of~\cite{AttiyaBDPR1990}.  One odd feature of
the algorithm is that, as written, it is not anonymous: processes
communicate using an atomic snapshot object and use their process IDs
to select which component of the snapshot array to write to.  But if
we think of the process IDs used in the algorithm as the inputs
$x_{i}$ rather than the actual process IDs $i$, then everything works.
The version given in Algorithm~\ref{alg-snapshot-renaming} makes this substitution explicit, by treating the
original name $i$ as the input.

\newData{\WFRview}{view}
\newFunc{\getName}{getName}
\newFunc{\releaseName}{releaseName}

\begin{algorithm}
    \Procedure{$\getName(i)$}{
        $s ← 1$ \;
        \While{\True}{
            $A[i] ← s$ \;
            $\WFRview ← \Snapshot(a)$ \;
            \eIf{$\WFRview[j] = s$ for some $j≠i$}{
                $r ← \card*{ \{ j : \WFRview[j] \ne ⊥ ∧ j ≤ i \} }$ \;
                \label{line-snapshot-renaming-rank}
                $s ← r$-th positive integer not in $\{ \WFRview[j] : j
                \ne i ∧ \WFRview[j] ≠ ⊥ \}$\;
                \label{line-snapshot-renaming-choose-name}
            }{
                \Return $s$\;
            }
        }
    }
\caption{Wait-free deterministic renaming}
\label{alg-snapshot-renaming}
\end{algorithm}

The array $a$ holds proposed names for each process (indexed by the
original names), or $⊥$ for processes that have not proposed a name
yet. If a process proposes a name and finds that no other process has
proposed the same name, it takes it; otherwise it chooses a new name
by first computing its id's rank $r$ among the active processes and
then choosing the $r$-th smallest name that hasn't been proposed by
another process. Because the rank is at most $n$ and there are at
most $n-1$ names proposed by the other processes, this always gives
proposed names in the range $[1\dots 2n-1]$. We also have anonymity,
since every process runs the same code (with the only difference in
behavior resulting from the input name $i$).

To show uniqueness, consider two process with original names
$i$ and $j$.  Suppose that $i$ and $j$ both decide on $s$. Then $i$
sees a view in which $a[i] = s$ and $a[j] \ne s$, after which it no
longer updates $a[i]$.  Similarly, $j$ sees a view in which $a[j] = s$
and $a[i] \ne s$, after which it no longer updates $a[j]$.  If $i$'s
view is obtained first, then $j$ can't see $a[i] \ne s$, but the same
holds if $j$'s view is obtained first.  So in either case we get a
contradiction, proving uniqueness.

Termination is a bit trickier. Here we argue that no process can run
forever without returning a name, by showing that if we have a set of
processes that are doing this, the one with smallest input name
eventually returns an output name, contradicting the assumption that
they all run forever.

Imagine some execution in which processes with input names $p_1 < p_2
< \dots < p_k$ take infinitely many steps, while the remaining
processes with input names $q_j$ do not.
Observe that the rank $r$ computed by each
$p_i$ eventually stabilizes, since it can only change if $p_i$
observes a new non-null entry $a[j]$ for $j≤p_i$, and this can only
happen a finite number of times. Suppose that we wait both for these
ranks to stabilize and for all the processes $q_j$ to perform their
last operations.

At this point, any name that appears in $a[q_j]$ for some $q_j$ is no
longer available any process $p_i$, either because $q_j$ has already
returned it (if we are lucky) or because $q_j$ has stopped (and thus
won't change $a[q_j]$ again). Let $z_1 < z_2 < \dots z_m$ be the names
that do \emph{not} appear in $a[q_j]$ for any $q_j$ after 
all $a[q_j]$ have stabilized.
Let $r_i$ be the final, stable rank of
process $p_i$. Then we can argue that after ranks and the $a[q_j]$
have stabilized, $p_i$ never picks a new name from
$\Set{z_1,\dots,z_{r-1}}$, because it picks the $r_i$-th smallest name
among those not already taken in its view, and these names are all
smaller.

We would like to argue that this means that $z_{r_1}$ is eventually
returned by $p_1$ (which will contradict the supposition that $p_1$
runs forever). This may not happen immediately, because even though
$z_1,\dots,z_{r_1}$ are not covered by any $a[q_j]$, they may be
covered by $a[p_i]$ for some $p_i ≠ p_1$. But any such $p_i$ takes
infinitely many steps, so it eventually chooses a new
name not in $z_1,\dots,z_{r-1}$. Once all the $p_i$ have picked names
outside this range, $z_{r_1}$ becomes the $r_1$-th smallest available
name, so $p_1$ chooses it, sees no conflict, and returns.

Note that we haven't proved any complexity bounds on this algorithm at
all, but we know that the snapshot alone takes at least $\Omega(N)$
time and space.  
With some tinkering this can be reduced.
Brodksy~\etal~\cite{BrodskyEW2011} cite a paper of Bar-Noy and
Dolev~\cite{Bar-NoyD1989}
as giving a shared-memory version of~\cite{AttiyaBDPR1990} with
complexity $O(n ⋅ 4^n)$; they also give algorithms and pointers to
algorithms with much better complexity.

\subsection{Long-lived renaming}

In 
\index{renaming!long-lived}
\concept{long-lived renaming} a process can release a name for later
use by other processes (or the same process, if it happens to run
choose-name again).  Now the bound on the number of names needed is
$2k-1$, where $k$ is the maximum number of concurrently active
processes.  Algorithm~\ref{alg-snapshot-renaming} can be converted to a long-lived
renaming algorithm by adding the \releaseName procedure given in
Algorithm~\ref{alg-snapshot-renaming-release-name}.  This just erases
the process's proposed name, so that some other process can claim it.

\begin{algorithm}
\Procedure{$\releaseName()$}{
    $a[i] ← ⊥$\;
}
\caption{Releasing a name}
\label{alg-snapshot-renaming-release-name}
\end{algorithm}

Here the termination requirement is weakened slightly, to say that
some process always makes progress in \getName.  It may be, however,
that there is some process that never successfully obtains a name,
because it keeps getting stepped on by other processes zipping in and
out of \getName and \releaseName.

\subsection{Renaming without snapshots}
\label{section-moir-anderson}

Moir and Anderson~\cite{MoirA1995} give a renaming protocol that is
somewhat easier to understand and doesn't require taking snapshots
over huge arrays.  A downside is that the basic version requires
$k(k+1)/2$ names to handle $k$ active processes.

\subsubsection{Splitters}
\label{section-splitters}

The Moir-Anderson renaming protocol uses a network of
\concept{splitters}, which we last saw providing a fast path for
mutual exclusion in §\ref{section-mutex-fast}.  Each splitter
is a widget, built from a pair of atomic registers, that assigns to
each processes that arrives at it the value \SplitterRight,
\SplitterDown, or \SplitterStop.  
As discussed previously, the useful properties of
splitters are that if at least one process arrives at a splitter, then
(a) at least one process returns \SplitterRight or \SplitterStop; and (b) at
least one process returns \SplitterDown or \SplitterStop; (c) at most
one process returns \SplitterStop; and (d) any process that runs by
itself returns \SplitterStop.  

We proved the last two properties in
§\ref{section-mutex-fast}; we'll prove the first two here.
Another way of describing these properties is that of all the
processes that arrive at a splitter, some process doesn't go down and
some process doesn't go right.  By arranging splitters in a grid, this
property guarantees that every row or column that gets at least one
process gets to keep it—which means that with $k$ processes, no
process reaches row $k+1$ or column $k+1$.

Algorithm~\ref{alg-splitter-again} gives the implementation of a
splitter (it's identical to Algorithm~\ref{alg-splitter}, but it will
be convenient to have another copy here).

\begin{algorithm}
\AlgSplitterBody{alg-splitter-again}
\caption{Implementation of a splitter}
\label{alg-splitter-again}
\end{algorithm}

\begin{lemma}
\label{lemma-splitter-right}
If at least one process completes the splitter, at least one process
returns $\SplitterStop$ or $\SplitterRight$.
\end{lemma}
\begin{proof}
Suppose no process returns \SplitterRight; then every process sees
\SplitterOpen in \SplitterDoor, which means that every process writes
its ID to \SplitterRace before any process closes the door.  Some
process writes its ID last: this process will see its own ID in
\SplitterRace and return \SplitterStop.
\end{proof}

\begin{lemma}
\label{lemma-splitter-down}
If at least one process completes the splitter, at least one process
returns $\SplitterStop$ or $\SplitterDown$.
\end{lemma}
\begin{proof}
First observe that if no process ever writes to \SplitterDoor, then no
process completes the splitter, because the only way a process can
finish the splitter without writing to \SplitterDoor is if it sees
\SplitterClosed when it reads \SplitterDoor (which must have been
written by some other process).  So if at least one process finishes,
at least one process writes to \SplitterDoor.  Let $p$ be any such
process.  From the code, having written \SplitterDoor, it has already
passed up the chance to return \SplitterRight; thus it either returns
\SplitterStop or \SplitterDown.
\end{proof}

\subsubsection{Splitters in a grid}
\label{section-splitters-grid}

\begin{figure}
\centering
\includegraphics[scale=0.4]{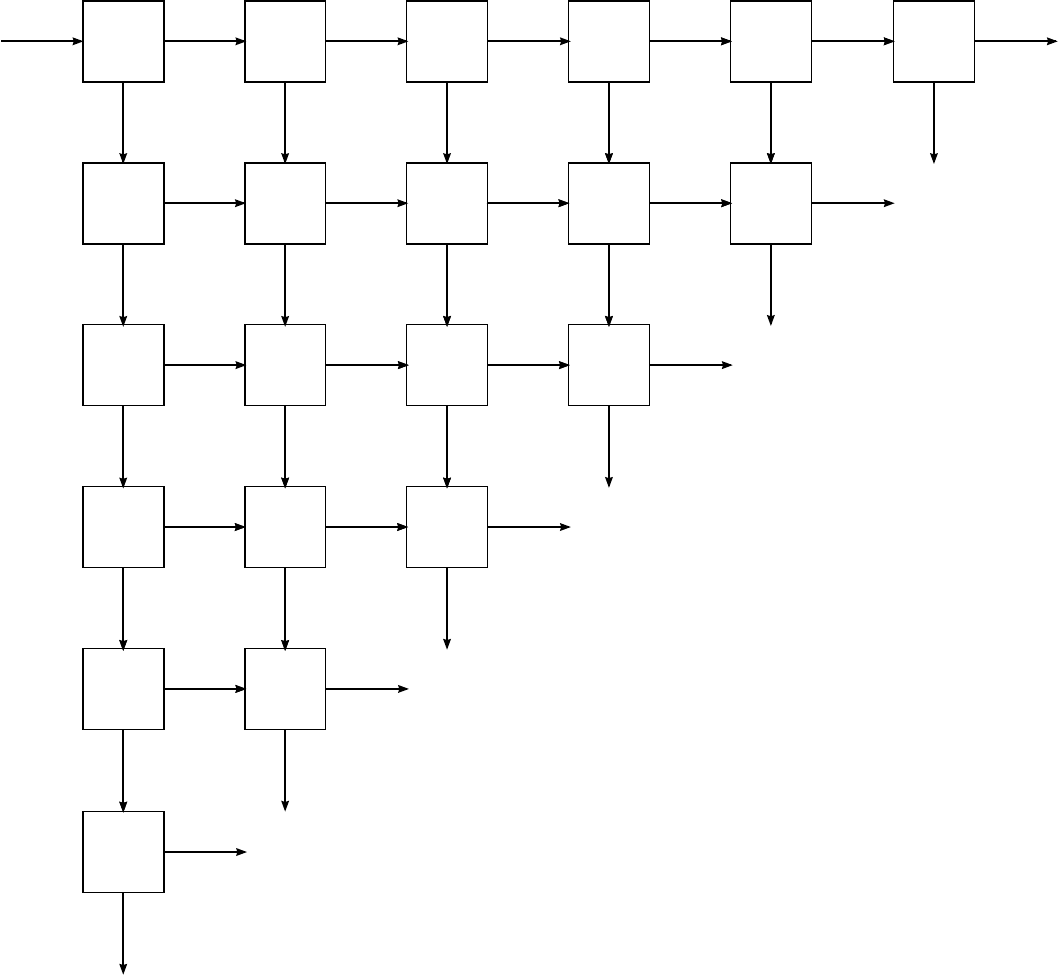}
    \caption[A $6 \times 6$ Moir-Anderson grid]
    {A $6 \times 6$ Moir-Anderson grid
    (From~\cite{Aspnes2010splitters}.)}

\label{figure-moir-anderson}
\end{figure}

Now build an $m$-by-$m$ triangular grid of splitters, arranged as rows $0\dots
m-1$ and columns $0\dots m-1$, where a splitter appears in each
position $(r,c)$ with $r+c ≤ m-1$ (see
Figure~\ref{figure-moir-anderson} for an example; this figure is taken
from~\cite{Aspnes2010splitters}).
Assign a distinct name to each of the $\binom{m}{2}$ splitters in this
grid.  
To obtain a
name, a process starts at $(r,c) = (0,0)$, and repeatedly executes the
splitter at its current position $(r,c)$.  If the splitter returns
\SplitterRight, it moves to $(r,c+1)$; if \SplitterDown, it moves to
$(r+1,c)$; if \SplitterStop, it stops, and returns the name of its
current splitter.  This gives each name to at most one process (by
Lemma~\ref{lemma-splitter-mutex}); we also have to show that if at
most $m$ processes enter the grid, every process stops at some
splitter.

\begin{figure}
\centering
\includegraphics[scale=0.4]{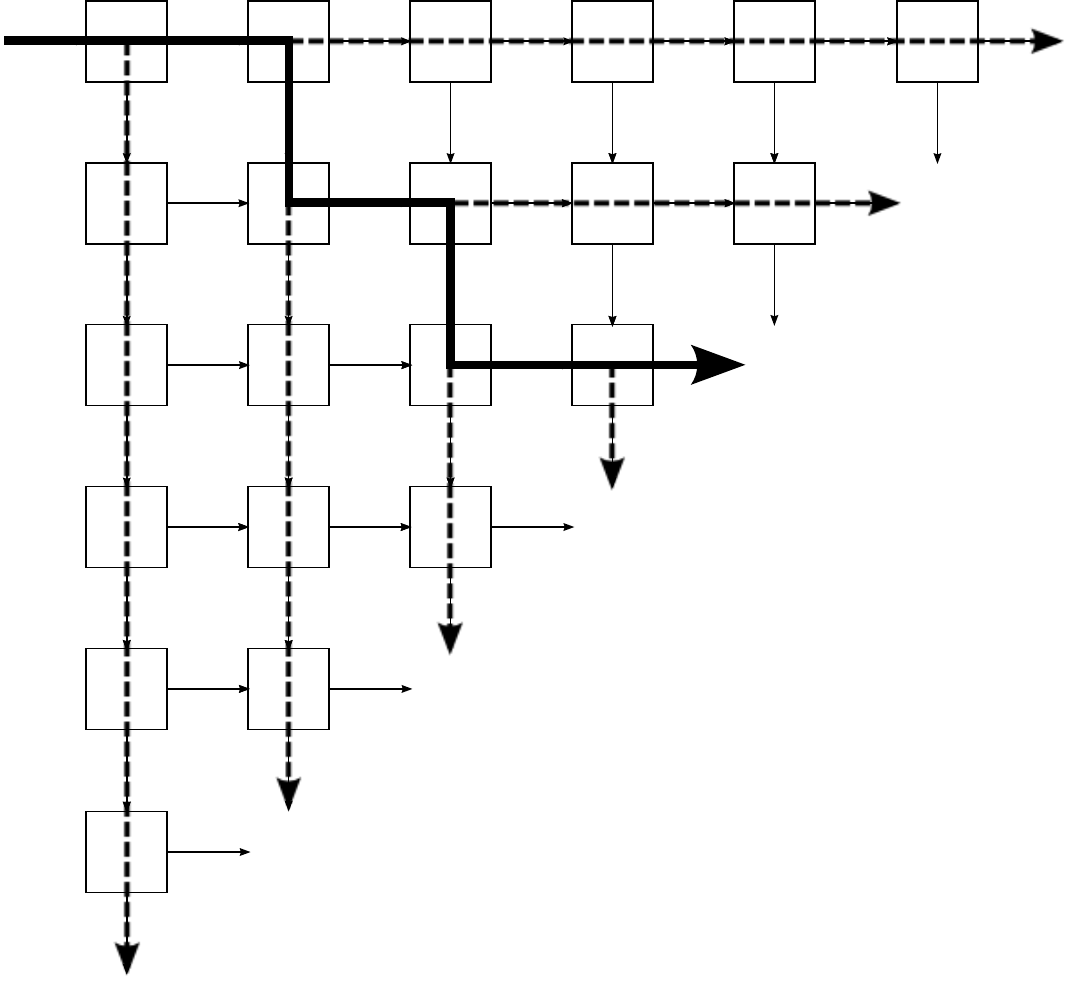}
\caption[Path through a Moir-Anderson grid]{Path taken by a single process through a $6 \times 6$
Moir-Anderson grid (heavy path), 
    and the 6 disjoint paths it spawns (dashed paths).
    (From~\cite{Aspnes2010splitters}.)}
\label{figure-disjoint-paths}
\end{figure}

The argument for this is simple.  Suppose some process $p$ leaves the grid
on one of the $2m$ output wires.
Look at the path it takes to get there (see
Figure~\ref{figure-disjoint-paths}, also taken
from~\cite{Aspnes2010splitters}).
Each splitter on this
path must handle at least two processes (or $p$ would have stopped at
that splitter, by Lemma~\ref{lemma-splitter-solo-wins}).  
So some other process leaves on the other output
wire, either $\SplitterRight$ or $\SplitterDown$.  If we draw a path from each of
these wires that continues $\SplitterRight$ or $\SplitterDown$ to the end of the grid,
then at every step along this path we either have a process stop or
continue in this same direction as long as there is a process left to
do so.
This means that on each of these $m$ disjoint paths, either some splitter stops
a process, or some process reaches a final output wire, each of which
is at a distinct splitter.  But this gives $m$ distinct processes in addition
to $p$, for a total of $m+1$ processes.  It follows that:
\begin{theorem}
\label{theorem-moir-anderson}
An $m\times m$ Moir-Anderson grid solves renaming for up to $m$
processes.
\end{theorem}

The time complexity of the algorithm is $O(m)$: Each process spends at
most 4 operations on each splitter, and no process goes through more
than $2m$ splitters.  In general, any splitter network will take at
least $n$ steps to stop $n$ processes, because the adversary can run
them all together in a horde that drops only one process at each
splitter.

If we don't know $k$ in advance, we can still guarantee names of size
$O(k^{2})$ by carefully arranging them so that each $k$-by-$k$ subgrid
contains the first $\binom{k}{2}$ names.  
This gives an adaptive renaming algorithm (although the namespace size
is pretty high).
We still have to choose our
grid to be large enough for the largest $k$ we might actually
encounter; the resulting space complexity is $O(n^2)$.

With a slightly more clever arrangement of the splitters, it is
possible to reduce the space complexity to
$O(n^{3/2})$~\cite{Aspnes2010splitters}.  Whether further reductions
are possible is an open problem.  Note however that linear time
complexity makes splitter networks uncompetitive with much faster
randomized algorithms (as we'll see in
§\ref{section-randomized-renaming}), so this may not be a very
important open problem.

\subsection{Getting to \texorpdfstring{$2n-1$}{2n-1} names in polynomial space}
\label{section-renaming-space}

From before, we have an algorithm that will get $2n-1$ names for $n$
processes out of $N$ possible processes when run using $O(N)$ space
(for the enormous snapshots).  To turn this into a bounded-space
algorithm, run Moir-Anderson first to get down to $Θ(k^{2})$
names, then run the previous algorithm (in $Θ(n^{2})$ space)
using these new names as the original names.

Since we didn't prove anything about time complexity of the
humongous-snapshot algorithm, we can't say much about the time
complexity of this combined one.
Moir and Anderson suggest instead
using an $O(Nk^{2})$ algorithm of Borowsky and Gafni to get $O(k^{4})$
time for the combined algorithm.  

This is close to the best known: a
later paper by Afek and Merritt~\cite{AfekM1999} holds the current record for
deterministic adaptive renaming into $2k-1$ names at $O(k^2)$
individual steps.  On the lower bound side, it is known that
$\Omega(k)$ is a lower bound on the individual steps of any
renaming protocol with a polynomial output
namespace~\cite{AlistarhAGG2011}.

\subsection{Renaming with test-and-set}
\label{section-renaming-with-test-and-set}

Moir and Anderson give a simple renaming algorithm based on
test-and-set that is 
\index{renaming!strong}
\indexConcept{strong renaming}{strong}
($k$ processes are assigned exactly the
names $1 \dots k$), 
\concept{adaptive} (the time complexity to acquire a name is
$O(k)$), and 
\indexConcept{long-lived renaming}{long-lived}, which means that a process can
release its name and the name will be available to processes that
arrive later.
In fact, the resulting algorithm gives
\index{renaming!long-lived strong}
\concept{long-lived strong renaming},
meaning
that the set of names in use will always be no larger than the set of
processes that have started to acquire a name and not yet finished
releasing one; this is a little stronger than just saying that the
algorithm is strong and that it is long-lived separately.

The algorithm is simple: we have a line of test-and-set bits $T[1]
\dots T[n]$.  To acquire a name, a process starts at $T[1]$ and
attempts to win each test-and-set until it succeeds; whichever $T[i]$
it wins gives it name $i$.  To release a name, a process releases the
test-and-set.

Without the releases, the same mechanism gives
fetch-and-increment~\cite{AfekWW1993}.  Fetch-and-increment by itself
solves tight renaming (although not long-lived renaming,
since there is no way to release a name).

\section{Randomized renaming}
\label{section-randomized-renaming}

With randomization, we can beat both the $2k-1$ lower bound on the
size of the output namespace from~\cite{HerlihyS1999} and the
$\Omega(k)$ lower bound on individual work
from~\cite{AlistarhAGG2011}, achieving strong renaming with $O(\log k)$
expected individual work~\cite{AlistarhACHGZ2011}.

The basic idea is that we can use randomization for 
\concept{load balancing}, where we avoid the problem of having an army
of processes marching together with only a few peeling off at a time
(as in splitter networks) by having the processes split up based on
random choices.  For example, if each process generates a random name
consisting of $2 \ceil{\lg n}$ bits, then it is reasonably likely that
every process gets a unique name in a namespace of size $O(n^2)$ (we
can't hope for less than $O(n^2)$ because of the \concept{birthday
paradox}).  But we want all processes to be guaranteed to have unique
names, so we need some more machinery.

We also need the processes to have initial names; if they don't, there
is always some nonzero probability that two identical processes will flip their
coins in exactly the same way and end up with the same name.  This
observation was formalized by Buhrman, Panconesi, Silvestri, and
Vitányi~\cite{BuhrmanPSV2006}.

\subsection{Randomized splitters}

Attiya~\etal~\cite{AttiyaKPWW2006}
suggested the use of
\index{splitter!randomized}
\indexConcept{randomized splitter}{randomized splitters} in the
context of another problem 
(\index{collect!adaptive}\concept{adaptive collect})
that is closely related to renaming.

A randomized splitter is just like a regular splitter, except that if
a process doesn't stop it flips a coin to decide whether to go right
or down.  Randomized splitters are nice because they usually split
better than deterministic splitters: if $k$ processes reach a
randomized splitter, with high probability no more than $k/2 +
O(\sqrt{k \log k})$ will leave on either output wire.

It's not hard to show that a binary tree of these things of
depth $2 \ceil{\lg n}$ stops all but a constant expected number of
processes on average;\footnote{The proof is to consider the expected
number of pairs of processes that flip their coins the same way for
all $2 \ceil{\lg n}$ steps.  This is at most $\binom{n}{2} n^{-2} <
1/2$, so on average at most 1 process escapes the tree, giving (by
symmetry) at most a $1/n$ chance that any particular process escapes.
Making the tree deeper can give any polynomial fraction of escapees
while still keeping $O(\log n)$ layers.}
processes that don't stop can be dropped into a backup
renaming algorithm (Moir-Anderson, for example) with only a
constant increase in expected individual work.  

Furthermore, the binary tree of randomized splitters is adaptive; if
only $k$ processes show up, we only need $O(\log k)$ levels
levels on average to split them up.
This gives renaming
into a namespace with expected size $O(k^2)$ in $O(\log k)$ expected individual
steps.

\subsection{Randomized test-and-set plus sampling}
\label{section-ratrace}
\label{section-ratrace-and-reshuffle}

Subsequent work by Alistarh~\etal~\cite{AlistarhAGGG2010} 
showed how some of the same ideas
could be used to get strong renaming, where the output namespace has
size exactly $n$ (note this is not adaptive; another result in the same
paper gives adaptive renaming, but it's not strong).
There are two pieces to this result: an implementation of randomized
test-and-set called \FuncSty{RatRace}, and a sampling procedure for
getting names called \FuncSty{ReShuffle}.

The \index{RatRace}\FuncSty{RatRace} protocol implements a randomized test-and-set
with $O(\log k)$ expected individual work.  The essential idea is to
use a tree of randomized splitters to assign names, then have
processes walk back up the same tree attempting to win a $3$-process
randomized test-and-set at each node (there are 3 processes, because in
addition to the winners of each subtree, we may also have a process
that stopped on that node in the renaming step); this test-and-set is
just a very small binary tree of 2-process test-and-sets implemented
using the algorithm of Tromp and Vitányi~\cite{TrompV2002}.  A \AWWgate bit is added
at the top as in the test-and-set protocol of
Afek~\etal~\cite{AfekGTV1992} to get linearizability.

Once we have test-and-set, we could get strong renaming using a linear
array of test-and-sets as suggested by Moir and
Anderson~\cite{MoirA1995}, but it's more efficient to use the
randomization to spread the processes out.  In the
\index{ReShuffle}\FuncSty{ReShuffle}
protocol, each process chooses a name in the range $[1 \dots n]$
uniformly at random, and attempts to win a test-and-set guarding that
name.  If it doesn't work, it tries again.  Alistarh~\etal{} show that
this method produces unique names for everybody in $O(n \log^4 n)$
total steps with high probability.  The individual step complexity of
this algorithm, however, is not very good: there is likely to be some
unlucky process that needs $\Omega(n)$ probes (at an expected cost of
$Θ(\log n)$ steps each) to find an empty slot.

\subsection{Renaming with sorting networks}

A later paper by Alistarh~\etal~\cite{AlistarhACHGZ2011} reduces the
cost of renaming still further, getting $O(\log k)$ expected individual step
complexity for acquiring a name.  
The resulting algorithm is both adaptive and strong: with $k$
processes, only names $1$ through $k$ are used.
We'll describe the non-adaptive
version here.

The basic idea is to build a 
\index{network!sorting}\concept{sorting network} out of test-and-sets;
the resulting structure, called a
\index{network!renaming}\concept{renaming network}, routes each
process through a sequence of test-and-sets to a unique output wire.
Unlike a splitter network, a renaming network uses the stronger
properties of test-and-set to guarantee that (once the dust settles)
only the lowest-numbered output wires are chosen; this gives
strong renaming.

\subsubsection{Sorting networks}

A \concept{sorting network}\index{network!sorting} 
is a kind of parallel sorting algorithm that
proceeds in synchronous rounds, where in each round the elements of an
array at certain fixed positions are paired off and swapped if they
are out of order. The main difference between a sorting network and a
standard comparison-based sort is that the choice of which positions
to compare at each step is static, and doesn't depend on the outcome
of previous comparisons; also, the only effect of a comparison is
possibly swapping the two values that were compared.

Sorting networks are drawn as in Figure~\ref{figure-sorting-network}.
Each horizontal line or \concept{wire} corresponds to a position in
the array.  The vertical lines are \concept{comparators} that compare
two values coming in from the left and swap the larger value to the
bottom.  A network of comparators is a sorting network if the
sequences of output values is always sorted no matter what the order
of values on the inputs is.

\begin{figure}
    \centering
    \begin{tikzpicture}[scale=1.5]
        \foreach \y in {0,...,3} {
            \draw (0,\y) -- (5,\y);
        }
        \foreach \x/\y/\yy in {1/1/3,2/0/2,3/0/1,3/2/3,4/1/2} {
            \draw (\x,\y) -- (\x,\yy);
            \draw[fill] (\x,\y) circle[radius=0.05];
            \draw[fill] (\x,\yy) circle[radius=0.05];
        }
    \end{tikzpicture}
    \caption{A sorting network}
    \label{figure-sorting-network}
\end{figure}

The \indexConcept{depth!sorting network}{depth} of a sorting network
is the maximum number of comparators on any path from an input to an
output.  The \indexConcept{width!sorting network}{width} is the number
of wires; equivalently, the number of values the network can sort.
The sorting network in Figure~\ref{figure-sorting-network} has depth 3
and width 4.

Explicit constructions of sorting networks with width $n$ and depth
$O(\log^2 n)$ are known~\cite{Batcher1968}.  It is
also known that sorting networks with depth $O(\log n)$
exist~\cite{AjtaiKS1983},
but no explicit construction of such a network is
known.

\subsubsection{Renaming networks}

To turn a sorting network into a renaming network, we replace the
comparators with test-and-set bits, and allow processes to walk
through the network asynchronously.  This is similar to an earlier
mechanism called a
\index{network!counting}
\concept{counting network}~\cite{AspnesHS1994}, which used certain
special classes of sorting networks as counters, but here any sorting
network works.

Each process starts on a separate input wire, and we maintain the
invariant that at most one process ever traverses a wire.  It follows
that each test-and-set bit is only used by two processes.  The first
process to reach the test-and-set bit is sent out the lower output,
while the second is sent out the upper output.  If we imagine each
process that participates in the protocol as a one and each process
that doesn't as a zero, the test-and-set bit acts as a comparator: if
no processes show up on either input (two zeros), no processes leave
(two zeros again); if processes show up on both inputs (two ones),
processes leave on both (two ones again); and if only one process ever
shows up (a zero and a one), it leaves on the bottom output (zero and
one, sorted).  Because the original sorting network sorts all the ones
to the bottom output wires, the corresponding renaming network sorts
all the processes that arrive to the bottom outputs.  Label these
outputs starting at $1$ at the bottom to get renaming.

Since each test-and-set involves at most two processes, we can carry
them out in $O(1)$ expected register operations using, for example,
the protocol of Tromp and Vitányi~\cite{TrompV2002}.  The expected
cost for a process to acquire a name is then $O(\log n)$ (using an
AKS~\cite{AjtaiKS1983}
sorting network).  A more complicated construction in the
Alistarh~\etal{} paper
shows how to make this adaptive, 
giving an expected cost of $O(\log k)$ instead.

The problem with using an AKS network is that the AKS result is
non-constructive: what Ajtai, Komlós, and Szemerédi show is that there
is a particular randomized construction of candidate sorting networks
that succeeds in producing a correct sorting network with nonzero (but
very small) probability.  Other disturbing features of this result are
that we have no efficient way to test candidate sorting networks
(determining if a network of comparators is in fact a sorting network
is co-\classNP-hard), and the constant in the big-O for AKS is quite
spectacularly huge.  So it probably makes more sense to think of
renaming networks as giving renaming in $O(\log^2 n)$ time, since this
is the most efficient practical sorting network we currently know
about.  This has led to efforts to produce $O(\log k)$-work tight renaming algorithms
that don't depend on AKS.  So far this has not worked out in the
standard shared-memory model, even allowing test-and-sets.\footnote{The
closest to this so far is an algorithm of 
Berenbrink~\etal~\cite{BerenbrinkBEFN2015}, who use an extended model
that incorporates an extra primitive called a \concept{$τ$-register},
which is basically a collection of $2 \log n$ test-and-set objects
that are restricted so that at most $τ < 2 \log n$ of them can be set
at a time.  Adding this primitive to the model is not entirely cheating, as the
authors make a case that it could be plausibly implemented in
hardware.  But it does mean that we don't know what happens if we
don't have this additional primitive.}

The use of test-and-sets to route processes to particular names is
similar to the line of test-and-sets proposed by Moir and
Anderson~\cite{MoirA1995} as described in
§\ref{section-renaming-with-test-and-set}.  Some differences
between that protocol and renaming networks is that renaming networks
do not by themselves give fetch-and-increment (although
Alistarh~\etal{} show how to build fetch-and-increment on top of
renaming networks at a small additional cost), 
and renaming networks do not provide any
mechanism for releasing names.
The question of whether it is possible to get cheap 
long-lived strong renaming is still open.

\subsection{Randomized loose renaming}

Loose renaming should be easier than strong renaming, and using
a randomized algorithm it essentially reduces to randomized load
balancing.  A basic approach is to use $2n$ names, and guard each with
a test-and-set; because less than half of the names are taken at any
given time, each process gets a name after $O(1)$ tries and the most
expensive renaming operation over all $n$ processes takes $O(\log n)$
expected steps.

A more sophisticated version of this strategy, which appears
in~\cite{AlistarhAGW2013}, uses $n(1+ε)$ output names to get
$O(\log \log n)$ maximum steps.  The intuition for why this works is
if $n$ processes independently choose one of $cn$ names uniformly at
random, then the expected number of collisions—pairs of processes
that choose the same name—is $\binom{n}{2}/cn$, or about $n/2c$.
This may seem like only a constant-factor improvement, but if we
instead look at the ratio between the survivors $n/2c$ and the number
of allocated names $cn$, we have now moved from $1/c$ to $1/2c^2$.
The $2$ gives us some room to reduce the number of names in the next
round, to $cn/2$, say, while still keeping a $1/c^2$ ratio of
survivors to names.  

So the actual renaming algorithm consists of allocating $cn/2^i$ names
to round $i$, and squaring the ratio of survivors to names in each
rounds.  It only takes $O(\log \log n)$ rounds to knock the ratio of
survivors to names below $1/n$, so at this point it is likely that all
processes will have finished.  At the same time, the sum over all
rounds of the allocated names forms a geometric series, so only $O(n)$
names are needed altogether.

Swept under the carpet here is a lot of careful analysis of the
probabilities.  Unlike what happens with sifters
(see~§\ref{section-sifters}), Jensen's inequality goes the wrong way
here, so some additional technical tricks are needed (see the paper
for details).  But the result is that only $O(\log \log n)$ rounds are
to assign every process a name with high probability, which is the
best value currently known.

There is a rather weak lower bound in the Alistarh~\etal{} paper that
shows that $Ω(\log \log n)$ steps are needed for some process in the
worst case, under the assumption that the renaming algorithm uses only
test-and-set objects and that a process acquires a name as soon as it
wins some test-and-set object.  This does not give a lower bound on
the problem in general, and indeed the renaming-network based
algorithms discussed previously do not have this property.  So the
question of the exact complexity of randomized loose renaming is still
open.

\myChapter{Software transactional memory}{2011}{If you are interested in software
    transactional memory from a theoretical perspective, there is a
    more recent survey on
    this material by Attiya~\cite{Attiya2014}, available at
\url{http://www.eatcs.org/images/bulletin/beatcs112.pdf}.}
\label{chapter-STM}

\index{transactional memory!software}
\indexConcept{software transactional memory}{Software transactional memory}, or \concept{STM} for short,
goes back to Shavit and Touitou~\cite{ShavitT1997} based on earlier
proposals for hardware support for transactions by Herlihy and
Moss~\cite{HerlihyM1993}.  Recently very popular in programming
language circles.  We'll give a high-level description of the Shavit and Touitou
results; for full details see the actual paper.

We start with the basic idea of a \concept{transaction}.  In a
transaction, I read a bunch of registers and update their values, 
and all of these operations appear to be \concept{atomic},
in the sense that the transaction either happens
completely or not at all, and serializes with other transactions as if each occurred instantaneously.  Our goal is to implement this with minimal hardware support, and use it for everything.

Generally we only consider 
\index{transaction!static}
\indexConcept{static transaction}{static transactions}
where the set of memory locations accessed is known in advance, as opposed to
\index{transaction!dynamic}
\indexConcept{dynamic transaction}{dynamic transactions} where it may
vary depending on what we read (for example, maybe we have to follow
pointers through some data structure).  Static transactions are easier
because we can treat them as multi-word read-modify-write.

Implementations are usually \concept{non-blocking}: some infinite
stream of transactions succeed, but not necessarily yours.  This
excludes the simplest method based on acquiring locks, since we have
to keep going even if a lock-holder crashes, but is weaker than
wait-freedom since we can have starvation.

\section{Motivation}

Some selling points for software transactional memory:
\begin{enumerate}
 \item We get atomic operations without having to use our brains much.
 Unlike hand-coded atomic snapshots, counters, queues, etc., we have a
 universal construction that converts any sequential data structure
 built on top of ordinary memory into a concurrent data structure.
 This is useful since most programmers don't have very big brains.  We
 also avoid burdening the programmer with having to remember to lock things.
 \item We can build large shared data structures with the possibility of concurrent access.  For example, we can implement atomic snapshots so that concurrent updates don't interfere with each other, or an atomic queue where enqueues and dequeues can happen concurrently so long as the queue always has a few elements in it to separate the enqueuers and dequeuers.
 \item We can execute atomic operations that span multiple data structures,
 even if the data structures weren't originally designed to work
 together, provided they are all implemented using the STM mechanism.
 This is handy in classic database-like settings, as when we
 want to take \$5 from my bank account and put it in yours.
\end{enumerate}

On the other hand, we now have to deal with the possibility that operations may fail.  There is a price to everything.

\section{Basic approaches}
\begin{itemize}
 \item Locking (not non-blocking).  Acquire either a single lock for
 all of memory (doesn't allow much concurrency) or a separate lock for
 each memory location accessed.  The second approach can lead to
 deadlock if we aren't careful, but we can prove that if every
 transaction acquires locks in the same order (e.g., by increasing memory address), then we never get stuck: we can order the processes by the highest lock acquired, and somebody comes out on top.  Note that acquiring locks in increasing order means that I have to know which locks I want before I acquire any of them, which may rule out dynamic transactions.
 \item Single-pointer compare-and-swap (called ''Herlihy's
 method'' in~\cite{ShavitT1997}, because of its earlier use for constructing concurrent data
 structures by Herlihy~\cite{Herlihy1993}). 
 All access to the data
 structure goes through a pointer in a CAS.  To execute a transaction,
 I make my own copy of the data structure, update it, and then attempt
 to redirect the pointer.  Advantages: trivial to prove that the
 result is linearizable (the pointer swing is an atomic action) and
 non-blocking (somebody wins the CAS); also, the method allows dynamic
 transactions (since you can do anything you want to your copy).
 Disadvantages: There's a high overhead of the many copies,\footnote{This
 overhead can be reduced in many cases by sharing components, a
 subject that has seen much work in the functional programming
 literature.  See for
 example~\cite{Okasaki1999}.} and the single-pointer bottleneck limits concurrency even when two transactions use disjoint parts of memory.
 \item Multiword RMW: This is the approach suggested by Shavit and
 Touitou, which most subsequent work follows.  As usually implemented, it only works for static transactions.  The idea is that I write down what registers I plan to update and what I plan to do to them.  I then attempt to acquire all the registers.  If I succeed, I update all the values, store the old values, and go home.  If I fail, it's because somebody else already acquired one of the registers.  Since I need to make sure that somebody makes progress (I may be the only process left alive), I'll help that other process finish its transaction if possible.  Advantages: allows concurrency between disjoint transactions.  Disadvantages: requires implementing multi-word RMW—in particular, requires that any process be able to understand and simulate any other process's transactions.  Subsequent work often simplifies this to implementing multi-word CAS, which is sufficient to do non-blocking multi-word RMW since I can read all the registers I need (without any locking) and then do a CAS to update them (which fails only if somebody else succeeded).
\end{itemize}

\section{Implementing multi-word RMW}

We'll give a sketchy description of
Shavit and Touitou's method~\cite{ShavitT1997}, which
essentially follows the locking approach but allows other processes to
help dead ones so that locks are always released.  

The synchronization primitive used is 
\index{load-linked/store-conditional}
\concept{LL/SC}: LL
(\concept{load-linked}) reads a
register and leaves our ID attached to it, SC
(\concept{store-conditional}) writes a register only
if our ID is still attached, and clears any other IDs that might also
be attached.  It's easy to build a 1-register CAS (CAS1) out of this,
though Shavit and Touitou exploit some additional power of LL/SC.

\newData{\STMstatus}{status}
\newData{\STMrec}{rec}
\newData{\STMversion}{version}
\newData{\STMfailure}{failure}
\newData{\STMsuccess}{success}
\newFunc{\LL}{LL}
\newFunc{\SC}{SC}

\subsection{Overlapping LL/SC}
\label{section-overlapping-LL/SC}

The particular trick that gets used in the Shavit-Touitou protocol is
to use two overlapping LL/SC pairs to do a CAS-like update on one
memory location while checking that another memory location hasn't
changed.  The purpose of this is to allow multiple processes to
work on the same transaction (which requires the first CAS to avoid
conflicts with other transactions) while making sure that slow
processes don't cause trouble by trying to complete transactions that
have already finished (the second check).

To see this in action, suppose we have a register $r$ that we want to
do a CAS on, while checking that a second register $\STMstatus$ is
$⊥$ (as opposed to $\STMsuccess$ or $\STMfailure$).  If we execute
the code fragment in Algorithm~\ref{alg-overlapping-LL/SC}, it will
succeed only if nobody writes to $\STMstatus$ between its \LL and \SC
and similarly for $r$; if this occurs, then at the time of $\LL(r)$,
we know that $\STMstatus = ⊥$, and we can linearize the write to
$r$ at this time if we restrict all access to $r$ to go through LL/SC.

\begin{algorithm}
\If{$\LL(\STMstatus) = ⊥$}{
    \If{$\LL(r) = \DataSty{oldValue}$}{
        \If{$\SC(\STMstatus, ⊥) = \True$}{
            $\SC(r, \DataSty{newValue})$\;
        }
    }
}
\caption{Overlapping LL/SC}
\label{alg-overlapping-LL/SC}
\end{algorithm}

\subsection{Representing a transaction}

\newData{\STMaddresses}{addresses}
\newData{\STMoldValues}{oldValues}
\newData{\STMowner}{owner}

Transactions are represented by records $\STMrec$.  Each such record
consists of a \STMstatus component that describes how far the
transaction has gotten (needed to coordinate cooperating processes), a
\STMversion component that distinguishes between versions that may
reuse the same space (and that is used to shut down the transaction
when complete), a \DataSty{stable} component that indicates when the
initialization is complete, an \DataSty{Op} component that describes the RMW to be
performance, an array $\STMaddresses[]$ of pointers to the arguments
to the RMW, and an array $\STMoldValues[]$ of old values at these
addresses (for the R part of the RMW).
These are all initialized by the initiator of the transaction, who
will be the only process working on the transaction until it starts
acquiring locks.

\subsection{Executing a transaction}

Here we give an overview of a transaction execution:
\begin{enumerate}
 \item Initialize the record \STMrec for the
 transaction.  (Only the initiator does this.)
 \item Attempt to acquire ownership of registers in
 $\STMaddresses[]$.  See the \FuncSty{AcquireOwnerships} code in the
 paper for details.  The essential idea is that we want to set the
 field $\STMowner[r]$ for each memory location $r$ that we need to
 lock; this is done using an overlapping LL/SC as described above so
 that we only set $\STMowner[r]$ if (a) $r$ is currently unowned, and
 (b) nothing has happened to $\STMrec.\STMstatus$ or
 $\STMrec.\STMversion$.  Ownership is acquired in order of increasing
 memory address; if we fail to acquire ownership for some $r$, our
 transaction fails.  In case of failure, we set $\STMrec.\STMstatus$
 to \STMfailure and release all the locks we've acquired (checking
 $\STMrec.\STMversion$ in the middle of each LL/SC so we don't release
 locks for a later version using the same record).  If we are the
 initiator of this transaction, we will also go on to attempt to
 complete the transaction that got in our way.
 \item Do a \LL on $\STMrec.\STMstatus$ to see if
 \FuncSty{AcquireOwnerships} succeeded.  If
 so, update the memory, store the old results in \STMoldValues, and release
 the ownerships.  If it failed, release ownership and help the next
 transaction as described above.
\end{enumerate}

Note that only an initiator helps; this avoids a long chain of
helping and limits the cost of each attempted transaction to the cost
of doing two full transactions, while (as shown below) still allowing
some transaction to finish.

\subsection{Proof of linearizability}
Intuition is:

\begin{itemize}
 \item Linearizability follows from the linearizability of the locking protocol: acquiring ownership is equivalent to grabbing a lock, and updates occur only when all registers are locked.
 \item Complications come from (a) two or more processes trying to
 complete the same transaction and (b) some process trying to complete
 an old transaction that has already terminated.  For the first part
 we just make sure that the processes don't interfere with each other,
 e.g. I am happy when trying to acquire a location if somebody else
 acquires it for the same transaction.  For the second part we have to
 check $\STMrec.\STMstatus$ and $\STMrec.\STMversion$ before doing just about anything.
 See the pseudocode in the paper for details on how this is done.
\end{itemize}

\subsection{Proof of non-blockingness}

To show that the protocol is non-blocking we must show that if an
unbounded number of transactions are attempted, one eventually
succeeds.  First observe that in order to fail, a transaction must be
blocked by another transaction that acquired ownership of a
higher-address location than it did; eventually we run out of
higher-address locations, so there is some transaction that doesn't
fail.  Of course, this transaction may not succeed (e.g., if its
initiator dies), but either (a) it blocks some other transaction, and
that transaction's initiator will complete it or die trying, or (b) it
blocks no future transactions.  In the second case we can repeat the
argument for the $n-1$ surviving processes to show that some of them complete transactions, ignoring the stalled transaction from case (b).

\section{Improvements}
One downside of the Shavit and Touitou protocol is that it uses LL/SC
very aggressively (e.g., with overlapping LL/SC operations) and uses
non-trivial (though bounded, if you ignore the ever-increasing version
numbers) amounts of extra space.  Subsequent work has aimed at
knocking these down; for example a paper by Harris, Fraser, and
Pratt~\cite{HarrisFP2002} builds multi-register CAS out of
single-register CAS with $O(1)$ extra bits per register.  The proof of
these later results can be quite involved; Harris~\etal , for example, base their algorithm on an implementation of 2-register CAS whose correctness has been verified only by machine (which may be a plus in some views).

\section{Limitations}

There has been a lot of practical work on STM designed to reduce
overhead on real hardware, but there's still a fair bit of overhead.
On the theory side, a lower bound of Attiya, Hillel, and
Milani~\cite{AttiyaHM2009} shows that any STM system that guarantees
non-interference between non-overlapping RMW transactions has the
undesirable property of making read-only transactions as expensive as
RMW transactions: this
conflicts with the stated goals of many practical STM implementations,
where it is assumed that most transactions will be read-only (and
hopefully cheap).  So there is quite a bit of continuing research on
finding the right trade-offs.

\myChapter{Obstruction-freedom}{2011}{In particular:
§\ref{section-boosting-obstruction-freedom-to-wait-freedom} has not
been updated to include some more recent
results~\cite{AlistarhCS2016,GiakkoupisHHW2013};
and §\ref{section-lock-free-lower-bounds} mostly follows the conference
version~\cite{FichHS2005} of the Ellen-Hendler-Shavit paper and omits
stronger results from the journal version~\cite{EllenHS2012}.}
\label{chapter-obstruction-freedom}

The gold standard for shared-memory objects is \index{wait-free}\concept{wait-freedom}:
I can finish my operation in a bounded number of steps no matter what
anybody else does.  Like the gold standard in real life, this can be
overly constraining.  So researchers have developed several weaker progress guarantees that are nonetheless useful.  The main ones are:
\begin{description}
 \item[Lock-freedom] An implementation is \concept{lock-free} if
 infinitely many operations finish in any infinite execution.  In
 simpler terms, somebody always makes progress, but maybe not you.
 (Also called \concept{non-blocking}.)
 \item[Obstruction-freedom] An implementation is
 \concept{obstruction-free} if, starting from any reachable
 configuration, any process can finish in a bounded number of steps if
 all of the other processes stop.  This definition was proposed in
 2003 by Herlihy, Luchangco, and Moir~\cite{HerlihyLM2003}.  In lower
 bounds (e.g., the Jayanti-Tan-Toueg bound described in
 Chapter~\ref{chapter-JTT}) essentially the same property is often called \concept{solo-terminating}.
\end{description}

Both of these properties exclude traditional lock-based algorithms,
where some process grabs a lock, updates the data structure, and then
release the lock; if this process halts, no more operations finish.
Both properties are also weaker than wait-freedom.  It is not hard to
show that lock-freedom is a stronger condition that
obstruction-freedom: given a lock-free implementation, if we can keep
some single process running forever in isolation, we get an infinite
execution with only finitely many completed operations.  So we have a
hierarchy: wait-free $>$ lock-free $>$ obstruction-free $>$ locking.

\section{Why build obstruction-free algorithms?}

The pitch is similar to the pitch for building locking algorithms: an
obstruction-free algorithm might be simpler to design, implement, and
reason about than a more sophisticated algorithm with stronger
properties.  Unlike locking algorithms, an obstruction-free algorithm
won't fail because some process dies holding the lock; instead, it
fails if more than one process runs the algorithm at the same time.
This possibility may be something we can avoid by building a
\concept{contention manager}, a high-level protocol that detects contention and delays some processes to avoid it (say, using randomized exponential back-off).

\section{Examples}

\subsection{Lock-free implementations}

Pretty much anything built using compare-and-swap or LL/SC ends up being lock-free.  A simple example would be a counter, where an increment operation does
\begin{algorithm}[h]
 $x ← \LL(C)$ \;
 $\SC(C, x+1)$ \;
\end{algorithm}

This is lock-free (the only way to prevent a store-conditional from succeeding is if some other store-conditional succeeds, giving infinitely many successful increments) but not wait-free (I can starve).  It's also obstruction-free, but since it's already lock-free we don't care about that.

\subsection{Double-collect snapshots}

Similarly, suppose we are doing atomic snapshots.  We know that there exist wait-free implementations of atomic snapshots, but they are subtle and confusing.  So we want to do something simpler, and hope that we at least get obstruction-freedom.

If we do double-collects, that is, we have updates just write to a register and have snapshots repeatedly collect until they get two collects in a row with the same values, then any snapshot that finishes is correct (assuming no updaters ever write the same value twice, which we can enforce with nonces).  This isn't wait-free, because we can keep a snapshot going forever by doing a lot of updates.  It \emph{is} lock-free, because we have to keep doing updates to make this happen.

We can make this merely obstruction-free if we work hard (there is no
reason to do this, but it illustrates the difference between
lock-freedom—good—and obstruction-freedom—not so good).  Suppose
that every process keeps a count of how many collects it has done in a
register that is included in other process's collects (but not its
own).  Then two concurrent scans can stall each other forever (the
implementation is not lock-free), but if only one is running it
completes two collects in $O(n)$ operations without seeing any changes (it is obstruction-free).

\subsection{Software transactional memory}

Similar things happen with software transactional memory (see
Chapter~\ref{chapter-STM}).  Suppose that I have an implementation of multiword compare-and-swap, and I want to carry out a transaction.  I read all the values I need, then execute an MCAS operation that only updates if these values have not changed.  The resulting algorithm is lock-free (if my transaction fails, it's because some update succeeded).  If however I am not very clever and allow some values to get written outside of transactions, then I might only be obstruction-free.

\subsection{Obstruction-free test-and-set}
\label{section-obstruction-free-test-and-set}

Algorithm~\ref{alg-obstruction-free-TAS} gives an implementation of
$2$-process test-and-set from atomic registers that is
obstruction-free; this demonstrates that obstruction-freedom lets us
evade the wait-free impossibility results implied by the consensus
hierarchy (\cite{Herlihy1991waitfree}, discussed in
Chapter~\ref{chapter-wait-free-hierarchy}).

The basic idea
goes back to the \concept{racing counters} technique used in consensus
protocols starting with Chor, Israeli, and Li~\cite{ChorIL1994}, and
there is some similarity to a classic randomized wait-free
test-and-set due to Tromp
and Vitányi~\cite{TrompV2002}.  Each process keeps a position $x$ in
memory that it also stores from time to time in its register $a[i]$.
If a process gets $2$ steps ahead of the other process (as observed by
comparing $x$ to $a[i-1]$, it wins the test-and-set; if a process
falls one or more steps behind, it (eventually) loses.  To keep space down and
guarantee termination in bounded time, all values are tracked modulo
$5$.

\begin{algorithm}
$x ← 0$ \\
\While{\True}{
    $δ ← x - a[1-i]$\;
    \uIf{$δ = 2 \pmod{5}$}{
        \Return $0$\;
    }
    \uElseIf{$δ = -1 \pmod{5}$}{
        \Return $1$\;
    }
    \Else{
        $x ← (x+1) \bmod{5}$ \;
        $a[i] ← x$\;
    }
}
\caption{Obstruction-free $2$-process test-and-set}
\label{alg-obstruction-free-TAS}
\end{algorithm}

Why this works: observe that whenever a process computes $δ$, $x$
is equal to $a[i]$; so $δ$ is always an instantaneous snapshot of
$a[i] - a[1-i]$.  If I observe $δ = 2$ and return $0$, your next
read will either show you $δ = -2$ or $δ = -1$ (depending on
whether you increment $a[1-i]$ after my read).  In the latter case,
you return $1$ immediately; in the former, you return after one more
increment (and more importantly, you can't return $0$).
Alternatively, if I ever observe $δ = -1$, your next read will
show you either $δ = 1$ or $δ = 2$; in either case, you will
eventually return $0$.
(We chose $5$ as a modulus because this is the smallest value that
makes the cases $δ = 2$ and $δ = -2$ distinguishable.)

We can even show that this is linearizable, by considering a solo
execution in which the lone process takes two steps and returns $0$
(with two processes, solo executions are the only interesting case for linearizability).

However, Algorithm~\ref{alg-obstruction-free-TAS} is not 
wait-free or even lock-free: if both processes run
in lockstep, they will see $δ = 0$ forever.  But it is
obstruction-free.  If I run by myself, then whatever value of $δ$
I start with, I will see $-1$ or $2$ after at most $6$
operations.\footnote{The worst case is where an increment by my fellow
process leaves $δ = -1$ just before
my increment.}

This gives an 
\index{complexity!obstruction-free step}
\index{step complexity!obstruction-free}
\concept{obstruction-free step complexity} of $6$, where the
obstruction-free step complexity is defined as the maximum number of
operations any process can take after all other processes stop.
Note that our usual wait-free measures of step complexity don't make a
lot of sense for obstruction-free algorithms, as we can expect a
sufficiently cruel adversary to be able to run them up to whatever
value he likes.

Building a tree of these objects as in
§\ref{section-test-and-set-from-two-process-consensus} gives
$n$-process test-and-set with obstruction-free step complexity
$O(\log n)$.

\subsection{An obstruction-free deque}

(We probably aren't going to do this in class.)

So far we don't have any good examples of why we would want to be
obstruction-free if our algorithm is based on CAS.  So let's describe
the case Herlihy~\etal{} suggested.

\newFunc{\OFrightPush}{rightPush}
\newFunc{\OFrightPop}{rightPop}
\newFunc{\OFleftPush}{leftPush}
\newFunc{\OFleftPop}{leftPop}
\newFunc{\OForacle}{oracle}
\newData{\OFtop}{top}
\newData{\OFright}{right}
\newData{\OFprev}{prev}
\newData{\OFcur}{cur}
\newData{\OFnext}{next}
\newData{\OFvalue}{value}
\newData{\OFversion}{version}
\newData{\OFRN}{RN}
\newData{\OFLN}{LN}

A \concept{deque} is a generalized queue that supports push and pop at both ends (thus it can be used as either a queue or a stack, or both).  A classic problem in shared-memory objects is to build a deque where operations at one end of the deque don't interfere with operations at the other end.  While there exist lock-free implementation with this property, there is a particularly simple implementation using CAS that is only obstruction-free.

Here's the idea: we represent the deque as an infinitely-long array of
compare-and-swap registers (this is a simplification from the paper,
which gives a bounded implementation of a bounded deque).  The middle
of the deque holds the actual contents.  To the right of this region
is an infinite sequence of \index{null!right}\concept{right null} (\OFRN) values, which are
assumed never to appear as a pushed value.  To the left is a similar
infinite sequence of \index{null!left}\concept{left null} (\OFLN)
values.  Some magical external mechanism (called an \concept{oracle} in the paper) allows processes to quickly find the first null value at either end of the non-null region; the correctness of the protocol does not depend on the properties of the oracle, except that it has to point to the right place at least some of the time in a solo execution.  We also assume that each cell holds a version number whose only purpose is to detect when somebody has fiddled with the cell while we aren't looking (if we use LL/SC, we can drop this).

Code for \OFrightPush and \OFrightPop is given in
Algorithm~\ref{alg-obstruction-free-deque} (the code for \OFleftPush
and \OFleftPop is symmetric).

\begin{algorithm}
\Procedure{$\OFrightPush(v)$}{
    \While{\True}{
        $k ← \OForacle(\OFright)$ \;
        $\OFprev ← a[k-1]$ \;
        $\OFnext ← a[k]$ \;
        \If{$\OFprev.\OFvalue \ne \OFRN$ \KwAnd $\OFnext.\OFvalue = \OFRN$}{
            \If{$\CAS(a[k-1], \OFprev, [\OFprev.\OFvalue, \OFprev.\OFversion+1])$}{
                \If{$\CAS(a[k], \OFnext, [v, \OFnext.\OFversion+1])$}{
                    we win, go home\;
                }
            }
        }
    }
}
\bigskip
\Procedure{$\OFrightPop()$}{
    \While{\True}{
        $k ← \OForacle(\OFright)$ \;
        $\OFcur ← a[k-1]$ \;
        $\OFnext ← a[k]$ \;
        \If{$\OFcur.\OFvalue \ne \OFRN$ \KwAnd $\OFnext.\OFvalue = \OFRN$}{
            \uIf{$\OFcur.\OFvalue = \OFLN$ \KwAnd $A[k-1] = \OFcur$}{
                \Return \DataSty{empty}\;
            }
            \ElseIf{$\CAS(a[k], \OFnext, [\OFRN, \OFnext.\OFversion+1])$}{
                \If{$\CAS(a[k-1], \OFcur, [\OFRN, \OFcur.\OFversion+1])$}{
                    \Return $\OFcur.\OFvalue$\;
                }
            }
        }
    }
}
\caption{Obstruction-free deque}
\label{alg-obstruction-free-deque}
\end{algorithm}

It's easy to see that in a solo execution, if the oracle doesn't lie,
either operation finishes and returns a plausible value after $O(1)$ operations.  So the implementation is obstruction-free.  But is it also correct?

To show that it is, we need to show that any execution leaves the deque in a sane state, in particular that it preserves the invariant that the deque consists of left-nulls followed by zero or more values followed by right-nulls, and that the sequence of values in the queue is what it should be.  

This requires a detailed case analysis of which operations interfere
with each other, which can be found in the original paper.  But we can
give some intuition here.  The two CAS operations in \OFrightPush or
\OFrightPop succeed only if neither register was modified between the
preceding read and the CAS.  If both registers are unmodified at the
time of the second CAS, then the two CAS operations act like a single
two-word CAS, which replaces the previous values $(\OFtop, \OFRN)$
with $(\OFtop, \OFvalue)$ in \OFrightPush or $(\OFtop, \OFvalue)$ with
$(\OFtop, \OFRN)$ in \OFrightPop; in
either case the operation preserves the invariant.  So the only way we
get into trouble is if, for example, a \OFrightPush does a CAS on
$a[k-1]$ (verifying that it is unmodified and incrementing the version
number), but then some other operation changes $a[k-1]$ before the CAS
on $a[k]$.  If this other operation is also a \OFrightPush, we are
happy, because it must have the same value for $k$ (otherwise it would
have failed when it saw a non-null in $a[k-1])$, and only one of the
two right-pushes will succeed in applying the CAS to $a[k]$.  If the
other operation is a \OFrightPop, then it can only change $a[k-1]$ after
updating $a[k]$; but in this case the update to $a[k]$ prevents the
original right-push from changing $a[k]$.  With some more tedious
effort we can similarly show that any interference from \OFleftPush
or \OFleftPop either causes the interfering operation or the original operation to fail.  This covers 4 of the 16 cases we need to consider.  The remaining cases will be brushed under the carpet to avoid further suffering.

\section{Boosting obstruction-freedom to wait-freedom}
\label{section-boosting-obstruction-freedom-to-wait-freedom}

Naturally, having an obstruction-free implementation of some object is
not very helpful if we can't guarantee that some process eventually
gets its unobstructed solo execution.  In general, we can't expect to
be able to do this without additional assumptions; for example, if we
could, we could solve consensus using a long sequence of adopt-commit
objects
with no randomization at all.\footnote{This fact was observed by
Herlihy~\etal~\cite{HerlihyLM2003} in their original obstruction-free paper; it also
implies that there exists a universal obstruction-free implementation
of anything based on Herlihy's universal construction.}  So we need to make some sort of assumption about timing, or find somebody else who has already figured out the right assumption to make.

Those somebodies turn out to be Faith Ellen Fich, Victor Luchangco,
Mark Moir, and Nir Shavit, who give an algorithm for boosting obstruction-freedom
to wait-freedom~\cite{FichLMS2005}.
The timing assumption is 
\index{semisynchrony!unknown-bound}\concept{unknown-bound
semisynchrony}, which means that in any execution there is some
maximum ratio $R$ between the shortest and longest time interval
between any two consecutive steps of the same non-faulty process, but
the processes don't know what this ratio is.\footnote{This is a much
older model, which goes back to a famous paper of Dwork, Lynch, and
Stockmeyer~\cite{DworkLS1988}.}  In particular, if I can execute more
than $R$ steps without you doing anything, I can reasonably conclude
that you are dead—the semisynchrony assumption thus acts as a 
failure detector.

The fact that $R$ is unknown might seem to be an impediment to using
this failure detector, but we can get around this.  The idea is to
start with a small guess for $R$; if a process is suspected but then
wakes up again, we increment the guess.  Eventually, the guessed value
is larger than the correct value, so no live process will be falsely
suspected after this point.  Formally, this gives an 
eventually perfect ($◇P$) failure detector, although the
algorithm does not specifically use the failure detector abstraction.

To arrange for a solo execution, when a process detects a conflict (because
its operation didn't finish quickly), it enters into a ``panic
mode'' where processes take turns trying to finish unmolested.  A
fetch-and-increment register is used as a timestamp generator, and only the
process with the smallest timestamp gets to proceed.  However, if this
process is too sluggish, other processes may give up and overwrite its
low timestamp with $\infty{}$, temporarily ending its turn.  If the sluggish process is in fact alive, it can restore its low timestamp and kill everybody else, allowing it to make progress until some other process declares it dead again.

The simulation works because eventually the mechanism for detecting dead processes stops suspecting live ones (using the technique described above), so the live process with the winning timestamp finishes its operation without interference.  This allows the next process to proceed, and eventually all live processes complete any operation they start, giving the wait-free property.

The actual code is in Algorithm~\ref{alg-obstruction-free-boosting}.  It's a rather long algorithm but most of the details are just bookkeeping.

\newData{\OFop}{op}
\newData{\OFpanic}{PANIC}
\newData{\OFmyTimestamp}{myTimestamp}
\newData{\OFminTimestamp}{minTimestamp}
\newData{\OFotherTimestamp}{otherTimestamp}
\newData{\OFwinnerTimestamp}{winnerTimestamp}
\newData{\OFwinner}{winner}

\begin{algorithm}
\If{$¬\OFpanic$}{
    execute up to $B$ steps of the underlying algorithm \;
    \lIf{we are done}{
        \Return
    }
}
$\OFpanic ← \True$   \tcp{enter panic mode}
$\OFmyTimestamp ← \FuncSty{fetchAndIncrement}()$ \;
$A[i] ← 1$  \tcp{reset my activity counter}
\While{\True}{
    $T[i] ← \OFmyTimestamp$ \;
    $\OFminTimestamp ← \OFmyTimestamp$;
    $\OFwinner ← i$ \;
    \For{$j ← 1 \dots n, j≠i$}{
        $\OFotherTimestamp ← T[j]$ \;
        \uIf{$\OFotherTimestamp < \OFminTimestamp$}{
            $T[\OFwinner] ← \infty$  \tcp{not looking so winning any more}
            $\OFminTimestamp ← \OFotherTimestamp$;
            $\OFwinner ← j$\;
        }
        \ElseIf{$\OFotherTimestamp < \infty$}{
             $T[j] ← \infty$\;
        }
    }
    \If{$i = \OFwinner$}{
        \Repeat{$T[i] = \infty{}$}{
            execute up to $B$ steps of the underlying algorithm \;
            \eIf{we are done}{
                $T[i] ← \infty{}$ \;
                $\OFpanic ← \False$ \;
                \Return\;
            }{
                $A[i] ← A[i] + 1$ \;
                $\OFpanic ← \True$\;
            }
        }
    }{
        \Repeat{$a = A[\OFwinner]$ \KwOr $\OFwinnerTimestamp \ne \OFminTimestamp$}{
            $a ← A[\OFwinner]$\;
            wait $a$ steps \;
            $\OFwinnerTimestamp ← T[\OFwinner]$\;
        }
        \If{\OFwinnerTimestamp = \OFminTimestamp}{
            $T[\OFwinner] ← \infty{}$ \tcp{kill winner for inactivity}
        }
    }
}
\caption{Obstruction-freedom booster from~\cite{FichLMS2005}}
\label{alg-obstruction-free-boosting}
\end{algorithm}

The preamble before entering \OFpanic mode is a fast-path computation
that allows a process that actually is running in isolation to skip
testing any timestamps or doing any extra work (except for the one
register read of \OFpanic).  The assumption is that the constant $B$ is
set high enough that any process generally will finish its operation
in $B$ steps without interference.  If there is interference, then the timestamp-based mechanism kicks in: we grab a timestamp out of the convenient fetch-and-add register and start slugging it out with the other processes.

(A side note: while the algorithm as presented in the paper assumes a
fetch-and-add register, any timestamp generator that delivers
increasing values over time will work.  So if we want to limit
ourselves to atomic registers, we could generate timestamps by taking snapshots of previous timestamps, adding 1, and appending process IDs for tie-breaking.)

Once I have a timestamp, I try to knock all the higher-timestamp
processes out of the way (by writing $\infty{}$ to their timestamp
registers).  If I see a smaller timestamp than my own, I'll drop out
myself ($T[i] ← \infty{}$), and fight on behalf of its
owner instead.  At the end of the $j$ loop, either I've decided I am
the winner, in which case I try to finish my operation (periodically
checking $T[i]$ to see if I've been booted), or I've decided somebody
else is the winner, in which case I watch them closely and try to shut
them down if they are too slow ($T[\OFwinner] ← \infty{}$).  I
detect slow processes by inactivity in $A[\OFwinner]$; similarly, I
signal my own activity by incrementing $A[i]$.  The value in $A[i]$ is
also used as an increasing guess for the time between increments of
$A[i]$; eventually this exceeds the $R(B+O(1))$ operations that I execute between incrementing it.

We still need to prove that this all works.  The essential idea is to
show that whatever process has the lowest timestamp finishes in a
bounded number of steps.  To do so, we need to show that other
processes won't be fighting it in the underlying algorithm.  Call a
process \emph{active} if it is in the loop guarded by the 
``if $i = \OFwinner$'' statement.  Lemma 1 from the paper states:
\begin{lemma}[\protect{\cite[Lemma 1]{FichLMS2005}}]
 If processes $i$ and $j$ are both active, then $T[i] =
 \infty{}$ or $T[j] = \infty{}$.
\end{lemma}
\begin{proof}
Assume without loss of generality that $i$ last set $T[i]$ to
\OFmyTimestamp in the main loop after $j$ last set $T[j]$.  In order to
reach the active loop, $i$ must read $T[j]$.  Either $T[j] = \infty{}$
at this time (and we are done, since only $j$ can set $T[j] <
\infty{}$), or $T[j]$ is greater than $i$'s timestamp (or else $i$
wouldn't think it's the winner).  In the second case, $i$ sets $T[j] =
\infty{}$ before entering the active loop, and again the claim holds.
\end{proof}

The next step is to show that if there is some process $i$ with a
minimum timestamp that executes infinitely many operations, it
increments $A[i]$ infinitely often (thus eventually making the failure detector stop suspecting it).  This gives us Lemma 2 from the paper:
\begin{lemma}[\protect{\cite[Lemma 2]{FichLMS2005}}]
\label{lemma-OF-booster-2}
  Consider the set of all processes that execute
 infinitely many operations without completing an operation.  Suppose
 this set is non-empty, and let $i$ hold the minimum timestamp of all
 these processes.  Then $i$ is not active infinitely often.
\end{lemma}
\begin{proof}
Suppose that from some time on, $i$ is active forever, i.e., it never
leaves the active loop.  Then $T[i] < \infty$ throughout this interval
(or else $i$ leaves the loop), so for any active $j$, $T[j] = \infty$
by the preceding lemma.  It follows that any active $T[j]$ leaves the
active loop after $B+O(1)$ steps of $j$ (and thus at most $R(B+O(1))$
steps of $i$).  Can $j$ re-enter?  If $j$'s timestamp is less than
$i$'s, then $j$ will set $T[i] = \infty$, contradicting our
assumption.  But if $j$'s timestamp is greater than $i$'s, $j$ will
not decide it's the winner and will not re-enter the active loop.  So
now we have $i$ alone in the active loop.  It may still be fighting
with processes in the initial fast path, but since $i$ sets \OFpanic
every time it goes through the loop, and no other process resets \OFpanic
(since no other process is active), no process enters the fast path
after some bounded number of $i$'s steps, and every process in the
fast path leaves after at most $R(B+O(1))$ of $i$'s steps.  So
eventually $i$ is in the loop alone forever—and obstruction-freedom
means that it finishes its operation and leaves.  This contradicts our
initial assumption that $i$ is active forever.
\end{proof}

So now we want to argue that our previous assumption that there exists
a bad process that runs forever without winning leads to a
contradiction, by showing that the particular $i$ from
Lemma~\ref{lemma-OF-booster-2}
actually finishes (note that Lemma~\ref{lemma-OF-booster-2} doesn't quite do this—we only
show that $i$ finishes if it stays active long enough, but maybe it doesn't stay active).  

Suppose $i$ is as in Lemma~\ref{lemma-OF-booster-2}.  Then $i$
leaves the active loop infinitely often.  So in particular it
increments $A[i]$ infinitely often.  After some finite number of
steps, $A[i]$ exceeds the limit $R(B+O(1))$ on how many steps some
other process can take between increments of $A[i]$.  For each other
process $j$, either $j$ has a lower timestamp than $i$, and thus
finishes in a finite number of steps (from the premise of the choice
of $i$), or $j$ has a higher timestamp than $i$.  Once we have cleared
out all the lower-timestamp processes, we follow the same logic as in
the proof of Lemma~\ref{lemma-OF-booster-2} to show that eventually (a) $i$ sets $T[i] <
\infty$ and \OFpanic = true, (b) each remaining $j$ observes $T[i] <
\infty$ and \OFpanic = true and reaches the waiting loop, (c) all such
$j$ wait long enough (since $A[i]$ is now very big) that $i$ can
finish its operation.  This contradicts the assumption that $i$ never finishes the operation and completes the proof.

\subsection{Cost}

If the parameters are badly tuned, the potential cost of this
construction is quite bad.  For example, the slow increment process
for $A[i]$ means that the time a process spends in the active loop
even after it has defeated all other processes can be as much as the
square of the time it would normally take to complete an operation
alone—and every other process may pay $R$ times this cost waiting.
This can be mitigated to some extent by setting $B$ high enough that a
winning process is likely to finish in its first unmolested pass
through the loop (recall that it doesn't detect that the other
processes have reset $T[i]$ until after it makes its attempt to
finish).  An alternative might be to double $A[i]$ instead of
incrementing it at each pass through the loop.  However, it is worth
noting (as the authors do in the paper) that nothing prevents the
underlying algorithm from incorporating its own 
\concept{contention management}
scheme to ensure that most operations complete in $B$ steps and \OFpanic mode is rarely entered.  So we can think of the real function of the construction as serving as a backstop to some more efficient heuristic approach that doesn't necessarily guarantee wait-free behavior in the worst case.

\section{Lower bounds for lock-free protocols}
\label{section-lock-free-lower-bounds}

So far we have seen that obstruction-freedom buys us an escape from
the impossibility results that plague wait-free constructions, while
still allowing practical implementations of useful objects under
plausible timing assumptions.  Yet all is not perfect: it is still
possible to show non-trivial lower bounds on the costs of these
implementations in the right model.  We will present one of these
lower bounds, the linear-contention lower bound of Ellen, Hendler, and
Shavit~\cite{EllenHS2012}.\footnote{The result first appeared in FOCS
    in 2005~\cite{FichHS2005}, with a small but easily fixed 
    bug in the definition of the class of objects the proof applies
to.  We'll use the corrected definition from the journal version.}
First we have to define what is meant by contention.

\subsection{Contention}
A limitation of real shared-memory systems is that physics generally
won't permit more than one process to do something useful to a shared
object at a time.  This limitation is often ignored in computing the
complexity of a shared-memory distributed algorithm (and one can make
arguments for ignoring it in systems where communication costs
dominate update costs in the shared-memory implementation), but it is
useful to recognize it if we can't prove lower bounds otherwise.
Complexity measures that take the cost of simultaneous access into
account go by the name of \concept{contention}.

The particular notion of contention used in the Ellen~\etal{} paper is an
adaptation of the contention measure of Dwork, Herlihy, and
Waarts~\cite{DworkHW1997}.  The idea is that if I access some shared object, I pay a price in 
\index{stall}\indexConcept{memory stall}{memory stalls} for all the
other processes that are trying to access it at the same time but got
in first.  In the original definition, given an execution
of the form
$A\phi_{1}\phi_{2}\dots{}\phi_{k}\phi{}A'$, where all operations
$\phi_{i}$ are applied to the same object as $\phi{}$, and the last
operation in $A$ is not, then $\phi_{k}$ incurs $k$ memory stalls.
Ellen~\etal{} modify this to only count
sequences of \emph{non-trivial} operations, where an operation is
non-trivial if it changes the state of the object in some states
(e.g., writes, increments, compare-and-swap—but not reads).  Note that this change only strengthens the bound they eventually prove, which shows that in the worst case, obstruction-free implementations of operations on objects in a certain class incur a linear number of memory stalls (possibly spread across multiple base objects).

\subsection{The class \texorpdfstring{$G$}{G}}

The Ellen~\etal{} bound is designed to be as general as possible, so the
authors define a class $G$ of objects to which it applies.  As is
often the case in mathematics, the underlying meaning of $G$ is ``a reasonably large class of objects for which this particular proof works,'' but the formal definition is given in terms of when certain operations of the implemented object are affected by the presence or absence of other operations—or in other words, when those other operations need to act on some base object in order to let later operations know they occurred.

\newData{\FHSop}{Op}

An object is in \concept{class $G$} if it has some operation \FHSop and
initial state $s$ such that for any two processes $p$ and $q$ and
every sequence of operations $A\phi{}A'$, where
\begin{enumerate}
\item  $\phi{}$ is an
instance of \FHSop executed by $p$, 
\item no operation in $A$ or $A'$ is
executed by $p$, 
\item no operation in $A'$ is executed by
$q$, and 
\item
no two operations in $A'$ are executed by the same process; 
\end{enumerate}
then there
exists a sequence of operations $Q$ by $q$ such that for every
sequence $H\phi{}H'$ where 
\begin{enumerate}
\item $HH'$ is an interleaving of $Q$ and the
sequences $AA'|r$ for each process $r$, 
\item
$H'$ contains no operations of $q$, and 
\item no two operations in $H'$
are executed by the same process; 
\end{enumerate}
then the return value of $\phi$ to
$p$ changes depending on whether it occurs after $A\phi$ or $H\phi$.

This is where ``makes the proof work'' starts looking like a much
simpler definition.  The intuition is that deep in the guts of the
proof, we are going to be injecting some operations of $q$ into an
existing execution (hence adding $Q$), and we want to do it in a way
that forces $q$ to operate on some object that $p$ is looking at
(hence the need for $A\phi{}$ to return a different value from 
$H\phi{})$, without breaking anything else that is going on (all the rest of the conditions).  The reason for pulling all of these conditions out of the proof into a separate definition is that we also want to be able to show that particular classes of real objects satisfy the conditions required by the proof, without having to put a lot of special cases into the proof itself.
\begin{lemma}
A mod-$m$ fetch-and-increment object, with $m ≥ n$, is in $G$.
\end{lemma}
\begin{proof}
This is a classic proof-by-unpacking-the-definition.  Pick some
execution $A\phi{}A'$ satisfying all the conditions, and let $a$ be
the number of fetch-and-increments in $A$ and $a'$ the number in $A'$.
Note $a' ≤ n-2$, since all operations in $A'$ are by different processes.

Now let $Q$ be a sequence of $n-a'-1$ fetch-and-increments by $q$, and
let $HH'$ be an interleaving of $Q$ and the sequences $AA'|r$ for each
$r$, where $H'$ includes no two operation of the same process and no
operations at all of $q$.  Let $h$, $h'$ be the number of
fetch-and-increments in $H$, $H'$, respectively.  Then $h+h' =
a+a'+(n-a'-1) = n+a-1$ and $h' ≤ n-2$ (since $H'$ contains at most
one fetch-and-increment for each process other than $p$ and $q$).
This gives $h ≥ (n+a+1)-(n-2) = a+1$ and $h ≤ n+a-1$, and the
return value of $\phi{}$ after $H\phi{}$ is somewhere in this range
mod $m$.  But none of these values is equal to $a$ mod $m$ (that's why
we specified $m ≥ n$, although as it turns out $m ≥ n-1$ would
have been enough), so we get a different return value from $H\phi{}$
than from $A\phi{}$.
\end{proof}

As a corollary, we also get stock fetch-and-increment registers, since
we can build mod-$m$ registers from them by taking the results mod
$m$.

A second class of class-$G$ objects is obtained from snapshot:
\begin{lemma}
Single-writer snapshot objects are in $G$.\footnote{For the purposes
of this lemma, ``single-writer'' means that each segment can be
written to by only one process, not that there is only one process
that can execute update operations.}
\end{lemma}
\begin{proof}
Let $A\phi{}A'$ be as in the definition, where $\phi{}$ is a scan
operation.  Let $Q$ consist of a single update operation by $q$ that
changes its segment.  Then in the interleaved sequence $HH'$, this
update doesn't appear in $H'$ (it's forbidden), so it must be in $H$.
Nobody can overwrite the result of the update (single-writer!), so it
follows that $H\phi$ returns a different snapshot from $A\phi$.
\end{proof}

\subsection{The lower bound proof}

\begin{theorem}[\protect{\cite[Theorem 5.2]{EllenHS2012}}]
For any obstruction-free implementation of some object in class $G$
from RMW base objects, there is an execution in which some operation
incurs $n-1$ stalls.
\end{theorem}

We can't do better than $n-1$, because it is easy to come up with
implementations of counters (for example) that incur at most $n-1$
stalls.  Curiously, we can even spread the stalls out in a fairly
arbitrary way over multiple objects, while still incurring at most
$n-1$ stalls.  For example, a counter implemented using a single
counter (which is a RMW object) gets exactly $n-1$ stalls if $n-1$
processes try to increment it at the same time, delaying the remaining
process.  At the other extreme, a counter implemented by doing a
collect over $n-1$ single-writer registers (also RMW objects) gets at
least $n-1$ stalls—distributed as one per register—if each
register has a write delivered to it while the reader waiting to read
it during its collect.  So we have to allow for the possibility that
stalls are concentrated or scattered or something in between, as long
as the total number adds up at least $n-1$.

The proof supposes that the theorem is not true and then shows how to
boost an execution with a maximum number $k < n-1$ stalls to an
execution with $k+1$ stalls, giving a contradiction.  (Alternatively,
we can read the proof as giving a mechanism for generating an $(n-1)$-stall execution by repeated boosting, starting from the empty execution.)

This is pretty much the usual trick: we assume that there is a class of bad executions, then look for an extreme member of this class, and show that it isn't as extreme as we thought.  In doing so, we can restrict our attention to particularly convenient bad executions, so long as the existence of some bad execution implies the existence of a convenient bad execution.

\newcommand{\FHSobj}{\mathcal{O}}

Formally, the authors define a \emph{$k$-stall execution} for process
$p$ as an execution $Eσ_{1}\dots{}σ_{i}$ where $E$ and
$σ_{i}$ are sequence of operations such that:
\begin{enumerate}
 \item $p$ does nothing in $E$,
 \item Sets of processes $S_{j}$, $j = 1\dots i$, whose union
 $S=\bigcup_{j=1}^{i} S_{j}$ has size $k$, are each covering objects $\FHSobj_{j}$
 after $E$ with pending non-trivial operations,
 \item Each $σ_{j}$ consists of $p$ applying events by itself
 until it is about to apply an event to $\FHSobj_{j}$, after which each
 process in $S_{j}$ accesses $\FHSobj_{j}$, after which $p$ accesses
 $\FHSobj_{j}$.
 \item All processes not in $S$ are idle after $E$,
 \item $p$ starts at most one operation of the implemented object in
 $σ_{1}\dots{}σ_{i}$, and
 \item In every extension of $E$ in which $p$ and the processes in $S$
 don't take steps, no process applies a non-trivial event to any base
 object accessed in $σ_{1}\dots{}σ_{i}$.  (We will call
 this the \concept{weird condition} below.)
\end{enumerate}

So this definition includes both the fact that $p$ incurs $k$ stalls
and some other technical details that make the proof go through.  The
fact that $p$ incurs $k$ stalls follows from observing that it incurs
$\card*{S_{j}}$ stalls in each segment $σ_{j}$, since all
processes in $S_{j}$ access $\FHSobj_{j}$ just before $p$ does.

Note that the empty execution is a
$0$-stall execution (with $i=0$) by the definition.
This shows that a $k$-stall execution exists for some $k$.

Note also that the weird condition is pretty strong: it claims not
only that there are no non-trivial operation on $\FHSobj{1} \dots
\FHSobj{i}$ in $\tau$, but also that there are no non-trivial
operations on \emph{any} objects accessed in $σ_1\dots
σ_{i}$, which may include many more objects accessed by
$p$.\footnote{\label{footnote-screwup-2011-11-14}And here is where I
screwed up in class on 2011-11-14, by writing the condition as the
weaker requirement that nobody touches $\FHSobj_1 \dots \FHSobj_i$.}

We'll now show that if a $k$-stall execution exists, for $k ≤ n-2$,
then a $(k+k')$-stall execution exists for some $k' > 0$.  Iterating
this process eventually produces an $(n-1)$-stall execution.

Start with some $k$-stall execution
$Eσ_{1}\dots{}σ_{i}$.  Extend this execution by a
sequence of operations $σ{}$ in which $p$ runs in isolation until
it finishes its operation $\phi{}$ (which it may start in $σ{}$
if it hasn't done so already), then each process in $S$ runs in
isolation until it completes its operation.  Now linearize the
high-level operations completed in
$Eσ_{1}\dots{}σ_{i}σ{}$ and factor them as
$A\phi{}A'$ as in the definition of class $G$.

Let $q$ be some process not equal to $p$ or contained in any $S_{j}$
(this is where we use the assumption $k ≤ n-2$).
Then there is some sequence of high-level operations $Q$ of $q$ such
that $H\phi{}$ does not return the same value as $A\phi{}$ for any
interleaving $HH'$ of $Q$ with the sequences of operations in $AA'$
satisfying the conditions in the definition.  We want to use this fact
to shove at least one more memory stall into
$Eσ_{1}\dots{}σ_{i}σ{}$, without breaking any of
the other conditions that would make the resulting execution a
$(k+k')$-stall execution.

Consider the extension $\tau{}$ of $E$ where $q$ runs alone until it
finishes every operation in $Q$.  Then $\tau{}$ applies no nontrivial
events to any base object accessed in
$σ_{1}\dots{}σ_{k}$, (from the weird condition on
$k$-stall executions) and the value of each of these base objects is
the same after $E$ and $E\tau{}$, and thus is also the same after
$Eσ_{1}\dots{}σ_{k}$ and
$E\tau{}σ_{1}\dots{}σ_{k}$.

Now let $σ'$ be the extension of
$E\tau{}σ_{1}\dots{}σ_{k}$ defined analogously to
$σ{}$: $p$ finishes, then each process in each $S_{j}$ finishes.
Let $H\phi{}H'$ factor the linearization of
$E\tau{}σ_{1}\dots{}σ_{i}σ'$.  Observe that $HH'$
is an interleaving of $Q$ and the high-level operations in $AA'$, that
$H'$ contains no operations by $q$ (they all finished in $\tau{}$,
before $\phi{}$ started), and that $H'$ contains no two operations by
the same process (no new high-level operations start after $\phi{}$
finishes, so there is at most one pending operation per process in $S$
that can be linearized after $\phi$).

Now observe that $q$ does some non-trivial operation in $\tau$ to some
base object accessed by $p$ in $σ$.  If not, then $p$ sees the
same responses in $σ'$ and in $σ$, and returns the same
value, contradicting the definition of class $G$.

So does $q$'s operation in $\tau$ cause a stall in $σ$?  Not
necessarily: there may be other operations in between.  Instead, we'll
use the existence of $q$'s operation to demonstrate the existence of
at least one operation, possibly by some other process we haven't even
encountered yet, that does cause a stall.  We do this by considering
the set $F$ of all finite extensions of $E$ that are free of $p$ and
$S$ operations, and look for an operation that stalls $p$ somewhere in this infinitely large haystack.

Let $\FHSobj_{i+1}$ be the first base object accessed by $p$ in $σ{}$
that is also accessed by some non-trivial event in some sequence in
$F$.  We will show two things: first, that $\FHSobj_{i+1}$ exists, and
second, that $\FHSobj_{i+1}$ is distinct from the objects
$\FHSobj_{1}\dots{}\FHSobj_{i}$.  The first part follows from the fact that
$\tau{}$ is in $F$, and we have just shown that $\tau{}$ contains a
non-trivial operation (by $q$) on a base object accessed by $p$ in
$σ$.  For the second part, we use the weird condition on
$k$-stall executions again: since every extension of $E$ in $F$ is
$(\{p\}\cup S)$-free, no process applies a non-trivial event to any
base object accessed in $σ_{1}\dotsσ_{i}$, which includes
all the objects $\FHSobj_{1}\dots \FHSobj_{i}$.

You've probably guessed that we are going to put our stalls in on
$\FHSobj_{i+1}$.  We choose some extension $X$ from $F$ that maximizes the
number of processes with simultaneous pending non-trivial operations
on $\FHSobj_{i+1}$ (we'll call this set of processes $S_{i+1}$ and let
$\card*{S_{i+1}}$ be the number $k'>0$ we've been waiting for), and let
$E'$ be the minimum prefix of $X$ such that these pending operations
are still pending after $EE'$.

We now look at the properties of $EE'$.  We have:
\begin{itemize}
 \item $EE'$ is $p$-free (follows from $E$ being $p$-free and
 $E'\in{}F$, since everything in $F$ is $p$-free).
 \item Each process in $S_{j}$ has a pending operation on $\FHSobj_{j}$
 after $EE'$ (it did after $E$, and didn't do anything in $E'$).
\end{itemize}

This means that we can construct an execution
$EE'σ_{1}\dots{}σ_{i}σ_{i+1}$ that includes
$k+k'$ memory stalls, by sending in the same sequences
$σ_{1}\dots{}σ_{i}$ as before, then appending a new
sequence of events where (a) $p$ does all of its operations in
$σ{}$ up to its first operation on $\FHSobj_{i+1}$; then (b) all the
processes in the set $S_{i+1}$ of processes with pending events on
$\FHSobj_{i+1}$ execute their pending events on $\FHSobj_{i+1}$; then (c) $p$ does
its first access to $\FHSobj_{i+1}$ from $σ{}$.  Note that in addition
to giving us $k+k'$ memory stalls, $σ_{i+1}$ also has the right
structure for a $(k+k')$-stall execution.  But there is one thing missing: we have to show that the weird condition on further extensions still holds.

Specifically, letting $S' = S\cup{}S_{i+1}$, we need to show that any
$(\{p\}\cup{}S')$-free extension $α$ of $EE'$ includes a
non-trivial access to a base object accessed in
$σ_{1}\dots{}σ_{i+1}$.  Observe first that since $α{}$
is $(\{p\}\cup{}S')$-free, then $E'α{}$ is $(\{p\}\cup{}S)$-free,
and so it's in $F$: so by the weird condition on
$Eσ_{1}\dots{}σ_{i}$, $E'α{}$ doesn't have any
non-trivial accesses to any object with a non-trivial access in
$σ_{1}\dots{}σ_{i}$.  So we only need to squint very closely
at $σ_{i+1}$ to make sure it doesn't get any objects in there either.

Recall that $σ_{i+1}$ consists of (a) a sequence of accesses by
$p$ to objects already accessed in $σ_1 \dots σ_i$ (already excluded); (b) an access
of $p$ to $\FHSobj_{i+1}$; and (c) a bunch of accesses by processes in
$S_{i+1}$ to $\FHSobj_{i+1}$.  So we only need to show that $α$
includes no non-trivial accesses to $\FHSobj_{i+1}$.  Suppose that it does:
then there is some process that eventually has a pending non-trivial
operation on $\FHSobj_{i+1}$ somewhere in $α$.  If we stop after this
initial prefix $α'$ of $α$, we get $k'+1$ processes with
pending operations on $\FHSobj_{i+1}$ in $EE'α'$.  But then
$E'α'$ is an extension of $E$ with $k'+1$ processes with a
simultaneous pending operation on $\FHSobj_{i+1}$.  This contradicts the
choice of $X$ to maximize $k'$.  So if our previous choice was in fact
maximal, the weird condition still holds, and we have just constructed
a $(k+k')$-stall execution.  This concludes the proof.

\subsection{Consequences}

We've just shown that counters and snapshots have $(n-1)$-stall
executions, because they are in the class $G$.  A further, rather
messy argument (given in the Ellen~\etal{} paper) extends the result to
stacks and queues, obtaining a slightly weaker bound of $n$ total
stalls and operations for some process in the worst case.\footnote{This is
    out of date: Theorem 6.2 of \cite{EllenHS2012} gives a stronger
result than what's in \cite{FichHS2005}.} In both cases, we can't expect to get a sublinear worst-case bound on time under the reasonable assumption that both a memory stall and an actual operation takes at least one time unit.  This puts an inherent bound on how well we can handle hot spots for many practical objects, and means that in an asynchronous system, we can't solve contention at the object level in the worst case (though we may be able to avoid it in our applications).

But there might be a way out for some restricted classes of objects.
We saw in Chapter~\ref{chapter-restricted-use} that we could escape from the
Jayanti-Tan-Toueg~\cite{JayantiTT2000} lower bound by considering
bounded objects.  Something similar may happen here: the
Fich-Herlihy-Shavit bound on fetch-and-increments requires executions
with $n(n-1)^{d}+n$ increments to show $n-1$ stalls for some
fetch-and-increment if each fetch-and-increment only touches $d$
objects, and even for $d = \log n$ this is already superpolynomial.
The max-register construction of a
counter~\cite{AspnesAC2012} doesn't help
here, since everybody hits the switch bit at the top of the max
register, giving $n-1$ stalls if they all hit it at the same time.  But there might be some better construction that avoids this.

\subsection{More lower bounds}

There are many more lower bounds one can prove on lock-free
implementations, many of which are based on previous lower bounds for
stronger models.  We won't present these in class, but if you are
interested, a good place to start is~\cite{AttiyaGHK2006}.

\section{Practical considerations}

Also beyond the scope of what we can do, there is a paper by Fraser
and Harris~\cite{FraserH2007} that gives some nice examples of the
practical trade-offs in choosing between multi-register CAS and
various forms of software transactional memory in implementing lock-free
data structures.

\myChapter{BG simulation}{2026}{}
\label{chapter-BG-simulation}

The \concept{Borowsky-Gafni simulation}~\cite{BorowskyG1993}, or \concept{BG simulation}
for short, is a deterministic, wait-free algorithm that allows $t+1$
processes to collectively construct a simulated execution of a system
of $n > t$ processes of which $t$ may crash.  For both the simulating
and simulated system, the underlying shared-memory primitives are
atomic snapshots; these can be replaced by atomic registers using 
any standard snapshot algorithm.  The main consequence of the BG
simulation is that the question of what decision tasks can be computed
deterministically by an asynchronous
shared-memory system that tolerates $t$ crash failures reduces to the
question of what can be computed by a wait-free system with exactly
$t+1$ processes.  This is an easier problem, and in principle can be
solved exactly using the topological approach described in
Chapter~\ref{chapter-topological-methods}.

The intuition for how this works is that the $t+1$ simulating processes
solve a sequence of agreement problems to decide what the $n$
simulated processes are doing; these agreement problems are structured
so that the failure of a simulator stops at most one agreement.  So if
at most $t$ of the simulating processes can fail, only $t$ simulated
processes get stuck as well.

We'll describe here a version of the BG simulation that appears in a
follow-up paper by Borowsky, Gafni, Lynch, and
Rajsbaum~\cite{BorowskyGLR2001}.  This gives a more rigorous
presentation of the mechanisms of the original Borowsky-Gafni paper,
and includes a few simplifications.

\section{High-level strategy}

To avoid having to simulate specific choices of operations, the BG
simulation assumes that all simulated processes alternate between
taking snapshots and doing updates.  This assumption is not very
restrictive, because two snapshots with no intervening update are
equivalent to two snapshots separated by an update that doesn't change
anything, and two updates with no intervening snapshot can be replaced
by just the second update, since the adversary could choose to
schedule them back-to-back anyway.

This approach means that we can determine the actions of some
simulated process by determining the sequence of snapshots that it
receives. So the goal will be to allow any of the real processes to
take a snapshot on behalf of any of the simulated processes, and then
coordinate these snapshots via weak consensus objects to enforce
consistency if more than one real process tries to simulate a step of
the same simulated process. The key tool for doing this is a
\concept{safe agreement}\index{agreement!safe} object, described in
§\ref{section-safe-agreement}.

\section{Safe agreement}
\label{section-safe-agreement}

\newFunc{\SApropose}{propose}
\newFunc{\SAsafe}{safe}
\newFunc{\SAagree}{agree}

A naive approach to simulate $n$ processes using $f+1$ processes would
be to lock each simulated process behind a mutex, and have the real
processes take turns grabbing a lock, simulating a step, and releasing
the lock. If we could somehow guarantee that processes never get stuck
waiting for a particular mutex just because some process died holding
the lock, then we could treat any blocked simulated process as dead,
and charge its death to the dead process holding the lock. This would
give the mapping of at most $f$ simulated failures to $f$ real
failures we are hoping for. But this depends on a lot of subtleties in
how we implement the mutexes, so the standard BG simulation goes
through a weakening of consensus instead.

The \index{agreement!safe}\concept{safe agreement} mechanism performs
agreement without running into the FLP bound, by providing a weaker
termination condition. It is guaranteed to terminate only if there are
no failures by any process during an initial, bounded, \concept{unsafe} section
of its execution, but if a process fails later, it can prevent
termination. Processes can detect when they leave the unsafe section
and have to wait for other processes only in the safe section. This
means that they can dovetail spinning in the safe sections of
multiple safe agreement objects without getting stuck entirely, even
if dead processes in the unsafe sections are blocking some of the
objects.

Each process $i$ starts
the agreement protocol with a $\SApropose_i(v)$ event for its input
value $v$.  At some point
during the execution of the protocol, the process receives a
notification $\SAsafe_i$, followed later (if the protocol finishes) 
by a second notification
$\SAagree_i(v')$ for some output value $v'$.  It is guaranteed that
the protocol terminates as long as all processes continue to take steps until they receive the
$\SAsafe$ notification, and that the usual validity (all outputs equal
some input) and agreement (all outputs equal each other) conditions
hold.
There is also a wait-free progress condition 
that the $\SAsafe_i$ notices do eventually
arrive for any process that doesn't fail, no matter what the other
processes do (so nobody gets stuck in their unsafe section).

\newData{\SAvalue}{value}
\newData{\SAlevel}{level}

Pseudocode for a safe agreement object is given in
Algorithm~\ref{alg-safe-agreement}.  This is a translation of the
description of the algorithm in~\cite{BorowskyGLR2001}, which is
specified at a lower level using I/O automata.\footnote{The I/O
automaton model is described in Appendix~\ref{appendix-IO-automata}.}

\begin{algorithm}
    \tcp{$\SApropose_i(v)$}
    $A[i] ← \Tuple{v,1}$\;
    \eIf{$\Snapshot(A)$ contains $\Tuple{j,2}$ for some $j≠i$}{
        \tcp{Back off}
        $A[i] ← \Tuple{v,0}$\;
    }{
        \tcp{Advance}
        $A[i] ← \Tuple{v,2}$\;
    }
    \tcp{$\SAsafe_i$}
    \Repeat{$s$ does not contain $\Tuple{j,1}$ for any $j$}{
        $s ← \Snapshot(A)$\;
    }
    \tcp{$\SAagree_i$}
    \Return $s[j].\SAvalue$ where $j$ is smallest index with $s[j].\SAlevel = 2$\;
    \caption{Safe agreement (adapted from~\protect{\cite{BorowskyGLR2001}})}
    \label{alg-safe-agreement}
\end{algorithm}

The communication mechanism is a snapshot object containing a pair
$A[i] = \Tuple{\SAvalue_i,\SAlevel_i}$ for each process $i$, initially
$\Tuple{⊥,0}$.  When a process carries out $\SApropose_i(v)$, it sets
$A[i]$ to $\Tuple{v,1}$, advancing to level 1.  It then looks around to see if anybody else
is at level 2; if so, it backs off to $0$, and if not, it advances to
$2$.  In either case it then spins until it sees a snapshot with
nobody at level 1, and agrees on the level-2 value with the smallest
index $i$.

The $\SAsafe_i$ transition occurs when the process leaves level $1$
(no matter which way it goes).  This satisfies the progress condition,
since there is no loop before this, and guarantees termination if all
processes leave their unsafe interval, because no process can then
wait forever for the last $1$ to disappear.

To show agreement, observe that at least one process advances to level
$2$ (because the only way a process doesn't is if some other process
has already advanced to level $2$), so any process $i$ that terminates
observes a snapshot $s$ that contains at least one level-$2$ tuple and no
level-$1$ tuples.  This means that any process $j$ whose value is
not already at level $2$ in $s$ can at worst reach level $1$
after $s$ is taken.  But then $j$ sees a level-$2$ tuples and backs
off.  It follows that any other process $i'$ that takes a later snapshot
$s'$ that includes no level-$1$ tuples sees the same level-$2$ tuples
as $i$, and computes the same return value.
(Validity also holds, for the usual trivial reasons.)

\section{The basic simulation algorithm}

The basic BG simulation uses a single snapshot object $A$ with $t+1$
components and an infinite array of safe agreement objects $S_{jr}$.

Each component $A[i]$ of $A$ belongs to one of the $t+1$ simulating
processes, and is a vector of values $A[i][j]$ that process $i$
believes process $j$ will have written at some point during the
simulated execution. These values are tagged with round numbers: each
$A[i][j]$ holds a tuple $\Tuple{v,r}$ representing the value $v$ that
process $i$ determines process $j$ would have written after taking $r$
snapshots.

The contents of these snapshots are obtained from the $S_{jr}$
objects. The inputs to $S_{jr}$ are simulated snapshots, and the
output $s_{jr}$ of $S_{jr}$ represents the value of the $r$-th
snapshot performed by simulated process $j$.

Each simulating process $i$ cycles through all simulated processes
$j$. Simulating one round of a particular process $j$ involves four
phases:
\begin{enumerate}
    \item Make an initial guess for $s_{jr}$ by taking a
        snapshot of $A$ and taking the value with the largest round
        number for each component $A[-][k]$.
    \item Initiate the safe agreement protocol $S_{jr}$
        using this guess.  It continues to run $S_{jr}$ until it
        leaves the unsafe interval.
    \item Attempt to finish $S_{jr}$, by performing one
        iteration of the loop from Algorithm~\ref{alg-safe-agreement}.
        If this iteration doesn't succeed, move on to simulating
        $j+1$ (but come back to this phase for $j$ eventually).
    \item If $S_{jr}$ terminates, compute a new value
        $v_{jr}$ for
        $j$ to write based on the simulated snapshot returned by
        $S_{jr}$, and update $A[i][j]$ with $\Tuple{v_{jr}, r}$.
\end{enumerate}

Actually implementing this while maintaining an abstraction barrier
around safe agreement is tricky.  One approach might be to have each
process $i$ manage a separate thread for each simulated process $j$,
and wrap the unsafe part of the safe agreement protocol inside a mutex
just for threads of $i$.  This guarantees that $i$ enters the unsafe
part of any safe agreement object on behalf of only one simulated $j$
at a time, while preventing delays in the safe part of $S_{jr}$ from
blocking it from finishing some other $S_{j'r'}$.

\section{Effect of failures}

So now what happens if a simulating process $i$ fails?  This won't
stop any other process $i'$ from taking snapshots on behalf of $j$, or
from generating its own values to put in $A[i'][j]$.  What it may do
is prevent some safe agreement object $S_{jr}$ from terminating.  The
termination property of $S_{jr}$ means that this can only occur if the
failure occurs while $i$ is in the unsafe interval for $S_{jr}$—but since
$i$ is only in the unsafe interval for at most one $S_{jr}$ at a time,
this stalls only one simulated process $j$.  It doesn't block any
$i'$, because any other $i'$ is guaranteed to leave its own unsafe
interval for $S_{jr}$ after finitely many steps, and though it may
waste some effort waiting for $S_{jr}$ to finish, once it is in the
safe interval it doesn't actually wait for it before moving on to other simulated $j'$.

It follows that each failure of a simulating process knocks out at
most one simulated process.  So a wait-free system with $t+1$
processes—and thus at most $t$ failures in the executions we care
about—will produces at most $t$ failures inside the simulation.

\section{Inputs and outputs}

Two details not specified in the description above are how $i$
determines $j$'s initial input and how $i$ determines its own outputs
from the outputs of the simulated processes.  For the basic BG
simulation, this is pretty straightforward: we use the safe agreement
objects $S_{j0}$ to agree on $j$'s input, after each $i$ proposes its
own input vector for all $j$ based on its own input to the simulator
protocol. For outputs, $i$ waits for at least $n-t$ of the simulated
processes to finish, and computes its own output based on what it
sees.

One issue that arises here is that we can only use the simulation to
solve \index{task!colorless}\indexConcept{colorless task}{colorless
tasks}, which are decision problems where any process can return the
output of any other process without causing trouble.\footnote{The term
    ``colorless'' here comes from use of colors to represent process
    IDs in the topological approach described in
    Chapter~\ref{chapter-topological-methods}.  These colors aren't
    really colors, but topologists like coloring nodes better than
assigning them IDs.}
This works for consensus or $k$-set agreement, but fails pretty
badly for renaming.  The \index{BG simulation!extended}\concept{extended BG
simulation}, due to Gafni~\cite{Gafni2009}, solves this problem by mapping each
simulating process $p$ to a specific simulated process $q_p$, and using a more
sophisticated simulation algorithm to guarantee that $q_p$ doesn't
crash unless $p$ does; details can be found in Gafni's paper.
There is also a later paper by Imbs and Raynal~\cite{ImbsR2009} 
that simplifies some details of the construction.  Here, we will limit
ourselves to the basic BG simulation.

\section{Correctness of the simulation}

To show that the simulation works, observe that we can extract a
simulated execution by applying the following rules:
\begin{enumerate}
    \item The round-$r$ write operation of $j$ is represented by the
        first write tagged with round $r$ performed for $j$.
    \item The round-$r$ snapshot operation of $j$ is represented by
        whichever snapshot operation wins $S_{jr}$.
\end{enumerate}

The simulated execution then consists of a sequence of write and
snapshot operations, with order of the operations determined by the
order of their representatives in the simulating execution, and the
return values of the snapshots determined by the return values of
their representatives.

Because all processes that simulate a write for $j$ in round $r$
use the same snapshots to compute the state of $j$, they all write the
same value.  So the only way we get into trouble is if the writes
included in our simulated snapshots are inconsistent with the
ordering of the simulated operations defined above.  Here the fact
that each simulated snapshot corresponds to a real snapshot makes
everything work: when a process performs a snapshot for $S_{jr}$, then
it includes all the simulated write operations that happen before this
snapshot, since the $s$-th write operation by $k$ will be represented
in the snapshot if and only if the first instance of the $s$-th write
operation by $k$ occurs before it.  The only tricky bit is that
process $i$'s snapshot for $S_{jr}$ might include some operations that
can't possibly be included in $S_{jr}$, like $j$'s round-$r$ write or
some other operation that depends on it.  But this can only occur if
some other process finished $S_{jr}$ before process $i$ takes its
snapshot, in which case $i$'s snapshot will not win $S_{jr}$ and will
be discarded.

\section{BG simulation and consensus}

BG simulation was originally developed to attack $k$-set agreement,
but (as pointed out by Gafni~\cite{Gafni2009}) it gives a particularly
simple proof of the impossibility of consensus with one faulty
process.  Suppose that we had a consensus protocol that solved
consensus for $n > 1$ processes with one crash failure, using only
atomic registers.  Then we could
use BG simulation to get a wait-free consensus protocol for two
processes.  But it's easy to show that atomic registers can't solve
wait-free consensus, because (following~\cite{LouiA1987}), we only
need to do the last step of FLP that gets a contradiction when moving
from a bivalent $C$ to $0$-valent $Cx$ or $1$-valent $Cy$.  We thus
avoid the complications that arise in the original FLP proof from
having to deal with fairness.

More generally, BG simulation means that increasing the number of
processes while keeping the same number of crash failures doesn't let
us compute anything we couldn't before.  This gives a formal
justification for the slogan that
the difference between distributed computing and parallel computing is
that in a distributed system, more processes can only make things
worse.

\myChapter{Topological methods}{2026}{}
\label{chapter-topological-methods}

Here we'll describe some results applying topology to
distributed computing, mostly following a classic paper
of Herlihy and Shavit~\cite{HerlihyS1999}.  This was
one of several
papers~\cite{BorowskyG1993,SaksZ2000}
that independently proved lower bounds on
\index{agreement!$k$-set}
\concept{$k$-set agreement}~\cite{Chaudhuri1993}, which is a relaxation of consensus where
we require only that there are at most $k$ distinct output values
(consensus is $1$-set agreement).  These lower bounds had failed to
succumb to simpler techniques.

\section{Basic idea}

The basic idea is to use tools from combinatorial topology to
represent indistinguishability proofs.  We've seen a lot of
indistinguishability proofs that involving showing that particular
pairs of executions are indistinguishable to some process, which means
that that process produces the same output in both executions.  In a
typical proof of this kind, we then construct a chain of executions
$Ξ_1,\dots,Ξ_k$ such that for each $i$, there is some $p$ with $Ξ_i|p
= Ξ_{i+1}|p$.  We've generally been drawing these with the executions
as points and the indistinguishability relation as an edge between two
executions.  In the topological method, we use the dual of this picture:
each process's view (the restriction of some execution to events
visible to that process) is represented as a point, and an execution
$Ξ$ is represented as a \concept{simplex} connecting all of the points
corresponding to views of $Ξ$ by particular processes.

A simplex is a generalization to arbitrary dimension of the sequence
that starts with a point (a $0$-simplex), an edge (a $1$-simplex), a
triangle (a $2$-simplex), or a tetrahedron (a $3$-simplex).  In
general, an $n$-simplex is a solid $n$-dimensional object with $n+1$
vertices and $n+1$ faces that are $(n-1)$-simplexes.  As a
combinatorial object, this is a fancy way of depicting the power set
of the set of vertices: each subset corresponds to a facet of the
original simplex.  A simplicial complex consists of a bunch of
simplexes pasted together by identifying vertices: this is similar to
the technique in graphics of representing the surface of a
three-dimensional object by decomposing it into triangles.  Topologists use
these to model continuous surfaces, and have many tools for deriving
interesting properties of those surfaces from a description of the
simplicial complex.

For distributed computing, the idea is that some of these topological
properties, when computed for the simplicial complex resulting from some protocol
or problem specification may sometimes useful to determine properties of the underlying protocol or problem.

\section{\texorpdfstring{$k$}{k}-set agreement}
\label{section-k-set-agreement}

The motivating problem for much of this work was getting impossibility
results for 
\index{agreement!$k$-set}
\concept{$k$-set agreement}, proposed by Chaudhuri~\cite{Chaudhuri1993}.
The $k$-set
agreement problem is similar to consensus, where each process starts
with an input and eventually returns a decision value that must be
equal to some process's input, but the agreement condition is relaxed
to require only that the set of decision values include at most $k$
values.

With $k-1$ crash failures, it's easy to build a $k$-set agreement
algorithm: wait until you have seen $n-k+1$ input values, then choose the
smallest one you see. This works because any value a process returns
is necessarily among the $k$ smallest input values (including the
$k-1$ it didn't see).

Chaudhuri conjectured that $k$-set agreement was not solvable with $k$
failures. Proving this is surprisingly difficult. Being able to solve
the problem with $k-1$ failures knocks out many standard
indistinguishability arguments that use only $1$ failure, and 
it is now known that a large class of bivalence-like arguments where the adversary
probes the future looking for a bad execution also can't work for this
problem~\cite{AlistarhAEGZ2023}. So the $k$-set agreement problem
quickly became a central test case for more general impossibility
results for computations with crash failures.

In her original paper,
Chaudhuri gave a proof of a partial
result (analogous to the existence of an initial bivalent
configuration for consensus) based on \index{Sperner's Lemma}
Sperner's Lemma~\cite{Sperner1928}.
This is a classic
result in topology that says that certain colorings of the vertices of
a graph in the form of a triangle that has been divided into smaller
triangles necessarily contain a small triangle with three different
colors on its corners.
This connection between $k$-set agreement and Sperner's Lemma became
the basic idea behind each the three independent proofs of the conjecture
that appeared shortly
thereafter~\cite{HerlihyS1999,BorowskyG1993,SaksZ2000}, all of which
adopted an approach that reduces decision problems in distributed
systems to the existence of certain structures in combinatorial topology.

Our plan is to give a sufficient high-level description of the
topological approach that the connection between $k$-set agreement and
Sperner's Lemma becomes obvious.  It is possible to avoid this by
approaching the problem purely combinatorially, as is done, for
example,
in Section 16.3 of~\cite{AttiyaW2004}.  The presentation there is
obtained by starting with a topological argument and getting rid of
the topology (in fact, the proof in~\cite{AttiyaW2004} contains a
proof of Sperner's Lemma with the serial numbers filed off).
The disadvantage of this approach is that it obscures what is really
going in and makes it harder to obtain insight into how topological
techniques might help for other problems.  The advantage is that
(unlike these notes) the resulting text includes actual proofs instead
of handwaving.

\section{Representing distributed computations using topology}
\label{section-topological-representations}

Topology is the study of properties of shapes that are preserved by
continuous functions between their points that have continuous inverses, which get the rather fancy name of 
\indexConcept{homeomorphism}{homeomorphisms}.  A continuous
function\footnote{Strictly speaking, this is the definition a continuous function between
metric spaces, which are spaces that have a consistent notion of
distance.  There is an even more general definition of continuity that holds for
spaces that are too strange for this.} is one that maps nearby points to nearby points.  A homeomorphism is continuous in both directions: this basically means that you can stretch and twist and otherwise deform your object however you like, as long as you don't tear it (which would map nearby points on opposite sides of the tear to distant points) or glue bits of it together (which turns into tearing when we look at the inverse function).  Topologists are particularly interested in showing when there is no homeomorphism between two objects; the classic example is that you can't turn a sphere into a donut without damaging it, but you can turn a donut into a coffee mug (with a handle).

Working with arbitrary objects embedded in umpteen-dimensional spaces
is messy, so topologists invented a finite way of describing certain
well-behaved objects combinatorially, by replacing ugly continuous
objects like spheres and coffee mugs with simpler objects pasted
together in complex ways.  The simpler objects are
\indexConcept{simplex}{simplexes}, and the more complicated pasted-together
objects are called 
\indexConcept{simplicial complex}{simplicial complexes}.  The nifty thing about
simplicial complexes is that they give a convenient tool for describing what states or outputs of processes in a distributed algorithm are ``compatible'' in some sense, and because topologists know a lot about simplicial complexes, we can steal their tools to describe distributed algorithms.

\subsection{Simplicial complexes and process states}
\label{section-topological-representation-of-process-states}

The formal definition of a $k$-dimensional \concept{simplex} is the
convex closure of $(k+1)$ points $\{ x_{1}\dots{}x_{k+1} \}$ in
general position; the convex closure part means the set of all points
$\sum{} a_{i}x_{i}$ where $\sum{} a_{i} = 1$ and each $a_{i} ≥ 0$,
and the general position part means that the $x_{i}$ are not all
contained in some subspace of dimension $(k-1)$ or smaller (so that
the simplex isn't squashed flat somehow).  What this gives us is a
body with $(k+1)$ corners and $(k+1)$ faces, each of which is a
$(k-1)$-dimensional simplex (the base case is that a $0$-dimensional simplex is a point).  Each face includes all but one of the corners, and each corner is on all but one of the faces.  So we have:
\begin{itemize}
 \item 0-dimensional simplex: point.\footnote{For consistency, it's sometimes
 convenient to define a point as having a single $(-1)$-dimensional
 face defined to be the empty set.  We won't need to bother with this,
 since $0$-dimensional simplicial complexes correspond to $1$-process
 distributed systems, which are amply covered in almost every other Computer
 Science class you have ever taken.}
 \item 1-dimensional simplex: line segment with 2 endpoints (which are both corners and faces).
 \item 2-dimensional simplex: triangle (3 corners with 3 1-dimensional simplexes for sides).
 \item 3-dimensional simplex: tetrahedron (4 corners, 4 triangular faces).
 \item 4-dimensional simplex: 5 corners, 5 tetrahedral faces.  It's probably best not to try to visualize this.
\end{itemize}

A simplicial complex is a bunch of simplexes stuck together; formally,
this means that we pretend that some of the corners (and any faces
that include them) of different simplexes are identical points.
There are ways to do this right using equivalence relations.  But it's
easier to abstract out the actual geometry and go straight to a
combinatorial structure.

An
\index{simplicial complex!abstract}
\index{abstract simplicial complex}
(abstract) simplicial complex is just a collection
of sets with the property that if $A$ is a subset of $B$, and $B$ is
in the complex, then $A$ is also in the complex (this means that if
some simplex is included, so are all of its faces, their faces, etc.).
This combinatorial version is nice for reasoning about simplicial
complexes, but is not so good for drawing pictures.  

The trick to using this for distributed computing problems is that we are going to build simplicial complexes by letting points be process states (or sometimes process inputs or outputs), each labeled with a process ID, and letting the sets that appear in the complex be those collections of states/inputs/outputs that are compatible with each other in some sense.  For states, this means that they all appear in some global configuration in some admissible execution of some system; for inputs and outputs, this means that they are permitted combinations of inputs or outputs in the specification of some problem.

Example: For 2-process binary consensus with processes 0 and 1, the
\index{complex!input}
\concept{input complex}, which describes all possible combinations of
inputs, consists of the sets 
\begin{displaymath}
\left\{ \{\}, \{p0\}, \{q0\}, \{p1\}, \{q1\},
\{p0,q0\}, \{p0,q1\}, \{p1,q0\}, \{p1,q1\} \right\},
\end{displaymath}
which we might draw like this:
\begin{center}
\begin{tikzpicture}[auto,node distance=2cm]
    \node (p0) {$p0$};
    \node (q0) [right of=p0] {$q0$};
    \node (q1) [below of=p0] {$q1$};
    \node (p1) [right of=q1] {$p1$};

    \path
        (p0) edge (q0)
             edge (q1)
        (p1) edge (q0)
             edge (q1)
             ;
\end{tikzpicture}
\end{center}

Note that there are no edges from $p0$ to $p1$ or $q0$ to $q1$: we can't have two different states of the same process in the same global configuration.

The 
\index{complex!output}
\concept{output complex}, which describes the permitted outputs, is
\begin{displaymath}
\left\{ \{\}, \{p0\}, \{q0\}, \{p1\}, \{q1\}, \{p0,q0\}, \{p1,q1\}
\right\}.
\end{displaymath}
As a picture, this omits two of the edges (1-dimensional simplexes) from the input complex:
\begin{center}
\begin{tikzpicture}[auto,node distance=2cm]
    \node (p0) {$p0$};
    \node (q0) [right of=p0] {$q0$};
    \node (q1) [below of=p0] {$q1$};
    \node (p1) [right of=q1] {$p1$};

    \path
        (p0) edge (q0)
        (p1) edge (q1)
             ;
\end{tikzpicture}
\end{center}

One thing to notice about this output complex is that it is not
\concept{connected}: there is no path from the $p0$--$q0$ component
to the $q1$--$p1$ component.

Here is a simplicial complex describing the possible states of two
processes $p$ and $q$, after each writes 1 to its own bit then reads
the other process's bit.  Each node in the picture is labeled by a
sequence of process IDs.  The first ID in the sequence is the process
whose view this node represents; any other process IDs are processes
this first process sees (by seeing a $1$ in the other process's
register).  So $p$ is the view of process $p$ running by itself, while
$pq$ is the view of process $p$ running in an execution where it reads
$q$'s register after $q$ writes it.
\begin{center}
\begin{tikzpicture}[auto,node distance=2cm]
    \node (p) {$p$};
    \node (qp) [right of=p] {$qp$};
    \node (pq) [right of=qp] {$pq$};
    \node (q) [right of=pq] {$q$};

    \path
        (p) edge (qp)
        (qp) edge (pq)
        (pq) edge (q)
    ;
\end{tikzpicture}
\end{center}

The edges express the constraint that if we both write before we read,
then if I don't see your value you must see mine (which is why there
is no $p$--$q$ edge), but all other combinations are possible.  Note that this complex \emph{is} connected: there is a path between any two points.

Here's a fancier version in which each process writes its input (and
remembers it), then reads the other process's register (i.e., a
one-round full-information protocol).  We now have final states that
include the process's own ID and input first, then the other process's
ID and input if it is visible.
For example, $p1$ means
$p$ starts with $1$ but sees a null and $q0p1$ means $q$ starts with $0$ but
sees $p$'s $1$.  The general rule is that two states are compatible if
$p$ either sees nothing or $q$'s actual input and similarly for $q$,
and that at least one of $p$ or $q$ must see the other's input.  This gives the following simplicial complex:

\begin{center}
\begin{tikzpicture}[auto,node distance=1.5cm]
    \node (p0) {$p0$};
    \node (q0p0) [right of=p0] {$q0p0$};
    \node (p0q0) [right of=q0p0] {$p0q0$};
    \node (q0) [right of=p0q0] {$q0$};
    \node (q1p0) [below of=p0] {$q1p0$};
    \node (p0q1) [below of=q1p0] {$p0q1$};
    \node (q1) [below of=p0q1] {$q1$};
    \node (p1q0) [below of=q0] {$p1q0$};
    \node (q0p1) [below of=p1q0] {$q0p1$};
    \node (p1q1) [right of=q1] {$p1q1$};
    \node (q1p1) [right of=p1q1] {$q1p1$};
    \node (p1) [right of=q1p1] {$p1$};

    \path
        (p0) edge (q0p0)
        (q0p0) edge (p0q0)
        (p0q0) edge (q0)
        (q0) edge (p1q0)
        (p1q0) edge (q0p1)
        (q0p1) edge (p1)
        (p1) edge (q1p1)
        (q1p1) edge (p1q1)
        (p1q1) edge (q1)
        (q1) edge (p0q1)
        (p0q1) edge (q1p0)
        (q1p0) edge (p0)
    ;
\end{tikzpicture}
\end{center}

Again, the complex is connected.

The fact that this looks like four copies of the
$p$--$qp$--$pq$--$q$ complex pasted into each edge of the input
complex is not an accident: if we fix a pair of inputs $i$ and $j$, we
get $pi$--$qjpi$--$piqj$--$qj$, and the corners are pasted together
because if $p$ sees only $p0$ (say), it can't tell if it's in the
$p0/q0$ execution or the $p0/q1$ execution.

The same process occurs if we run a two-round protocol of this form,
where the input in the second round is the output from the first
round.  Each round subdivides one edge from the previous round into
three edges:

\begin{align*}
&p-q \\
 \\
&p-qp-pq-q \\
 \\
&p-(qp)p-p(qp)-qp-(pq)(qp)-(qp)(pq)-pq-q(pq)-(pq)q-q \\
\end{align*}

Here $(pq)(qp)$ is the view of $p$ after seeing $pq$ in the first
round and seeing that $q$ saw $qp$ in the first round.

\subsection{Subdivisions}

In the simple write-then-read protocol above, we saw a single input
edge turn into 3 edges.  Topologically, this is an example of a
\concept{subdivision}, where we represent a simplex using several new simplexes pasted together that cover exactly the same points.

Certain classes of protocols naturally yield subdivisions of the input
complex.  The \concept{iterated immediate snapshot}
(\index{IIS}IIS) model, defined by Borowsky and
Gafni~\cite{BorowskyG1997},
considers executions made up of a sequence of rounds
(the iterated part) where each round is made up of one or more
mini-rounds in which some subset of the processes all write out their
current views to their own registers and then take snapshots of all
the registers (the immediate snapshot part).  
The two-process protocols of the previous section are
special cases of this model.

Within each round, each process $p$ obtains a view $v_p$ that contains the
previous-round views of some subset of the processes.  
We can represent the views as a subset of the processes, which we will
abbreviate in pictures by putting the view owner first: $pqr$ will be
the view $\{p, q, r\}$ as seen by $p$, while $qpr$ will be the same
view as seen by $q$.
The
requirements on these views are that (a) every process sees its own
previous view: $p \in v_p$ for all $p$; (b) all views are
comparable: $v_p \subseteq v_q$ or $v_q \subseteq v_p$; and (c) if I
see you, then I see everything you see: $q \in v_p$ implies $v_q
\subseteq v_p$.  This last requirement is called \concept{immediacy}
and follows from the assumption that writes and snapshots are done in
the same mini-round: if I see your write, then I see all the values
you do, because your snapshot is either in an earlier mini-round than
mine or in the same mini-round.  Note this depends on the peculiar
structure of the mini-rounds, where all the writes precede all the snapshots.

The IIS model does not correspond exactly to a standard shared-memory
model (or even a standard shared-memory model augmented with cheap
snapshots).  There are two reasons for this: standard snapshots don't
provide immediacy, and standard snapshots allow processes to go back
and perform more than one snapshot on the same object.  The first issue
goes away if we are looking at impossibility proofs, because the
adversary can restrict itself only to those executions that satisfy
immediacy; alternatively, we can get immediacy from the
\concept{participating set} protocol of~\cite{BorowskyG1997}, which we
will describe in 
§\ref{section-participating-set}.  The second issue is more delicate, but Borowsky and Gafni
demonstrate that any decision protocol that runs in the standard model
can be simulated in the IIS model, using a variant of the BG
simulation algorithm described in Chapter~\ref{chapter-BG-simulation}.

\begin{figure}
\centering
\includegraphics[scale=0.8]{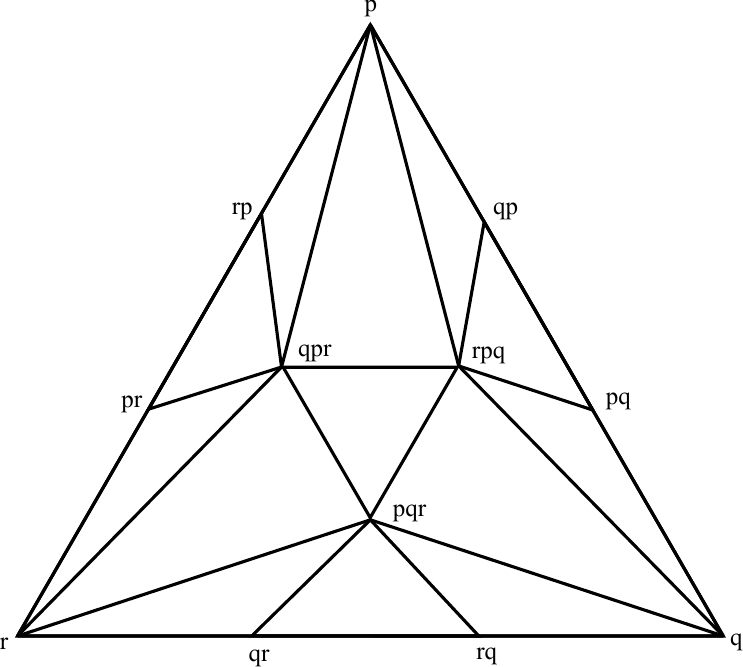}
\caption{Subdivision corresponding to one round of immediate snapshot}
\label{figure-immediate-snapshot}
\end{figure}

For three processes, one round of immediate snapshots gives rise to
the simplicial complex depicted in
Figure~\ref{figure-immediate-snapshot}.  The corners of the big
triangle are the solo
views of processes that do their snapshots before anybody else shows
up.  Along the edges of the big triangle are views corresponding to
2-process executions, while in the middle are complete views of
processes that run late enough to see everything.  Each little
triangle corresponds to some execution.  For example, the triangle
with corners $p$, $qp$, $rpq$ corresponds to a sequential 
execution where $p$
sees nobody, $q$ sees $p$, and $r$ sees both $p$ and $q$.  The
triangle with corners $pqr$, $qpr$, and $rpq$ is the
maximally-concurrent execution where all three processes write before
all doing their snapshots: here everybody sees everybody.  It is not
terribly hard to enumerate all possible executions and verify that the
picture includes all of them.  In higher dimension, the picture is
more complicated, but we still get a subdivision that preserves the
original topological structure~\cite{BorowskyG1997}.

\begin{figure}
\centering
\includegraphics[scale=0.8]{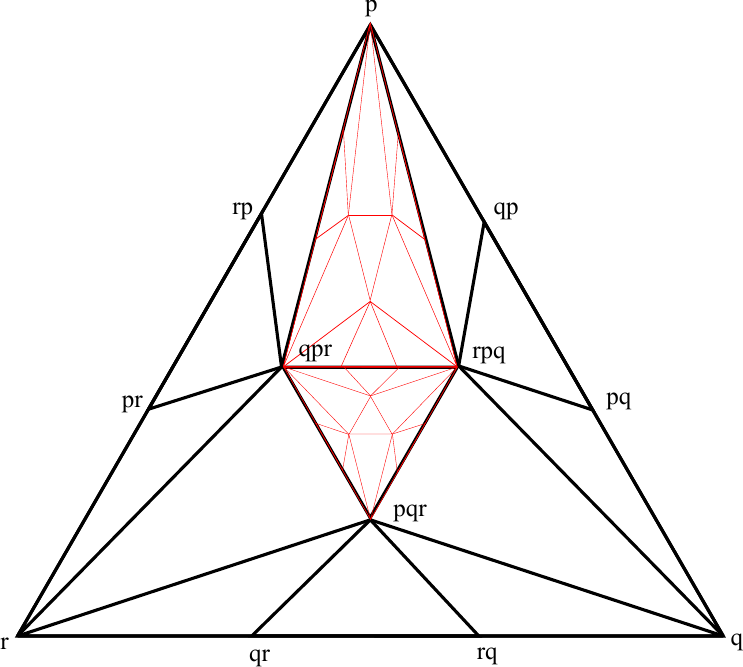}
\caption{Subdivision corresponding to two rounds of immediate snapshot}
\label{figure-immediate-snapshot-2}
\end{figure}

Figure~\ref{figure-immediate-snapshot-2} shows (part of) the next step
of this process: here we have done two iterations of immediate
snapshot, and filled in the second-round subdivisions for the
$p$--$qpr$--$rpq$ and $pqr$--$qpr$--$rpq$ triangles.  (Please imagine
similar subdivisions of all the other triangles that I was too lazy to
fill in by hand.)  The structure is recursive, with each first-level
triangle mapping to an image of the entire first-level complex.
As in the two-process case, adjacent triangles overlap because the
relevant processes don't have enough information; for example, the
points on the $qpr$--$rpq$ edge correspond to views of $q$ or $r$ that
don't include $p$ in round 2 and so can't tell whether $p$ saw $p$ or
$pqr$ in round 1.

The important feature of the round-2 complex (and the round-$k$
complex in general) is that it's a \concept{triangulation} of the
original outer triangle: a partition into little triangles where each
corner aligns with corners of other little triangles.

(Better pictures of this process in action can be found in Figures 25
and 26 of~\cite{HerlihyS1999}.)

\section{Impossibility of \texorpdfstring{$k$}{k}-set agreement}
\label{section-k-set-agreement-impossible}

Now let's show that there is no way to do $k$-set agreement with
$n=k+1$ processes in the IIS model.

\begin{figure}
\centering
\includegraphics[scale=0.8]{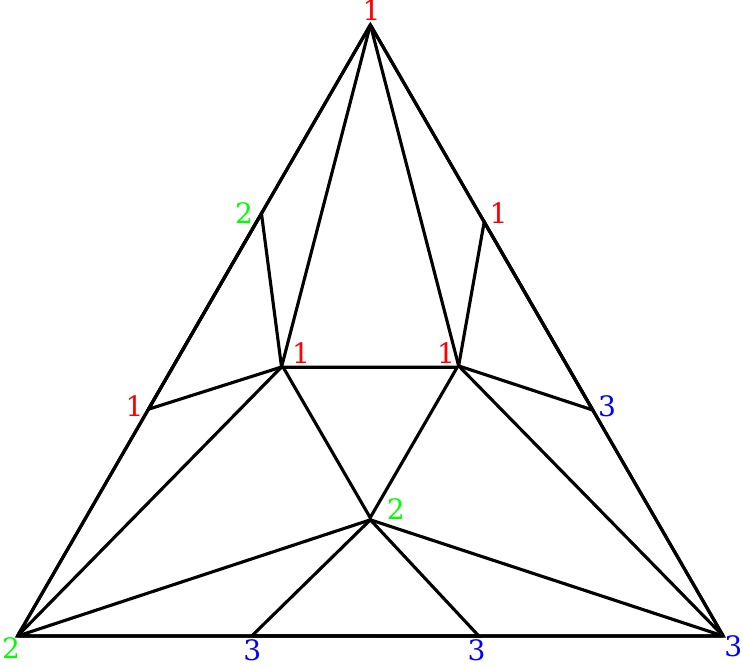}
\caption{An attempt at $2$-set agreement}
\label{figure-immediate-snapshot-outputs}
\end{figure}

Suppose that after some fixed number of
rounds, each process chooses an output
value.  This output can only depend on the view of the process, so is
fixed for each vertex in the subdivision.  Also, the validity
condition means that a process can only choose an output that it can
see among the inputs in its view.  This means that at the corners of
the outer triangle (corresponding to views where the process thinks
it's alone), a process must return its input, while along the outer
edges (corresponding to views where two processes may see each other
but not the third), a process must return one of the two inputs that
appear in the corners incident to the edge.  Internal corners
correspond to views that include—directly or indirectly—the
inputs of all processes, so these can be labeled arbitrarily.
An example is given in Figure~\ref{figure-immediate-snapshot-outputs},
for a one-round protocol with three processes.

We now run into Sperner's Lemma~\cite{Sperner1928}, which says that,
for any subdivision of a simplex into smaller simplexes, if each
corner of the original simplex has a different color, and each corner
that appears on some face of the original simplex has a color equal to
the color of one of the corners of that face, then within the
subdivision there are an odd number of simplexes whose corners are all
colored differently.\footnote{The proof of Sperner's Lemma is not
hard, and is done by induction on the dimension $k$.  For $k=0$, any
subdivision consists of exactly one zero-dimensional simplex whose
single corner covers all $k+1=1$ colors.  For $k+1$, suppose that the
colors are $\Set{1, \dots, k+1}$, and construct a graph
with a vertex for each little simplex in the subdivision and an extra
vertex for the region outside the big simplex.  Put an edge in this
graph between each pair of regions that share a $k$-dimensional face
with colors $\Set{1, \dots, k}$.  The induction hypothesis tells us
that there are an odd number of edges between the outer-region vertex
and simplexes on the $\Set{1,\dots,k}$-colored face of the big
simplex.  The Handshaking Lemma from graph theory says that the sum of
the degrees of all the nodes in the graph is even.  But this can only
happen if there are an even number of nodes with odd degree, implying
that the are are an odd number of simplexes in the subdivision with an
odd number of faces colored $\Set{1,\dots,k}$, because the extra node
for the outside region has exactly one face colored $\Set{1,\dots,k}$.
Since zero is even, this means there is at least one simplex in the subdivision with an
odd number of faces colored $\Set{1,\dots,k}$.

Now suppose we have a simplex with an odd number of faces colored
$\Set{1,\dots,k}$.  Let $f$ be one such face.  If the corner $v$ not
contained in $f$ is colored $c \ne k+1$, then
our simplex has exactly two faces colored $\Set{1,\dots,k}$: $f$, and the face
that replaces $f$'s $c$-colored corner with $v$.  So
the only way to get an odd number of $\Set{1,\dots,k}$-colored faces is
to have all $k+1$ colors.  It follows that there are an odd number of
$(k+1)$-colored simplexes.}

How this applies to $k$-set agreement: Suppose we have $n = k+1$
processes in a wait-free system (corresponding to allowing up to $k$
failures).  With the cooperation of the adversary, we can restrict
ourselves to executions consisting of $\ell$ rounds of iterated
immediate snapshot for some $\ell$ (termination comes in here to show
that $\ell$ is finite).  This gives a subdivision of a simplex, where
each little simplex corresponds to some particular execution and each
corner some process's view.  Color
all the corners of the little simplexes in this subdivision with the
output of the process holding the corresponding view.  Validity means
that these colors satisfy the requirements of Sperner's Lemma.
Sperner's Lemma then says that some little simplex has all $k+1$
colors, giving us a bad execution with more than $k$ distinct output
values.

The general result says that we can't do $k$-set agreement with $k$
failures for any $n > k$. This follows immediately from the $n=k+1$
version using BG simulation (Chapter~\ref{chapter-BG-simulation}).

\section{Simplicial maps and specifications}

Let's step back and look at consensus again.

One thing we could conclude from the fact that the output complex for
consensus was not connected but the ones describing our simple
protocols were was that we can't solve consensus (non-trivially) using
these protocols.  The reason is that to solve consensus using such a
protocol, we would need to have a mapping from states to outputs (this is just whatever rule tells each process what to decide in each state) with the property that if some collection of states are consistent, then the outputs they are mapped to are consistent.

In simplicial complex terms, this means that the mapping from states to
outputs is a 
\index{map!simplicial}
\concept{simplicial map}, a function $f$ from points in one simplicial
complex $C$ to points in another simplicial complex $D$ such that for
any simplex $A \in C$, $f(A) = \{ f(x) | x \in{} A \}$ gives a simplex
in $D$.  (Recall that consistency is represented by including a
simplex, in both the
state complex and the output complex.)  A mapping from states to
outputs that satisfies the consistency requirements encoded in the
output complex $s$ always a simplicial map, with the additional
requirement that it preserves process IDs (we don't want process $p$
to decide the output for process $q$).  Conversely, any id-preserving simplicial map gives an output function that satisfies the consistency requirements.

Simplicial maps are examples of 
\index{function!continuous}
\indexConcept{continuous function}{continuous functions}, which have
all sorts of nice topological properties.  One nice property is that a
continuous function can't separate a path-connected space (
one in which there is a path between any two points) into path-disconnected
components.  We can prove this directly for simplicial maps: if there
is a path of 1-simplexes $\{x_{1},x_{2}\}, \{x_{2},x_{3}\}, \dots{}
\{x_{k-1},x_{k}\}$ from $x_1$ to $x_k$ in $C$, and $f:C\rightarrow{}D$
is a simplicial map, then there is a path of 1-simplexes
$\{f(x_{1}),f(x_{2})\}, \dots{}$ from $f(x_{1})$ to $f(x_{k})$.  Since
being path-connected just means that there is a path between any two
points, if $C$ is connected we've just shown that $f(C)$ is as well.

Getting back to our consensus example, it doesn't matter what
simplicial map $f$ you pick to map process states to outputs; since
the state complex $C$ is connected, so is $f(C)$, so it lies entirely
within one of the two connected components of the output complex.
This means in particular that everybody always outputs $0$ or $1$: the protocol is trivial.

\subsection{Mapping inputs to outputs}

For general decision tasks, it's not enough for the outputs to be
consistent with each other.  They also have to be consistent with the
inputs.  This can be expressed by a relation $Δ$ between input simplexes and output simplexes.

Formally, a decision task is modeled by a triple $(I, O, Δ)$,
where $I$ is the input complex, $O$ is the output complex, and $(A,B)
\in{} Δ$ if and only if $B$ is a permissible output given input
$A$.  Here there are no particular restrictions on $Δ$ (for example, it doesn't have to be a simplicial map or even a function), but it probably doesn't make sense to look at decision tasks unless there is at least one permitted output simplex for each input simplex.

\section{The asynchronous computability theorem}

Given a decision task specified in this way, there is a topological
characterization of when it has a wait-free solution.  This is given
by the \concept{Asynchronous Computability Theorem} (Theorem 3.1
in~\cite{HerlihyS1999}), which says:
\begin{theorem}
A decision task $(I,O,Δ{})$ has a wait-free protocol using shared
memory if and only if there exists a chromatic subdivision $σ$ of
$I$ and a color-preserving simplicial map $\mu{}: σ{}(I)
\rightarrow{} O$ such that for each simplex $S$ in $σ{}(I)$,
$\mu{}(S) \in{} Δ{}(\carrier(S, I))$.
\end{theorem}

To unpack this slightly, a 
\index{subdivision!chromatic}
\concept{chromatic subdivision} is a subdivision where each vertex is
labeled by a process ID (a color), and no simplex has two vertices
with the same color.  A color-preserving simplicial map is a
simplicial map that preserves IDs.  The carrier of a simplex in a
subdivision is whatever original simplex it is part
of.  So the theorem says that I can only solve a task if I can find a
simplicial map from a subdivision of the input complex to the output
complex that doesn't do anything strange to process IDs and that is
consistent with $Δ$.

Looking just at the theorem, one might imagine that the proof consists
of showing that the 
\index{complex!protocol}
\concept{protocol complex} defined by the state complex after running the protocol to completion is a subdivision of the input complex, followed by the same argument we've seen already about mapping the state complex to the output complex.  This is almost right, but it's complicated by two inconvenient facts: (a) the state complex generally isn't a subdivision of the input complex, and (b) if we have a map from an arbitrary subdivision of the input complex, it is not clear that there is a corresponding protocol that produces this particular subdivision.

So instead the proof works like this:

\begin{description}
 \item[Protocol implies map] Even though we don't get a subdivision
     with the full protocol, there is a restricted set of executions
     that does give a subdivision.  So if the protocol works on this
     restricted set of executions, an appropriate map exists.  There
     are two ways to prove this: Herlihy and Shavit do so directly, by
     showing that this restricted set of executions exists, and
     Borowksy and Gafni~\cite{BorowskyG1997} do so indirectly, by
     showing that the IIS model (which produces exactly the standard
     chromatic subdivision used in the ACT proof) can simulate an ordinary snapshot
     model.  Both methods are a bit involved, so we will skip over
     this part.
 \item[Map implies protocol] This requires an algorithm.
     The idea here is that that \concept{participating set} algorithm,
     originally developed to solve $k$-set
     agreement~\cite{BorowskyG1993}, produces precisely the standard
     chromatic subdivision used in the ACT proof.  In particular, it
     can be used to solve the problem of
 \index{agreement!simplex}
 \concept{simplex agreement}, the problem of getting the processes to
 agree on a particular simplex contained within the subdivision of
 their original common input simplex.  This is a little easier to
 explain, so we'll do it.
\end{description}

\subsection{The participating set protocol}
\label{section-participating-set}

\newData{\PSlevel}{level}

Algorithm~\ref{alg-participating-set} depicts the participating set
protocol; this first appeared in~\cite{BorowskyG1993}, although the
presentation here is heavily influenced by the version in Elizabeth Borowsky's
dissertation~\cite{Borowsky1995}.  
The shared data consists of a snapshot object $\PSlevel$,
and processes start at a high level and float down until they reach a
level $i$ such that there are already $i$ processes at this level or
below.  The set returned by a process consists of all processes it
sees at its own level or below, and it can be shown that this in fact
implements a one-shot immediate snapshot.  Since immediate snapshots
yield a standard subdivision, this gives us what we want for
converting a color-preserving simplicial map to an actual protocol.

\begin{algorithm}
    Initially, $\PSlevel[i] = n+2$ for all $i$.\;

    \Repeat{$\card*{S} ≥ \PSlevel[i]$}{
        $\PSlevel[i] ← \PSlevel[i] - 1$\;
        $v ← \Snapshot(\PSlevel)$\;
        $S ← \SetWhere{j}{v[j] ≤ \PSlevel[i]}$\;
    }
    \Return $S$\;
    \caption{Participating set}
    \label{alg-participating-set}
\end{algorithm}

The following theorem shows that the return values from participating
set have all the properties we want for iterated immediate snapshot:
\begin{theorem}
    \label{theorem-participating-set}
    Let $S_i$ be the output of the participating set algorithm for
    process $i$.  Then all of the following conditions hold:
    \begin{enumerate}
        \item For all $i$, $i ∈ S_i$.  (Self-containment.)
        \item For all $i,j$, $S_i ⊆ S_j$ or $S_j ⊆ S_i$.  (Atomic
            snapshot.)
        \item For all $i,j$, if $i ∈ S_j$, then $S_i ⊆ S_j$.
            (Immediacy.)
    \end{enumerate}
\end{theorem}
\begin{proof}
    Self-inclusion is trivial, but we will have to do some work for
    the other two properties.

    We will show that
    Algorithm~\ref{alg-participating-set} neatly sorts the processes
    out into levels, where each process that returns at level $\ell$ returns
    precisely the set of processes at level $\ell$ and below.

    For each process $i$, let $S_i$ be the set of process IDs that $i$
    returns, let
    $\ell_i$ be the final value of $\PSlevel[i]$ when $i$ returns,
    and let $S'_i = \SetWhere{j}{\ell_j ≤ \ell_i}$.  Our goal is to show
    that $S'_i = S_i$, justifying the above claim.

    Because no process ever increases its level, if process $i$
    observes $\PSlevel[j] ≤ \ell_i$ in its last snapshot, then $\ell_j
    ≤ \PSlevel[j] ≤ \ell_i$.  So $S'_i$ is a superset of $S_i$.  We
    thus need to show only that no extra processes sneak in; in
    particular, we will to show that $\card*{S_i} = \card*{S'_i}$, by showing that
    both equal $\ell_i$.

    The first step is to show that $\card*{S'_i} ≥ \card*{S_i} ≥
    \ell_i$.  The first inequality follows from the fact that $S'_i ⊇
    S_i$; the second follows from the code (if not,
    $i$ would have stayed in the loop).

    The second step is to show that $\card*{S'_i} ≤ \ell_i$.
    Suppose not; that is, suppose that 
    $\card*{S'_i} > \ell_i$.  Then there are at least $\ell_i+1$ processes
    with level $\ell_i$ or less, all of which take a snapshot on level
    $\ell_i+1$.  Let $i'$ be the last of these processes to take a snapshot
    while on level $\ell_i+1$.  Then $i'$ sees at least $\ell_i+1$ processes at
    level $\ell_i+1$ or less and exits, contradicting the assumption that
    it reaches level $\ell_i$.  So $\card*{S'_i} ≤ \ell_i$.

    The atomic snapshot property follows immediately from the fact
    that if $\ell_i ≤ \ell_j$, then $\ell_k ≤ \ell_i$ implies
    $\ell_k ≤ \ell_j$, giving $S_i = S'_i ⊆ S'_j = S_j$.
    Similarly, for immediacy we have that if $i ∈ S_j$, then $\ell_i ≤
    \ell_j$, giving $S_i ≤ S_j$ by the same argument.
\end{proof}

The missing piece for turning this into IIS is that in
Algorithm~\ref{alg-participating-set}, I only learn the identities of
the processes I am supposed to include but not their input values.
This is easily dealt with by the usual trick of adding an extra register for each
process, to which it writes its input before executing participating
set.

\section{Proving impossibility results}

To show something is impossible using the ACT, we need to show that
there is no color-preserving simplicial map from a subdivision of $I$
to $O$ satisfying the conditions in $Δ$.  This turns out to be
equivalent to showing that there is no continuous function from $I$ to
$O$ with the same properties, because any such simplicial map can be
turned into a continuous  function (on the geometric version of $I$, which includes the intermediate points in addition to the corners).  Fortunately, topologists have many tools for proving non-existence of continuous functions.

\subsection{\texorpdfstring{$k$-connectivity}{k-connectivity}}

Define the $m$-dimensional \concept{disk} to be the set of all points
at most 1 unit away from the origin in $\mathbb{R}^{m}$, and the
$m$-dimensional \concept{sphere} to be the surface of the
$(m+1)$-dimensional disk (i.e., all points exactly 1 unit away from
the origin in $\mathbb{R}^{m+1}$).  Note that what we usually think of
as a sphere (a solid body), topologists call a disk, leaving the term
sphere for just the outside part.

An object is 
\indexConcept{$k$-connectivity}{$k$-connected} if any continuous image of an
$m$-dimensional sphere can be extended to a continuous image of an
$(m+1)$-dimensional disk, for all $m ≤ k$.\footnote{This
definition is for the topological version of $k$-connectivity.  It
is not related in any way to the definition of $k$-connectivity in graph
theory, where a graph is $k$-connected if there are $k$ disjoint paths
between any two points.} This is a roundabout
way of saying that if we can draw something that looks like a deformed
sphere inside our object, we can always include the inside as well:
there are no holes that get in the way.  The punch line is that
continuous functions preserve $k$-connectivity: if we want to map an
object with no holes continuously into some other object, the image had better not have any holes either.

Ordinary path-connectivity is the special case when $k = 0$; here, the
$0$-sphere consists of two points and the $1$-disk is the path between
them.  So $0$-connectivity says that for any two points, there is a path between them.

For $1$-connectivity, if we draw a loop (a path that returns to its
origin), we can include the interior of the loop somewhere.  One way
to thinking about this is to say that we can shrink the loop to a
point without leaving the object (the technical term for this is that
the path is \concept{null-homotopic}, where a \concept{homotopy} is a way to transform one thing continuously into another thing over time and the 
\index{path!null}\concept{null path} sits on a single point).  An
object that is $1$-connected is also called 
\index{connected!simply}
\concept{simply connected}.

For 2-connectivity, we can't contract a sphere (or box, or the surface
of a 2-simplex, or anything else that looks like a sphere) to a point.

The important thing about $k$-connectivity is that it is possible to
prove that any subdivision of a $k$-connected simplicial complex is
also $k$-connected (sort of obvious if you think about the pictures,
but it can also be proved formally), and that $k$-connectivity is
preserved by simplicial maps (if not, somewhere in the middle of all
the $k$-simplexes representing our surface is a $(k+1)$-simplex in the domain that maps to a hole in the range, violating the rule that simplicial maps map simplexes to simplexes).  So a quick way to show that the Asynchronous Computability Theorem implies that something is not asynchronously computable is to show that the input complex is $k$-connected and the output complex isn't.

\subsection{Impossibility proofs for specific problems}
Here are some applications of the Asynchronous Computability Theorem and $k$-connectivity:

\begin{description}
 \item[Consensus] There is no nontrivial wait-free consensus protocol
 for $n ≥ 2$ processes.  Proof: The input complex is 1-connected, but
        the output complex is not, and we need a map that covers the
        entire output complex (by non-triviality).
 \item[$k$-set agreement] There is no wait-free $k$-set agreement for
 $n ≥ k+1$ processes.   Proof: The output complex for $k$-set agreement is not
 $k$-connected, because buried inside it are lots of
 $(k+1)$-dimensional holes corresponding to missing simplexes where
 all $k+1$ processes choose different values.  But these holes aren't
 present in the input complex—it's OK if everybody starts with
 different inputs—and the validity requirements for $k$-set
 agreement force us to map the surfaces of these non-holes around
 holes in the output complex.  (This proof actually turns into the
 Sperner's Lemma proof if we fully expand the claim about having to
 map the input complex around the hole.)
 \item[Renaming] There is no wait-free renaming protocol with less
 than $2n-1$ output names for all $n$.  The general proof of this
 requires showing that with fewer names we get holes that are too big
 (and ultimately reduces to Sperner's Lemma); for the special case of
 $n=3$ and $m=4$, see Figure~\ref{fig-renaming-output-complex}, which shows how
 the output complex of renaming folds up into the surface of a torus.
 This means that renaming for $n=3$ and $m=4$ is \emph{exactly the same} as trying to stretch a basketball into an inner tube.
\end{description}

\begin{figure}
    \centering
    \begin{tikzpicture}[auto,node distance=1.5cm]
        \node (a1) {$a1$};
        \node (b2) [right of=a1] {$b2$};
        \node (c3) [above right of=a1] {$c3$};
        \node (a4) [right of=c3] {$a4$};
        \node (c1) [right of=b2] {$c1$};
        \node (b1') [above right of=c3] {$b1$};
        \node (c2') [right of=b1'] {$c2$};
        \node (b3) [right of=a4] {$b3$};
        \node (a2) [right of=c1] {$a2$};
        \node (c4) [right of=b3] {$c4$};
        \node (b1) [right of=a2] {$b1$};
        \node (a1') [right of=c2'] {$a1$};
        \node (a3) [right of=c4] {$a3$};
        \node (b2') [right of=a1'] {$b2$};
        \node (c2) [right of=b1] {$c2$};
        \node (c1') [right of=b2'] {$c1$};
        \node (b4) [right of=a3] {$b4$};
        \node (a1'') [right of=c2] {$a1$};
        \node (a2') [right of=c1'] {$a2$};
        \node (c3') [right of=b4] {$c3$};
        \node (b1'') [right of=a2'] {$b1$};

\newcommand{\s}[3]{\filldraw[fill=blue!20,draw=black!35] 
            (#1.center) -- (#2.center) -- (#3.center) -- (#1.center);
        }

        \begin{pgfonlayer}{background}
            \s{a1}{b2}{c3}
            \s{a1'}{b2'}{c4}
            \s{a1'}{b3}{c2'}
            \s{a1'}{b3}{c4}
            \s{a1''}{b4}{c2}
            \s{a1''}{b4}{c3'}
            \s{a2'}{b1''}{c3'}
            \s{a2}{b1}{c4}
            \s{a2}{b3}{c1}
            \s{a2}{b3}{c4}
            \s{a2'}{b4}{c1'}
            \s{a2'}{b4}{c3'}
            \s{a3}{b1}{c2}
            \s{a3}{b1}{c4}
            \s{a3}{b2'}{c1'}
            \s{a3}{b2'}{c4}
            \s{a3}{b4}{c1'}
            \s{a3}{b4}{c2}
            \s{a4}{b1'}{c2'}
            \s{a4}{b1'}{c3}
            \s{a4}{b2}{c1}
            \s{a4}{b2}{c3}
            \s{a4}{b3}{c1}
            \s{a4}{b3}{c2'}
        \end{pgfonlayer}
    \end{tikzpicture}
\caption[Output complex for renaming with $n=3$, $m=4$]{Output complex
    for renaming with $n=3$, $m=4$.  Each vertex is labeled by a
    process ID $(a,b,c)$ and a name $(1,2,3,4)$.  Observe that the left and right
    edges of the complex have the same sequence of labels, as do the top and
    bottom edges; the complex thus folds up into a (twisted) torus. 
    (This is a poor imitation of part of \cite[Figure 9]{HerlihyS1999}.)}
\label{fig-renaming-output-complex}
\end{figure}

\myChapter{Approximate agreement}{2011}{}
\label{chapter-approximate-agreement}

The
\index{agreement!approximate}
\concept{approximate agreement}~\cite{DolevLPSW1986}
or
\index{agreement!$ε$-}
\concept{$ε$-agreement}
problem is another relaxation of consensus where input and output
values are real numbers, and a protocol is required to satisfy
modified validity and agreement conditions.

Let $x_i$ be the input of process $i$ and $y_i$ its output.  Then a
protocol satisfies approximate agreement if it satisfies:
\begin{description}
\item[Termination] Every non-faulty process eventually decides.
\item[Validity] Every process returns an output within the range of
inputs.  Formally, for all $i$, it holds that
$(\min_j x_j) ≤ y_i ≤ (\max_j x_j)$.
\item[$ε$-agreement] For all $i$ and $j$, 
$\abs*{i-j} ≤ ε$.
\end{description}

Unlike consensus, approximate agreement has wait-free algorithms for
asynchronous shared memory, which we'll see in
§\ref{section-approximate-agreement-upper-bounds}).  But a curious property of approximate
agreement is that it has no 
\index{wait-free!bounded}
\concept{bounded wait-free} algorithms, even for two processes (see
§\ref{section-approximate-agreement-lower-bound})

\section{Algorithms for approximate agreement}
\label{section-approximate-agreement-upper-bounds}

Not only is approximate agreement solvable, it's actually easily
solvable, to the point that there are many known algorithms for
solving it.

\newFunc{\AAsnapshot}{snapshot}
\newcommand{\AAmaxRound}{r_{\max}}

We'll use the algorithm of Moran~\cite{Moran1995}, mostly as presented
in~\cite[Algorithm 54]{AttiyaW2004} but with a slight bug fix;\footnote{The original algorithm from~\cite{AttiyaW2004}
does not include the test $\AAmaxRound ≥ 2$.  This allows for
bad executions in which process $1$ writes its input of $0$ in
round $1$ and takes a snapshot that includes only its own
input, after which process $2$ runs the algorithm to completion with
input $1$.  Here process $2$ will see $0$ and $1$ in round $1$, and
will write $(1/2, 2, 1)$ to $A[2]$; on subsequent iterations, it will
see only the value $1/2$ in the maximum round, and after
$\ceil{\log_2(1/ε)}$ rounds it will decide on $1/2$.  But if we
now wake process $1$ up, it will decided $0$ immediately based on its
snapshot, which includes only its own input and gives $\spread(x) =
0$.  Adding the extra test prevents this from happening, as new values
that arrive after somebody writes round $2$ will be ignored.}
pseudocode appears
in Algorithm~\ref{alg-approximate-agreement}.\footnote{Showing that
    this particular algorithm works takes a lot of effort.  If I were
    to do this over, I'd probably go with a different algorithm due to
Schenk~\cite{Schenk1995}.}

The algorithm carries out a
sequence of asynchronous
rounds in which processes adopt new values, such that the
\concept{spread} 
 of the vector of all 
values $V_r$ in round $r$, defined as
$\spread V_r = \max V_r - \min V_r$, drops by a factor of $2$ per
round.
This is done by having each process choose a new value in each round
by taking the midpoint (average of min
and max) of all the values it sees in the previous round.  Slow
processes will jump to the maximum round they see rather than
propagating old values up from ancient rounds; this is enough to
guarantee that latecomer values that arrive after some process writes in round
$2$ are ignored.

The algorithm uses a single
snapshot object $A$ to communicate, and each process stores its
initial input and a round number along with its current preference. 
We assume that the initial values in this object all have round
number $0$, and that $\log_2 0 = -\infty$ (which avoids a
special case in the termination test).
\begin{algorithm}
$A[i] ← \langle x_i, 1, x_i \rangle$ \;
\Repeat{$\AAmaxRound ≥ 2$ \KwAnd $\AAmaxRound ≥
\log_2(\spread(\{x'_j\})/ε)$}{
    $\langle x'_1, r_1, v_1 \rangle \dots \langle x'_n, r_n, v_n \rangle
        ← \AAsnapshot(A)$ \;
    $\AAmaxRound ← \max_j r_j$ \;
    $v ← \midpoint \{ v_j \,|\, r_j = \AAmaxRound \}$ \;
    $A[i] ← 
      \langle
        x_i, 
        \AAmaxRound + 1, 
        v
      \rangle$\;
}
\Return $v$\;
\caption{Approximate agreement}
\label{alg-approximate-agreement}
\end{algorithm}

To show this works, we want to show 
that the midpoint operation guarantees that the
spread shrinks by a factor of $2$ in each round.
Let
$V_r$ be the set of all values $v$ that are ever written to the
snapshot object with round number $r$.
Let $U_r \subseteq V_r$ be the set of values that are ever written to
the snapshot object with round number $r$ before some process writes a
value with round number $r+1$ or greater; the intuition here is that
$U_r$ includes only those values that might contribute to the
computation of some round-$(r+1)$ value.
\begin{lemma}
\label{lemma-approximate-agreement-spread}
For all $r$ for which $V_{r+1}$ is nonempty, 
\begin{align*}
\spread(V_{r+1}) &≤ \spread(U_r)/2.
\end{align*}
\end{lemma}
\begin{proof}
Let $U_r^i$ be the set of round-$r$ values observed by a process $i$
in the iteration in which it sees
$\AAmaxRound = r$ in some iteration, if such an iteration exists.
Note that $U_r^i \subseteq U_r$, because if some value with round
$r+1$ or greater is written before $i$'s snapshot, then $i$ will
compute a larger value for $\AAmaxRound$.

Given two processes $i$ and $j$, we can argue from the properties of
snapshot that either $U_r^i \subseteq U_r^j$ or $U_r^j \subseteq
U_r^i$.  The reason is that if $i$'s snapshot comes first, then $j$
sees at least as many round-$r$ values as $i$ does, because the only way
for a round-$r$ value to disappear is if it is replaced by a value in a
later round.  But in this case, process $j$ will compute a larger
value for $\AAmaxRound$ and will not get a view for round $r$.  The
same holds in reverse if $j$'s snapshot comes first.

Observe that if $U_r^i \subseteq U_r^j$, then 
\begin{align*}
\abs*{\,\midpoint(U_r^i) - \midpoint(U_r^j)} &≤ \spread(U_r^j)/2.
\end{align*}
This holds because $\midpoint(U_r^i)$ lies within the interval 
$\left[\min U_r^j, \max U_r^j\right]$, and every point in this interval is within
$\spread(U_r^j)/2$ of $\midpoint(U_r^j)$.  The same holds if $U_r^j
\subseteq U_r^i$.  So any two values written in round $r+1$ are
within $\spread(U_r)/2$ of each other.

In particular, the minimum and maximum values in $V_{r+1}$ are within
$\spread(U_r)/2$ of each other, so
$\spread(V_{r+1}) ≤ \spread(U_r)/2$.
\end{proof}

\begin{corollary}
\label{corollary-approximate-agreement}
For all $r ≥ 2$ for which $V_{r}$ is nonempty, 
\begin{align*}
\spread(V_{r}) &≤ \spread(U_1)/2^{r-1}.
\end{align*}
\end{corollary}
\begin{proof}
By induction on $r$.  For $r=2$, this is just
Lemma~\ref{lemma-approximate-agreement-spread}.  For larger $r$,
use the fact that $U_{r-1} \subseteq V_{r-1}$ and thus
$\spread(U_{r-1}) ≤ \spread(V_{r-1})$ to compute
\begin{align*}
\spread(V_{r})
&≤ \spread(U_{r-1})/2
\\
&≤ \spread(V_{r-1})/2
\\
&≤ (\spread(U_1)/2^{r-2})/2
\\
&= \spread(U_1)/2^{r-1}.
\end{align*}
\end{proof}

Let $i$ be some process that finishes in the fewest number of rounds.
Process $i$ can't finish until it reaches round
$\AAmaxRound+1$, where $\AAmaxRound ≥ \log_2(\spread(\{x'_j\})/ε)$
for a vector of input
values $x'$ that it reads after some process writes round $2$ or
greater.  We have $\spread(\{x'_j\}) ≥ \spread(U_1)$, because
every value in $U_1$ is included in $x'$.  
So $\AAmaxRound ≥ \log_2\left(\spread(U_1)/ε\right)$
and 
$\spread(V_{\AAmaxRound+1})
    ≤ \spread(U_1)/2^{\AAmaxRound}
    ≤ \spread(U_1)/(\spread(U_1)/ε)
    = ε$.
Since any value returned is either included in
$V_{\AAmaxRound+1}$ or some later $V_{r'} \subseteq
V_{\AAmaxRound+1}$, this gives us that the spread of all the outputs
is less than $ε$: Algorithm~\ref{alg-approximate-agreement}
solves approximate agreement.

The cost of Algorithm~\ref{alg-approximate-agreement} depends on the
cost of the snapshot operations, on $ε$, and on the initial
input spread $D$.  For linear-cost snapshots, this works out to $O(n
\log (D/ε))$.

\section{Lower bound on step complexity}
\label{section-approximate-agreement-lower-bound}

The dependence on $D/ε$ is necessary, at least for
deterministic algorithms.  Here we give a lower
bound due to Herlihy~\cite{Herlihy1991pram}, which shows that any
deterministic approximate agreement algorithm takes at least $\log_3 (D/ε)$
total steps even with just two processes.

Define the \concept{preference} of a process in some configuration as
the value it will choose if it runs alone starting from this
configuration.  The preference of a process $p$ is well-defined because
the process is deterministic; it also can only change as a result of a
write operation by another process $q$ (because no other operations are
visible to $p$, and $p$'s own operations can't change its preference).  The
validity condition means that in an initial state, each process's
preference is equal to its input.

Consider an execution with two processes $p$ and $q$, where $p$ starts
with preference $p_0$ and $q$ starts with preference $q_0$.
Run $p$ until it is about to perform a write that would change $q$'s
preference.  Now run $q$ until it is about to change $p$'s preference.
If $p$'s write no longer changes $q$'s preference, start $p$ again and
repeat until both $p$ and $q$ have pending writes that will change the
other process's preference.  Let $p_1$ and $q_1$ be the new
preferences that result from these operations.  The adversary can now
choose between running $P$ only and getting to a configuration with
preferences $p_0$ and $q_1$, $Q$ only and getting $p_1$ and $q_0$, or
both and getting $p_1$ and $q_1$; each of these choices incurs at
least one step.  By the triangle inequality,
$\abs*{p_0 - q_0}
≤
\abs*{p_0 - q_1}
+ \abs*{q_1 - p_1}
+ \abs*{p_1 - q_0}$,
so at least on of these configurations has a spread between
preferences that is at least $1/3$ of the initial spread.  It follows
that after $k$ steps the best spread we can get is $D/3^k$, requiring
$k ≥ \log_3 (D/ε)$ steps to get $ε$-agreement.

Herlihy uses this result to show that there are decisions problems that have
wait-free but not bounded wait-free deterministic solutions using
registers.  Curiously, the lower bound says nothing about the
dependence on the number of processes; it is conceivable that there is
an approximate agreement protocol with running time that depends only
on $D/ε$ and not $n$.

\part{Other communication models}
\label{part-other-models}

\myChapter{Overview}{2026}{}

In this part, we consider models that don't fit well into the standard
message-passing or shared-memory models.  These includes models where
processes can directly observe the states of nearby processes
(Chapter~\ref{chapter-self-stabilization});
where computation is inherently local and the emphasis is on computing
information about the communication graph
(Chapter~\ref{chapter-distributed-graph-algorithms}); 
where processes
(in the form of robots) communicate only by observing each others'
locations and movements (Chapter~\ref{chapter-mobile-robots});
where processes can transmit only beeps, and are able to
observe only whether at least one nearby process beeped
(Chapter~\ref{chapter-beeping});
and where processes wander about and exchange
information only with processes they physically encounter
(Chapter~\ref{chapter-population-protocols});

Despite the varying
communication mechanisms, these models all share the usual features of
distributed systems, where processes must contend with nondeterminism
and incomplete local information.

\myChapter{Self-stabilization}{2026}{}
\label{chapter-self-stabilization}

A \index{self-stabilization}\concept{self-stabilizing} algorithm has the property that, starting
from any arbitrary configuration, it eventually reaches a
\concept{legal} configuration, and this property is \concept{stable}
in the sense that it remains in a legal configuration
thereafter.  The notion of which configurations are legal depends on
what problem we are trying to solve, but the overall intuition is that
an algorithm is self-stabilizing if it can recover from arbitrarily
horrible errors, and will stay recovered as long as no new errors
occur.

It's generally not possible to detect whether the algorithm is in a
legal configuration from the inside: if a process has a bit that says
that everything is OK, the adversary can set that bit in the initial
configuration, even if everything is in fact broken.  So
self-stabilizing algorithms don't actually terminate: at best, they
eventually converge to a configuration where the necessary ongoing paranoid
consistency checks produce no further changes to the configuration (a property
called \index{self-stabilization!silent}\concept{silent
self-stabilization}.

The idea of self-stabilization first appeared in a paper by
Edsger Dijkstra~\cite{Dijkstra1974}, where he considered the problem
of building robust token-ring networks.  In a token-ring network,
there are $n$ nodes arranged in a directed cycle, and we want a single
token to circulate through the nodes, as a mechanism for enforcing
mutual exclusion: only the node currently possessing the token can
access the shared resource.

The problem is: how do you get the token started?  Dijkstra worried
both about the possibility of starting with no tokens or with more
than one token, and he wanted an algorithm that would guarantee that,
from any starting state, eventually we would end up with exactly one
token that would circulate as desired.  He called such an algorithm
\concept{self-stabilizing}, and gave three examples, the simplest of
which we will discuss in §\ref{section-self-stabilizing-token-ring}
below.  These became the
foundation for the huge field of self-stabilization,
which spans thousands of papers, at least one
textbook~\cite{Dolev2000}, a specialized conference (SSS, the
\emph{International Symposium on Stabilization, Safety, and Security
in Distributed Systems)}, and its own domain name
\url{http://www.selfstabilization.org/}.  We won't attempt to
summarize all of this, but will highlight a few results to give a
sampling of what self-stabilizing algorithms look like.

\section{Model}

Much of the work in this area, dating back to Dijkstra's
original paper, does not fit well in either the message-passing or
shared-memory models that we have been considering in this class, both
of which were standardized much later.  Instead, Dijkstra assumed that
processes could, in effect, directly observe the states of their
neighbors.  A self-stabilizing program would consist of a collection
of what he later called \index{command!guarded}\indexConcept{guarded
command}{guarded commands}~\cite{Dijkstra1975}, statements of the form
``if [some condition is true] then [update my state in this way].''
In any configuration of the system, one or more of these guarded
commands might have the if part (the \concept{guard}) be true; these
commands are said to be \concept{enabled}.

A step consists of one or more of these enabled commands being
executed simultaneously, as chosen by an adversary scheduler, called
the \index{daemon!distributed}{distributed daemon}. The usual fairness
condition applies: any process that has an enabled command eventually
gets to execute it. If no commands are enabled, nothing happens. With
the \index{daemon!central}\concept{central daemon} variant of the
model, only one step can happen at a time.  With the
\index{daemon!synchronous}\concept{synchronous daemon}, every enabled
step happens at each time. Note that both the central and synchronous
daemons are special cases of the distributed daemon.

More recent work has tended to assume a distinction between the part
of a process's state that is visible to its neighbors and the part
that isn't.  This usually takes the form of explicit
\index{register!communication}
\indexConcept{communication register}{communication registers}
or
\index{register!link}
\indexConcept{link register}{link registers},
which allow a process to write a specific message for a specific
neighbor.  This is still not quite the same as standard
message-passing or shared-memory, because a process is often allowed to read
and write multiple link registers atomically.

\section{Token ring circulation}
\label{section-self-stabilizing-token-ring}

For example, let us consider Dijkstra's token ring circulation
algorithm.  There are several versions of this in Dijkstra's
paper~\cite{Dijkstra1974}.  We will do the unidirectional $(n+1)$-state version, which is the
simplest to describe and analyze.

For this algorithm, the processes are numbered as elements $0 \dots
n-1$, with all arithmetic on process IDs being done modulo
$n$.\footnote{In Dijkstra's paper, there are $n+1$ processes numbered
$0\dots n$, but this doesn't really make any difference.}
Each process $i$ can observe both its own state and that of its
predecessor at $(i-1) \bmod n$.

Process $0$ has a special role and has different code from the
others, but the rest of the processes are symmetric.  
Each process $i$ has a variable $\ell_i$ that takes on values in the
range $0\dots n$, interpreted as elements of $ℤ_{n+1}$.
The algorithm is
given in Algorithm~\ref{alg-dijkstra-token-ring}.

\begin{algorithm}
    Code for process 0:\\
    \lIf{$\ell_{0} = \ell_{n-1}$}{$\ell'_0 ← (\ell_{n-1} + 1) \bmod (n+1)$}
    Code for process $i≠0$:\\
    \lIf{$\ell_{i} ≠ \ell_{i-1}$}{$\ell'_i ← \ell_{i-1}$}
    \caption{Dijkstra's large-state token ring algorithm~\cite{Dijkstra1974}}
    \label{alg-dijkstra-token-ring}
\end{algorithm}

In this algorithm, the nonzero processes just copy the state of the
process to their left.  The zero process increments its state if it
sees the same state to its left.  Note that the nonzero processes have
guards on their commands that might appear useless at first glance,
but these are there ensure that the adversary can't waste steps
by getting nonzero processes to carry out operations that have no effect.

What does this have to with tokens?  The algorithm includes an
additional interpretation of the state, which says that:
\begin{enumerate}
    \item If $\ell_0 = \ell_{n-1}$, then $0$ has a token, and
    \item If $\ell_i ≠ \ell_{i-1}$, for $i≠0$, then $i$ has a token.
\end{enumerate}

Like the update rule, the token rule can be evaluated by a node that can
only see its predecessor. This allows it to do detect when it acquires
the token and do whatever leaderly things it needs to before applying
an update to pass it on to the next process.

Using the token rule instantly guarantees that there is at least one token: if none of
the nonzero processes have a token, then all the $\ell_i$ variables
are equal.  But then $0$ has a token.  It remains though to show that
we eventually converge to a configuration where at most one process
has a token.

Define a configuration $\ell$ as legal if there is some value $j$ such
that $\ell_i = \ell_j$ for all $i≤j$ and $\ell_i = \ell_j - 1
\pmod{n+1}$ for all $i > j$. When $j = n-1$, this makes all $\ell_i$
equal, and $0$ has the only token. When $j < n-1$, then $\ell_0 ≠
\ell_{n-1}$ (so $0$ does not have a token), $\ell_j ≠ \ell_{j+1}$ (so
$j+1$ has a token), and $\ell_i = \ell_{i+1}$ for all $i∉{j,n-1}$ (so
nobody else has a token). That each legal configuration has exactly
one token partially justifies our definition of legal configurations.

If a configuration $\ell$ is legal, then when $j = n-1$, the only
enabled step is $\ell'_0 ← (\ell_{n-1} + 1) \bmod (n+1)$; when $j <
n-1$, the only enabled step is $\ell'_{j+1} ← \ell_j$. In either case,
we get a new legal configuration $\ell'$. So the property of being a
legal configuration is stable, which is the other half of justifying
our definition.

Now we want to show that we eventually converge to a legal
configuration. Fix some initial configuration $\ell^0$, and let $c$ be
some value such that $\ell^0_i ≠ c$ for all $i$.
(There is at least one such $c$ by the Pigeonhole Principle.) We will
argue that there is a sequence of configurations with $c$ as a prefix
of the values that forms a bottleneck forcing us into a legal
configuration.

\begin{lemma}
    \label{lemma-Dijkstra-bottleneck}
    Let $\ell^0,\ell^1,\dots$ be the sequence of configurations in
    some execution of Dijkstra's token ring circulation algorithm.
    Let $0≤c≤n$ be such that $\ell^0_i ≠ c$ for all $i$.
    Then for any configuration $\ell^t$, either $t$ is legal, or there
    is some $0≤j<n$ such that $\ell^t_i = c$ if and only if $i < j$.
\end{lemma}
\begin{proof}
    By induction on $t$. For the base case, $\ell^0$ satisfies
    $\ell^0_i = c$ if and only if $i < j$ when $j = 0$.

    If $\ell^t$ is legal, $\ell^{t+1}$ is also legal. So the
    interesting case is when $\ell^t$ is not legal. In this case,
    there is some $0≤j<n$ such that $\ell^t_i = c$ if and only if
    $i < j$.

    If $j = 0$, then $\ell^t_i ≠ c$ for all $i$. Then the only way to
    get $\ell^{t+1}_i = c$ is if $i = 0$. But then $\ell^{t+1}$
    satisfies the condition with $j'=1$.

    If $0 < j < n$, then $\ell^t_i = c$ for at least one $i < j$, and
    $\ell^t_{n-1} ≠ c$ since $n-1 \not< j$. So we may get a transition
    that sets $\ell^{t+1}_j = \ell^t_{j-1} = c$, giving a new
    configuration $\ell^{t+1}$ that satisfies the induction hypothesis
    with $j' = j+1$, or we may get a transition that does not create
    or remove any copies of $c$. In either case the induction goes
    through.
\end{proof}

To show that we eventually hit this bottleneck, we use a potential
function.
Starting in some initial configuration $\ell^0$, let $c$ be some
missing value in $\ell^0$ as defined above.
For any configuration $\ell$, define 
$g(\ell) = (c-\ell_0) \bmod
(n+1)$ to be the gap between $\ell_0$ and $c$.
For each $i ∈ \Set{0,\dots,n-2}$, define $u_i(\ell) = [\ell_i ≠ \ell_{i+1}]$ to be
the indicator variable for whether $i$ is \emph{unhappy} with its successor,
because its successor has not yet agreed to adopt its
value.\footnote{The notation $[P]$, where $P$ is some logical
predicate, is
called an \concept{Iverson bracket} and means the function that is $1$
when $P$ is true and $0$ when $P$ is false.}
The idea is that unhappiness moves right when some $i ≠ 0$ copies its
predecessor and that the gap drops when $0$ increments its value. By
weighting these values appropriately, we can arrange for a function
that always drops.

Let
\begin{equation}
    \label{eq-dijkstra-potential}
    Φ(\ell) = ng(\ell)  + ∑_{i=0}^{n-2} (n-1-i) u_i(\ell).
\end{equation}

Most of the work here is being done by the first two terms. The
$g$ term tracks the gap between $\ell_0$ and $c$, weighted by $n$. The sum tracks unhappiness, weighted by distance to position
$n-1$.

In the initial configuration $\ell^0$, $g$ is at most $n$, and each $u_i$ is at
most $1$, so $Φ(\ell^0) = O(n^2)$. We also have that $Φ≥0$ always; if
$Φ=0$, then $g=0$ and $u_i=0$ for all $i$ implies we are in an all-$c$
configuration, which is legal. So we'd like to 
argue that every step of the algorithm in a
non-legal configuration reachable from $\ell^0$ reduces $Φ$ by
at least $1$, forcing us into a legal configuration after $O(n^2)$
steps.

Consider any step of the algorithm starting from a non-legal
configuration $\ell^t$ with
$Φ(\ell^t) >0$ that satisfies the condition in
Lemma~\ref{lemma-Dijkstra-bottleneck}:
\begin{itemize}
    \item 
If it is
a step by $i≠0$, then $u_{i-1}$ changes from $1$ to $0$, reducing $Φ$
by $(n-1-(i-1)) = n-i$.  It may be that $u_i$ changes from $0$ to $1$,
increasing $Φ$ by $n-i-1$, but the sum of these changes is at most
        $-1$.
\item If it is a step by $0$, then $u_0$ may increase from $0$ to $1$,
    increasing $Φ$ by $n-1$. 
        But $g$ drops by $1$ as long as $\ell^t_0 ≠ c$, reducing $Φ$
        by $n$, for a total change of at most $-1$.
        The case $\ell^t = c$ is excluded by the assumption that
        $\ell^t$ is non-legal and satisfies the conditions of
        Lemma~\ref{lemma-Dijkstra-bottleneck}, as the only way for $0$
        to change its value away from $c$ is if $\ell^t_{n-1}$ is also $c$.
\end{itemize}

Since the condition of Lemma~\ref{lemma-Dijkstra-bottleneck} holds for
any reachable $\ell^t$, as long as we are in a non-legal
configuration, $Φ$ drops by at least $1$ per step. If we do not reach
a legal configuration otherwise, $Φ$ can only drop
$O(n^2)$ times before hitting $0$, giving us a legal configuration.
Either way, the configuration stabilizes in $O(n^2)$ steps.

\section{Synchronizers}

Self-stabilization has a curious relationship with failures: the
arbitrary initial state corresponds to an arbitrarily bad initial
disruption of the system, but once we get past this there are no
further failures.  So it is not surprising that many of the things we
can do in a failure-free distributed system we can also do in a
self-stabilizing system.  One of these is to implement a synchronizer,
which will allow us to pretend that our system is synchronous even if
it isn't.

The self-stabilizing synchronizer we will describe here, due to
Awerbuch~\etal~\cite{AwerbuchKMPV1993,AwerbuchKMPV1997}, is a variant of the alpha
synchronizer.  It assumes that each process can observe the states of
its neighbors and that we have a central daemon (meaning that one
process takes a step at a time).

To implement this synchronizer in a self-stabilizing system, each
process $v$ has a variable $P(v)$, its current pulse.  
We also give each process a rule for adjusting $P(v)$ when it takes a
step.
Our goal is to
arrange for every $v$ to increase its pulse infinitely often
while staying at most one ahead of its neighbors $N(v)$.
Awerbuch~\etal{} give several possible rules for achieving this goal,
and consider the effectiveness of each.

The simplest rule is taken directly from the alpha synchronizer.  When
activated, $v$ sets
\begin{equation*}
    P(v) ← \min_{u∈N(v)} (P(u)+1)
\end{equation*}

This rule works find as long as every process starts synchronized.
But it's not self-stabilizing.  A counterexample, given in the paper,
assumes we have $10$ processes organized in a ring.  By carefully
choosing which processes are activated at each step, we can go through
the following sequence of configurations, where in each configuration
the updated node is shown in boldface:
\begin{center}
    1234312343\\
    1234\textbf{2}12343\\
    12342\textbf{3}2343\\
    123423\textbf{4}343\\
    1234234\textbf{5}43\\
    123423454\textbf{2}\\
    \textbf{3}234234542\\
    3\textbf{4}34234542\\
    34\textbf{5}4234542
\end{center}
Here the final configuration is identical to the original if we
increment each value by one and shift the values to the left one
position.  So we can continue this process indefinitely.  But at the
same time, each configuration has at least one pair of adjacent nodes
whose values differ by more than one.

The problem that arises in the counterexample is that sometimes values
can go backwards.
A second rule proposed by Awerbuch~\etal{} avoids this problem by
explicitly 
forbidding $P(v)$ to drop, using the rule:
\begin{equation*}
P(v) ← \max\parens*{P(v), \min_{u∈N(v)} (P(u) + 1)}
\end{equation*}

This turns out to be self-stabilizing, but the time to stabilize is
unbounded even in small networks.  One counterexample is a network
consisting of just three nodes:
\begin{center}
\begin{tikzpicture}
    \node[labeled] (a) at(0,0) {$1$};
    \node[labeled] (b) at(1.5,0) {$1$};
    \node[labeled] (c) at(3,0) {$10^{50}$};
    \path
        (a) edge (b)
        (b) edge (c)
        ;
\end{tikzpicture}
\end{center}
If we run the nodes in round-robin order, the left two nodes will
eventually catch up to the rightmost, but it will take a while.

After some further tinkering, the authors present their optimal rule,
which they call \concept{max minus one}:
\begin{equation*}
    P(v) ←
    \begin{cases}
        \min_{u ∈ N(v)} (P(u)+1) & \text{if $P(v)$ looks legal,} \\
        \max_{u ∈ N(v)} (P(u)-1) & \text{otherwise.}

    \end{cases}
\end{equation*}
Here $P(v)$ looks legal if it is within $±1$ of all of its neighbors.

The intuition for why this works is that the most backward node pulls
the rest down to its level in $O(D)$ time\footnote{Defining a time
unit as a minimum interval in which every process takes at least one
step.} using the max-minus-one rule, after which we just get the alpha
synchronizer since everybody's local values look legal.

The actual proof uses a potential function at each node $v$ given by
\begin{equation*}
    \label{eq-max-minus-one-potential}
    φ(v) = \max_u \parens*{P(u) - P(v) - d(u,v)},
\end{equation*}
where $d(u,v)$ is the distance between $u$ and $v$ in the graph.
This is zero if the skew between any pair of nodes is equal to the
distance, which is the most we can expect from a synchronizer. The
proof shows that applying the max-minus-one rule never increases
$φ(v)$, and decreases it by at least $1$ whenever a node $v$ with positive
$φ(v)$ changes $P(v)$. Because this only gives a bound of $∑ φ(v)$,
which can be arbitrarily big,
the rest of the proof uses a second potential function
\begin{equation*}
    Φ(v) = \min_u \SetWhere{ d(u,v) }{P(u)-P(v)-d(u,v) = φ(v)},
\end{equation*}
which measures the distance from $v$ to the nearest node $u$ that
supplies the maximum in $φ(v)$. It is shown that $Φ(v)$ drops by $1$
per time unit. When it reaches $0$, then $φ(v) = P(v) - P(v) - d(v,v)
= 0$. Since $Φ(v)$ can never start at more than the diameter $D$, this
implies convergence in $D$ time units.

The intuition for why this works is that if the closest node $u$ to
$v$ with $P(u)$ too high is at distance $d$, then max-minus-one will
pull $P(w)$ up for some node $w$ at distance $d-1$ the next time $w$
takes a step. The full set of cases is more complicated, and we'll
skip over the details of the argument here. If you are interested, the
presentation in the paper is not too hard to follow.

The important part is that once we have a synchronizer, we can
effectively assume synchrony in other self-stabilizing algorithms.
We just run the synchronizer underneath our main protocol, and when
the synchronizer stabilizes, that gives us the initial starting point
for the main protocol.  Because the main protocol itself should
stabilize starting from an arbitrary configuration, any insanity
produced while waiting for the synchronizer to converge is eventually
overcome.

\section{Spanning trees}

\newData{\BFroot}{root}
\newData{\BFdist}{dist}

The straightforward way to construct a spanning tree in a graph is to
use Bellman-Ford~\cite{Bellman1958,Ford1956} to compute a
breadth-first search tree rooted at the node with lowest ID.  This has
a natural implementation in the self-stabilizing model: each process
maintains $\BFroot$ and $\BFdist$, and when a process takes
a step, it sets $\BFroot$ to the minimum of its own ID and the minimum
$\BFroot$ among its neighbors, and sets $\BFdist$ to $0$ if it has the
minimum ID, or to one plus the minimum distance to the root among its
neighbors otherwise.
It is not hard to show that in the absence of errors, this
converges to a configuration where every node knows the ID of the root
and its distance to the root in $O(D)$ time, where $D$ is the diameter
of the network.  A spanning tree can then be extracted by the usual
trick of having each node select as parent some neighbor closer to the
root.

But it fails badly if the system starts in an arbitrary state, because
of the \index{root!ghost}\concept{ghost root} problem.  Suppose that
some process wakes up believing in the existence of a distant, fake
root with lower ID than any real process.  This fake root will
rapidly propagate through the other nodes, with distances increasing
without bound over time.  For most graphs, the algorithm will never
converge to a single spanning tree.

Awerbuch~\etal~\cite{AwerbuchKMPV1993} solve this problem by assuming
a known upper bound on the maximum diameter of the graph.  Because the
distance to a ghost root will steadily increase over time, eventually
only real roots will remain, and the system will converge to a correct
BFS tree.

It's easiest to show this if we assume synchrony, or at least some
sort of asynchronous round structure.  Define a round as
the minimum time for every node to update at least once.  Then the
minimum distance for any ghost root rises by at least one per round,
since any node with the minimum distance either has no neighbor with
the ghost root (in which case it picks a different root), or any neighbor that
has the ghost root has at least the same distance (in which case it
increases its distance)
Once the minimum distance exceeds the upper bound
$D'$, all the ghost roots will have been eliminated, and only real
distances will remain.  This gives a stabilization time (in rounds)
linear in the upper bound on the diameter.

\section{Self-stabilization and local algorithms}

In Chapter~\ref{chapter-distributed-graph-algorithms}, we will look at
algorithms in the \concept{LOCAL} model, where named processes in a
synchronous network, organized as an unknown graph, can send a
polynomial-sized message to each neighbor in each round and perform
arbitrary computation locally. The goal is usually to compute some
property of the graph quickly, often in significantly fewer rounds
than the diameter of the graph.

There is a close connection between self-stabilizing algorithms and
the LOCAL model. The idea is that if we have a local algorithm that runs
in $f(n)$ rounds, each process can propagate its information in a
self-stabilizing way to all nodes at distance at most $f(n)$, and we
can reconstruct the output of the local algorithm whenever this
information changes.

For each node $u$, let $x_u$ be its input value; we assume that this
is fixed once the system stabilizes.
The state of $u$ will be a table $T_u$, where $T_u$ is a
partial function from sequences of nodes of length at most $f(n)$ to
input values. 
We can represent this partial function as a set of ordered pairs
$T_u = \Tuple{w,x}$, where we write $T_u(w) = x$ if $x$ is the unique
value such that $\Tuple{w,x} ∈ T_u$, or $T_u(w) = ⊥$ if there is no
such value.
We have one rule at each node $u$, which we can imagine is guarded so that it fires
only if it changes $T_u$:
\begin{equation}
    T_u \gets
    \Set{\Tuple{u, x_u}}
    ∪ \bigcup_{v∈δ(u)} \SetWhere{ \Tuple{uw,x} }{ \card{uw} \le f(n),
    T_v(w) = x}
    \label{eq-self-stabilizing-local}
\end{equation}

We can now argue that, after stabilization, this process
eventually converges to $T_u$ consisting precisely of 
the set of all pairs $\Tuple{w,x_v}$ where $w$ is a $u$–$v$ path of
length at most $f(n)$ and
$x_v$ is the input to $v$.
Indeed, this works under almost any reasonable assumption about
scheduling. The relevant lemma:
\begin{lemma}
    \label{lemma-self-stabilizing-local}
    Starting from any initial configuration, for any sequence $w$ of
    at most $f(n)$ vertices starting at $u$ and ending at $v$,
    if \eqref{eq-self-stabilizing-local} fires for each node in $w$ in
    reverse order, then $T_u(w) = x_v$ if $w$ is a $u$–$v$ path,
    and $T_u(w) = ⊥$ otherwise.
\end{lemma}
\begin{proof}
The proof is by induction on the length of $w$. The base case is when
    $\card{w} = 1$, implying $w = u = v$. Here rule
    \eqref{eq-self-stabilizing-local} writes 
    $\Tuple{u,x_u}$ to $T_u$, giving $T_u(u) = x_u$ as claimed.

    For a sequence $w = uw'$ where $w'$ is a nonempty path from some node
    $u'$ to $v$, if $u'$ is a neighbor of $u$, then firing rule
    \eqref{eq-self-stabilizing-local} at $u$ after firing the rule for
    each node in $w'$ has $T_u(uw') \gets T_{u'}(w) = x_v$ by the
    induction hypothesis. If $uw'$ is not a path from $u$ to $v$, then
    either $u'$ is not a neighbor of $u$, or
    $w'$ is not a path from $u'$ to $v$ and $T_{u'}(w') = ⊥$ by the
    induction hypothesis. In either case, $T_u(uw') \gets ⊥$.
\end{proof}

What does this buy us? Suppose we have a deterministic synchronous
algorithm that runs in $f(n)$ rounds. Starting from a stable
configuration, Lemma~\ref{lemma-self-stabilizing-local} tells us 
that any fair daemon will eventually leave us in a configuration where
each node $u$ stores in $T_u$ both the inputs of all nodes within distance
$f(n)$ and enough information to reconstruct how they are connected.
So $u$ can simulate the execution of any node at distance $d$ for up
to $f(n)-d$ rounds. In particular, it can simulate its own execution
for $f(n)$ rounds, computing the same output as it would produce in
the LOCAL model.

\myChapter{Distributed graph algorithms}{2026}{}
\label{chapter-distributed-graph-algorithms}

In Chapter~\ref{chapter-self-stabilization}, we saw that certain
classes of ``local'' algorithms have a straightforward conversion to
self-stabilizing algorithms. In this chapter, we'll look more closely
at what kinds of problems can be solved with this kind of locality,
where each process is limited in how much information it can acquire
quickly by distance or constraints on message size.

Often we will be trying to compute some property of the communication
graph, giving us a \concept{distributed graph algorithm}. The field of
distributed graph algorithms is very active, and we will only be able
to touch on a few highlights. A more comprehensive introduction can be
found in the on-line textbook of Hirvonen and
Suomela~\cite{HirvonenS2025}, which also informed some of the presentation
in this chapter (particularly §\ref{section-CONGEST-APSP}).

\section{The LOCAL and CONGEST models}

The LOCAL and CONGEST models were defined by Peleg~\cite{Peleg2000} to
formalize the idea of local distributed computation. Similar models
had been considered previously without being specifically
named~\cite{Linial1992}, but these names are now standard.

The LOCAL model is a synchronous message-passing model where the
processes are organized into a graph, all run the same code,
and can communicate only with their neighbors in the graph. 
To break symmetry, each process starts
with a unique ID that is polynomial in the number of processes $n$.
The processes may also start with local inputs, but often we are
interested simply in computing some property of the graph itself.
There is no bound on the size of messages.

The CONGEST model is like the local model, but messages are limited to
$O(\log n)$ bits. More generally, the $\text{CONGEST}(b)$ model allows
messages of size $b$, making $\text{LOCAL} = \text{CONGEST}(∞)$ and
$\text{CONGEST} = \text{CONGEST}(O(\log n))$. The choice of $O(\log
n)$ as the default bound allows each message to contain $O(1)$ process
ids (and perhaps other information).

In both models, we usually assume that the processes do \emph{not}
know the structure of the graph or their place in it. But for specific
problems, we might require the graph to be from some restricted class
(for example, a ring, a tree, or a clique).

We'll mostly focus on the LOCAL model in this chapter, using
the problem of graph coloring as our primary example.

\section{Local graph coloring}

One of the first problems studied in the LOCAL model is local
graph coloring~\cite{Linial1992}, where we wish to assign each node in
the graph a small label distinct from its neighbors. Because the nodes
initially start with large distinct labels, graph coloring in the
LOCAL model shares some similarities with renaming
(Chapter~\ref{chapter-renaming}), since we will use
the unique IDs as a starting point for generating the colors.

\subsection{Coloring graphs with out-degree 1}
\label{section-Cole-Vishkin}

Let us start by describing a classic local algorithm for
$3$-coloring a directed graph with maximum out-degree 1, a class of
graphs that includes both cycles and rooted trees. The algorithm we
will use is ultimately due to Cole and Vishkin~\cite{ColeV1986},
although the application to local graph coloring was given by
Linial~\cite{Linial1992}, and the version given here incorporates some
additional features from Peleg's textbook~\cite{Peleg2000}.

The core idea from the Cole and Vishkin algorithm is to treat each identity
$x$ as a long bit-string $x_{k} x_{k-1} \dots x_0$, where $k =
\floor{\lg N}$ and $x = ∑ 2^i x_i$ and repeatedly apply an operation that
to shorten these IDs while maintaining the property that neighbors have distinct IDs.

At each synchronous round, each process adopts a new identity based on
its old identity $x$ and the identity $y$ of its successor. We look
for the smallest index $i$ for which $x_i ≠ y_i$.  We then generate a
new identity $2i + x_i$; this is the same as taking the bit-vector
representation of $i$ and shifting it one position to the left so we
can append $x_i$ to the end of it. 

In the case of a node with no successor, we pretend that it has a
successor with $y_0 ≠ x_0$.
This will knock $x$ down to just its last bit $x_0$.

We now argue that this never produces two adjacent identities that are
the same. Consider three consecutive identities $x$, $y$, and $z$. Let
$i$ be the smallest index with $x_i ≠ y_i$, and let $j$ be the
smallest index with $x_j ≠ y_j$.  If $j ≠ i$, then my successor's new
identity $2j + y_{j}$ will not equal my new identity $2i + x_i$,
because the initial bits will be different.  But if $j = i$, then my
successor's new identity is $2i+y_i ≠ 2i+x_i$ because $y_i ≠ x_i$.

Assuming that the largest initial color is $N$, the largest possible
value for $i$ is $\floor{\lg N}$, and so the largest possible value for
$2i+x_i$ is $2\floor{\lg N}+1$.  
Iterating the function $2\floor{\lg N}+1$ converges to at most $5$ after $O(\log^* N)$ rounds, which gives
us six colors $0,\dots,5$, where no two adjacent processes have the
same color.

To reduce this to three colors, add a phase for each $c∈\Set{3,4,5}$
to eliminate $c$. In each phase, we carry out a two-stage process. The
first stage cleans up the neighborhood around each node, and the
second stage replaces all copies of $c$ with some color in
$\Set{0,1,2}$.

In the first stage, we shift all
colors down, by having each node switch its color to that of its
successor (or some new color chosen from $\Set{0,1,2}$ if it doesn't
have a successor). The reason for doing this is that it guarantees
that each node's predecessors will all share the same color, meaning
that that node now has at most two colors represented among its
predecessors and successor. 
At the same time, it doesn't create any new pair of adjacent nodes
with the same color.

For the second stage, each node $v$ that
currently has color $c$ chooses a new color from $\Set{0,1,2}$
that is the smallest color that doesn't appear in its neighborhood.
Since none of $v$'s neighbors change color during this stage (they don't have color
$c$), this replaces all instances of $c$ with a color from
$\Set{0,1,2}$ while keeping all edges two-colored.
After doing this for all $c∈\Set{3,4,5}$, the only colors left are in
$\Set{0,1,2}$.

Doing the $6$ to $3$ reduction in the obvious way takes an additional
$6$ rounds, which is (asymptotically) dominated by the $O(\log^* N)$
rounds of reducing from initial IDs with values up to $N$.

Because the reduction to $6$ colors technically requires more than
constant time, it's theoretically necessary for the nodes to have an
upper bound on $O(\log^* N)$ to know when to switch to the $6→3$ step.
In practice, $\log^* N ≤ 7$ for any $N$ that can be
represented by bits encoded using 
subatomic particles contained in the visible universe,
so we may be able to get away with fixing a constant.
Despite this useful property
of $\log^*$ in practice, we can't get rid of it in theory, because of
an $Ω(\log^* n)$ bound on coloring rings shown in the next section.

\subsection{Lower bound for rings}

Using a Ramsey-theoretic argument, Linial~\cite{Linial1992} showed
that $Ω(\log^* n)$ is a lower bound on the time to color a directed
ring with $n$ nodes in the LOCAL model, which implies that the
algorithm of the previous section is optimal up to constants, since a
directed ring is a special case of a graph with out-degree $1$. We'll
describe here a simplified version of Linial's original proof given by
Laurinharju and Suomela~\cite{LaurinharjuS2014}. (The Laurinharju and
Suomela paper is only two pages long, so it may be worth skipping the
rest of this section and just reading it in the original.)

The idea is that any coloring algorithm in the local model that runs
in time $T$ assigns a color to each node based only on the initial
IDs of the $2T+1$ nodes that are within $T$ hops. So we can represent any
possible deterministic coloring algorithm by specifying the mapping
from these $2T+1$ IDs to colors.

Define a \concept{$k$-ary $c$-coloring
function}\index{function!$k$-ary $c$-coloring}\index{coloring
function!$k$-ary $c$-}
as a function $A: [n]^k → [c]$ where
$[n] = \Set{0,\dots,n-1}$ is the ID space and $[c] = \Set{0,\dots,c-1}$ is a
set of $c$ colors, with the property that
\begin{equation}
    \label{eq-coloring-function}
    A(x_1,x_2,\dots,x_k) ≠ A(x_2,\dots,x_k,x_{k+1})
\end{equation}
for any $0 ≤ x_1 < x_2 < \dots < x_{k+1} ≤
n-1$.

The restriction to increasing sequences and values in $[n]$ rather
than $[N]$ is more restrictive that a general $c$-coloring algorithm,
but if we have a successful $3$-coloring algorithm that runs in time
$T$, we can extract from it
a $(2T+1)$-ary $3$-coloring function, and condition
\eqref{eq-coloring-function} will hold given that the original
algorithm never assigns the same color to adjacent nodes. Taking the
contrapositive,
if condition \eqref{eq-coloring-function} fails for some sequence 
$(x_1,x_2, \dots x_{k+1})$, then we can supply this sequence as the IDs
for the first $k+1$ nodes in the ring and show the algorithm fails.
This implies that a $3$-coloring algorithm that runs in
time $T$ can exist only if there is a $(2T+1)$-ary $3$-coloring
function. The lower bound proof works by showing that $T$ needs to be
$Ω(\log^* n)$ for this to be possible.

It holds trivially that any $1$-ary $c$-coloring function requires
$c≥n$. The proof works by showing how to transform any $k$-ary
$c$-coloring function into a $(k-1)$-ary $2^c$-coloring function,
which hits the trivial bound after $k-1$ steps.

\begin{lemma}[\protect{\cite[Lemma 2]{LaurinharjuS2014}}]
    \label{lemma-coloring-function-amplification}
    For $k>1$, given a $k$-ary $c$-coloring function $A$, it is possible to
    construct a $(k-1)$-ary $2^c$-coloring function $B$.
\end{lemma}
\begin{proof}
    Let 
    $B'(x_1,\dots,x_{k-1}) = \SetWhere{ A(x_1,x_2,\dots,x_{k-1},x_k)
    }{x_k > x_{k-1}}$. In other words, we fill in the missing
    parameter $x_k$ with all possible values $x_k > x_{k-1}$, and
    return the set of colors that we obtain from $A$. Since there are
    exactly $2^c$ possible sets, we can obtain $B: [N]^{k-1} → [2^c]$ by encoding
    each set as a distinct number in $[2^c] = \Set{1,\dots,2^c}$.

    We will now prove that $B$ satisfies \eqref{eq-coloring-function}
    whenever $A$ does, by showing the contraposition that if $B$ does
    not satisfy \eqref{eq-coloring-function}, then $A$ doesn't either.

    Suppose now that \eqref{eq-coloring-function} does not hold for
    $B$, that is, there is some increasing sequence $(x_1,\dots,x_k)$ such
    that $B(x_1,\dots,x_{k-1}) = B(x_2,\dots,x_k)$, or equivalently
    $B'(x_1,\dots,x_{k-1}) = B'(x_2,\dots,x_k)$.

    We will feed this bad sequence to $A$ and see what happens.
    Let $α = A(x_1,\dots,x_k)$. Since $x_k$ is one of the possible
    extensions of $(x_1,\dots,x_{k-1})$ used to generate
    $B'(x_1,\dots,x_{k-1})$, we get $α ∈ B'(x_1,\dots,x_{k-1})$.
    But then $α$ is also contained in $B'(x_2\dots,x_k) =
    B'(x_1,\dots,x_{k-1})$. From the definition of
    $B'(x_2,\dots,x_k)$, this implies that there is some $x_{k+1} >
    x_k$ such that $α = A(x_2,\dots,x_k,x_{k+1}) =
    A(x_1,x_2,\dots,x_k)$. But then $A$ is not a $k$-ary $c$-coloring
    function.
\end{proof}

To get the $Ω(\log^* n)$ lower bound, start with a $k$-ary
$3$-coloring function and iterate
Lemma~\ref{lemma-coloring-function-amplification} to get a $1$-ary
$f(k-1)$-coloring function where $f(k)$ is the result of iteratively applying the
function $2^x$ to $3$, $k-1$ times.
Then $f(k-1) ≥ n$, which implies $k
= Ω(\log^* n)$.

\subsection{Coloring bounded-degree graphs}

The $O(\log^* n)$-time $3$-coloring algorithm for out-degree $1$
digraphs can be used to get a simple $O(Δ^2 + \log^* n)$ time
algorithm for
$(Δ+1)$-coloring any graph with maximum degree $Δ$, using an algorithm
of Panconesi and Rizzi~\cite{PanconesiR2001}.

This algorithm has three steps:
\begin{enumerate}
    \item First, partition the original graph $G$ into $Δ$ directed
        graphs $G_1,\dots,G_Δ$, each with
        maximum out-degree $1$. We can do this in $O(1)$ rounds: each
        process collects the IDs of its neighbors, and assigns each a
        \concept{port number}\index{number!port} in $\Set{1,\dots,Δ}$
        in increasing order of ID, while also orienting each edge to
        point to the neighbor with larger ID. Each directed graph
        $G_i$ then consists of all edges for which the source node
        assigns port number $1$.
    \item Next, use Cole-Vishkin (§\ref{section-Cole-Vishkin}) to
        $3$-color each $G_i$.
    \item To color the original graph $G$, start with $H_1 = G_1$ and
        repeatedly merge each $H_i$ with the next unmerged $G_{i+1}$
        to get $H_{i+1}$. Each $H_i$ will have at most $Δ+1$ colors,
        which we will show by induction on $i$.\footnote{We are
        assuming here that $Δ≥2$ to get the induction going,
        but the cases $Δ=0$ and $Δ=1$ are easy to handle using a
        simpler algorithm.}

        The merging process consists of assigning
        each node a color in $[3(Δ+1)]$ by taking an ordered pair of
        its color in $G_{i+1}$ (3 choices) and its color in $H_i$. 
        Then for each $c ∈ \Set{Δ+2,\dots,3(Δ+1)}$, have each node
        with color $c$ choose the smallest color not represented among
        its neighbors. This is the same color-reduction scheme used
        to go from six to three colors in §\ref{section-Cole-Vishkin},
        except without the shifting, and just like there we don't
        create any new conflicts because no two nodes with the same
        color $c$ are adjacent to begin with.

        Each merging step costs $O(Δ)$ rounds (mostly for polling the
        neighbors to see what colors they currently have).
        There are $O(Δ)$ total merges, so it takes $O(Δ^2)$ rounds to
        complete them all and get $Δ+1$ colors.
\end{enumerate}

Since we are using Cole-Vishkin as a subroutine, we do need an upper
bound on $\log^* n$, but this shouldn't be too hard to obtain in
practice.

If we go back and check all the steps, we find that the largest value
we are transmitting in any message is a color, which we can do in
$O(\log n)$ bits (assuming that the initial colors are all polynomial
in $n$). So in fact this gives us an $O(Δ^2 + \log^*n)$ round
algorithm in both the LOCAL and CONGEST models.

The Panconesi and Rizzi algorithm has the advantage of simplicity, but
there are faster algorithms. An algorithm of Ghaffari and
Kuhn~\cite{GhaffariK2021} obtains a $(Δ+1)$-coloring of a graph with
maximum degree $Δ$ in $O(\log^2 Δ \log n)$ rounds.

\section{All-pairs shortest paths in CONGEST}
\label{section-CONGEST-APSP}

Here we're going to roughly follow~\cite[Chapter 5]{HirvonenS2025},
which introduces the CONGEST model by starting with constructing a
breadth-first-search
tree from a single source and then works up to computing all-pairs
shortest paths (where each process learns its distance from every
other process) using a rather clever algorithm of Holzer and
Wattenhofer~\cite{HolzerW2012}.

The tricky part this is showing that each step of this process 
works within the constraints of the CONGEST model, 
which only allows each process to send $O(\log n)$ bits
of information to each of its neighbors in each round. 

\subsection{BFS with fixed starting root}

To construct a BFS tree starting from a known root, we can just use
flooding as in Algorithm~\ref{alg-flooding-parents}; the spanning tree
constructed by this algorithm will be BFS in a synchronous model like
CONGEST. As written, the algorithm uses only constant-size messages,
so it easily fits within the message-size bound, and it finishes in
$O(D)$ rounds where $D$ is the diameter of the network.

By supplementing the outgoing messages with distance information, we
can also arrange for each process to learn its distance from the root.
We'll need this later to compute all-pairs shortest paths.

\subsection{Leader election in CONGEST}

To elect a single root, we just pick the node with smallest id. As
described in §\ref{section-leader-election-lower-bounds}, we can
do this by running $n$ copies of flooding in parallel, each
transmitting its root's id, and have each
process propagate only the smallest id it has seen so far. This again
fits in the CONGEST model because we only need $O(\log n)$ bits to
represent ids and each process sends at most one message per round to
each neighbor.

By combining each flooding algorithm with convergecast (see
Algorithm~\ref{alg-flooding-with-convergecast}), we can have the
winning leader find out that it has won after $O(D)$ rounds.

\subsection{All-pairs shortest paths}

Now we want to have every process learn its distance from every other
process.

For single-source shortest paths, we can just have the single source
do flooding, with an increasing distance field in the message (see
Algorithm~\ref{alg-sssp}).

\begin{algorithm}
    \Initially{
        \eIf{$\Pid = \Root$}{
            $\DataSty{dist} ← 0$ \;
            send $\DataSty{dist}$ to all neighbors\;
        }{
            $\DataSty{dist} ← ⊥$\;
        }
    }
    \UponReceiving{$d$}{
        \If{$\DataSty{dist} = ⊥$}{
            $\DataSty{dist} ← d+1$\;
            send $\DataSty{dist}$ to all neighbors\;
        }
    }
    \caption{Single-source shortest paths using flooding}
    \label{alg-sssp}
\end{algorithm}

A straightforward induction on distance from the root shows that this
algorithm sets $\DataSty{dist}$ at each process to the correct value
within $D$ rounds.

For all-pairs shortest paths, it's tempting to just run $n$ copies of
Algorithm~\ref{alg-sssp}. In the standard synchronous message-passing model, this
will solve the problem in $D$ rounds. But it is likely to violate the
message-size bounds from CONGEST since (a) each process may send as many
as $n$ messages to each of its neighbors across all $n$ protocols,
resulting in $Ω(n/D)$ messages to some neighbor in some round; and (b)
even if we are clever about combining these messages, because we have
to label messages from different floods by their roots, we will still
need to transmit $Ω(n/D)$ process ids at the cost of $Ω((n/d) \log n)$
bits in some message.

So instead we are going to delay starting each copy of
Algorithm~\ref{alg-sssp} so that they don't interfere with each other.
We assume that we have already constructed elected a leader and
constructed a BFS tree rooted at the leader as described in the
previous sections. Following~\cite{HolzerW2012}, we now start a token
at the leader and have the token do a \emph{depth-first} traversal of
the BFS tree, slowed down so that the token waits one round at each node before
moving to the next node. Whenever the token arrives at a node for the
first time, that node starts its instance of Algorithm~\ref{alg-sssp}.

Let's show that this prevents excessive traffic across any edge.
In particular, we'll argue for any distinct nodes $x$, $y$, and $z$, it is
never the case that a message starting at $x$ and a message starting
at $y$ arrive for the first time at $z$ in the same round. The proof
uses the triangle inequality $d(y,z) ≤ d(x,y) + d(x,z)$; if we assume
that $d(x,z) ≤ d(y,z)$ (the other case is symmetric), then this gives
$d(x,y) ≥ d(y,z) - d(x,z)$. Since $d(y,z)$ is the time for the flood
starting at $y$ to reach $z$, and $d(x,z)$ is similarly the time for
the flood starting at $x$ to reach $z$, this means that $x$ started
its flood exactly $d(y,z) - d(x,z)$ rounds before $y$ did.
But then the token made it from $x$ to $y$ in at most $d(y,z) - d(x,z)
≤ d(x,y)$ rounds. Even if the path it takes in the DFS traversal is a
shortest path, the built-in delay means that it will take at least
$2d(x,y) > d(x,y)$ rounds to get from $x$ to $y$. So assuming that two
distinct floods arrive at the same node at the same round gives a
contradiction.

For the CONGEST bound, since a process only sends messages for a
particular instance of Algorithm~\ref{alg-sssp} when it first receives
a message from this instance, we get that within any one round each
process only sends messages from one instance. These all have size
$O(\log n)$ (instance id plus a distance), so they fit in the bound.

We can also argue that the protocol as a whole takes $O(n+D) = O(n)$
rounds since the initial leader election step finishes in $O(D)$ rounds,
the depth-first traversal takes $O(n)$ rounds, and finishing each
flood takes $O(D)$ rounds from when it starts. So we can compute
all-pairs shortest paths in CONGEST in $O(n)$ rounds.

\myChapter{Mobile robots}{2026}{}
\label{chapter-mobile-robots}

Mobile robots are a model of distributed computation where the agents
(robots) are located in a plane, and have no ability to communicate
except by observing each others' positions. Typical problems are to
get a group of robots to gather on a single point, or more generally
to arrange themselves in some specified pattern. This is complicated
by the usual issues of asynchrony and failures, as well as a common
assumption that the robots have no memory and can only base their
movements on what they can observe at the moment.

\section{Model}
\label{section-mobile-robots-model}

We will start by describing the original
\concept{Suzuki-Yamashita model}~\cite{SuzukiY1999},
and some variants. We'll follow the naming conventions used by Agmon and
Peleg~\cite{AgmonP2006} and Flocchini~\etal~\cite{FlocchiniNPPS2026}.

The basic model involves a collection of robots 
    represented by points in the plane $ℝ^2$.
    Each robot repeatedly executes a \concept{look-compute-move} cycle
    consisting of three phases:
        \begin{itemize}
            \item Look phase: obtain snapshot of relative positions of
                all the other robots.
            \item Compute phase: pick a new point to move to.
            \item Move phase: move to that point.
        \end{itemize}

The robots are limited in the capabilities. Some common limitation:
        \begin{itemize}
            \item \indexConcept{anonymity}{Anonymity}: Any two robots
                that see the same view take the same action.
            \item \indexConcept{oblivious!mobile robots}{Oblivious}:
                The output of the compute phase is based \emph{only}
                on the results of the last
                look phase, and not on any previous observations.
                Robots have no memory!
            \item No \concept{absolute coordinates}: Translations of
                the space don't change the behavior of the robots.
            \item No \indexConcept{sense of direction!mobile
                robots}{sense of direction}: Robots don't know which
                way is north.
            \item No \indexConcept{sense of scale!mobile robots}{sense
                of scale}: Robots don't have a consistent linear
                measure.
            \item No \indexConcept{chirality}{sense of chirality}:
                robots can't tell counter-clockwise from clockwise.
            \item No ability to detect
                \indexConcept{multiplicity!mobile
                robots}{multiplicities}: the view of other robots is a
                set of points (rather than a multiset), so if two
                robots are on the same point, they look like one
                robot.
            \item \index{robot!fat}\indexConcept{fat robot}{Fat
                robots}: robots block the view of the robots behind
                them.
        \end{itemize}

        Many of these limitations can be described in terms of
        symmetries of the space the robots are moving in. No absolute
        coordinates means that if we translate a configuration of
        robot positions $C = \Tuple{c_1,c_2,\dots,c_n}$ to $C+x =
        \Tuple{c_1+x,c_2+x,\dots,c_n+x}$, then a robot that tries to
        move to $c'$ in $C$ will try to move to $c'+x$ in $C+x$. No
        sense of direction applies the same rule to rotations; no
        sense of scale to scaling; and no sense of chirality to
        reflections. A minor complication is what to do when a
        configuration is itself symmetric: in this case, we don't
        require the robot to stay put (which might be the only move
        that is invariant with respect to possible transformations);
        instead, we'll let the adversary pick which transformed
        version of the configuration the robot sees when doing its
        look phase.

    The adversary can also interfere directly in the computation in
    various ways. One restriction is that robots are guaranteed to be
    able to any distance less than some bound $δ>0$, but any attempt
    to move more than $δ$ can be interrupted by the adversary. An
    interrupted move leaves the robot at a location somewhere of the
    adversary's choosing on the line between the starting point and
    ending point of the move, so long as it gets to move at least
    distance $δ$.

    The other power of the adversary is control over timing.
    Look-compute-move phases are asynchronous, and
    adversary can schedule robots subject to various
    constraints. 
    In the
    \indexConcept{asynchronous!mobile
    robots}{asynchronous} model (\concept{ASYNC}),
    The adversary can delay a robot between look and move
    phases, so that robots might be moving based
    on out-of-date information.
    In a \indexConcept{synchronous!mobile robots} model, the robots
    operate in a sequence of rounds, and each robot that participates
    in a round is guaranteed to complete its entire look-compute-move
    cycle during that round, with all robots doing a look before
    any robot moves. However, it may not be the case that every robot
    participates in every round. There are three main variants in the
    literature:
\begin{itemize}
\item \indexConcept{semi-synchronous!mobile
    robots}{Semi-synchronous} model (\concept{SSYNC}):
    The adversary may schedule more
    one or more robots to do their
    look-compute-move in each round. Also known
    as the \concept{ATOM}
    model. This was the original model given by Suzuki and
    Yamashita~\cite{SuzukiY1999}.
\item \indexConcept{fully synchronous!mobile
    robots}{Fully synchronous} model (\concept{FSYNC}):
    Like SSYNC, but every robot is active in every
    round.
\item \indexConcept{sequential!mobile robots}{Sequential} mode
    (\concept{SQ}) Like SSYNC, but exactly one robot is active in
        every round.
                \end{itemize}
                In each of these models, the adversary is required to
                obey a fairness condition and schedule every robot
                infinitely often (this is only a concern for SSYNC
                and SQ).

                We may also have faults:
        \begin{itemize}
            \item \indexConcept{Byzantine fault!mobile
                robots}{Byzantine faults}: Byzantine robots
                will move wherever the adversary chooses.
            \item \indexConcept{Crash fault!mobile
                robots}{Crash faults}: Crashed robots don't
                move even when they are supposed to.
        \end{itemize}

The simplest goal is to gather the non-faulty robots together on a
single point despite all these possible disasters. Other goals might
be formation of particular shapes. An additional source of variation
here is whether we want exact gathering (every robot eventually gets
to exactly where it should be) or just convergence (over time, robots
get closer and closer to where they should be). We'll start by looking
at gathering in §\ref{section-mobile-robot-gathering} and then talk a
bit about the more general problem of pattern formation in
§\ref{section-mobile-robot-pattern-formation}.

\section{Gathering}
\label{section-mobile-robot-gathering}

Here, we will mostly be looking at the semi-synchronous model, with
the assumption that robots are anonymous and oblivious, and have no
absolute coordinates, sense of direction, or sense of scale. However,
we will usually let robots detect multiplicity. Depending on the
results we are describing, we may or may not assume chirality.

\subsection{Two robots, no faults}
\label{section-mobile-robot-gathering-two-robots}

Suzuki and Yamashita~\cite{SuzukiY1999} showed that it's impossible to
get two deterministic, oblivious robots to the same point in the semi-synchronous
model assuming no absolute coordinates and no sense of direction, 
although they can converge.  The convergence algorithm is
simple: have each robot move to the midpoint of the two robots
whenever it is activated.  This always reduces the distance between
the robots by $\min(δ, d/2)$, giving convergence to within $ε$ for two
robots that start at distance $D$ in $O(D/δ + \ln δ/ε)$ rounds.
But it doesn't solve gathering if only one robot moves at a time.

This turns out to be true in general~\cite[Theorem 3.1]{SuzukiY1999}.
The idea is this: Suppose we have an oblivious algorithm for
gathering. Consider two robots at distinct points $p$ and $q$, and suppose
that after one round they both move to $r$ (any algorithm that works
eventually takes such a step).  There are two cases:
\begin{enumerate}
    \item Both robots move.  By symmetry, $r = (p+q)/2$.  So now
        construct a different execution in which only one robot moves
        (say, the one that moved least recently, to avoid running into
        fairness).
    \item Only one robot moves.  Without loss of generality, suppose
        the robot at $p$ moves to $q$.  Then there is a different
        execution where $q$ also moves to $p$ and the robots switch
        places.
\end{enumerate}

In either case two robots in the modified execution don't reach the
same point, which means that the algorithm doesn't finish. Note that
this works even if the adversary can't stop a robot in mid-move.

Both obliviousness and the lack of coordinates and sense of direction
are necessary.  If the robots are not oblivious, then they can try
moving to the midpoint, and if only one of them moves then it stays
put until the other one catches up.  If the robots have absolute
coordinates or a sense of direction, then we can deterministically
choose one of the two initial positions as the ultimate gathering
point (say, the northernmost position, or the westernmost position if both
are equally far north).  But if we don't have any of this we are in
trouble.

Like the 3-process impossibility result for Byzantine agreement, the
2-process impossibility result for robot gathering extends to any
even number of robots where half of them are on one point and half on
the other.  Anonymity then means that each group of robots acts the
same way a single robot would if we activate them all together.
Later work (e.g.,~\cite{BouzidDT2012}) refers to this as
\indexConcept{bivalent!mobile robots}{bivalent} configuration, and it turns out to be the only
initial configuration for which it is not possible to solve gathering
absent Byzantine faults.

\subsection{Three robots}

Agmon and Peleg~\cite{AgmonP2006} show that with three robots, it is
possible to solve gathering in the SSYNC model with one crash fault
but not with one Byzantine fault.  We'll start with the crash-fault
algorithm.  Given a view $v = \Set{p_1,p_2,p_3}$, this sends each robot to the ``goal'' point $p_G$
determined according to the following rules:
\begin{enumerate}
\item If $v$ has a point $p$ with more than one robot, set $p_G =
    p$.
\item If $p_1,p_2,$ and $p_3$ are collinear, set $p_G$ to the middle
    point.
\item If $p_1,p_2,$ and $p_3$ form an obtuse triangle (one with a
    corner whose angle is $≥π/2$, set $p_G$ to the obtuse corner.
\item If $p_1,p_2,$ and $p_3$ form an acute triangle (one with no
    angles $≥π/2$), set $p_G$ to the intersection of the angle
    bisectors.
\end{enumerate}

Here is a sketch of why this works.  For the real proof
see~\cite{AgmonP2006}.

\begin{enumerate}
    \item If we are in a configuration with multiplicity $>1$, any non-faulty robot not
        on the multiplicity point eventually gets there.
    \item If we are in a collinear configuration, we stay collinear
        until eventually one of
        the outer robots gets to the middle point, giving a
        configuration with multiplicity $>1$.
    \item If we are in an obtuse-triangle configuration, we stay in an
        obtuse-triangle configuration until eventually one of the
        acute-corner robots gets to the obtuse corner, again giving a
        configuration with multiplicity $>1$.
    \item If we are in an acute-triangle configuration, then a
        somewhat messy geometric argument shows that if at least one
        robot moves at least $δ$ toward the intersection of the angle
        bisectors, then the circumference of the triangle drops by
        $cδ$ for some constant $c>0$.  This eventually leads either to
        the obtuse-triangle case (if we happen to open up one of the
        angles enough) or the multiplicity $>1$ case (if the
        circumference drops to zero).
\end{enumerate}

However, once we have a Byzantine fault, this blows up.  This is
shown by considering a lot of cases, and giving a strategy for the
adversary and the Byzantine robot to cooperate to prevent the other
two robots from gathering in each case.  This applies to both
algorithms for gathering and convergence: the bad guys can arrange so
that the algorithm eventually makes no progress at all.

The first trick is to observe that any working algorithm for the
$n=3,f=1$ case must be \concept{hyperactive}: every robot attempts to
move in every configuration with multiplicity $1$.  If not, the
adversary can (a) activate the non-moving robot (which has no effect);
(b) stall the moving non-faulty robot if any, and (c) move the
Byzantine robot to a symmetric position relative to the first two so
that the non-moving robot becomes the moving robot in the next round
and vice versa.  This gives an infinite execution with no progress.

The second trick is to observe that if we can ever reach a
configuration where two robots move in a way that places them further
away from each other (a \concept{diverging} configuration), then we
can keep those two robots at the same or greater distance forever.
The idea is that when the robots start to move closer together, we can
stop them at their original distance, and then move the Byzantine
robot to restore the diverging configuration.
This depends on the adversary being able to stop a robot in the middle
of its move, which in turn depends on the robot moving at least $δ$
before the adversary stops it. The paper argues that since the
algorithm has to work for any choice of $δ$, we can always assume $δ$
small enough that this is not a problem.\footnote{This claim depends
on a mildly fishy assumption that the algorithm doesn't know $δ$.
An alternative is just to assume no sense of scale.}

Here is the full argument: Suppose that from positions $p_0$ and $q_0$
there is a step in which the non-faulty robots move to $p_1$ and $q_1$
with $d(p_1,q_1) > d(p,q)$.  Starting from $p_1$ and $q_1$, run both
robots until they are heading for states $p_2$ and $q_2$ with
$d(p_2,q_2) ≤ d(p_0,q_0)$.  By continuity, somewhere along the paths
$p_1 p_2$ and $q_1 q_2$ there are intermediate points $p'_2$ and
$q'_2$ with $d(p'_2,q'_2) = d(p_0,q_0)$.  Stop the robots at these
points, move the Byzantine robot $r$ to the appropriate location to
make everything look like the initial $p_0,q_0$ configuration, and we
are back where we started.

So now we know that (a) we have to move every robot, and (b) we can't
move any two robots away from each other. So now construct a
configuration where the robots are collinear (this is easily arranged
by moving the Byzantine robot in line with the other ones). Since the
middle robot has to move, it has to move away from one of the other
robots. But now we have diverging robots and we're doomed.

\subsection{Many robots, with crash failures}

It turns out that we can solve the gathering problem even if we have
many robots and some of them can crash, as long as the robots do not
start on the same line. The reason for this is that any set of
non-collinear points $x_1, \dots, x_n$ in $ℝ^2$ has a unique
\index{median!geometric}\concept{geometric median}, defined as the
point $m$ that minimizes $∑_{i=1}^n d(m,x_i)$, and the geometric
median is unchanged if we move any of the points towards
it. 

So the algorithm, as suggested by Bramas and
Tixeuil~\cite{BramasT2015} is simply for all the robots to walk toward
this point.\footnote{In the Bramas and Tixeuil paper, the geometric
media is referred to as the \concept{Weber point}\index{point!Weber},
but it's the same thing.} It doesn't matter if some of the robots
don't move, or don't move at the same speed, because the median
doesn't change. Eventually, all the non-faulty processes will reach
it.

There is one drawback to this approach, which is that even though very
good approximation algorithms exist~\cite{CohenLMPS2016}, the
geometric median appears to be difficult to compute exactly.  We could
declare that we are willing to assume that our robots have infinite
computational power, but this may not an easy assumption to justify.
The alternative described by Bramas and Tixeuil
is to build an algorithm that marches toward the
geometric median in certain cases where it is straightforward to
compute, and does something more sophisticated otherwise.

\section{Pattern formation}
\label{section-mobile-robot-pattern-formation}

A \concept{pattern} is specified by a set of points $\hat{P}$ in
$ℝ^2$, with the goal of placing at least one robot on each point in
the pattern. Since the robots are usually assumed not to have absolute
coordinates, a send of direction, a sense of scale, or a sense of
chirality, placing the robots on the points of any equivalent pattern
up to rotation, scaling, and reflection will be treated as meeting
this goal.

A \concept{universal pattern formation} (\concept{UPF}) protocol
allows robots to form any given pattern. Unfortunately, universal
pattern formation is trivially unsolvable in FSYNC, and thus also in
SSYNC and ASYNC, since these models can act like FSYNC. The reason for
this (as shown by Flocchini~\etal~\cite{FlocchiniNPPS2026}) is that
FYSNC can't break multiplicities: if we stack $n$ anonymous robots on
a single point, they will move together forever in a fully synchronous
model. For this reason, Flocchini~\etal study universal pattern
formation in the \concept{sequential model}, which they abbreviate as
\concept{SQ}. This model is a special case of SSYNC where exactly one
robot, chosen by the adversary, moves in each round, breaking
symmetry.

Flocchini~\etal give a general result showing that universal pattern
formation is possible in SQ for every pattern except one consisting of
a single point, giving a problem they call UPF*: essentially,
universal pattern formation minus gathering.

Gathering is excluded because it is unsolvable in SQ without
multiplicity detection. In a configuration where two or more robots
are placed on two points $p$ and $q$, a protocol that has a robot at
$p$ move to $q$ is defeated by placing two robots on $p$, and
alternating between moving one of these robots to $q$ and moving one
of robots on $q$ back to $p$; while a protocol that moves a robot at
$p$ anywhere else succumbs to Zeno's paradox if there are only two
robots.\footnote{By contrast, in a system that has even weak
multiplicity detection, where a robot can observe that a point is
occupied by multiple robots but not count exactly how many,
Flocchini~\etal demonstrate SQ can solve gathering using a simple
protocol that moves all robots to the unique multiple point if there
is one, and if there is not, creates a unique multiple point by either
moving excess robots to empty spots if there is more than one multiple
point and moving one robot onto another if there are none. But for
forming any pattern that is not a single point, they don't need even
weak multiplicity detection, hence the exclusion of gathering.}

For any pattern with more than one point, the paper gives a rather
complicated protocol based on electing a leader from all the robots
on the boundary of the smallest circle that encloses all the robots,
then using the position of this leader to match up the rest of the
robots with points in the pattern.

We won't try to describe this protocol here, but some of the ideas can
be found in a more specialized protocol that is given for the case
where $1 < \card{\hat{P}} < 5$.
For this protocol, a robot that is asked to move looks at the set of
points $Q$ that contain robots, and applies the first these rules that
works:
\begin{enumerate}
    \item If $\card{Q} < \card{\hat{P}}$, try to separate from a
        possible multiple point by moving to an empty space.
    \item If there is not a unique pair $q_1,q_2$ at maximum
        distance, move to create such a pair.
    \item If $\card{Q} > \card{P}$, move to the closer of $q_1,q_2$.
    \item If I am not at $q_1$ or $q_2$, map $q_1,q_2$ to the most
        distant points $\hat{p}_1,\hat{p}_2$ in $\hat{P}$, then move
        to the first of points $\hat{p}_3$ or $\hat{p}_4$ that exists and is unoccupied.
\end{enumerate}

The idea is that $q_1$ and $q_2$ form a spine for the rest of the
pattern. There are some annoying special cases when the pattern
contains more than one pair of points at the same maximum distance,
but these can be dealt with by choosing the order in which the
remaining points are filled carefully. Details can be found in the
paper.

\myChapter{Beeping}{2016}{}
\label{chapter-beeping}

The (discrete) \index{model!beeping}\concept{beeping model} was introduced by
Cornejo and Kuhn~\cite{CornejoK2010} to study what can be computed in
a wireless network where communication is limited to nothing but
carrier sensing.  According to the authors, the model is inspired in
part by some earlier work on specific
algorithms based on carrier sensing due to
Scheideler~\etal~\cite{ScheidelerRS2008} and Flury and
Wattenhofer~\cite{FluryW2010}.  It has in turn spawned a significant
literature, not only in its original domain of wireless networking,
but also in analysis of biological systems, which often rely on very limited
signaling mechanisms.  Some of this work extends or adjusts the
capabilities of the processes in various ways, but the essential idea of tightly limited
communication remains.

In its simplest form, the model consists of synchronous processes
organized in an undirected graph.  Processes wake up at arbitrary
rounds chosen by the adversary, and do not know which round they are
in except by counting the number of rounds since they woke.
Once awake, a process chooses in each
round to
either send (beep) or listen.  A process that sends learns nothing in
that round.  A process that listens learns whether any of its
neighbors sends, but not how many or which one(s).  

From a practical perspective, the justification for the model is that carrier sensing
is cheap and widely available in radio networks.
From a theoretical perspective, the idea is to make the communication
mechanism as restrictive as possible while still allowing some sort of
distributed computing.  The assumption of synchrony both adds to and
limits the power of the model.  With no synchrony at all, it's
difficult to see how to communicate anything with beeps, since each
process will just see either a finite or infinite sequence of beeps
with not much correlation to its own actions.  With continuous time,
subtle changes in timing can be used to transmit arbitrarily detailed
information.  So the assumption of a small number of synchronous
rounds is a compromise between these two extremes.  The assumption
that processes wake at different times and have no common sense of
time prevents synchronization on rounds, for example by reserving
certain rounds for beeps by processes with particular IDs.  It is up
to the protocol to work around these limitations.

\section{Interval coloring}
\label{section-interval-coloring}

One way to get around the problem of not having a common global clock
is to solve 
\index{coloring!interval}
\concept{interval coloring}, the main problem considered by Cornejo
and Kuhn.  This is related to TDMA multiplexing in
cell phone networks, and involves partitioning a repeating interval of
$T$ rounds in a network with maximum degree $Δ$ into subintervals of
length $Ω(T/Δ)$ such that each process is assigned a subinterval and
no two adjacent processes are assigned overlapping subintervals.  The
idea is that these intervals can then be used to decide when each
process is allowed to use its main radio to
communicate.\footnote{We may really want \index{coloring!2-hop}
    \concept{2-hop coloring} here, where no two of my neighbors get
    the same color, because this is what (a) allows me to  tell my
    neighbors apart, and (b) allows my neighbors not to interfere with
    each other, but that is a subject for later papers (see, for
example, \cite{MetivierRZ2015}.}

Cornejo and Kuhn give an algorithm for interval coloring that assigned
a subinterval of length $Ω(T/Δ)$ to each process assuming
that the size of the interval $T$ is known to all processes and that $T$ is
at least a constant multiple of $Δ$.
However, the processes do not
know anything about the structure of the graph, and in particular do
not know $Δ$.  This requires each process to get an estimate of the
size of its neighborhood (so that it knows how large a subinterval to
try to acquire) and to have a mechanism for collision detection that
keeps it from grabbing an interval that overlaps with a neighbor's
interval.  The process is complicated by the fact that my length-$T$
intervals and your length-$T$ intervals may be offset from each other,
and that I can't detect that you are beeping if you and I are beeping
at the same time.

To simplify things a bit, the presentation below will assume that the
graph is regular, so that $d(v)$ equals the maximum degree $Δ$ for all
nodes in the graph.  The paper~\cite{CornejoK2010} gives an analysis
that does not need this assumption.  We'll also wave our hands around
a lot instead of doing actual algebra in many places.

\subsection{Estimating the degree}

The basic idea is to have each process beep once in every $T$
consecutive slots.  Each process divides time into
\indexConcept{period}{periods} of length
$T$, starting when it wakes up.  Because processes wake up at
different times, my period might overlap with up to two of yours.
This means that if $S$ is the set of times during my period where I
hear beeps, then $S$ includes
at most two beeps per process, so $\card{S}$ is at most twice my actual
degree.  This gives an upper bound on $d(v)$, and indeed each process
will just use the maximum number of beeps it heard in the last period
as the basis for its estimate of $d(v)$.

For the lower bound side, we want to argue that if processes choose
slots at random in a way that is more likely to avoid collisions than
create them, and there are enough slots, then we expect to get few
enough collisions that $\card{S}$ is also $Ω(Δ)$.  The details of this
depend on the mechanism for choosing slots, but if we imagine slots
are chosen uniformly, then $\Exp{\card{S}} ≥ Δ(1-Δ/T)$, which is
$Ω(Δ)$ under our assumption that $T ≥ cΔ$ for some sufficiently large
$c$.  We can compensate by the error by inserting a fudge factor $η$,
chosen so that $(1/η) \card{S}$ is very likely to be an upper bound on the
degree.

\subsection{Picking slots}

Each process will try to grab a subinterval of size $b = η
\frac{T}{\card{S} + 1}$, where $η$ is the fudge factor mentioned
above.
If it has not already picked a position $p$, then it chooses one
uniformly at random from the set of all positions $p$ such that
$S[p-b-2, p+b+1]$ from the most recent period includes no beeps.  
Because this selection criterion knocks out up to $(2b+4) \card{S}$
possible choices, it
does tend to concentrate uncolored processes on a smaller range of
positions than a uniform pick from the entire interval, increasing
the likelihood of collisions.  But we can choose $η$ to make $2(b+4)
\card{S} = 2ηT \frac{\card{S}}{\card{S}+1}$ a small enough fraction of
$T$ that this is not a problem.

\subsection{Detecting collisions}

The basic idea for detecting a collision is that I will abandon my
color $p$ if I hear any beeps in $[p-b-2,p+b+1]$ during the next
period.  This works great as long as nobody chooses exactly the same
$p$ as me.  To avoid this, each process flips a coin and beeps at
either $p$ or $p+1$.  So even if I choose the same slot as one or more
of my neighbors, there is a $1/2$ chance per period that I detect
this and pick a new color next time around.

What this means is that in each round (a) I have a constant
probability of getting a good estimate of my degree (which means I set
$b$ correctly); (b) I have a constant probability of detecting a
collision with a neighbor if there is one (which means I pick a new
position if I have to); and (c) I have a constant probability that if
I pick a new position it is a good one.  If we repeat all these
constant-probability wins for $O(\log n)$ periods, then all $n$
processes win, and we are done.

\section{Maximal independent set}

A high-impact early result in the beeping model was a paper by
Afek~\etal.~\cite{AfekABHBB2011} that showed that a biological
mechanism used in fruit-fly sensory organ development to choose a
subset of cells that are not too close to each other can be viewed as
an implementation of 
maximal independent set using beeps.  As a
distributed algorithm, this algorithm is not so good, so instead we
will talk about a follow-up paper~\cite{AfekABCHK2011} by some of the same authors on
more effective beeping algorithms for MIS.

Recall that a subset of the vertices of a graph is
\index{independent set}
\indexConcept{set!independent}{independent} if no two vertices in the
set are adjacent.  A 
\index{independent set!maximal}
\index{set!maximal independent}
\concept{maximal independent set} (\concept{MIS})
is an independent set of vertices that can't be increased without
including adjacent vertices.  Equivalently, it's an independent set
where every non-member is adjacent to some member.

Afek~\etal{} give a couple of algorithms for beeping MIS that require
either special knowledge of the graph or extensions to the beeping
model.  The justification for this is a lower bound, which they also
give, that shows that without any knowledge of the graph, computing an
MIS in the standard beeping model takes $Ω(\sqrt{n/\log n})$ time with 
constant probability.  We'll describe the lower bound and then show how to compute
MIS in $O(\log^3 n)$ time given a polynomial upper bound on $n$.

\subsection{Lower bound}
\label{section-beeping-MIS-lower-bound}

For the lower bound, the idea is to exploit the fact that the
adversary can wake up nodes over time.  To avoid allowing the
adversary to delay the algorithm from finishing indefinitely by just
not waking up some nodes for a while, the running time is computed as
the maximum time from when any particular node $p$ wakes up to when
$p$ converges to being in the MIS or not.

The cornerstone of the proof is the observation that if a process
doesn't know the size of the graph, then it has to decide whether to
beep or not within a constant number of rounds.  Specifically, for any
fixed sequence of beeps $b_0, b_1, \dots$, where $b_i$ is an indicator
variable for whether
the process hears a beep in round $i$ after it wakes up, either the
process never beeps or there are constant $\ell$ and $p$ such that the
process beeps in round $\ell$ with probability $p$.  This follows
because if the process is ever going to beep, there is some first
round $\ell$ where it might beep, and the probability that it does so
is constant because it depends only on the algorithm and the sequence
$b$, and not on $n$.

If an algorithm that hears only silence remains silent, then nobody
ever beeps, and nobody learns anything about the graph.  
Without knowing anything, it's impossible to correctly compute an MIS
(consider a graph with only two nodes that might or might not have an
edge between them).
This means
that in any working algorithm, there is some round $\ell$ and
probability $p$ such that each process beeps with probability $p$
after $\ell$ rounds of silence.

We can now beep the heck out of everybody by assembling groups of
$Θ(\frac{1}{p}\log n)$ processes and waking up each one $\ell$ rounds before
we want them to deliver their beeps.  But we need to be a little bit
careful to keep the graph from being so connected that the algorithm
finds an
MIS despite this.

There are two cases, depending on what a process that hears only beeps
does:
\begin{enumerate}
    \item If a process that hears only beeps stays silent forever,
        then we build a graph with $k-1$ cliques $C_1,\dots,C_{k-1}$
        of size $Θ(\frac{k}{p} \log n)$ each, and a set of $k$ cliques
        $U_1,\dots,U_k$ of size $Θ(\log n)$ each.  Here $k \gg \ell$ is
        a placeholder that will be filled in later (foreshadowing: it's the
        biggest value that doesn't give us more than $n$ processes).  Each $C_i$ clique
        is further partitioned into subcliques $C_{i1},\dots,C_{ik}$
        of size $Θ(\frac{1}{p}\log n)$ each.  Each $C_{ij}$ is
        attached to $U_j$ by a complete bipartite graph.

        We wake up each clique $C_i$ in round $i$, and wake up all the
        $U$ cliques in round $\ell$.  We can prove by induction on
        rounds that with high probability, at
        least one process in each $C_{ij}$ beeps in round $i+\ell$,
        which means that every process in every $U_i$ hears a beep in
        the first $k-1$ rounds that it is awake, and remains silent,
        causing the later $C$ cliques to continue to beep.

        Because each $C_i$ is a clique, each contains at most one
        element of the $MIS$, and so between them they contain at
        most $k-1$ elements of the MIS.  But there are $k$ $U$
        cliques, so one of them is not adjacent to any MIS element in
        a $C$ clique.  This means that one of the $U_j$ must contain
        an MIS element.

        So now we ask when this extra $U_j$ picks an MIS element.  If
        it's in the first $k-1$ rounds after it wakes up, then all
        elements have seen the same history, so if any of them attempt
        to join
        the MIS then all of them do with independent constant probability
        each.  This implies that we can't converge to the MIS until at
        least $k$ rounds.

        Now we pick $k$ to be as large as possible so that the total number of processes $Θ(k^2
        \log n) = n$.  This gives $k=Ω(\sqrt{n/\log n})$ as claimed.
    \item If a process starts beeping with probability $p'$ after
        hearing beeps for $m$ rounds, then we can't apply the
        silent-beeper strategy because the $C$ cliques will stop
        hearing silence.  Instead, we replace the $C$ cliques with new
        cliques $S_1,\dots,S_{m-1}$ of size $Θ(\frac{1}{p} \log n)$
        each.  We start the process by having the $S$ cliques shout at
        the $U$ cliques for the first $m-1$ rounds.  After this, we
        can start having the $U$ cliques shout at each other: 
        each clique $U_j$ is connected to $q$ earlier cliques, 
        consisting of up to $q$ $U_{j'}$ for $j < j$ and enough $S_i$
        to fill out the remainder.

        We now argue that if a process that hears only beeps chooses to join the MIS with
        constant probability after $q$ rounds, then every $U$
        clique gets at least two processes joining with high
        probability, which is trouble.  Alternatively, if no process
        in a $U$ clique tries to join the MIS for at least $q$ rounds,
        then for $q = O(n / \log n)$, there are $U$ cliques that are
        connected only to other $U$ cliques, which again means we
        don't get an MIS.  So in this case we get a lower bound of
        $Ω(n/\log n)$ on the time for each node to converge.
\end{enumerate}

The lower bound in the paper is actually a bit stronger than this,
since it allows the processes to send more detailed messages than beeps as long
as there are no collisions.  Reducing this back to beeping 
means tuning the constants so we get at least two messages out of every clique.

\subsection{Upper bound with known bound on \texorpdfstring{$n$}{n}}
\label{section-beeping-MIS-upper-bound}

Algorithm~\ref{alg-beeping-MIS}~\cite{AfekABCHK2011}
converges to a maximal independent set in $O(\log^2 N \log n)$ rounds,
from any initial state,
given an upper bound $N$ on the number of processes $n$.

\begin{algorithm}
    Leave MIS and restart the algorithm here upon hearing a beep\;
    \For{$c \lg^2 N$ rounds}{listen\;}
    \For{$i ← 1$ \KwTo $\lg N$}{
        \For{$c \lg N$ rounds}{
            \eWithProbability{$\frac{2^i}{8N}$}{
                beep\;
            }{
                listen\;
            }
        }
    }
    Join MIS\;
    \While{I don't hear any beeps}{
        \eWithProbability{$\frac{1}{2}$}{
            beep\;
            listen\;
        }{
            listen\;
            beep;
        }
    }
    \caption{Beeping a maximal independent set (from \cite{AfekABCHK2011})}
    \label{alg-beeping-MIS}
\end{algorithm}

The proof that this works is a bit involved, so if you want to see all
the details, you should look at the paper.  The intuition runs like
this:
\begin{enumerate}
    \item At least one of any two adjacent processes that both think
        they are in the MIS will eventually notice the other during the final phase, causing it
        to restart.
    \item If I start at the beginning of the protocol, and I have a
        neighbor already in the MIS, then I will hear them during my
        initial listening phase and restart.
    \item If two adjacent nodes both execute the middle phase of
        increasing-probability beeps, then one of them will go through
        a phase where it listens with probability at least $1/2$ while
        the other beeps with probability at least $1/2$ (note that
        this might not be the same phase for both, because the nodes
        might not start at the same time).  This gives at least a $1/4$ chance
        per round that the likely listener drops out, for at least a
        $1-n^{-c/2}$ chance that it drops out during the $c \lg n$ rounds that it
        listens with this probability, assuming its neighbor does not
        drop out.  This means that by tuning $c$ large enough, we can
        make it highly improbable that any pair of neighbors both
        enter the MIS (and if they do, eventually at least one drops
        out).  So we eventually get a set that is independent, but
        maybe not maximal.
    \item The hard part: After $O(\log^2 N \log n)$ rounds, it holds
        with high probability that every node is either in the MIS or
        has a neighbor in the MIS.  This will give us that the alleged
        MIS is in fact maximal.

        The bad case for termination is when some node $u$ hears a
        neighbor $v$ that is then knocked out by one of its
        neighbors $w$.  So now $u$ is not in the MIS, but neither is
        its (possibly only) neighbor $v$.  The paper gives a rather
        detailed argument that this can't happen too often, which we
        will not attempt to reproduce here.  The basic idea is that
        if one of $v$'s neighbors were going
        to knock $v$ shortly after $v$ first beeps, then the sum of
        the probabilities of those neighbors beeping must be pretty
        high (because at least one of them has to be beeping instead
        of listening when $v$ beeps).  But they don't increase their
        beeping probabilities very fast, so if this is the case, then
        with high probability one of them would have beeped in the previous $c \log N$ rounds
        before $v$ does.  So the most likely scenario is that $v$
        knocks out $u$ and knocks out the rest of its neighbors at the
        same time, causing it to enter the MIS and remain there
        forever.  This doesn't happen always, so we might have to have
        some processes go through the whole $O(\log^2 N)$ initial
        rounds of the algorithm more than once before the MIS
        converges.  But $O(\log n)$ attempts turn out to be enough to
        make it work in the end.
\end{enumerate}

\myChapter{Population protocols}{2026}{}
\label{chapter-population-protocols}

Here are four mostly-equivalent models:
\begin{description}
    \item[Population protocols] A
        \index{protocol!population}\concept{population
        protocol}~\cite{AngluinADFP2006}
        consists of a collection of agents with states in some state
        space $Q$.  At each
        step, the adversary picks two of the agents to interact, and
        both get to update their state according to a joint transition
        function $δ:Q×Q→Q×Q$.  
        A \index{fairness!global}\concept{global fairness} condition
        requires that if some global configuration $C$ of the system
        occurs infinitely often, and there is a step that transforms
        $C$ to $C'$, then this step eventually occurs.

        Computation by population protocols usually consists of
        computing some function of the initial states of the
        population and propagating the output of this function to all
        agents.  As in a self-stabilizing system, termination is not
        detected; instead, we hope to converge to the correct answer
        eventually.

        In some versions of the model, interactions between agents are
        limited by an \index{graph!interaction}{interaction graph}
        (only adjacent agents can interact), or are assumed to be
        chosen randomly instead of adversarially.  These assumptions
        may in some cases increase the power of the model.
    \item[Chemical reaction networks]  In a
        \index{network!chemical reaction}
        \index{reaction network!chemical}
        \index{chemical reaction network} (\concept{CRN} for short), we have a collection of
        molecules representing different \concept{species}.  These
        molecules may undergo chemical reactions that transform one or
        more inputs into one or more outputs, as in this bit of
        classic rocketry:
        \begin{equation*}
            H_2 + O_2 → H_2 O + O
        \end{equation*}

        Computation by a chemical reaction network consists of putting
        some appropriate mix of molecules into a test tube, stirring
        it up, and hoping to learn something from the final product.

        Unlike population protocols, chemical reaction networks do not
        necessarily conserve the number of molecules in each
        interaction, and (in reality at least) require some source of
        energy to keep the reactions going.
    \item[Petri nets] A \index{net!Petri}\concept{Petri
        net}~\cite{Petri1962} is a
        collection of 
        of \indexConcept{place}{places}
        and
        \indexConcept{transition (Petri net)}{transitions},
        in the form of a bipartite graph, with
        \indexConcept{token (Petri net)}{tokens} wandering around
        through the places.  A transition 
        \indexConcept{fire (Petri net)}{fires} by consuming one token
        from each place in its in-neighborhood and adding one token to
        each place in its out-neighborhood, assuming there is at least
        one token on each place in its in-neighborhood.  Various
        conditions are assumed on which transitions fire in which
        order.

        Petri nets were invented to model chemical reaction networks,
        so it's not surprising that they do so.  Pretty much any
        result in population protocols or CRNs can be translated to
        Petri nets or vice versa, by the mapping:

        \begin{center}
            \begin{tabular}{ccc}
                agent & molecule & token \\
                state & species & place \\
                transition & reaction & transition
            \end{tabular}
        \end{center}

        We will not talk much about Petri nets, but there
        has been a fair bit of cross-pollenization between the
        population protocol, CRN, and Petri net literature.
    \item[Vector addition systems] You have a non-negative integer vector $x$.  There is a
        set of rules $-a+b$, where $a$ and $b$ are both non-negative
        integer vectors, and you are allowed to replace $x$ by
        $x-a+b$ if $x-a≥0$.  These are basically Petri nets without
        jargon.
\end{description}

Of these models, population protocols are currently the most popular
in the theory of distributed computing community, with chemical
reaction networks moving up fast.  So we'll talk about population
protocols.

\section{Definition of a population protocol}

Let us begin by giving a formal definition of a population protocol,
following the original definition of
Angluin~\etal~\cite{AngluinADFP2006}.

A \index{protocol!population}\concept{population protocol}
is a tuple $\Tuple{X,Y,Q,I,O,δ}$, where $X$ is the input alphabet, $Y$
is the output alphabet, $Q$ is the state space, $I:X→Q$ maps inputs to
initial states, $O:Q→Y$ maps states to outputs, and $δ:Q×Q→Q×Q$ is the
transition function.

A \concept{population} consists of $n$ agents, taken to be the
vertices of a directed graph called the
\index{graph!interaction}{interaction graph}.  Most of the time we
will assume the interaction graph is a complete graph, but the model
allows for more restrictive assumptions.  A \concept{configuration} is
a mapping $C$ from agents to $Q$.  A \concept{transition}
involves choosing two agents $x$ and $y$ such that $xy$ is an edge in
the interaction graph, and updating the configuration $C$ to a new
configuration $C'$ with $\Tuple{C'_x,C'_y} = δ(\Tuple{C_x,C_y})$ and
$C'_z = C_z$ for all $z ∉ \Set{x,y}$.

The first agent in an
interaction is called the \concept{initiator} and the second agent the
\concept{responder}.  Note that this means that the model breaks
symmetry for us.

With a complete interaction graph, we can will often not bother with
the identities of specific agents and just treat the configuration $C$ as
a multiset of states.

The main difference between population protocols and similar models is
the input and output mappings, and the notion of stable computation,
which gets its own section.

\section{Stably computable predicates}

A predicate $P$ on a vector of initial inputs is \concept{stably
computable} if there exists a population protocol such that it
eventually holds forever that every agent correctly outputs whether $P$ is
true or false.  
Stably computable functions are defined similarly.

One of the big early results on population protocols
was an exact characterization of stably computable predicates for the
complete interaction graph.  We will give a sketch of this result
below, after giving some examples of protocols that compute
particular predicates.

\subsection{Time complexity}

The original population protocol did not define a notion of time,
since the fairness condition allows arbitrarily many junk transitions
before the system makes progress.  More recent work has tended to
compute time complexity by assuming random scheduling, where the pair
of agents to interact is determined by choosing an edge uniformly from
the interaction graph (which means uniformly from all possible pairs
when the interaction graph is complete). 

Assuming random scheduling (and allowing for a small probability of
error) greatly increases the power of population protocols.  So when
using this time measure we have to be careful to mention whether we
are also assuming random scheduling to improve our capabilities.  Most
of the protocols in this section are designed to work as long as the
scheduling satisfies global fairness—they don't exploit random
scheduling—but we will discuss running time
in the random-scheduling case as well.

\subsection{Examples}

These examples are mostly taken from the original paper of
Angluin~\etal~\cite{AngluinADFP2006}.

\subsubsection{Leader election}
\label{section-population-protocol-leader-election}

Most stably computable predicates can be computed as a side-effect of
\indexConcept{leader election!for population protocols}{leader
election}, so we'll start with a leader election protocol.  The state
space consists of $L$ (leader) and $F$ (follower); the input map makes
every process a leader initially.  Omitting transitions that have no
effect, the transition relation is given by
\begin{align*}
    L,L &→ L,F.
\end{align*}

It is easy to see that in any configuration with more than one leader,
there exists a transition that eventually reduces the number of
leaders.  So global fairness says this happens eventually, which
causes us to converge to a single leader after some finite number of
interactions.

If we assume random scheduling, the expected number of transitions to get down to one leader
is exactly
\begin{align*}
∑_{k=2}^{n} \frac{n(n-1)}{k(k-1)}
&= n(n-1) ∑_{k=2}^{n} \frac{1}{k(k-1)}
\\&= n(n-1) ∑_{k=2}^{n} \parens*{\frac{1}{k-1} - \frac{1}{k}}
\\&= n(n-1) \parens*{1 - \frac{1}{n}}
\\&= n^2.
\end{align*}

\subsubsection{Distributing the output}

The usual convention in a population protocol is that we want every
process to report the output.  It turns out that this is equivalent to
the leader reporting the output.

Given a protocol $A$ with states of the form $\Tuple{\ell,x}$ where $\ell
∈ \Set{L,F}$ is the leader bit and $x$ is whatever the protocol is
computing, define a new protocol $A'$ with states $\Tuple{\ell,x,y}$
where $y = O(x)$ when $\ell = L$ and $y$ is the output of the last
leader the agent met when $\ell = F$.

Now as soon as the leader has converged on an output, it only needs to
meet each other agent once to spread it to them.  This takes an
additional $n H_{n-1} / 2 = O(n^2 \log n)$ interactions on average.

\subsubsection{Remainder mod \texorpdfstring{$m$}{m}}
\label{section-population-protocol-remainder}

We can now give an example of a protocol that stably computes a
function: we will count the number of agents in some special initial
state $A$, modulo a constant $m$.  (We can't count the exact total
because the agents are finite-state.)

Each agent has a state $\Tuple{\ell,x}$, where $\ell ∈ \Set{L,F}$ as
in the leader election protocol, and $x ∈ ℤ_m$.  The input mapping
sends $A$ to $\Tuple{L,1}$ and everything else to $\Tuple{L,0}$.  The
non-trivial transitions are given by
\begin{align*}
    \Tuple{L,x}, \Tuple{L,y} &→ \Tuple{L,(x + y) \bmod m}, \Tuple{F,0}
\end{align*}

This protocol satisfies the invariant that the sum over all agents of
the second component, mod $m$, is unchanged by any transition.  Since
the components for any is follower is zero, this means that when we
converge to a unique leader, it will contain the count of initial
$A$'s mod $m$.

\subsubsection{Linear threshold functions}
\label{section-population-protocol-linear-threshold}

Remainder mod $m$ was one of two tools
in~\cite{AngluinADFP2006} that form the foundation for computing all stably computable
predicates.  The second
computes linear threshold predicates, of the form 
\begin{equation}
    ∑ a_i x_i ≥ b,
    \label{eq-linear-threshold}
\end{equation}
where the $x_i$ are the counts of various possible inputs and the
$a_i$ and $b$ are integer constants.  This includes comparisons like
$x_1 > x_2$ as a special case.

The idea is to compute a truncated version of the left-hand side
of~\eqref{eq-linear-threshold} as a side-effect of leader election.

Fix some $k > \max\parens*{\abs{b}, \max_i \abs{a_i}}$.  In addition
to the leader bit, each agent stores an integer in the range $-k$
through $k$.
The input map sends each $x_i$ to the corresponding coefficient $a_i$,
and the transition rules cancel out positive and negative $a_i$, and
push any remaining weight to the leader as much as possible subject to
the limitation that values lie within $[-k,k]$.

Formally, define a truncation function $t(x) = \max(-k, \min(k,
r))$, and a remainder function $r(x) = x - t(x)$.  These have the
property that if $\abs{x} ≤ 2k$, then $t(x)$ and $r(x)$ both have
their absolute value bounded by $k$.  If we have the stronger
condition $\abs{x} ≤ k$, then $t(x) =
x$ and $r(x) = 0$.

We can now define the transition rules:
\begin{align*}
    \Tuple{L,x}, \Tuple{-,y} &→ \Tuple{L, t(x+y)}, \Tuple{F, r(x+y)}\\
    \Tuple{F,x}, \Tuple{F,y} &→ \Tuple{F, t(x+y)}, \Tuple{F, r(x+y)}
\end{align*}

These have the property that the sum of the second components is
preserved by all transitions.  Formally, if we write $y_i$ for the second
component of agent $i$, then $∑ y_i$ does not change through the
execution of the protocol.

When agents with positive and negative values meet, we get
cancellation.  This reduces the quantity $∑ \abs{y_i}$.  Global
fairness implies that this quantity will continue to drop until
eventually all nonzero $y_i$ have the same sign.

Once this occurs, and there is a unique leader, then the leader will
eventually absorb as much of the total as it can.  This will leave the
leader with $y = \min\parens*{k, \max\parens*{-k, ∑ y_i}}$.  By
comparing this quantity with $b$, the leader can compute the threshold
predicate.

\subsection{Presburger arithmetic and semilinear sets}

\index{arithmetic!Presburger}
\concept{Presburger arithmetic}~\cite{Presburger1929} is the first-order theory (in the logic sense) of
the natural numbers with addition, equality, $0$, and $1$.  This allows expressing
ideas like ``$x$ is even:''
\begin{align*}
∃y: x = y+y
\intertext{or ``$x$ is greater than $y$'':}
∃z: x = y+z+1
\end{align*}
but not ``$x$ is prime'' or even $x = y⋅z$.''

Presburger arithmetic has various amazing properties, including
\concept{decidability}–there is an algorithm that will tell you if any
statement in Presburger arithmetic is true or not (in
doubly-exponential time~\cite{FischerR1998})—and \concept{quantifier
elimination}—a formula using any
combination of $∀$ and $∃$ quantifiers can be converted to a
formula with no quantifiers, using the predicates $<$ and $≡_k$ for
constant values of $k$, where $x ≡_k y$ if $x$ and $y$ have the same
remainder mod $k$.

There is also a one-to-one correspondence between predicates in
Presburger arithmetic and
\index{set!semilinear}
\indexConcept{semilinear set}{semilinear sets}, which are finite unions of
\index{set!linear}\indexConcept{linear set}{linear sets} of the form $\Set{b + ∑ a_i x_i}$ where $b$ is a non-negative
integer vector, the
$a_i$ are non-negative integer coefficients, and the $x_i$ are
non-negative integer vectors, and there are only finitely many terms.

(We will not attempt to prove any of this.)

It turns out that Presburger arithmetic (alternatively, semilinear
sets) captures exactly what can and can't be stably computed by a
population protocol.  For example, no semilinear set contains all and
only primes (because any infinite semilinear set on one variable is an
arithmetic progression), and primes aren't recognizable by a
population protocol.  An intuitive and not entirely incorrect
explanation is that in both cases we can't do multiplication because
we can't do nested loops.  In population protocols this is because
even though we can do a single addition that turns exactly $A$ many
blank tokens into $B$'s,
using the rule
\begin{align*}
    A,- &→ A',B
\end{align*}
we can't multiply by repeated addition, because we can't
detect that the first addition step addition has ended to start the
next iteration of the outer loop.

Below we'll describe the correspondence between semilinear sets and
stably-computable predicates.  For full details
see~\cite{AngluinADFP2006,AngluinAE2006semilinear}.

\subsubsection{Semilinear predicates are stably computable}

This part is easy.  We have that any Presburger formula can be
represented as a logical combination of $<$, $+$, and $≡_k$
operators.  We can implement any formula of the form $∑ a_i x_i < b$,
where $a_i$ and $b$ are integer constants, using the linear threshold
function from §\ref{section-population-protocol-linear-threshold}.
We can implement any formula of the form $∑ a_i x_i ≡_k b$ using a
straightforward extension of the mod-$k$ counter from
§\ref{section-population-protocol-remainder}.  If we run these in parallel
for each predicate in our formula, we can then apply any logical
connectives to the result.

For example, if we want to express the
statement that ``$x$ is an odd number greater than $5$'', we have out
agents compute separately $x ≡_2 1$ and $x > 5$; if the leader
computes true for both of these, it assigns true to its real output.

\subsubsection{Stably computable predicates are semilinear}

This is the hard direction, because we have to exclude any possible
algorithm for computing a non-semilinear set.  The full proof is
pretty involved, and can be found in~\cite{AngluinAE2006semilinear}.
A survey paper of Aspnes and Ruppert~\cite{AspnesR2009} gives a
simplified proof of the weaker result (modeled on an
introductory argument in~\cite{AngluinAE2006semilinear}) that any 
stably-computable
set is a finite union of \indexConcept{monoid}{monoids}.  Like
linear sets, monoids are of the form $\Set{b + ∑ a_i x_i}$, but the
number of terms in the sum might be infinite.

We won't do either of these proofs.

\section{Random interactions}

An alternative to assuming worst-case scheduling is to assume random
scheduling: at each step, a pair of distinct agents is chosen
uniformly at random to interact.  This gives the population protocol
substantially more power, and (with some tweaks to allow for different
reactions to occur at different rates) is the standard assumption in
chemical reaction networks.

An example of an algorithm that exploits the assumption of random
scheduling is the \concept{approximate majority} protocol of Angluin,
Aspnes, and Eisenstat~\cite{AngluinAE2008majority}, which was
also independently discovered by Perron, Vasudevan, and
Vojnovic~\cite{PerronVV2009}.  This protocol starts
with a mix of agents in states $x$ and $y$, and uses a third state $b$
(for blank) to allow the initial majority value to quickly take over
the entire population.  The non-trivial transition rules are:
\begin{align*}
    xy &→ xb\\
    yx &→ yb\\
    xb &→ xx\\
    bx &→ xx\\
    yb &→ yy\\
    by &→ yy
\end{align*}

If two opposite agents meet, one becomes blank, depending on
which initiates the reaction (this is equivalent to flipping a coin
under the random-scheduling assumption).  These reactions produce a
supply of blank agents, drawing equally from both sides.  But when a
blank agent meets a non-blank agent, it adopts the non-blank agent's
state.  This is more likely to be the majority state, since there are
more agents to meet in the majority state.  So if we consider only
transitions that change the net number of $x$ agents minus $y$
agents, we get a random walk biased toward the majority value with an
absorbing barrier in the state where all agents are equal.  However,
the rate at which these special transitions occur depends on how
quickly blank agents are created, which in turn depends on the
relative numbers of $x$ and $y$ agents.

Analysis of the full process is difficult, but Angluin~\etal~show that
with high probability all agents end up in the initial majority state
in $O(n \log n)$ interactions, provided the initial majority is large
enough ($Ω(\sqrt{n} \log n)$, later improved to $Ω(\sqrt{n \log n}$ by
Condon~\etal~\cite{CondonHKM2019}).  Curiously, a later paper by
Cardelli and Csikász-Nagy~\cite{CardelliC2012} showed that the cell
cycle controlling mitosis in all living organisms uses a chemical
switch that looks suspiciously like the approximate majority
algorithm, making this algorithm roughly three billion years old.

But we can do better than this.  With random scheduling, we have much
more control over how a computation evolves, and this can be used to
simulate (with high probability) a register machine, which in turn can
be used to simulate a Turing machine.  The catch is that the state of
a population protocol with $n$ agents can be described using $O(\log
n)$ bits, by counting the number of agents in each state.  So the most
we can simulate is a machine that has $O(\log n)$ space.

The original population protocol paper included a simulation of an
$O(\log n)$-space Turing machine, but the time overhead per operation
was very bad, since most operations involved a controller agent
personally adjusting the state of some other agent, which requires 
$Θ(n)$ time on average before the controller meets its target.

A better construction was given by
Angluin~\etal~\cite{AngluinAE2008fast}, under the assumption
that the population starts with a single agent in a special leader
state.  The main technique used in this paper it to propagate
a message $m$ using an epidemic protocol $mb → mm$.  The time for an
epidemic to spread through a population of $n$ individuals through
random pairwise interactions is well-understood, and has the property that (a)
the time to infect everybody is $Θ(\log n)$ with high
probability, and (b) it's still $Θ(\log n)$ with high probability
if we just want to infect a polynomial fraction $n^ε$ of the agents.

So now the idea is that if the leader, for example, wants to test if
there is a particular state $x$ in the population, it can spread a
message $x?$ using an epidemic, and any agent with $x$ can respond by
starting a counter-epidemic $x!$.  So if there is an $x$, the leader finds
out about it in $O(\log n)$ time, the time for the first epidemic to
go out plus the time for the second epidemic to come back.

What if there is no $x$ agent?  Then the query goes out but nothing
comes back.  If the leader can count off $Θ(\log n)$ time units (with
an appropriate constant, it can detect this.  But it does not have
enough states by itself to count to $Θ(\log n)$.

The solution is to take advantage of the known spreading time for
epidemics to build a \index{clock!phase}\concept{phase clock} out of
epidemics.  The idea here is that the leader will always be in some
\concept{phase} $0\dots m-1$.  Non-leader agents try to catch up with
the leader by picking up on the latest rumor of the leader's phase,
which is implemented formally by transitions of the form $\Tuple{x,i}
\Tuple{F,j} → \Tuple{x,i} \Tuple{F,i}$ when $0 < i-j < m/2 \pmod{m}$.
The leader on the other hand is a hipster and doesn't like it when
everybody catches up; if it sees a follower in the same phase, it
advances to the next phase to maintain its uniqueness:
$\Tuple{L,i}\Tuple{F,i} → \Tuple{L,i+1} \Tuple{F,i}$.

Because the current phase spreads like an epidemic, when the leader
advances to $i+1$, every agent catches up in $a \log n$ time w.h.p.
This means both that the leader doesn't spend too much time in $i+1$
before meeting a same-phase follower and that followers don't get too
far behind.  (In particular, followers don't get so far behind that
they start pulling other followers forward.)  But we also have that it
takes at least $b \log n$ time w.h.p.~before more than $n^ε$ followers
catch up.  This gives at most an $n^{ε-1} \ll 1$ probability that the
leader advances twice in $b \log n$ time.  By making $m$ large enough,
the chances that this happens enough to get all the way around the
clock in less than, say $b (m/2) \log n)$ time can be made at most
$n^{-c}$ for any fixed $c$.  So the leader can now count of $Θ(\log
n)$ time w.h.p., and in particular can use this to time any other
epidemics that are propagating around in parallel with the phase
clock.

Angluin~\etal~use these techniques to implement various basic arithmetic
operations such as addition, multiplication, division, etc., on the
counts of agents in various states, which gives the register machine
simulation.  The simulation can fail with nonzero probability, which
is necessary because otherwise it would allow implementing
non-semilinear operations in the adversarial scheduling model.

The assumption of an initial leader can be replaced by a leader
election algorithm, but at the time of the Angluin~\etal{} paper, no
leader election algorithm better than the $Θ(n)$-time
fratricide protocol described
§\ref{section-population-protocol-leader-election} was known, and even
using this protocol requires an additional polynomial-time cleaning step
before we can run the main algorithm,
to be sure that there are no leftover phase clock remnants from
deposed leaders to cause trouble.  So the question of whether this
could be done faster remained open.

Hopes of finding a better leader election protocol without changing
the model ended
when Doty and Soloveichek~\cite{DotyS2015} proved a matching
$Ω(n)$
lower bound on the expected time to convergence for
any leader election algorithm in the more general model
of chemical reaction networks.  This results holds assuming constant
states and a \concept{dense} initial population where any state that
appears is represented by a constant fraction of the agents.

Because of this and related lower bounds, recent work on fast
population protocols has tended to assume more states.  This is a
fast-moving area of research, so I will omit trying to summarize the
current state of the art here. For an introduction to this
work see~\cite{AlistarhG2018,ElsasserR2018}.

\part*{Appendix}
\phantomsection
\addcontentsline{toc}{part}{Appendix}

\appendix

\chapter{Sample assignments from Spring 2026}

\section{Assignment 1: due Thursday 2026-01-29, at 23:59 Eastern US time}

    \subsection{Summing over a hypercube}

    Suppose we have an asynchronous network with $n=2^d$ processes in
    the form of a hypercube. Each process $p_i$ has a location $i =
    i_1 \dots i_n ∈ \Set{0,1}^d$, and can send messages to any process
    $p_j$ such that $i_k = j_k$ for all but one coordinate $k$. Any
    message sent is delivered in at most one time unit as usual, and
    there are no failures.

    Let us consider two versions of this network. In the first
    version, each process is aware of its location. In the second, 
    no process is aware of its location, and all run exactly the same
    code, although each can tell which of its $d$ neighbors
    differ in which coordinate.

    Initially, each process starts with a value $v_i ∈ \Set{0,1}$.
    We'd like all processes to compute $∑_i v_i$.

    \begin{enumerate}
        \item Find an algorithm that solves this problem in the
            location-aware model, and show that it has asymptotically
            optimal time and message complexity.
        \item Do the same for the location-oblivious model.
    \end{enumerate}

        \subsubsection*{Solution}

        \begin{enumerate}
            \item Construct a tree of depth $d$ on the hypercube in
                some reasonable way. One possibility is assigning as
                parent to each node $p_i$ the node $p_j$ where $j$
                replaces the rightmost $1$ in $i$ with $0$. This tree
                will have $p_{0^d}$ as its root and has depth exactly
                $d$. There is no cost to constructing this tree,
                because each process can just compute its parent and
                children locally from its id.

                Run convergecast on this tree in $d$ time using $n-1$
                messages to compute $∑ v_i$ at $p_{0^d}$. Then run
                broadcast in the same time to distribute the sum to
                all nodes. This gives a solution to the problem with
                time complexity $O(d)$ and message complexity $O(n)$.

                It is straightforward to show that no protocol can do
                better than $d$ time using an indistinguishability
                argument. Consider two synchronous executions of
                length $d-1$ in which each $v_i = 0$ except $v_{1^d}$,
                which is either $0$ or $1$. We can easily show by
                induction on $t$ that after $t$ time units, any node
                at distance more than $t$ from $p_{1^d}$ has the same
                state and sends the same messages in both executions.
                So it is not possible for $p_{0^d}$ in particular to
                distinguish these two inputs (and thus report a
                correct sum) after this time.

                For message complexity, we'd like to argue that any
                correct protocol must have each process send at least
                one message in any execution. That this is necessary
                is not immediately obvious: in a synchronous protocol,
                a process could signal if its input is $1$ or not by
                choosing whether or not to send a message. But in an
                asynchronous protocol this doesn't work.

                Fix some protocol and suppose it has an execution $Ξ$ in
                which some process $p_i$ sends no messages. Let $v_i$
                be the input to $p_i$ in this execution, and consider
                an execution $Ξ'$ in which $p_i$ instead has input
                $v'_i = ¬v_i$. Schedule delivery of all messages by
                processes other than $p_i$ in the same order in $Ξ'_i$
                as in $Ξ_i$ while delaying any messages sent by $p_i$.
                Because the prefix of $Ξ'_i$ in which $p_i$'s messages
                are delayed is indistinguishable to all other
                processes from $Ξ_i$, whatever sum the other processes
                decided on in $Ξ_i$, they will decide on the same sum
                in $Ξ'_i$. But this sum is incorrect for
                $Ξ'_i$.\footnote{The other processes may learn of
                their error when $p_i$'s messages are finally
                delivered, but having made their decision, their tears
                will be too late to wash it away.}

            \item Here we solve the problem recursively for
                increasingly large subcubes of the full cube.
                
                Each process $p_i$ initially sets a local variable
                $s_i$ to $v_i$.
                In each of $d$ asynchronous rounds $r=1,\dots,d$, 
                it (a) sends $s_i$ to its neighbor $p_j$ that differs in the
                $r$-th coordinate, (b) waits to receive $s_j$ from
                that neighbor, then (c) sets $s_i$ to $s_i + s_j$. As
                usual, messages from later rounds are deferred until
                that round. At the end of $d$ rounds, $p_i$ returns
                $s_i$ as the sum.

                To show that this works, we will argue an invariant
                that at the end of round $r$, $s_i$ will be equal to
                $∑_x x v_{i_{r+1} i_{r+2} \dots i_d}$, where $x$ ranges over
                all bit-vectors of length $d-r$. This holds trivially
                after $0$ rounds. For $r>0$, assuming it holds after
                $r-1$ rounds, we have that $s_i$ starts as 
                $∑_x x v_{i_{r} i_{r+1} \dots i_d}$,
                to which is
                added $s_j = ∑_x x v_{¬i_{r} i_{r+1} \dots i_d}$,
                giving 
                $∑_x x v_{i_{r+1} \dots i_d}$, where in each sum $x$
                ranges over all bit-vectors of appropriate length.
                After $d$ rounds, this gives $s_i = ∑_x v_x$.

                The total time complexity of this algorithm is exactly
                $d$ and the total message complexity is exactly $nd$,
                since each process sends exactly one message in each
                of $d$ rounds.

                The same $Ω(d)$ lower bound for time complexity for the
                location-aware case applies \emph{a fortiori} to the
                location unaware case, so time complexity is
                asymptotically optimal.

                For message complexity, consider a synchronous
                execution starting with $v_i = 0$ for all $i$. 
                By symmetry, ever process sends the same sequence of
                messages in the same directions and passes through the
                same states before ultimately deciding. We will argue
                that for a correct protocol, this makes every process
                send at least one message in each of the $d$
                directions, giving an $nd$ lower bound on message
                complexity.

                Suppose that there is some direction $j$ in this
                execution such that no process sends a message in
                direction $j$. Assuming the protocol is correct, every
                process decides $0$. Now consider a second execution
                in which every process $p_i$ has $v_i = i_j$, and run
                all the processes with $i_j = 0$ until they all
                decide, while delaying any messages from the processes
                with $i_j = 1$. Since none of the $i_j=0$ processes ever
                receive a message from a process with $i_j = 1$ in
                either execution, the two executions are
                indistinguishable, and so the processes with $i_j = 0$
                all decide $0$. This is the wrong value in the second
                execution, which contradicts our assumption that there
                is a correct protocol in which some process doesn't
                send in direction $i_j$ in the synchronous execution
                with all $v_i = 0$.
        \end{enumerate}

    \subsection{A broken ring}

    A manufacturer of low-budget asynchronous one-way ring networks
    has hired you to construct a protocol for finding a faulty edge.
    The idea is that we have $n≥2$ processes $p_0,\dots,p_{n-1}$,
    each process $p_i$ can send messages to process $p_{i+1}$ (where
    all arithmetic on indices is done mod $n$), and
    every message is eventually delivered unless $i=b$ where $b$ is
    the starting position of a unique broken edge. For this edge, no
    messages are ever delivered from $p_b$ to $p_{b+1}$.

    The processes will each take infinitely many steps in an infinite
    execution, but they are deterministic and anonymous: every process
    runs the same code and no process knows its position in the ring
    or the position of the bad edge $b→b+1$.
    The processes \emph{do} know $n$.

    We'd like to find a protocol that eventually sets off an alarm at exactly
    one of $p_b$ or $p_{b+1}$. This protocol should use as little
    time and as few messages as possible, and should never produce a
    false alarm where both of $p_b$ or $p_{b+1}$ set off the alarm or
    some process other than $p_b$ or $p_{b+1}$ does so.

    Describe such a protocol, prove that it is correct, and prove that
    its time complexity and message complexity are both asymptotically
    optimal.

        \subsubsection*{Solution}

        A protocol is given in Algorithm~\ref{alg-hw-broken-ring}. It
        sets off an alarm at $p_b$ after $O(n)$ time using $O(n^2)$
        messages. The idea is that each message acts like a token counting how many
        edges it has traversed. So when $p_b$ receives $n-1$, it can
        tell that none of the $n-1$ edges to its left are broken,
        implying that the edge to its right is.

        \begin{algorithm}
            \Initially{
                send $0$\;
                \label{line-hw-broken-ring-send-zero}
            }

            \UponReceiving{$m$}{
                send $m+1$\;
                \label{line-hw-broken-ring-send-plus}
                \If{$m=n-2$}{
                    set off alarm\;
                    \label{line-hw-broken-ring-alarm}
                }
            }
            \caption{Detecting the broken edge}
            \label{alg-hw-broken-ring}
        \end{algorithm}

        Since the protocol is anonymous we can without loss
        of generality assume that $b=n-1$. We'll show that $p_{n-1} =
        p_b$ sets off the alarm, and is the only process to do so.

        We start with a lemma showing exactly which messages are sent
        in any execution.

        \begin{lemma}
            \label{lemma-hw-broken-ring-messages}
            Each $p_i$ sends one copy of each $m∈\Set{0,\dots,i}$ by
            time $m$.
        \end{lemma}
        \begin{proof}
            By induction on $i$. 

            For $i=0$, $p_0$ sends $0$ once in
            Line~\ref{line-hw-broken-ring-send-zero} at time $0$.
            Since $b=n-1$, no message from $p_{n-1}$ is every
            delivered to $p_0$, so it never sends a message in
            Line~\ref{line-hw-broken-ring-send-plus}. This makes
            the set of messages sent by $p_0$ exactly $\Set{0}$.

            For $i>0$, the induction hypothesis says that $p_{i-1}$
            sends one copy of each $m∈\Set{0,\dots,i-1}$ no later than
            time $m$. Since $i-1 ≠
            n-1 = b$, each of these messages is eventually delivered
            no later than time $m+1$.
            This causes $p_i$ to send one copy of each message $m+1$ in
            $\Set{1,\dots,m}$ in
            Line~\ref{line-hw-broken-ring-send-plus} no later than
            time $m+1$.
            Adding the
            $0$ sent in Line~\ref{line-hw-broken-ring-send-zero} at
            time $0$ gives the full claim.
        \end{proof}

        Since only $p_{n-2}$ and $p_{n-1}$ send $n-2$, and no message
        from $p_{n-1}$ is every delivered, it follows that only
        $p_{n-1}$ receives $n-2$ and sets off the alarm in
        Line~\ref{line-hw-broken-ring-alarm}.

        It is immediate from the lemma that the time complexity of
        this protocol is $O(n)$ and the message complexity is
        $O(n^2)$. We now need to show that this is optimal for any
        anonymous protocol.

        Again we will assume that $b=n-1$ so that we don't have to
        keep adding in $b$ everywhere.

        First let's rule out protocols that set off the alarm at
        $p_0$. Suppose we have such a protocol. Since $p_0$ never
        receives any messages, it must set off the alarm after some
        finite number of steps. Run $p_1$ for the same finite number
        of steps without delivering any messages: now $p_1$ sets off
        the alarm too.

        Now fix some protocol and consider an execution of
        that protocol organized into a sequence of rounds, where each
        process receives in round $r+1$ the messages sent by its
        neighbor in round $r$.

        This execution maintains partial symmetry between the
        processes for a while:
        \begin{lemma}
            \label{lemma-hw-broken-ring-symmetry}
            In the execution described above, in every round $r$,
            every $p_i$ with $r≤i≤n-1$ 
            enters the same state and sends the same messages.
        \end{lemma}
        \begin{proof}
            By induction on $r$.
            For $r=0$, this is immediate from determinism and
            anonymity. For $r>0$, since every $p_i$ with $r-1≤i≤n-1$ sent
            the same messages in round $r-1$, every $p_i$ with $r≤i≤n-1$
            receives the same messages at $r$, updates its state to the
            same value, and sends the same messages.
        \end{proof}

        For the time complexity lower bound,
        the lemma tells us that $p_{n-2}$ and $p_{n-1}$ have the same
        state at $r=n-2$. So if $p_{n-1}$ sets off its alarm at or
        before this time, so does $p_{n-2}$. So any correct protocol
        must take at least $n-1$ rounds, translating into $n-1$ time,
        exactly equal to the upper
        bound given by Algorithm~\ref{alg-hw-broken-ring}.

        This construction doesn't say much about message complexity,
        because there is no guarantee that every process sends a
        message in every round. We'll construct a different execution
        to get an $Ω(n^2)$ message complexity lower bound. In this
        execution, it will help that we don't need to think about its
        time complexity.

        Here we run the processes sequentially. 
        For each process $p_i$, we deliver the messages sent by the previous
        process (if there is one) one at a time and run $p_i$ after
        each message until it stops sending messages. In the event
        that some $p_i$ attempts to send infinitely many messages, we
        truncate the protocol after $n^2$ messages total, which
        already gets us our desired lower bound. So below we will
        assume that this doesn't happen, and every $p_i$ gets its turn to
        run.

        Because each $p_i$ is anonymous and deterministic, letting
        $M_i$ be the sequence of messages received by $p_i$, it holds
        that (a) there is a function $f$ such that $M_{i+1} = f(M_i)$;
        (b) $f$ is monotone in the sense that if $M$ is a prefix of
        $M'$, $f(M)$ is a prefix of $f(M')$; and (c) if $M_i = M_j$,
        $p_i$ sets off the alarm if and only if $p_j$ does.

        We will use this to argue that each $M_i$ is a \emph{proper}
        prefix of $M_{i+1}$ for all $i < n-1$. Suppose otherwise. Then
        $M_i = M_{i+1}$ for some $i$. But then $M_{i+2} = f(M_{i+1}) =
        f(M_i) = M_{i+1} = M_i$ as well, and iterating this equation
        gives $M_j = M_i$ for all $j≥i$. But then $p_{n-2}$ and
        $p_{n-1}$ either both set off the alarm or neither does.

        But if $M_i$ is a proper prefix of $M_{i+1}$, we have
        $\card{M_i} < \card{M_{i+1}}$. Starting with $\card{M_0} = 0$,
        this gives $\card{M_i} ≥ i$ for all $i ≤ n-1$. So the total
        number of messages is at least $∑_{i=0}^{n-1} \card{M_i} ≥
        ∑_{i=0}^{n-1} i = Ω(n^2)$.

\section{Assignment 2: due Thursday 2026-02-12, at 23:59 Eastern US time}

    \subsection{Misdirected mail}

    Somehow we acquired a network router that doesn't understand
    recipient addresses. The result is an asynchronous message-passing
    system where messages are never lost or duplicated, but any
    message sent may be delivered to any machine, including the
    original sender. The only thing keeping this network from being
    completely useless is a rather weak fairness guarantee: in any
    execution, (a) every process takes infinitely many steps, and (b)
    if infinitely many messages are sent, then every process receives
    infinitely many messages.

    We'd like to perform a broadcast operation in this terrible
    network. Let a leader process $p$ start with some
    initial value $m∈\Set{0,1}$. In a successful execution of broadcast
    protocol, $n>0$ follower processes $q_1,\dots,q_n$ eventually
    decide on $m$. A follower process can continue to send and receive
    messages after deciding, but it can only decide once. All
    processes have unique identities and know $n$.

    Prove or disprove: Any broadcast protocol in which every follower
    eventually decides correctly in all executions sends infinitely
    many messages in some execution.

        \subsubsection*{Solution}
            Disproof: We'll construct an working protocol that sends only
            finitely many messages in any execution.

            This protocol will satisfy an invariant
            that at any step, (a) either all followers have decided
            $m$ and there are no messages in transit, or there will be
            exactly one message in transit, of the form $\Tuple{m,S}$
            where $m$ is the leader's initial value, $S$ is the set
            of followers that have already decided $m$, and $\card{S}
            < n$.

            Initially, the leader sends $\Tuple{m,∅}$. Upon receiving
            $\Tuple{m,S}$, process $q_i$ decides $m$ if it has not
            done so already, and sends $\Tuple{m,S∪\Set{q_i}}$ if
            $\card{S∪\Set{q_i}}<n$. If $p$ receives $\Tuple{m,S}$, it
            just resends it.

            Now let us check the invariant:
            \begin{description}
                \item[One message in transit before termination:] True in the initial
                    state. If a follower or leader sends a message,
                    it does so only after receiving one, leaving the
                    total at one. If a follower $q_i$ doesn't resend a
                    message after receiving $\Tuple{m,S}$, the protocol is done:
                    since every process is in $S∪\Set{q_i}$, each
                    $q_j$ has already decided $m$ because of the
                    $q_j∈S$ and the second clause of the invariant
                    applies; or it decides $m$ at this step because
                    $q_j = q_i$.
                \item[Any $q_i∈S$ has decided $m$:] When $q_i$ adds
                    itself to $S$ for the first time, it also decides
                    $m$.
                \item[In any message $\Tuple{m,S}$, $\card{S} < n$:] 
                    If it's the initial message from $p$, we have
                    $\card{S} = 0 < n$.
                    If it's a message sent by a follower, this is
                    immediate from the protocol. If sent by the
                    leader, it follows from the fact that the leader
                    is resending a message $\Tuple{m,S}$ in which
                    $\card{S} < n$ because of the invariant holding in
                    the previous step.
            \end{description}

            We now need to show that the protocol sends only finitely
            many messages. Suppose otherwise. Since infinitely many
            messages are sent, every process receives infinitely many
            messages. So there is some step at which every
            follower process has received at least one message and
            decided $m$. But then any message $\Tuple{m,S}$ sent
            after this step has $\card{S} = n$, contradicting the 
            invariant.

            We also get that the protocol successfully broadcasts to
            all processes: if only finitely many messages are sent,
            there is a step at which there are no messages in transit,
            and the first clause of the invariant tells us that all
            processes have decided $m$ by this step.

        \subsection{Synchronous agreement with staggered start}

        Let us define a variant of the synchronous model where the $n$
        processes do not all start at the same time. Instead, a
        process $i$ either wakes up when it first receives one or more
        messages, or wakes up in round $s_i$ with a delivery event that
        includes no messages if it has not done so already, where $s$
        is a bijective mapping between process ids and
        $\Set{0,\dots,n-1}$. The processes do not have access to $s$
        and cannot directly compute round numbers, but as in the usual
        synchronous model, it is guaranteed that any message sent in
        round $r$ is received in round $r+1$ and that a non-faulty
        process, once started, is given the opportunity to take a step
        in every subsequent round whether it receives any messages or
        not.

        We will also assume crash failures as described in
        Chapter~\ref{chapter-synchronous-agreement}. When a faulty
        process crashes in some round $r$, some or all of its outgoing
        messages are unsent, and the process takes no steps in round
        $r+1$ or later. We assume that at most $f<n$ processes are
        faulty. The value of $f$ is known to the processes.

        Our goal is to solve agreement by constructing a protocol that
        satisfies the termination, agreement, and validity conditions
        given in §\ref{section-synchronous-agreement-problem}.

        Prove or disprove: $Θ(f)$ rounds are necessary and sufficient
        to reach agreement in the worst case in this model.

            \subsubsection*{Solution}

            We'll prove this by showing that both the upper and lower
            bounds from Dolev-Strong~\cite{DolevS1983} still work
            after a little adjustment.

            For the upper bound, we'll argue that all processes start
            no later than round $f+1$, and then if they then run the
            standard Dolev-Strong algorithm (see
            §\ref{section-synchronous-agreement-flooding} for at least
            another $f+1$ rounds, there is some round in which every
            non-faulty process transmits its current set, after which
            every non-faulty process maintains the same set forever as
            shown in
            Lemma~\ref{lemma-synchronous-agreement-flooding-gives-same-set}.

            In detail, each process $p$ keeps a set $S$ of pairs
            $\Tuple{i,v}$, which initially contains only the
            process's own id and input.
            If the process $i$ wakes up on its own, it sends this set
            $\Set{\Tuple{i,v}}$ to all processes. If it wakes up by
            receiving some collection of sets $S_j$, it updates its
            own set $S_i ← S_i ∪ \bigcup_j S_j$ and sends the updated
            set to all processes. In either case it continues
            taking the union of incoming sets with its own and
            broadcasting the result for $2f+2$ rounds, using a local
            counter to keep track. At the end of these $2f+2$ rounds,
            it returns the smallest value in its set.

            For the wake-up part of this, observe that at least one
            process $p$ with $s(p) ∈ \Set{0,\dots,f}$ is non-faulty.
            Since $p$ sends messages in round $s(p)$ to all processes,
            every process starts no later than round $s(p)+1 ≤ f+1$
            (and no earlier than round $0$).

            Now consider what happens in rounds $f+1$ through $2f+1$.
            Since these include $f+1$ rounds, there is some round $r$
            in which no fault occurs. We also have that in this round
            $r$ all non-faulty processes are (a) awake and (b) still
            running, since every non-faulty process starts no later
            than round $f+1$ and finishes no earlier than round
            $2f+2$. Applying the same reasoning as in
            Lemma~\ref{lemma-synchronous-agreement-flooding-gives-same-set},
            every process $i$ stores the same set $S$ starting in
            round $r+1 ≤ 2f+2$. While the processes do not necessarily
            all finish at the same round, they each return the same
            value, the minimum in this common set. This gives
            agreement.

            Validity follows from the invariant that every value
            in any $S^r_i$ is some process's input. Termination occurs
            in at most $3f+3 = O(f)$ rounds, since every process wakes
            up by round $f+1$ and finishes at most $2f+2$ rounds
            later.

            For the lower bound, suppose that we have an algorithm
            that works in fewer than $f+1$ rounds in the
            staggered-start model. We can simulate this algorithm
            in the standard synchronous model by assignment our own
            starting rounds $s_i$ and having each process send nothing
            in any round where it would not have woken up yet. This
            gives an algorithm for the standard model that satisfies
            validity and agreement, and that terminates in fewer than
            $f+1$ rounds, contradicting the original Dolev-Strong lower bound.

\section{Assignment 3: due Thursday 2026-02-26, at 23:59 Eastern US time} 

    \subsection{Measuring unreliably}

    Consider a synchronous message-passing system with
    $m$ thermometers $t_1,\dots,t_m$ and $n$ deterministic processes
    $p_1,\dots,p_n$. Under normal conditions, each thermometer sends a
    reading $0$ (cold) or $1$ (hot) as a message to each process at
    time $0$. The processes then need to agree on a common decision
    value based on these readings.

    Sadly, up to $d$ of the thermometers are
    defective, meaning that they can send arbitrary values to each
    process (including different values to different processes); the
    remaining non-defective processes each send the same value to all
    processes, although it is possible that two different non-defective
    thermometers send different values. It is also the case that up to
    $f$ of the processes can suffer Byzantine faults.

    We'd like a protocol that satisfies the following requirements:
    \begin{description}
        \item[Termination] Every non-faulty process outputs a decision
            value in a fixed number of rounds.
        \item[Agreement] Every non-faulty process outputs the same
            value.
        \item[Validity] If every non-defective thermometer sends the
            same value $v$, all non-faulty processes output $v$.
    \end{description}

    Let's call a protocol that satisfies these conditions a
    \conceptFormat{measurement agreement} protocol.

    \begin{enumerate}
        \item Show that there is measurement agreement protocol
            that works when $m≥2d+1$ and $n≥3f+1$.
        \item Show that no such protocol works when $1 < m < 2d+1$
            and $n=3f+1$.
        \item Show that no such protocol works when $3 ≤ m = 2d+1$
            and $3 ≤ n < 3f+1$.
    \end{enumerate}

        \subsubsection*{Solution}

        \begin{enumerate}
            \item For this case we can reduce to any standard
                Byzantine agreement protocol that works when $n≥3f+1$.

                Have each process $p_i$ collect $m≥2d+1$ readings from the
                thermometers and compute a simulated input $v_i$ as the
                majority of the readings it sees. Then run Byzantine
                agreement on these inputs.

                This satisfies termination and agreement because the
                Byzantine agreement protocol does.

                For validity, suppose that all $m-d$ non-defective
                thermometers report the same value $v$. Since $m-d ≥
                2d+1-d = d+1 > d$, each non-faulty process $p_i$ sees a
                majority of $v$ readings and sets $v_i=v$. This means
                that the standard Byzantine agreement protocol sees
                only $v$ as an input from non-faulty processes, so by
                the validity condition on it, every non-faulty process
                outputs $v$.

            \item Here we can construct two executions that differ
                only in which thermometers are labeled as defective.
                This will violate validity in at least one execution
                no matter what the protocol does as long as there is
                at least one non-faulty process, covering all cases
                where $f<n$, including $n=3f+1$.

                Divide the thermometers into two non-empty sets $T_0$
                and $T_1$ of size at most $d$ each. Making both sets
                non-empty uses $1<d$; making each set have size at
                most $d$ uses $m<2d+1$ which implies $m≤2d$.

                In the first execution, all thermometers in $T_0$ are
                defective; in the second, all thermometers in $T_1$ are
                defective. In both executions all thermometers in each
                set $T_b$ report value $b$ to all processes. Since
                these executions are indistinguishable to the
                processes, if we assume termination and agreement
                hold, each non-faulty process will output the same
                value $v$ in both executions. But in the first
                execution, validity requires that all non-faulty
                processes output $1$, the common value of all
                non-defective thermometers, while in the second,
                validity requires that they all output $0$. So
                validity is violated in at least one of these
                executions.

            \item Here we reduce from standard Byzantine agreement.

                Suppose that $m=2d+1$.
                To solve Byzantine agreement, have each process $p_i$ simulate
                receiving $0$ from $t_1,\dots,t_d$, $1$ from
                $t_{d+1},\dots,t_{2d}$, and its input $v_i$ from
                $t_{2d+1}$. Then run the measurement agreement
                protocol on these simulated values.
                This gives termination and agreement if the
                measurement agreement protocol does. For validity, if
                every non-faulty process has the same input $v$,
                then every non-faulty process simulates receiving
                $d+1$ copies of $v$ from the same set of $d+1$
                thermometers. Since there exists an execution in which
                these $d+1$ thermometers are non-defective, validity for
                the measurement agreement protocol requires that all
                non-faulty processes output $v$, satisfying the
                validity condition for Byzantine agreement.

                This tells us that if we can solve
                measurement agreement with $m=2d+1$ and some values
                for $f$ and $n$, we can solve Byzantine agreement
                under the same conditions. But since we can't solve
                Byzantine agreement when $n<3f+1$, we can't solve
                measurement agreement when $m=2d+1$ and $n<3f+1$
                either.
        \end{enumerate}

        It's probably worth noting that these three cases do not
        actually cover all the possible values of $m$, $d$, $n$, and
        $f$. For example, $m=d=1$ and $n=3f+1$ turns into standard
        Byzantine agreement if each process takes the value reported
        by the single thermometer as its input; this is solvable
        even though $m < 2d+1$. We have also not excluded the
        possibility that measurement agreement might be solvable when
        $d$ is much smaller than $g$, even if $n < 3f+1$.

    \subsection{Measuring unreliability}

    Suppose we have access to a failure detector that, instead of
    reporting a set of suspect processes, reports the size of
    this set. We'll describe a class of such \conceptFormat{counting failure
    detectors} by prefixing the name of the underlying failure detector
    with a $\#$ symbol. For example, a counting failure detector
    that returns the number of processes suspected by $P$ will be
    written as $\#P$ (and neither of these failure detector classes should be
    confused with the computational complexity class of the same name).

    We would like to use these counting failure detectors to solve
    consensus in an asynchronous message-passing system with $n$
    processes and up to $f$ crash failures.

    \begin{enumerate}
        \item What is the maximum value of $f<n$ for which we can solve
            consensus using $\#S$?
        \item What is the maximum value of $f<n$ for which we can solve
            consensus using $\#P$?
    \end{enumerate}

    Prove your answer in each case.

        \subsubsection*{Solution}

        \begin{enumerate}
            \item The maximum value of $f$ for which we can solve
                consensus using $\#S$ is $0$.

                Proof: We'll show that a message-passing system
                equipped with no failure detector at all can simulate
                $\#S$. Recall that $S$ is strongly complete and weakly
                accurate, meaning that (a) all faulty processes are
                eventually permanently suspected, and (b) there exists
                some non-faulty process that is never suspected.

                Pick some non-faulty $c$, and let a failure detector
                never suspect $c$ and always suspect every $p≠c$. This
                satisfies the requirements for $S$. The size of the
                set of suspect processes is always $n-1$, so there is
                an implementation of $\#S$ that simply always returns
                $n-1$ regardless of the failure pattern.

                Since the FLP impossibility proof
                (\cite{FischerLP1985}, or see
                Chapter~\ref{chapter-FLP}) tells use that an
                asynchronous message-passing system with one crash
                failure can't solve consensus, an asynchronous
                message-passing system with one crash failure where
                every process has a sticker that says $n-1$ glued to
                it can't solve consensus either.
            \item The maximum value of $f$ for which we can solve
                consensus using $\#S$ is $n$.

                In this case, we can reduce $P$ to $\#P$, then use
                Algorithm~\ref{alg-strong-failure-detector-consensus},
                which requires only $S$ (which reduces to $P$) 
                to solve consensus for any number of failures $f$.

                \newData{\Heartbeat}{heartbeat}

                The $P$ construction works like this:
                \begin{enumerate}
                    \item Whenever the output of my instance of
                        $\#P$ increases to
                        $k$ for the first time,
                        send a message $\Heartbeat_k$ to
                        all processes (including myself).
                    \item Whenever I receive $\Heartbeat_k$
                        from $n-k$ processes, mark any process I
                        didn't receive this message from as suspect.
                \end{enumerate}

                To show that this construction is strongly accurate,
                observe that when the output of $\#P$ reaches $k$ at
                some process $p$, then $k$ processes have already
                crashed (since the instance of $P$ inside $\#P$ is
                strongly accurate). So at most $n-k$ processes send
                $\Heartbeat_k$, and any process that doesn't has
                already crashed when the first such message goes out.
                It follows that any process that receives $n-k$ of
                these messages and suspects the non-senders does so
                correctly.

                For strong completeness, let $k$ be the maximum number
                of processes that crash. Then eventually $\#P$ reports
                $k$ at all $n-k$ non-faulty processes, these processes
                all send $\Heartbeat_k$, and eventually all $n-k$
                non-faulty processes receive these messages and
                suspect the $k$ faulty processes that didn't.
        \end{enumerate}

\section{Assignment 4: due Thursday 2026-03-26, at 23:59 Eastern US time}

    \subsection{Filling in the blanks}

    We'd like to implement a structure similar to a distributed ledger
    in an asynchronous message-passing system with deterministic
    processes and up to one crash failure. The idea is to avoid the
    FLP impossibility result (we hope) by weakening the requirements
    on this data structure.

    The interface for a \conceptFormat{weak ledger}
    object will consist of an $\Append(v)$ operation that adds a new
    entry with value $v$ to the ledger, and a $\Read()$ operation that
    returns a state of the ledger in the form of a sequence $x = x_1
    x_2 \dots x_k$, where each entry is either a value supplied by an
    $\Append$ operation or a null value corresponding to an incomplete
    $\Append$ operation. The intent is to allow incomplete $\Append$
    operations to be represented by blank entries in the ledger, which
    will eventually be filled in if no crashes prevent the $\Append$
    operation from being finished.

    A correct implementation must guarantee that in any admissible
    execution, every operation by a non-faulty process eventually
    finishes.

    We'll say that an execution of the object is \conceptFormat{consistent}
    if there is an ordering $α_1, α_2, \dots$ of all
    $\Append$ operations that start during the execution such that
    \begin{itemize}
        \item In the sequence $x$ returned by a $\Read$ operation $ρ$,
            each $x_i$ is either $⊥$ or the value $v_i$ supplied to
            $α_i$.
        \item If $α_i$ finishes before $ρ$ starts, then $\card{x} ≥ i$
            and $x_i = v_i$.
        \item If $ρ$ finishes before $α_i$ starts, then $\card{x} <
            i$.
    \end{itemize}

    An implementation of the object is consistent if all of its
    executions are consistent.

    Prove or disprove: There is a minimum value $n_0$ such that for
    any $n≥n_0$, there exists a consistent implementation of this
    object in an asynchronous message-passing system with $n$
    deterministic processes and up to one crash failure.

        \subsubsection*{Solution}

        Disproof: We'll show that despite the limitations of this
        object, it can still solve consensus for $n$ processes and one
        crash failure for any value of $n$. So an implementation in an
        asynchronous message-passing system would contradict the FLP
        impossibility result.

        The protocol is given in
        Algorithm~\ref{alg-hw-weak-ledger-consensus}.

        \begin{algorithm}
            \Procedure{$\FuncSty{consensus}(v)$}{
                send $\Tuple{\Id,v}$ to all processes \;
                $\Append(\Id)$ \;
                \While{\True}{
                    $x ← \Read()$ \;
                    \uIf{$x_1 ≠ ⊥$} {
                        \label{line-hw-weak-ledger-consensus-first}
                        \tcp{$x_1$ gives winner}
                        wait to receive $\Tuple{x_1,v'}$ \;
                        \Return $v'$\;
                    }
                    \ElseIf{$x_1 = ⊥$ and $\card{x} = n$ and each $x_2,\dots,x_n ≠ ⊥$}{
                        \label{line-hw-weak-ledger-consensus-full}
                        \tcp{missing $x_1$ gives winner}
                        $w ←$ id not in $x_2,\dots,x_n$ \;
                        \label{line-hw-weak-ledger-consensus-missing-id}
                        wait to receive $\Tuple{w,v'}$ \;
                        \Return $v'$
                    }
                    \tcp{else try again}
                }
            }
            \caption{Consensus using a weak ledger}
            \label{alg-hw-weak-ledger-consensus}
        \end{algorithm}

        The idea is that whichever process gets the first slot in the
        ledger will supply the winning consensus value, whether that
        process successfully completes its $\Append$ operation or not.
        Each process can identify the winner either by observing
        directly in the first slot returned by $\Read$, or by a
        process of elimination when the $n-1$ subsequent slots are
        filled.

        To show this solves consensus, observe that at least $n-1$
        non-faulty processes will eventually complete their $\Append$
        operations. Any subsequent $\Read$ will either see $x_1 ≠ ⊥$,
        or will see $x_1 = ⊥$ and $x_2,\dots,x_n ≠ ⊥$. A reader that
        sees either of these conditions will wait for a message from
        whichever process $p$ started the $\Append$ operation assigned
        position $α_1$ in the consistency condition. Such a message
        exists because $p$ will have sent it before starting $α_1$, so
        we get termination. Each such message also contains the input
        to $p$, giving validity, and since every non-faulty process
        uses the message from the same $p$, we have agreement.

    \subsection{A matchmaking protocol}

    Suppose we have $n$ processes in a shared-memory system that we'd
    like to repeatedly pair off. Each process may enter a matchmaking
    protocol, and as long as at least two processes have entered the
    matchmaking protocol and not yet left, eventually some pair of
    processes $p$ and $q$ will be matched in the sense that both leave the
    protocol with a return value $\Tuple{p,q}$. This matching is
    exclusive: once a match $\Tuple{p,q}$ is made, neither process can
    be matched again unless it re-enters the protocol.

    We'd like the protocol to be starvation-free in the sense that
    every time a process enters the protocol, it will eventually be
    matched and leave, unless we reach a configuration where this
    process is the only process in the protocol and no new process
    ever arrives.

    Show how to implement such a protocol such that each time a
    process enters it incurs $O(\log n)$ RMR complexity in the
    cache-coherent model, and prove your protocol works.

        \subsubsection*{Solution}

        See Algorithm~\ref{alg-hw-matchmaking} for pseudocode.
        This uses two registers $r_0$ and $r_1$, both initialized to
        $⊥$, and a starvation-free
        mutex $\DataSty{lock}$ with
        $O(\log n)$ RMR complexity, as can be constructed by using
        a Peterson-Fischer tree~\cite{PetersonF1977}
        (see §\ref{section-mutex-tournament}).

        \begin{algorithm}
            \Procedure{$\FuncSty{match}()$}{
                acquire $\DataSty{lock}$\;
                \label{line-hw-matchmaking-acquire-lock}
                \eIf{$r_0 = ⊥$}{
                    \tcp{I am first}
                    $r_0 ← \Id$\;
                    \tcp{let second in}
                    release $\DataSty{lock}$\;
                    \tcp{wait for a match}
                    \While{$r_1 = ⊥$}{
                        \tcp{try again}
                    }
                    $v ← \Tuple{\Id, r_1}$\;
                    $r_0 ← ⊥$\;
                }{
                    \tcp{I am second}
                    $v ← \Tuple{r_0, \Id}$\;
                    \tcp{notify first}
                    $r_1 ← \Id$\;
                    \tcp{wait for first to leave}
                    \While{$r_0 ≠ ⊥$}{
                        \tcp{try again}
                    }
                    \tcp{reset presence}
                    $r_1 ← ⊥$\;
                    release $\DataSty{lock}$\;
                }
                \Return $v$\;
            }
            \caption{A matchmaking algorithm}
            \label{alg-hw-matchmaking}
        \end{algorithm}

        To show that this works as claimed, observe first that since
        each process acquires $\DataSty{lock}$ each time it enters the
        protocol, we can construct a sequence of process ids $p_1,
        p_2, \dots,$ where $p_i$ is the $i$-th process to acquire
        $\DataSty{lock}$. Starvation-freedom of $\DataSty{lock}$ means
        that every process appears in this list once for each time it
        enters the protocol.

        Process $p_1$ will see $r_0 = ⊥$ and take the first branch of
        the if statement. This will cause it to assign its id to $r_0$
        then release the lock and wait for another process to set $r_1
        ≠ ⊥$.

        Process $p_2$ will do this; since it acquires $\DataSty{lock}$
        only after $p_1$ releases it, it will see $r_0 ≠ ⊥$ and take
        the second branch of the if statement. Its first step is to
        record its return value $\Tuple{p_1,p_2}$; it then writes its
        id to $r_1$, which allows $p_1$ to escape its busy-waiting
        loop. Process $p_2$ will wait for $p_1$ to read $r_1$ to
        compute its own output $\Tuple{p_1,p_2}$ since it is only
        after this step that $p_1$ resets $r_0$. Having observed $r_0
        = ⊥$, $p_1$ then resets $r_1$, restoring the initial state,
        before finally releasing the lock. At this point both $p_1$
        and $p_2$ eventually return $\Tuple{p_1,p_2}$: they have been
        successfully matched.

        This describes the behavior of $p_1$ and $p_2$ in isolation.
        But any later process can only modify $r_0$ or $r_1$ after
        acquiring the lock, which happens only after $p_2$
        releases it, and thus after $p_1$ and $p_2$ are done reading
        or writing $r_0$ and $r_1$.

        Since $p_1$ and $p_2$ leave the registers in their initial
        state, the same reasoning applies to any subsequent pair
        $p_{2k+1}$ and $p_{2k+2}$. So we get starvation-free matching
        as long as such pairs continue to arrive, though it is
        possible that some process $p_{2k+1}$ gets stuck waiting for a
        $p_{2k+2}$ that never arrives (which is permitted by the
        specification of the protocol).

        To compute RMR complexity, observe that the total RMR cost to
        acquire and release $\DataSty{lock}$ is $O(\log n)$. Each
        process occurs at most and addition $O(1)$ RMRs for operations
        on $r_0$ and $r_1$ outside the while loops; the reads inside
        each while loop incur a total of only $1$ RMR, because the
        loop exits as soon as the value in the register changes. So
        the entire protocol has a cost of $O(\log n)$ RMRs.

\section{Assignment 5: due Thursday 2026-04-09, at 23:59 Eastern US time}

    \subsection{A set data type}

    Consider a \concept{set} data type that holds a finite subset $S$ of the natural
    numbers $ℕ$, initially the empty set. The data type has $\Read$
    operation that returns $S$, and has an $\Insert(T)$ operation that
    sets $S$ to $S∪T$, and a $\Delete(T)$ operation that sets $S$ to
    $S∖T$.

    Let's imagine that there are two versions of this data type.
    A set with singleton operations only allows sets $T$ with
    one element to be the arguments of the $\Insert$ or $\Delete$
    operations. A set with general operations allows any finite subset
    $T$ of $ℕ$.

    For each of the following objects, prove or disprove whether there
    exists a deterministic wait-free linearizable implementation of
    that object from one set object with singleton operations (and no
    other objects).
    \begin{enumerate}
        \item A set that has a $\Read$ operation and $\Insert(T)$
            operation for any finite $T⊆ℕ$, but that does not have a
            $\Delete$ operation.
        \item A set that has a $\Read$ operation and both $\Insert(T)$
            and $\Delete(T)$ operations for any finite $T⊆ℕ$.
        \item A single-writer snapshot array holding natural numbers,
            where each process $p_i∈\Set{p_1,\dots,p_n}$ can write any
            $v∈ℕ$ to its location $A[i]$, and a $\Snapshot$ operation
            returns the entire contents of $A$. You should assume that
            each $A[i]$ is initially $0$, and that $n$ is arbitrary
            (meaning that an implementation should work for any fixed
            $n$, while an impossibility proof only needs to find one
            bad $n$).
    \end{enumerate}

        \subsubsection*{Solution}

        \begin{enumerate}
            \item This we can implement. The trick is to observe that
                there is a bijective map between the set of all finite
                subsets of $ℕ$ and elements of $ℕ$ given by the
                bit-vector encoding $f(T) = ∑_{i∈T} 2^i$. We can use
                this to implement $\Insert(T)$ by applying
                $\Insert(\Set{f(T)})$ to the singleton object $S$ and
                $\Read()$ by letting $U = \Read(S)$ and returning $T =
                \bigcup_{i∈U} f^{-1}(i)$.

                Since each operation on the general set object maps to
                a single operation on the singleton set object, this
                implementation is trivially wait-free and
                linearizable.

            \item This we can't implement.

                First we'll show that the singleton-operation set
                object has consensus number $1$. This holds because:
                \begin{itemize}
                    \item $\Read$ is overwritten by either $\Insert$
                        or $\Delete$,
                    \item $\Insert$ or $\Delete$ operations on
                        different values $i$ commute.
                    \item $\Delete(\Set{i})$ overwrites
                        $\Insert(\Set{i})$,
                    \item Two copies of $\Insert(\Set{i})$ or
                        $\Delete(\Set{i})$ commute.
                \end{itemize}
                (Here as usual, $x$ overwrites $y$ if, for any
                configuration $C$, $Cyx$ is
                indistinguishable to some process from $Cy$, and $x$
                commutes with $y$ if $Cxy$ is indistinguishable from
                $Cyx$.)

                Next, let's show that the general-operation set has
                consensus number at least $2$. The idea is similar to
                the argument for multi-register write (see
                §\ref{section-wait-free-multi-register-writes}), with
                the complication that we can effectively only write
                single bits $0$ ($\Delete$) or $1$ ($\Insert$).

                We'll start with a set object initialized to $\Set{2}$.

                Given two processes $p$ and $q$, we'll have each write its
                input to some auxiliary register $r_p$ or $r_q$, then
                fight it out over $S$ by having $p$ do
                $\Insert(\Set{0,1})$ and $q$ do $\Delete(\Set{1,2})$.
                The output of any subsequent $\Read(S)$ then tells us
                which of these operations happened and in which order:
                \begin{itemize}
                    \item $∅  ⇒ \Delete(\Set{1,2})$.
                    \item $\Set{0} ⇒ \Insert(\Set{0,1})\,\Delete(\Set{1,2})$.
                    \item $\Set{0,1} ⇒ \Delete(\Set{1,2})\,\Insert(\Set{0,1})$.
                    \item $\Set{0,1,2} ⇒  \Insert(\Set{0,1})$.
                \end{itemize}
                In each case we can have each process return the value
                $r_p$ or $r_q$ based on which of $p$ or $q$ did the
                first operation on $S$, solving $2$-process wait-free consensus.

                This shows that any number of singleton-operation set
                objects supplemented with any number of atomic
                registers can't implement a general-operation set
                object. The even stronger claim that just one
                singleton-operation set object isn't enough follows.

            \item Here we can implement a snapshot array. The trick,
                as for the implementation of the insert-only set
                object, is to come up with an appropriate encoding of
                snapshot write operations that allows us to decode $A$
                from $\Read(S)$.

                Pick some bijective pairing function $π:ℕ×ℕ→ℕ$ and
                define $f(p,t,v) = π(p,π(t,v))$. Then $f:ℕ×ℕ×ℕ→ℕ$ is
                also bijective.

                Let $p$'s $t$-th write to $A[p]$ have value $v$.
                Simulate this operation by executing
                $\Insert(\Set{f(t,p,v})$.

                To take a snapshot of $A$, perform a $\Read$
                operation, and extract the list of all previous writes
                by computing $f^{-1}(S)$. Return an array $A$ in which
                $A[p] = v$ where $v$ is the value in the tuple
                $\Tuple{p,t,v} ∈ f^{-1}(S)$ with maximum $t$ out of
                all tuples for $p$, or $A[p] = 0$ if
                there is no such tuple.

                Wait-freedom and linearizability are immediate from
                the fact that we are mapping each operation of the
                snapshot array to exactly one operation of the set.
        \end{enumerate}

    \subsection{A backup register}

    A \concept{backup register}\index{register!backup} $r$ has two
    components $r_0$ (the primary copy) and $r_1$ (the backup copy),
    each of which can hold arbitrary values. Normal $\Read$ and
    $\Write$ operations act on the primary copy $r_0$. A
    $\FuncSty{backup}$ operation is provided to copy $r_0$ to the
    backup copy $r_1$, and any value backed up in this way can be
    recovered either by a $\FuncSty{readBackup}$ operation that
    returns the value of $r_1$ directly or a $\FuncSty{restore}$
    operation that copies $r_1$ back to $r_0$. Each of these
    operations is atomic. Pseudocode for these operations
    is given in Algorithm~\ref{alg-hw-backup-register}.

    \begin{algorithm}
        \lProcedure{$\Read(r)$:}{ \Return $r_0$ }
        \lProcedure{$\Write(r,v)$:}{ $r_0 ← v$ }
        \lProcedure{$\FuncSty{backup}(r)$:}{ $r_1 ← r_0$ }
        \lProcedure{$\FuncSty{restore}(r)$:}{ $r_0 ← r_1$ }
        \lProcedure{$\FuncSty{readBackup}(r)$:}{ \Return $r_1$ }
        \caption{Operations on a backup register}
        \label{alg-hw-backup-register}
    \end{algorithm}

    \begin{enumerate}
        \item Suppose we have a backup register that provides the
            operations $\Read$,
            $\Write$, $\FuncSty{backup}$, and $\FuncSty{readBackup}$, but
            not $\FuncSty{restore}$. What is the consensus number of
            this object?
        \item Suppose we have a backup register that provides the
            operations $\Read$,
            $\Write$, $\FuncSty{backup}$, and $\FuncSty{restore}$, but
            not $\FuncSty{readBackup}$. What is the consensus number of
            this object?
    \end{enumerate}

        \subsubsection*{Solution}

        The consensus number of both versions of the object is
        infinite. We'll show this by showing that we can solve
        consensus if we have $\FuncSty{readBackup}$, then show how to
        replace $\FuncSty{readBackup}$ if only have
        $\FuncSty{restore}$.

        Let's start by exploring what affect the three update
        operations $\Write(1)$, $\FuncSty{backup}$, and
        $\FuncSty{restore}$ have on the state of the object starting
        in an initial configuration $01$ representing $r_0 = 0$ and
        $r_1 = 1$. The transition graph is shown in
        Figure~\ref{fig-hw-backup-register}.

        \begin{figure}
            \centering
            \begin{tikzpicture}[scale=2]
                \foreach \n/\x/\y in {01/0/2,00/-1/1,11/1/1,10/0/0} {
                    \node[draw,circle] (\n) at (\x,\y) { $\n$ };
                }
                \draw[->] (01) edge node[above left] { $b$ } (00);
                \draw[->] (01) edge node[above right] { $w$, $r$ } (11);
                \draw[->] (00) edge[loop left] node[left] { $b$, $r$ } (00);
                \draw[->] (00) edge[bend right] node[below left] { $w$ } (10);
                \draw[->] (11) edge[loop right] node[right] { $b$, $w$, $r$ } (11);
                \draw[->] (10) edge[loop below] node[below] { $w$ } (10);
                \draw[->] (10) edge[bend right] node[above right] { $r$ } (00);
                \draw[->] (10) edge[red] node[below right] { $b$ } (11);
            \end{tikzpicture}
            \caption[Backup register transitions]{Backup register
            transitions. Each node shows $r_0$ and $r_1$. Operations
            $\FuncSty{backup}$, $\Write(1)$, and $\FuncSty{restore}$
            are abbreviated as $b$, $w$, and $r$. Transition
            $10\stackrel{b}{\rightarrow} 11$ (marked in red) is only
            transition from $\Set{00,10}$ to $\Set{11}$.}
            \label{fig-hw-backup-register}
        \end{figure}

        To solve consensus, we need to find two classes of
        operations leaving our starting state such that we can always tell
        which class went first. Looking at the transitions out of $01$, we
        see that doing $\FuncSty{backup}$ sends us to $00$ while
        $\Write(1)$ and $\FuncSty{restore}$ both send us to $10$.
        Unfortunately there is a path back from $00$ to $11$ using 
        $\Write(1)$ followed by $\FuncSty{backup}$.

        We can avoid this path by preventing the transition from $10$ to
        $11$, which we can do by only allowing one call to
        $\FuncSty{backup}$. With this restriction, applying
        $\FuncSty{backup}$ to the initial state $10$ makes $r_1 = 0$
        forever, while applying $\Write(1)$ makes $r_1 = 1$ forever.
        If we have $\FuncSty{readBackup}$, we can just read $r_1$ and
        determine which operation went first.

        This is enough to solve a one-vs-many consensus problem where all
        but one process have the same input.
        This gives general consensus for any fixed $n$
        using the recursive construction of
        Ruppert\cite{Ruppert2000} (see
        Lemma~\ref{lemma-n-discerning-possiblity}).

        For the case where we have $\FuncSty{restore}$ instead of
        $\FuncSty{readBackup}$,
        we'll implement a destructive version of $\FuncSty{readBackup}$
        using $\FuncSty{restore}$ followed by $\Read$. This version is destructive
        in the sense that it will overwrite $r_0$, but since both $00$ and
        $10$ give the same decision value, we don't care about this. A
        bigger issue is that a $\Write(1)$ that slips in between the
        $\FuncSty{restore}$ and $\Read$ might cause the $\Read$ to return
        the wrong value. The solution is to repeat the
        $\FuncSty{restore}$-then-$\Read$ $n+1$ times and observe that
        since each process only does at most one $\Write(1)$, if any of
        these iterations reads a $0$, we can return $0$.

        Algorithm~\ref{alg-hw-backup-register-consensus} gives the full
        protocol for one-vs-many consensus. It includes the destructive
        implementation of $\FuncSty{readBackup}$ from $\FuncSty{restore}$
        if $\FuncSty{readBackup}$ is not available.

        \begin{algorithm}
            \tcp{assumes all processes except $0$ have same input}
            \Procedure{$\FuncSty{consensus(v)}$}{
                \eIf{I am process $0$}{
                    $\Input[0] ← v$ \;
                    $\FuncSty{backup}(r)$\;
                }{
                    $\Input[1] ← v$\;
                    $\Write(r,1)$\;
                }
                \tcp{figure out who went first}
                $\DataSty{winner} ← \FuncSty{readBackup}(r)$\;
                \Return $\Input[\DataSty{winner}]$\;
            }
            \tcp{replacement for $\FuncSty{readBackup}$ if not provided by object}
            \tcp{does not preserve $r_0$}
            \tcp{assumes $r_1$ is initially $0$}
            \tcp{assumes at most $n$ $\Write(1)$ operations and no other $\Write$s}
            \Procedure{$\FuncSty{readBackup}(r)$}{
                \For{$i ← 1$ \KwTo $n+1$}{
                    $\FuncSty{restore}(r)$\;
                    \lIf{$\Read(r) = 0$}{\Return $0$}
                }
                \Return $1$\;
            }
            \caption{One-vs-many consensus using a backup register}
            \label{alg-hw-backup-register-consensus}
        \end{algorithm}

        \paragraph*{A better solution for the $\FuncSty{restore}$-only
        case.} Boyang
        Huang was kind enough to show me an easier way to solve
        consensus using 
        $\FuncSty{backup}$ and
        $\FuncSty{restore}$ 
        without needing to read from 
        from $r_1$, either directly or indirectly.
        Starting from state $01$ ($r_0=0$, $r_1=1$), have any
        process with input $0$ execute $\FuncSty{backup}$ and any process with
        input $1$ execute $\FuncSty{restore}$. Now the state is either $00$ or
        $11$ forever no matter how many additional $\FuncSty{backup}$ or
        $\FuncSty{restore}$ operations are done, and these states can be
        distinguished with $\Read$. This
        solves binary consensus for any $n$, giving this version of
        the backup register consensus number $∞$.

\section{Assignment 6: due Thursday 2026-04-23, at 23:59 Eastern US time}

    \subsection{Cooperative work without waste}

    Suppose we have an asynchronous shared-memory system with $n$
    deterministic processes $p_1\dots p_n$, up to $n-1$ of which can fail;
    $m$ counters $c_1 \dots c_m$; and as many additional atomic read/write
    registers as we might need. We'd like a protocol that increments as many of the
    counters as it can exactly once. The protocol should be wait-free
    (every process eventually terminates) and should never 
    increment a counter twice. The quality of the protocol is
    measured by its \emph{waste}, the number of counters that are never
    incremented.

    Show that there is a protocol that solves this problem with $O(n)$
    waste.

        \subsubsection*{Solution}

        We can solve this by adapting the deterministic $2n-1$
        renaming protocol of Attiya~\etal~\cite{AttiyaBDPR1990}, as
        previously described in Algorithm~\ref{alg-snapshot-renaming}. Each
        process will store in its cell in a snapshot array the set of
        counters it has incremented, plus one target counter it
        intends to increment. If it observes that no other process has
        claimed its target counter, it will go ahead and increment
        that counter. Otherwise it will pick a new target from the
        unclaimed counters with a rank corresponding to the process's
        id. The process terminates when the number of claimed counters
        reaches $m-n$; at this point at least $m-2n$ counters will
        have been incremented, giving the desired waste bound.

        Details of the protocol are given in
        Algorithm~\ref{alg-hw-cooperative-counting}. Competing claims
        are tracked in a single-writer snapshot array $A$.
        \begin{algorithm}
            \Procedure{\FuncSty{incrementCounters}}{
                \tcp{counters we have already incremented}
                $C ← ∅$\;
                \tcp{counter we are trying to claim}
                $s ← 1$\;
                \While{\True}{
                    $A[i] ← C ∪ \Set{s}$\;
                    $\DataSty{view} ← \FuncSty{snapshot}(A)$\;
                    \uIf{$\card*{\bigcup_{j=1}^{n} \WFRview[j]} ≥ m-n$}{
                        \label{line-hw-cooperative-counting-termination}
                        \Return
                    }
                    \uElseIf{$s ∈ \WFRview[j]$ for some $j≠i$}{
                        \tcp{pick a new candidate}
                        $s ← i$-th positive integer not in $\bigcup_{j≠i} \WFRview[j]$\;
                        \label{line-hw-cooperative-counting-choose-name}
                    }
                    \Else{
                        \tcp{nobody else has claimed our candidate}
                        increment $c_{s}$\;
                        $C ← C ∪ \Set{s}$\;
                    }
                }
            }
            \caption{Low waste cooperative counting}
            \label{alg-hw-cooperative-counting}
        \end{algorithm}

        Proof of correctness of the algorithm is similar to the proof
        for Algorithm~\ref{alg-snapshot-renaming}.

        The termination test in
        Line~\ref{line-hw-cooperative-counting-termination} ensures
        that a process that has not finished can always choose a new
        candidate in
        Line~\ref{line-hw-cooperative-counting-choose-name} without
        running out of counters.

        The safety property that every counter is incremented at most
        once follows from the fact that if process $i$ increments some
        counter $c_k$, then there is a time before the increment at
        which $k ∈ A[i]$ and $k ∉ A[j]$ for any $j≠i$, and since $i$
        always writes a set containing $k$ to $A[i]$ after this point,
        any $j≠i$ will see $k ∈ A[i]$.

        The liveness property that no process runs forever without
        incrementing a counter follows from a modified version of the
        contradiction proof for the renaming algorithm. Suppose that
        some process does run forever. Let $P$ be the set of indices
        of all process that do this. Wait until all processes not in
        $P$ have stopped and all processes in $P$ have incremented a
        counter for the last time. Let $i$ be the smallest index in
        $P$, and let $z_1 < z_2 < z_3 \dots < z_k$ be the sequence of
        indices of counters that don't appear in $C_j$ for any $j$ and
        aren't equal to $s_j$ for any $j∉P$. Then no process in $P ∖
        \Set{i}$ can choose a candidate in $\Set{z_1,\dots,z_i}$, and
        since every process in $P$ runs forever, eventually each of
        the processes $j∈P∖\Set{i}$ will choose some candidate outside
        this range and write its new candidate to $A[j]$, leaving
        $z_i$ open for $i$ to take, a contradiction.

        The waste of this algorithm is bounded by $2n$. When all
        processes have either crashed or finished, any process that
        has finished has seen $\card*{\bigcup_{j=1}^{n} A[j]} ≥ m-n$.
        Each $A[j]$ is equal to some past value of $C'_j ∪ \Set{s'_j}$.
        Since each $C_j$ can only increase over time, it holds for the
        current values of $C_j$ that
        $\bigcup_{j=1}^{n} C_j ⊇ \bigcup_{j=1}^n (A[j] ∖ \Set{s'_j})$
        and thus $\card*{\bigcup_{j=1}^{n} C_j} ≥ m-2n$.

    \subsection{Self-stabilizing clique detection}

    Given a 5-regular connected graph $G$ with diameter $D$, we'd like
    to identify all the vertices that are part of at least one clique
    of 5 nodes, by having each such vertex set a flag locally
    indicating its status. We will be doing so in the context of a synchronous
    self-stabilizing system where each node has a port number
    $0,\dots,4$ for each of its 5 neighbors, and in each round updates
    its state based on its previous state and a vector of the states
    of its neighbors ordered by port number. Each node has no other
    information about the graph or the identity of itself or its
    neighbors except as specified below.

    For each of the cases below, show one of: (a) there is a protocol
    for detecting members of 5-cliques that stabilizes in $O(1)$
    rounds for every graph; (b) there is no such protocol that
    stabilizes in $O(1)$ rounds for all graphs, but there is one that
    stabilizes in $O(D)$ rounds; or (c) there is no protocol that
    eventually stabilizes to a correct configuration for every graph.

    \begin{enumerate}
        \item The nodes are anonymous.
        \item The nodes are anonymous, except that exactly one
            process is marked as a leader.
        \item The nodes have unique identities.
    \end{enumerate}
    
        \subsubsection*{Solution}

        \begin{enumerate}
            \item Here we show impossibility by constructing two graphs that are
                indistinguishable to the nodes, where one contains
                several $5$-cliques and the other contains none. This
                construction will be designed to also work in the
                marked-leader case.

                Figure~\ref{fig-hw-graph-with-cliques} depicts the
                clique version, which we will call $G$. A central node
                $p$ is adjacent to five middle nodes $q_0,\dots,q_4$, and
                each node $q_i$ is adjacent to four outer nodes
                $r_{4i},\dots,r_{4i+4}$, organized into four $5$-cliques.

                \newcommand{\hwCliqueGraphCore} {
                    \node (p) at (0,0) { $p$ };
                    \foreach \i in {0,...,4} {
                        \node (q\i) at (-72*\i+90:2) { $q_\i$ };
                        \draw (p) edge (q\i);
                    }
                    \foreach \i in {0,...,19} {
                        \node (r\i) at (-18*\i+117:6) { $r_{\i}$ };
                    }
\foreach \i in {0,...,4} {
                        \pgfmathsetmacro{\jstart}{int(4*\i)}
                        \pgfmathsetmacro{\jend}{int(\jstart+3)}
                        \foreach \j in {\jstart,...,\jend} {
                            \draw (q\i) edge (r\j);
                        }
                    }
                }

                \begin{figure}
                    \centering
                    \begin{tikzpicture}
                        \hwCliqueGraphCore
\foreach \i in {0,...,3} {
                            \foreach \j in {0,...,3} {
                                \pgfmathsetmacro{\jj}{int(\j+1)}
                                \foreach \k in {\jj,...,4} {
                                    \pgfmathsetmacro{\ij}{int(5*\i+\j)}
                                    \pgfmathsetmacro{\ik}{int(5*\i+\k)}
                                    \draw[blue] (r\ij) edge (r\ik);
                                }
                            }
                        }
                    \end{tikzpicture}
                    \caption{Graph $G$ (with cliques)}
                    \label{fig-hw-graph-with-cliques}
                \end{figure}

                Figure~\ref{fig-hw-graph-without-cliques} depicts the
                non-clique version, which we will call $H$. Again a
                central node $p$ is adjacent to five middle nodes
                $q_0,\dots,q_4$, and each $q_i$ is adjacent to four
                outer nodes $r_{4i},\dots,r_{4i+3}$, but now each
                outer node $r_j$ is adjacent to four other outer nodes
                $r_{j-2},r_{j-1},r_{j+1},$ and $r_{j+2}$, where all
                index computation is done mod $20$. Any set of three
                outer-ring neighbors of $r_i$ is missing an edge,
                since two of these neighbors must be at least $3$
                positions distant. So the outer ring contains no
                $4$-cliques, which means that even after including the
                $q$ nodes we still have no $5$-cliques.

                \begin{figure}
                    \centering
                    \begin{tikzpicture}
                        \hwCliqueGraphCore
\foreach \i in {0,...,19} {
                            \foreach \j in {1,2} {
                                \pgfmathsetmacro{\jj}{int(Mod(\i+\j,20)}
                                \draw[blue] (r\i) edge (r\jj);
                            }
                        }
                    \end{tikzpicture}
                    \caption{Graph $H$ (without cliques)}
                    \label{fig-hw-graph-without-cliques}
                \end{figure}

                In both graphs, each node $r_j$ assigns port $0$ to
                its neighbor $q_i$, and each node $q_i$ assigns port
                $0$ to $p$. The remaining port number assignments are
                arbitrary.

                Suppose we start both $G$ and $H$ in a configuration
                where all nodes are in the same state $s_0$. Each node
                sees five neighbors all in $s_0$, so all nodes update
                to a new state $s_1$. Iterating this process gives a
                sequence of states $s_0, s_1, \dots$ such that every
                node is in state $s_t$ after $t$ rounds in either
                graph. If some node $r_i$ decides it is in a clique or
                not in a clique at some time $t$, it will have
                computed the wrong value for one of $G$ or $H$.

            \item Unfortunately, adding a leader doesn't help. 
                In each of the graphs $G$ and $H$, start in a
                configuration where $p$ is the leader and starts in
                some state $s^0_p$, all $q$ nodes
                start in the same state $s^0_q$, and all $r$ nodes
                start in the same state $s^0_r$. Suppose that after
                $t$ rounds, $p$ is in state $s^t_p$, all $q$ nodes are in the same state
                $s^t_q$ and all $r$ nodes are in the same state
                $s^t_r$. Then each $q$ node observes $s^t_p$ on port
                $0$ and $s^t_r$ on ports $1$ through $4$, and computes
                the same new state $s^{t+1}_q$ as every other $q$
                node. Similarly, all the $r$ nodes compute the same
                new state $s^{t+1}_r$. So again any $r$ node that
                decides it is in or not in a clique after $t$ rounds
                will be wrong for one of $G$ or $H$.

            \item In this case, the nodes can stabilize to the correct
                answer in $O(1)$.
                rounds in any graph. Each node $v$ observes the
                identities of all of its
                neighbors in every round and stores the result in a
                local variable $N_v$. It also observes $N_u$ for each
                of its neighbors $u$, and reports that it is in a
                clique of size $k$ if it sees a set $S$ of $k-1$
                neighbors such that for each distinct pair of
                neighbors $u,u'∈S$, $u∈N_{u'}$ and $u'∈N_u$. Since the
                $N_u$ variables will be correct after at most one round, this
                protocol stabilizes in at most two rounds.
        \end{enumerate}

\chapter{Sample assignments from Spring 2025}

\section{Assignment 1: due Tuesday 2025-01-28, at 23:59 Eastern US time}

    \subsection{Local agreement}

    \begin{algorithm}
        \Procedure{\FuncSty{agreement}}{
            \While{\True}{
                $\ell ← r ← ⊥$\;
                \label{line-alg-local-agreement-on-a-ring-reset}
                Send $\DataSty{query}$ to both neighbors.\;
                Wait until $\ell≠⊥$ and $r≠⊥$.\;
                \label{line-alg-local-agreement-on-a-ring-waiting}
                $\DataSty{opinion} ← \maj(\ell, \DataSty{opinion}, 
                r)$\;
                \label{line-alg-local-agreement-on-a-ring-new-opinion}
            }
        }
        \UponReceiving{\DataSty{query} from $j$}{
            Send $\DataSty{response}(\DataSty{opinion})$ to $j$\;
            \label{line-alg-local-agreement-on-a-ring-receive-respond}
        }
        \UponReceiving{$\DataSty{response}(b)$ from $i-1$}{
            $\ell ← b$\;
        }
        \UponReceiving{$\DataSty{response}(b)$ from $i+1$}{
            $r ← b$\;
            \label{line-alg-local-agreement-on-a-ring-receive-response-right}
        }
        \caption[Local agreement algorithm: code for process $i$]{Local agreement algorithm}
        \label{alg-local-agreement-on-a-ring}
    \end{algorithm}

    Algorithm~\ref{alg-local-agreement-on-a-ring} runs on an
    asynchronous message-passing ring of $n$ processes labeled
    $0,\dots,n-1$, where each process $i$ can send messages only to
    its immediate neighbors $i-1$ and $i+1 \pmod{n}$.

    Each process starts with an $\DataSty{opinion}$ that is either $0$
    or $1$. The main loop
    repeatedly polls both neighbors for their current opinions, and in
    each iteration the process adopts the majority opinion among
    itself and its two neighbors. The hope is that this procedure will
    eventually lead to some level of agreement among all of the processes.

    Unfortunately this does not happen in most executions, and indeed
    when $n$ is even it is not hard to construct executions where the
    nodes' opinions flip back and forth forever. In an effort to
    salvage this otherwise magnificent algorithm, the designer
    declares that (a) it should be used only with odd $n$; and (b) the
    algorithm will guarantee only that each process's opinion
    stabilizes in the sense that it eventually stops changing.

    Prove or disprove: When $n$ is odd, every fair execution of
    Algorithm~\ref{alg-local-agreement-on-a-ring}
    eventually reaches a configuration where every process's
    $\DataSty{opinion}$ value never changes again.

        \subsubsection*{Solution}

        Here is a proof.

        Let $C_0 C_1 \dots$ be the sequence of configurations in some
        fair execution $Ξ$ of
        Algorithm~\ref{alg-local-agreement-on-a-ring}.
        Call a process $i$ \conceptFormat{stable}
        at step $t$ if its opinion is the same in all configurations
        $C_{t'}$ with $t'≥t$.
        We'll show that in an odd ring, some pair of
        process $i$ and $i+1$ are both stable at $0$, and that
        stability propagates in the sense that if some $i$ is stable at 
        $t$, then $i+1$ becomes stable at some subsequent $t'$.

        For the first part, the intuition is that if any pair of
        adjacent processes start
        the same opinion, that opinion is stable, since these
        processes will see at least two out of three votes for that
        opinion forever. 

        We can prove this by writing out an
        invariant.
        A complication is that it's not enough to
        look at the $\DataSty{opinion}$ values alone, because
        out-of-date information might cause confusion. So we
        must use a stronger invariant that also includes $\ell$, $r$, and
        the contents of messages in transit:
        \begin{lemma}
            \label{lemma-local-agreement-adjacent-pairs}
            Let $C$ be a configuration where
            \begin{enumerate}
                \item\label{item-local-agreement-invariant-opinions}
                    $\DataSty{opinion}_i = \DataSty{opinion}_{i+1} = b$,
                \item\label{item-local-agreement-invariant-variables} 
                    $\Set{r_i, \ell_{i+1}} ⊆ \Set{⊥,b}$, and
                \item\label{item-local-agreement-invariant-messages} 
                    Any $\DataSty{query}(b')$ messages in transit
                    between $i$ and $i+1$ have $b' = b$.
            \end{enumerate}
            Then any
            configuration reachable from $C$ in one step also has
            these properties.
        \end{lemma}
        \begin{proof}
            We will prove the invariant is preserved by events
            at $i$; the case for events at $i+1$ is symmetric.
            \begin{enumerate}
                \item The only place where $\DataSty{opinion}_i$
                    changes is in
                    Line~\ref{line-alg-local-agreement-on-a-ring-new-opinion}.
                    Since the process waits in
                    Line~\ref{line-alg-local-agreement-on-a-ring-waiting}
                    for both $\ell_i$ and $r_i$ to be non-null,
                    \iref{item-local-agreement-invariant-variables}
                    tells us that $r_i = b$.
                    We also have $\DataSty{opinion}_i = b$ from
                    \iref{item-local-agreement-invariant-opinions}.
                    This gives us two out of three inputs to the
                    majority, so the new value of
                    $\DataSty{opinion}_i$ is still $b$.
                \item If $r_i$ changes, it is either set to $⊥$ in
                    Line~\ref{line-alg-local-agreement-on-a-ring-reset}
                    or set to some value $b'$ in
                    Line~\ref{line-alg-local-agreement-on-a-ring-receive-response-right}.
                    But $b'=b$ from
                    \iref{item-local-agreement-invariant-messages}.
                    So in either case the new value of $r_i$ is in
                    $\Set{⊥,b}$.
                \item Any new $\DataSty{response}(b')$ message from $i$
                    to $i+1$ is generated in
                    Line~\ref{line-alg-local-agreement-on-a-ring-receive-respond}.
                    From
                    \iref{item-local-agreement-invariant-opinions}
                    this will have $b'=b$.
            \end{enumerate}
        \end{proof}

        In an odd ring, we must have two adjacent processes $i$ and
        $i+1$ that start with the same opinion $b$. So these processes are
        stable at $0$ by
        Lemma~\ref{lemma-local-agreement-adjacent-pairs}, since
        \iref{item-local-agreement-invariant-opinions} holds by
        assumption and
        \iref{item-local-agreement-invariant-variables}
        and
        \iref{item-local-agreement-invariant-messages} both hold
        trivially in the initial state.

        We now show that stability propagates:
        \begin{lemma}
            \label{lemma-local-agreement-stability-propagates}
            Let $C_0 C_1 \dots$ be the sequence of configurations in some
            fair execution $Ξ$ of
            Algorithm~\ref{alg-local-agreement-on-a-ring}.
            Let $i$ be stable at $t$.
            Then there is a step $t'$ such that $i+1$ is stable at
            $t'$.
        \end{lemma}
        \begin{proof}
            Let $i$ have opinion $b$ in $C_t$.
            Suppose $i+1$ is not stable at any $t'≥t$. Then
            for every $t' ≥ t$, there is some $t'' ≥ t'$
            where $i+1$ reaches
            Line~\ref{line-alg-local-agreement-on-a-ring-new-opinion}
            and sets its opinion to $b$.
            Because $i$ is
            stable, any subsequent message $\DataSty{response}(b')$
            that $i+1$ receives from $i$ will have $b'=b$.
            It follows that $i+1$ will always compute
            $\maj(\ell_{i+1}, \DataSty{opinion}_{i+1}, r_{i+1}
            = b$, making it stable with opinion $b$,
            contradicting the assumption.
        \end{proof}

        From Lemma~\ref{lemma-local-agreement-adjacent-pairs}, we started with two stable processes, so applying
        Lemma~\ref{lemma-local-agreement-stability-propagates}
        inductively tells us that all processes eventually become
        stable, which is what we wanted.

    \subsection{Finding your place}

    Consider an asynchronous message-passing system 
    in the form of an $n×n$ torus,
    where each process $p_{ij}$ has neighbors 
    $p_{i,j-1}$ (up),
    $p_{i,j+1}$ (down), 
    $p_{i-1,j}$ (left), and
    $p_{i+1,j}$ (right),
    where all arithmetic is in $ℤ_n$.
    The processes do not know $n$ or their coordinates; instead, each
    process sends outgoing messages to a neighbor in a particular
    direction (up, down, left, or right), and similarly receives
    messages labeled by the direction of the neighbor they came from.

    An exception to this rule is $p_{00}$, which does not know
    $n$, but does initially know its position $\Tuple{0,0}$. We would like to use this to
    build a distributed algorithm that allows each process to compute
    its coordinates. This means that each process $p_{ij}$ has an
    internal register $c_{ij}$ that starts with $⊥$ but should
    eventually be set, once and only once, to the tuple 
    $\Tuple{i,j}$.

    Give an algorithm that does this using $O(n)$ time and $O(n^2)$
    messages, and show that it is correct.

        \subsubsection*{Solution}

        It's possible to do this without computing $n$, but the proof
        of correctness seems to be easier if we just compute $n$
        first.

        This give a protocol that runs in two stages:
        \newData{\CountN}{count}
        \begin{enumerate}
            \item First, $p_{00}$ computes $n$. This involves sending
                a message $\CountN(0)$ to the right, and having each
                process $p_{i0}$ forward the $\CountN(i)$ message it receives
                to $p_{i+1,0}$ as $\CountN(i+1)$. After $n$ time units
                and $n$ messages, $p_{00}$
                receives $\CountN(n)$.

                We can prove correctness of this stage by a
                straightforward
                induction showing that each process $p_{i0}$ sends
                exactly one message $\CountN(i)$ at time no later than
                $i$.
            \item Knowing $n$, $p_{00}$ can now initiate a flooding
                protocol by sending $\Tuple{0,0,n}$ to all of its
                neighbors, and having each process $p_{ij}$, upon
                receiving a message for the first time, calculate and
                store its coordinates $\Tuple{i,j}$ based on the
                message it received and then send $\Tuple{i,j,n}$ to its
                neighbors. If we do this naively, we send $4$ message
                per process for a total of $4n^2$ messages, and each
                process $p_{ij}$ receives its message at time no later than
                $\min(i,n-i)+\min(j,n-j) ≤ n$, giving $n+1$ time for
                this stage if we take into account the time to deliver 
                extra messages sent
                at the end to processes that have already been
                recruited. With some careful pruning, we can arrange
                that each process other than the root only receives
                one message and does so by time equal to its distance,
                reducing the message complexity for this stage to $n^2-1$ and the
                time complexity to $n$.

                The proof of correctness for this stage is pretty much the same as
                for standard flooding, with the additional invariant
                that each process stores the correct value
                $\Tuple{i,j}$ and that each message
                $\Tuple{i,j,n}$ in transit also has the correct value
                for $n$ and correctly reports the coordinates of its
                sender. This is again a straightforward
                induction, although it may require some caution to
                check the specific rules used to calculate each
                receiver's coordinates based on the relative position
                of the sender.
        \end{enumerate}

        Even without the second-stage optimizations, this algorithm
        uses $O(n)$ time and $O(n^2)$ message. With the second-stage
        optimizations, the total comes to exactly $2n$ time and
        $n^2+n-1$ messages. I don't know if a more clever algorithm
        can do better than this.

\section{Assignment 2: due Thursday 2025-02-06, at 23:59 Eastern US time}

    \subsection{Leader election in a ring with bounded ids}

    Suppose we have a synchronous bidirectional ring with $n$
    processes that start with unique ids in the range
    $\Set{1,\dots,n^2}$. We'd like to solve leader election, where one
    and only one process marks itself as leader. We assume that the
    processes are deterministic and that the algorithm is uniform,
    meaning that the processes do not know $n$.

    Prove or disprove: Any uniform deterministic algorithm for this
    problem requires at least $Ω(n \log n)$ messages in the worst
    case.

        \subsubsection*{Solution}

        Disproof: We'll construct an algorithm that uses only $O(n
        \log \log n)$ messages by exploiting synchrony.
        The idea is to run LCR to elect the minimum-id process,
        but delay each
        process's start until round $\Id⋅n$, with each process
        starting only if it has not already received a message
        containing an id smaller than its own.

        Since the processes do not
        know $n$, we can't just have each process start at round
        $\Id⋅n$; instead, we will have each process start at round
        $2^{2^\Id}$ and argue that not too many processes start before
        the lowest-id process successfully sends a message all the way
        round the ring.

        Let $m$ be the lowest id of any process, and let $k$ be the id
        of some other process. Then $k$ transmits its id only if
        $2^{2^k} < 2^{2^m} + n$.

        Suppose now that $k ≥ m + \lg \lg n$.
        Then $2^{2^k} ≥ 2^{2^{m + \lg \lg n}} = 2^{2^m \lg n} =
        2^{2^m} ⋅ n = 2^{2^m} + 2^{2^m} (n-1) ≥ 2^{2^m} + 4(n-1) >
        2^{2^m} + n$ when $n≥2$. It follows that at most $O(\lg \lg
        n)$ processes with ids $m ≤ k < m + \lg \lg n$ transmit an initial
        message, accounting for $O(n \log \log n)$ messages total.

        Fortunately the problem does not ask about time complexity,
        which for this protocol is $O\parens*{2^{2^{n^2}}}$.
        A better algorithm that uses only $O(n)$ messages and
        $O\parens*{n⋅2^{n^2}}$ time (for this case)
        is given by Frederickson and Lynch~\cite[Lemma
        1]{FredericksonL1987}; this modifies LCR to propagate each
        $\Id$ $i$ only every $2^i$ rounds.

    \subsection{A covering problem}

    Suppose we have an asynchronous message-passing system with an
    arbitrary connected bidirectional communication graph $G=(V,E)$. Each process in
    $V$ starts with the
    same distance bound $\ell$, and a Boolean value \DataSty{anchor} that
    identifies some of the processes as anchors and the rest as
    non-anchors. Processes do not have any other identity beyond being
    an anchor or not.

    A process is \emph{covered} if it is at distance at most
    $\ell$ from the nearest anchor. 

    Prove or disprove: There is a protocol in this model that allows
    each process to determine if is covered in $O(\ell)$ time using $O(\ell \card{E})$
    messages.

        \subsubsection*{Solution}

        Proof: We'll construct a synchronous protocol with the desired
        properties, then argue that it can be made asynchronous using
        the $α$ synchronizer.

        For the synchronous protocol, have each anchor initiate a
        flooding protocol, which will stop after $\ell$ rounds.
        Each process that receives the message after
        $\ell$ rounds marks itself as covered, and each process that
        does not do so marks itself as not covered. It is immediate
        from the behavior of synchronous flooding that this marks only
        the processes at distance $\ell$ or less from some anchor.

        To make this asynchronous, apply the $α$ synchronizer,
        terminating after round $\ell$. Because the time complexity of
        the $α$ synchronizer is equal to the round complexity of the
        simulated synchronous algorithm, this gives $O(\ell)$ time.
        Because each process sends one message across each of its
        outgoing edges in each round, this also gives $O(\ell
        \card{E})$ messages.

\section{Assignment 3: due Thursday 2025-02-20, at 23:59 Eastern US time}

    \subsection{Coalition government}

    Algorithm~\ref{alg-hw-coalition} gives an outline of a multivalued
    consensus algorithm for an asynchronous message-passing system
    with up to $f$ crash failures. Each process starts with an input
    that is an arbitrary natural number, and we'd like the protocol to
    satisfy the usual requirements of agreement (no processes decide
    on different values), termination (all non-faulty processes
    decide), and validity (any decision value is equal to some
    process's input).

    \begin{algorithm}
        \Procedure{$\FuncSty{consensus}(v)$}{
            $c ← \FuncSty{joinCoalition}(v)$ \;
            send $c$ to all processes\;
            wait to receive values $\Set{c_i}$ from $n-f$ processes\;
            let $a,b$ be the values appearing in $\Set{c_i}$\;
            let $S_a = \SetWhere{i}{c_i = a}$\;
            let $S_b = \SetWhere{i}{c_i = b}$\;
            \uIf{$\card{S_a} > \card{S_b}$}{
                decide $a$
            } \uElseIf {$\card{S_b} > \card{S_a}$} {
                decide $b$
            } \Else {
                decide $\min(a,b)$
            }
        }
        \caption{Coalition consensus}
        \label{alg-hw-coalition}
    \end{algorithm}
    
    The algorithm proceeds in two stages:
    \begin{enumerate}
        \item An unspecified
            subprotocol \FuncSty{joinCoalition} assigns each
            non-faulty process to one of at most two coalitions,
            each of which is identified by
            one of the original input values.
        \item The coalitions vote on the common output value.
            The larger coalition wins, or, if both coalitions have
            equal size, the smaller value wins.
    \end{enumerate}

    Let $f=1$. Do \emph{one} of the following:
    \begin{enumerate}
        \item Show that it is not possible to implement
            \FuncSty{joinCoalition}.
        \item Show that it is possible to implement
            \FuncSty{joinCoalition}, but give an execution where
            \FuncSty{joinCoalition} works but
            Algorithm~\ref{alg-hw-coalition}
            nonetheless
            violates one of agreement, termination, or validity.
    \end{enumerate}

        \subsubsection*{Solution}

        \begin{algorithm}
            \Procedure{\FuncSty{joinCoalition}}{
                \If{$\MyId = 1$ or $\MyId = 2$}{
                    send $\Input$ to all processes
                }
                wait to receive a value $v$ from some process\;
                \Return $v$
            }
            \caption{Implementation of \FuncSty{joinCoalition} for
            $f=1$.}
            \label{alg-hw-coalition-solution}
        \end{algorithm}

        Algorithm~\ref{alg-hw-coalition-solution} gives an
        implementation of $\FuncSty{joinCoalition}$. Each process
        adopts the input value of either process $1$ or $2$, depending
        on which arrives first. Since at most one process can fail, at
        least one of these two messages will arrive eventually, giving
        termination, and every value returned by
        \FuncSty{joinCoalition} is one of at most two actual input
        values, satisfying the other conditions.

        To get a bad execution, let $n$ be even, and have
        \FuncSty{joinCoalition} split the processes into two equal
        groups $A$ and $B$ of size $n/2$ with different values $a$ and $b$.
        All $n$ processes send out their values. Pick some process
        $p$, and deliver to $p$ $n/2$ copies of $a$ and $n/2-1$ copies
        of $b$, causing $p$ to decide $a$. Now pick a second process
        $q$, and deliver to $q$ $n/2-1$ copies of $a$ and $n/2$ copies
        of $b$, causing $q$ to decide $b$. 
        Let the remaining processes finish however they like; whatever
        they do, $p$ and $q$ have already violated agreement.

    \subsection{Stronger Byzantine agreement}

    For this problem we consider binary Byzantine agreement, where the
    inputs and outputs are in $\Set{0,1}$.

    Recall that the validity condition for Byzantine agreement with
    $f$ faulty processes among $n$ total processes says
    that if all $n-f$ faulty processes have the same input value, they all
    agree on that value.

    What if the non-faulty processes don't have the same input?
    An obvious modification would be to ask the protocol to decide on
    a common input of sufficiently many non-faulty processes.

    For $n≥3f+1$,
    give an explicit threshold $s$ as a function of $n$ and $f$,
    and prove that for this choice of $s$:
    \begin{enumerate}
        \item 
            There is a protocol for synchronous Byzantine
            agreement that satisfies agreement and termination in all
            executions, such that whenever at least $s$ non-faulty
            processes start with the same input value $v$, all
            non-faulty processes decide $v$.
        \item For any protocol that satisfies agreement and termination,
            there exists an execution where at least $s-1$ non-faulty processes
            start with some input value $v$, but all non-faulty
            processes decide on some $v'≠v$.
    \end{enumerate}

        \subsubsection*{Solution}

        Let $s = \ceil{\frac{n+1}{2}}$; this means that we require
        that the non-faulty processes with the common input represent
        a strict majority of all processes.

        Consider the following protocol, which adds an extra round at
        the start of a Byzantine agreement protocol using the usual
        notion of validity:
        \begin{enumerate}
            \item Each process sends its input to all other processes
                (including itself).
            \item If a process receives at least $s = \ceil{\frac{n+1}{2}}$
                copies of a single value $v$, it replaces its input
                with $v$.
            \item The processes then execute a standard Byzantine
                agreement protocol tolerating $f < n/3$ faults
                using their new inputs.
        \end{enumerate}

        Suppose $k≥s$ non-faulty processes start with the same input
        $v$. Then each non-faulty process receives at least $s$ copies
        of $v$ and so sets its input to $v$. Validity of the embedded
        BA protocol thus guarantees that all processes decide $v$.
        The embedded BA protocol also enforces agreement and
        termination even if this condition does not hold. This
        demonstrates that $s=\ceil{\frac{n+1}{2}}$ is sufficient to
        agree on the majority input.

        To show that $\ceil{\frac{n+1}{2}}$ is necessary, consider an
        initial configuration with $\ceil{n/2}$ inputs $0$ and
        $\floor{n/2}$ inputs $1$. Given some protocol, run it with
        only non-faulty processes until all processes decide some
        value $v'$.
        Now replace $f$ of the processes with input $v'$ with faulty
        processes that behave exactly like non-faulty processes. We
        can do this because $\ceil{n/2} ≥ \floor{n/2} ≥ \frac{n-1}{2} 
        ≥ \frac{n-1}{3} ≥ f$. Then input $v' = 1-v$ is shared among at
        least $\floor{n/2}$ processes. When $n$ is even, this is
        exactly $n/2$ and $s$ is exactly $\ceil{\frac{n+1}{2}} = n/2 +
        1$. When $n$ is odd, this is $\frac{n-1}{2}$ and $s$ is
        $\frac{n+1}{2} = \frac{n-1}{2} + 1$. In either case, at least
        $s-1$ non-faulty processes start with input $v≠v'$.

\section{Assignment 4: due Thursday 2025-03-06, at 23:59 Eastern US time}

    \subsection{A primary-backup clock}

    \newcommand{\ReadClock}{\FuncSty{readClock}}

    Primary-backup replication is a special case of replicated state
    machines where the clients interact with a primary server that
    maintains some shared object, but switch over to a backup server
    if the primary server fails.

    Algorithm~\ref{alg-hw-primary-backup} gives an implementation of a
    fault-tolerant clock for an asynchronous message-passing system
    with failure detectors. The algorithm uses two servers, the primary and the
    backup, where the primary is responsible for updating the clock
    and the backup handles queries from clients if the primary fails.
    We assume that at most one of these two servers can fail.

    \begin{algorithm}
        \tcp{update procedure for primary}
        \Procedure{$\FuncSty{primaryMainLoop}()$}{
            \While{\True}{
                send $\Update(t+1)$ to backup\;
                wait to receive $\Ack(t+1)$ from backup or to suspect
                backup\;
                $t ← t+1$\;
            }
        }
        \tcp{update procedure for backup}
        \Procedure{$\FuncSty{backupMainLoop}()$}{
            \UponReceiving{$\Update(s)$ from primary}{
                $t ← s$\;
                send $\Ack(s)$ to primary\;
            }
        }
        \tcp{responder thread for both primary and backup}
        \Procedure{$\FuncSty{responder}()$}{
            \UponReceiving{$\Query$ from $p$}{
                send $\Response(t)$ to $p$\;
            }
        }
        \Procedure{$\ReadClock()$}{
            send $\Query$ to primary\;
            wait to receive $\Response(s)$ from primary or to suspect primary\;
            \eIf{received $\Response(s)$}{
                \Return $s$\;
            }{
                send $\Query$ to backup\;
                wait to receive $\Response(s)$ from backup\;
                \Return $s$\;
            }
        }
        \caption{Primary-backup clock}
        \label{alg-hw-primary-backup}
    \end{algorithm}

    The primary and backup processes both implement separate threads
    for updating their copy of the clock (the
    $\FuncSty{primaryMainLoop}$ and $\FuncSty{backupMainLoop}$
    procedures) and for responding to client queries (the
    $\FuncSty{responder}$ procedure. These operate on a local variable
    $t$ in each server that is implemented as an atomic register.

    Each of an unbounded number of client processes uses the procedure $\ReadClock$ to
    obtain a recent clock value from either the primary or backup
    server, depending on what messages it receives and what it is told
    by its failure detector.

    We would like the algorithm to satisfy the following properties:
    \begin{description}
        \item[Availability] Every call by a client
            process to $\ReadClock$ eventually returns.
        \item[Linearizability] Given any concurrent
            history of calls to $\ReadClock$, there is a
            linearization $π_1 π_2 \dots$ consistent with the
            observable execution order such that for any two
            $\ReadClock$ operations $π_i, π_j$ with $i < j$,
            the return values $s_i$ of $π_i$ and $s_j$ of $π_j$
            satisfy $s_i ≤ s_j$.
    \end{description}

    The algorithm does not specify what failure detector is used. For
    each of the failure detectors $P, ◇P, S,$ and $◇S$, show whether
    or not the algorithm satisfies both of
    the above two properties using this failure detector.

        \subsubsection*{Solution}

        We'll start by knocking out $S$ and $◇S$. These are only
        weakly accurate, meaning that some non-faulty process is never
        suspected, but it is possible to suspect any other process
        whether it has crashed or not.

        Let's abbreviate the primary server as $p$, the backup server
        as $b$, and one of the clients as $c$.

        Consider an execution using $S$ where $b$ is the non-faulty
        process that is never suspected, but $p$ can be suspected or
        not at any time. Starting in a configuration where $t_p = t_b
        = 0$:
        \begin{enumerate}
            \item Let $p$ send $\Update(1)$ to $b$.
            \item Let $b$ receive $\Update(1)$ and set $t_b = 1$.
                Delay its acknowledgment for now.
            \item Let $c$ execute $π_1 = \ReadClock$ while
                suspecting $p$, so that it sends $\Query$ to $p$ but
                does not wait for a response, instead sending a second
                $\Query$ to $b$ that receives a response
                $\Response(1)$. This causes $c$ to return $1$.
            \item Let $c$ execute a second operation
                $π_2 = \ReadClock$ where it does not suspect $p$.
                Now $c$ will wait for a response from $p$, which
                will be a response to either its first or second
                query to $p$; in either case it will be $\Response(0)$
                since $t_p = 0$ throughout the execution. So $c$
                returns $0$, violating linearizability.
        \end{enumerate}

        Since this fails for $S$, it also fails for the weaker
        $◇S$.\footnote{I was worrying to much about server failures to
        notice this when writing the sample solutions, but there are
        even easier counterexamples that make some client $c$ the
        non-faulty never-suspected process, and both servers are
        suspected as much as needed whenever it is needed. Weak
        accuracy can be very, very weak.}

        What if we have $◇P$? Here we eventually reach a point where
        the failure detector suspects all crashed processes and only
        crashed processes, but until that point, its choice of which
        processes to suspect is arbitrary. So we can use the same
        counterexample above for $◇P$ to show that it doesn't work either.

        Fortunately, $P$ saves us. Using a perfect failure detector,
        we can show:
        \begin{description}
            \item[Availability] Consider a call to
                $\ReadClock$ by some client $c$.

                If $p$ does not fail during this
                call, then $c$ sends $\Query$ to $p$, this message is
                eventually received, $p$ sends $\Response(t)$ to $c$
                for some value $t$, and this message is also
                eventually received. Because $P$ correctly reports $p$
                as non-suspect throughout the call, $c$ waits for the
                $\Response(t)$ message and returns.

                Alternatively, if $p$ does fail during this call, then
                either (a) $c$ receives $\Response(t)$ anyway and
                returns, or (b) $c$'s failure detector eventually
                permanently suspects $p$. This cases $c$ to stop
                waiting for $p$ and send a $\Query$ to $b$. Because
                only one of $p$ or $b$ is faulty, $b$ must be
                non-faulty. So $b$ sends $\Response(t)$ to $c$ for
                some value $t$ and $c$'s operation returns.
            \item[Linearizability] Consider some concurrent history
                $H$ consisting of $\Invoke$ and $\Respond$ events for
                various calls to $\ReadClock$. In general $H$ may
                include incomplete operations, but thanks to availability we can do the usual
                trick of running each $\ReadClock$ to completion to
                obtain an extension $H'$ in which all operations are
                complete. We will show how to linearize $H'$ and thus
                $H$.

                For each operation $π$ in $H'$, let $t_π$ be the value
                returned by $π$. Order the operations first by $t_π$,
                next by the observable execution order $<_{H'}$, and
                finally by breaking ties consistently. Because we are
                ordering by $t_π$ first, this automatically satisfies
                the requirement that for $i < j$, $t_{π_i} ≤ t_{π_j}$.
                But it remains to show that this ordering is
                consistent with $<_{H'}$. The only part of the
                ordering rule that might conflict with $<_{H'}$ is the
                first part, so this can only occur if there
                are events $π_i <_{H'} π_j$ with $t_{π_i} > t_{π_j}$.

                Let $t_p$ and $t_b$ be the values of the local
                variables in $p$ and $b$ respectively. Initially,
                $t_p=t_b=0$. Let us prove the following invariant,
                which covers configurations in which $b$ has not
                crashed:

                \begin{lemma}
                    \label{lemma-alg-hw-clock-invariant}
                    If $b$ has not crashed, then either $t_b =
                    t_p$ and there is an $\Update(t_p+1)$ message in
                    transit from $p$ to $b$, or $t_b = t_p+1$ and there is
                    an $\Ack(t_p+1)$ message in transit from $b$ to $p$.
                \end{lemma}
                \begin{proof}
                    Initially we have $t_p = t_b = 0$ and $p$ sends
                    $\Update(0)$.

                    If the first branch of the invariant holds, then
                    the only way for $t_p$ or $t_b$ to change is when
                    $b$ receives $\Update(t_p+1)$. This sets $t_b =
                    t_p+1$ and adds $\Ack(t_p+1)$ to the buffer,
                    making the second branch of the invariant hold.

                    Similarly, if the second branch of the invariant
                    holds, the variables only change when $p$ receives
                    $\Ack(t_p+1)$, which sets up the first branch.
                \end{proof}

                If $b$ crashes, then the invariant is not useful; but
                in this case, $p$ never crashes, so all values
                obtained by $\ReadClock$ are obtained from $t_p$. Since
                these values only increase over time, if $π_i <_{H'}
                π_j$ then $t_{π_i} ≤ t_{π_j}$ and so these events are
                not linearized out of order.

                If $b$ never crashes, it is possible that $p$ is
                faulty. We have already analyzed the case where $p$
                does not crash, so let's look at what happens if it
                does.

                For each operation $π_i$, either $π_i$ returns a value
                obtained from $p$, or $π_i$ suspects $p$ and returns a
                value obtained from $b$. If $π_i <_{H'} π_j$ and both
                return values obtained from $p$ or both return values
                obtained from $b$, then the fact that $t_p$ and $t_b$
                are both non-decreasing over time shows $t_{π_i} ≤
                t_{π_j}$.

                If $π_i$ returns some value $t_p$ and $π_j$ returns
                some value $t_b$, then the invariant tells us that at
                the time $t_{π_i}$' is sent, $t_p ≤ t_b$; so
                $t_{π_j}$ equals some later $t_b ≥ t_{π_i}$.

                If $π_i$ returns some value $t_b$ and $π_j$ returns
                some value $t_p$, then $π_i$ suspects $p$, and $p$ has
                crashed, before $π_j$ starts. So $π_j$ can't get a
                response from $p$, and we get a contradiction.

                For the cases that can occur, we have $π_i <_{H'} π_j$
                implies $t_{π_i} ≤ t_{π_j}$. This tells us that our
                linearization is consistent with $<_{H'}$ and thus
                that linearizability holds.
        \end{description}

    \subsection{Simulating an atomic register with churn}

    \begin{algorithm}
        \Procedure{$\Write(v)$}{
            $t ← t+1$\;
            choose quorum $Q_t$ such that $\card*{Q_t} = n$ and $\card*{Q_t ∩ Q_{t-1}} ≥
            n-1$\;
            send $\Write(\Tuple{t,Q,v})$ to all processes in $Q$\;
            \label{line-alg-hw-ABD-dynamic-write-broadcast}
            wait for $\Ack(t)$ from $n-f$ processes\;
        }
        \Procedure{$\FuncSty{responder}()$}{
            \Initially{
                $\Tuple{t^i,Q^i,v^i} ← \Tuple{0,Q_{0},⊥}$
            }
            \UponReceiving{$\Write(\Tuple{t,Q,v})$ from $p$}{
                \If{$t > t^i$}{
                    $\Tuple{t^i,Q^i,v^i} ← \Tuple{t,Q,v}$
                }
                send $\Ack(t)$ to $p$
            }
            \UponReceiving{$\Read(\Nonce)$ from $p$}{
                send $\Respond(\Nonce, \Tuple{t^i,Q^i,v^i})$ to $p$
            }
        }
        \Procedure{$\Read()$}{
            \tcp{new local timestamp used to construct nonces}
            $\ell ← \ell+1$\;
            $t ← 0$\;
            $Q ← Q_0$\;
            \For{$r ← 1 \dots ∞$}{
                $\Nonce ← \Tuple{\ell,r}$\;
                send $\Read(\Nonce)$ to each process in $Q$\;
                wait to receive $\Respond(\Nonce,\Tuple{t^i,Q^i,v^i})$ from $n-f$ processes\;
                let $i$ maximize $t^i$\;
                \eIf{$t^i = t$}{
                    \tcp{we have converged on some quorum}
                    \Return $v^i$\;
                }{
                    \tcp{we found a newer quorum}
                    $Q ← Q^i$\;
                    $t ← t^i$\;
                    \tcp{ABD-style completion of possible partial write}
                    send $\Write(\Tuple{t,Q,v^i})$ to all processes in $Q$\;
                    \label{line-alg-hw-ABD-dynamic-rewrite-broadcast}
                    wait for $\Ack(t)$ from $n-f$ processes\;
                }
            }
        }
        \caption{Simulating a register with variable quorums}
        \label{alg-hw-ABD-dynamic}
    \end{algorithm}

    Algorithm~\ref{alg-hw-ABD-dynamic} gives an implementation
    of a single-writer atomic register in an asynchronous message passing system tolerating up to
    $f < \frac{n-1}{2}$ crash failures, based on the Attiya-Bar-Noy-Dolev
    protocol (see §\ref{section-ABD}). Instead of using a fixed
    collection of $n$ processes to hold the register value, this
    implementation uses a dynamic quorum, where $Q_0$ is some set of
    $n$ processes initially known to all processes but later quorums
    $Q_t$ may be chosen by the unique writer process from some
    arbitrarily large collection of processes running
    $\FuncSty{responder}$, subject to a
    constraint that prevents the quorum from changing too
    quickly.

    This allows processes to be slowly
    swapped out of the quorum (perhaps to reduce load on these
    processes), although we assume that processes that leave the
    current quorum still respond to queries about the most recent
    values they have received.

    \begin{enumerate}
        \item Show that any $\Read$ operation is guaranteed to
            terminate eventually in any execution in which
            only finitely many $\Write$ operations occur.
        \item Show that any execution of this protocol is linearizable.
    \end{enumerate}

        \subsubsection*{Solution}

        \begin{enumerate}
            \item For termination, if only finitely many $\Write$
                operations occur, then there is a finite sequence of
                quorums $Q_0,Q_1,\dots,Q_m$.

                Each time a $\Read$
                operation executes its loop without returning, it
                updates $Q$ to a new quorum $Q^i = Q_t$ for some $t$.
                It then gets $\Ack(t)$ from $n-f > \frac{n+1}{2}$
                processes. In the next iteration, the reader gets
                $\Respond$ messages from $n-f > \frac{n+1}{2}$
                processes in $Q_t$, so there is at least one process
                that sent both and $\Ack$ and a $\Respond$. So the new
                maximum $t^i$ is at least $t$, and either the new
                iteration returns (if $t^i = t$) or obtains a new
                $Q = Q_{t'}$ where $t' > t$. But since there are only
                finitely many such $t'$, eventually the second case can
                no longer occur and the $\Read$ returns.
            \item 
                For linearizability, we'd like to do the usual
                argument where we order each operation by the
                timestamp of the last quorum it writes to, but there
                is a complication in that we can no longer guarantee
                that all quorums overlap. Instead we will show that
                there is always a trail of sufficiently-populated
                quorums leading to the most recent one.

                Define a quorum $Q_t$ as \conceptFormat{visible} in
                some configuration if (a) the writer has chosen $Q_t$
                already and (b) at least $n-f$ processes in $Q_t$
                store a timestamp that is $t$ or greater.

                Claim: For any $t≥0$, if $Q_{t+1}$ is visible, so is
                $Q_t$. Proof: When $t=0$, $Q_0$ is trivially visible.
                For larger $t$, if the writer has chosen $Q_{t+1}$, it
                must have previously completed a $\Write$ operation
                in which it chose $Q_t$ (which gives (a)) and received an $\Ack(t)$
                message from $n-f$ processes in $Q_t$ (which gives
                (b)).

                Applying induction to the claim shows that for any
                visible $Q_{t}$, all $Q_{t'}$ with $t' < t$ are also
                visible.

                Given an execution $H$, assume without loss of
                generality that all operations in $H$ are complete.
                Assign each operation $π$ a timestamp $t_π$ equal to
                the timestamp used in
                Line~\ref{line-alg-hw-ABD-dynamic-write-broadcast}
                or the last call to
                Line~\ref{line-alg-hw-ABD-dynamic-rewrite-broadcast}.
                Construct a linearization $S$ 
                as in the original ABD proof
                by ordering the operations first
                by increasing $t_π$, then by
                putting $\Write$ operations before $\Read$ operations,
                then by $<_H$, then by breaking ties consistently.
                Because each $\Read$ returns the value associated with
                the last preceding $\Write$ in $S$ (or $⊥$ if there is
                no such $\Write$), $S$ looks like a sequential atomic
                register execution. It remains only to show that the
                $<_S ⊇ <_H$.

                Let $π_1 <_H π_2$. Since $π_1$ receives
                $\Ack(t_{π_1})$ from $n-f$ processes, $Q_{t_{π_1}}$,
                and thus also any $Q_{t'}$ with $t'<t_{π_1}$, is
                visible when $π_1$ finishes in $H$. Since $π_2$ starts
                after $π_1$ finishes, each time it receives $n-f >
                \frac{n+1}{2}$
                responses from a quorum $Q_{t'}$ with $t'<t_{π_1}$,
                visibility of $Q_{t'+1}$ says that
                there are at least $n-f>\frac{n+1}{2}$ processes in $Q_{t'+1}$ that
                store timestamps that are at least $t'+1$. Since at
                most one process in $Q_{t'+1}$ is not in $Q_{t'+1}$,
                at least $n-f-1 > \frac{n-1}{2}$ processes in $Q_{t'}$
                store a time step that is at least $t'+1$. 
                At least one of
                those processes is among the $n-f$ processes
                responding to $π_2$, so $π_2$ learns about some quorum
                $Q_{t'+1}$ or higher, and doesn't stop at $Q_{t'}$. It
                follows that $t_{π_2} ≥ t_{π_1}$, and the ordering of
                these operations by timestamp is consistent with
                $<_H$. The same argument as in the ABD proof shows
                that the remaining ordering rules are also
                consistent with $<_H$ when they apply, so $<_S$ is consistent with
                $<_H$.
        \end{enumerate}

\section{Assignment 5: due Thursday 2025-04-03, at 23:59 Eastern US time} 

    \subsection{Fetch-and-max from fetch-and-add}

    Suppose you are given atomic registers, and fetch-and-add registers
    that store values in $ℕ$, and want to build a fetch-and-max register that
    also stores values in $ℕ$. For the purposes of this problem, assume
    that a fetch-and-add supports a single operation
    $\FetchAndAdd(x)$ that adds $x$ to the value of the
    register and returns the previous value before the add; similarly,
    a fetch-and-max supports a single operation $\FetchAndMax(x)$ that
    replaces the value $v$ of the register with $\max(v,x)$ and
    returns the old value $v$.

    Prove or disprove: For any fixed $n$, there exists a wait-free
    linearizable implementation of a fetch-and-max register using atomic
    registers and fetch-and-add registers.

        \subsubsection*{Solution}

        Algorithm~\ref{alg-hw-fetch-and-max} gives an implementation
        that works for a given fixed $n$. It uses a single
        fetch-and-add object $r$ that encodes the max value in the
        largest digit of a number represented in base $(n+1)$.

        \begin{algorithm}
            \Procedure{$\FetchAndMax(x)$}{
                \tcp{make sure nobody has already stored $x$}
                $v ← r$\;
                \label{line-hw-fetch-and-max-read}
                \If{$v < (n+1)^x$}{
                    $v ← \FetchAndAdd\parens*{r,(n+1)^x}$
                    \label{line-hw-fetch-and-max-fetch-and-add}
                }
                \Return largest $y∈N$ such that $v ≥ (n+1)^y$
            }
            \caption{Fetch-and-max from fetch-and-add}
            \label{alg-hw-fetch-and-max}
        \end{algorithm}

        The idea is to store a number of the form $r=∑_{i=0}^{∞} r_i
        (n+1)^i$, where $0≤r_i≤n$ is the number of times some process
        calls $\FetchAndAdd\parens*{(n+1)^i}$. We will let the
        of the fetch-and-max object be the largest $i$ for which $r_i$
        is nonzero.

        To avoid overflow, the $\FetchAndMax$ procedure first checks
        if $r$ is already at least $(n+1)^x$, and skips changing $r$
        if it is. Since each process can execute at most one
        $\FetchAndAdd\parens*{(n+1)^x}$ before seeing $r≥(n+1)^x$,
        this implies that we never do more than $n$ calls to
        $\FetchAndAdd\parens*{(n+1)^x}$ in any execution.
        So in the unique expansion $r=∑_{i=0}^{∞} r_i (n+1)^i$, the
        value of $r_i$ accurately counts the number of previous calls to 
        $\FetchAndAdd\parens*{(n+1)^i}$, since smaller increments
        don't add up to $(n+1)^i$ and larger increments don't
        contributed to $r_i (n+1)^i$.

        To show linearizability, we'll assign linearization points to
        each call to $\FetchAndMax$. If a call executes the
        $\FetchAndAdd$ in
        Line~\ref{line-hw-fetch-and-max-fetch-and-add}, linearize it
        there. If not, linearize it at at the read operation in
        Line~\ref{line-hw-fetch-and-max-read}. Note that in either
        case the value $v$ used to compute the return value is
        obtained from an operation that takes place at the
        linearization point.

        Because we are using linearization points, we don't need to
        argue consistency of the resulting schedule with the observed
        execution ordering. But we do need to argue that the
        sequential schedule gives the correct return values for
        $\FetchAndMax$. Most of this follows from the fact that the
        largest $y$ such that $v ≥ (n+1)^y$ will be precisely the
        most significant nonzero digit $r_i$, corresponding to the
        largest previous increment $(n+1)^i$ to $r$. This will always
        be the result of some previous execution of
        Line~\ref{line-hw-fetch-and-max-fetch-and-add}, which will be
        the linearization point of a $\FetchAndMax(i)$ operation.

        In the other direction, at the time of linearization of a
        $\FetchAndMax(i)$ operation, either $r_i$ is already nonzero
        (Line~\ref{line-hw-fetch-and-max-read}); or $r_i$ becomes
        nonzero if it is not already nonzero
        (Line~\ref{line-hw-fetch-and-max-fetch-and-add}).
        In either case, $\FetchAndMax$ that are linearized later will
        read a value with $r_i > 0$ and return at least $i$. So each
        $\FetchAndMax$ will correctly return the value of the largest
        argument to any $\FetchAndMax$ that linearizes before it.

        \subsubsection*{Alternate solution}

        If you happen to have read Chapter~\ref{chapter-common2},
        which we didn't happen to cover this semester, you may have found an
        easier solution. Update operations on fetch-and-max commute,
        so fetch-and-max objects are in the common2 class as defined
        by Afek~\etal~\cite{AfekWW1993}.  We also know that
        fetch-and-increment has consensus number 2. So applying the
        results of Afek~\etal~\cite{AfekWW1993} shows that
        fetch-and-increment plus registers implements fetch-and-max.

        \subsubsection*{Solution with unbounded number of processes}

        We can use a different encoding that works for an unbounded
        number of processes, provided each has a unique id in $ℕ$.
        Fix some pairing function $\Tuple{⋅,⋅}: ℕ × ℕ → ℕ$ then run
        Algorithm~\ref{alg-hw-fetch-and-max-pairing}. For each process
        $i$ and input $x$, we will use the bit in position
        $\Tuple{i,x}$ in the fetch-and-add to record if $i$ has done
        $\FetchAndMax(x)$ at least once.

        \begin{algorithm}
            \Procedure{$\FetchAndMax(x)$}{
                \tcp{make sure I have not already stored $x$}
                $v ← r$\;
                \If{$v < 2^{\Tuple{i,x}}$}{
                    \tcp{set bit $\Tuple{i,x}$ in $r$}
                    $v ← \FetchAndAdd\parens*{r,2^{\Tuple{i,v}}}$
                }
                \Return largest $y∈N$ such that bit $\Tuple{j,y}$ is
                set for some $y$
            }
            \caption{Fetch-and-max from fetch-and-add with unbounded
            $n$}
            \label{alg-hw-fetch-and-max-pairing}
        \end{algorithm}

        The proof of correctness is essentially the same as for
        Algorithm~\ref{alg-hw-fetch-and-max}.

        It's possible to generalize this solution further to implement
        any object with a countable number of commuting RMW
        operations: encode the $i$-th execution of operation $π$ by
        process $p$ as $\Tuple{\Tuple{p,π},i}$ and implement it by
        setting bit $i$ using
        $\FetchAndAdd(r,2^i)$. This shows that unbounded integer
        fetch-and-add is already powerful enough to implement
        countable instances of the \concept{generalized
        fetch-and-add}\index{fetch-and-add!generalized} defined
        by~\cite{AfekWW1993}. The moral of this story is that natural
        numbers can encode a lot of information.

    \subsection{Read-modify-write consensus}

    Algorithm~\ref{alg-hw-RMW-consensus} gives an partial implementation of
    wait-free binary consensus for an unbounded number of processes
    from a single RMW object $r$. In this algorithm, each
    process applies a function $f_0$ or $f_1$ to the object (depending
    on its input), then reads the state $q$ of the object as a separate
    operation and returns $g(q)$. The algorithm does not specify the
    state space $Q$ of the object, its initial state $q_0∈Q$, or the functions $f_0:Q→Q$,
    $f_1:Q→Q$, and $g:Q→\Set{0,1}$.

    \begin{algorithm}
        \Procedure{$\FuncSty{RMW}(r,f)$}{
            \Atomically{
                $r ← f(r)$
            }
        }
        \Procedure{$\FuncSty{consensus}(v)$}{
            $\FuncSty{RMW}(r,f_v)$\;
            \Return $g(r)$
        }
        \caption{Consensus from a single RMW object}
        \label{alg-hw-RMW-consensus}
    \end{algorithm}

    \begin{enumerate}
        \item Prove or disprove: There is a choice of $Q, q_0, f_0, f_1,$
            and $g$ that makes Algorithm~\ref{alg-hw-RMW-consensus}
            satisfy the usual requirements of agreement, termination,
            and validity, where $Q$ is finite.
        \item Prove or disprove: There is a choice of $Q, q_0, f_0, f_1,$
            and $g$ that makes Algorithm~\ref{alg-hw-RMW-consensus}
            satisfy the usual requirements of agreement, termination,
            and validity, where $Q$ is finite and $f_0$ and $f_1$ are
            invertible.
    \end{enumerate}

        \subsubsection*{Solution}

        \begin{enumerate}
            \item Proof: Let $Q = \Set{⊥,0,1}$, let $f_v$ map $⊥$ to
                $v$ and any $x≠⊥$ to $x$, and let $g$ be the identity
                function. Then whichever process does the first RMW
                operation sets $r$ to its input $v$, which is then
                returned by all processes, satisfying agreement and
                validity. Termination is immediate since each process
                only does two operations.
            \item Disproof: We will show that if $f_0$ and $f_1$ both
                have inverses, then $Q$ cannot be finite. Consider an
                execution of Algorithm~\ref{alg-hw-RMW-consensus} with
                $n = 2k$ processes. Observe that in the initial
                configuration $C$, applying an operation
                $π_v = \FuncSty{RMW}(r,f_v)$
                operation yields a $v$-valent configuration, since
                $Cπ_v$ is indistinguishable to all other processes
                from a configuration where the process applying $π_v$
                subsequently reads $r$ and decides $v$.

                Now consider an execution where all $n$ processes have
                input $1$. Let $q_0$ be the initial state of $r$, and
                let $q_i = f^{(i)}_1(q_0)$ be the result of applying
                $f_1$ to $q_0$ $i$ times.
                We will abuse notation a bit and use $q_i$ to refer
                both to the element of $Q$ and to
                the configuration in which $r$ contains $q_i$.
                Because $q_0$ is bivalent
                and $q_1 = f_1(q_0)$ is $1$-valent, we have $q_0 ≠ q_1$. 
                More generally, since for any $i>0$, $q_i$ is $1$-valent, $q_i ≠
                q_0$ unless $i=0$.

                We can now use the fact that $f_1$ has an inverse
                function $f^{-1}_1$ to show that $q_i ≠ q_j$ for any
                $0≤i<j≤n$. The proof is by induction on $j$. If $i=0$,
                we have already shown that $q_0 ≠ q_j$ If $i>0$, then
                the induction hypothesis tells us that 
                $f^{-1}_1(q_i) = q_{i-1} ≠ q_{j-1} = f^{-1}(q_j)$.
                But then $q_i ≠ q_j$.

                We now have states $q_0, \dots, q_n$ that are all
                distinct, implying $\card{Q} ≥ n+1$. But since $n$ was
                arbitrary, $\card{Q}$ can't be finite.
        \end{enumerate}

\section{Assignment 6: due Thursday 2025-04-17, at 23:59 Eastern US time}

    \subsection{A mod-2 counter}

    A \concept{mod-2 counter}\index{counter!mod-2} stores a single
    bit. It has an increment operation $\Inc$ that flips the bit and
    returns nothing and a read operation $\Read$ that returns the current
    value of the bit.

    We would like to have a solo-terminating linearizable
    implementation of a mod-2 counter that uses as few base objects as
    possible.

    \begin{enumerate}
        \item Suppose our base objects are atomic registers. As a
            function of the number of processes $n$, what is the
            minimum number of registers needed to implement a
            solo-terminating linearizable mod-2 counter?
        \item Suppose our base objects are swap registers, which
            support an operation $\FuncSty{swap}(x)$ that replaces the
            value in the register and returns the old value, and an
            operation
            $\Read$ that just returns the old value without changing
            it. What is the minimum number of these objects needed to
            implement a solo-terminating linearizable mod-2 counter?
    \end{enumerate}

    For each case, prove the correctness of your answer. (You may
    assume $n≥2$ to avoid the trivial $n=1$ case.)

        \subsubsection*{Solution}

        \begin{enumerate}

            \item We need $n$ registers exactly to implement a mod-2
                counter.

                To show that $n$ are needed, we'll do a covering
                argument. Unfortunately we can't just apply the
                Jayanti-Tan-Toueg
                bound~\cite{JayantiTT2000}\footnote{Also discussed in
                Chapter~\ref{chapter-JTT}.} directly because (a) it only gives
                an $n-1$ lower bound, and (b) mod-2 counters aren't
                perturbable.

                Instead, we'll do a JTT-like covering argument that
                ends up with $n$ registers covered, showing the
                claimed space lower bound. This is a bit simpler than
                stock JTT because we are only interested in the space
                bound, and can re-use one of the processes doing
                increments as the reader.

                The induction hypothesis is that we can construct a
                schedule of the form $Λ_k Σ_k$, where $Λ_k$ consists
                of $k$ incomplete $\Inc$ operations by processes
                $p_1,\dots,p_k$ and $Σ_k$ consists of $k$ $\Write$
                operations to distinct atomic registers
                $r_1,\dots,r_k$ by the same processes.

                For the base case, let $p_1$ start an $\Inc$
                operation, and let $Λ_1 Σ_1$ be the prefix of its solo
                execution of this operation ending with its first
                attempt to write to some register, which we will call
                $r_1$.

                For the induction step, start with $Λ_k Σ_k$ and
                consider an execution $Λ_k γ Σ_k Δ ρ$ where $γ$ is a
                complete $\Inc$ operation by $p_{k+1}$ running alone,
                $Δ$ consists of $p_1,\dots,p_k$ finishing their $\Inc$
                operations (each running alone, so
                solo-termination applies) and $ρ$ consists of a
                $\Read$ operation by $p_1$.

                If $γ$ does not include a write to some atomic
                register not covered by some write in $Σ_k$, this
                execution is indistinguishable to $p_1,\dots,p_k$ from
                the execution $Λ_k Σ_k Δ ρ$ where the $\Inc$ by
                $p_{k+1}$ does not occur. In this case $ρ$ returns the
                same value in both executions but it is the wrong
                value in one of them, since there are $k+1$ complete
                $\Inc$ operations before it in the first execution and
                only $k$ in the second, and $k+1≠k \pmod{2}$.

                It follows that $γ$ includes a
                write to some register $r_{k+1}$ not in
                $r_1,\dots,r_k$. Expand $γ = λσδ$ where $σ$ is the
                first write to such a register and let $Λ_{k+1} =
                Λ_k λ$ and $Σ_{k+1} = σ Σ_k$.

                Repeating the induction step eventually yields an
                execution $Λ_n Σ_n$ in which $Σ_n$ covers $n$ distinct
                registers $r_1,\dots,r_n$, showing that $n$ registers
                exist.

                For the matching upper bound, use a snapshot (which
                can be implemented using $n$ registers) to track
                the total number of increments by each process, and
                have a $\Read$ operation return the sum of these
                quantities mod 2.
            \item We need only one swap register to implement a
                solo-terminating mod-2 counter. The implementation is
                given in
                Algorithm~\ref{alg-hw-mod-2-counter-from-swap}.

                \begin{algorithm}
                    \Procedure{$\Inc(r)$}{
                        \tcp{first try to update counter to $0$}
                        $v ← 0$\;
                        \While{$\FuncSty{swap}(r,v) = v)$}{
                            \label{line-hw-mod-2-counter-from-swap-while}
                            \tcp{try again with a different value}
                            $v ← ¬v$\;
                        }
                    }
                    \Procedure{$\Read(r)$}{
                        \Return $r$\;
                        \label{line-hw-mod-2-counter-from-swap-return}
                    }
                    \caption{Solo-terminating mod-2 counter from swap}
                    \label{alg-hw-mod-2-counter-from-swap}
                \end{algorithm}

                This algorithm is not wait-free, since the test in
                Line~\ref{line-hw-mod-2-counter-from-swap-while} could
                fail every time if the value of the counter keeps
                changing. But it is solo-terminating: if some process
                runs alone, each call to $\FuncSty{swap}$ returns the
                same value $v'$, but the process switches $v$ to $¬v$
                after the first failed test, making the next test succeed.

                For linearizability, use linearization points. For an
                $\Inc$ operation the linearization point is the
                $\FuncSty{swap}$ operation that causes the test in
                Line~\ref{line-hw-mod-2-counter-from-swap-while} to
                succeed, which corresponds to switching the value in
                $r$ from $v$ to $¬v$. For a $\Read$ operation, the
                linearization point is the (implicit) swap register read in
                Line~\ref{line-hw-mod-2-counter-from-swap-return}.

                That these linearization points give a correct
                sequential execution is immediate from the fact that
                only successful swaps change the value of the swap
                register, making the value of the swap register
                exactly equal to the number of successful swaps mod 2.
        \end{enumerate}

    \subsection{A linear splitter network}

    \begin{algorithm}
        \Procedure{$\FuncSty{renaming}(\Id)$}{
            $i ← \Id \bmod m$\;
            \While{\True}{
                \Switch{$S_i(\Id)$}{
                    \Case{\SplitterStop}{
                        \Return $i$
                    }
                    \Case{\SplitterRight}{
                        $i ← i+1$\;
                    }
                    \Case{\SplitterDown}{
                        $i ← i+n$\;
                    }
                }
            }
        }
        \caption{Renaming with a linear splitter network}
        \label{alg-hw-linear-splitter-network}
    \end{algorithm}

    Algorithm~\ref{alg-hw-linear-splitter-network} gives an
    implementation of renaming based on a network of $2m$ splitters
    (as defined in Algorithm~\ref{alg-splitter-again})
    organized in a long line. Each of $n$ processes with unique
    initial identity $\Id ∈ \Set{0,\dots,N-1}$
    starts at the splitter at position $i = \Id \bmod m$. At each
    splitter $S_i$, a process returns $i$ if it stops, advances to
    $i+1$ if it goes right, and advances to $i+n$ if it goes down.
    The algorithm fails if any process goes past $S_{2m-1}$.

    Prove or disprove: It is possible to choose
    $m$ polynomial in $n$ such that
    Algorithm~\ref{alg-hw-linear-splitter-network} implements
    wait-free renaming where each process obtains a name in the range
    $\Set{0,\dots,2m-1}$.

        \subsubsection*{Solution}
        For each $i$, let $A_i$ be the set of processes that enter splitter $S_i$. 
        Let $t$ be the maximum over all processes of $\Id \bmod m$,
        which is the largest $t$ for which $A_t$ includes some process
        that reaches $S_t$ without first going through $S_{t-1}$ or
        $S_{t-n}$.

        We start by showing that $n$ consecutive splitters are enough to
        stop at least one process:
        \begin{lemma}
            \label{lemma-alg-hw-linear-splitter-network-stop}
            If $A_i ≠ ∅$, at least one process
            stops in some splitter in the range $S_i,\dots,S_{i+n-1}$.
        \end{lemma}
        \begin{proof}
            Suppose no processes stop in this range.
            Then each of the $n$ splitters $S_j$ with $i≤j<n$ for
            which $\card*{A_j} > 0$ sends at least one process right
            and at least one process down. But for any $S_j$ that
            sends a process right, we have $\card*{A_{j+1}} > 0$, so
            by induction on $j$, every $S_j$ in this range sends at
            least one process right and at least one process down.
            Process that go down from $S_j$ next enter $S_{j+n}$,
            which is not in the range; so these processes are all
            distinct. But there is also a process that goes
            right from $S_{i+n-1}$, for a total of $n+1$ processes, a
            contradiction.
        \end{proof}

        We also note that processes can't jump too far:
        \begin{lemma}
            \label{lemma-alg-hw-linear-splitter-network-gap}
            For any $i>t$, if $A_j = ∅$ for all $j$ with $i≤j<i+n$, then $A_j=∅$ for
            all $j≥i$.
        \end{lemma}
        \begin{proof}
            Proof is by induction on $j$. For $i≤j≤i+n-1$, $A_j=∅$ is
            given. For larger $j$, suppose the claim holds for all
            $j'$ with $i≤j'<j$. Then $A_j$ is the union of (a) the set
            of all
            processes that start at position $j$,
            (b) the set of all processes that move right from
            $j-1$, and (c) the set of all processes that move down
            from $j-n$.
            Case (a) includes no processes because $j ≥ i+n > t$.
            Cases (b) and (c) include no processes because $A_{j-1}$
            and $A_{j-n}$ are both empty by the induction hypothesis.
        \end{proof}

        Now let us argue that the maximum name returned is not too
        high. Since $A_t ≠ ∅$,
        Lemma~\ref{lemma-alg-hw-linear-splitter-network-stop} shows that
        some process stops at position $t_0 ≤ t + n-1$. After $t_0$,
        Lemma~\ref{lemma-alg-hw-linear-splitter-network-gap} shows that
        either all processes have already stopped, or we have $A_j ≠
        ∅$ for some $j ≤ t_0 + n$. Applying
        Lemma~\ref{lemma-alg-hw-linear-splitter-network-stop} gives us a
        new position $t_1 ≤ j + n-1 ≤ t_0 + n + (2n-1)$ at which
        at least one additional processes has stopped. Iterating this
        argument gives us $n$ stopped processes in the worst case by
        $t_{n-1} ≤ t_0 + n + (n-1)(2n-1) ≤ t + (n-1) + n + (n-1)(2n-1)
        ≤ (m-1) + n + 2n(n-1) = m + 2n^2 - n - 2$.

        Setting $m = 2n^2$, which is polynomial in $n$, gives a
        maximum name bounded by $2m - n - 2 ≤ 2m-1$ as required.

\chapter{Sample assignments from Fall 2023}

\section{Assignment 1: due Thursday 2023-09-21, at 23:59 Eastern US time}

    \subsection{Maximal independent set in a ring}

    Given a graph, a \concept{maximal independent set} (MIS) is a subset $S$ of
    the vertices that is an \concept{independent set} (no two vertices
    in $S$ have an edge between them) that is \concept{maximal} (no superset
    of $S$ is also an independent set). We will say that a distributed
    algorithm computes a maximal independent set if every process
    eventually returns $0$ or $1$, and the set of processes that
    return $1$ form an MIS.

    Let's suppose we have an asynchronous bidirectional ring of
    unknown size with deterministic processes. For each of the
    following assumptions, show either (a) no algorithm correctly
    computes an MIS in the worst case, or (b) there is an algorithm
    that computes an MIS in $O(f(n))$ time in the worst case,
    \emph{and} there is a matching lower bound showing no algorithm
    can do better than $Ω(f(n))$ in the worst case.

    \begin{enumerate}
        \item The network is anonymous.
        \item All processes have distinct ids, but the algorithm is
            comparison-based.
    \end{enumerate}

        \subsubsection*{Solution}

        \begin{enumerate}
            \item This case is impossible. The proof is the same as
                for leader election in an anonymous ring. In a
                synchronous execution, symmetry is never broken, and
                so if any process returns, all processes return the
                same value. This either yields $S=∅$ (not maximal) or
                $S = V$ (not independent).
            \item Possible, with $Θ(n)$ time both necessary and
                sufficient.
                \begin{itemize}
                    \item For the algorithm, elect a leader using a
                        comparison-based
                        $O(n)$-time leader election algorithm (LCR
                        works).
                        Relay a message clockwise from the leader to
                        count off the position of each node (see
                        Algorithm~\ref{alg-mis-solution-counting}).
                        Send a single message counterclockwise from
                        the leader to notify node $n-1$ of its special
                        position.
                        The time to complete both of these steps is at
                        most $O(n)$.

                        Now have each node return $1$ if and only if
                        (a) it has an even position and (b) it is not
                        in position $n-1$.

                        This is an independent set since no two
                        even-position nodes other than $n-1$ are
                        adjacent. It is maximal because adding any
                        other node $i$ creates two adjacent nodes ($i$
                        and $i-1$ in the case of an odd node, $n-1$
                        and $0$ in the case of $n-1$). So we get an
                        MIS in $O(n)$ time.
                    \item For the lower bound, adapt the
                        Frederickson-Lynch lower bound for leader
                        election. As for the upper bound we need to be
                        a little careful about odd vs even rings.

                        Consider a synchronous execution,
                        then after $k$ rounds, two nodes will return
                        the same value (if any) if their
                        $k$-neighborhoods are order-equivalent. Now
                        observe that in a ring of size $n$ i
                        with $\Id_i = i$ for all $i$, nodes
                        $\floor{\frac{n}{2}}-1$, $\floor{\frac{n}{2}}$, 
                        and $\floor{\frac{n}{2}}+1$
                        have order-equivalent
                        $\parens*{\floor{\frac{n}{2}}-1}$-neighborhoods, as in both
                        cases the ids in these neighborhoods are
                        strictly increasing. It follows that if any of
                        of these nodes returns a value after
                        $\floor{\frac{n}{2}}-1$ or fewer rounds, either all three
                        return $0$ (meaning that the computed
                        independent set is not maximal, since
                        node $\floor{\frac{n}{2}}$ can be added
                        without creating two adjacent nodes); or all three
                        return $1$ (meaning that the computed set is
                        not independent). This gives the desired matching $Ω(n)$
                        worst-case lower bound.
                \end{itemize}
        \end{enumerate}

        \begin{algorithm}
            \Initially{
                \If{I am the leader}{
                    $\DataSty{position} ← 0$\;
                    send $0$ clockwise
                }
            }
            \UponReceiving{$m$}{
                \If{I am not the leader}{
                    $\DataSty{position} ← m+1$\;
                    send $m+1$ clockwise
                }
            }
            \caption{Counting off nodes in a ring}
            \label{alg-mis-solution-counting}
        \end{algorithm}

    \subsection{Deanonymization}

    Suppose you have an asynchronous bidirectional message-passing
    network in the form of an arbitrary connected graph, which is
    mostly anonymous in the sense that every node but one runs the
    same code, and that each node can only identify its neighbors by a
    local \concept{port number}, an element of $ℕ$, that is only
    meaningful to that node and is not correlated with any other
    node's port numbers. This means that when a message is delivered
    from a node $i$ to a node $j$, $j$ sees that the message came from
    a particular port $p$ that it uniquely associates with $i$; it can
    similarly send messages to port $p$ that will be delivered to $i$.
    You may assume that each node has a complete list of its
    neighbors' port numbers, so it can tell, for example, if it has a
    neighbor that it hasn't received any messages from.

    The one non-anonymous node is marked as the initiator and can run
    special code, but is subject to the same port-number limitations
    as all the other nodes. None of the nodes know the size of the
    graph $n$ or its diameter $D$.

    We would like to assign a unique id to every node in the system in
    the range $1$ through $n$. Prove or disprove: There exists an
    algorithm that does this in time $O(D)$.

        \subsubsection*{Solution}

        We'll show that it is possible by constructing an algorithm.

        First observe that we can apply the alpha synchronizer to this
        this system, since the alpha synchronizer only requires that a
        node be able to detect when it has received a message (or
        $\NoMsg$ from each of its neighbors, and the assumptions
        on port numbers are sufficient to do this. We also don't care
        about message complexity. So we can simplify
        our life by assuming that the model is synchronous.
        (Alternatively we can replace the synchronous breadth-first
        search protocol in the first step below with an asynchronous
        breadth-first search protocol, but the end result is pretty
        much the same either way.)

        Run a synchronous breadth-first search protocol to construct a
        shortest-path tree rooted at the initiator. This takes $O(D)$
        time and yields a tree with depth at most $D$. Note that the
        parent pointers in the usual protocol will now have port
        numbers rather than ids, but this doesn't affect the
        algorithm.

        Using convergecast, compute the size of every subtree and have
        each node pass this information on to its parent. This takes
        an additional $O(D)$ time.

        We can now recursively assign ids through the tree. The
        initiator starts the process by sending itself a message
        containing the id range $\Set{1\dots n}$. Each node that
        receives an id range $\Set{i\dots j}$ assigns $i$ to itself
        and then partitions the remaining range $\Set{i+1\dots j}$
        into subranges $\Set{i_1\dots j_1}, \dots \Set{i_k \dots
        j_k}$, where $k$ is the number of children it has and each
        range has length equal to the number of nodes at the subtree
        rooted at the corresponding child when sorted by port number.
        Now send each child its range. A straightforward induction
        argument shows that this assigns a unique identifier to every
        node. The time to perform this broadcast-like operation is
        proportional to the depth of the tree, giving another $O(D)$
        time. So the total time for all steps is $O(D)$.

\section{Assignment 2: due Thursday 2023-10-05, at 23:59 Eastern US time}

    \subsection{Synchronous agreement in a bipartite network}

    Let $n≥2$, and suppose you have a synchronous network with $2n$ processes
    $p_1,\dots,p_n$ and $q_1,\dots,q_n$.
    The network is bipartite:
    each $p_i$ can send and receive messages from each $q_j$, but no
    pair of processes $p_i$ and $p_j$ or $q_i$ and $q_j$ can
    communicate directly. The processes are subject to crash failures,
    where as usual a process that crashes in a particular round may send any
    subset of the messages it intended to send in that round. Our goal
    is to solve synchronous agreement, as defined in
    §\ref{section-synchronous-agreement-problem}.

    \begin{enumerate}
        \item As a function of $n$, what is the largest number of
            potential crash failures $t$ for which it is possible to solve
            agreement? (Give an exact value.)
        \item As a function of $n$ and $t$, what is the best possible
            asymptotic worst-case running time for synchronous agreement,
            assuming $t$ is small enough to make synchronous agreement
            possible. (Give an asymptotic expression in $t$ and/or $n$.)
    \end{enumerate}

    You should justify your answers with matching upper and lower
    bounds.
    For the upper bound side, you may find it helpful to give a single
    algorithm that applies to both cases.

        \subsubsection*{Solution}

        For our algorithm, we'll run Dolev-Strong (see
        §\ref{section-synchronous-agreement-flooding}) largely
        unmodified, except that we will run for $2t+2$ rounds and each process
        will send messages only to its neighbors.

        Observe that (a) if $t≤n-1$, there is at least one
        non-faulty $p_i$ and at least one non-faulty $q_j$; and (b)
        since we can divide the rounds into $t+1$ phases of two rounds
        each,
        there are at least two consecutive rounds $2s$ and $2s+1$ with no new crash
        failures in either round. Let $\Tuple{k,v}$ appear in
        $S_{p_i}$ at the beginning of round $2s$, where $p_i$ has not
        yet crashed. Then $\Tuple{k,v}$ is
        transmitted to all surviving $q_j$ in round $2s$, and at least
        one such $q_j$ forwards $\Tuple{k,v}$ to all surviving $p_i$
        in round $2s+1$. Similarly, any $\Tuple{k,v}$ that appears in
        $S_{q_j}$ at the beginning of round $2s$ is transmitted to all
        $p_i$ and $q_{j'}$ by the end of round $2s+1$. It follows that
        $S_{p_i}^{2s+1} = S_{q_j}^{2s+1}$ for all $p_i$ and $q_j$ that
        do not crash in round $2s+1$ or earlier. The same argument as
        used for the original algorithm shows that this continues to
        hold for all subsequent rounds, and so all processes choose
        the same value from the same set at the end of the protocol,
        giving agreement. Termination is trivial as usual, and
        validity follows from the same argument as for the original
        algorithm.

        This shows that consensus is possible in $O(t)$ time when
        $t≤n-1$. Now we just need the corresponding lower bounds.

        \begin{enumerate}
            \item To show that $t≤n-1$ is necessary, suppose $t≥n$.
                Then the adversary can crash all processes $q_j$
                immediately, leaving each $p_i$ with no live
                neighbors. If some $p_i$ decides on a value that is
                not its input, it violates validity in the execution
                where all other processes have the same input. But if
                each $p_i$ decides its own input and $n≥2$, then we
                violate agreement if the inputs don't all agree.

                (There is an annoying special case when $n=1$. In this
                case the protocol can tolerate one crash failure,
                because the survivor will agree with itself.
                Fortunately the problem statement excludes this case.)
            \item For the time bound, suppose that we can solve the
                problem in $t$ rounds. Then in a
                complete network we can also solve synchronous
                agreement with $t$ failures in $t$ rounds, since
                nothing prevents the processes in the complete network
                from choosing only to communicate using a bipartite
                subnetwork. But this violates the Dolev-Strong lower
                bound (see §\ref{section-synchronous-agreement-lower-bound}).
                So we get matching upper and lower bounds of $Θ(t)$ on
                the time complexity for this problem.
        \end{enumerate}

    \subsection{Leader rotation}

    We are given an asynchronous bidirectional network of $n$
    processes in the form of an arbitrary connected graph with
    diameter $D$. Each process has access to a \concept{special
    action}\index{action!special}, a local operation it uses to claim
    temporary leadership of the network. We'd like this leader role to
    repeatedly rotate through the $n$ processes, in the sense that
    there is an assignment $0,\dots n-1$ of positions to the processes
    such that the $i$-th special action is always carried out by the
    process in position $i \bmod n$. (Note that these positions can be
    chosen by the protocol and do not necessarily have any meaning
    outside of showing that the protocol satisfies this requirement.)

    We assume that the processes are not anonymous and that every
    process in the network knows the entire structure of the graph,
    including all process identities.

    Since we are considering infinite executions, we can't talk about
    the time complexity of the execution has a whole, so instead we
    will define the \concept{responsiveness} of an execution of the
    protocol as the maximum time between any two consecutive special
    actions.

    Show that there is a function $f(n,D)$ such that any protocol for
    this problem has
    responsiveness $Ω(f(n,D))$ in the worst case, and that some such
    protocol has responsiveness $O(f(n,D))$ in the worst case.

        \subsubsection*{Solution}

        It turns out that the diameter is not important. There exists
        a protocol with responsiveness $Θ(1)$, which is also the lower
        bound.

        We'll start by showing that no protocol with $n≥2$
        can have responsiveness less than $1$.

        Consider a synchronous execution $Ξ$, and suppose that 
        there are two consecutive special actions $s_i$ and $s_{i+1}$
        such that the time between $s_i$ and $s_{i+1}$ is less than
        $1$. Then $s_i$ and $s_{i+1}$ are not causally ordered, and
        there is a causal shuffle $Ξ'$ of $Ξ$ in which $s_{i+1}$
        occurs before $s_i$ but the other special actions occur in the
        same order as before. Let $p_i$ and $p_{i+1}$ be the processes
        execution $s_i$ and $s_{i+1}$. Then in $Ξ'$, $p_i$ executes
        both the $(i+1)$-th special action and the $(i+n)$-th special
        action, which is requires $p_i$ to have both positions $(i+1)
        \bmod n$ and $(i+n) \bmod n = i \bmod n$, which is
        inconsistent withe requirement of distinction positions when
        $n≥2$.

        For the upper bound, we need to show that for any graph $G$
        there is a protocol that rotates through the special actions
        as described above with a gap of at most $O(1)$ time between
        any consecutive special actions. We can do this by adapting a
        depth-first traversal of a spanning tree $T$ of $G$, circulating a token
        along the $2n$ edges of the tour so that it reaches every node
        at least once every $2n$ steps. A node will execute its
        special action on exactly one of these occasions where it
        receives the token, carefully chosen so that the token doesn't
        travel too far without triggering a special action.

        \begin{algorithm}
            \For{ever}{
                \If{I am not the root}{
                    wait to receive token from my parent
                }
                \If{My depth is even}{
                    perform special action
                }
                \Foreach{child $c$ in increasing order by id}{
                    send token to $c$\;
                    wait to receive token from $c$
                }
                \If{My depth is odd}{
                    perform special action
                }
            }
            \caption{Leader rotation algorithm}
            \label{alg-leader-rotation}
        \end{algorithm}

        The algorithm is given as Algorithm~\ref{alg-leader-rotation}.
        We assume that a rooted spanning tree has already been
        constructed and that each node knows its parent (if any), its
        children, and its depth in the tree. (Each process can easily
        compute this at the start of the tree based on its knowledge
        of the graph; so long as the processes use the same algorithm
        to construct the tree, they will all behave consistently.)

        This protocol repeatedly carries out a depth-first traversal
        of the tree by passing a single token along the edges of the
        tree. Each even-depth node (including the root, at depth $0$)
        performs its special action when the token enters its subtree;
        each odd-depth node performs its special action when the token
        leaves. Since exactly one of these events occurs during each
        traversal and each traversal visits the nodes in the same
        fixed order, we satisfy the requirements that the special
        actions rotate among the nodes.

        To show a bound on the gap between special actions, consider
        four consecutive events in the execution, and look at the
        messages sent by the first three events along edges of the tree.
        Classify these messages as $D$ or $U$ depending on whether the
        message goes down the tree (is sent to a child) or goes up (is
        sent to a parent). There are eight possible patterns $DDD$,
        $DDU$, $DUD$, \dots, $UUU$ for the three messages.

        \begin{enumerate}
            \item If $DD$ appears in the pattern, then one of the
                receivers of these messages has even depth and performs
                the special action. This covers $DDD$, $DDU$, and
                $UDD$.
            \item Similarly, if $UU$ appears in the pattern, then one
                of the senders of these messages has odd depth and
                performs the special action. This covers $DUU$, $UUD$,
                and $UUU$.
            \item The remaining patterns are $DUD$ and $UDU$. In both
                cases, the process in the middle of $DU$
                that receives the $D$ message and
                sends the $U$ message is a leaf. No matter what its
                depth, it performs the special action.
        \end{enumerate}

        It follows that any sequence of four consecutive send events
        involves at least one special action. So the maximum time
        between special actions is $4$.

        This gives a matching $Ω(1)$ lower bound and $O(1)$ upper
        bound on the responsiveness of this protocol.

\section{Assignment 3: due Thursday 2023-10-26, at 23:59 Eastern US time}

    \subsection{Evil twins}

    Suppose we have a system where a process $p$ can be paired with an
    \concept{evil twin}, a Byzantine process $\evil{p}$ that can
    send messages that appear to come from $p$. Messages from
    $\evil{p}$ enter the same buffer as messages from $p$, and cannot
    be distinguished by the recipient from legitimate messages from
    $p$. The existence of the evil twin does not otherwise affect the
    execution of $p$, which continues to behave normally.

    Prove or disprove: There exists a constant $c>0$ such that it is
    possible to solve binary consensus in
    an asynchronous message-passing system with deterministic
    processes, as long as the number of evil twins $t$ is less than
    $cn$.

    Here binary consensus is defined as a protocol that satisfies
    the usual requirements of agreement (all processes decide on the
    same value), termination (all processes eventually decide), and
    validity (if all processes start with the same input, they all
    decide on this input)?

        \subsubsection*{Solution}

        We can solve the problem for $t < n/3$, by simulating any
        standard synchronous Byzantine agreement algorithm with optimal fault
        tolerance, with an extra round at the end to handle processes
        that have evil twins but still need to decide on the common
        value.

        To enforce synchrony, we use the alpha synchronizer. Since
        every good process sends a message to every other process in
        every simulated round, nobody gets stuck waiting for messages
        from all other processes, and the worst that happens is that
        some process might receive a round-$r$ message from
        $\evil{p}_i$ instead of $p_i$. In this case we treat $p_i$ as
        Byzantine for the simulated execution.

        We will also assume that any $p_i$ with an evil twin behaves arbitrarily
        during the main protocol. This absolves us from worrying about
        good processes sending bad messages, and again bad messages
        from a twinned $p_i$ are indistinguishable from bad messages
        from a Byzantine $p_i$ in the simulated execution.

        Running EIG or a similar algorithm then gives agreement among
        all the processes that do not have evil twins. We add one more
        round where each process announces its decision value, and all
        good processes (including twinned processes) wait to receive
        decision values from all $n$ processes and decide on the
        majority. Since at least $\frac{2}{3}n$ processes agree coming
        out of the Byzantine agreement protocol, all good processes
        will see the same majority value and reach the same decision.

    \subsection{Crash failures with recovery}

    Consider an asynchronous message-passing model with deterministic
    processes, where a process can crash, losing all of its state
    (including its input), but
    then recovers to a default state from which it can continue its
    execution.  We would like to solve binary consensus in this model,
    characterized by agreement (all processes eventually decide the
    same value), validity (if all processes start with the same input,
    they all decide this input), and termination (every process
    eventually decides on some value). Note that while defining the
    problem in this model we do not necessarily think of processes as
    being faulty or non-faulty; any process can crash, possibly more
    than once, but we still require that it eventually makes a
    decision on the same value as all the others.

    As a function of $n$, what the largest number of possible crash
    failures $t$ for which it is consensus as defined above can be
    solved in this model?

        \subsubsection*{Solution}

        The largest number of crash failures we can tolerate is
        $t=n-1$. At $t=n$, it is possible for every process to crash
        immediately, erasing all inputs. Since this gives the same
        configuration in both an all-$0$-input and all-$1$-input
        execution, whatever the processes decide will violate validity in one
        of these executions.

        To solve consensus with $t=n-1$, we'll first show how to
        simulate a system with standard crash failures and a perfect
        failure detector, then adapt the consensus protocol from
        Chandra and Toueg~\cite{ChandraT1996} for the strong failure
        detector (see
        Algorithm~\ref{alg-strong-failure-detector-consensus})
        to solve the problem in the crash-with-recovery model.

        The idea is that whenever a process recovers, it will send a
        message \DataSty{failed} to all other processes, and otherwise
        act like a crashed process by no longer participating in the
        simulated consensus protocol. A never-crashed process that
        receives a \DataSty{failed} message from some process $p$ will
        (a) add $p$ to its list of suspect processes; and (b) send $p$
        its decision value, if it has already decided, or add $p$ to a
        list of processes to be notified of its decision value when it
        decides, if it has not already decided. A previously-crashed
        process that is notified of a decision value decides on that
        value.  Other than these changes, the never-crashed processes
        run Algorithm~\ref{alg-strong-failure-detector-consensus}
        essentially unmodified.

        Agreement follows from the fact that all never-crashed
        processes agree in
        Algorithm~\ref{alg-strong-failure-detector-consensus} and all
        crashed processes that decide choose a value sent to them by a
        never-crashed process. Validity follows from validity of
        Algorithm~\ref{alg-strong-failure-detector-consensus} and the
        same argument.

        Termination is a bit trickier since we have to allow for the
        possibility that a process might crash more than once. Any
        process that doesn't crash decides at the end of
        Algorithm~\ref{alg-strong-failure-detector-consensus} (but
        note that it may still need to respond to \DataSty{failed}
        messages). For a process $p$ that does crash, consider what
        happens when it recovers for the last time. At this point the
        process sends \DataSty{failed} to all processes, including at
        least one process $q$ that does not crash. Eventually $q$
        sends a value to $p$ (either immediately in response to $p$'s
        message or eventually when it decides). This value is sent
        after $p$'s last crash, so eventually $p$ receives it and
        decides.

\section{Assignment 4: due Thursday 2023-11-09, at 23:59 Eastern US time}

    \subsection{A one-object mutex}

    The Deadlock-Free Lock Company has hired you as a consultant for
    its project to built a new fetch-and-add-based mutex that works
    for any number of processes and uses no
    extra registers. Their starting point is the ticket algorithm for
    simulating a queue using a RMW object as described in
    §\ref{section-mutex-RMW}, but rather than use a general RMW
    object, they wish to use a fetch-and-add object that supports a
    single operation $\FuncSty{FAA}(r, v)$ that adds $v$ to the
    current value of $r$ and returns the old value. Both $v$ and the
    contents of the register may be arbitrary integers (including
    negative integers) of any size.

    The intern who previously worked on the project suggested the
    implementation in Algorithm~\ref{alg-bad-faa-lock}. Here $K$ is a
    large constant. The intuition is that $r \bmod K$ is used to track
    which tickets will be given out next and $\floor{r/K}$ stores which
    ticket can be used to enter the critical section. Each process calls
    $\FuncSty{acquire}(r)$ in its entry section and
    $\FuncSty{release}(r)$ in its exit section. The fetch-and-add
    register starts with value $0$.

    \begin{algorithm}
        \tcp{acquire the lock}
        \Procedure{$\FuncSty{acquire}(r)$}{
            \tcp{take a ticket}
            $t ← \FuncSty{FAA}(r, 1) \bmod K$\nllabel{line-bad-faa-take-ticket}\;
            \tcp{spin until I am at the front of the line}
            \While{$\floor{\FuncSty{FAA}(r,0)/K} ≠ t$}{
                \nllabel{line-bad-faa-while}
                spin
            }
        }
        \tcp{release the lock}
        \Procedure{$\FuncSty{release}(r)$}{
            \tcp{advance the front of the line}
            $\FuncSty{FAA}(r, K)$
        }
        \caption{Candidate fetch-and-add mutex}
        \label{alg-bad-faa-lock}
    \end{algorithm}

    \begin{enumerate}
        \item Show that Algorithm~\ref{alg-bad-faa-lock} 
            can violate both mutual exclusion and deadlock-freedom.
        \item Prove or disprove: For \emph{any} algorithm, if
            (a) it uses
            only one fetch-and-add object and no other objects and
            (b) it works for an arbitrarily large unknown number of processes,
            then there exists an execution in which it 
            eventually violates at least one of mutual exclusion or 
            deadlock-freedom.
    \end{enumerate}

        \subsubsection*{Solution}

        \begin{enumerate}
            \item Since I am lazy I will give a single execution that
                violates both mutual exclusion and deadlock-freedom.

                Send in $K+2$ processes $p_0 \dots p_{K+1}$, and have all of them
                execute Line~\ref{line-bad-faa-take-ticket} in order.
                Then each process $p_i$ gets ticket $i \bmod K$, and
                in particular $p_1$ and $p_{K+1}$ both get $1$. This
                is unfortunate, because $\floor{r/K} = 1$, so both of
                these processes leave the loop in
                Line~\ref{line-bad-faa-while} and enter the critical
                section together. Mutex is violated!

                Even worse, since $r$ never decreases, poor process
                $p_0$ can never see $\floor{r/K} = 0$ and thus remains
                stuck at Line~\ref{line-bad-faa-while} forever. This
                is true even if every other process runs to completion
                and makes no attempt to re-enter the critical section.
                We haven't actually shown that every other process
                \emph{can} run to completion, but we eventually reach
                some configuration where either (a)
                every remaining process is stuck, or (b) $p_0$ is
                alone and stuck. In either case, deadlock-freedom is
                violated.

            \item We'll disprove the claim by showing that a working
                mutex is possible.

                Here condition (b) makes things difficult, because
                even if we could tweak the calculation of
                $\floor{r/K}$ to make it wrap around like $r \bmod K$, for
                $n>K$ we still have the issue of two processes getting the same
                ticket. So we will need to abandon
                Algorithm~\ref{alg-bad-faa-lock} and do something
                else.

                Algorithm~\ref{alg-good-faa-lock} gives a mutex
                algorithm using a single fetch-and-add object, which
                we assume is initialized to $0$.
                The idea is similar to the mutex using test-and-set
                given in Algorithm~\ref{alg-mutex-TAS}. Each process
                will attempt to acquire the lock by incrementing the
                fetch-and-add object, and only a process that sees $0$
                will win. But since we can't reset the object we'll
                have the winner decrement the object on its way out,
                and have each loser decrement the object once to remove
                its excess increment and then spin until it sees a $0$
                before attempting to increment again.

                \begin{algorithm}
                    \tcp{acquire the lock}
                    \Procedure{$\FuncSty{acquire}(r)$}{
                        \While{$\FuncSty{FAA}(r,1) ≠ 0$}{
                            \nllabel{line-good-faa-outer-while}
                            $\FuncSty{FAA}(r,-1)$\nllabel{line-good-faa-decrement}\;
                            \While{$\FuncSty{FAA}(r,0) ≠ 0$}{
                                \nllabel{line-good-faa-inner-while}
                                spin
                            }
                        }
                    }
                    \tcp{release the lock}
                    \Procedure{$\FuncSty{release}(r)$}{
                        $\FuncSty{FAA}(r,-1)$\nllabel{line-good-faa-release}
                    }
                    \caption{Improved fetch-and-add mutex}
                    \label{alg-good-faa-lock}
                \end{algorithm}

                Let's prove that Algorithm~\ref{alg-good-faa-lock}
                works. We'll write that a process $p$ is in the
                critical section if it has escaped the loop by seeing
                $0$ in Line~\ref{line-good-faa-outer-while} and has
                not yet performed the decrement in
                Line~\ref{line-good-faa-release}.

                We can now state an invariant: The value of $r$ is
                equal to the number of processes $c$ in the critical
                section plus the number of processes $d$ at
                Line~\ref{line-good-faa-decrement}. To prove this,
                start by noting that in the initial configuration,
                $r=c+d=0$. The value of $r$ changes only when a
                process executes a fetch-and-add in
                Line~\ref{line-good-faa-outer-while},
                Line~\ref{line-good-faa-decrement}, or
                Line~\ref{line-good-faa-release}, so we need to show that
                $r=c+d$ continues to hold in each
                of these cases:
                \begin{itemize}
                    \item In
                        Line~\ref{line-good-faa-outer-while}, $r$ increasing
                        by $1$ and exactly one of $c$ or $d$ increases by $1$,
                        depending on whether the process sees $0$ and enters
                        the critical section or sees $1$ and moves to
                        Line~\ref{line-good-faa-decrement}.
                    \item In Line~\ref{line-good-faa-decrement}, $r$
                        and $d$ both drop by $1$.
                    \item In Line~\ref{line-good-faa-release}, $r$
                        and $c$ both drop by $1$.
                \end{itemize}
                Conversely, these three lines are also the only places
                where $c$ or $d$ change. Since we have already shown
                that they preserve $r=c+d$, the invariant holds
                throughout any execution of the algorithm.

                The invariant directly gives mutual exclusion: If in
                some configuration there is already a process in the
                critical section, then $r = c+d ≥ c ≥ 1$ and so no
                process can observe $r=0$ in
                Line~\ref{line-good-faa-outer-while} and enter the
                critical section.

                For deadlock-freedom we want to show that if there is
                at least one process in the entry section, $r$
                eventually reaches $0$ and stays there long enough for
                some process to see it in
                Line~\ref{line-good-faa-outer-while}. Start in any
                reachable configuration. If $c=1$, then we can run
                until the process in the critical section leaves,
                reducing $c$ to $0$. Suppose that $c$ remains $0$
                forever (if not, some process entered the critical
                section and we are done). If $r$
                never reaches $0$, every process in the entry section
                eventually gets stuck at
                Line~\ref{line-good-faa-inner-while}. But then $d=0$
                implies $r=0$, a contradiction. If instead $r$ reaches
                $0$, then in that configuration no process is in
                Line~\ref{line-good-faa-decrement}, so every process
                is either at Line~\ref{line-good-faa-inner-while} or
                Line~\ref{line-good-faa-outer-while}. Processes in
                Line~\ref{line-good-faa-inner-while} see $r=0$ and
                move to Line~\ref{line-good-faa-outer-while}; this
                does not change $r$. So eventually some process
                executes Line~\ref{line-good-faa-outer-while}, sees
                $r-0$, and
                enters the critical section.

                \paragraph*{A more general solution.} Here's an
                alternative approach that is a bit more general.
                Let $r = ∑_{i=0}^{∞} 2^i r_i$ be the value of the
                fetch-and-add register. Assign a countably infinite
                sequence $b_{p0}, b_{p1}, \dots$ of bit positions to
                each process $p$, so that no two processes' bits
                overlap. (We can do this for countably many processes
                using Cantor's pairing
                function.) Observe that (a) any process can take a
                snapshot of all the bits of all processes using
                $\FuncSty{FAA}(r,0)$, and (b) any process $p$ can update its own
                bits atomically by doing $\FuncSty{FAA}(r,δ)$ where $δ
                = ∑ 2^{b_{pj}} δ_j$ with $δ_j ∈ \Set{-1,0,1}$ being
                the desired change in $p$'s $j$-th bit. This gives an
                implementation of snapshot over single-writer 
                registers of unbounded size using a single $\FuncSty{FAA}$.

                Since unbounded single-writer registers are 
                enough to implement Lamport's bakery
                algorithm for starvation-free mutex (see
                §\ref{section-Lamport-bakery-algorithm}), we are done.

                A curious feature of this construction is that we
                don't actually need full-blown fetch-and-add, since we
                are effective only doing reads and generalized
                increments. So an unbounded generalized counter by
                itself is enough to simulate unbounded single-writer
                snapshot for any finite number of processes.
        \end{enumerate}

    \subsection{A locker object}

    The Wait-Free Locker Company has hired you as a consultant to
    evaluate the strength of its new locker object. This object,
    intended for delivery of licensed digital content to 
    subscribing consumer processes, 
    stores at most one value. It guarantees that data is not
    lost by ignoring writes to a non-empty locker, and preserves the
    licensor's valuable intellectual property rights by emptying the locker when
    it is read.

    Specifically, a $\Write$ operation inserts a value
    into the locker if none is present already; otherwise it discards
    the new value. A $\Read$ operation removes and returns any value 
    in the locker, returning $⊥$ if the locker is empty. Pseudocode
    describing these operations is given in
    Algorithm~\ref{alg-locker-problem}.

    \begin{algorithm}
        \Procedure{$\Write(\ell, v)$}{
            \Atomically{
                \lIf{$\ell = ⊥$}{$\ell ← v$}
            }
        }
        \Procedure{$\Read(\ell)$}{
            \Atomically{
                $v ← \ell$\;
                $\ell ← ⊥$\;
                \Return $v$
            }
        }
        \caption{Locker operations}
        \label{alg-locker-problem}
    \end{algorithm}

    What is the consensus number of this object?

        \subsubsection*{Solution}

        The consensus number of this object is $2$.

        To solve consensus for $n=2$, initialize the locker with some
        non-null default value, say $1$, and have each process attempt
        to read the locker after writing its input to a register. Then
        whichever process gets $1$ has won and can return its own
        input, while the other process can read the winning input from
        the winner's register as usual.

        To show we can do consensus for $n=3$, we'll use an argument
        similar to that for queues without peek.
        Consider an alleged three-consensus protocol using
        locker objects and atomic registers. Do the usual thing to get
        to a bivalent configuration $C$ with pending operations $x$
        and $y$ on the same locker object $\ell$ by processes $p$ and $q$
        such that $Cx$ is $0$-valent and $Cy$ is $1$-valent. Let $z$
        be a pending operation by the third process $r$. We have that
        $Cr$ is univalent but we don't care about this for the purpose
        of the argument.

        We want to show that for any choice of $x$ and $y$, we can
        construct an execution in which $r$ can't tell which of $x$
        and $y$ went first. As usual we know that $x$ and $y$ must be
        operations on the same object and that this object must be a
        locker.

        If $x$ and $y$ are both $\Read$ operations, then $Cxy$ and
        $Cyx$ both leave an empty locker and are indistinguishable to
        $r$.

        If $x$ is a $\Write$ and $y$ is a $\Read$, then we need to
        consider two cases depending on whether the locker is empty in
        $C$ or not. If the locker is empty, then $Cyx \sim_r Cx$,
        since in either case only $q$ knows if $y$ occurred or not.
        If the locker is not empty, then $Cxy \sim_r Cy$ since $x$ has
        no effect on a non-empty locker and only $p$ knows whether it
        occurred or not.

        If $x$ and $y$ are both $\Write$s, then we have to put in some
        effort to destroy the evidence of which went first. We can
        assume that the locker is empty in $C$, because otherwise $x$
        and $y$ are both no-ops. Configurations $Cxy$ and $Cyx$ now
        differ in the value in the locker. Run $p$ solo starting from
        either of these configurations. To decide, it
        must be able to distinguish between them, which requires
        reading the locker. Let $α$ be the sequence of operations done
        by $p$ up to and including its first read of the locker. Then
        $Cxyα \sim_r Cyxα$ since the locker is now empty and only $p$
        knows its value.

\section{Assignment 5: due Thursday 2023-11-30, at 23:59 Eastern US time}
    
    \subsection{Writable max registers}

    Consider a
    \concept{writable max register}\index{max register!writable}
    object $r$ that supports operations
    $\Read(r)$, $\Write(r,v)$ and $\WriteMax(r,v)$, where $\Read(r)$ returns
    the current value of $r$, $\Write(r,v)$ replaces
    the value of $r$ with $v$, and $\WriteMax(r,v)$ replaces the
    value of $r$ with $v$ only if $v$ is larger than the current value.

    Since this object implements an unbounded max register (just don't
    do any $\Write$ operations), the Jayanti-Tan-Toueg bound shows
    that any possible solo-terminating linearizable implementation of
    a writable max register from atomic registers requires at least
    $Ω(n)$ steps for some operation in the worst case. So let us consider a writable max register
    restricted by the following constraints:

    \begin{enumerate}
        \item The register holds only $m$ possible values $0 \dots
            m-1$, where $m$ is polynomial in $n$.
        \item At most $w$ $\Write$ operations can safely be applied to the
            register, where $w$ is also polynomial in $n$. Any
            additional $\Write$ operations have an unpredictable
            effect.
    \end{enumerate}

    Note that the limited-use restriction only applies to $\Write$
    operations. There is no limit on the number of $\Read$ or
    $\WriteMax$ operations.

    Prove or disprove: There exists a wait-free linearizable
    implementation of a restricted writable max register as defined
    above from atomic registers that uses $o(n)$ steps for any
    operation in the worst case.

        \subsubsection*{Solution}

        To implement a writable max register $r$, 
        we'll use a standard bounded max register $m_r$ to store
        lexicographically-ordered tuples
        $\Tuple{g,i,v}$ where $g$ is a generation number in $\Set{0\dots
        w}$, $i$ is a process id in the range $0 \dots n-1$, and $v$ is a value in $\Set{0 \dots m-1}$. We can do this
        by encoding $\Tuple{g,i,v}$ as $mn⋅g+m⋅i+v$, which is both bijective
        and order-preserving. To simplify the presentation of the
        algorithm, we will treat this encoding as happening
        implicitly. We assume that $m_r$ starts with its minimum value
        $0$, corresponding to the tuple $\Tuple{0,0,0}$.

        We can then increment the generation to reset the register
        in response to $\Write$ operations, and use the max-register
        property within a generation to implement $\WriteMax$.
        Pseudocode for the resulting algorithm is given in
        Algorithm~\ref{alg-writable-max-register}.

        \begin{algorithm}
            \Procedure{$\Read(r)$}{
                $\Tuple{-,-,v} ← \Read(m_r)$\;
                \label{line-writable-max-register-read-read}
                \Return $v$
            }
            \Procedure{$\Write(r,v)$}{
                $\Tuple{g,-,-} ← \Read(m_r)$\;
                $\WriteMax(m_r, \Tuple{g+1,\MyId,v})$
                \label{line-writable-max-register-write-write}
            }
            \Procedure{$\WriteMax(r,v)$}{
                $\Tuple{g,i,-} ← \Read(m_r)$\;
                $\WriteMax(m_r, \Tuple{g,i,v})$
                \label{line-writable-max-register-writemax-write}
            }
            \caption{Writable max register}
            \label{alg-writable-max-register}
        \end{algorithm}

        We assume that the number of calls to $\Write$ is bounded by
        $w$; this avoids overflow in
        Line~\ref{line-writable-max-register-write-write}.
        Under this assumption,
        the embedded max register $m_r$ takes on values in the range
        $\Set{0 \dots mnw + m(n-1) + (m-1)}$. So we can implement it with the
        standard construction of~\cite{AspnesAC2012} (see
        §\ref{section-max-register-implementation}) using $O(\log mnw)
        = O(\log n) = o(n)$ steps per operation. This gives a wait-free
        implementation that uses $o(n)$ steps per operation, since
        Algorithm~\ref{alg-writable-max-register} uses only a constant
        number of operations on $m_r$ for each operation of $r$.

        To linearize a concurrent execution,
        the intuition is that the generation and id (used as a
        tie-breaker) gives an increasing sequence of intervals, each
        consisting of a $\Write$ operation, followed by zero or more $\Read$ and
        $\WriteMax$ operations. But we need to be a little careful to
        deal with out-of-date $\Write$ and $\WriteMax$ operations that have no
        effect on $m_r$.

        Call a $\Write$ or $\WriteMax$ operation 
        \conceptFormat{punctual} if it writes a $\Tuple{g,i,v}$ where
        $\Tuple{g,i}$ is at least as big as the corresponding
        components of $m_r$, and \conceptFormat{delayed} otherwise.
        Assign linearization points as follows:
        \begin{enumerate}
            \item A $\Read$ operation is linearized at the time of its
                $\Read(m_r)$ operation in
                Line~\ref{line-writable-max-register-read-read}.
            \item A punctual $\Write$ operation is linearized at the
                time of its $\WriteMax$ operation in
                Line~\ref{line-writable-max-register-write-write}.
            \item A punctual $\Write$ operation is linearized at the
                time of its $\WriteMax$ operation in
                Line~\ref{line-writable-max-register-writemax-write}.
            \item A late $\Write$ or $\WriteMax$ operation that writes
                $\Tuple{g,v}$ is linearized just before the first
                $\Write$ operation that writes $\Tuple{g',i',v'}$ where
                $\Tuple{g',i'} > \Tuple{g,i}$. Ties between such late
                operations are broken
                arbitrarily.
        \end{enumerate}

        First let us show that each linearization point lies within
        the interval of its operation. For the first three cases, this
        is trivial. For the last case, in order for an operation $π$
        to be late, it must read a pair $\Tuple{g,i}$ from $m_r$ and
        then write $m_r$ while $m_r$ holds some pair $\Tuple{g',i'} >
        \Tuple{g,i}$. Since the only operation that changes this pair
        in $m_r$ is a $\Write$, the first such $\Write$ writes to
        $m_r$ between $π$'s $\Read$ and $\WriteMax$ operations, and
        thus within the interval of $π$. So the sequential execution
        order is consistent with the observed execution order.

        To show that this gives a correct sequential execution $S$,
        observe that we can organize $S$ as a sequence of intervals.
        The first interval consists only of zero or more $\Read$ and $\WriteMax$
        operations with initial pair $\Tuple{0,0}$, followed by zero
        or more delayed operations; subsequent intervals are similar
        but start with a $\Write$ that writes some $\Tuple{g,i,v_0}$,
        Within
        each such interval, $m_r$ starts with some value $\Tuple{g,i,v_0}$, and all
        operations that precede a $\Read$ operation have the same
        initial pair $\Tuple{g,i}$. So a $\Read$ operation within the
        interval returns the largest of the value $v_0$ supplied
        by the most recent write $\Write$ or any value
        written in the same interval by a $\WriteMax$. This
        matches the specification of the writable max register, so we
        are done.

    \subsection{Approximate vector agreement}

    Given two vectors $x$ and $y$, the \concept{Hamming
    distance}\index{distance!Hamming} between $x$ and $y$ is the
    number of positions $i$ such that $x_i ≠ y_i$.
    
    Consider the following vector agreement problem. Each process $p$
    has an input vector $x^p$ with $m$ components, where $m$ is
    typically much larger than $n$. We would like a
    protocol that gives to each process $p$ an output $y^p$,
    satisfying the following conditions, for some choice of $k$:

    \begin{description}
        \item[Wait-free termination] Each process obtains an output
            after a finite number of its own steps.
        \item[Validity] For each position $i$ and process $p$, $y^p_i$ is equal to some $x^q_i$.
        \item[Maximum distance] The Hamming distance between
            any two outputs $y^p$ and $y^q$ is at most $k$.
    \end{description}

    For example, the following might be an example of inputs and
    outputs that satisfy these constraints for $n=3$ and $k=3$:

    \begin{center}
    \begin{tabular}{ll}
        \texttt{saffron} & \texttt{sanding} \\
        \texttt{evening} & \texttt{winding} \\
        \texttt{windows} & \texttt{winning} \\
    \end{tabular}
    \end{center}

    Show that there is wait-free deterministic solution to this problem
    using atomic registers for some $k = O(n)$, where $n$ is the number of
    processes.

        \subsubsection*{Solution}

        We'll use a safe agreement object~\cite{BorowskyGLR2001} (see
        §\ref{section-safe-agreement}) for each position $i$. Since it
        takes a distinct failure to knock out each safe agreement
        object, at most $n-1$ of these objects will get stuck. So when
        a process $p$ sees return values from $m-(n-1)$ objects, it will
        combine these with its own inputs for the missing positions to
        produce its output $y^p$.

        To avoid a lot of handwaving about how the safe agreement
        objects interact, we'll break the abstraction barriers around
        their implementations and build an explicit loop for
        managing the unsafe phases. This also allows us to skip
        looping in the safe phase. Pseudocode is given in
        Algorithm~\ref{alg-vector-agreement-Hamming}.

        \begin{algorithm}
            \Procedure{\FuncSty{vectorAgreement}(x)}{
                \tcp{unsafe phase of safe agreement for each $i$}
                \For{$i ← 1$ \KwTo $m$}{
                    \tcp{propose $x_i$ at level $1$ as in safe agreement}
                    $a[p]_i ← \Tuple{1,x_i}$\;
                    $s ← \Snapshot(a)$\;
                    \eIf{$s$ contains $a[q]_i$ with level $2$}{
                        \nllabel{line-vector-agreement-Hamming-advance-test}
                        \tcp{back off}
                        $a[p]_i ← \Tuple{0,x_i}$\;
                    }{
                        \tcp{advance}
                        $a[p]_i ← \Tuple{2,x_i}$\;
                        \nllabel{line-vector-agreement-Hamming-advance}
                    }
                }
                \tcp{safe phase of safe agreement}
                \nllabel{line-vector-agreement-Hamming-last-snapshot}
                $s ← \Snapshot(a)$\;
                \For{$i ← 1$ \KwTo $m$}{
                    \eIf{$s$ contains a proposal at level $1$ for $i$}{
                        $y_i ← x_i$\;
                        \nllabel{line-vector-agreement-Hamming-use-input}
                    }{
                        $y_i ←$ some level $2$ proposal for $i$\;
                        \nllabel{line-vector-agreement-Hamming-use-proposal}
                    }
                }
                \Return $y$
            }
            \label{alg-vector-agreement-Hamming}
            \caption{Solution to vector agreement problem}
        \end{algorithm}

        We claim that this satisfies all three requirements for
        $k=2n-3$.

        Validity is easy. Any $y^p_i$ is either $x^p_i$ or a
        proposal derived from some $x^q_i$.

        Termination is also easy, since the algorithm contains no
        unbounded loops.

        For maximum distance, observe that the final snapshot $s$
        always contains at least one level $2$ proposal for each
        position, since every process that reaches this line either
        observes a level $2$ proposal in
        Line~\ref{line-vector-agreement-Hamming-advance-test} or
        writes one in
        Line~\ref{line-vector-agreement-Hamming-advance}.
        We can argue that any two such level $2$ proposals that are
        used in Line~\ref{line-vector-agreement-Hamming-use-proposal}
        are equal, because if I take a snapshot that includes a level
        $2$ proposal in position $1$ and no level $1$ proposal, any
        process working on position $i$ that has not yet written a
        level $1$ proposal will see the level $2$
        proposal and back off instead of writing a new one.  
        So the only places where $y^p$ and $y^q$ can differ are
        locations where at least one of $p$ or $q$ sees a level $1$
        proposal in its last snapshot. Suppose $p$ does the last
        snapshot first. Then there are at most $n-1$ level $1$
        proposals in $p$'s snapshot, since each process has at most
        one level $1$ proposal at a time, and $p$ has already removed
        any of its level $1$ proposals. For $q$, there are at most
        $n-2$ level $1$ proposals, since both $p$ and $q$ have left
        the unsafe phase when $q$ does its snapshot. This gives the
        claimed bound of $k≤(n-1)+(n-2) = 2n-3 = O(n)$.

        There is a much simpler solution that I did not come up with
        myself, but which was
        suggested by several people during office hours.
        Construct a multi-writer snapshot array
        $A$ with $m$ entries, initially blank. Have each process
        repeatedly take a snapshot, and if the snapshot contains a
        blank position $A[i]$, write the process's value $x_i$ to $A[i]$.
        If not, return the snapshot.

        When some process sees a full snapshot and returns,
        there are at most $n-1$ pending write operations that together
        can change at most $n-1$ positions in $A$ before all processes
        see a full snapshot and return. Since any two return values
        can disagree only in one of these $n-1$ positions, this gives
        $k=n-1=O(n)$.

\chapter{Sample assignments from Fall 2022}

\section{Assignment 1: due Thursday 2022-09-22, at 23:59 Eastern US time}

    \subsection{Leader election using broadcast}

    In the usual asynchronous message-passing model, each process can
    choose to send a message to any of its neighbors. To make our
    system super-anonymous, suppose that we eliminate the need for a
    process to know what neighbors it has by replacing these
    point-to-point channels with a
    \index{channel!broadcast}\concept{broadcast channel} where any
    message that is sent is eventually delivered to every process
    (including the sender). This is equivalent to requiring in the
    standard model that whenever a process sends a message, it sends
    $n$ copies of the message, one to each possible recipient.  As in
    the standard model, we assume that every copy of a message is
    delivered after at most $1$ time unit, but by default impose no
    other constraints on the time at which each copy of a message is
    delivered.

    We would like to solve leader election in this model, under
    various assumptions. By leader election, we mean a protocol in
    which exactly one process eventually sets its $\Leader$ bit to
    $1$. For each of the conditions below, give an algorithm for
    solving leader election, prove its correctness, and compute its
    message complexity and running time; or prove that no such
    algorithm is possible.

    \begin{enumerate}
        \item An anonymous system in which all processes run the same
            code and do not have unique IDs.
        \item A uniform system with IDs, where uniformity means that
            the code for each process depends only on its ID and not
            on the size of the system.
        \item A non-uniform system with IDs, where the processes
            know $n$.
        \item A uniform system with IDs, but where the broadcast
            channel is replaced by an \concept{ordered
            broadcast}\index{broadcast!ordered}
            channel that guarantees for each pair of messages $m_1$
            and $m_2$, that if $m_1$ is sent before $m_2$, each
            process receives $m_1$ before it receives $m_2$.
    \end{enumerate}

        \subsubsection*{Solution}

        For computing message complexity, there is an ambiguity in the
        problem description: does sending a single broadcast count as
        $n$ messages or one message? Below, we assume a broadcast
        counts as $n$ messages, but one message is also a reasonable
        interpretation, so either assumption is acceptable as long as
        it is clear.

        \begin{enumerate}
            \item Not possible. Construct a synchronous execution in
                which we alternate between having all $n$ processes
                take steps until each sends a message then
                having all $n^2$ messages delivered.
                The usual symmetry argument shows that each process
                updates to the same state and sends the same messages
                in each round, so either no process ever declares
                itself the leader, or they all do.
            \item Not possible. Consider a system with two processes
                $p_1$ and $p_2$. Run $p_1$ but do not deliver any of
                its messages to $p_2$. Since this execution is
                indistinguishable from an execution in which $p_1$ is
                the only process, it must eventually set its
                $\Leader$ bit. Now run $p_2$ without delivering any of
                its messages to or from $p_1$. It also must eventually set
                its $\Leader$ bit. We can now satisfy admissibility by
                delivering all the undelivered messages, but it's too
                late: we already have two leaders.
            \item Possible. Have each process broadcast its ID then
                wait to collect $n$ IDs. The process with the smallest
                ID among these $n$ IDs sets its $\Leader$ bit. Message
                complexity is $n^2$ and time complexity is $1$.
            \item Possible. Have each process broadcast its ID. If a
                process receives its own ID before any others, it sets
                its leader bit. Since the broadcast channel is
                ordered, only the first process to do a broadcast
                wins. Message complexity is $n^2$ and time complexity
                is $1$.
        \end{enumerate}

    \subsection{Discovery by flooding}

    In the usual message-passing model, it is assumed that every
    process has the ability to communicate directly only with its
    immediate neighbors in the communication graph. For this problem
    we will consider model closer to the current Internet, where (in
    principle) any machine in the network can send a message to any
    other machine, provided it knows the other machine's IP address.

    For each process $p_i$, let $S_i$ be the set of processes $p_j$
    such that $p_i$ knows $p_j$'s address, and let $G = (V,E)$ be the
    directed graph whose vertices $V$ are all processes and which contains an
    edge $ij ∈ E$ for each pair $p_i, p_j$ such that $p_j ∈ S_i$ in
    the initial configuration. Assume that $p_i$ knows about itself,
    so that $G$ includes all the self-loops $ii$.

    We'd like the processes to exchange messages until this graph is
    complete, with an edge for every pair of processes.
    The protocol is simple: 
    In each (synchronous) round, every process $p_i$ sends its current list $S_i$ to every
    process in $S_i$, then updates $S_i$ to be the union of every
    message it receives.

    Show that if the initial graph $G$ is weakly-connected, 
    then after at most $O(\log n)$ rounds,
    this protocol reaches a configuration
    where $S_i = V$ for all $i$.

        \subsubsection*{Solution}

        For each $r$, let $S^r_i$ be the value of $S_i$ after
        $r$ rounds of messages.
        Define $G^r =
        (V, E^r)$ as the graph where $V$ is the set of processes and
        $ij ∈ E^r$ if and only if $p_j ∈ S_i$.
        From the definition we have $G^0 = G$.

        It is convenient to work with undirected graphs. Let $H^r$ be
        the \emph{undirected} graph that contains an edge $ij$ if and
        only if $ij$ and $ji$ are both edges in $G^r$. Note that
        $H^r$ is always a subgraph of $G^r$.

        Claim: $H^1$ is connected. Proof: For each edge $ij ∈
        G^0$, $p_i$ sends $p_i ∈ S_i$ to $p_j$, so $p_j$ updates $S^1_j$
        to include $ji$. So $H^1$ contains the undirected version of
        $G^0$ as a subgraph. Since $G^0$ is weakly connected, $H^0$ is
        connected.

        Because $H^1$ is connected, there is a path in $H^1$ between
        any two nodes, and the diameter $d(H^1)$ of $H^1$ is at most
        $n-1$. We now show that each round of the
        protocol reduces the diameter of $H$ by roughly half.

        Claim: If $uv$ and $vw$ are both edges in $H^r$, then $uw$ is
        an edge in $H^{r+1}$. Proof: From the definition of $H^r$, we
        have $\Set{u,w} ⊆ S^r_v$. So both of $u$ and $w$ add the other
        upon receiving $S^r_v$ from $v$.

        Now consider arbitrary $u,v ∈ H^r$ with $d(u,v) = m$. This
        means that there is a path $u = u_0 u_1 \dots u_m = v$ in
        $H^r$. From the claim, we have that $u = u_0 u_2 u_4 \dots
        u_m = v$ is a path in $H^{r+1}$ if $m$ is even, and $u = u_0
        u_2 u_4 \dots u_{m-1} u_m = v$ is a path in $H^{r+1}$ if $m$
        is odd. In either case we have $d_{H_{r+1}}(u,v) ≤ \ceil{m/2}$.
        It follows that $d(H^{r+1}) = \max_{u,v} d_{H_{r+1}}(u,v) ≤
        \max_{u,v} \ceil{d_{H_r}(u,v)/2} ≤ \ceil{d(H^r)/2}$.

        A simple induction on $r$ shows that if $d(H^1) ≤ 2^k$, then
        $d(H^r) ≤ \min(1, 2^{k-r+1})$. In particular for $r =
        \ceil{\lg n}+1$ we have $d(H^r) ≤ 1$, which shows that there
        is an edge between every pair of nodes in $H^r$. Since $H^r$
        is defined to contain $ij$ if and only if $ij$ and $ji$ are
        edges in $G^r$, it follows that $G^r$ is complete for
        $r=\ceil{\lg n} + 1 = O(\log n)$.

\section{Assignment 2: due Thursday 2022-10-06, at 23:59 Eastern US time}

    \subsection{Maximum consensus}

    Suppose you have a synchronous message-passing system with $n$
    processes that may experience up to $f$ crash failures.
    Each process $p_i$ starts with an input $x_i$ that is an arbitrarily-large
    natural number. What is the minimum number of rounds needed to
    solve each of the following problems in the worst case as a
    function of $f$? In each case, provide
    matching upper and lower bounds for sufficiently large $n$.

    \begin{enumerate}
        \item Each non-faulty process $p_i$ outputs a value $y_i$ such
            that (a) $y_i = x_j$ for some process $p_j$, and (b) $y_i
            ≥ x_j$ for all non-faulty processes $p_j$.
        \item As above, but in addition $y_i = y_j$ for all non-faulty
            processes $i$ and $j$.
    \end{enumerate}

        \subsubsection*{Solution}

        \begin{enumerate}
            \item One round is enough. Each process sends $x_i$ to all
                processes (including itself), and each process returns $y_i$ equal
                to the largest of all $x_j$ it received.
                
                Condition (a) follows immediately from $y_i$ being
                equal to some $x_j$. For (b), if $p_j$ is non-faulty,
                $p_i$ receives $x_j$ from $p_j$, so it returns either
                $x_j$ or some larger $x_{j'}$.

                For the lower bound, if a protocol uses zero rounds,
                then no messages are sent. If process $p_i$ decides
                $x_i$ in some execution, then for $n≥2$ there exists
                an execution indistinguishable to $p_i$ from this one,
                where
                some non-faulty $p_j$ with $j≠i$ has $x_j > x_i$,
                violating (b).
                Similarly,
                if $p_i$ decides a value $y_i ≠ x_i$, there
                exists an indistinguishable
                execution where no process has $y_i$ as its
                input value, violating (a).

            \item Here we need $f+1$ rounds. For the lower bound we
                can reduce from synchronous consensus and apply
                Dolev-Strong
                (\cite{DolevS1983}; see also §\ref{section-synchronous-agreement-lower-bound}).
                To solve consensus using this problem, have each
                process $p_i$ decide on $y_i$. This satisfies validity
                from (a) and agreement from the added condition that
                $y_i = y_j$ for all non-faulty $i$ and $j$. So if we have an
                algorithm that uses less than $f+1$ rounds,
                we get an algorithm for consensus that also uses less
                than $f+1$ rounds, contradicting the known lower bound
                for consensus.

                For the upper bound, we can use the flooding mechanism
                from Dolev-Strong
                (\cite{DolevS1983}; see also §\ref{section-synchronous-agreement-flooding}). This
                guarantees that after $f+1$ rounds, every non-faulty
                process obtains the same set $S$ of input values,
                which includes the inputs of all non-faulty processes.
                So taking $\max S$ gives a common return value for all
                non-faulty processes that satisfies both (a) and (b).
        \end{enumerate}

    \subsection{Colorful Byzantine agreement}

    Consider a synchronous system with $n$ processes, each of which is
    labeled with one of four colors: red, green, blue, or yellow. The
    processes have unique IDs that are known to all the other
    processes, and all processes know which processes have which
    color.

    The adversary can turn as many processes as it likes
    Byzantine, provided that all the processes corrupted by the
    adversary are of the same color.

    Prove or disprove: It is possible to solve Byzantine agreement in
    this system for any number of processes $n ≥ 4$ using any
    assignment of colors that gives at least one process of each
    color.

        \subsubsection*{Solution}

        Possible. The idea is to reduce the problem to four processes
        of which at most one is Byzantine, then use any Byzantine
        agreement algorithm that tolerates $f < n/3$ Byzantine faults
        to solve agreement. One possibility would be exponential
        information gathering~\cite{PeaseSL1980} (see
        §\ref{section-Byzantine-exponential-information-gathering}),
        since we don't particularly care about anything but fault
        tolerance and $4$ is a constant anyway.

        For each color group, let the process with maximum ID
        represent the group (this does not require any rounds of
        communication under the assumption that all IDs are known to
        all processes). We then have four representatives that can
        execute EIG in $f+1 = 2$ rounds to solve Byzantine agreement among
        themselves. Each representative then broadcasts its decision
        value to all $n$ processes, and each non-faulty process
        decides on the value broadcast by the majority of
        representatives. (Note that it is not enough for a process to
        follow its own representative, because there may be non-faulty
        processes within the faulty group.)

        We would like to show that this algorithm solves Byzantine
        agreement for all $n$ processes. Termination is immediate. For
        validity, if all non-faulty processes have the same input $v$,
        then so do the three non-faulty representatives; validity in
        the four-process protocols implies that all three non-faulty
        representatives broadcast this value and thus all non-faulty
        processes decide it. Agreement is similar: because all three
        non-faulty representatives agree on the same value $v$, each
        non-faulty process will see a majority for $v$ and decide on
        $v$.

\section{Assignment 3: due Thursday 2022-10-27, at 23:59 Eastern US time}

    \subsection{A census of failure}

    Suppose we have an asynchronous message passing system with crash
    failures, and we want to implement an oracle that returns a count
    of the number of processes that haven't crashed yet. Define a
    \concept{census protocol} to be a protocol that stores at every
    point in the execution a value $c_i$ at each process $p_i$, 
    such that (a) $c_i ≥ n-f$ always, where $n$ is the number of
    processes in the system and $f$ is the number of processes that
    have crashed so far, and (b) once $f$ converges to a fixed value,
    $c_i$ eventually converges to $n-f$. These properties should hold
    for every non-crashed process $p_i$.

    Prove or disprove each of the following statements. In each case
    assume that we have an asynchronous message-passing system with
    a complete communication graph, deterministic processes, and 
    crash failures modeled as explicit crash events, and that any
    implementation must work for arbitrarily large $n$ (which is known
    to the processes).

    \begin{enumerate}
        \item It is possible to implement a census protocol without
            using a failure detector.
        \item It is possible to implement a census protocol using an
            eventually perfect ($◇P$) failure detector.
        \item It is possible to implement a census protocol using a
            perfect ($P$) failure detector.
    \end{enumerate}

        \subsubsection*{Solution}

        \begin{enumerate}
            \item Disproof: With no failure detector, consider two
                executions of a two-process system.
                In one execution, process $p_1$ takes no steps
                because it crashes immediately.
                In the other, $p_1$ takes no steps for a very long
                time.

                If $p_2$ eventually sets $c_2$ to $1$, this violations
                $c_2 ≥ n-f$ in the execution where $p_1$ has not
                crashed.

                If $p_2$ does not eventually set $c_2$ to $1$, this
                violations $c_2$ converging to $n-f$ in the execution
                where $p_1$ has crashed.
            \item Disproof: Consider the two executions in the
                previous case, and suppose that $◇P$ correctly
                suspects $p_1$ throughout the crash execution and
                incorrectly suspects $p_1$ in the no-crash execution.

                If $p_2$ sets $c_2$ to $1$, it violates (a) again in
                the no-crash execution, and
                afterwards we can both wake up $p_1$ and have $◇P$
                stop suspecting $p_1$.

                If $p_2$ doesn't set $c_2$ to $1$, it violates (b) in
                the crash execution.
            \item Proof: Recall that $P$ eventually permanently
                suspects every crashed process and never suspects a
                process before it crashes. So have each process $p_i$
                set $c_i$ to $n-f_i$, where $f_i$ is the number of
                processes that $p_i$'s instance of $P$ currently
                suspects. Because $P$ only suspects crashed processes,
                $f_i ≤ f$ and thus $c_i = n-f_i ≥ n-f$, satisfying
                (a). Because $P$ eventually permanently suspects all
                crashed processes, once every process that will crash
                has crashed, $P$ will eventually suspect all of them
                at each $p_i$. This gives $f_i = f$ and $c_i = n-f_i =
                n-f$.
        \end{enumerate}

    \subsection{Distributed shared memory with Byzantine servers}

    Consider the following modification to the usual asynchronous
    message-passing model:
    \begin{enumerate}
        \item There are $m$ clients, and any of them may crash at any
            time.
        \item There are $n$ servers. These do not crash, but up to $f$
            of them may be Byzantine.
    \end{enumerate}

    We would like to have a linearizable implementation of a single-writer
    multi-reader register in this model, where the single writer and
    multiple readers are all clients, and any operation by a
    non-faulty client eventually finishes. Show that there is a constant
    $c$ such that this is possible for $n ≥ cf+1$.
    
        \subsubsection*{Solution}

        We can do this when $n ≥ 4f+1$
        by modifying ABD (see §\ref{section-ABD}).

        To make things easier, we will assume that the honest servers
        keep track of every timestamp-value pair $\Tuple{t,v}$ they
        have every received, instead of just the one with the maximum
        timestamp. Upon receiving a $\Read(u)$ message, the server
        responds with its entire list (including $\Tuple{t,v}$ if it
        wasn't there already).

        To perform a write operation with value $v$, the writer
        increments its local timestamp $t$, sends $\Write(t,v)$ to all
        servers, and waits for $n-f$ acknowledgments.

        To perform a read operation, a reader sends $\Read(u)$ to all
        servers, waits for $n-f$ replies, and then chooses a pair
        $\Tuple{t,v}$ that (a) is sent by at least $f+1$ servers, and
        (b) has the largest $t$ out of all such pairs. If there is no
        pair sent by $f+1$ servers, the reader returns the default
        initial register value $⊥$. Otherwise, it sends $\Write(t,v)$ to all
        servers, waits for $n-f$ acknowledgments, then returns $v$.

        To show this gives a linearizable implementation of a
        single-writer multi-reader register, we will largely follow
        the original proof for ABD, constructing an explicit
        linearization of any complete execution. We start with a
        simple invariant:

        \begin{lemma}
            \label{lemma-Byzantine-server-ABD-validity}
            Let $\Tuple{t,v}$ be a pair that is (a) in some honest server's list, (b) in
            a $\Write(t,v)$ message, or (c) adopted by a reader. Then
            $\Tuple{t,v}$ was previously sent by the writer.
        \end{lemma}
        \begin{proof}
            It is easy to see that if (b) and (c) hold in some
            configuration, then (a) and (b) hold in any successor
            configuration, since we can only add a tuple to an honest
            server if it was in a $\Write(t,v)$ message and we can
            only generate a $\Write(t,v)$ message if $\Tuple{t,v}$ is
            sent by the writer or was previously adopted by a reader.
            To show that (c) holds, observe that if a reader adopts
            $\Tuple{t,v}$, it must first receive it from $f+1$
            servers. At least one of these servers is honest, so (a)
            applies.
        \end{proof}

        For any operation $a$, let $t(a)$ be the timestamp of the pair
        $\Tuple{t,v}$ that $a$ sends in its $\Write(t,v)$ messages.
        Observe that if $a$ finishes, then it receives
        acknowledgments from $n-f$ servers of which at least $n-2f$
        are not faulty: this implies that by the time $a$ finishes, at
        least $n-f$ servers have $\Tuple{t,v}$ in their lists. If $b$
        is a read operation with $a <_H b$, then $b$ receives
        responses from at least $n-3f$ of these servers. With $n ≥
        4f+1$, this is at least $f+1$. So $b$ either adopts
        $\Tuple{t,v}$ or adopts some other $\Tuple{t',v'}$ with $t' >
        t$. So whenever $a <_H b$, $t(a) ≤ t(b)$.

        To define $<_S$, $a$ before $b$ if (1) $t(a) <
        t(b)$ (which we've just shown is consistent with $<_H$); or
        (2) $t(a) = t(b)$, $a$ is a write, and $b$ is a read (which is
        consistent with $<_H$ by
        Lemma~\ref{lemma-Byzantine-server-ABD-validity}); or (3)
        $t(a)=t(b)$, both operations are reads, and $a <_H b$
        (definitely consistent with $<_H$!). Then extend the resulting
        partial order to a total order. As in the original ABD
        algorithm, we get a sequence of blocks of operations where all
        operations in a block have the
        same $\Tuple{t,v}$ pair, and the first operation in each
        block (except possibly the first block) is a write of $v$ and the
        rest are reads that return $v$. So the resulting sequential
        execution is consistent both with $H$ and the specification of
        a register, and we have shown that the implementation is
        linearizable.

\section{Assignment 4: due Thursday 2022-11-10, at 23:59 Eastern US time}

    \subsection{Arithmetic registers}

    An \concept{arithmetic register} holds an integer value and
    supports operations $\Read()$, $\FuncSty{add}(x)$, and
    $\FuncSty{multiply}(x)$, where $\Read()$ returns the current value
    of the register; $\FuncSty{add}(x)$ updates the current value by
    adding $x$ to it; and $\FuncSty{multiply}(x)$ updates the current
    value by multiplying it by $x$. The $\FuncSty{add}$ and
    $\FuncSty{multiply}$ operations do not return a value.

    Suppose that arithmetic registers come in two flavors: a
    \conceptFormat{signed} arithmetic register can hold any integer value and
    allows any integer argument to $\FuncSty{add}$ or
    $\FuncSty{multiply}$, while an \conceptFormat{unsigned} arithmetic
    register holds only non-negative integer values and allows only
    non-negative integer arguments.

    Prove or disprove: There exists a deterministic, wait-free,
    linearizable implementation of a signed arithmetic register from
    unsigned arithmetic registers and ordinary atomic registers.

        \subsubsection*{Solution}

        Proof: We'll show that an unsigned arithmetic register
        implements consensus for any fixed number of processes $n$,
        then use universality of consensus to get an implementation of
        a signed arithmetic register.

        The consensus construction follows a similar argument of
        Ellen~\etal~\cite{EllenGSZ2020} for registers supporting
        multiplication and decrement, but we have to be a little
        careful to only use non-negative values. Start with a single
        unsigned arithmetic register $r$ initialized to $1$. A process
        with input $0$ applies $\FuncSty{add}(1)$ to $r$. A process
        with input $1$ applies $\FuncSty{multiply}(n+2)$ to $r$, where
        $n$ is the number of processes.

        Consider some sequence of operations $s$, and let $a_i$ be the
        number of calls to $\FuncSty{add}(1)$ in $s$ that
        are followed by exactly $i$ calls to $\FuncSty{multiply}(n+2)$
        in $s$. Let $k$ be the total number of calls to
        $\FuncSty{multiply}(n+2)$ in $s$. Then it is easily shown by
        induction on the length of $s$ that the value of the
        register at the end of $s$ is given by 
        $r = (a_k+1)(n+2)^k + ∑_{i=0}^{k-1} a_i (n+2)^i$.

        Since each coefficient in this expansion is at most $n+1$, we
        can recover the expansion uniquely from $r$. The value $a_k$
        will be nonzero if and only if the first operation on the
        register was $\FuncSty{add}(1)$. Since this holds for any
        sequence of operations, any process reading the register 
        can determine whether an adder or multiplier went first, and
        so all processes can return $0$ in the first case and $1$ in
        the second.

        Now apply Herlihy's universal construction to implement a
        signed arithmetic register.

        (With some tinkering, we can even drop the requirement for
        atomic registers by showing that they can be implemented from
        unsigned arithmetic registers, but this is not required by the problem.)
    
    \subsection{Counting to two}

    Let us say that we can count to $k$ with $m$ registers for $n$
    processes if there is a deterministic, wait-free, linearizable,
    one-shot implementation of a $k$-bounded counter from $m$
    registers that works for $n$ processes. A $k$-bounded register
    starts at $0$, has a read operation that returns its current value, and has an increment
    operation that increases the value by $1$ unless it
    is already $k$. It is one-shot if each process is only allowed to
    call the increment operation at most once.

    It is easy to show that we can count to $1$ for any number of
    processes using $1$ register: start with a $0$ in the register,
    and implement an increment by writing $1$. It is also
    straightforward to count to any value $k$ for $n$ processes using
    $n$ registers: give a register to each process; implement
    increment by writing $1$ to my register; and sum over a collect to
    get a number of increments $s$, returning $\min(k,s)$ to enforce
    $k$-boundedness.
    
    Prove or disprove: We can count to $2$ with $3$ registers for $4$
    processes.

        \subsubsection*{Solution}

        Proof: In fact, we can do this for any $n$, not just $n=4$.

        Use two of the three registers to build a splitter
        (Algorithm~\ref{alg-splitter}). The third
        register, initially $0$, will be a flag indicating at least
        two increments.

        To do an increment: Try to win the splitter. If I win, I am
        done. If not, write $1$ to the flag.

        To do a read: Check $\SplitterDoor$. If it's open, assume no
        increments have finished yet and return $0$. If it's closed,
        use the flag to decide whether to return $1$ or $2$.

        Code is given in Algorithm~\ref{alg-2-bounded-counter-3}.

        \begin{algorithm}
\SharedData\\
atomic register $\SplitterRace$, big enough to hold an ID, initially $⊥$\\
atomic register $\SplitterDoor$, big enough to hold a bit, initially
$\SplitterOpen$\\
            atomic register $\DataSty{flag}$, big enough to hold a
            bit, initially $0$\\
            \Procedure{$\FuncSty{increment}(\Id)$}{
                $\SplitterRace ← \Id$\;
                \If{$\SplitterDoor = \SplitterClosed$}{
                    $\DataSty{flag} ← 1$\;
                }
    $\SplitterDoor ← \SplitterClosed$\;
    \If{$\SplitterRace ≠ \Id$}{
        $\DataSty{flag} ← 1$\;
    }
}

            \Procedure{\Read}{
                \uIf{$\SplitterDoor = \SplitterOpen$}{
                    \Return $0$\;
                }
                \uElseIf{$\DataSty{flag} = 0$}{
                    \Return $1$\;
                }
                \Else{
                    \Return $2$\;
                }
            }
            \caption{Counting to $2$ with a splitter}
            \label{alg-2-bounded-counter-3}
        \end{algorithm}

        Since each operation does at most a constant number of steps,
        this is clearly wait-free. But we need to show that it is
        linearizable. We'll use linearization points.

        For a read that returns $0$: linearize it at the point where
        it reads $\SplitterDoor$ and sees $\SplitterOpen$.

        For any other read: linearize it at the point where it reads
        $\DataSty{flag}$.

        This orders all reads that return $0$ (when the door
        is still open) before all reads that return $1$ or $2$; and
        orders all reads that return $1$ (when the flag is not yet
        set) before all reads that return $2$. So now we just need to
        fit in some increments to justify the changes.

        If there is an increment $I_1$ that wins the splitter and does
        not set the flag, assign its linearization point to the step
        where the door closes (whether $I_1$ closes the door or not).
        Then $I_1$ linearizes between all reads that return $0$ and
        all reads that return $1$ or $2$.
        Because every other increment loses the splitter, every other
        increment sets the flag; make each such increment's linearization
        point be the step where it sets the flag. The first such
        increment $I_2$ linearizes between all reads that return $0$
        or $1$ and all reads that return $2$.

        If no increment wins the splitter, then no increment finishes
        before setting the flag, at the point where the flag is first
        set there are at least two increments in progress and none
        have already finished. Let $I_1$ be one of these increments
        that starts before the door closes, and assign its
        linearization point to the step where the door closes. Let
        $I_2$ be any other increment in progress when the flag is
        first set, and assign its linearization point to when the flag
        is first set. Assign the linearization points of any other
        increment anywhere during its execution interval that is after
        $I_2$'s.  Again we get one increment linearized between the
        $0$ and $1$ reads, and at least one between the $1$ and $2$
        reads. We are done.

\section{Assignment 5: due Monday 2022-12-05, at 23:59 Eastern US time}

    \subsection{A hidden counter}

    Consider a system with $n$ processes; $n$ single-writer
    multi-reader atomic registers, one for each process; and a counter
    that can be incremented by any process but that can be read by
    nobody. We would like a wait-free protocol that results in the
    counter being incremented by at least $f(n)$ using as few total
    operations, across all processes, as possible, counting both
    increment operations on the counter and read and write operations on the
    registers.

    In this context, wait-freedom means that a process can only return
    when it is sure that $f(n)$ increments have been done, which may,
    in the worst case, require it to do all $f(n)$ increments by
    itself. A process that returns is scheduled for no more
    operations.

    \begin{enumerate}
        \item Show that $O(n^2)$ total operations are sufficient to
            increment the counter at least $n^2$ times.
        \item Show that $T(n)$ total operations are sufficient to
            increment the counter at least $n$ times, for some $T(n) =
            o(n^2)$.
    \end{enumerate}

        \subsubsection*{Solution}

        \begin{enumerate}
            \item We'll have each process $i$ alternate between
                incrementing the counter and writing out the total
                number of increments it has done so far to its
                register $r_i$.

                After $n$ increments, the process will read all the
                registers $r_j$, and if $∑ r_j ≥ n^2$, return.

                This gives an amortized cost of $3$ operations per
                increment, so as long as we only do $O(n^2)$
                increments, we are fine. To show this, observe that
                once the total value in the registers exceeds $n^2$,
                each process does at most $n$ increments before it
                re-reads the registers, for at most $n^2$ extra
                increments.

            \item There are a number of ways to do this. One simple
                approach is to divide the processes into groups of
                size $k = \sqrt{n}$, and have each group independently do
                at least $n = k^2$ increments using the algorithm from the
                previous case. This costs $O(n)$ operations per group,
                or $O(n^{3/2})$ operations total.
        \end{enumerate}

    \subsection{One register to rule them all}

    \emph{This problem was nearly identical to
    Problem~\ref{problem-randomized-consensus-with-one-max-register}
    from 2020 and has been withdrawn. Any submission for this
    assignment will be graded as if a complete solution to this
    problem had been provided.}

\chapter{Sample assignments from Spring 2020}

\section{Assignment 1: due Wednesday, 2020-09-23, at 5:00pm Eastern US time}

    \subsection{A token-passing game}

    \newData{\HasToken}{hasToken}

    Suppose we have an asynchronous bidirectional message-passing
    network in the form of a connected graph, where initially $m$ of
    the $n$ nodes possess a token, represented by a local variable
    $\HasToken$ being set to true. We'd like to be able to move the
    tokens around, while preserving the total number of tokens.

    \begin{enumerate}
        \item Show that no algorithm that allows tokens to move can
            guarantee that there are exactly $m$ tokens in any
            reachable configuration.
        \item Give an algorithm that satisfies the following two
            properties, starting with a configuration with $m$ tokens:
            \begin{enumerate}
                \item \textit{Safety:} In any reachable configuration,
                    there are at most $m$ tokens. You should give an
                    explicit invariant that implies this, and show
                    that any transition of your algorithm preserves
                    the invariant.
                \item \textit{Liveness:} From any reachable
                    configuration $C_0$, for any subset $S$ of the
                    processes with $\card{S} = m$,
                    there exists an execution
                    starting in $C_0$ that ends with a configuration
                    in which every process in $S$ has a
                    token.\footnote{Strictly speaking, this is a lot
                    weaker than the usual definition of liveness,
                    because it effectively assumes that the adversary
                    is cooperating with us. In retrospect I should
                    have written this as ``for any admissible
                    adversary strategy, there is a sequence of nondeterministic
                    choices by the algorithm that causes the execution
                    to reach a desired configuration.'' But I didn't
                    write this, and so it's fine to answer the problem
                    I did write.}
            \end{enumerate}
                    To keep things simple, you may assume that the processes can make
                    non-deterministic choices. For example, a process
                    $p$ might choose arbitrarily between sending a
                    message to a neighbor $q$ or to a different
                    neighbor $r$, and each choice leads to a different
                    possible execution.
    \end{enumerate}

        \subsubsection*{Solution}

        \newData{\TakeThis}{takeThis}
        
        \begin{enumerate}
            \item Suppose that we are preserving total tokens.
                Consider some transition between configurations
                $C_1$ and $C_2$. 
                If some process switches $\HasToken$ from $1$
                to $0$ between these configurations, then some other
                process must switch $\HasToken$ from $0$ to $1$. But
                the definition of delivery events in the asynchronous
                message-passing model only allows one process at a
                time to change its state. It follows that no process
                can change $\HasToken$ from $1$ to $0$ in any
                transition, so tokens can't move.
            \item 
        Consider the following algorithm:
        \begin{itemize}
            \item At any time, a process with $\HasToken = 1$ may send
                a message $\TakeThis$ to any of its neighbors and set
                $\HasToken = 0$.
            \item A process that receives $\TakeThis$ when $\HasToken
                = 1$ sends $\TakeThis$ to any of its neighbors. 
                A process that receives $\TakeThis$ when $\HasToken =
                0$ may either set $\HasToken = 1$ or send $\TakeThis$
                to any of its neighbors.

                Let us show that this has the desired properties:
                \begin{enumerate}
                    \item \textit{Safety:} Our invariant will be that
                        the sum of the number of processes with
                        $\HasToken = 1$ plus the number of $\TakeThis$
                        messages in transit will be $m$.
                        
                        The invariant holds in
                        the initial configuration because there are
                        exactly $m$ processes with $\HasToken = 1$ and
                        no message in transit.

                        It is preserved by transitions, because in
                        each possible transition, either:
                        \begin{enumerate}
                            \item Some process changes $\HasToken = 1$
                                to $\HasToken = 0$ and generates a
                                $\TakeThis$ message;
                            \item Some process changes $\HasToken = 0$
                                to $\HasToken = 1$ while consuming a
                                $\TakeThis$ message; or
                            \item Some process consumes a $\TakeThis$
                                message but generates a new
                                $\TakeThis$ message.
                        \end{enumerate}
                        In each case, the total number of tokens plus
                        messages is preserved.
                    \item \textit{Liveness:}
                        For any configuration $C$, let $T(C)$ be the
                        set of processes with $\HasToken = 1$.
                        We will argue that if $T(C) ≠ S$, there exists
                        a partial execution that increases $\card{T(C)
                        ∩ S}$ by $1$. First pick some $p ∈ S ∖ T(C)$.
                        Now consider two cases:
                        \begin{enumerate}
                            \item If there is at least one $\TakeThis$
                                message $t$ in transit, apply the
                                following strategy. Deliver $t$. If
                                the recipient of $t$ is $p$, set
                                $p.\HasToken = 1$. If not, have the
                                recipient send $\TakeThis$ to some
                                neighbor that is closer to $p$ than it
                                is. Repeat this process until a
                                $\TakeThis$ message reaches $p$.
                            \item If there is no $\TakeThis$ message
                                in transit, generate a $\TakeThis$
                                message at some $q ∈ T(C) ∖ S$ while
                                setting $q.\HasToken$ to $0$. Then
                                apply the previous case.
                        \end{enumerate}
                        For any configuration with $T(C) ≠ S$, at
                        least one of these two conditions will hold
                        because of the safety property.

                        Each partial execution defined above
                        increases $\card{T(C) ∩ S}$ by one, and we can
                        only increase $\card{T(C) ∩ S}$ at most $m$
                        times because $\card{S} = m$, so
                        after at most $m$ of these partial executions 
                        we reach a configuration with
                        $T(C) = S$.
                \end{enumerate}
        \end{itemize}
        \end{enumerate}

    \subsection{A load-balancing problem}
    \label{section-problem-load-balancing}

    Consider a two-way message-passing ring with $n = mk$ nodes, where
    $m>1$ and $k$ is odd. Nodes at positions $0, k, 2k, \dots, (m-1)k$ are
    initially marked as leaders, while nodes at other positions are
    followers. All nodes have a sense of direction, and can
    distinguish their left neighbor from their right, but they do not
    have any other ID information.

    Algorithm~\ref{alg-problem-load-balancing} is intended to allow
    the leaders to recruit followers.
    It is not hard to
    show that every follower eventually adds itself to a tree of
    parent pointers rooted at some leader.
    We would like all of these trees to contain roughly the same
    number of nodes.

    \newData{\Recruit}{recruit}

    \begin{algorithm}
        \Initially{
            \eIf{I am a leader}{
                $\Parent ← \Id$\;
                send \Recruit{} to both neighbors\;
            }{
                $\Parent ← ⊥$\;
            }
        }
        \UponReceiving{\Recruit{} from $p$}{
            \If{$\Parent = ⊥$}{
                $\Parent ← p$\;
                send \Recruit{} to my neighbor who is not $p$\;
            }
        }
        \caption{Recruiting algorithm for Problem
        \ref{section-problem-load-balancing}.}
        \label{alg-problem-load-balancing}
    \end{algorithm}

    \begin{enumerate}
        \item Suppose we run this algorithm in a synchronous system.
            What is the minimum and maximum possible size of a tree?
        \item Suppose instead we run the algorithm in an asynchronous
            system. Now what is the minimum and maximum possible size
            of a tree?
        \item Give an algorithm for the asynchronous version of this
            model that guarantees that all trees are the same size.
    \end{enumerate}

        \subsubsection*{Solution}

        \begin{enumerate}
            \item In a synchronous execution, we can prove by
                induction that for each $t$ with $0≤t≤\frac{k-1}{2}$,
                and each $0≤i≤m-1$,
                each node at position $ik±t$ joins the tree rooted at $ik$ at
                time $t$. This puts exactly $k$ nodes in each tree.
            \item In an asynchronous execution, by rushing messages
                from $0$, we can recruit all nodes in the range
                $[-k+1,k-1]$ to the $0$ tree before any other messages
                are delivered. Conversely, each leader $ik$ can't
                recruit nodes $(i-1)k$ or $(i+1)k$, because these are
                leaders.
                So the maximum size of any tree is $2k-1$.

                For the minimum size, suppose we rush all messages
                from nodes $k$ and $(m-1)k$. Then nodes $1$ and $m-1$
                are recruited into the trees rooted at these nodes
                before either message from $0$ is delivered. This
                shows that there are executions with a minimum tree
                size of $1$.
            \item The easiest fix may be to have each leader initially send just one
                $\Recruit$ message to the right. For each $i$, this recruits all agents
                $ik,\dots,ik+(k-1)$ to a tree of size $k$ rooted at $ik$.
        \end{enumerate}

\section{Assignment 2: due Wednesday, 2020-10-07, at 5:00pm Eastern US time}

\subsection{Synchronous agreement with limited broadcast}

Suppose that we are given a synchronous message passing system on a
complete network in which messages are replaced by $k$-way broadcasts,
where the recipient is replaced by a recipient list of up to $k$
processes.
Suppose further that when a process crashes in round $r$,
each of its round-$r$ messages is either delivered to all of the
processes on the message's recipient list or to none of them.
A process can send as many messages as it likes to as many 
groups as it likes, but if it crashes in some round, any subset of the
messages sent in that round may be lost.

Show that for it is possible to solve agreement in this model in
$O(f/k)$ rounds, assuming $n > f$.

    \subsubsection*{Solution}

    We'll use the flooding algorithm of Dolev and
    Strong~\cite{DolevS1983} (see
    §\ref{section-synchronous-agreement-flooding}), but replace sending
    $S$ to all $n$ processes in each round with sending $S$ to all
    $\binom{n}{k}$ possible recipient lists. As in the original
    algorithm, we want to prove that after some round with few
    enough failures, all the non-faulty processes have the same set.

    Let $S^r_i$ be the set stored by process $i$ after $r$ rounds.
    Suppose there is some round $r+1$ in which fewer than $k$ processes
    fail. Then every recipient list in round $r$ includes a process
    that does not fail in round $r+1$. Let $L$ be the set of processes
    that successfully deliver a message to at least one recipient list
    in round $r$, and let $S = ∪_{i∈L} S^r_i$.
    Then for each value $v ∈ S$, there is
    some process that receives $v$ during round $r$, does not crash in
    round $r+1$, and so retransmits $v$ to all processes in round
    $r+1$, causing it to be added to $S^{r+2}_i$.
    On the other hand, for any $v ∉ S$, $v$ is not
    transmitted to any recipient list in round $r$, which means that
    no non-faulty process $i$ includes $v$ in $S^{r+1}_i$. 
    So $S ⊆ S^{r+2}_i ⊆ ∪_j S^{r+1}_j ⊆ S$ for all $i$, and the usual
    induction argument shows that $S^{r'}_i$ continues to equal $S$
    for all non-faulty $i$ and all $r' ≥ r+2$.

    We can have at most $\floor{f/k}$ rounds with $≥k$ crashes before
    we run out, so the latest possible round in which we have fewer
    than $k$ crashes is is $r=\floor{f/k}+1$, giving agreement after
    $\floor{f/k}+2$ rounds (since we don't need to send any messages
    in round $r+2$).

    (With some tinkering, it is not too hard to adapt the Dolev-Strong
    lower bound to get a $\floor{f/k}+1$ lower bound for this model.
    The main issue is now we have to crash $k$ processes fully in
    round $r+1$ before we can remove one outgoing broadcast from a
    process in round $r$, which means we need to budget $tk$ failures
    to break a $t$-round protocol. The details are otherwise pretty
    much the same as described in
    §\ref{section-synchronous-agreement-lower-bound}.)

\subsection{Asynchronous agreement with limited failures}

Algorithm~\ref{alg-asynchronous-agreement-bogus} describes an
algorithm for asynchronous agreement with $f$ crash failures in a
fully-connected message-passing network.
The idea is to collect values from $n-f$ other processes in each of
$m$ rounds, and then decide on the smallest value collected.

\begin{algorithm}
    $\Preference \gets \Input$\;
    \For{$i \gets 1$ \KwTo $m$}{
        send $\Tuple{i,\Preference}$ to all processes\;
        wait to receive $\Tuple{i,v}$ from $n-f$ processes\;
        \Foreach{$\Tuple{i,v}$ received}{
            $\Preference \gets \min(\Preference, v)$\;
        }
    }
    decide $\Preference$\;
    \caption{Candidate algorithm for asynchronous agreement}
    \label{alg-asynchronous-agreement-bogus}
\end{algorithm}

The value $m$ is a parameter of the algorithm and may depend on $n$
and $f$.

As usual, when waiting for messages from round $i$, any messages
delivered for with other round numbers will be buffered internally and
processed when the algorithm is ready for them.

Note that when a process sends a message to all process, that includes
itself.

Show that, for any $n$ and $0 < f < n/2$, there exists a value of $m$ such that
Algorithm~\ref{alg-asynchronous-agreement-bogus} satisfies agreement,
termination, and validity; or show how to construct an execution for
any $n$, $0<f<n/2$, and $m$ that causes
Algorithm~\ref{alg-asynchronous-agreement-bogus} to fail at least one
of these requirements.

    \subsubsection*{Solution}

    We know from the FLP bound (\cite{FischerLP1985}, Chapter \ref{chapter-FLP}) that
    Algorithm~\ref{alg-asynchronous-agreement-bogus} can't work. So
    the only question is how to find an execution that shows it
    doesn't work.

    It's not too hard to see that
    Algorithm~\ref{alg-asynchronous-agreement-bogus} satisfies both
    termination and validity. So we need to find a problem with
    agreement.

    The easiest way I can see to do this is to pick a patsy process
    $p$ and give it input $0$, while giving all the other processes
    input $1$. Now run
    Algorithm~\ref{alg-asynchronous-agreement-bogus} while delaying
    all outgoing messages $\Tuple{i,v}$ from $p$ until after the
    receiver has finished the protocol. Because each other process is
    waiting for $n-f ≤ n-1$ messages, this will not prevent the other
    processes from finishing. But all the other processes have input
    $1$, so we have an invariant that messages in transit from
    processes other than $p$ and preferences of processes other than
    $p$ will be $1$ that holds as long as no messages from $p$ are
    delivered. This results in the non-$p$ processes all deciding $1$.
    We can then run $p$ to completion, at which point it will decide
    $0$.

\section{Assignment 3: due Wednesday, 2020-10-21, at 5:00pm Eastern US time}

\subsection{Too many Byzantine processes}

The phase king algorithm (Algorithm~\ref{alg-phase-king}) described in
§\ref{section-Byzantine-phase-king} solves Byzantine agreement for $f
< n/4$ processes. For larger values of $f$, it may fail by violating
one or more of the properties of termination, validity, or agreement.

For this algorithm:
\begin{enumerate}
    \item How big does $f$ need to be to prevent termination?
    \item How big does $f$ need to be to prevent validity?
    \item How big does $f$ need to be to prevent agreement?
\end{enumerate}

Assume that the processes know the new bound on $f$, and
any thresholds in the algorithm that use $f$ are adjusted to
correspond to this new bound.

        \subsubsection*{Solution}

                \begin{enumerate}
                        \item Termination: The algorithm always
                            terminates in $f+1$ synchronous rounds, so
                            $f$ doesn't matter.
                        \item Validity: To violate validity, we need
                            to convince some non-faulty process to
                            decide on the wrong value when all
                            non-faulty processes have the same input.

                            Suppose all the non-faulty processes have
                            input $0$, and we want to introduce a $1$
                            somewhere. Each process updates its
                            preference in each round to be either the
                            majority value it sees, if this value has
                            multiplicity greater than $n/2+f$, or the
                            $\PKkingMajority$ broadcast by the phase
                            king otherwise.

                            If $f < n/2$, it's going to be hard to
                            show a process a bogus majority. But a
                            Byzantine phase king gives us more
                            options. Suppose that all the $f$
                            Byzantine processes send out $1$ in all
                            rounds. Then for $f ≥ n/4$, the
                            multiplicity of the correct value $0$ will
                            be $n-f ≤ (3/4)n$, while the required
                            multiplicity to ignore the phase king will
                            be strictly greater than $n/2+f ≥ (3/4)n$.
                            So at $f=n/4$, all non-faulty processes
                            adopt the phase king's bad value $1$. In
                            any subsequent round, we can just run the
                            algorithm with the Byzantine agents
                            pretending to be non-faulty processes with
                            preference $1$, and eventually all
                            processes incorrectly decide $1$.
                        \item Agreement: Now we need to get two
                            non-faulty processes to decide different
                            values. Wait to the last
                            round, and use $f = n/4$ Byzantine
                            processes to prevent the non-faulty
                            processes from seeing a high enough
                            multiplicity on any majority value to
                            accept it, and use a Byzantine phase king
                            to transmit different $\PKkingMajority$
                            values to different non-faulty processes.
                            So again, the algorithm fails at $f=n/4$.
                \end{enumerate}

    \subsection{Committee election}

    Consider the following \conceptFormat{committee election} problem
    in an asynchronous message-passing system with $f < n/2$ crash
    failures. Each process runs a committee election protocol, at the
    end of which it receives a value $1$ (elected) or $0$ (not
    elected). The requirements of the protocol are:
    \begin{enumerate}
        \item Nonempty committee: If no processes fail, at least one process receives $1$.
        \item No latecomers: In any execution, if some process $p$ finishes the
            protocol before another process $q$ starts the protocol,
            then $q$ receives $0$.
    \end{enumerate}

    Give an algorithm that solves this problem, and show that it
    satisfies these requirements.

    (For the purpose of defining when a process starts or ends the
    protocol, imagine that it uses explicit invoke and respond
    events. Your protocol should have the property that all non-faulty
    processes eventually terminate.)

        \subsubsection*{Solution}

        The easiest way to do this may be to use ABD
        (see §\ref{section-ABD}).
        Algorithm~\ref{alg-problem-committee-election-ABD} has each
        process read the simulated register, which we assume is
        initialized to $1$, then write a $0$ before returning the
        value it read.

        \begin{algorithm}
            Let $r$ be an ABD register initialized to $1$.\;
            \Procedure{\FuncSty{elect}}{
                $\DataSty{onCommittee} \gets r$\;
                $r \gets 0$\;
                \Return $\DataSty{onCommittee}$\;
            }
            \caption{Committee election using ABD}
            \label{alg-problem-committee-election-ABD}
        \end{algorithm}

        This satisfies nonempty committee, because the first operation
        in the linearization of the register must be a read operation
        that returns $1$. It satisfies no latecomers, because if $p$
        finishes before $q$ starts, then $p$'s write finishes before
        $q$'s read starts, and linearizability of ABD implies $q$
        reads a $0$.

        This takes 3 round-trips to finish (2 for the ABD read and
        1 for the ABD write). It is not too hard to reduce this
        to 2 round-trips by replacing the embedded write in the ABD
        read operation with a write of $1$, but this requires a more
        detailed correctness argument. 

\section{Assignment 4: due Wednesday, 2020-11-04, at 5:00pm Eastern US time}

    \subsection{Counting without snapshots}

    Algorithm~\ref{alg-generalized-counter-collect-only} gives a
    wait-free implementation of a generalized counter using a collect. The
    $\Inc(v)$ procedure adjusts the value of the counter by $v$: if it
    was $x$ before $\Inc(v)$, it should be $x+v$ after. The $\Read$
    procedure returns the current value of the counter. Assume that
    the initial value of the counter is $0$, as are the initial values
    of the registers $A[i]$ that implement it.

    \begin{algorithm}
        \Procedure{$\Inc(v)$}{
            $A[i] \gets A[i] + v$\;
        }
        \Procedure{$\Read()$}{
            $s \gets 0$\;
            \For{$j \gets 1$ \KwTo $n$}{
                $s \gets s + A[j]$\;
            }
            \Return $s$\;
        }
        \caption{An alleged counter. Code for process $i$.}
        \label{alg-generalized-counter-collect-only}
    \end{algorithm}

    This counter implementation is not linearizable in all executions,
    but it may be linearizable if we restrict the allowed values $v$
    that can be supplied as arguments to an $\Inc$ operations. For
    each of the following sets $V$, show that any execution in which
    all increments are elements of $V$ is linearizable, or show that
    there exists an execution with increments in $V$ that is not.

    \begin{enumerate}
        \item $V=\Set{0,1}$.
        \item $V=\Set{-1,1}$.
        \item $V=\Set{1,2}$.
    \end{enumerate}

        \subsubsection*{Solution}

        \begin{enumerate}
            \item The $\Set{0,1}$ case is linearizable. Given an
                execution $S$ of
                Algorithm~\ref{alg-generalized-counter-collect-only},
                we assign to a linearization point to each $\Inc$
                operation at the step where 
                it writes to $A$, and assign a linearization point to
                each $\Read$ operation
                $ρ$ that returns $s$ at the later of the first step 
                that leaves $∑_j A[j] = s$ or the first step of $ρ$.
                Since this may assign the same linearization point to
                some write operation $π$ and one or more read
                operations $ρ_1,\dots,ρ_k$, when this occurs, we order
                the write before the reads and the reads arbitrarily.

                Observe that:
                \begin{enumerate}
                    \item The value of each $A[j]$ individually is
                        non-decreasing over time, and increases by at
                        most one at each step.
                    \item The same holds for $∑_{j=1}^{n} A[j]$.
                \end{enumerate}
                These are easily shown by induction on the steps of
                the execution, since each $\Inc$ operation only
                changes at most one $A[j]$ and only changes it by
                increasing it by $1$.

                The first condition implies that the value $v_j$ of $A[j]$
                used by a particular $\Read$ operation $ρ$ lies somewhere between
                the minimum and maximum values of $A[j]$ during the
                operation's interval, which implies the same about the
                total $∑_j A[j]$. In particular, if $ρ$ returns $s$
                the value of $∑_j A[j]$ is no greater than $s$, and it
                reaches $s$ no later than the end of $ρ$. 
                
                Because $∑_j A[j]$ increases by at most one per step,
                this means that either $∑_j A[j] = s$ at the first
                step of $ρ$, or $∑_j A[j] = s$ at some step within the
                execution interval of $ρ$. In either case, $ρ$ is
                assigned an execution point within its interval that
                follows exactly $s$ non-trivial increments. This means
                that the return values of all $\Read$ operations are
                consistent with a sequential generalized counter
                execution, and because both $\Read$ and $\Inc$ operations 
                are ordered consistently with the execution ordering
                in $S$, we have a linearization of $S$.
            \item For increments in $\Set{-1,1}$, there are executions
                of
                Algorithm~\ref{alg-generalized-counter-collect-only}
                that are not linearizable. We will construct a
                specific bad execution for $n=3$. Let $p_1$ perform
                $\Inc(1)$ and $p_2$ perform $\Inc(2)$, where $p_1$
                finishes its operation before $p_2$ starts. Because
                the $\Inc(1)$ must be linearized before the
                $\Inc(-1)$, the values of the counter in any
                linearization will be $0,1,0$ in this order.

                Now add a $\Read$ operation by $p_3$ that is
                concurrent with both $\Inc$ operations. Suppose that
                in the execution, the follow operations are performed
                on the registers $A[1]$ through $A[3]$:
                \begin{enumerate}
                    \item $p_3$ reads $0$ from $A[1]$.
                    \item $p_1$ writes $1$ to $A[1]$.
                    \item $p_2$ writes $-1$ to $A[2]$.
                    \item $p_3$ reads $-1$ from $A[2]$.
                    \item $p_3$ reads $0$ from $A[3]$.
                \end{enumerate}

                Now $p_3$ returns $-1$. There is no point in the
                sequential execution at which this is the correct
                return value, so there is no linearization of this
                execution.
            \item For increments in $\Set{1,2}$, essentially the same
                counterexample works. Here we let $p_1$ do $\Inc(1)$
                and $p_2$ do $\Inc(2)$, while $p_3$ again does a
                concurrent read. The bad execution is:
                \begin{enumerate}
                    \item $p_3$ reads $0$ from $A[1]$.
                    \item $p_1$ writes $1$ to $A[1]$.
                    \item $p_2$ writes $2$ to $A[2]$.
                    \item $p_3$ reads $2$ from $A[2]$.
                    \item $p_3$ reads $0$ from $A[3]$.
                \end{enumerate}
                Now $p_3$ returns $2$, but in any linearization of the
                two write operations, the values in the counter are
                $0,1,3$.
        \end{enumerate}

    \subsection{Rock-paper-scissors}

    Define a \concept{rock-paper-scissors object} as having three
    states $0$ (rock), $1$ (paper), and $2$ (scissors), with a $\Read$
    operation that returns the current state and a $\FuncSty{play}(v)$
    operation for $v∈\Set{0,1,2}$ that changes the state from $s$ to $v$ if $v = (s+1)
    \pmod{3}$ and has no effect otherwise.

    Prove or disprove: There exists a deterministic wait-free linearizable
    implementation of a rock-paper-scissors object from atomic
    registers.

        \subsubsection*{Solution}

        Proof: We will show how to implement a rock-paper-scissors
        object using an unbounded max register, which can be built
        from atomic registers using snapshots. The idea is to store a
        value $v$ such that $v \bmod 3$ gives the value of the
        rock-paper-scissors object. Pseudocode for both operations is
        given in
        Algorithm~\ref{alg-rock-paper-scissors-implementation}.

        \begin{algorithm}
            Let $m$ be a shared max register.\;
            \Procedure{$\FuncSty{play}(v)$}{
                $s \gets m$\;
                \If{$v = ((s+1) \bmod 3)$}{
                    $m \gets s+1$\;
                }
            }
            \Procedure{$\Read()$}{
                \Return $(m \bmod 3)$\;
            }
            \caption{Implementation of a rock-paper-scissors object}
            \label{alg-rock-paper-scissors-implementation}
        \end{algorithm}

        Linearize each $\FuncSty{play}$ operation that does not write
        $m$ at the step at which it reads $m$.

        Linearize each $\FuncSty{play}$ operation that writes $s+1$ to
        $m$ at the first step at which $m ≥ s+1$. If this produces
        ties, break first in order of increasing $s+1$ and then
        arbitrarily. Since each such operation has $m
        ≤ s$ when the operation starts and $m ≥ s+1$ when it finishes,
        these linearization points fit within the intervals of their
        operations.

        Linearize each $\Read()$ operation at the step where it reads
        $m$.

        Since each of these linearization points is within the
        corresponding operation's interval, this preserves the
        observed execution ordering.

        Observe that the $\FuncSty{play}$ operations that write are linearized in
        order of increasing values written, and there are no gaps in
        this sequence because no process writes $s+1$ without first
        seeing $s$. (This actually shows there is no to break
        ties by value.) So the sequence of values in the max register,
        taken mod $3$, iterates through the values $0,1,2,0,\dots$ in
        sequence, with each value equal mod $3$ to some argument to a
        $\FuncSty{play}$ operation. 
        So we can take these values mod $3$ as the actual value
        of the register for the purposes of $\Read$ operations,
        meaning the
        $\Read$ operations all return correct values.
        The $\FuncSty{play}$ operations that don't write are
        linearized at a point where they would have no effect on the
        state of the rock-paper-scissors object, which is also
        consistent with the sequential specification.

        It follows that
        Algorithm~\ref{alg-rock-paper-scissors-implementation} is a
        linearizable implementation of a rock-paper-scissors object
        from max registers.
        It is also wait-free, since each operation is implemented
        using a constant number of max-register operations.
        By implementing max registers using snapshots, we get a
        wait-free linearizable implementation
        from atomic registers.

\section{Assignment 5: due Wednesday, 2020-11-18, at 5:00pm Eastern US time}

    \subsection{Randomized consensus with one max register}
    \label{problem-randomized-consensus-with-one-max-register}

        Prove or disprove: A single max register, with no other
            objects, is sufficient to solve randomized wait-free
            binary consensus for two processes against an oblivious adversary.

            \subsubsection*{Solution}

            We'll disprove it.

            Let $p_0$ and $p_1$ be the two processes.  The idea is to
            consider, for each $i∈\Set{0,1}$ some nonzero-probability
            solo terminating execution $ξ_i$ of $p_i$ with input $i$, then show that
            $ξ_0$ and $ξ_1$ can be interleaved to form a two-process
            execution $ξ$ that is indistinguishable by each $p_i$ from
            $ξ_i$.

            The oblivious adversary will simply choose to schedule the
            processes for $ξ$. Since the processes flip a finite
            number of coins in this execution, there is a nonzero
            chance that the adversary gets lucky and they flip their
            coins exactly the right way.

            Fix $ξ_0$ and $ξ_1$ as above. Partition each $ξ_i$ as
            $α_iβ_{i1}β_{i2}\dots β_{ik_i}$ where $α_i$ contains only
            read operations and each $β_{ij}$ starts with a write
            operation of a value $v_{ij}$ strictly larger than any previous
            write operation.

            Let $ξ = α_0 α_1 β_{i_1 j_1} β_{i_2 j_2} \dots β_{i_k
            j_k}$ where $k=k_0+k_1$ and the blocks $β_{i_\ell j_\ell}$
            are the blocks $\Set{β_{0j}}$ and $\Set{β_{1j}}$ sorted in
            order of non-decreasing $v_{ij}$. Then each block
            $β_{i_\ell j_\ell}$ in $ξ$ starts with a write of a value no
            smaller than the previous value in the max register,
            causing each read operation within the block to return the
            value of this write, just as in the solo execution
            $ξ_{i_\ell}$. Assuming both processes flip their coins as
            in the solo executions, they both perform the same
            operations and return the same values. These values will
            either violate agreement in $ξ$ or validity in at least
            one of $ξ_0$ or $ξ_1$.

        \subsection{A plurality object}

        Consider a shared-memory object with operations
        $\FuncSty{vote}(v)$ and $\FuncSty{winner}()$, where
        $\FuncSty{winner}()$ returns the value $v$ that appeared in
        the largest number of previous $\FuncSty{vote}$ operations, or
        $⊥$ if there is no such unique $v$. For example, in a
        sequential execution with votes $a,b,b,c,c,c,a,a,a$, the value
        returned by a $\FuncSty{winner}$ operation following each vote
        will be $a, ⊥, b, b, ⊥, c, c, ⊥, a$.

        Pick one of these statements, and show that it is true:
        \begin{enumerate}
            \item There is a deterministic wait-free linearizable implementation of
                this object for $n$ processes that uses $o(n)$
                registers.
            \item\label{line-plurality-object-JTT} There is such an implementation that uses $O(n)$
                registers, but not $o(n)$ registers.
            \item There is no such implementation using $O(n)$
                registers.
        \end{enumerate}

            \subsubsection*{Solution}

            Case (\ref{line-plurality-object-JTT}) holds. 
            
            To implement
            the object, use a snapshot array to hold the total votes
            from each process, and have the $\FuncSty{winner}$
            operation take a snapshot, add up all the votes and return
            the correct result. This can be done using $n$ registers.

            To show that it can't be done with $o(n)$ registers, use
            the JTT bound (see Chapter~\ref{chapter-JTT}).
            We need to argue that the object is perturbable.
            Let $ΛΣπ$ be an execution that needs to be perturbed, and
            let $m$ be the maximum number of $\FuncSty{vote}(v)$
            operations that start in $Λ$ for any value $v$. Then a
            sequence $γ$ of $m+1$ votes for some $v'$ that does not
            appear in $Λ$  will leave the object with $v'$ as the
            plurality value, no matter how the remaining operations
            are linearized. Since $v'$ did not previously appear in
            $Λ$, this gives a different return value for $π$ in $ΛγΣπ$
            from $ΛΣπ$ as required. The JTT bound now implies that any
            implementation of the object requires at least $n-1$
            registers.

\chapter{Sample assignments from Spring 2019}

\section{Assignment 1: due Wednesday, 2019-02-13, at 5:00pm}

    \subsection{A message-passing bureaucracy}

    Alice and Bob are communicating with each other by alternately
    exchanging messages.  But Bob finds Alice's messages alarming,
    and whenever he responds to Alice, he also forwards a copy of
    Alice's message to his good friend Charlie 1, a secret policeman.
    Charlie 1 reports to Charlie 2, but following the rule that ``once
    is happenstance, twice is coincidence, the third time it's enemy
    action,''~\cite{Fleming1959} Charlie 1 only sends a report to Charlie 2 after receiving
    three messages from Bob.  Similarly, Charlie 2 only sends a
    message to his supervisor Charlie 3 after receiving three messages
    from Charlie 2, and so on up until the ultimate boss Charlie $n$.
    Pseudocode for each participant is given in
    Algorithm~\ref{alg-three-times-enemy-action}.

    \begin{algorithm}
        Alice:\\
        \Initially{
            send message to Bob\;
        }
        \UponReceiving{message from Bob}{
            send message to Bob\;
        }
        Bob:\\
        \UponReceiving{message from Alice}{
            send message to Alice\;
            send message to Charlie 1\;
        }
        Charlie $i$, for $i<n$:\\
        \Initially{
            $c ← 0$\;
        }
        \UponReceiving{message from Bob or Charlie $i-1$}{
            $c ← c+1$\;
            \If{$c = 3$}{
                $c ← 0$\;
                send message to Charlie $i+1$\;
            }
        }

        \caption{Reporting Alice's alarming messages}
        \label{alg-three-times-enemy-action}
    \end{algorithm}

    Assuming we are in a standard asynchronous message-passing system,
    that Alice sends her first message at time $0$,
    and that the protocol finishes as soon as Charlie $n$ receives a
    message, what is the worst-case time and message complexity of
    this protocol as a function of $n$?

    \subsubsection*{Solution}

    \paragraph{Time complexity}
    Observe that Alice sends at least $k$
    messages by time $2k-2$.  This is easily shown by induction on
    $k$, because Alice sends at least $1$ message by time $0$, and if
    Alice has sent at least $k-1$ message by time $2k-4$, the last of
    these is received by Bob no later than time $2k-3$, and Bob's response
    is received by Alice no later than time $2k-2$.

    Because each message from Alice prompts a message from Bob at most
    one time unit later, this implies that Bob sends at least $k$
    messages by time $2k-1$.

    Write $T_0(k) = 2k-1$ for the maximum time for Bob to send $k$
    messages.  Write $T_i(k)$ for the maximum time for Charlie $i$ to
    send $k$ messages, for each $0 < i < n$.  In order for Charlie $i$
    to send $k$ messages, it must receive $3k$ messages from Bob or
    Charlie $i-1$ as appropriate.  These messages are sent no later
    than $T_{i-1}(3k)$, and the last of them is received no later than
    $T_{i-1}(3k)+1$.  So we have the recurrence
    \begin{align*}
        T_i(k) &= T_{i-1}(3k) + 1 \\
        T_0(k) &= 2k-1 \\
        \intertext{with the exact solution}
        T_i(k) &= (2⋅3^i⋅k-1) + k.
    \end{align*}

    For $i=n-1$ and $k=1$, this is $2⋅3^{n-1}-1+n-1 = 2⋅3^{n-1} + n =
    O(3^n)$.  We can get the exact time to finish by adding one more
    unit to account for the delay in delivering the message from
    Charlie $n-1$ to Charlie $n$.  This gives $2⋅3^{n-1}+n+1$ time
    exactly in the worst case, or $O(3^n)$ if we want an asymptotic
    bound.

    \paragraph{Message complexity}
    Message complexity is easier: there is no bound on the number of
    messages that may be sent before Charlie $n$ receives his first
    message.  This is because in an asynchronous system, Alice and Bob
    can send an unbounded (though finite) number of messages to each
    other even before
    Bob's first message to Charlie $0$ is delivered, without violating
    fairness.

    \subsection{Algorithms on rings}

    In Chapter~\ref{chapter-leader-election}, we saw several leader
    election algorithms for rings.  But nobody builds rings.  However,
    it may be that an algorithm for a ring can be adapted to other
    network structures.

    \begin{enumerate}
        \item Suppose you have a network in the form of a
            $d$-dimensional hypercube $Q^d$.  This means we have $n=2^d$
            nodes, where each node is labeled by a $d$-bit coordinate
            vector, and two nodes are adjacent if their vectors differ
            in exactly one coordinate.
            We also assume that each node knows its own coordinate
            vector and those of its neighbors.

            Show that any algorithm for an asynchronous
            ring can be adapted to an asynchronous $d$-dimensional
            hypercube with no increase in its time or message complexity.
        \item 
            What difficulties arise if we try to generalize this to an
            arbitrary graph $G$?
    \end{enumerate}

    \subsubsection*{Solution}

    \begin{enumerate}
        \item The idea is to embed the ring in the hypercube, so that
            each node is given a clockwise and counterclockwise
            neighbors, and any time the ring algorithm asks to send a
            message clockwise or counterclockwise, we send to the
            appropriate neighbor in the hypercube.  
            We can then argue that for any execution of the hypercube
            algorithm there is a corresponding execution of the ring
            algorithm and vice versa; this implies that the worst-case
            time and message-complexity in the hypercube is the same
            as in the ring.

            It remains only to construct an embedding.  For $d=0$,
            $d=1$, and $d=2$, the ring and hypercube are the same graph, so it's easy.
            For larger $d$, split the hypercube into two
            subcubes $Q^{d-1}$, consisting of nodes with coordinate vectors of
            the form $0x$ and $1x$.  Use the previously constructed
            embedding for $d-1$ to embed a ring on each subcube, using
            the same embedding for both.  Pick a pair of matching
            edges $(0x,0y)$ and $(1x,1y)$ and remove them, replacing
            them with $(0x,1x)$ and $(0y,1y)$.  We have now
            constructed an undirected Hamiltonian cycle on $Q^d$.  Orient the
            edges to get a directed cycle, and we're done.
        \item There are a several problems that may come up:
            \begin{enumerate}
                \item Maybe $G$ is not Hamiltonian.
                \item Even if $G$ is Hamiltonian, finding an
                    Hamiltonian cycle in an arbitrary graph is
                    $\classNP$-hard.  This could be trouble for a
                    practical algorithm.
                \item Even if we can find a Hamiltonian cycle for $G$
                    (maybe because $G$ is a nice graph of some kind,
                    or maybe by taking advantage of the unbounded
                    computational power of processes assumed in the
                    standard message-passing model), the processes
                    don't necessarily know what $G$ looks like at the
                    start.  So they would need some initial start-up
                    cost to map the graph, adding to the time and
                    message complexity of the ring algorithm.
            \end{enumerate}
    \end{enumerate}

    \subsection{Shutting down}

    \newFunc{\Stop}{stop}

    Suppose we want to be able to stop a running protocol in an
    asynchronous message-passing system prematurely.
    Define a \concept{shutdown mechanism} as a modification to an
    existing protocol in which any process can nondeterministically 
    issue a $\Stop$ order
    that eventually causes all processes to stop sending messages.
    We would like such a shutdown mechanism to satisfy two properties:
    \begin{enumerate}
        \item \conceptFormat{Termination.}  If some process issues a
            $\Stop$ order at time $t$, no process sends a message 
            at time $t+Δ$ or later, for some finite bound
            $Δ$ that may depend on the structure of the network.
        \item \conceptFormat{Non-interference.}  If no process issues
            a $\Stop$ order, the protocol carries out an
            execution identical to some execution of the underlying
            protocol without a shutdown mechanism.
    \end{enumerate}

    Show how to implement a shutdown mechanism, and prove tight upper
    and lower bounds on $Δ$ as a function of the structure of the
    network.

    \subsubsection*{Solution}

    This is pretty much the same as a Chandy-Lamport
    snapshot~\cite{ChandyL1985}, as described in
    §\ref{section-chandy-lamport}.  The main difference is that
    instead of recording its state upon receiving a
    $\Stop$ message, a process shuts down the underlying protocol.
    Pseudocode is given in Algorithm~\ref{alg-shutdown}.
    We assume that the initial $\Stop$ order takes the form of a
    $\Stop$ message delivered by a process to itself.
    
    \begin{algorithm}
        \newData{\Stopped}{stopped}
        \Initially{$\Stopped ← \False$\;}
        \UponReceiving{$\Stop$}{
            \If{$¬\Stopped$}{
                $\Stopped ← \True$\;
                send $\Stop$ to all neighbors\;
                replace all events in underlying protocol with no-ops\;
            }
        }
        \caption{Shutdown mechanism based on Chandy-Lamport}
        \label{alg-shutdown}
    \end{algorithm}

    An easy induction argument shows that if $p$ receives a $\Stop$
    message at time $t$, then any process $q$ at distance $d$ from $p$
    receives a $\Stop$ message no later than time $t+d$.  
    It may be that $q$ sends $\Stop$ messages in responds to this $\Stop$
    message, but these are the last messages $q$ ever sends.
    It follows that no process sends a message later than time $t+D$, where $D$
    is the diameter of the graph.  This gives an upper bound on $Δ$.

    For the lower bound, we can apply an indistinguishability
    argument.  Let $p$ and $q$ be processes at distance $D$ from each other,
    and suppose that the underlying protocol involves processes
    sending messages to their neighbors at every opportunity.
    Consider two synchronous executions: $X$, an execution in which no
    $\Stop$ order is ever issued, and $X_t$, an execution in which
    $p$ delivers a $\Stop$ message to itself at time $t$.

    We can show by induction on $d$ that any process $r$ at distance
    $d$ from $p$ carries out the same steps in both $X$ and $X_t$ up
    until time $t+d-1$.  The base case is when $d=0$, and we are
    simply restating that $p$ runs the underlying protocol before time
    $t$.  For the induction step, we observe for any time $t'<t+d-1$,
    any message sent to $r$ from some neighbor $s$ was sent at time
    $t'-1 < t+d-2$, and since $d(p,s) ≥ d-1$, the induction hypothesis
    gives that $s$ sends the same messages at $t'-1$ in both $X$ and
    $X_t$.

    It follows that $q$ sends the same message in $X$ and $X_t$ at
    time $t+D-1$.  If it sends a message, then we have $Δ > D-1$.  If
    it does not send a message, then the mechanism violates the
    non-interference condition.  So any correct shutdown mechanism
    requires exactly $Δ = D$ time to finish in the worst case.

\section{Assignment 2: due Wednesday, 2019-03-06, at 5:00pm}

    \subsection{A non-failure detector}

    Consider the following vaguely monarchist leader election
    mechanism for an asynchronous message-passing system with crash
    failures.  Each process has access to an oracle that starts with
    the value $0$ and may increase over time.  The oracle guarantees:
    \begin{enumerate}
        \item No two processes ever see the same nonzero value.
        \item Eventually some non-faulty process is given a fixed
            value that is larger than the values for all other
            processes for the rest of the execution.
    \end{enumerate}

    As a function of the number of processes $n$, what is the largest
    number of crash failures $f$ for which it is possible to solve
    consensus using this oracle?

    \subsubsection*{Solution}

    We need $f < n/2$.
    
    To show that $f < n/2$ is sufficient, observe that we
    can use the oracle to construct an eventually strong ($◇S$) failure detector.

    Recall that $◇S$ has the property that there is some non-faulty process that
    is eventually never suspected, and every fault process is
    eventually permanently suspected.  Have each process broadcast the
    current value of its leader oracle whenever it increases; when a
    process $p$ receives $i$ from some process $q$, it stops suspecting
    $q$ if $i$ is greater than any value $p$ has previously seen, and
    starts suspecting all other processes.  The guarantee that
    eventually some non-faulty $q$ gets a maximum value that never
    changes ensures that eventually $q$ is never suspected, and all
    other processes (including faulty processes) are suspected.  We
    can now use Algorithm~\ref{alg-strong-failure-detector-consensus}
    to solve consensus.

    To show that $f < n/2$ is necessary, apply a partition argument.
    In execution $Ξ_0$, processes $n/2+1$ through $n$ crash, and
    processes $1$ through $n/2$ run with input $0$ and 
    with the oracle assigning value $1$ to process $1$ (and no
    others).  In execution $Ξ_1$, processes $1$ through $n/2$ crashes,
    and processes $n/2+1$ through $n$ run with input $1$ and with the
    oracle assigning value $2$ to process $n$ (and no others).
    In each of these executions, termination and validity require that eventually
    the processes all decide on their respective input values $0$ and
    $1$.

    Now construct an execution $Ξ_2$, in which both groups of
    processes run as in $Ξ_0$ and $Ξ_1$, but no messages are exchanged
    between the groups until after both have decided (which must occur
    after a finite prefix because this execution is indistinguishable
    to the processes from $Ξ_0$ or $Ξ_1$).  We now violate agreement.

    \subsection{Ordered partial broadcast}

    Define \index{broadcast!ordered partial}\concept{ordered partial broadcast} 
    as a protocol that allows any process to broadcast a message, with
    the guarantees that, for messages sent through the broadcast
    mechanism:
    \begin{enumerate}
        \item Any message sent by a non-faulty process is received by
            at least one process;
        \item Any message that is received by at least one process is
            received by at least $k$ processes; and
        \item If two processes $p$ and $q$ both receive messages $m_1$
            and $m_2$ from the protocol, then either $p$ and $q$ both receive $m_1$
            before $m_2$, or they both receive $m_2$ before $m_1$.
    \end{enumerate}

    Give an implementation of ordered partial broadcast with 
    $k=3n/4$ that works for sufficiently large $n$ in a
    fully-connected asynchronous message-passing system with
    up to $f=n/6$ crash failures, or show that no such implementation
    is possible.

    \subsubsection*{Solution}

    No such implementation is possible.  The proof is by showing that
    if some such implementation could work, we could solve
    asynchronous consensus with $1$ crash failure, contradicting the
    Fischer-Lynch-Patterson bound~\cite{FischerLP1985} (see
    Chapter~\ref{chapter-FLP}).

    An implementation of consensus based on totally-ordered partial
    broadcast for $k=3n/4$ is given in
    Algorithm~\ref{alg-totally-ordered-partial-broadcast-consensus}.
    In fact, $k=3n/4$ is overkill when $f=1$; $k > n/2+f$ is enough.
   
    \newFunc{\OPBreceived}{received}
    \newData{\OPBfirst}{first}
    \newData{\OPBcount}{count}
    \newData{\OPBvalue}{value}

    \begin{algorithm}
        $\OPBfirst ← ⊥$\;
        \For{$i ← 1 \KwTo n$}{
            $\OPBcount[i] ← 0$\;
            $\OPBvalue[i] ← ⊥$\;
        }

        broadcast $\Tuple{i,\Input}$\;
        \UponReceiving{$\Tuple{j,v}$}{
            \If{$\OPBfirst = ⊥$}{
                $\OPBfirst ← \Tuple{j,v}$\;
                send $\OPBreceived(\Tuple{j,v})$ to all processes\;
            }
        }
        \UponReceiving{$\OPBreceived(\Tuple{j,v})$}{
            $\OPBcount[j] ← \OPBcount[j] + 1$\;
            $\OPBvalue[j] ← v$\;
            \If{$\OPBcount[j] = k-f$}{
                decide $\OPBvalue[j]$\;
            }
        }
        \caption[Consensus from totally-ordered partial broadcast]{
            Consensus from totally-ordered partial broadcast.  Code for process $i$.}
        \label{alg-totally-ordered-partial-broadcast-consensus}
    \end{algorithm}

    The idea of the algorithm is to use the broadcast mechanism to
    chose a decision value, by looking at which value is delivered
    first.  Since not every process will see the same value delivered
    first, this requires a second round of communication in which
    processes retransmit their first incoming message.  The following
    lemma shows that this is enough to get agreement:
    \begin{lemma}
        \label{lemma-OPB-consensus}
        In any execution of
        Algorithm~\ref{alg-totally-ordered-partial-broadcast-consensus}
        with $k > n/2+f$,
        there is a unique pair $\Tuple{j,v}$ such that at least $k-f$
        non-faulty processes resend $\OPBreceived(\Tuple{j,v})$.
    \end{lemma}
    \begin{proof}
        Because all processes that receive messages $m_1$ and $m_2$
        through the broadcast mechanism receive them in the same
        order, we can define a partial order on messages by letting
        $m_1 < m_2$ if any process receives $m_1$ before $m_2$.
        
        There are only finitely many messages, so there is at least one pair
        $\Tuple{j,v}$ that is minimal in this partial order.  This
        message is received by at least $k$ processes, of which at
        least $k-f$ are non-faulty.  Each such process receives
        $\Tuple{j,v}$ before any other broadcast messages, so it sets
        $\OPBfirst$ to $\Tuple{j,v}$ and resends
        $\OPBreceived(\Tuple{j,v})$.

        To show that $\Tuple{j,v}$ is unique, observe that $k-f > n/2$
        implies that if there is some other pair $\Tuple{j',v'}$ that
        is resent by $k-f$ non-faulty processes, then there is some
        process that resends both $\Tuple{j,v}$ and $\Tuple{j',v'}$.
        But each process resends at most one pair.
    \end{proof}

    Lemma~\ref{lemma-OPB-consensus} immediately gives agreement,
    because a process only decides on a value $v$ after receiving
    $\OPBreceived(\Tuple{j,v})$ from $k-f$ processes, and only one
    such pair is sent by so many.  Termination follows from the
    existence of such a pair: eventually every non-faulty process
    receives $\Tuple{j,v}$ from $k-f$ processes.  Validity is
    immediate from the fact that $v$ is $j$'s input.

    It follows that
    Algorithm~\ref{alg-totally-ordered-partial-broadcast-consensus}
    solves consensus whenever $k
    > n/2 + f$, which includes the case $k=3n/4$ and $f=n/6$.  If an
    implementation of ordered partial broadcast with these parameters
    exists in the standard message-passing model, this would give a
    protocol for asynchronous consensus with $f = n/6 ≥ 1$ when $n≥6$.
    This contradicts FLP, showing that such an implementation is
    impossible.
    
    \subsection{Mutual exclusion using a counter}

\newData{\CMTwaiting}{waiting}
\newData{\CMTcount}{count}

    Algorithm~\ref{alg-mutex-Peterson-modified} gives a modified
    version of Peterson's two-process mutual exclusion algorithm
    (§\ref{section-mutex-Peterson}) that replaces the $\Present$
    bits with an atomic counter $\CMTcount$.  This object supports read,
    increment, and decrement operations, where increment and decrement
    increase and decrease the value in the counter by one,
    respectively.  Unlike the $\Present$ array, $\CMTcount$ doesn't
    depend on the number of processes $n$.  So in principle this algorithm
    might work for arbitrary $n$.

\begin{algorithm}
\SharedData\\
$\CMTwaiting$, atomic register, initially arbitrary\\
$\CMTcount$, atomic counter, initially $0$\\

Code for process $i$:\\
\While{\True}{
        \tcp{trying}
        increment $\CMTcount$
        \nllabel{line-alg-modified-Peterson-present}\;
        $\CMTwaiting ← i$
        \nllabel{line-alg-modified-Peterson-waiting}\;

        \While{\True}{
            \If{$\CMTcount = 1$}{
                \nllabel{line-alg-modified-Peterson-break-1}
                \Break\;
            }
\If{$\CMTwaiting = i + 1 \pmod{n}$}{
                \nllabel{line-alg-modified-Peterson-break-2}
                \Break\;
            }
        }

        \tcp{critical}
        (do critical section stuff)
        \nllabel{line-alg-modified-Peterson-critical}\;
        \tcp{exiting}
        decrement $\CMTcount$
        \nllabel{line-alg-modified-Peterson-reset-present}\;
        \tcp{remainder}
        (do remainder stuff) 
        \nllabel{line-alg-modified-Peterson-remainder}\;
}
\caption{Peterson's mutual exclusion algorithm using a counter}
\label{alg-mutex-Peterson-modified}
\end{algorithm}

    Show that Algorithm~\ref{alg-mutex-Peterson-modified} provides starvation-free mutual exclusion
    for two processes, but not for three processes.

    \subsubsection*{Solution}

    The proof that this works for two processes is essentially
    the same as in the original algorithm.  The easiest way to see
    this is to observe that process $p_i$ sees $\CMTcount = 1$ in
    Line~\ref{line-alg-modified-Peterson-break-1} under exactly the
    same circumstances as it sees $\Present[¬i] = 0$ in
    Line~\ref{line-alg-Peterson-break-1} in the original algorithm;
    and similarly with two processes $\CMTwaiting$ is always set to
    the same value as $\Waiting$ in the original algorithm.  So we
    can map any execution of
    Algorithm~\ref{alg-mutex-Peterson-modified} for two processes
    to an execution of Algorithm~\ref{alg-mutex-Peterson}, and all of
    the properties of the original algorithm carry over to the
    modified version.

    To show that the algorithm doesn't work for three processes, we
    construct an explicit bad execution:

    \begin{enumerate}
        \item $p_0$ increments $\CMTcount$
        \item $p_1$ increments $\CMTcount$
        \item $p_2$ increments $\CMTcount$
        \item $p_0$ writes $0$ to $\CMTwaiting$
        \item $p_1$ writes $1$ to $\CMTwaiting$
        \item $p_2$ writes $2$ to $\CMTwaiting$
        \item $p_0$ reads $3$ from $\CMTcount$
        \item $p_1$ reads $3$ from $\CMTcount$
        \item $p_2$ reads $3$ from $\CMTcount$
        \item $p_1$ reads $2$ from $\CMTwaiting$ and enters the
            critical section.
        \item $p_1$ leaves the critical section and decrements
            $\CMTcount$.
    \end{enumerate}

    At this point we have $\CMTcount = 2$ and $\CMTwaiting = 2$, with
    both $p_0$ and $p_2$ at the start of the loop to check these
    variables.  Suppose that $p_1$ doesn't come back.  Because neither
    $p_0$ nor $p_2$ changes $\CMTcount$ or $\CMTwaiting$, both
    variables remain at $2$ forever.  But then neither $p_0$ nor $p_2$
    enters the critical section, because $\CMTcount$ is never $1$ and
    $\CMTwaiting$ is never equal to $0+1$ or $2+1 \pmod{3}$.

\section{Assignment 3: due Wednesday, 2019-04-17, at 5:00pm}

    \subsection{Zero, one, many}

    Consider a counter supporting $\Inc$ and $\Read$ operations
    that is capped at $2$.  This means that after the first two calls
    to $\Inc$, any further calls to $\Inc$ have no effect:
    a $\Read$
    operation will return $0$ if it follows no calls to $\Inc$, $1$ if
    it follows exactly one call to $\Inc$, and $2$ if it follows two
    or more calls to $\Inc$.

    There is a straightforward implementation
    of this object using snapshot.  This
    requires $O(n)$ space and $O(n)$ steps per operation in the worst
    case.

    Is it possible to do better?  That is, can one give a
    deterministic, wait-free, linearizable implementation of a
    $2$-bounded counter from atomic registers that uses $o(n)$ space
    and $o(n)$ steps per operation in the worst case?

    \subsubsection*{Solution}

    One possible implementation is given in
    Algorithm~\ref{alg-2-bounded-counter}.  This requires $O(1)$
    space and $O(1)$
    steps per call to $\Inc$ or $\Read$.

    \begin{algorithm}
        \Procedure{\Inc}{
            \eIf{$c[1] = 1$} {
                \tcp{somebody already did $\Inc$}
                $c[2] ← 1$\;
            }{
                $c[1] ← 1$\;
                \tcp{maybe somebody else is doing $\Inc$}
                \If{$\Splitter$ returns $\SplitterRight$ or $\SplitterDown$}{
                     $c[2] ← 1$\;
                }
            }
        }
        \Procedure{\Read}{
            \uIf{$c[2] = 1$}{
                \Return $2$\;
            }
            \ElseIf{$c[1] = 1$}{
                \Return $1$\;
            }
            \Else{
                \Return $0$\;
            }
        }
        \caption{A $2$-bounded counter}
        \label{alg-2-bounded-counter}
    \end{algorithm}

    The implementation uses two registers $c[1]$ and $c[2]$ to
    represent the value of the counter.  Two additional registers
    implement a splitter object as in
    Algorithm~\ref{alg-splitter}.\footnote{It may be possible shave
    off a register by breaking the splitter abstraction and using the
    $\SplitterRace$ or $\SplitterDoor$ register in place of $c[1]$,
    but I haven't worked out the linearizability for this case.}

    Claim: For any two calls to $\Inc$, at least one sets $c[2]$
    to $1$.  Proof: Suppose otherwise.  Then both calls are by
    different processes $p$ and $q$ (or else the second call would see $c[1] = 1$)
    and both execute the splitter.  
    Since a splitter returns $\SplitterStop$ to at most one process,
    one of the two processes gets $\SplitterRight$ or $\SplitterDown$,
    and sets $c[2]$.

    It is also straightforward to show that a single $\Inc$
    running alone will set $c[1]$ but not $c[2]$, since in this case
    the splitter will return $\SplitterStop$.

    Now we need to argue linearizability.  We will do so by assigning
    linearization points to each operation.

    If some $\Inc$ does not set $c[2]$, assign it the step at
    which it sets $c[1]$.  Assign each other $\Inc$ the step at
    which it first sets $c[2]$.

    If every $\Inc$ sets $c[2]$, assign the first $\Inc$
    to set $c[1]$ the step at which it does so, and assign all others
    the first point during its execution interval at which $c[2]$ is
    nonzero.

    For a $\Read$ operation that returns $2$, assign the step at which
    it reads $c[2]$.  For a $\Read$ operation that returns $1$, assign
    the first point in the execution interval after it reads $c[2]$ at
    which $c[1] = 1$.  For a $\Read$ operation that returns $0$,
    assign the step at which it reads $c[1]$.

    This will assign the same linearization point to some operations;
    in this case, put $\Inc$s before $\Read$s and otherwise
    break ties arbitrarily.

    These choices create a linearization which consists of (a) a
    sequence of $\Read$ operations that return $0$, all of which are
    assigned linearization points before the first step at which $c[1]
    = 1$; (b) the first $\Inc$ operation that sets $c[1]$; (c) a
    sequence of $\Read$ operations that return $1$, all of which are
    linearized after $c[1] = 1$ but before $c[2] = 1$; (c) some
    $\Inc$ that is either the first to set $c[2]$ or 
    spans the step that sets $c[2]$; and (d) additional $\Inc$
    operations together with  
    $\Read$ operations that all return $2$.  Since each $\Read$
    returns the minimum of $2$ and the number of $\Inc$s that precede it, this is a
    correct linearization.

    \subsection{A very slow counter}

    Consider a \index{counter!slow}\concept{slow counter} object with
    operations $\Inc$ and $\Read$, where the value $v$ of a counter
    starts at $0$ and increases by $1$ as the result of each call to
    $\Inc$, but $\Read$ returns $\log^* v$ instead of $v$.

    Suppose we want a deterministic, wait-free, linearizable
    implementation of a slow counter as defined above from atomic
    registers.  Give tight bounds on the worst-case step complexity of
    operations on such an implementation.

    \subsubsection*{Solution}

        The worst-case step complexity of an operation is $Θ(n)$.

            For the upper bound, implement a counter on top of
            snapshots (or just collect), and have $\Read$ compute
            $\log^*$ of whatever value is read.

            For the lower bound, observe that a slow counter has
            the perturbability property needed for the JTT proof.
            Given an execution of the form $Λ_k Σ_k π$ as described in
            Chapter~\ref{chapter-JTT}, we can always
            insert some sequence of $\Inc$ operations between $Λ_k$
            and $Σ_k$ that will change the return value of $π$.
            The number of $\Inc$s needed will be the number needed to
            raise $\log^* v$, plus an extra $n$ to overcome the
            possibility of pending $\Inc$s in $Σ_k$ being linearized before
            or after $π$.  Since this object is perturbable, and the
            atomic registers we are implementing it from are
            historyless, JTT applies and gives an $Ω(n)$ lower bound
            on the cost of $\Read$ in the worst case.

    \subsection{Double-entry bookkeeping}

    \newFunc{\DEBtransfer}{transfer}
    \newFunc{\DEBclose}{close}

    Consider an object that implements an unbounded collection of
    accounts $A_1, A_2, \dots$, each of which holds an integer value,
    and that provides three operations:
    \begin{itemize}
        \item The operation $\Read(i)$ returns the current value of $A_i$.
        \item The operation $\DEBtransfer(i,j,n)$ moves $n$ units from $A_i$ to $A_j$;
            if $A'_i$ and $A'_j$ are the new values, then $A'_i = A_i
            - n$ and $A'_j = A_j + n$.
        \item The operation $\DEBclose(i,j)$ sets $A_i$ to zero and
            adds the previous value to $A_j$.  It is equivalent to atomically executing
            $\DEBtransfer(i,j,\Read(i))$.
    \end{itemize}

    \begin{enumerate}
        \item What is the consensus number of this object?
        \item What is the consensus number of a restricted version of
            this object that provides
            only the $\Read$ and $\DEBtransfer$ operations?
    \end{enumerate}

    \subsubsection*{Solution}

    \begin{enumerate}
        \item The consensus number of the object is infinite.
            Initialize $A_0$ to $1$ and the remaining $A_i$ to $0$.
            We can solve ID consensus by having process $i$ (where $i
            > 0$ execute $\DEBclose(0,i)$ and then applying $\Read$ to
            scan all the $A_j$ values for itself and other processes.
            Whichever process gets the
            $1$ wins.
        \item The consensus number without $\DEBclose$ is $1$.  Proof:
            Observe that $\DEBtransfer$ operations commute.
    \end{enumerate}

\newcommand{\MMXIXproblem}[1]{\subsection{{#1} (20 points)}}

\section{CS465/CS565 Final Exam, May 7th, 2019}
\label{appendix-final-exam-2019}

Write your answers in the blue book(s).  Justify your answers.  Work
alone.  Do not use any notes or books.  

There are four problems on this exam, each worth 20
points, for a total of 80 points.
You have approximately three hours to complete this
exam.

\MMXIXproblem{A roster}

\newFunc{\RosterAnnounce}{announce}

A \concept{roster} object has operations \RosterAnnounce and \Read,
where \Read returns a list of the identities of all processes that
have previously called \RosterAnnounce at least once.

Suppose you want a wait-free, linearizable implementation of this
object from multiwriter atomic registers.  As a function of the number
of processes $n$, give the best upper and lower bound you can on the
number of registers you will need.

You may assume that the set of process identities is fixed for each
$n$ and that each process knows its own identity.

\subsubsection*{Solution}

You will need exactly $n$ registers ($Θ(n)$ is also an acceptable
answer).

For the upper bound, have each process write its ID to its own
register, and use a double-collect snapshot to read all of them.
This uses exactly $n$ registers.  The double-collect snapshot is
wait-free because after each process has called $\RosterAnnounce$
once, the contents of the registers never change, so $\Read$ finishes
after $O(n)$ collects or $O(n^2)$ register reads.  It's linearizable
because double-collect snapshot returns the exact contents of the
registers at some point during its execution.\footnote{If we only do a
single collect, the implementation is not linearizable.  An example of
a bad execution is one where a reader reads $r_1$, then $p_1$ writes
to $r_1$ (starting and finishing its $\RosterAnnounce$ operation), then $p_2$
writes to $r_2$ (starting an finishing its $\RosterAnnounce$
operation), and finally the reader reads $r_2$.  In this execution the
reader will return $\Set{p_2}$ only, which is inconsistent with the
observed ordering that puts $\RosterAnnounce(p_1)$ before
$\RosterAnnounce(p_2)$.}

For the lower bound, use a covering argument.\footnote{It's tempting
to use JTT~\cite{JayantiTT2000} here, but the roster object is not
perturbable.  Once all IDs of $p_1$ through $p_{n-1}$ have been registered, subsequent
$\RosterAnnounce$ operations by $p_1$ through $p_{n-1}$ have no effect.  

The subtlety here is that in the JTT argument, we probably won't
choose $γ$ when perturbing $Λ_kΣ_k$ to include more than one new
$\RosterAnnounce$, but replacing $γ$ by $γ'$ to hit the first possible
uncovered register in $π$ might involve an arbitrary sequence of
operations by $p_1$ through $p_{n-1}$, including a sequence where all
of them call $\RosterAnnounce$ at least once.  Once this happens, we
get a $Λ_{k+1} Σ_{k+1}$ execution that can no longer be perturbed.}

Have the processes
$p_1,\dots,p_n$ run $\RosterAnnounce$ in order, stopping each process
when it covers a new register.
This will give sequence of partial executions $Ξ_i$ where at the end
of $Ξ_i$, there is a
set of $i$ registers $r_1\dots r_i$ that are covered by $p_1\dots
p_i$, and no other operations are in progress.

To show this works, we need to argue that each $p_{i+1}$ does in fact
cover a register $r_{i+1} ∉ \Set{r_1,\dots,r_i}$.  If not, then we can
extend $Ξ_i$ by running $p_{i+1}$'s $\RosterAnnounce$ operation to
completion, then delivering all the covering writes by $p_1\dots p_i$.
Now any subsequent $\Read$ will fail to return $p_{i+1}$, violating
the specification.  (If we have a spare process, we can have it do the
bad $\Read$; otherwise we can run $p_1$ to completion and let it do
it.)

At the end of $Ξ_n$, we have covered $n$ distinct registers, proving
the lower bound.

\MMXIXproblem{Self-stabilizing consensus}

Consider a model with $n$ processes $p_0,\dots,p_{n-1}$ organized in a
ring, where $p_i$ can directly observe both its own state and that of
$p_{(i-1) \bmod n}$.  Suppose that the state $x_i$ of each process
$p_i$ is always a natural number.

At each step, an adversary chooses a process $p_i$ to run.  This
process then updates its own state based on its previous state $x_i$ and
the state $x_{(i-1) \bmod n}$ of its counterclockwise neighbor $p_{(i-1) \bmod n}$.  The
adversary is required to run every process infinitely often but is not
otherwise constrained.

Is there a protocol for the processes that guarantees that, starting
from an arbitrary initial configuration, they eventually reach a
configuration where (a) all processes have the same state $x∈ℕ$; and
(b) no process ever changes its state as the result of taking
additional steps?

Give such a protocol and prove that it works, or show that no such
protocol is possible.

\subsubsection*{Solution}

It turns out that this problem is a good example of what happens
if you don't remember to include some sort of validity condition.  As
pointed in several student solutions, having each process pick a
fixed constant $x_i$ the first time it updates works.

Here is a protocol that also works, and satisfies the validity
condition that the common output was some process's input (which was
not required in the problem statement).  When $p_i$ takes a step, it sets $x_i$
to $\max(x_i, x_{(i-1) \bmod n})$.

To show that this works, we argue by induction that the maximum value
eventually propagates to all processes.  Let $x = x_i$ be the initial
maximum value.  The induction hypothesis is that for each
$j∈\Set{0,\dots,n-1}$, eventually all processes in the range $i$
through $i+j \pmod n$ hold value $x$ forever.

Suppose that the hypothesis holds for $j$; to show that it holds for
$j+1$, start in a configuration where $x_i$ through $x_{i+j}$ are all
$x$.  No transition can change any of these values, because taking the
max of $x$ and any other value yields $x$.  Because each process is
scheduled infinitely often, eventually $p_{i+j+1}$ takes a step; when
this happens, $x_{i+j+1}$ is set to $\max(x, x_{i+j+1}) = x$.

Since the hypothesis holds for all $j∈\Set{0,\dots,n-1}$, it holds for
$j=n-1$; but this just says that eventually all $n$ processes hold $x$ forever.

\MMXIXproblem{All-or-nothing intermittent faults}

Recall that in the standard synchronous message-passing model with
crash failures, a faulty process runs correctly up until the round in
which it crashes, during which it sends out some subset of the correct
messages, and after which it sends out no messages at all.

Suppose instead we have intermittent faults, where any process may
fail to send outgoing messages in a particular round, but these are
all-or-nothing faults in the sense that a process either sends all of
its messages in a given round or no messages in that round.  To avoid
shutting down a protocol completely, we require that in every
round, there is at least one process that sends all of its messages.
We also allow a process to send a message to itself.

If we wish to solve agreement (that is, get agreement, termination,
and validity) in this model, what is the minimum number of rounds we
need in the worst case?

\subsubsection*{Solution}

We need one round.  Every process transmits its input to all
processes, including itself.  From the all-or-nothing property,
all processes receive the same set of messages.  From the assumption
that some process is not faulty in this round, this set is nonempty.
So the processes can reach agreement by applying any consistent rule
to choose an input from the set.

\MMXIXproblem{A tamper-proof register}

Consider a \index{register!tamper-proof}\concept{tamper-proof register}, which is a
modified version of a standard multiwriter atomic register
for which the $\Read$ operation returns $⊥$ if no $\Write$ operation
has occurred, $v$ if exactly one $\Write(v)$ operation has occurred,
and $\DataSty{fail}$ if two or more $\Write$ operations have occurred.

What is the consensus number of this object?

\subsubsection*{Solution}

The consensus number is $1$.

Proof: We can implement it from atomic snapshot, which can be
implemented from atomic registers, which have consensus number $1$.

For my first $\Write(v)$ operation, write $v$ to my component of the snapshot; for subsequent
$\Write(v)$ operations, write $\DataSty{fail}$.  For a $\Read$
operation, take a snapshot and return (a) $⊥$ if all components are
empty; (b) $v$ if exactly one component is non-empty and has value
$v$; and (c) $\DataSty{fail}$ if more than one component is non-empty
or any component contains $\DataSty{fail}$.

\chapter{Sample assignments from Spring 2016}

\section{Assignment 1: due Wednesday, 2016-02-17, at 5:00pm}

\subsection*{Bureaucratic part}

Send me email!  My address is
\mailto{james.aspnes@gmail.com}.

In your message, include:

\begin{enumerate}
\item Your name.
\item Your status: whether you are an undergraduate, grad student, auditor, etc.
\item Whether you are taking the course as CPSC 465 or CPSC 565.
\item Anything else you'd like to say.
\end{enumerate}

(You will not be graded on the bureaucratic part, but you should do it anyway.)

\subsection{Sharing the wealth}

A kindergarten consists of $n$ children in a ring, numbered $0$
through $n-1$, with all arithmetic on positions taken mod $n$.

In the initial configuration, child $0$ possesses $n$ cookies.
The children take steps asynchronously, and whenever child $i$ takes a
step in a configuration where they have a cookie but child $i+1$ does
not, child $i$ gives one cookie to child $i+1$.  If child $i+1$
already has a cookie, or child $i$ has none, nothing happens.
We assume that a fairness condition guarantees that even though some
children are fast, and some are slow, each of them takes a step
infinitely often.

\begin{enumerate}
    \item Show that after some finite number of steps, every child has
        exactly one cookie.
    \item Suppose that we define a measure of time in the usual way by
        assigning each step the largest possible time consistent with
        the assumption that that no child ever waits more than one time unit to
        take a step.  Show the best asymptotic upper bound you can, as a
        function of $n$, on the
        time until every child has one cookie.
    \item Show the best asymptotic lower bound you can, as a function
        of $n$, on the worst-case time until every child has one cookie.
\end{enumerate}

    \subsection*{Solution}

    \begin{enumerate}
        \item First observe that in any configuration reachable from
            the initial configuration, child $0$ has $k$ cookies,
            $n-k$ of the remaining children have one cookie each, and
            the rest have zero cookies.  Proof: Suppose we are in a
            configuration with this property, and consider some
            possible step that changes the configuration.  Let
            $i$ be the child that takes the step.  If $i=0$, then
            child $i$ goes from $k$ to $k-1$ cookies, and child $1$
            goes from $0$ to $1$ cookies, increasing the number of
            children with one cookie to $n-k+1$.  If $i>0$, then child
            $i$ goes from $1$ to $0$ cookies and child $i+1$ from $0$
            to $1$ cookies, with $k$ unchanged.  In either case, the
            invariant is preserved.

            Now let us show that $k$ must eventually drop as long as
            some cookie-less child remains.  Let $i$
            be the smallest index such that the $i$-th child has no
            cookie.  Then after finitely many steps, child $i-1$ takes
            a step and gives child $i$ a cookie.  If $i-1=0$, $k$
            drops.  If $i-1 > 0$, then the leftmost $0$ moves one
            place to the left.  It can do so only finitely many times
            until $i=1$ and $k$ drops the next time child $0$ takes a
            step.  It follows that after finitely many steps, $k=1$,
            and by the invariant all $n-1$ remaining children also
            have one cookie each.
        \item Number the cookies $0$ through $n-1$.  When
    child $0$ takes a step, have it give the largest-numbered cookie
    it still possesses to child $1$.  For each cookie $i$, let $x_i^t$
    be the position of the $i$-th cookie after $t$ asynchronous
    rounds, where an asynchronous round is the shortest interval in
    which each child takes at least one step.

    Observe that no child $j>0$ ever gets more than one cookie, since
    no step adds a cookie to a child that already has one.  It follows
    that cookie $0$ never moves, because if child $0$ has one cookie,
    so does everybody else (including child $1$).  We can thus ignore
    the fact that the children are in a cycle and treat them as being
    in a line $0 \dots n-1$.

    We will show by induction on $t$ that, for all $i$ and $t$,
    $x_i^t ≥ y_i^t = \max(0, \min(i, z_i^t))$ where $z_i^t = t + 2(i-n+1)$.

    Proof: The base case is when $t=0$.  Here $x_i^t = 0$ for all $i$.
    We also have $z_i^t = 2(i-n+1) ≤ 0$ so $y_i^t = \max(0, \min(i,
    z_i^t)) = \max(0, z_i^t) = 0$.  So the induction hypothesis holds
    with $x_i^t = y_i^t = 0$.

    Now suppose that the induction hypothesis holds for $t$.  For each
    $i$, there are several cases to consider:
    \begin{enumerate}
        \item $x_i^t = x_{i+1}^t = 0$.  In this case
            cookie $i$ will not move, because it's not at the top of
            child $0$'s stack.  But from the induction hypothesis we
            have that $x_{i+1}^t = 0$ implies $z_{i+1}^t = t +
            2(i+1-n+1) ≤ 0$, which gives $z_i^t = z_{i+1}^t - 2 ≤ -2$.
            So $z_i^{t+1} ≤ z_{i+1}^t + 1 ≤ -1$ and $y_i^{t+1} = 0$,
            and the induction hypothesis holds for $x_i^{t+1}$.
        \item $x_i^t = i$.  Then even if cookie $i$ doesn't move (and
            it doesn't), we have $x_i^{t+1} ≥ x_i^t ≥ \min(i, z_i^t)$.
        \item $x_i^t < i$ and $x_{i+1}^t = x_i^t+1$.  Again, even if
            cookie $i$ doesn't move, we still have $x_i^{t+1} ≥ x_i^t
            = x_{i+1}^t - 1 ≥ y_{i+1}^t - 1 ≥ t + 2(i+1-n+1)-1 = t +
            2(i-n+1) + 1 > y_i^t$.
        \item $x_i^t < i$ and $x_{i+1}^t > x_i^t+1$.  Nothing is
            blocking cookie $i$, so it moves: $x_i^{t+1} = x_i^t + 1 ≥
            t + 2(i-n+1) + 1 = (t+1) + 2(i-n+1) = y_i^{t+1}$.
    \end{enumerate}

    It follows that our induction hypothesis holds for all $t$.  In
    particular, at $t=2n-2$ we have $z_i^t = 2n-2 + 2(i-n+1) = 2i-1 ≥
    i$ for all $i > 0$.  So at time $2n-2$, $x_i^t ≥ y_i^t = i$ for
    all $i$ and every child has one cookie.  This gives an asymptotic
    upper bound of $O(n)$.
        \item There is an easy lower bound of $n-1$ time.  Suppose we run
            the processes in round-robin order, i.e., the $i$-th step
            is taken by process $i \bmod n$.  Then one time unit goes
            by for every $n$ steps, during which each process takes
            exactly one step.  Since process $0$ reduces its count by
            at most $1$ per step, it takes at least $n-1$ time to get
            it to $1$.  This gives an asymptotic lower bound of
            $Ω(n)$, which is tight.

            I believe it should be possible to show an exact lower bound of
            $2n-2$ time by considering a schedule that runs in reverse
            round-robin order $n-1,n-2,\dots,0,n-1,n-2,\dots$, but
            this is more work and gets the same bound up to constants.
    \end{enumerate}

\subsection{Eccentricity}

Given a graph $G=(V,E)$, the \concept{eccentricity} $ε(v)$ of a vertex
$v$ is the maximum distance $\max_{v'} d(v,v')$ from $v$ to any 
vertex in the graph.

Suppose that you have an anonymous\footnote{Clarification added
2016-02-13: Anonymous means that processes don't have global IDs, but
they can still tell their neighbors apart.  If you want to think of
this formally, imagine that each process has a
\index{identifier!local}\concept{local identifier} for each of its neighbors: a process with
three neighbors might number them $1, 2, 3$ and when it receives or
sends a message one of these identifiers is attached.  But the local
identifiers are arbitrary, and what I call you has no
relation to what you call me or where either of us is positioned in the network.}
asynchronous message-passing system
with no failures whose network forms a tree.

\begin{enumerate}
    \item Give an algorithm that allows each node in the network to
        compute its eccentricity.
    \item Safety: Prove using an invariant
        that any value computed by a node using your
        algorithm is in fact equal to its eccentricity.  (You should
        probably have an explicit invariant for this part.)
    \item Liveness: Show that every node eventually computes its
        eccentricity in your algorithm, and that the worst-case message complexity and time complexity 
        are both within a constant factor of optimal for
        sufficiently large networks.
\end{enumerate}

\subsubsection*{Solution}

\newData{\Notified}{notified}
\newFunc{\Notify}{notify}

\begin{algorithm}
    \LocalData{
        $d[u]$ for each neighbor $u$, initially $⊥$\;
        $\Notified[u]$ for each neighbor $u$, initially $\False$\;
    }

    \bigskip

    \Initially{
        \Notify()\;
    }

    \UponReceiving{$d$ from $u$}{
        $d[u] ← d$\;
        \Notify()\;
    }

    \Procedure{\Notify()}{
        \ForEach{neighbor $u$}{
            \If{$¬\Notified[u]$ and $d[u'] ≠ ⊥$ for all $u' ≠ u$}{
                Send $1 + \max_{u'≠u} d[u']$ to $u$\;
                $\Notified[u] ← \True$\;
            }
        }
        \If{$\Notified[u] = \True$ for all neighbors $u$}{
            $ε ← \max_{u} d[u]$\;
        }
    }
    \caption{Computing eccentricity in a tree}
    \label{alg-tree-eccentricity}
\end{algorithm}

\begin{enumerate}
    \item Pseudocode is given in
        Algorithm~\ref{alg-tree-eccentricity}.  For each edge $vu$,
        the algorithm sends a message $d$ from $v$ to $u$, where $d$ is the maximum
        length of any simple path starting with $uv$.
        This can be computed as soon as $v$ knows the maximum
        distances from all of its other neighbors $u'≠u$.
    \item We now show correctness of the values computed by
        Algorithm~\ref{alg-tree-eccentricity}.
        Let $d_v[u]$ be the value of $d[u]$ at $v$.  
        Let $\ell_v[u]$ be the maximum length of any simple path starting
        with the edge $vu$.
        To show that the algorithm computes the correct values,
        we will prove the
        invariant that $d_v[u] ∈ \Set{⊥, \ell_v[u]}$ always, and for
        any message $d$ in transit from $u$ to $v$, $d = \ell_v[u]$.

        In the initial configuration, $d_v[u] = ⊥$ for all $v$ and
        $u$, and there are no messages in transit.  So the invariant
        holds.

        Now let us show that calling $\Notify$ at some process $v$
        preserves the invariant.
        Because $\Notify()$ does not change $d_v$, we need only show
        that the messages it sends contain the correct distances.

        Suppose $\Notify()$ causes $v$ to send a message $d$ to $u$.
        Then $d = 1 + \max_{u'≠u} d_v[u'] = 1 + \max_{u'≠u}
        \ell_v[u']$, because $d_v[u'] ≠ ⊥$ for all neighbors $u'≠u$
        by the condition on the if statement and thus $d_v[u'] =
        \ell_v[u']$ for all $u' ≠ u$ by the invariant.

        So the invariant will continue to hold in this case provided
        $\ell_u[v] = 1 + \max_{u'≠u} \ell_v[u']$.  The longest simple path
        starting with $uv$ either consists of $uv$ alone, or is of the
        form $uvw\dots$ for some neighbor $w$ of $v$ with $w≠u$.
        In the former case, $v$ has no other
        neighbors $u'$, in which case $d = 1+\max_{u'≠u} \ell_v[u'] = 1+ 0
        = 1$, the correct answer.  In the latter case, $d = 1 +
        \max_{u'≠u} \ell_v[u'] = 1 + \ell_v[w]$, again the length of
        the longest path starting with $uv$.

        This shows that $\Notify$ preserves the invariant.  We must
        also show that assigning $d_v[u] ← d$ upon receiving $d$ from
        $u$ does so.  But in this case we know from the invariant that
        $d = \ell_v[u]$, so assigning this value to $d_v[u]$ leaves
        $d_v[u] ∈ \Set{⊥,\ell_v[u]}$ as required.
    \item 
        First let's observe that at most one message is sent in each
        direction across each edge, for a total of $2\card{E}=2(n-1)$
        messages.  This is optimal, because if in some execution we do
        not send a message across some edge $uv$, then we can replace the
        subtree rooted at $u$ with an arbitrarily deep path, and
        obtain an execution indistinguishable to $v$ in which its
        eccentricity is different from whatever it computed.

        For time complexity (and completion!) we'll argue by induction
        on $\ell_v[u]$ that we send a message across $uv$ by time
        $\ell_v[u] - 1$.

        If $\ell_v[u] = 1$, then $u$ is a leaf; as soon as $\Notify$
        is called in its initial computation event (which we take as
        occurring at time $0$), $u$ notices it has no neighbors other
        than $v$ and sends a message to $v$.

        If $\ell_v[u] > 1$, then since $\ell_v[u] = 1 + \max_{v'≠v}
        \ell_u[v']$, we have $\ell_u[v]' ≤ \ell_v[u] - 1$ for all
        neighbors $v'≠v$ of $u$, which by the induction hypothesis
        means that each such neighbor $v'$ sends a message to $u$ no
        later than time $\ell_v[u]-2$.  These messages all arrive at
        $u$ no later than time $\ell_v[u]-1$; when the last one is
        delivered, $u$ sends a message to $v$.

        It follows that the last time a message is sent is no later
        than time $\max_{uv} (\ell_v[u] - 1)$, and so the last
        delivery event occurs no later than time $\max_{uv}
        \ell_v[u]$.  This is just the diameter $D$ of the tree, giving
        a worst-case time complexity of exactly $D$.

        To show that this is optimal, consider an execution of some
        hypothetical algorithm that terminates by time $D-1$ in the worst
        case.  Let $u$ and $v$ be nodes such that $d(u,v) = D$.  Then
        there is an execution of this algorithm in no chain of
        messages passes from $u$ to $v$, meaning that no event of $u$
        is causally related to any event of $v$.  So we can replace
        $u$ with a pair $uw$ of adjacent nodes with $d(w,v) =
        d(u,v)+1$, which changes $ε(v)$ but leaves an execution that
        is indistinguishable to $v$ from the original.  It follows
        that $v$ returns an incorrect value in some executions, and
        this hypothetical algorithm is not correct.  So time
        complexity $D$ is the best possible in the worst case.
\end{enumerate}

\subsection{Leader election on an augmented ring}

Suppose that we have an asynchronous ring where each process has a
distinct identity, but the processes do not know the size $n$ of the
ring.  Suppose also that each process $i$ can send messages not only to
its immediate neighbors, but also to the processes at positions at
positions $i-3$ and $i+3$ (mod $n$) in the ring.

Show that $Θ(n \log n)$ messages are both necessary and sufficient in
the worst case to elect a unique leader in this system.

\subsubsection*{Solution}

For sufficiency, ignore the extra edges and use
Hirschberg-Sinclair~\cite{HirschbergS1980} (see
§\ref{section-hirschberg-sinclair}).

For necessity, we'll show that an algorithm that solves leader
election in this system using at most $T(n)$ messages can be modified
to solve leader election in a standard ring without the extra edges
using at most $3T(n)$ messages.  The idea is that whenever a process
$i$ attempts to send to $i+3$, we replace the message with a sequence
of three messages relayed from $i$ to $i+1$, $i+2$, and then $i+3$,
and similarly for messages sent in the other direction.  Otherwise the
original algorithm is unmodified.  Because both systems are
asynchronous, any admissible execution in the simulated system has a
corresponding admissible execution in the simulating system (replace
each delivery event by three delivery events in a row for the relay
messages) and vice versa (remove the initial two relay delivery events
for each message and replace the third delivery event with a direct
delivery event).  So in particular if there exists an execution in the
simulating system that requires $Ω(n \log n)$ messages, then there is
a corresponding execution in the simulated system that requires at
least $Ω(n \log n / 3) = Ω(n \log n)$ messages as well.

\section{Assignment 2: due Wednesday, 2016-03-09, at 5:00pm}

    \subsection{A rotor array}

    Suppose that you are given an object that acts as described in
    Algorithm~\ref{alg-rotor-array}.  A \Write operation on this
    object writes to location $A[r]$ and increments $r$ mod $n$.  A
    \Read operation by process $i$ (where $i ∈ \Set{0\dots n-1}$)
    returns $A[i]$.  Initially, $r=0$ and $A[i] = ⊥$ for all $i$.

    \begin{algorithm}
        \Procedure{$\Write(A,v)$}{
            \Atomically{
                $A[r] ← v$;
                $r ← (r+1) \bmod n$\;
            }
        }
        \Procedure{$\Read(A)$}{
            \Return $A[i]$\;
        }
        \caption[Rotor array]{Rotor array: code for process $i$}
        \label{alg-rotor-array}
    \end{algorithm}

    What is the consensus number of this object?

        \subsubsection*{Solution}

        First let's show that it is at least $2$, by exhibiting an
        algorithm that uses a single rotor array plus two atomic
        registers to solve $2$-process wait-free consensus.

        \begin{algorithm}
            \Procedure{$\FuncSty{consensus}(v)$}{
                $\Input[i] ← v$\;
                $\Write(A,i)$\;
                $i' ← \Read(A)$\;
                \eIf{$i' = i$}{
                    \tcp{Process $0$ wrote first}
                    \Return $\Input[0]$\;
                }{
                    \tcp{Process $1$ wrote first}
                    \Return $\Input[1]$\;
                }
            }
            \caption{Two-process consensus using a rotor array}
            \label{alg-rotor-array-consensus}
        \end{algorithm}

        The algorithm is given as
        Algorithm~\ref{alg-rotor-array-consensus}.  Each process $i$ first
        writes its input value to a single-writer register
        $\Input[i]$.  The process then writes its ID to the rotor
        array.  There are two cases:
        \begin{enumerate}
            \item If process $0$ writes first, then process $0$ reads
                $0$ and process $1$ reads $1$.  Thus both processes
                see $i' = i$ and return $\Input[0]$, which gives
                agreement, and validity because $\Input[0]$ is then
                equal to $0$'s input.
            \item If process $1$ writes first, then process $0$ reads
                $1$ and process $1$ reads either $0$ (if $0$ wrote
                quickly enough) or $⊥$ (if it didn't).  In either
                case, both processes see $i'≠i$ and return
                $\Input[1]$.
        \end{enumerate}

        Now let us show that a rotor array can't be used to solve
        wait-free consensus with three processes.  We will do the
        usual bivalence argument, and concentrate on some bivalent
        configuration $C$ and pair of operations $π_0$ and $π_1$ such
        that $Cπ_i$ is $i$-valent for each $i$.

        If $π_0$ and $π_1$ are operations on different objects or
        operations on an atomic register, then they either commute or
        the usual analysis for atomic registers gives a contradiction.
        So the interesting case is when $π_0$ and $π_1$ are both
        operations on a single rotor array object $A$.

        If either operation is a $\Read$, then only the process that
        carries out the $\Read$ knows whether it occurred.  The same
        argument as for atomic registers applies in this case.  So the
        only remaining case is when both operations are $\Write$s.

        Consider the configurations $Cπ_0π_1$ (which is $0$-valent)
        and $Cπ_1π_0$ (which is $1$-valent).  These
        differ in that there are two locations $j$ and $(j+1) \bmod n$
        (which we will just write as $j+1$) that contain values $v_0$
        and $v_1$ in the first configuration and $v_1$ and $v_0$ in
        the second.  Suppose that we stop processes $j$ and $j+1$, and
        let some other process run alone until it decides.  Because
        this third process can't observe either locations $j$ or
        $j+1$, it can't distinguish between $Cπ_0π_1$ and $Cπ_1π_0$,
        and thus decides the same value starting from either
        configuration.  But this contradicts the assumption that
        $Cπ_i$ is $i$-valent.  It follows that there is no escape from
        bivalence with three processes, and the rotor array plus
        atomic registers cannot be used to solve three-process
        wait-free consensus.

        The consensus number of this object is $2$.

    \subsection{Set registers}

    \newFunc{\SetRegisterInsert}{insert}

    Suppose we want to implement a 
    \index{register!set}\concept{set register} $S$ in a
    message-passing system, where a set register provides operations
    $\SetRegisterInsert(S,v)$, which inserts a new element $v$ in $S$,
    and $\Read(S)$, which returns the set of all elements previously
    inserted into $S$.  So, for example, after executing
    $\SetRegisterInsert(S, 3)$, $\SetRegisterInsert(S, 1)$, and
    $\SetRegisterInsert(S, 1)$; $\Read(S)$ would return $\Set{1,3}$.

    \begin{enumerate}
        \item Give an algorithm for implementing a linearizable set
            register where all operations terminate in finite time,
            in a deterministic asynchronous message-passing system with $f < n/2$ crash
            failures and no failure detectors, or show that no such algorithm is possible.
        \item Suppose that we change the $\Read(S)$ operation to
            return a list of all the elements of $S$ in the order they
            were first inserted (e.g., $[3,1]$ in the example above).
            Call the resulting object an 
            \index{register!ordered set}
            \index{set register!ordered}\concept{ordered set register}.

            Give an algorithm for implementing a 
            linearizable ordered set
            register under the same conditions as above, or show that no such algorithm is possible.
    \end{enumerate}

        \subsubsection*{Solution}

        \begin{enumerate}
            \item It's probably possible to do this with some variant
                of ABD, but getting linearizability when there are
                multiple concurrent $\SetRegisterInsert$ operations
                will be tricky.

                Instead, we'll observe that it is straightforward to
                implement a set register using a shared-memory
                snapshot: each process writes to $A[i]$ the set of all
                values it has ever inserted, and a $\Read$ consists of
                taking a snapshot and then taking the union of the
                values.  Because we can implement snapshots using
                atomic registers, and we can implement atomic
                registers in a message-passing system with $f < n/2$
                crash failures using ABD, we can implement this
                construction in a message-passing system with $f <
                n/2$ failures.
            \item This we can't do.  The problem is that an ordered
                set register can solve agreement: each process inserts
                its input, and the first input wins.  But FLP says we
                can't solve agreement in an asynchronous
                message-passing system with one crash failure.
        \end{enumerate}
    
    \subsection{Bounded failure detectors}

    Suppose you have a deterministic asynchronous message-passing system equipped
    with a failure detector that is eventually weakly accurate and
    $k$-bounded strongly complete, meaning that at least $\min(k,f)$ 
    faulty processes are eventually permanently suspected by all
    processes, where $f$ is the number of faulty processes.

    For what values of $k$, $f$, and $n$ can this system solve
    agreement?

        \subsubsection*{Solution}

        We can solve agreement using the $k$-bounded failure detector
        for $n ≥ 2$ processes if and only if $f ≤ k$ and $f < n/2$.

        Proof:

        If $k ≥ f$, then every faulty process is eventually
        permanently suspected, and the $k$-bounded failure detector is
        equivalent to the $◇S$ failure detector.
        The Chandra-Toueg protocol~\cite{ChandraT1996} then solves
        consensus for us provided $f < n/2$.

        If $f ≥ n/2$, the same partitioning argument used to show
        impossibility with $◇P$ applies to the $k$-bounded detector as
        well (as indeed it applies to any failure detector that is
        only eventually accurate).

        If $k < f$, then if we have an algorithm that solves
        agreement for $n$ processes, then we can turn it into an
        algorithm that solves agreement for $n-k$ processes with $f-k$
        failures, using no failure detector at all.  The idea is that
        the $n-k$ processes can pretend that there are an extra $k$
        faulty processes that send no messages and that are
        permanently suspected.  But this algorithm runs in a standard
        asynchronous system with $f-k$ failures, and FLP says we can't solve agreement
        in such a system with $n ≥ 2$ and $f ≥ 1$.  So this rules out solving agreement
        in the original system if $k < f$ and $k ≤ n-2$.

        There is one remaining case, where $k = n-1$ and $f = n$.
        Here we can actually solve consensus when $n=1$ (because we
        can always solve consensus when $n=1$).  For larger $n$, we
        have $f ≥ n/2$.  So there is only one exception to the general
        rule that we need $f ≤ k$ and $f < n/2$.

\section{Assignment 3: due Wednesday, 2016-04-20, at 5:00pm} 

    \subsection{Fetch-and-max}

\begin{algorithm}
\Procedure{$\FetchAndMax(r, 0:x)$}{
    \eIf{$\MRswitch = 0$}{
       \Return $0 : \FetchAndMax(\MRleft, x)$\;
   }{
       \Return $1 : \FetchAndMax(\MRright, 0)$\;
   }
}

\Procedure{$\FetchAndMax(r, 1:x)$}{
    $v ← \FetchAndMax(\MRright, x)$ \;
    \eIf{$\TAS(\MRswitch) = 0$}{
        \Return $0 : \FetchAndMax(\MRleft, 0)$\;
    }{
        \Return $1 : v$\;
    }
}
\caption{Max register modified to use a test-and-set bit}
\label{alg-fetch-and-max-bogus}
\end{algorithm}

    Algorithm~\ref{alg-fetch-and-max-bogus} replaces the \MRswitch bit
    in the max register implementation from
    Algorithm~\ref{alg-max-register-write} with a test-and-set, and
    adds some extra machinery to return the old value of the register
    before the write.

    Define a fetch-and-max register as a RMW object that supports a single
    operation $\FetchAndMax(x)$ that, as an atomic action, (a) replaces the old value $v$ in
    the register with the maximum of $x$ and $v$; and (b) returns the
    old value $v$.  

    Suppose that $\MRleft$ and $\MRright$ are both
    linearizable wait-free $k$-bit fetch-and-max registers.
    Show that Algorithm~\ref{alg-fetch-and-max-bogus}
    implements a linearizable wait-free $(k+1)$-bit fetch-and-max register, or give an
    example of an execution that violates linearizability.

        \subsubsection*{Solution}

        Here is a bad execution (there are others).  Let $k=1$, and let $π_1$ do
        $\FetchAndMax(01)$ and $π_2$ do $\FetchAndMax(10)$.  Run these
        operations concurrently as follows:
        \begin{enumerate}
            \item $π_1$ reads $\MRswitch$ and sees $0$.
            \item $π_2$ does $\FetchAndMax(\MRright, 0)$.
            \item $π_2$ does $\TAS(\MRswitch)$ and sees $0$.
            \item $π_2$ does $\FetchAndMax(\MRleft, 0)$ and sees $0$.
            \item $π_1$ does $\FetchAndMax(\MRleft, 1)$ and sees $0$.
        \end{enumerate}

        Now both $π_1$ and $π_2$ return $00$.  But in the sequential
        execution $π_1 π_2$, $π_2$ returns $01$; and in the
        sequential execution $π_2 π_1$, $π_1$ returns $10$.  Since
        $π_1$ and $π_2$ 
        return the values they return in the concurrent execution in
        neither sequential execution,
        the concurrent execution is not linearizable.

    \subsection{Median}

    \newFunc{\AddSample}{addSample}
    \newFunc{\ComputeMedian}{computeMedian}

    Define a \index{register!median}\concept{median register} as an
    object $r$ with two operations $\AddSample(r,v)$, where $v$ is
    any integer, and
    $\ComputeMedian(r)$.  The $\AddSample$ operation adds a sample to
    the multiset $M$ of integers stored in the register, which is
    initially empty.  The $\ComputeMedian$ operations returns a median
    of this multiset, defined as a value $x$ with the property that
    (a) $x$ is in the multiset; (b) at least $\card{M}/2$ values $v$ in the
    multiset are less than or equal to $x$; (c) at least $\card{M}/2$ values
    $v$ in the multiset are greater than or equal to $x$.

    For example, if we add the samples $1, 1, 3, 5, 5, 6$, in any
    order, then a subsequent $\ComputeMedian$ can return either $3$ or
    $5$.

    Suppose that you wish to implement a linearizable wait-free median register using
    standard atomic registers and resettable test-and-set bits.  Give
    tight (up to constants) asymptotic upper and lower bounds on the
    number of such objects you would need.  You may assume that the
    atomic registers may hold arbitrarily large values.

        \subsubsection*{Solution}

        For the upper bound, we can do it with $O(n)$ registers using
        any linear-space snapshot algorithm (for example,
        Afek~\etal~\cite{AfekADGMS1993}).  Each process stores in its
        own segment of the snapshot object the multiset of all samples
        added by that process; $\AddSample$ just adds a new sample to
        the process's segment.  For $\ComputeMedian$, take a
        snapshot, then take the union of all the multisets, then
        compute the median of this union.  Linearizability and
        wait-freedom of both operations are immediate from the
        corresponding properties of the snapshot object.

        For the lower bound, use JTT~\cite{JayantiTT2000}.  Observe
        that both atomic registers and resettable test-and-sets are
        historyless: for both types, the new state after an operation
        doesn't depend on the old state.  So JTT applies if we can
        show that the median register is perturbable.

        Suppose that we have a schedule $Λ_k Σ_k π$ in which $Λ_k$
        consists of an arbitrary number of median-register operations
        of which at most $k$ are incomplete, $Σ_k$ consists of $k$ pending
        base object operations (writes, test-and-sets, or test-and-set
        resets) covering $k$ distinct base objects, and $π$ is a read
        operation by a process not represented in $Λ_k Σ_k$.  We need
        to find a sequence of operations $γ$ that can be inserted
        between $Λ_k$ and $Σ_k$ that changes the outcome of $π$.

        Let $S$ be the multiset of all values appearing as arguments
        to $\AddSample$ operations that start in $Λ_k$ or $Σ_k$.  Let
        $x = \max S$ (or $0$ if $S$ is empty), and let $γ$ consist of
        $\card{S} + 1$
        $\AddSample(r,x+1)$ operations.  Write $T$ for the multiset of
        $\card{S}+1$ copies of $x+1$.  Then in any linearization of
        $Λ_k γ Σ_k π$, the multiset $U$ of samples contained in $r$
        when $π$ executes includes at least all of $T$ and at most all
        of $S$; this means that a majority of values in $U$ are equal
        to $x+1$, and so the median is $x+1$.  But $x+1$ does not
        appear in $S$, so $π$ can't return it in $Λ_k Σ_k π$.  It
        follows that a median register is in fact perturbable, and JTT
        applies, which means that we need at least $Ω(n)$ base objects
        to implement a median register.

    \subsection{Randomized two-process test-and-set with small registers}

    Algorithm~\ref{alg-small-two-TAS} gives an implementation of a
    randomized one-shot test-and-set for two processes, each of which
    may call the procedure at most once, with its process ID ($0$ or
    $1$) as an argument.

    The algorithm uses two registers, $a_0$ and
    $a_1$, that are both initialized to $0$ and hold values in
    the range $0\dots m-1$, where $m$ is a positive integer.
    Unfortunately, whoever wrote it forgot to specify the value of
    $m$.

    \newData{\TwoTASmyPosition}{myPosition}
    \newData{\TwoTASotherPosition}{otherPosition}

    \begin{algorithm}
        \Procedure{$\TAS(i)$}{
            $\TwoTASmyPosition ← 0$\;
            \While{\True}{
                $\TwoTASotherPosition ← \Read(a_{¬i})$\;
                $x ← \TwoTASmyPosition - \TwoTASotherPosition$ \;
                \uIf{$x ≡ 2 \pmod{m}$}{
                    \Return 0\;
                }
                \uElseIf{$x ≡ -1 \pmod{m}$}{
                    \Return 1\;
                }
                \uElseIf{fair coin comes up heads}{
                    $\TwoTASmyPosition ← (\TwoTASmyPosition + 1) \bmod m$\;
                }
                $\Write(a_i, \TwoTASmyPosition)$\;
            }
        }
        \caption{Randomized one-shot test-and-set for two processes}
        \label{alg-small-two-TAS}
    \end{algorithm}

    For what values of $m$ does
    Algorithm~\ref{alg-small-two-TAS} correctly implement a one-shot,
    probabilistic wait-free, linearizable test-and-set, assuming:
    \begin{enumerate}
        \item An oblivious adversary?
        \item An adaptive adversary?
    \end{enumerate}

        \subsubsection*{Solution}

        For the oblivious adversary, we can quickly rule out $m < 5$,
        by showing that there is an execution in each case where both
        processes return $0$:
        \begin{itemize}
            \item When $m=1$ or $m=2$, both processes immediately return $0$,
                because the initial difference $0$ is congruent to $2$
                mod $m$.
            \item When $m=3$, there is an execution in which $p_0$
                writes $1$ to $a_0$, $p_1$ reads this $1$ and computes
                $x = -1 ≡ 2 \pmod{3}$ and returns $0$, then $p_0$
                reads $0$ from $a_0$, computes $x=1$, advances $a_0$ to $2$, then
                re-reads $0$ from $a_0$, computes $x=2$, and returns
                $0$.
            \item When $m=4$, run $p_0$ until it writes $2$ to $a_0$.
                It then computes $x = 2$ and returns $0$.  If we now
                wake up $p_1$, it computes $x=-2≡2 \pmod{4}$ and also
                returns $0$.
        \end{itemize}

        When $m≥5$ and the adversary is oblivious, the implementation works.  We need to show both
        linearizability and termination.  We'll start with
        linearizability.

        Observe that in a sequential execution,
        first process to perform \TAS returns $0$ and the
        second $1$.  So we need to show (a) that the processes between
        them return both values, and (b) that if one process finishes
        before the other starts, the first process returns $0$.

        It is immediate from Algorithm~\ref{alg-small-two-TAS} that in
        any reachable configuration, $\TwoTASmyPosition_i ∈
        \Set{a_i,a_i+1}$, because process $i$ can only increment
        $\TwoTASmyPosition$ at most once before writing its value to $a_i$.

        Below we will assume without loss of generality that $p_0$ is
        the first process to perform its last read before returning.
        
        Suppose that $p_0$ returns $0$.  This means that $p_0$
        observed $a_1 ≡ a_0 - 2 \pmod{m}$.  So at the time $p_0$ last
        read $a_1$, $\TwoTASmyPosition_1$ was congruent to either $a_0 - 1$
        or $a_0 - 2$.  This means that on its next read of $a_0$,
        $p_1$ will compute $x$ congruent to either $-1$ or $-2$.
        Because $m$ is at least $5$, in neither case will it mistake
        this difference for $2$.
        If it computed $x ≡ -1$, it returns $1$; if it computed $x ≡
        -2$, it does not return immediately, but eventually it will
        flip its coin heads, increment $\TwoTASmyPosition_1$, and return
        $1$.  In either case we have that exactly one process returns
        each value.

        Alternatively, suppose that $p_0$ returns $1$.  Then $p_0$
        reads $a_1 ≡ a_0 + 1$, and at the time of this read,
        $\TwoTASmyPosition_1$ is either congruent to $a_0 + 1$ or $a_0 + 2$.
        In the latter case, $p_1$ returns $0$ after its next read; in
        the former, $p_1$ eventually increments $\TwoTASmyPosition_1$ and
        then returns $0$.  In either case we again have that exactly
        one process returns each value.

        Now suppose that $p_0$ runs to completion before $p_1$ starts.
        Initially, $p_0$ sees $a_0 ≡ a_1$, but eventually $p_0$
        increments $a_0$ enough times that $a_0 - a_1 ≡ 2$; $p_0$
        returns $0$.

        To show termination (with probability $1$), consider any
        configuration in which neither process has returned.  During
        the next $2k$ steps, at least one process takes $k$ steps.
        Suppose that during this interval, this fast process increments $\TwoTASmyPosition$ at every
        opportunity, while the other process does not increment
        $\TwoTASmyPosition$ at all (this event occurs with nonzero
        probability for any fixed $k$, because the coin-flips are
        uncorrelated with the oblivious adversary's choice of which
        process is fast).  Then for $k$ sufficiently
        large, the fast process eventually sees $a_0 - a_1$ congruent
        to either $2$ or $-1$ and returns.  Since this event occurs
        with independent nonzero probability in each interval of
        length $2k$, eventually it occurs.\footnote{The fancy way to
            prove this is to invoke
            the second Borel-Cantelli lemma of
            probability theory.  Or we can just argue that the
            probability that we don't terminate in the first $\ell$
            intervals is at most $(1-ε)^\ell$, which goes to zero in
        the limit.}

        Once one process has terminated, the other increments
        $\TwoTASmyPosition$ infinitely often, so it too eventually sees a
        gap of $2$ or $-1$.
        
        For the adaptive adversary, the adversary can prevent the
        algorithm from terminating.  Starting from a state in which
        both processes are about to read and $a_0 = a_1 = k$, run
        $p_0$ until it is about to write $(k+1) \bmod m$ to $a_0$
        (unlike the oblivious adversary, the adaptive adversary can
        see when this will happen).  Then run $p_1$ until it is about
        to write $(k+1) \bmod m$ to $a_1$.  Let both writes go
        through.  We are now in a state in which both processes are
        about to read, and $a_0 = a_1 = (k+1) \bmod m$.  So we can
        repeat this strategy forever.

\section{Presentation (for students taking CPSC 565): due Wednesday,
2016-04-27}

    Students taking CPSC 565, the graduate version of the class, are
    expected to give a 15-minute
    presentation on a recent paper in the
    theory of distributed computing.

    The choice of paper to present should be made in consultation with
    the instructor.  To a first approximation, any paper from PODC,
    DISC, or a similar conference in the last two or three years (that
    is not otherwise covered in class)
    should work.

    Because of the limited presentation time, you are not required to
    get into all of the technical details of the paper, but your
    presentation should include\footnote{Literary theorists will recognize this as a three-act structure
    (preceded by a title card): introduce the main character, make
    their life difficult, then resolve their problems in time for the
    final curtain.  This is not the only way to organize a talk, but
if done right it has the advantage of keeping the audience awake.}
    \begin{enumerate}
        \item Title, authors, and date and venue of publication of the paper.
        \item A high-level description of the main result.  Unlike a
            talk for a general audience, you can assume that your
            listeners know at least everything that we've talked about
            so far in the class.
        \item A description of where this result fits into the
            literature (e.g., solves an open problem previously
            proposed in [...], improves on the previous best running
            time for an algorithm from [...], gives a lower bound or
            impossibility result for a problem previously proposed by
            [...], opens up a new area of research for studying
            [...]), and why it is interesting and/or hard.
        \item A description (also possibly high-level) of the main
            technical mechanism(s) used to get the main result.
    \end{enumerate}

    You do not have to prepare slides for your presentation if you
    would prefer to use the blackboard, but you should make sure to
    practice it in advance to make sure it fits in the
    allocated time.  The instructor will be happy to offer feedback on
    draft versions if available far enough before the actual
    presentation date.

    Relevant dates:
    \begin{description}
        \item[2016-04-13] Paper selection due.
        \item[2016-04-22] Last date to send draft slides or arrange for a
            practice presentation with the instructor if you want
            guaranteed feedback.
        \item[2016-04-27] Presentations, during the usual class time.
    \end{description}

\newcommand{\MMXVIproblem}[1]{\subsection{{#1} (20 points)}}

\section{CS465/CS565 Final Exam, May 10th, 2016}

Write your answers in the blue book(s).  Justify your answers.  Work
alone.  Do not use any notes or books.  

There are four problems on this exam, each worth 20
points, for a total of 80 points.
You have approximately three hours to complete this
exam.

\MMXVIproblem{A slow register}

Define a \index{register!second-to-last}\concept{second-to-last
register} as having a read operation that always returns the
second-to-last value written to it.  For example, after  
\Write(1), \Write(2), \Write(3), a subsequent $\Read$ operation will
return $2$.  If fewer that two $\Write$ operations have occurred, a
$\Read$ will return $⊥$.

What is the consensus number of this object?

\subsubsection*{Solution}

The consensus number of this object is $2$.

For two processes, have each process $i$ write its input to a standard
atomic register $r[i]$, and then write its ID to a shared
second-to-last-value register $s$.  We will have whichever process
writes to $s$ first win.  After writing, process $i$ can detect which
process wrote first by reading $s$ once, because it either sees $⊥$ (meaning the other
process has not written yet) or it sees the identity of the process
that wrote first.  In either case it can return the winning process's
input.

For three processes, the usual argument gets us to a configuration
$C$ where all three processes are about to execute operations $x$, $y$,
and $z$ on the same object, where each operation moves from a bivalent
to a univalent state.  Because we know that this object can't be a
standard atomic register, it must be a second-to-last register.
We can also argue that all of $x$, $y$, and $z$ are writes, because if
one of them is not, the processes that don't perform it can't tell if
it happened or not.

Suppose that $Cx$ is $0$-valent and $Cy$ is 1-valent.  Then $Cxyz$ is
$0$-valent and $Cyz$ is $1$-valent.  But these configurations are
indistinguishable to any process but $x$.  It follows that the
second-to-last register can't solve consensus for three processes.

\MMXVIproblem{Two leaders}

Assume that you are working in an asynchronous message-passing system
organized as a connected graph, where all processes run the same code
except that each process starts with an ID and the knowledge of the IDs
of all of its neighbors.  Suppose that all of these IDs are unique,
except that the smallest ID (whatever it is) might appear on either one or two
processes.

Is it possible in all cases to detect which of these situations hold?
Either give an algorithm that allows all processes to eventually
correctly return whether there are one or two minimum-id processes in
an arbitrary connected graph, or show that no such algorithm is
possible.

\subsubsection*{Solution}

Here is an algorithm.

If there are two processes $p$ and $q$ with the same ID that are
adjacent to each other, they can detect this in the initial
configuration, and transmit this fact to all the other processes by
flooding.

If these processes $p$ and $q$ are not adjacent, we will need some
other mechanism to detect them.
Define the extended ID of a process as its own
ID followed by a list of the IDs of its neighbors in some fixed order.
Order the extended IDs lexicographically, so that a process with a
smaller ID also has a smaller extended ID.

Suppose now that $p$ and $q$ are not adjacent and have the same
extended ID.  Then they share the same neighbors, and each of these
neighbors will see that $p$ and $q$ have duplicate IDs.  So we can do
an initial round of messages where each process transmits its extended
ID to its neighbors, and if $p$ and $q$ observe that their ID is a
duplicate, they can again notify all the processes to return that
there are two leaders by flooding.

The remaining case is that $p$ and $q$ have distinct extended IDs, or
that only one minimum-process ID exists.  In either case we can run
any standard broadcast-based leader-election algorithm, using the
extended IDs, which will leave us with a tree rooted at whichever
process has the minimum extended ID.  This process can then perform
convergecast to detect if there is another process with the same ID,
and perform broadcast to inform all processes of this fact.

\MMXVIproblem{A splitter using one-bit registers}

Algorithm~\ref{alg-one-bit-register-splitter} implements a
splitter-like object using one-bit registers.  
It assumes that each process has a unique ID $\DataSty{ID}$ consisting of
$k = \ceil{\lg n}$ bits 
$\DataSty{ID}_{k-1} \DataSty{ID}_{k-2} \dots \DataSty{ID}_0$.
We would like this
object to have the properties that (a) if exactly one process executes
the algorithm, then it wins; and (b) in any execution, at most one
process wins.

\begin{algorithm}
    \SharedData\\
    one-bit atomic registers $A[i][j]$ for $i=0\dots\ceil{\lg n}-1$
    and $j∈\Set{0,1}$, all initially $0$\;
    one-bit atomic register $\DataSty{door}$, initially $0$\;
    \Procedure{$\FuncSty{splitter}(\DataSty{ID})$}{
        \For{$i ← 0$ \KwTo $k-1$}{
            $A[i][\DataSty{ID}_i] ← 1$\;
        }
        \If{$\DataSty{door} = 1$}{
            \Return \DataSty{lose}\;
        }
        $\DataSty{door} ← 1$\;
        \For{$i ← 0$ \KwTo $k-1$}{
            \If{$A[i][¬\DataSty{ID}_i] = 1$}{
                \Return \DataSty{lose}\;
            }
        }
        \Return \DataSty{win}\;
    }
    \caption{Splitter using one-bit registers}
    \label{alg-one-bit-register-splitter}
\end{algorithm}

Show that the algorithm has these properties, or give an example of an
execution where it does not.

\subsubsection*{Solution}

The implementation is correct.

If one process runs alone, it sets $A[i][\DataSty{ID}_i]$ for each
$i$, sees $0$ in $\DataSty{door}$, then sees $0$ in each location
$A[i][¬\DataSty{ID}_i]$ and wins.  So we have property (a).

Now suppose that some process with ID $p$ wins in an execution that may involve
other processes.  Then $p$ writes $A[i][p_i]$ for all $i$ before
observing $0$ in $\DataSty{door}$, which means that it sets all these
bits before any process writes $1$ to $\DataSty{door}$.  If some other
process $q$ also wins, then there is at least one position $i$ where
$p_i = ¬q_i$, and $q$ reads 
$A[i][p_i]$ after writing $1$ to $\DataSty{door}$.  But then $q$ sees
$1$ in this location and loses, a contradiction.

\MMXVIproblem{Symmetric self-stabilizing consensus}

Suppose we have a synchronous system consisting of processes organized
in a connected graph.  The state of each process is a single bit, and
each process can directly observe the number of neighbors that it has
and how many of them have $0$ bits and how many have $1$ bits.  At
each round, a process counts the number of neighbors $k_0$ with zeros,
the number $k_1$ with ones, and its own bit $b$, and chooses a new bit
for the next round $f(b, k_0, k_1)$ according to some 
rule $f$ that is the same for all processes.  The goal of the
processes is to reach consensus, where all processes have the same
bit forever, starting from an arbitrary initial configuration.  An
example of a rule that has this property is for $f$ to output $1$ if
$b=1$ or $k_1 > 0$.

However, this rule is not symmetric with respect to bit values: if we
replace all ones by zeros and vice versa, we get different behavior.

Prove or disprove: There exists a rule $f$ that is symmetric, by which
we mean that $f(b, k_0, k_1) = ¬f(¬b, k_1, k_0)$ always, such that
applying this rule starting from an arbitrary configuration in an
arbitrary graph eventually converges to all processes having the same
bit forever.

\subsubsection*{Solution}

Disproof by counterexample: Fix some $f$, and consider a graph with
two processes $p_0$ and $p_1$ connected by an edge.  Let $p_0$ start
with $0$ and $p_1$ start with $1$.  Then $p_0$'s next state is
$f(0,0,1) = ¬f(1,1,0) ≠ f(1,1,0)$, which is $p_1$'s next state.  So
either $p_0$ still has $0$ and $p_1$ still has $1$, in which case we
never make progress; or they swap their bits, in which case we can
apply the same analysis with $p_0$ and $p_1$ reversed to show that
they continue to swap back and forth forever.  In either case the
system does not converge.

\chapter{Sample assignments from Spring 2014}

\section{Assignment 1: due Wednesday, 2014-01-29, at 5:00pm}

\subsection*{Bureaucratic part}

Send me email!  My address is
\mailto{james.aspnes@gmail.com}.

In your message, include:

\begin{enumerate}
\item Your name.
\item Your status: whether you are an undergraduate, grad student, auditor, etc.
\item Anything else you'd like to say.
\end{enumerate}

(You will not be graded on the bureaucratic part, but you should do it anyway.)

\subsection{Counting evil processes}

A connected bidirectional 
asynchronous network of $n$ processes with identities has
diameter $D$ and may contain zero or more evil processes.  Fortunately,
the evil processes, if they exist, are not Byzantine, fully conform to
RFC 3514~\cite{rfc3514}, and will correctly execute any code we
provide for them.

Suppose that all processes wake up at time $0$ and start whatever
protocol we have given them.  Suppose 
that each process initially knows whether it is evil, and knows the identities of
all of its neighbors.
However, the processes do not know the number of processes $n$ or the
diameter of the network $D$.

Give a protocol that allows every process to correctly return the
number of evil processes no later than time $D$.  Your protocol should only
return a value once for each process (no converging to the correct answer after an
initial wrong guess).

    \subsubsection*{Solution}

    There are a lot of ways to do this.  Since the problem doesn't ask
    about message complexity, we'll do it in a way that optimizes for
    algorithmic simplicity.

    At time $0$, each process initiates a separate copy of the
    flooding algorithm (Algorithm~\ref{alg-flooding}).  The message 
    $\Tuple{p,N(p),e}$
    it
    distributes consists of its own identity, the identities of all of
    its neighbors, and whether or not it is evil.

    In addition to the data for the flooding protocol, each process
    tracks a set $I$ of all processes it has seen that initiated a
    protocol and a set $N$ of all processes that have been mentioned
    as neighbors.  
    The initial values of these sets for process $p$ are $\Set{p}$ and
    $N(p)$, the neighbors of $p$.

    Upon receiving a message $\Tuple{q,N(q),e}$, a process
    adds $q$ to $I$ and $N(q)$ to $N$.  As soon as $I=N$, the process
    returns a count of all processes for which $e=\True$.

    Termination by $D$: Follows from the same analysis as flooding.  
    Any process at
    distance $d$ from $p$ has $p∈I$ by time $d$, so $I$ is complete by
    time $D$.

    Correct answer: Observe that $N = \bigcup_{i∈I} N(i)$
    always.  Suppose that there is some process
    $q$ that is not in $I$.  Since the graph is connected, there is a path from $p$ to
    $q$.  Let $r$ be the last node in this path in $I$, and let $s$ be
    the following node.  Then $s∈N∖I$ and $N≠I$.  By contraposition,
    if $I=N$ then $I$ contains all nodes in the network, and so the
    count returned at this time is correct.

\subsection{Avoiding expensive processes}

Suppose that you have a bidirectional but not necessarily complete
asynchronous message-passing 
network represented by a graph $G=(V,E)$ where each node in $V$
represents a process and each edge in $E$ connects two processes that
can send messages to each other.  Suppose further that each process is
assigned a weight $1$ or $2$.  Starting at some initiator process,
we'd like to construct a shortest-path tree, where each process points
to one of its neighbors as its parent, and following the parent
pointers always gives a path of minimum total weight to the
initiator.\footnote{Clarification added 
2014-01-26: The actual number of hops is not relevant for the
construction of the shortest-path tree.
By
shortest path, we mean path of minimum total weight.}

Give a protocol that solves this problem with reasonable time,
message, and bit complexity, and show that it works.

    \subsubsection*{Solution}

    There's an ambiguity in the definition of total weight: does it
    include the weight of the initiator and/or the initial node in the
    path?  But since these values are the same for all paths to the
    initiator from a given process, they don't affect which is
    lightest.

    If we don't care about bit complexity, there is a
    trivial solution: Use an existing BFS algorithm followed by
    convergecast to gather the entire structure of the network at the
    initiator, run your favorite single-source shortest-path algorithm
    there, then broadcast the results.  This has time complexity
    $O(D)$ and message complexity $O(DE)$ if we use the BFS algorithm
    from §\ref{section-distributed-BFS-synchronizer}.  But the last
    couple of messages in the convergecast are going to be pretty big.

    A solution by reduction: Suppose that we construct a new graph
    $G'$ where each weight-$2$ node $u$ in $G$ is replaced by a clique
    of nodes $u_1, u_2, \dots u_k$,
    with each node in the clique attached to a different neighbor of $u$.
    We then run any breadth-first search protocol of our choosing on
    $G'$, where each weight-$2$ node simulates all members of the
    corresponding clique.  Because any path that passes through a
    clique picks up an extra edge, each path in the breadth-first
    search tree has a length exactly equal to the sum of the weights
    of the nodes other than its endpoints.

    A complication is that if I am simulating $k$ nodes, between them
    they may have more than one parent pointer.  So we define
    $u.\Parent$ to be $u_i.\Parent$ where $u_i$ is a node at minimum
    distance from the initiator in $G'$.  We also re-route any
    incoming pointers to $u_j ≠ u_i$ to point to $u_i$ instead.
    Because $u_i$ was chosen to have minimum distance, 
    this never increases the length of any path, and the resulting
    modified tree is a still a shortest-path tree.

    Adding nodes blows up $\card*{E'}$, but we don't need to actually
    send messages between different nodes $u_i$ represented by the
    same process.  So if we use the
    §\ref{section-distributed-BFS-synchronizer} algorithm again, we
    only send up to $D$ messages per real edge, giving $O(D)$ time and
    $O(DE)$ messages.

    If we don't like reductions, we could also tweak one of our
    existing algorithms.  Gallager's layered BFS
    (§\ref{section-distributed-BFS-layered}) is easily modified by
    changing the depth bound for each round to a total-weight bound.
    The synchronizer-based BFS can also be modified to work, but the
    details are messy.

\section{Assignment 2: due Wednesday, 2014-02-12, at 5:00pm}

    \subsection{Synchronous agreement with weak failures}

    Suppose that we modify the problem of synchronous agreement with
    crash failures from Chapter~\ref{chapter-synchronous-agreement} so
    that instead of crashing a process forever, the adversary may jam
    some or all of its outgoing messages for a single round.  The
    adversary has limited batteries on its jamming equipment, and can
    only cause $f$ such one-round faults.  There is no restriction on
    when these one-round jamming faults occur: the adversary might jam
    $f$ processes for one round, one process for $f$ rounds, or
    anything in between, so long as the sum over all rounds of the
    number of processes jammed in each round is at most $f$.
    For the purposes of agreement and validity, assume that a process
    is non-faulty if it is never jammed.\footnote{
    Clarifications added 2014-02-10: We assume that processes don't
    know that they are being jammed or which messages are lost (unless
    the recipient manages to tell them that a message was not delivered).
    As in the original model, we assume a complete network and that
    all processes have known identities.}

    As a function of $f$ and $n$, how many rounds does it take to reach
    agreement in the worst case in this model, under the usual
    assumptions that processes are deterministic and the algorithm
    must satisfy agreement, termination, and validity?  Give the best upper
    and lower bounds that you can.

        \subsubsection*{Solution}

        The par solution for this is an $Ω(\sqrt{f})$ lower bound and
        $O(f)$ upper bound.  I don't know if it is easy to do better
        than this.

        For the lower bound, observe that the adversary can simulate
        an ordinary crash failure by jamming a process in every round
        starting in the round it crashes in.  This means that in an
        $r$-round protocol, we can simulate $k$ crash failures with
        $kr$ jamming faults.  From the Dolev-Strong lower
        bound~\cite{DolevS1983} (see also
        Chapter~\ref{chapter-synchronous-agreement}), we
        know that there is no $r$-round protocol with $k=r$ crash
        failures faults, so
        there is no $r$-round protocol with $r^2$ jamming faults.
        This gives a lower bound of $\floor{\sqrt{f}}+1$ on the number
        of rounds needed to solve synchronous agreement with $f$
        jamming faults.\footnote{Since Dolev-Strong only needs to
            crash one process per round, we don't really need the full
            $r$ jamming faults for processes that crash late.  This
        could be used to improve the constant for this argument.}

        For the upper bound, have every process broadcast its input
        every round.  After $f+1$ rounds, there is at least one round
        in which no process is jammed, so every process learns all the
        inputs and can take, say, the majority value.  
    
    \subsection{Byzantine agreement with contiguous faults}

    Suppose that we restrict the adversary in Byzantine agreement to
    corrupting a connected subgraph of the network; the
    idea is that the faulty nodes need to coordinate, but can't
    relay messages through the non-faulty nodes to do so.

    Assume the usual model for Byzantine agreement with a network in
    the form of an $m×m$ torus.  This means that each node has a
    position $(x,y)$ in $\Set{0,\dots,m-1}×\Set{0,\dots,m-1}$, and 
    its neighbors are the four nodes
    $(x+1 \bmod m, y)$,
    $(x-1 \bmod m, y)$,
    $(x, y+1 \bmod m)$, and
    $(x, y-1 \bmod m)$.

    For sufficiently large $m$,\footnote{Problem modified 2014-02-03.
        In the original version, it asked to compute $f$ for all $m$,
    but there are some nasty special cases when $m$ is small.}
        what is the largest number of faults $f$;
    that this system can tolerate and still solve Byzantine agreement?

        \subsubsection*{Solution}
        
        The relevant bound here is the requirement that the network
        have enough connectivity that the adversary can't take over
        half of a vertex cut (see
        §\ref{section-Byzantine-minimum-connectivity}).  This is
        complicated slightly by the requirement that the faulty nodes
        be contiguous.

        The smallest vertex cut in a sufficiently large torus consists of the four
        neighbors of a single node; however, these nodes are not
        connected.  But we can add a third node to connect two of
        them (see Figure~\ref{fig-assignment-Byzantine-contiguous}).

\begin{figure}
    \centering
\begin{tikzpicture}
    \node[circle,draw] (00) {};
    \node[circle,draw] (01) [right of=00] {};
    \node[circle,draw] (02) [right of=01] {};
    \node[circle,draw] (10) [below of=00] {};
    \node[circle,fill,draw,color=blue] (11) [right of=10] {};
    \node[circle,fill,draw,color=red] (12) [right of=11] {};
    \node[circle,draw] (20) [below of=10] {};
    \node[circle,fill,draw,color=red] (21) [right of=20] {};
    \node[circle,fill,draw,color=red] (22) [right of=21] {};

    \path
    (00) edge[dotted] (01) edge[dotted] (10)
    (01) edge[dotted] (02) edge (11)
    (02) edge[dotted] (12)
    (10) edge (11) edge[dotted] (20)
        (11) edge (12) edge (21)
        (12) edge[color=red,dashed] (22)
        (20) edge[dotted] (21)
        (21) edge[color=red,dashed] (22)
    ;
\end{tikzpicture}
\caption{Connected Byzantine nodes take over half a cut}
\label{fig-assignment-Byzantine-contiguous}
\end{figure}

    By adapting the usual lower bound we can use this construction to
    show that $f=3$ faults
    are enough to prevent agreement when $m≥3$.  The question then is
    whether $f=2$ faults is enough.

    By a case analysis, we can show that any two nodes in a
    sufficiently large torus are either adjacent themselves or can be
    connected by three paths, where no two paths have adjacent
    vertices.  Assume without loss of generality that one of the nodes
    is at position $(0,0)$.  Then any other node is covered by one of
    the following cases:
    \begin{enumerate}
        \item Nodes adjacent to $(0,0)$.  These can communicate
            directly.
        \item Nodes at $(0,i)$ or $(i,0)$.  These cases are symmetric,
            so we'll describe the solution for $(0,i)$.  Run one path
            directly north: $(0,1), (0,2), \dots, (0,i-1)$.  Similarly,
            run a second path south: $(0, -1), (0,-2), \dots (0,i+1)$.
            For the third path, take two steps east and then run
            north and back west: $(1,0), (2,0), (2,1), (2,2), \dots,
            (2,i), (1,i)$.  These paths are all non-adjacent as long
            as $m≥4$.
        \item Nodes at $(±1,i)$ or $(i,±1)$, where $i$ is not $-1$,
            $0$, or $1$.  Suppose the node is at $(1,i)$.  Run one
            path east then north through
            $(1,0), (1,1), \dots, (1,i-1)$.  The other
            two paths run south and west, with a sideways jog in the
            middle as needed.  This works for $m$ sufficiently large
            to make room for the sideways jogs.
        \item Nodes at $(±1,±1)$ or $(i,j)$ where neither of $i$ or
            $j$ is $-1$, $0$, or $1$.  Now we can run one path north
            then east, one east then north, one south then west, and
            one west then south, creating four paths in a figure-eight
            pattern centered on $(0,0)$.
    \end{enumerate}

\section{Assignment 3: due Wednesday, 2014-02-26, at 5:00pm}

    \subsection{Among the elect}

    The adversary has decided to be polite and notify each non-faulty
    processes when he gives up crashing it.  Specifically, we have
    the usual asynchronous message-passing system with up to $f$
    faulty processes, but every non-faulty process is eventually told
    that it is non-faulty.  (Faulty processes are told nothing.)

    For what values of $f$ can you solve consensus in this model?

        \subsubsection*{Solution}

        We can tolerate $f<n/2$, but no more.

        If $f<n/2$, the following algorithm works: Run Paxos, where
        each process $i$ waits to learn that it is non-faulty, then
        acts as a proposer for proposal number $i$.  The
        highest-numbered non-faulty process then carries out a
        proposal round that succeeds because no higher proposal is
        ever issued, and both the proposer (which is non-faulty) and a
        majority of accepters participate.

        If $f≥n/2$, partition the processes into two groups of
        size $\floor{n/2}$, with any leftover process crashing
        immediately.  Make all of the processes in both groups
        non-faulty, and tell each of them this at the start of the
        protocol.  Now do the usual partitioning argument:
        Run group $0$ with inputs $0$ with no messages
        delivered from group $1$ until all processes
        decide $0$ (we can do this because the processes can't
        distinguish this execution from one in which the group $1$
        processes are in fact faulty).  Run group $1$ similarly until
        all processes decide $1$.  We have then violated agreement,
        assuming we didn't previously violate termination of validity.

    \subsection{Failure detectors on the cheap}

    Suppose we do not have the budget to equip all of our machines
    with failure detectors.  Instead, we order an 
    eventually strong failure detector for $k$ machines, 
    and the remaining $n-k$
    machines get fake failure detectors that never suspect anybody.
    The choice of which machines get the real failure detectors and
    which get the fake ones is under the control of the adversary.

    This means that every faulty process is eventually permanently
    suspected by every non-faulty process with a real failure
    detector, and there is at least one non-faulty process that is
    eventually permanently not suspected by anybody.  Let's call the
    resulting failure detector $◇S_k$.

    Let $f$ be the number of actual failures.  Under what conditions
    on $f$, $k$, and $n$ can you still solve consensus in the usual
    deterministic asynchronous message-passing model using $◇S_k$?

        \subsubsection*{Solution}

        First observe that $◇S$ can simulate $◇S_k$ for any $k$ by
        having $n-k$ processes ignore the output of their failure
        detectors.  So we need $f < n/2$ by the usual lower bound on
        $◇S$.

        If $f≥k$, we are also in trouble.  The $f>k$ case is easy: If
        there exists a consensus protocol for $f>k$, then we can
        transform it into a consensus protocol
        for $n-k$ processes and $f-k$
        failures, with no failure detectors at all,
        by pretending that there are an extra $k$ processes
        with real failure detectors that crash immediately.
        The FLP impossibility result rules this out.

        If $f=k$, we have to be a little more careful.
        By immediately crashing $f-1$ processes with real failure
        detectors, we can reduce to the $f=k=1$ case.  Now the
        adversary runs the FLP strategy.  If no processes crash, then
        all $n-k+1$ surviving process report no failures; if it
        becomes necessary to crash a process, this becomes the one
        remaining process with the real failure detector.  In either
        case the adversary successfully prevents consensus.

        So let $f < k$.  Then we have weak completeness,
        because every faulty process is eventually permanently
        suspected by at least $k-f > 0$ processes.  We also have weak
        accuracy, because it is still the case that some process is
        eventually permanently never suspected by anybody.  By
        boosting weak completeness to strong completeness as described
        in §\ref{section-failure-detector-boosting-completeness}, we
        can turn out failure detector into $◇S$, meaning we can solve
        consensus precisely when $f < \min(k, n/2)$.

\section{Assignment 4: due Wednesday, 2014-03-26, at 5:00pm} 

    \subsection{A global synchronizer with a global clock}

    Consider an asynchronous message-passing system with $n$ processes in a
    bidirectional ring with no failures.  Suppose that the
    processes are equipped with a global clock, which causes a local
    event to occur simultaneously at each process every $c$ time
    units, where as usual $1$ is the maximum message delay.  We would
    like to use this global clock to build a global synchronizer.
    Provided $c$ is at least $1$, a trivial approach is to have every
    process advance to the next round whenever the clock pulse hits.
    This gives one synchronous round every $c$ time units.

    Suppose that $c$ is greater than $1$ but still $o(n)$.  Is it
    possible to build a global synchronizer in this model that runs
    more than a constant ratio faster than 
    this trivial global synchronizer in the worst case?

        \subsubsection*{Solution}

        No.  We can adapt the lower bound on the session problem from
        §\ref{section-session-problem} to apply in this model.

        Consider an execution of an algorithm for the session problem
        in which each message is delivered exactly one time unit after
        it is sent.
        Divide it as in the previous proof into a prefix $β$ containing
        special actions and a suffix $δ$ containing no special
        actions.  Divide $β$ further into segments $β_1, β_2, β_3, \dots, β_k$,
        where each segment ends with a clock pulse.  
        Following the standard argument, because each segment has
        duration less than the diameter of the network, there is no
        causal connection between any special actions done by
        processes at opposite ends of the network that are in the same
        segment $β_i$.  So we can causally shuffle each $β_i$ to get a
        new segment $β'_i$ where all special actions of process $p_0$
        (say) occur before all special actions of process $p_{n/2}$.
        This gives at most one session per segment, or at most one
        session for every $c$ time units.

        Since a globally 
        synchronous system can do one session per round, this
        means that our global synchronizer can only produce one
        session per $c$ time units as well.

    \subsection{A message-passing counter}
    \label{problem-message-passing-counter}

    A \concept{counter} is a shared object that support operations
    $\Inc$ and $\Read$, where $\Read$ returns the number of
    previous $\Inc$ operations.

    Algorithm~\ref{alg-problem-broadcast-counter} purports to
    implement a counter in an asynchronous
    message-passing system subject to $f < n/2$ crash failures. 
    In the algorithm, each process $i$ maintains a vector $c_i$ of
    contributions to the counter from all the processes, as well as a
    nonce $r_i$ used to distinguish responses to
    different read operations from each other.  All of these values
    are initially zero.

    Show
    that the implemented counter is
    linearizable, or give an example of an execution where it isn't.

    \begin{algorithm}
        \Procedure{\Inc}{
            $c_i[i] ← c_i[i] + 1$ \;
            Send $c_i[i]$ to all processes.\;
            Wait to receive $\Ack(c_i[i])$ from a majority of
            processes.\;
        }

        \UponReceiving{$c$ from $j$}{
            $c_i[j] ← \max(c_i[j], c)$ \;
            Send $\Ack(c)$ to $j$.\;
        }

        \Procedure{\Read}{
            $r_i ← r_i+1$\;
            Send $\Read(r_i)$ to all processes.\;
            Wait to receive $\Respond(r_i, c_j)$ from a majority of processes $j$.\;
            \Return $∑_k \max_j c_j[k]$
        }

        \UponReceiving{$\Read(r)$ from $j$}{
            Send $\Respond(r, c_i)$ to $j$\;
        }
    \caption{Counter algorithm for
    Problem~\ref{problem-message-passing-counter}.}
    \label{alg-problem-broadcast-counter}
    \end{algorithm}

        \subsubsection*{Solution}

        This algorithm is basically implementing an array of ABD
        registers~\cite{AttiyaBD1995}, but it
        omits the second phase on a $\Read$ where any information the
        reader learns is propagated to a majority.  So we expect it to
        fail the same way ABD would without this second round, by
        having two $\Read$ operations return values that are out of
        order with respect to their observable ordering.

        Here is one execution that produces this bad outcome:
        \begin{enumerate}
            \item Process $p_1$ starts an $\Inc$ by updating $c_1[1]$
                to $1$.
            \item Process $p_2$ carries out a $\Read$ operation in
                which it receives responses from $p_1$ and $p_2$, and
                returns $1$.
            \item After $p_2$ finishes, process $p_3$ carries out a
                $\Read$ operation in which it receives responses from
                $p_2$ and $p_3$, and returns $0$.
        \end{enumerate}

        If we want to be particularly perverse, we can exploit the
        fact that $p_2$ doesn't record what it learns in its first $\Read$ to
        have $p_2$ do the second $\Read$ that returns $0$ instead of
        $p_3$.  This shows that
        Algorithm~\ref{alg-problem-broadcast-counter} isn't even
        sequentially consistent.

        The patch, if we want to fix this, is to include the missing
        second phase from ABD in the $\Read$ operation: after
        receiving values from a majority, I set $c_i[k]$ to $\max_j
        c_j[k]$ and send my updated values to a majority.  That the
        resulting counter is linearizable is left as an exercise.

\section{Assignment 5: due Wednesday, 2014-04-09, at 5:00pm}

\subsection{A concurrency detector}
\label{section-assignment-concurrency-detector}

\newFunc{\CDenter}{enter}
\newFunc{\CDexit}{exit}

Consider the following optimistic mutex-like object, which we will
call a \concept{concurrency detector}.  A concurrency detector
supports two operations for each process $i$, $\CDenter_i$ and
$\CDexit_i$.  These operations come in pairs: a process enters a
critical section by executing $\CDenter_i$, and leaves by executing
$\CDexit_i$.  The behavior of the object is undefined if a process
calls $\CDenter_i$ twice without an intervening $\CDexit_i$, or calls
$\CDexit_i$ without first calling $\CDenter_i$.

Unlike mutex, a concurrency detector does not
enforce that only one process is in the critical section at a time;
instead, $\CDexit_i$ returns $1$ if the interval between it and the previous
$\CDenter_i$ overlaps with some interval between
a $\CDenter_j$ and corresponding $\CDexit_j$ for some $j≠i$, and
returns $0$ if there is no overlap.

Is there a deterministic linearizable
wait-free implementation of a concurrency detector from atomic
registers?  If there is, give an implementation.  If there is not,
give an impossibility proof.

    \subsubsection*{Solution}

    It is not possible to implement this object using atomic
    registers.

    Suppose that there were such an implementation.
    Algorithm~\ref{alg-assignment-concurrency-detector} implements
    two-process consensus using a two atomic registers and a single
    concurrency detector, initialized to the state following
    $\CDenter_1$.

    \begin{algorithm}
        \newFunc{\CDConsensus}{consensus}
        \Procedure{$\CDConsensus_1(v)$}{
            $r_1 ← v$\;
            \eIf{$\CDexit_1() = 1$}{
                \Return $r_2$\;
            }{
                \Return $v$\;
            }
        }
        \Procedure{$\CDConsensus_2(v)$}{
            $r_2 ← v$\;
            $\CDenter_2()$\;
            \eIf{$\CDexit_2() = 1$}{
                \Return $v$\;
            }{
                \Return $r_1$\;
            }
        }
        \caption{Two-process consensus using the object from
        Problem~\ref{section-assignment-concurrency-detector}}
        \label{alg-assignment-concurrency-detector}
    \end{algorithm}

    Termination is immediate from the absence of loops in the code.

    To show validity and termination, observe that one of two cases
    holds:
    \begin{enumerate}
        \item Process $1$ executes $\CDexit_1$ before process $2$
            executes $\CDenter_2$.  In this case there is no overlap
            between the interval formed by the implicit $\CDenter_1$
            and $\CDexit_1$ and the interval formed by $\CDenter_2$
            and $\CDexit_2$.  So the $\CDexit_1$ and $\CDexit_2$
            operations both return $0$, causing process $1$ to return
            its own value and process $2$ to return the contents of
            $r_1$.  These will equal process $1$'s value, because
            process $2$'s read follows its call to $\CDenter_2$, which
            follows $\CDexit_1$ and thus process $1$'s write to $r_1$.
        \item Process $1$ executes $\CDexit_1$ after process $2$
            executes $\CDenter_2$.  Now both $\CDexit$ operations
            return $1$, and so process $2$ returns its own value while
            process $1$ returns the contents of $r_2$, which it reads
            after process $2$ writes its value there.
    \end{enumerate}

    In either case, both processes return the value of the first
    process to access the concurrency detector, satisfying both
    agreement and validity.  This would give a consensus protocol for
    two processes implemented from atomic registers, contradicting the
    impossibility result of Loui and Abu-Amara~\cite{LouiA1987}.

    \subsection{Two-writer sticky bits}

    A \index{sticky bit!two-writer}
    \concept{two-writer sticky bit} is a sticky bit that can be read
    by any process, but that can only be written to by two specific
    processes.

    Suppose that you have an unlimited collection of two-writer sticky
    bits for each pair of processes, plus as many ordinary atomic
    registers as you need.  What is the maximum number of processes
    for which you can solve wait-free binary consensus? 

        \subsubsection*{Solution}

        If $n=2$, then a two-writer sticky bit is equivalent to a
        sticky bit, so we can solve consensus.

        If $n≥3$, 
        suppose that we maneuver our processes as usual to a
        bivalent configuration $C$ with no bivalent successors.  
        Then there are three pending operations $x$, $y$, and $z$, that among them
        produce both $0$-valent and $1$-valent configurations.
        Without loss of generality, suppose that $Cx$ and $Cy$ are
        both $0$-valent and $Cz$ is $1$-valent.  We now consider what
        operations these might be.

        \begin{enumerate}
            \item If $x$ and $z$ apply to different objects, then
                $Cxz=Czx$ must be both $0$-valent and $1$-valent, a
                contradiction.  Similarly if $y$ and $z$ apply to
                different objects.  This shows that all three
                operations apply to the same object $O$.
            \item If $O$ is a register, then the usual case analysis
                of Loui and Abu-Amara~\cite{LouiA1987} gives us a
                contradiction.
            \item If $O$ is a two-writer sticky bit, then we can split
                cases further based on $z$:
                \begin{enumerate}
                    \item If $z$ is a read, then either:
                        \begin{enumerate}
                            \item At least one of $x$ and $y$ is a
                                read.  But then $Cxz=Czx$ or
                                $Cyz=Czy$, and we are in trouble.
                            \item Both $x$ and $y$ are writes.  But
                                then $Czx$ ($1$-valent) is indistinguishable from
                                $Cx$ ($0$-valent) by the two processes
                                that didn't perform $z$: more trouble.
                        \end{enumerate}
                    \item If $z$ is a write, then at least one of $x$
                        or $y$ is a read; suppose it's $x$.  Then
                        $Cxz$ is indistinguishable from $Cz$ by the
                        two processes that didn't perform $x$.
                \end{enumerate}
        \end{enumerate}

        Since we reach a contradiction in all cases, it must be that
        when $n≥3$,
        every bivalent configuration has a bivalent successor, which
        shows that we can't solve consensus in this case.  
        The
        maximum value of $n$ for which we can solve consensus is $2$.

\section{Assignment 6: due Wednesday, 2014-04-23, at 5:00pm}

    \subsection{A rotate register}

    \newFunc{\RotateLeft}{RotateLeft}

    Suppose that you are asked to implement a concurrent $m$-bit register that
    supports in addition to the usual \Read and \Write operations
    a \RotateLeft operation that rotates all the bits to the
    left; this is equivalent to doing a left shift (multiplying the
    value in the register by two) followed by replacing the
    lowest-order bit with the previous highest-order bit.

    For example, if the register contains $1101$, and we do
    \RotateLeft, it now contains $1011$.

    Show that if $m$ is sufficiently large as a function of the number
    of processes $n$,
    $Θ(n)$ steps per operation in the worst case are necessary and sufficient
    to implement a linearizable wait-free $m$-bit shift register from atomic registers.

        \subsubsection*{Solution}

        The necessary part is easier, although we can't use JTT
        (Chapter~\ref{chapter-JTT}) directly because having write
        operations means that our rotate register is not perturbable.
        Instead, we argue that if we initialize the register to $1$,
        we get a mod-$m$ counter, where increment is implemented by
        $\RotateLeft$ and read is implemented by taking the log of the
        actual value of the counter.  Letting $m ≥ 2n$ gives the
        desired $Ω(n)$ lower bound, since a mod-$2n$ counter \emph{is}
        perturbable.

        For sufficiency, we'll show how to implement the rotate
        register using snapshots.  This is pretty much a standard
        application of known
        techniques~\cite{AspnesH1990waitfree,AndersonM1993}, but it's
        not a bad exercise to write it out.

        Pseudocode for one possible solution is given in
        Algorithm~\ref{alg-assignment-rotate-register}.

        The register is implemented using a single snapshot array $A$.
        Each entry in the snapshot array holds four values: a
        timestamp and process ID indicating which write the process's most recent
        operations apply to, the initial write value corresponding to
        this timestamp, and the number of rotate operations this
        process has applied to this value.  A write operation
        generates a new timestamp, sets the written value to its
        input, and resets the rotate count to $0$.  A rotate operation
        updates the timestamp and associated write value to the most recent that
        the process sees, and adjusts the rotate count as appropriate.
        A read operation combines all the rotate counts associated
        with the most recent write to obtain the value of the
        simulated register.

        \newData{\RRtimestamp}{timestamp}

        \begin{algorithm}

            \Procedure{$\Write(A,v)$}{
                $s ← \Snapshot(A)$ \;
                $A[\Id] ← \Tuple{\max_i s[i].\RRtimestamp + 1, \Id, v, 0}$ \;
            }

            \Procedure{$\RotateLeft(A)$}{
                $s ← \Snapshot(A)$ \;
            Let $i$ maximize $\Tuple{s[i].\RRtimestamp, s[i].\DataSty{process}}$ \;
                \eIf{
                    $s[i].\RRtimestamp = A[\Id].\RRtimestamp$ 
                    \KwAnd
                    $s[i].\DataSty{process} = A[\Id].\DataSty{process}$}{
                        \tcp{Increment my rotation count}
                        $A[\Id].\DataSty{rotations} ←
                        A[\Id].\DataSty{rotations} + 1$ \;
                    }{
                        \tcp{Reset and increment my rotation count}
                        $A[\Id] ← \Tuple{s[i].\RRtimestamp,
                            s[i].\DataSty{process},
                            s[i].\DataSty{value},
                    1}$\;
                }
            }

            \Procedure{$\Read(A)$}{
                $s ← \Snapshot(A)$ \;
                Let $i$ maximize $\Tuple{s[i].\RRtimestamp, s[i].\DataSty{process}}$ \;
                Let $r = ∑_{j, s[j].\RRtimestamp = s[i].\RRtimestamp ∧
                    s[j].\DataSty{process} = s[i].\DataSty{process}}
                    s[j].\DataSty{rotations}$\;
                \Return $s[i].\DataSty{value}$ rotated $r$ times.\;
                }

            \caption{Implementation of a rotate register}
            \label{alg-assignment-rotate-register}
        \end{algorithm}

        Since each operation requires one snapshot and at most one
        update, the cost is $O(n)$ using the linear-time snapshot
        algorithm of Inoue~\etal~\cite{InoueMCT1994}.  Linearizability
        is easily verified by observing that ordering all operations
        by the maximum timestamp/process tuple that they compute
        and then by the total number of rotations that
        they observe produces an ordering consistent with the concurrent execution
        for which all return values of reads are correct.

    \subsection{A randomized two-process test-and-set}
    \label{section-assignment-2TAS}

    Algorithm~\ref{alg-assignment-2TAS} gives pseudocode for a
    protocol for two processes $p_0$ and $p_1$.  It uses two shared unbounded single-writer
    atomic registers $r_0$ and $r_1$, both initially $0$.  Each
    process also has a local variable $s$.

    \begin{algorithm}
        \Procedure{$\FuncSty{TAS}_i()$}{
            \While{\True}{
                \eWithProbability{$1/2$}{
                    $r_i ← r_i+1$\;
                }{
                    $r_i ← r_i$\;
                }
                $s ← r_{¬i}$\;
                \uIf{$s > r_i$}{
                    \Return 1\;
                }
                \ElseIf{$s < r_i - 1$}{
                    \Return 0\;
                }
            }
        }

        \caption{Randomized two-process test-and-set
        for~\protect{\ref{section-assignment-2TAS}}}
    \label{alg-assignment-2TAS}
    \end{algorithm}

    \begin{enumerate}
        \item Show that any return values of the protocol are
            consistent with a linearizable, single-use test-and-set.
        \item Will this protocol always terminate with probability $1$
            assuming an oblivious adversary?
        \item Will this protocol always terminate with probability $1$
            assuming an adaptive adversary?
    \end{enumerate}

        \subsubsection*{Solution}

        \begin{enumerate}
            \item To show that this implements a linearizable
                test-and-set, we need to show that exactly one process
                returns $0$ and the other $1$, and that if one process
                finishes before the other starts, the first process to
                go returns $1$.

                Suppose that $p_i$ finishes before $p_{¬i}$ starts.  Then
                $p_i$ reads only $0$ from $r_{¬i}$, and cannot observe
                $r_i < r_{¬i}$: $p_i$ returns $0$ in this case.

                We now show that the two processes cannot return the
                same value.  Suppose that both processes terminate.
                Let $i$ be such that $p_i$ reads $r_{¬i}$ for the last
                time before $p_{¬i}$ reads $r_i$ for the last time.
                If $p_i$ returns $0$, then it observes $r_i ≥
                r_{¬i}+2$ at the time of its read; $p_{¬i}$ can
                increment $r_{¬i}$ at most once before reading $r_i$
                again, and so observed $r_{¬i} < r_i$ and returns $1$.

                Alternatively, if $p_i$ returns $1$, it observed $r_i
                < r_{¬i}$.  Since it performs no more increments on
                $r_i$, $p_i$ also observes $r_i < r_{¬i}$ in all
                subsequent reads, and so cannot also return $1$.

            \item Let's run the protocol with an oblivious adversary,
                and track the value of $r_0^t - r_1^t$ over time,
                where $r_i^t$ is the value of $r_i$ after $t$ writes
                (to either register).  Each
                write to $r_0$ increases this value by $1/2$ on
                average, with a change of $0$ or $1$ equally likely,
                and each write to $r_1$ decreases it by $1/2$ on
                average.  
                
                To make things look symmetric, let $Δ^t$ be
                the change caused by the $t$-th write and write $Δ^t$
                as $c^t + X^t$ where $c^t = ±1/2$ is a constant
                determined by whether $p_0$ or $p_1$ does the $t$-th
                write and $X^t = ±1/2$ is a random variable with
                expectation $0$.  Observe that the $X^t$ variables are
                independent of each other and the constants $c^t$
                (which depend only on the schedule).

                For the protocol to run forever, at every time $t$ it
                must hold that $\abs*{r_0^t-r_1^t} ≤ 3$; otherwise,
                even after one or both processes does its next write, we will
                have $\abs*{r_0^{t'}-r_1^{t'}}$ and the next process to
                read will terminate.  But
                \begin{align*}
                    \abs*{r_0^t - r_1^t}
                    &= \abs*{∑_{s=1}^{t} Δ^s}
                    \\&= \abs*{∑_{s=1}^{t} (c_s + X_s)}
                    \\&= \abs*{∑_{s=1}^{t} c_s + ∑_{s=1}^{t} X_s}.
                \end{align*}

                The left-hand sum is a constant, while the right-hand
                sum has a binomial distribution.  For any fixed
                constant, the probability that a binomial distribution
                lands within $±2$ of the constant goes to zero in the
                limit as $t→∞$, so with probability $1$ there is some
                $t$ for which this event does not occur.
                
            \item For an adaptive adversary, the following strategy
                prevents agreement:
                \begin{enumerate}
                    \item Run $p_0$ until it is about to increment
                        $r_0$.
                    \item Run $p_1$ until it is about to increment
                        $r_1$.
                    \item Allow both increments to proceed and repeat.
                \end{enumerate}

                The effect is that both processes always observe $r_0
                = r_1$ whenever they do a read, and so never finish.
                This works because the adaptive adversary can see the
                coin-flips done by the processes before they act on
                them; it would not work with an oblivious adversary or
                in a model that supported probabilistic writes.
        \end{enumerate}

\newcommand{\MMXIVproblem}[1]{\subsection{{#1} (20 points)}}

\section{CS465/CS565 Final Exam, May 2nd, 2014}
\label{appendix-final-exam-2014-solutions}

Write your answers in the blue book(s).  Justify your answers.  Work
alone.  Do not use any notes or books.  

There are four problems on this exam, each worth 20
points, for a total of 80 points.
You have approximately three hours to complete this
exam.

\MMXIVproblem{Maxima}

Some deterministic 
processes organized in an anonymous, synchronous ring are each given an integer
input (which may or may not be distinct from other processes' inputs),
but otherwise run the same code and do not know the size of the ring.
We would like the processes to each compute the maximum input.  As
usual, each process may only return an output once, and must do
so after a finite number of rounds, although it
may continue to participate in the protocol (say, by relaying
messages) even after it returns an output.

Prove or disprove: It is possible to solve this problem in this model.

\subsubsection*{Solution}

It's not possible.

Consider an execution with $n=3$ processes, each with input $0$.
If the protocol is correct, then
after some finite number of rounds $t$, each process returns $0$.
By symmetry, the processes all have the same states and send the same
messages throughout this execution.

Now consider a ring of size $2(t+1)$ where every process has input $0$,
except for one process $p$ that has input $1$.  Let 
$q$ be the process at
maximum distance from $p$.  By induction on
$r$, we can show that after $r$ rounds of communication, every process
that is more than $r+1$ hops away from $p$ has the same state as all
of the processes in
the $3$-process execution above.  So in particular, after $t$ rounds,
process $q$ (at distance $t+1$) is in the same state as it would be in the $3$-process
execution, and thus it returns $0$.  But—as it learns to its horror,
one round too late—the correct maximum is $1$.

\MMXIVproblem{Historyless objects}

Recall that a shared-memory object is \concept{historyless} if any
operation on the object either (a) always leaves the object in the
same state as before the operation, or (b) always leaves the object in
a new state that doesn't depend on the state before the operation.

What is the maximum possible consensus number for a historyless
object?  That is, for what value $n$ is it possible to solve wait-free consensus
for $n$ processes using some particular historyless object but not
possible to solve wait-free consensus for $n+1$ processes using any historyless
object?

    \subsubsection*{Solution}

    Test-and-sets are (a) historyless, and (b) have consensus number
    2, so $n$ is at least $2$.

    To show that no historyless object can solve wait-free $3$-process
    consensus, consider an execution that starts in a bivalent
    configuration and runs to a configuration $C$ with two pending
    operations $x$ and $y$ such that $Cx$ is $0$-valent and $Cy$ is
    $1$-valent.  By the usual arguments $x$ and $y$ must both be
    operations on the same object.  If either of $x$ and $y$ is a read
    operation, then ($0$-valent) $Cxy$ and ($1$-valent) $Cyx$ 
    are indistinguishable to a third
    process $p_z$ if run alone, because the object is left in the same
    state in both configurations; whichever way $p_z$ decides, it will
    give a contradiction in an execution starting with one of these
    configurations.  If neither of $x$ and $y$ is a read, then $x$
    overwrites $y$, and $Cx$ is indistinguishable from $Cyx$to $p_z$
    if $p_z$ runs alone; again we get a contradiction.

\MMXIVproblem{Hams}

\newFunc{\LaunchHam}{launch}

Hamazon, LLC, claims to be the world's biggest delivery service for
canned hams, with guaranteed delivery of a canned ham to your home
anywhere on Earth via suborbital trajectory from
secret launch facilities at the North and South Poles.  Unfortunately,
these launch facilities may be subject to crash failures due to
inclement weather, trademark infringement actions, or military
retaliation for misdirected hams.

For this problem, you are to evaluate Hamazon's business model from
the perspective of distributed algorithms.  Consider a system
consisting of a client process and two server processes (corresponding
to the North and South Pole facilities) that communicate by means of
asynchronous message passing.  In addition to the usual
message-passing actions, each server also has an irrevocable
\LaunchHam action that launches a ham at the client. 
As with messages, hams are delivered
asynchronously: it is impossible for the client to tell if a ham has
been launched until it arrives.

A ham protocol is correct provided (a) a client that orders no ham
receives no ham; and (b) a client that orders a ham receives exactly
one ham.  Show that there can be no correct deterministic protocol for this
problem if one of the servers can crash.

    \subsubsection*{Solution}

    Consider an execution in which the client orders ham.  Run the
    northern server together with the client until the server is about
    to issue a \LaunchHam action (if it never does so, the
    client receives no ham when the southern server is faulty).

    Now run the client together with the southern server.  There are
    two cases:
    \begin{enumerate}
        \item If the southern server ever issues \LaunchHam, execute
            both this and the northern server's \LaunchHam
            actions: the client gets two hams.
        \item If the southern server never issues \LaunchHam, 
            never run the northern server again: the client gets
            no hams.
    \end{enumerate}

    In either case, the one-ham rule is violated, and the protocol is
    not correct.\footnote{It's tempting to try to solve this problem
        by reduction from a known impossibility result, like Two
        Generals or FLP.  For these specific problems, direct
        reductions don't appear to work.  Two Generals assumes message
        loss, but
        in this model, messages are not lost.  FLP needs any
        process to be able to fail, but in this model, the client
        never fails.  Indeed, we can solve consensus in the Hamazon
        model by just having the client transmit its input to both
    servers.}

\MMXIVproblem{Mutexes}

\newFunc{\MMXIVswap}{swap}

A swap register $s$ has an operation $\MMXIVswap(s,v)$ that returns the
argument to the previous call to $\MMXIVswap$, or $⊥$ if it is the first such
operation applied to the register.  It's easy to build a mutex from a
swap register by treating it as a test-and-set: to grab the mutex, I
swap in $1$, and if I get back $⊥$ I win (and otherwise try again); 
and to release the mutex, I put back $⊥$.

Unfortunately, this implementation is not starvation-free: some other
process acquiring the mutex repeatedly
might always snatch the $⊥$ away just before I try to swap it out.
Algorithm~\ref{alg-MMXIV-alleged-mutex} uses a swap object $s$ along
with an atomic register $r$ to try to fix this.

\begin{algorithm}
    \Procedure{$\FuncSty{mutex}()$}{
        $\DataSty{predecessor} ← \MMXIVswap(s,\MyId)$\;
        \While{$r ≠ \DataSty{predecessor}$}{
            try again\;
        }
        \tcp{Start of critical section}
        $\ldots$\;
        \tcp{End of critical section}
        $r ← \MyId$\;
    }
    \caption{Mutex using a swap object and register}
    \label{alg-MMXIV-alleged-mutex}
\end{algorithm}

Prove that Algorithm~\ref{alg-MMXIV-alleged-mutex} gives a
starvation-free mutex, or give an example of an execution where it
fails.  You should assume that $s$ and $r$ are both initialized to
$⊥$.

\subsubsection*{Solution}

Because processes use the same ID if they try to access the mutex
twice, the algorithm doesn't work.

Here's an example of a bad execution:
\begin{enumerate}
    \item Process $1$ swaps $1$ into $s$ and gets $⊥$, reads $⊥$
        from $r$, performs its critical section, and writes $1$ to
        $r$.
    \item Process $2$ swaps $2$ into $s$ and gets $1$, reads $1$ from
        $r$, and enters the critical section.
    \item Process $1$ swaps $1$ into $s$ and gets $2$, and spins
        waiting to see $2$ in $r$.
    \item Process $3$ swaps $3$ into $s$ and gets $1$.  Because $r$ is
        still $1$, process $3$ reads this $1$ and enters the critical
        section.  We now have two processes in the critical section,
        violating mutual exclusion.
\end{enumerate}

I believe this works if each process adopts a new ID every time it
calls \FuncSty{mutex}, but the proof is a little tricky.\footnote{The
    simplest proof I can come up with is to apply an invariant that
    says that (a) the processes that have executed $\MMXIVswap(s,\MyId)$
    but have not yet left the while loop have $\DataSty{predecessor}$
    values that form a linked list, with the last pointer either equal
    to $⊥$ (if no process has yet entered the critical section) or the
    last process to enter the critical section; (b) $r$ is $⊥$ if
    no process has yet left the critical section, or the last process to
    leave the critical section otherwise; and (c) if there is a
    process that is in the critical section, its
    $\DataSty{predecessor}$ field points to the last process to leave
    the critical section.  Checking the effects of each operation shows that this
    invariant is preserved through the execution, and $(a)$ combined
    with $(c)$ show that we can't have two processes in the
    critical section at the same time.  Additional work is still
    needed to show starvation-freedom.  It's a good thing this
algorithm doesn't work as written.}

\chapter{Sample assignments from Fall 2011}

\section{Assignment 1: due Wednesday, 2011-09-28, at 17:00} 

\subsection*{Bureaucratic part}

Send me email!  My address is
\mailto{aspnes@cs.yale.edu}.

In your message, include:

\begin{enumerate}
\item Your name.
\item Your status: whether you are an undergraduate, grad student, auditor, etc.
\item Anything else you'd like to say.
\end{enumerate}

(You will not be graded on the bureaucratic part, but you should do it anyway.)

\subsection{Anonymous algorithms on a torus} 

An $n \times m$ \concept{torus} is a two-dimensional version of a ring, where a node
at position $(i,j)$ has a neighbor to the north at $(i,j-1)$, the east
at $(i+1,j)$, the south at $(i,j+1)$, and the west at $(i-1,j)$.
These values wrap around modulo $n$ for the first coordinate and
modulo $m$ for
the second; so $(0,0)$ has neighbors $(0,m-1)$, $(1,0)$, $(0,1)$, and
$(n-1,0)$.

Suppose that we have a synchronous message-passing network in the form
of an $n\times m$ torus, consisting of anonymous, identical processes
that do not know $n$, $m$, or their own coordinates, but do have a
sense of direction (meaning they can tell which of their neighbors is
north, east, etc.).

Prove or disprove: Under these conditions, there is a
deterministic\footnote{Clarification added 2011-09-28.} algorithm that
computes whether $n > m$.

\subsubsection*{Solution}

Disproof: Consider two executions, one in an $n\times m$ torus and one
in an $m \times n$ torus where $n > m$ and both $n$ and $m$ are at
least $2$.\footnote{This last assumption is not strictly necessary, but
it avoids having to worry about what it means when a process sends a
message to itself.}  Using the same argument as in
Lemma~\ref{lemma-symmetry}, show by induction on the round
number that, for each round $r$, all processes in both executions have
the same state.  It follows that if the processes correctly detect
$n>m$ in the $n\times m$ execution, then they incorrectly report $m>n$
in the $m\times n$ execution.

\subsection{Clustering} 

Suppose that $k$ of the nodes in an asynchronous message-passing
network are designated as cluster heads, and we want to have each
node learn the identity of the nearest head.  Given the most efficient
algorithm you can for this problem, and compute its worst-case time
and message complexities.

You may assume that processes have
unique identifiers and that all processes know how many neighbors they
have.\footnote{Clarification added 2011-09-26.} 

\subsubsection*{Solution}

The simplest approach would be to 
run either of the efficient distributed breadth-first search
algorithms from Chapter~\ref{chapter-distributed-BFS} simultaneously
starting at all cluster heads, and have each process learn the
distance to all cluster heads at once and pick the nearest one.  This
gives $O(D^2)$ time and $O(k(E+VD))$ messages if we use layering and
$O(D)$ time and $O(kDE)$ messages using local synchronization.

We can get rid of the dependence on $k$ in the local-synchronization
algorithm by running it almost unmodified, with the only difference
being the attachment of a cluster head ID to the \BFSexactly messages.
The simplest way to show that the resulting algorithm works is to
imagine coalescing all cluster heads into a single initiator; the
clustering algorithm effectively simulates the original algorithm
running in this modified graph, and the same proof goes through.  The
running time is still $O(D)$ and the message complexity $O(DE)$.

\subsection{Negotiation}

Two merchants $A$ and $B$ are colluding to fix the price of some valuable
commodity, by sending messages to each other for $r$ rounds in a synchronous
message-passing system.  To avoid the attention of antitrust
regulators, the merchants are transmitting their messages via carrier
pigeons, which are unreliable and may become lost.  Each merchant has
an initial price $p_A$ or $p_B$, which are integer values satisfying
$0 ≤ p ≤ m$ for some known value $m$, and their goal is to choose new
prices $p'_A$ and $p'_B$, where $\abs*{p'_A - p'_B} ≤ 1$.
If $p_A = p_B$ and no messages are lost, they want the stronger
goal that $p'_A = p'_B = p_A = p_B$.

Prove the best lower bound you can on $r$, as a function of $m$, for
all protocols that achieve these goals.

\subsubsection*{Solution}

This is a thinly-disguised version of the Two Generals Problem from
Chapter~\ref{chapter-two-generals}, with the agreement condition $p'_A
= p'_B$ replaced by an 
\index{agreement!approximate}\concept{approximate agreement} condition
$\abs*{p'_A - p'_B} ≤ 1$.  We can use a proof based on the
indistinguishability argument in
§\ref{section-two-generals-impossible} to show that $r ≥ m/2$.

Fix $r$, and suppose that in a failure-free execution both processes
send messages in all rounds (we can easily modify an algorithm that
does not have this property to have it, without increasing $r$).
We will start with a sequence of executions with $p_A = p_B = 0$.  Let
$X_0$ be the execution in which no messages are lost, $X_1$ the
execution in which $A$'s last message is lost, $X_2$ the execution in
which both $A$ and $B$'s last messages are lost, and so on, with
$X_k$ for $0 ≤ k ≤ 2r$ losing $k$ messages split evenly between the two processes,
breaking ties in favor of losing messages from $A$.

When $i$ is even, $X_i$ is indistinguishable from $X_{i+1}$ by
$A$; it follows that $p'_A$ is the same in both executions.  Because
we no longer have agreement, it may be that $p'_B(X_i)$ and
$p'_B(X_{i+1})$ are not the same as $p'_A$ in either execution; but
since both are within $1$ of $p'_A$, the difference between them is at
most $2$.  Next, because $X_{i+1}$ to $X_{i+2}$ are
indistinguishable to $B$, we have $p'_B(X_{i+1}) = p'_B(X_{i+2})$,
which we can combine with the previous claim to get
$\abs*{p'_B(X_i)-p'_B(X_{i+2})}$.  A simple induction then gives
$p'_B(X_{2r}) ≤ 2r$, where $X_{2r}$ is an
execution in which all messages are lost.

Now construct executions $X_{2r+1}$ and $X_{2r+2}$ by changing
$p_A$ and $p_B$ to $m$ one at a time.  Using essentially the same argument as
before, we get $\abs*{p'_B(X_{2r}) - p'_B(X_{2r+2})} ≤ 2$ and thus
$p'_B(X_{2r+2}) ≤ 2r+2$.

Repeat the initial $2r$ steps backward to get to an execution
$X_{4r+2}$ with $p_A = p_B = m$ and no messages lost.  Applying the
same reasoning as above shows $m = p'_B(X_{4r+2}) ≤ 4r+2$ or $r ≥
\frac{m-2}{4} = \Omega(m)$.

Though it is not needed for the solution, it is not too hard to unwind
the lower bound argument to extract an algorithm that matches the
lower bound up to a small constant factor.  For simplicity, let's
assume $m$ is even.

The protocol is to send my input in the
first message and then use $m/2-1$ subsequent acknowledgments,
stopping immediately if I ever fail to receive a message in some
round; the total number of rounds $r$ is exactly $m/2$. 
If I receive $s$
messages in the first $s$ rounds, I decide on
$\min(p_A, p_B)$ if that value lies in $[m/2-s, m/2+s]$ and the
nearest endpoint otherwise.  (Note that if $s=0$, I don't need to
compute $\min(p_A, p_B)$, and if $s > 0$, I can do so because I know
both inputs.)

This satisfies the approximate agreement condition because if I see only
$s$ messages, you see at most $s+1$, because I stop sending once I
miss a message.  So either we both decide $\min(p_A, p_B)$ or we
choose endpoints $m/2\pm s_A$ and $m/2 \pm s_B$ that are within $1$ of
each other.  It also satisfies the validity condition $p'_A = p'_B =
p_A = p_B$ when both inputs are equal and no messages are lost (and
even the stronger requirement that $p'_A = p'_B$ when no messages are
lost), because in this case $[m/2-s, m/2+s]$ is exactly $[0,m]$ and
both processes decide $\min(p_A, p_B)$.

There is still a factor-of-$2$ gap between the upper and lower bounds.
My guess would be that the correct bound is very close to $m/2$ on both
sides, and that my lower bound proof is not quite clever enough.

\section{Assignment 2: due Wednesday, 2011-11-02, at 17:00} 

\subsection{Consensus with delivery notifications}

The FLP bound (Chapter~\ref{chapter-FLP}) shows that we can't solve
consensus in an asynchronous system with one crash failure.  Part of
the reason for this is that only the recipient can detect when a
message is delivered, so the other processes can't distinguish between
a configuration in which a message has or has not been delivered to a
faulty process.

Suppose that we augment the system so that senders are notified
immediately when their messages are delivered.  
We can model this by
making the delivery of a single message an event that updates the
state of both sender and recipient, both of which may send additional
messages in response.  Let us suppose that this includes attempted 
deliveries to
faulty processes, so that any non-faulty process that sends a message
$m$ is eventually notified that $m$ has been delivered (although it
might not have any effect on the recipient if the recipient has already
crashed).

\begin{enumerate}
\item Show that this system can solve consensus with one faulty
process when $n=2$.
\item Show that this system cannot solve consensus with two faulty
processes when $n=3$.
\end{enumerate}

\subsubsection*{Solution}

\begin{enumerate}
\item To solve consensus, each process sends its input to the other.
Whichever input is delivered first becomes the output value for both
processes.
\item 
To show impossibility with $n=3$ and two faults,
run the usual FLP proof until we get to a configuration $C$ with
events $e'$ and $e$ such that $Ce$ is $0$-valent and $Ce'e$ is
$1$-valent (or vice versa).  
Observe that $e$ and $e'$ may involve two processes each (sender and
receiver), for up to four processes total, but only a process that is
involved in both $e$ and $e'$ can tell which happened first.  There
can be at most two such processes.  Kill both, and get that $Ce'e$ is
indistinguishable from $Cee'$ for the remaining process, giving the
usual contradiction.
\end{enumerate}

\subsection{A circular failure detector}

Suppose we equip processes $0\dots n-1$ in an asynchronous
message-passing system with $n$ processes
subject to crash failures with a failure
detector that is strongly accurate (no non-faulty process is ever
suspected) and causes process $i+1 \pmod{n}$ to eventually permanently suspect
process $i$ if process $i$ crashes.  Note that this failure detector
is not even weakly complete (if both $i$ and $i+1$ crash, no
non-faulty process suspects $i$).  Note also that the ring structure
of the failure detector doesn't affect the actual network: even
though only process $i+1 \pmod{n}$ may suspect process $i$, any
process can send messages to any other process.

Prove the best upper and lower bounds you can on 
the largest number of failures $f$ that allows solving
consensus in this system.

\subsubsection*{Solution}

There is an easy reduction to FLP that shows $f ≤ n/2$ is necessary
(when $n$ is even),
and a harder reduction that shows $f < 2\sqrt{n}-1$ is necessary.  The
easy reduction is based on crashing every other process; now no
surviving process can suspect any other survivor, and we are back in
an asynchronous message-passing system with no failure detector and
$1$ remaining failure (if $f$ is at least $n/2+1$).

The harder reduction is to crash every $(\sqrt{n})$-th process.  This
partitions the ring into $\sqrt{n}$ segments of length $\sqrt{n}-1$
each, where there is no failure detector in any segment that suspects
any process in another segment.  If an algorithm exists that solves
consensus in this situation, then it does so even if (a) all processes
in each segment have the same input, (b) if any process in one segment
crashes, all $\sqrt{n}-1$ process in the segment crash, and (c) if any
process in a segment takes a step, all take a step, in some
fixed order.  Under this additional conditions, each segment can be
simulated by a single process in an asynchronous system with no
failure detectors, and the extra $\sqrt{n}-1$ failures in
$2\sqrt{n}-1$ correspond to one failure in the simulation.  But we
can't solve consensus in the simulating system (by FLP), so we can't
solve it in the original system either.

On the other side, let's first boost completeness of the failure
detector, by having any process that suspects another transmit this
submission by reliable broadcast.  So now if any non-faulty process
$i$ suspects $i+1$, all the non-faulty processes will suspect $i+1$.
Now with up to $t$ failures, whenever I learn that process $i$ is faulty
(through a broadcast message passing on the suspicion of the
underlying failure detector, I will suspect processes $i+1$ through
$i+t-f$ as well, where $f$ is the number of failures I have heard
about directly.  I don't need to suspect process $i+t-f+1$ (unless
there is some intermediate process that has also failed), because the
only way that this process will not be suspected eventually is if
every process in the range $i$ to $i+t-f$ is faulty, which can't
happen given the bound $t$.

Now if $t$ is small enough that I can't cover the entire ring with
these segments, then there is some non-faulty processes that is far
enough away from the nearest preceding faulty process that it is never
suspected: this gives us an eventually strong failure detector, and we
can solve consensus using the standard Chandra-Toueg $◇S$
algorithm from §\ref{section-Chandra-Toueg}
or~\cite{ChandraT1996}.
The inequality I am looking for is $f(t-f) < n$, where the
left-hand side is maximized by setting $f = t/2$, which gives $t^2/4 <
n$ or $t < \sqrt{2n}$.  This leaves a gap of about $\sqrt{2}$ between
the upper and lower bounds; I don't know which one can be improved.

I am indebted to Hao Pan for suggesting the $Θ(\sqrt{n})$ upper
and lower bounds, which corrected an error in my original draft solution
to this problem.

\subsection{An odd problem}

Suppose that each of $n$ processes in a message-passing system with a
complete network is attached to a sensor.  Each sensor has two states,
\emph{active} and \emph{inactive}; initially, all sensors are
off.  When the sensor changes state, the
corresponding process is notified immediately, and can update its
state and send messages to other processes in response to this event.
It is also guaranteed that if a sensor changes state, it does not
change state again for at least two time units.
We would like to detect when an odd number of sensors are
active, by having at least one process update its state to set off an alarm at a
time when this condition holds.  

A correct protocol for this problem should satisfy two conditions:
\begin{description}
\item[No false positives] If a process sets of an alarm, then an odd
number of sensors are active.
\item[Termination] If at some time an odd number of sensors are
active,
and from that point on no sensor changes its state, then some process
eventually sets off an alarm.
\end{description}

For what values of $n$ is it possible to construct such a protocol?

\subsubsection*{Solution}

It is feasible to solve the problem for $n < 3$.

For $n=1$, the unique process sets off its alarm as soon as its sensor
becomes active.

For $n=2$, have each process send a message to the other containing
its sensor state whenever the sensor state changes.  Let $s_1$ and
$s_2$ be the state of the two process's sensors, with $0$ representing
inactive and $1$ active, and let $p_i$ set off its alarm if it
receives a message $s$ such that $s⊕ s_i = 1$.  This satisfies
termination, because if we reach a configuration with an odd number of
active sensors, the last sensor to change causes a message to be sent
to the other process that will cause it to set off its alarm.  It
satisfies no-false-positives, because if $p_i$ sets off its alarm,
then $s_{¬i} = s$ because at most one time unit has elapsed since
$p_{¬i}$ sent $s$; it follows that $s_{¬i} ⊕ s_i = 1$ and
an odd number of sensors are active.

No such protocol is possible for $n ≥ 3$.  Make $p_1$'s sensor
active.  Run the protocol until some process $p_i$ is about to enter
an alarm state (this occurs eventually because otherwise we violate
termination).  Let $p_j$ be one of $p_2$ or $p_3$ with $j\ne i$, 
activate $p_j$'s sensor (we can do this without violating the
once-per-time-unit restriction because it has never previously been
activated) and then let $p_i$ set off its alarm.  We have now violated
no-false-positives.

\section{Assignment 3: due Friday, 2011-12-02, at 17:00}

\subsection{A restricted queue}

Suppose you have an atomic queue $Q$ that supports operations $\Enq$
and
$\Deq$, restricted so that:
\begin{itemize}
\item $\Enq(Q)$ always pushes the identity of the current process onto
the tail of the queue.
\item $\Deq(Q)$ tests if the queue is nonempty and its head is equal
to the identity of the current process.  If so, it pops the head and
returns \True.  If not, it does nothing and returns \False.
\end{itemize}

The rationale for these restrictions is that this is the minimal
version of a queue needed to implement a starvation-free mutex using
Algorithm~\ref{alg-mutex-queue}.

What is the consensus number of this object?

\subsubsection*{Solution}

The restricted queue has consensus number 1.

Suppose we have 2 processes, and consider all pairs of operations on
$Q$ that might get us out of a bivalent configuration $C$.  Let $x$ be an
operation carried out by $p$ that leads to a $b$-valent state, and $y$
an operation by $q$ that leads to a $(¬b)$-valent state.  There
are three cases:
\begin{itemize}
\item Two \Deq operations.  If $Q$ is empty, the operations commute.
If the head of the $Q$ is $p$, then $y$ is a no-op and $p$ can't
distinguish between $Cx$ and $Cyx$.  Similarly for $q$ if the head is
$q$.
\item One \Enq and one \Deq operation.  Suppose $x$ is an \Enq and $y$
a \Deq.  If $Q$ is empty or the head is not $q$, then $y$ is a no-op:
$p$ can't distinguish $Cx$ from $Cyx$.  If the head is $q$, then $x$
and $y$ commute.  The same holds in reverse if $x$ is a \Deq and $y$
an \Enq.
\item Two \Enq operations.  This is a little tricky, because $Cxy$ and
$Cyx$ are different states.  However, if $Q$ is nonempty in $C$,
whichever process isn't at the head of $Q$ can't distinguish them,
because any \Deq operation returns false and never reaches the
newly-enqueued values.  This leaves the case where $Q$ is empty in
$C$.  Run $p$ until it is poised to do $x' = \Deq(Q)$ (if this never
happens, $p$ can't distinguish $Cxy$ from $Cyx$); then run $q$ until
it is poised to do $y' = \Deq(Q)$ as well (same argument as for $p$).  Now
allow both \Deq operations to proceed in whichever order causes them
both to succeed.  Since the processes can't tell which \Deq happened
first, they can't tell which \Enq happened first either.  Slightly
more formally, if we let
$α$ be the sequence of operations leading up to the two \Deq
operations, we've just shown $Cxyα x'y'$ is indistinguishable
from $Cyxα y'x'$ to both processes.
\end{itemize}
In all cases, we find that we can't escape bivalence.  It follows that
$Q$ can't solve $2$-process consensus.

\subsection{Writable fetch-and-increment}

Suppose you are given an unlimited supply of atomic registers and
\index{fetch-and-increment}fetch-and-increment objects, where the
fetch-and-increment objects are all initialized to $0$ and 
supply \emph{only} a fetch-and-increment
operation that increments the object and returns the old value.
Show how to use these objects to construct a wait-free, linearizable
implementation of 
an augmented fetch-and-increment that also supports a $\Write$
operation that sets the value of the fetch-and-increment and returns
nothing.

\subsubsection*{Solution}

\newFunc{\FAIfai}{FetchAndIncrement}
\newData{\FAItimestamp}{timestamp}
\newData{\FAIbase}{base}

We'll use a snapshot object $a$ to control access to an infinite array $f$ of
fetch-and-increments, where each time somebody writes to the
implemented object, we switch to a new fetch-and-increment.
Each cell in $a$ holds $(\FAItimestamp, \FAIbase)$, where $\FAIbase$
is the starting value of the simulated fetch-and-increment.
We'll also use an extra fetch-and-increment $T$ to hand out timestamps.

Code is in Algorithm~\ref{alg-resettable-fetch-and-increment}.

\begin{algorithm}
\Procedure{$\FAIfai()$}{
    $s ← \FuncSty{snapshot}(a)$\;
    $i ← \argmax_i(s[i].\FAItimestamp)$ \;
    \Return $f[s[i].\FAItimestamp] + s[i].\FAIbase$\;
}
\bigskip
\Procedure{$\Write(v)$}{
    $t ← \FAIfai(T)$ \;
    $a[\MyId] ← (t, v)$\;
}
\caption{Resettable fetch-and-increment}
\label{alg-resettable-fetch-and-increment}
\end{algorithm}

Since this is all straight-line code, it's trivially wait-free.

Proof
of linearizability is by grouping all operations by timestamp, using
$s[i].\FAItimestamp$ for \FAIfai operations and $t$ for $\Write$
operations, then putting $\Write$ before $\FAIfai$, then ordering $\FAIfai$
by return value.  Each group will consist of a $\Write(v)$ for some
$v$ followed by zero or more
\FAIfai operations, which will return increasing values starting at
$v$ since they are just returning values from the underlying \FAIfai
object; the implementation thus meets the specification.

To show consistency with the actual execution order,
observe that timestamps only increase over time and
that the use of snapshot means that any process that observes or
writes a timestamp $t$ does so at a time later than any process
that observes or writes any $t' < t$; this shows the group order is
consistent.  Within each group, the \Write writes $a[\MyId]$ before
any \FAIfai reads it, so again we have consistency between the \Write
and any \FAIfai operations.  The \FAIfai operations are linearized in
the order in which they access the underlying $f[\dots]$ object, so we
win here too.

\subsection{A box object}

Suppose you want to implement an object representing a $w \times h$
box whose width ($w$) and height ($h$) can be increased if needed.
Initially, the box is $1\times 1$, and the coordinates can be
increased by $1$ each using $\FuncSty{IncWidth}$ and
$\FuncSty{IncHeight}$ operations.  There is also a $\FuncSty{GetArea}$
operation that returns the area $w\cdot h$ of the box.

Give an obstruction-free deterministic implementation of this object from atomic
registers that optimizes
the worst-case individual step complexity of $\FuncSty{GetArea}$, and
show that your implementation is optimal by this measure up to
constant factors.

\subsubsection*{Solution}

Let $b$ be the box object.  Represent $b$ by a snapshot object $a$,
where $a[i]$ holds a pair $(\Delta w_i, \Delta h_i)$ representing the
number of times process $i$ has executed $\FuncSty{IncWidth}$ and
$\FuncSty{IncHeight}$; these operations simply increment the
appropriate value and update the snapshot object.  Let
$\FuncSty{GetArea}$ take a snapshot and return
$\left(\sum_i \Delta w_i\right)
\left(\sum_i \Delta h_i\right)$; 
the cost of the snapshot is $O(n)$.

To see that this is optimal, observe that we can use
$\FuncSty{IncWidth}$ and $\FuncSty{GetArea}$ to represent
$\FuncSty{inc}$ and \Read for a standard counter.  The
Jayanti-Tan-Toueg bound applies to counters, giving a worst-case cost of
$\Omega(n)$ for $\FuncSty{GetArea}$.

\section{CS465/CS565 Final Exam, December 12th, 2011}
\label{section-final-exam-2011}

\newcommand{\MMXIproblem}[1]{\subsection{{#1} (20 points)}}

Write your answers in the blue book(s).  Justify your answers.  Work
alone.  Do not use any notes or books.  

There are four problems on this exam, each worth 20
points, for a total of 80 points.
You have approximately three hours to complete this
exam.

\paragraph*{General clarifications added during exam} Assume all processes
have unique IDs and know $n$.  Assume that the network is complete in the
message-passing model.

\MMXIproblem{Lockable registers}

\newFunc{\FinalMMXIlock}{lock}

Most memory-management units provide the ability to control access to
specific memory pages, allowing a page to be marked (for example)
read-only.  Suppose that we model this by a
\index{register!lockable}
\concept{lockable register}
that has the usual register operations $\Read(r)$ and $\Write(r,v)$ plus an
additional operation $\FinalMMXIlock(r)$.  The behavior of the register is
just like a normal atomic register until somebody calls
$\FinalMMXIlock(r)$; after this, any call to $\Write(r)$ has no effect.

What is the consensus number of this object?

\subsubsection*{Solution}
The consensus number is $\infty$; a single lockable register solves
consensus for any number of processes.  Code is in
Algorithm~\ref{alg-lockable-register-consensus}.
\begin{algorithm}
$\Write(r, \DataSty{input})$\;
$\FinalMMXIlock(r)$ \;
\Return $\Read(r)$ \;
\caption{Consensus using a lockable register}
\label{alg-lockable-register-consensus}
\end{algorithm}

Termination and validity are trivial.  Agreement follows from the fact
that whatever value is in $r$ when $\FinalMMXIlock(r)$ is first called will
never change, and thus will be read and returned by all processes.

\MMXIproblem{Byzantine timestamps}

Suppose you have an asynchronous message passing system with exactly one
Byzantine process.

You would like the non-faulty processes to be able
to acquire an increasing sequence of timestamps.  A process
should be able to execute the timestamp protocol as often as it likes,
and it should be guaranteed that when a process is non-faulty, 
it eventually obtains a timestamp
that is
larger than any timestamp returned in any execution of the
protocol by a non-faulty process
that finishes before the current process's execution started.

Note that there is no bound on the size of a timestamp, so having the
Byzantine process run up the timestamp values is not a problem, as
long as it can't cause the timestamps to go down.

For what values of $n$ is it possible to solve this problem?

\subsubsection*{Solution}

It is possible to solve the problem for all $n$ except $n=3$.  For $n=1$, there are
no non-faulty processes, so the specification is satisfied trivially.
For $n=2$, there is
only one non-faulty process: it can just keep its own counter and
return an increasing sequence of timestamps without talking to the
other process at all.

For $n=3$, it is not possible.  Consider an execution in which
messages between non-faulty processes $p$ and $q$ are delayed
indefinitely.  If the Byzantine process $r$ acts to each of $p$ and
$q$ as it would if the other had crashed, this execution is
indistinguishable to $p$ and $q$ from an execution in which $r$ is
correct and the other is faulty.  Since there is no communication
between $p$ and $q$, it is easy to construct and execution in which
the specification is violated.

For $n ≥ 4$, the protocol given in
Algorithm~\ref{alg-Byzantine-timestamps} works.

\newFunc{\FinalMMXIgetTimestamp}{getTimestamp}
\newFunc{\FinalMMXIprobe}{probe}
\newFunc{\FinalMMXIresponse}{response}
\newFunc{\FinalMMXInewTimestamp}{newTimestamp}

\begin{algorithm}
\Procedure{$\FinalMMXIgetTimestamp()$}{
    $c_i ← c_i+1$ \;
    send $\FinalMMXIprobe(c_i)$ to all processes \;
    wait to receive $\FinalMMXIresponse(c_i, v_j)$ from $n-1$ processes \;
    $v_i ← (\max_j v_j) + 1$ \;
    send $\FinalMMXInewTimestamp(c_i, v_i)$ to all processes \;
    wait to receive $\Ack(c_i)$ from $n-1$ processes \;
    \Return $v_i$ \;
}
\bigskip
\UponReceiving{$\FinalMMXIprobe(c_j)$ \From $j$}{
    send $\FinalMMXIresponse(c_j, v_i)$ to $j$ \;
}
\bigskip
\UponReceiving{$\FinalMMXInewTimestamp(c_j, v_j)$ \From $j$}{
    $v_i ← \max(v_i, v_j)$ \;
    send $\Ack(c_j)$ to $j$ \;
}
\caption{Timestamps with $n ≥ 3$ and one Byzantine process}
\label{alg-Byzantine-timestamps}
\end{algorithm}

The idea is similar to the Attiya, Bar-Noy, Dolev distributed shared
memory algorithm~\cite{AttiyaBD1995}.  A process that needs a
timestamp polls $n-1$ other
processes for the maximum values they've seen and adds 1 to it; before
returning, it sends the new timestamp to all other processes and waits
to receive $n-1$ acknowledgments.  The Byzantine process may choose
not to answer, but this is not enough to block completion of the
protocol.

To show the timestamps are increasing, observe that
after the completion of any call by $i$ to
$\FinalMMXIgetTimestamp$, at least $n-2$ non-faulty processes $j$ have a value
$v_j ≥ v_i$.  Any call to $\FinalMMXIgetTimestamp$ that starts later
sees at least $n-3 > 0$ of these values, and so computes a $\max$ that
is at least as big as $v_i$ and then adds $1$ to it, giving a larger
value.

\MMXIproblem{Failure detectors and \texorpdfstring{$k$}{k}-set agreement}

Recall that in the 
\index{$k$-set agreement}
\index{agreement!$k$-set}
$k$-set agreement problem we want each of $n$ processes to choose a
decision value, with the property that the set of decision values has
at most $k$ distinct elements.  It is known that $k$-set agreement
cannot be solved deterministically in an asynchronous message-passing
or shared-memory system with $k$ or more crash failures.

Suppose that you are working in an asynchronous message-passing system
with an 
\index{eventually strong failure detector}
\index{failure detector!eventually strong}
eventually strong ($◇S$) failure detector.  Is it possible to
solve $k$-set agreement deterministically with $f$ crash failures, when $k ≤ f < n/2$?

\subsubsection*{Solution}

Yes.  With $f < n/2$ and $◇S$, we can solve consensus using
Chandra-Toueg~\cite{ChandraT1996}.  Since this gives a unique decision
value, it solves $k$-set agreement for any $k ≥ 1$.

\MMXIproblem{A set data structure}

\newFunc{\MMXIadd}{add}
\newFunc{\MMXIsize}{size}

Consider a data structure that represents a set $S$, with an operation
$\MMXIadd(S, x)$ that adds $x$ to $S$ by setting $S ← S \cup
\{x\}$), and an operation $\MMXIsize(S)$ that returns the number of
distinct\footnote{Clarification added during exam.}
elements $\card*{S}$ of $S$.  There are no restrictions on
the types or sizes of elements that can be added to the set.

Show that any deterministic wait-free implementation of this object
from atomic registers has individual step complexity $\Omega(n)$ for
some operation in the worst case.

\subsubsection*{Solution}

Algorithm~\ref{alg-set-based-counter} implements a counter from a set
object, where the counter read consists of a single call to
$\MMXIsize(S)$.  The idea is that each increment is implemented by
inserting a new element into $S$, so $\card*{S}$ is always equal to the
number of increments.

\newData{\MMXInonce}{nonce}
\begin{algorithm}
\Procedure{$\FuncSty{inc}(S)$}{
    $\MMXInonce ← \MMXInonce + 1$\;
    $\MMXIadd(S, \langle\MyId, \MMXInonce\rangle)$. \;
}
\bigskip
\Procedure{$\Read(S)$}{
    \Return $\MMXIsize(S)$ \;
}
\caption{Counter from set object}
\label{alg-set-based-counter}
\end{algorithm}

Since the Jayanti-Tan-Toueg lower bound~\cite{JayantiTT2000} gives a
lower bound of $\Omega(n)$ on the worst-case cost of a counter read,
there exists an execution in which $\MMXIsize(S)$ takes $\Omega(n)$
steps.

(We could also apply JTT directly by showing that the set object is
perturbable; this follows because adding an element not added by
anybody else is always visible to the reader.)

\chapter{Additional sample final exams}
\label{appendix-past-final-exams}

This appendix contains final exams from previous times the course was
offered, and is intended to give a rough guide to the typical format
and content of a final exam.  Note that the topics covered in past
years were not necessarily the same as those covered this year.

\newcommand{\MMVproblem}[1]{\subsection{{#1} (20 points)}}

\section{CS425/CS525 Final Exam, December 15th, 2005}

Write your answers in the blue book(s).  Justify your answers.  Work
alone.  Do not use any notes or books.  

There are three problems on this exam, each worth 20
points, for a total of 60 points.
You have approximately three hours to complete this
exam.

\MMVproblem{Consensus by attrition}

Suppose you are given a 
\index{fetch-and-subtract!bounded}
\concept{bounded fetch-and-subtract} register that holds
a non-negative integer value and supports an operation
fetch-and-subtract($k$) for each $k > 0$ that (a) sets the value of the
register to the previous value minus $k$, or zero if this result would
be negative, and (b) returns the previous value of the register.

Determine the consensus number of bounded fetch-and-subtract under the
assumptions that you can use arbitrarily many such objects, that you
can supplement them with arbitrarily many multiwriter/multireader
read/write registers, that you can initialize all registers of both
types to initial values of your choosing, and that the design of the
consensus protocol can depend on the number of processes $N$.

\subsubsection*{Solution}
The consensus number is $2$.

To implement 2-process wait-free consensus, use a single
fetch-and-subtract register initialized to 1 plus two auxiliary
read/write registers to hold the input values of the processes.  Each
process writes its input to its own register, then performs a
fetch-and-subtract(1) on the fetch-and-subtract register.  Whichever
process gets 1 from the fetch-and-subtract returns its own input; the
other process (which gets 0) returns the winning process's input
(which it can read from the winning process's read/write register.)

To show that the consensus number is at most $2$, observe that any two
fetch-and-subtract operations commute: starting from state $x$, after
fetch-and-subtract($k_1$) and fetch-and-subtract($k_2$) the value in
the fetch-and-subtract register is $\max(0, x - k_1 - k_2)$ regardless
of the order of the operations.

\MMVproblem{Long-distance agreement}

Consider an asynchronous message-passing model consisting of $N$
processes $p_1 \ldots p_N$ arranged in a line, so that each process $i$ can send
messages only to processes $i-1$ and $i+1$ (if they exist).  Assume
that there are no failures, that local computation takes zero time,
and that every message is delivered at most 1 time unit after it is
sent no matter how many messages are sent on the same edge.

Now suppose that we wish to solve agreement in this model, where the
agreement protocol is triggered by a local \emph{input} event at one
or more processes and it terminates when every process executes a
local \emph{decide} event.  As with all agreement problems, we want
Agreement (all processes decide the same value), Termination (all
processes eventually decide), and Validity (the common decision value
previously appeared in some input).  We also want no false starts: the
first action of any process should either be an \emph{input} action or
the receipt of a message.

Define the time cost of a protocol for this problem as the worst-case
time between the first \emph{input} event and the last \emph{decide}
event.  Give the best upper and lower bounds you can on this time as
function of $N$.  Your upper and lower bounds should be \emph{exact}:
using no asymptotic notation or hidden constant factors.  Ideally, they
should also be equal.

\subsubsection*{Solution}

\subsubsection*{Upper bound}

Because there are no failures, we can appoint a leader and have it
decide.  The natural choice is some process near the middle, say
$p_{\lfloor (N+1)/2 \rfloor}$.  Upon receiving an input, either directly
through an \emph{input} event or indirectly from another process, the
process sends the input value along the line toward the leader.  The
leader takes the first input it receives and broadcasts it back out in
both directions as the decision value.  The worst case is when the
protocol is initiated at $p_N$; then we pay $2(N - \lfloor (N+1)/2
\rfloor)$ time to send all messages out and back, which is $N$ time
units when $N$ is even and $N-1$ time units when $N$ is odd.

\subsubsection*{Lower bound}

Proving an almost-matching lower bound of $N-1$ time units is trivial:
if $p_1$ is the only initiator and it starts at time $t_0$, then by an
easy induction argument,in the worst case $p_i$ doesn't learn of any
input until time $t_0 + (i-1)$, and in particular $p_N$ doesn't find
out until after $N-1$ time units.  If $p_N$ nonetheless decides early,
its decision value will violate validity in some executions.

But we can actually prove something stronger than this: that $N$ time
units are indeed required when $N$ is odd.  Consider two slow executions
$\Xi_0$ and $\Xi_1$, where (a) all messages are delivered after
exactly one time unit in each execution; (b) in $\Xi_0$ only $p_1$ receives an input and
the input is $0$; and (c) in $\Xi_1$ only $p_N$ receives an input and the
input is $1$.  For each of the executions, construct a causal ordering
on events in the usual fashion: a send is ordered before a receive,
two events of the same process are ordered by time, and other events
are partially ordered by the transitive closure of this relation.

Now consider for $\Xi_0$ the set of all events that precede the
\emph{decide(0)} event of $p_1$ and for $\Xi_1$ the set of all events
that precede the \emph{decide(1)} event of $p_N$.  Consider further
the sets of processes $S_0$ and $S_1$ at which these events occur; if these two sets
of processes do not overlap, then we can construct an execution in
which both sets of events occur, violating Agreement.

Because $S_0$ and $S_1$ overlap, we must have $\card*{S_0}+\card*{S_1} ≥ N+1$,
and so at least one of the two sets has size at least $\lceil (N+1)/2
\rceil$, which is $N/2 + 1$ when $N$ is even.  Suppose that it is
$S_0$.  Then in order for any event to occur at $p_{N/2 + 1}$ at all
some sequence of messages must travel from the initial
input to $p_1$ to process $p_{N/2+1}$ (taking $N/2$ time units), and
the causal ordering implies that an additional sequence of messages
travels back from $p_{N/2+1}$ to $p_1$ before $p_1$ decides (taking
and additional $N/2$ time units).  The total time is thus $N$.

\MMVproblem{Mutex appendages}

An \concept{append} register supports standard read operations plus an
append operation that appends its argument to the list of values
already in the register.  An \concept{append-and-fetch} register is
similar to an append register, except that it returns the value in the
register after performing the append operation.  Suppose that you have
an failure-free asynchronous system with anonymous deterministic 
processes (i.e., deterministic processes that
all run exactly the same code).
Prove or disprove each of the
following statements:

\begin{enumerate}
\item It is possible to solve mutual exclusion using only append
registers.
\item It is possible to solve mutual exclusion using only
append-and-fetch registers.
\end{enumerate}

In either case, the solution should work for arbitrarily many
processes—solving mutual exclusion when $N=1$ is not interesting.
You are also not required in either case to guarantee starvation-freedom.

\subsubsection*{Clarification given during exam}

\begin{enumerate}
\item If it helps, you may assume that the processes know $N$.  (It
probably doesn't help.)
\end{enumerate}

\subsubsection*{Solution}

\begin{enumerate}
\item Disproof: With append registers only, it is not possible to solve 
mutual exclusion.  To prove this, construct a failure-free execution
in which the processes never break symmetry.  In the initial
configuration, all processes have the same state and thus execute
either the same read operation or the same append operation; in either
case we let all $N$ operations occur in some arbitrary order.  If the
operations are all reads, all processes read the same value and move
to the same new state.  If the operations are all appends, then no
values are returned and again all processes enter the same new state.
(It's also the case that the processes can't tell from the register's
state which of the identical append operations went first, but we
don't actually need to use this fact.)

Since we get a fair failure-free execution where all processes move
through the same sequence of states, if any process decides it's in
its critical section, all do.  We thus can't solve mutual exclusion in this model.

\item Since the processes are anonymous, any solution that depends
on them having identifiers isn't going to work.  But there is a simple
solution that requires only appending single bits to the register.

Each process trying to enter a critical section 
repeatedly executes an append-and-fetch operation with argument $0$;
if the append-and-fetch operation returns either a list consisting
only of a single $0$ or a list whose second-to-last element is $1$,
the process enters its critical section.  To leave the critical
section, the process does append-and-fetch(1).

\end{enumerate}

\newcommand{\MMVIIIproblem}[1]{\subsection{{#1} (20 points)}}
\newenvironment{MMVIIIsolution}{\subsubsection*{Solution}}{}

\section{CS425/CS525 Final Exam, May 8th, 2008}

Write your answers in the blue book(s).  Justify your answers.  Work
alone.  Do not use any notes or books.  

There are four problems on this exam, each worth 20
points, for a total of 80 points.
You have approximately three hours to complete this
exam.

\MMVIIIproblem{Message passing without failures}

Suppose you have an asynchronous message-passing system with a
complete communication graph, unique node identities, and no failures.
Show that any deterministic atomic shared-memory object can be
simulated in this model, or give an example of a shared-memory object
that can't be simulated.

\begin{MMVIIIsolution}
Pick some leader node to implement the object.  To execute an
operation, send the operation to the leader node, then have the leader
carry out
the operation (sequentially) on its copy of the object and send the
results back.
\end{MMVIIIsolution}

\MMVIIIproblem{A ring buffer}
\label{section-MMVIII-exam-ring-buffer}

Suppose you are given a 
\index{object!ring buffer}
\concept{ring buffer object} that consists of 
$k ≥ 1$
memory locations $a[0]\ldots a[k-1]$ with an atomic
\emph{shift-and-fetch} operation that takes an argument $v$ and (a)
shifts $v$ into the buffer, so that $a[i] ←
a[i+1]$ for each $i$ less than $k-1$ and $a[k-1] ← v$; and
(b) returns a snapshot of the new contents of the array (after the
shift).

What is the consensus number of this object as a function of $k$?

\begin{MMVIIIsolution}
We can clearly solve consensus for at least $k$ processes: each
process calls shift-and-fetch on its input, and returns the first
non-null value in the buffer.

So now we want to show that we can't solve consensus for $k+1$
processes.  Apply the usual FLP-style argument to get to a bivalent
configuration $C$ where each of the $k+1$ processes has a pending
operation that leads to a univalent configuration.  Let $e_0$ and
$e_1$ be particular operations leading to $0$-valent and $1$-valent
configurations, respectively, and let $e_2 \ldots e_k$ be the
remaining $k-1$ pending operations.

We need to argue first that no two distinct operations $e_i$ and $e_j$
are operations of different objects.  Suppose that $Ce_i$ is
$0$-valent and $Ce_j$ is $1$-valent; then if $e_i$ and $e_j$ are on
different objects, $Ce_ie_j$ (still $0$-valent) is indistinguishable
by all processes from $Ce_je_i$ (still $1$-valent), a contradiction.
Alternatively, if $e_i$ and $e_j$ are both $b$-valent, there exists
some $(1-b)$-valent $e_k$ such that $e_i$ and $e_j$ both operate on the
same object as $e_k$, by the preceding argument.  So all of $e_0
\ldots e_k$ are operations on the same object.

By the usual argument we know that this object can't be a register.
Let's show it can't be a ring buffer either.  Consider the
configurations $Ce_0e_1\ldots e_k$ and $Ce_1\ldots e_k$.  These are
indistinguishable to the process carrying out $e_k$ (because its sees
only the inputs to $e_1$ through $e_k$ in its snapshot).  So they must
have the same valence, a contradiction.

It follows that the consensus number of a $k$-element ring buffer is
exactly $k$.
\end{MMVIIIsolution}

\MMVIIIproblem{Leader election on a torus}

An $n \times n$ torus is a graph consisting of $n^2$ nodes, where each
node $(i,j)$, $0 ≤ i,j ≤ n-1$, is connected to nodes $(i-1,j),
(i+1,j), (i,j-1),$ and $(i,j+1)$, where all computation is done mod
$n$.

Suppose you have an asynchronous message-passing system with a
communication graph in the form of an $n\times n$ torus.  Suppose
further that each node has a unique identifier (some large natural
number) but doesn't know the value of $n$.  Give an algorithm for
leader election in this model with the best message complexity you can
come up with.

\begin{MMVIIIsolution}
First observe that each row and column of the torus is a bidirectional
ring, so we can run e.g. Hirschbirg and Sinclair's $O(n \log
n)$-message protocol within each of these rings to find the smallest
identifier in the ring.  We'll use this to construct the following
algorithm:

\begin{enumerate}
\item Run Hirschbirg-Sinclair in each row to get a local leader for
each row; this takes $n \times O(n \log n) = O(n^2 \log n)$ messages.
Use an additional $n$ messages per row to distribute the identifier
for the row leader to all nodes and initiate the next stage of the
protocol.
\item Run Hirschbirg-Sinclair in each column with each node adopting
the row leader identifier as its own.  This costs another $O(n^2 \log
n)$ messages; at the end, every node knows the minimum identifier of
all nodes in the torus.
\end{enumerate}

The total message complexity is $O(n^2 \log n)$.  (I suspect this is
optimal, but I don't have a proof.)
\end{MMVIIIsolution}

\MMVIIIproblem{An overlay network}

A collection of $n$ nodes—in an asynchronous message-passing system
with a connected, bidirectional communications graph with $O(1)$ links per
node—wish to engage in some strictly legitimate file-sharing.  Each
node starts with some input pair $(k, v)$, where $k$ is a key and $v$
is a value, and the search problem is to find the value $v$
corresponding to a particular key $k$.

\begin{enumerate}
\item Suppose that we can't do any preparation ahead of time.  Give an
algorithm for searching with the smallest asymptotic worst-case message
complexity you can find as a function of $n$.  You may assume that
there are no limits on time complexity, message size, or storage space
at each node.

\item Suppose now that some designated leader node can initiate a
protocol ahead of time to pre-process the data in the nodes before any
query is initiated.  Give a pre-processing algorithm (that does not
depend on which key is eventually searched for) and associated search
algorithm such that the search algorithm minimizes the asymptotic
worst-case message complexity.  Here you may assume that there are no
limits on time complexity, message size, or storage space for either
algorithm, and that you don't care about the message complexity of the
pre-processing algorithm.

\item Give the best lower bound you can on the total message
complexity of the pre-processing and search algorithms in the case
above.
\end{enumerate}

\begin{MMVIIIsolution}
\begin{enumerate}
\item Run depth-first search to find the matching key and return the
corresponding value back up the tree.  Message
complexity is $O(\card*{E}) = O(n)$ (since each node has only $O(1)$ links).

\item Basic idea: give each node a copy of all key-value pairs, then
searches take zero messages.  To give each node a copy of all
key-value pairs we could do convergecast followed by broadcast ($O(n)$
message complexity) or just flood each pair $O(n^2)$.  Either is fine
since we don't care about the message complexity of the pre-processing
stage.

\item Suppose the total message complexity of both the pre-processing stage
and the search protocol is less than $n-1$.  Then there is some node
other than the initiator of the search that sends no messages at any
time during the protocol.  If this is the node with the matching
key-value pair, we don't find it.  It follows that any solution to the
search problem.
requires a total of $\Omega(n)$ messages in the pre-processing and
search protocols.
\end{enumerate}
\end{MMVIIIsolution}

\newcommand{\MMXproblem}[1]{\subsection{{#1} (20 points)}}

\section{CS425/CS525 Final Exam, May 10th, 2010}

Write your answers in the blue book(s).  Justify your answers.  Work
alone.  Do not use any notes or books.  

There are four problems on this exam, each worth 20
points, for a total of 80 points.
You have approximately three hours to complete this
exam.

\MMXproblem{Anti-consensus}

A wait-free \concept{anti-consensus} protocol satisfies the conditions:

\begin{description}
\item[Wait-free termination] Every process decides in a bounded number
of its own steps.
\item[Non-triviality] There is at least one process that decides different
values in different executions.
\item[Disagreement] If at least two processes decide, then some
processes decide on different values.
\end{description}

Show that there is no deterministic wait-free anti-consensus protocol
using only atomic registers for two processes and two possible output
values, but there is one for three processes and three possible output
values.

\paragraph*{Clarification:} You should assume processes have distinct
identities.

\subsubsection*{Solution}
No protocol for two: turn an anti-consensus protocol with outputs in
$\{0,1\}$ into a consensus protocol by having one of the processes
always negate its output.

A protocol for three: Use a splitter.

\MMXproblem{Odd or even}

Suppose you have a protocol for a synchronous message-passing ring
that is anonymous (all processes run the same code) and uniform (this
code is the same for rings of different sizes).  Suppose also that the
processes are given inputs marking some, but not all, of them as
leaders.  Give an algorithm for
determining if the size of the ring is odd or even, or show that no
such algorithm is possible.

\paragraph*{Clarification:} Assume a bidirectional, oriented ring and a
deterministic algorithm.

\subsubsection*{Solution}
Here is an impossibility proof.  Suppose there is such an algorithm,
and let it correctly decide ``odd'' on a ring of size $2k+1$ for some
$k$ and some set of leader inputs.
Now construct a ring of size $4k+2$ by pasting two such rings
together (assigning the same values to the leader bits in each copy)
and run the algorithm on this ring.
By the usual symmetry argument,
every corresponding process sends the same messages and
makes the same decisions in both rings, implying that the processes
incorrectly decide the ring of size $4k+2$ is odd.

\MMXproblem{Atomic snapshot arrays using message-passing}

Consider the following variant of Attiya-Bar-Noy-Dolev for obtaining
snapshots of an array instead of individual register values, in an
asynchronous message-passing system with $t < n/4$ crash failures.
The data structure we are simulating is an array $a$ consisting of an
atomic register $a[i]$ for each process $i$, with the ability to
perform atomic snapshots.

Values are written by sending a set of 
$\langle i,v,t_i \rangle$ values to all processes, where $i$ specifies the
segment $a[i]$ of the array to write, $v$ gives a value for this
segment, and $t_i$ is an increasing timestamp used to indicate more
recent values.  We use a set of values because (as in ABD) some values
may be obtained indirectly.

To update segment $a[i]$ with value $v$, process $i$ generates a new
timestamp $t_i$, sends $\{\langle i,v,t_i \rangle\}$ to all processes, and
waits for acknowledgments from at least $3n/4$ processes.  

Upon receiving a message containing one or more $\langle i,v,t_i
\rangle$ triples, a process updates its copy of $a[i]$ for any $i$
with a higher timestamp than previously seen, and responds with an
acknowledgment (we'll assume use of nonces so that it's unambiguous
which message is being acknowledged).

To perform a snapshot, a process sends \textsc{snapshot} to all
processes, and waits to receive responses from at least $3n/4$
processes, which will consist of the most recent values of each $a[i]$
known by each of these processes together with their timestamps
(it's a set of triples as above).
The snapshot process then takes the most recent versions of $a[i]$ for
each of these responses and updates its own copy, then sends its
entire snapshot vector to all processes and waits to receive at least
$3n/4$ acknowledgments.  When it has received these acknowledgments,
it returns its own copy of $a[i]$ for all $i$.

Prove or disprove: The above procedure implements an atomic snapshot
array in an asynchronous message-passing system with $t < n/4$
crash failures.

\subsubsection*{Solution}
Disproof: Let $s_1$ and $s_2$ be processes carrying out snapshots and
let $w_1$ and $w_2$ be processes carrying out writes.  Suppose that
each $w_i$ initiates a write of $1$ to $a[w_i]$, but all of its
messages to other processes are delayed after it updates its own copy
$a_{w_i}[w_i]$.  Now let each $s_i$ receive responses from $3n/4-1$
processes not otherwise mentioned plus $w_i$.  Then $s_1$ will return
a vector with $a[w_1] = 1$ and $a[w_2] = 0$ while $s_2$ will return a
vector with $a[w_1] = 0$ and $a[w_2] = 1$, which is inconsistent.
The fact that these vectors
are also disseminated throughout at least $3n/4$ other processes is a
red herring.

\MMXproblem{Priority queues}

Let $Q$ be a priority queue whose states are multisets of natural
numbers and that has operations
$\Enq(v)$ and
$\Deq()$, where $\Enq(p)$ adds a new value $v$ to
the queue, and $\Deq()$ removes and returns the smallest value in the
queue, or returns null if the queue is empty.  (If there is more
than one copy of the smallest value, only one copy is removed.)

What is the consensus number of this object?

\subsubsection*{Solution}
The consensus number is 2.  The proof is similar to that for a queue.

To show we can do consensus for $n=2$, start with a priority queue with a single
value in it, and have each process attempt to dequeue this value.  If
a process gets the value, it decides on its own input; if it gets
null, it decides on the other process's input.

To show we can't do consensus for $n=3$, observe first that starting
from any states $C$ of the queue, given any two operations $x$
and $y$ that are both enqueues or both dequeues, the states $Cxy$ and
$Cyx$ are identical.  This means that a third process can't tell which
operation went first, meaning that a pair of enqueues or a pair of
dequeues can't get us out of a bivalent configuration in the FLP
argument.  We can also exclude any split involving two operations on
different queues (or other objects) 
But we still need to consider the case of a dequeue
operation $d$ and an enqueue operation $e$ on the same queue $Q$.
This splits into several
subcases, depending on the state $C$ of the queue in some bivalent
configuration:
\begin{enumerate}
\item $C = \{ \}$.  Then $Ced = Cd = \{ \}$, and a third process can't
tell which of $d$ or $e$ went first.
\item $C$ is nonempty and $e = \Enq(v)$, where $v$ is greater than
or equal to the smallest value in $C$.  Then $Cde$ and $Ced$ are
identical, and no third process can tell which of $d$ or $e$ went
first.
\item $C$ is nonempty and $e = \Enq(v)$, where $v$ is less than any value in $C$.
Consider the configurations $Ced$ and $Cde$.  Here the process $p_d$
that performs $d$ can tell which operation went first, because it either
obtains $v$ or some other value $v' \ne v$.  Kill this process.  No
other process in $Ced$ or $Cde$ can distinguish the two states without
dequeuing whichever of $v$ or $v'$ was not dequeued by $p_d$.  So
consider two parallel executions $Cedσ$ and $Cdeσ$ where
$σ$ consists of an arbitrary sequence of operations ending with a
$\Deq$ on $Q$ by some process $p$ (if no process ever attempts to dequeue
from $Q$, then we have already won, since the survivors can't
distinguish $Ced$ from $Cde$).  Now the state of all objects is
the same after $Cedσ$ and $Cdeσ$, and only $p_d$ and $p$ 
have different states in 
these two configurations.  So any third process is out of luck.
\end{enumerate}

\chapter{I/O automata}
\label{appendix-IO-automata}

\section{Low-level view: I/O automata}

\newFunc{\IOAtrans}{trans}
\newFunc{\IOAstates}{states}
\newFunc{\IOAacts}{acts}
\newFunc{\IOAtask}{task}
\newFunc{\IOAin}{in}
\newFunc{\IOAout}{out}
\newFunc{\IOAint}{int}
\newFunc{\IOAtrace}{trace}
\newFunc{\IOAtraces}{traces}
\newFunc{\IOAsig}{sig}
\newFunc{\IOAstart}{start}

An \concept{I/O automaton}~\cite{LynchT1987} is an automaton where transitions are
labeled by \indexConcept{action}{actions}, which come in three
classes: \index{action!input}\indexConcept{input action}{input
actions}, triggered by the outside world;
\index{action!output}\indexConcept{output action}{output actions}
triggered by the automaton and visible to the outside world; and
\index{action!internal}\indexConcept{internal action}{internal
actions}, triggered by the automaton but not visible to the outside
world.  These classes correspond to inputs, outputs, and internal
computation steps of the automaton; the latter are provided mostly to
give merged input/output actions a place to go when automata are
composed together.  A \concept{transition relation} $\IOAtrans(A)$
relates $\IOAstates(A)\times{}\IOAacts(A)\times{}\IOAstates(A)$; if
$(s,a,s')$ is in $\IOAtrans(A)$, it means that $A$ can move from state
$s$ to state $s'$ by executing action $a$.

There is also an equivalence relation $\IOAtask(A)$ on the output and internal actions, which is used for enforcing fairness conditions—the basic idea is that in a fair execution some action in each equivalence class must be executed eventually (a more accurate definition will be given below).

The I/O automaton model carries with it a lot of specialized jargon.
We'll try to avoid it as much as possible.  One thing that will be
difficult to avoid in reading~\cite{Lynch1996} is the notion of a
\concept{signature}, which is just the tuple $\IOAsig(A) = (\IOAin(A), \IOAout(A),
\IOAint(A))$ describing the actions of an automaton $A$.

\subsection{Enabled actions}
An action $a$ is \concept{enabled} in some state $s$ if $\IOAtrans(A)$
contains at least one transition $(s,a,s')$.  Input actions are
\emph{always} enabled—this is a requirement of the model.  Output
and internal actions—the ``locally controlled'' actions—are not
subject to this restriction.  A state $s$ is \concept{quiescent} if
only input actions are enabled in $s$.

\subsection{Executions, fairness, and traces}
An \concept{execution} of A is a sequence $s_{0} a_{0} s_{1} a_{1}
\dots{}$ where each triple $(s_{i}, a_{i} s_{i+1})$ is in
$\IOAtrans(A)$.  Executions may be finite or infinite; if finite, they must end in a state.  

A \concept{trace} of $A$ is a subsequence of some execution consisting
precisely of the external (i.e., input and output) actions, with
states and internal actions omitted.  If we don't want to get into the
guts of a particular I/O automaton—and we usually don't, unless we
can't help it because we have to think explicitly about states for
some reason—we can describe its externally-visible behavior by just giving its set of traces.

\subsection{Composition of automata}
Composing a set of I/O automata yields a new super-automaton whose
state set is the Cartesian product of the state sets of its components
and whose action set is the union of the action sets of its
components.  A transition with a given action $a$ updates the states
of all components that have $a$ as an action and has no effect on the states of other components.  The classification of actions into the three classes is used to enforce some simple compatibility rules on the component automata; in particular:

\begin{enumerate}
 \item An internal action of a component is never an action of another component—internal actions are completely invisible.
 \item No output action of a component can be an output action of another component.
 \item No action is shared by infinitely many
 components.\footnote{Note that infinite (but countable) compositions
 \emph{are} permitted.}  In practice this means that no action can be an input action of infinitely many components, since the preceding rules mean that any action is an output or internal action of at most one component.
\end{enumerate}

All output actions of the components are also output actions of the composition.  An input action of a component is an input of the composition only if some other component doesn't supply it as an output; in this case it becomes an output action of the composition.  Internal actions remain internal (and largely useless, except for bookkeeping purposes).  

The \IOAtask equivalence relation is the union of the \IOAtask
relations for the components: this turns out to give a genuine
equivalence relation on output and internal actions precisely because the first two compatibility rules hold.

Given an execution or trace $X$ of a composite automaton that includes
$A$, we can construct the corresponding execution or trace $X|A$ of
$A$ which just includes the states of $A$ and the actions visible to
$A$ (events that don't change the state of $A$ drop out).  The
definition of composition is chosen so that $X|A$ is in fact an
execution/trace of $A$ whenever $X$ is.

\subsection{Hiding actions}
Composing $A$ and $B$ continues to expose the outputs of $A$ even if
they line up with inputs of $B$.  While this may sometimes be desirable, often we want to shove such internal communication under the rug.  The model lets us do this by redefining the signature of an automaton to make some or all of the output actions into internal actions.

\subsection{Fairness}
I/O automata come with a built-in definition of \index{execution!fair}\indexConcept{fair
execution}{fair executions}, where an execution of $A$ is
\concept{fair} if, for each equivalence class $C$ of actions in
$\IOAtask(A)$, 

\begin{enumerate}
 \item the execution is finite and no action in $C$ is enabled in the final state, or
 \item the execution is infinite and there are infinitely many
 occurrences of actions in $C$, or
 \item the execution is infinite and there are infinitely many states
 in which no action in $C$ is enabled.
\end{enumerate}

If we think of $C$ as corresponding to some thread or process, this
says that $C$ gets infinitely many chances to do something in an
infinite execution, but may not actually do them if it gives ups and
stops waiting (the third case).  The finite case essentially says that
a finite execution isn't fair unless nobody is waiting at the end.
The motivation for this particular definition is that it guarantees
(a) that any finite execution can be extended to a fair execution and
(b) that the restriction $X|A$ of a fair execution or trace $X$ is also fair.

Fairness is useful e.g. for guaranteeing message delivery in a message-passing system: make each message-delivery action its own task class and each message will eventually be delivered; similarly make each message-sending action its own task class and a process will eventually send every message it intends to send.  Tweaking the task classes can allow for possibilities of starvation, e.g. if all message-delivery actions are equivalent then a spammer can shut down the system in a ``fair'' execution where only his (infinitely many) messages are delivered.

\subsection{Specifying an automaton}
The typical approach is to write down preconditions and effects for
each action (for input actions, the preconditions are empty).  An
example would be the spambot in Algorithm~\ref{alg-spambot}.

\newData{\IOAspamAction}{spam}
\newData{\IOAsetMessageAction}{setMessage}
\SetKwFor{IOAinputAction}{input action}{}{end action}
\SetKwFor{IOAoutputAction}{output action}{}{end action}
\SetKwBlock{IOAprecondition}{precondition}{end precondition}
\SetKwBlock{IOAeffects}{effects}{end effects}

\newData{\IOAspambotState}{state}

\begin{algorithm}
\IOAinputAction{$\IOAsetMessageAction(m)$}{
\IOAeffects{
    $\IOAspambotState ← m$ \;
}
}
\IOAoutputAction{$\IOAspamAction(m)$}{
 \IOAprecondition{$\IOAspamAction = m$}
 \IOAeffects{none (keep spamming)}
}
\caption{Spambot as an I/O automaton}
\label{alg-spambot}
\end{algorithm}

(Plus an initial state, e.g. $\IOAspambotState = ⊥$, where $⊥$ is not a possible message, and a task partition, of which we will speak more below when we talk about liveness properties.)

\section{High-level view: traces}

When studying the behavior of a system, traces are what we really care about, and we want to avoid talking about states as much as possible.  So what we'll aim to do is to get rid of the states early by computing the set of traces (or fair traces) of each automaton in our system, then compose traces to get traces for the system as a whole.  Our typical goal will be to show that the resulting set of traces has some desirable properties, usually of the form (1) nothing bad happens (a \concept{safety property}); (2) something good eventually happens (a \concept{liveness property}); or (3) the horribly complex composite automaton representing this concrete system acts just like that nice clean automaton representing a specification (a \concept{simulation}).

Very formally, a \concept{trace property} specifies both the signature of the automaton and a set of traces, such that all traces (or perhaps fair traces) of the automata appear in the set.  We'll usually forget about the first part.

Tricky detail: It's OK if not all traces in $P$ are generated by $A$
(we want $\IOAtrace(A) \subseteq P$, but not necessarily $\IOAtrace(A)
= P$).  But $\IOAtrace(A)$ will be pretty big (it includes, for
example, all finite sequences of input actions) so hopefully the fact
that $A$ has to do something with inputs will tell us something useful.

\subsection{Example}

A property we might demand of the spambot above (or some other
abstraction of a message channel) is that it only delivers messages
that have previously been given to it.  As a trace property this says
that in any trace $t$, if $t_{k} = \IOAspamAction(m)$, then $t_{j} =
\IOAsetMessageAction(m)$ for some $j < k$.  (As a set, this is just
the set of all sequences of external spambot-actions that have this
property.)  Call this property $P$.

To prove that the spambot automaton given above satisfies $P$, we
might argue that for any execution $s_{0}a_{0}s_{1}a_{1}\dots{}$, that
$s_{i} = m$ in the last $\IOAsetMessageAction$ action preceding
$s_{i}$, or $⊥$ if there is no such action.  This is easily proved by
induction on $i$.  It then follows that since $\IOAspamAction(m)$ can
only transmit the current state, that if $\IOAspamAction(m)$ follows
$s_{i} = m$ that it follows some earlier $\IOAsetMessageAction(m)$ as claimed.

However, there are traces that satisfy $P$ that don't correspond to
executions of the spambot; for example, consider the trace
$\IOAsetMessageAction(0) \IOAsetMessageAction(1) \IOAspamAction(0)$.
This satisfies $P$ (0 was previously given to the automaton
$\IOAspamAction(0)$), but the automaton won't generate it because the
0 was overwritten by the later $\IOAsetMessageAction(1)$ action.  Whether this is indicates a problem with our automaton not being nondeterministic enough or our trace property being too weak is a question about what we really want the automaton to do.

\subsection{Types of trace properties}
\subsubsection{Safety properties}
$P$ is a \concept{safety property} if

\begin{enumerate}
 \item $P$ is nonempty.
 \item $P$ is \concept{prefix-closed}, i.e. if $xy$ is in $P$ then $x$
 is in $P$.
 \item P is \concept{limit-closed}, i.e. if $x_{1}, x_{1}x_{2},
 x_{1}x_{2}x_{3}, \dots{}$ are all in $P$, then so is the infinite sequence obtained by taking their limit.
\end{enumerate}

Because of the last restrictions, it's enough to prove that $P$ holds
for all finite traces of $A$ to show that it holds for all traces (and
thus for all fair traces), since any trace is a limit of finite
traces.  Conversely, if there is some trace or fair trace for which
$P$ fails, the second restriction says that $P$ fails on any finite
prefix of $P$, so again looking at only finite prefixes is enough.  The spambot property mentioned above is a safety property.

Safety properties are typically proved using \concept{invariants}, properties that are shown by induction to hold in all reachable states.

\subsubsection{Liveness properties}

$P$ is a \concept{liveness property} of $A$ if any finite sequence of
actions in $\IOAacts(A)$ has an extension in $P$.  Note that liveness
properties will in general include many sequences of actions that
aren't traces of $A$, since they are extensions of finite sequences
that $A$ can't do (e.g. starting the execution with an action not
enabled in the initial state).  If you want to restrict yourself only
to proper executions of $A$, use a safety property.  (It's worth
noting that the same property $P$ can't do both: any $P$ that is both a liveness and a safety property includes all sequences of actions because of the closure rules.)

Liveness properties are those that are always eventually satisfiable;
asserting one says that the property is eventually satisfied.  The
typical way to prove a liveness property is with a \concept{progress
function},
a function $f$ on states that (a) drops by at least 1 every time
something that happens infinitely often happens (like an action from
an always-enabled task class) and (b) guarantees $P$ once it reaches 0.

An example would be the following property we might demand of our
spambot: any trace with at least one $\IOAsetMessageAction(\dots{})$
action contains infinitely many $\IOAspamAction(\dots{})$ actions.
Whether the spambot automaton will satisfy this property (in
fair traces) depends on its task partition.  If all
$\IOAspamAction(\dots{})$ actions are in the same equivalence class,
then any execution with at least one \IOAsetMessageAction will have
some \IOAspamAction(\dots{}) action enabled at all times thereafter,
so a fair trace containing a \IOAsetMessageAction can't be finite
(since spam is enabled in the last state) and if infinite contains
infinitely many spam messages (since spam messages of some sort are
enabled in all but an initial finite prefix).  On the other hand, if
$\IOAspamAction(m_1)$ and $\IOAspamAction(m_2)$ are not equivalent in
$\IOAtask(A)$, then the spambot doesn't satisfy the liveness property:
in an execution that alternates $\IOAsetMessageAction(m_1)
\IOAsetMessageAction(m_2) \IOAsetMessageAction(m_1)
\IOAsetMessageAction(m_2) \dots$ there are infinitely many states in
which $\IOAspamAction(m_1)$ is not enabled, so fairness doesn't
require doing it even once, and similarly for $\IOAspamAction(m_2)$.

\subsubsection{Other properties}

Any other property $P$ can be expressed as the intersection of a
safety property (the
closure of $P)$ and a liveness property (the union of $P$ and the set
of all finite
sequences that aren't prefixes of traces in $P$).  The intuition is
that the safety property prunes out the excess junk we threw into the
liveness property to make it a liveness property, since any sequence
that isn't a prefix of a trace in $P$ won't go into the safety
property.  This leaves only the traces in $P$.

Example: Let $P = \{ 0^{n}1^{\infty} \}$ be the set of traces where we
eventually give up on our pointless 0-action and start doing only
1-actions forever.  Then $P$ is the intersection of the safety
property $S = \{ 0^{n}1^{m} \} \cup P$ (the extra junk is from
prefix-closure) and the liveness property $L = \{ 0^{n}11^{m}0x | x in
\{0,1\}^{*} \} \cup P$.  Property $S$ says that once we do a 1 we
never do a 0, but allows finite executions of the form $0^{n}$ where
we never do a 1.  Property $L$ says that we eventually do a 1-action, but that we can't stop unless we later do at least one 0-action.

\subsection{Compositional arguments}

The \concept{product} of trace properties $P_{1}, P_{2} \dots{}$ is
the trace property $P$ where $T$ is in $P$ if and only if $T|\IOAsig(P_{i})$ is
in $P_{i}$ for each $i$.  If the $\{A_{i}\}$ satisfy corresponding
propertties $\{P_{i}\}$ individually, then their composition satisfies
the product property.  (For safety properties, often we prove
something weaker about the $A_{i}$, which is that each $A_{i}$
individually is not the first to violate $P$—i.e., it can't leave
$P$ by executing an internal or output action.  In an execution where
inputs by themselves can't violate $P$, $P$ then holds.)

Product properties let us prove trace properties by smashing together
properties of the component automata, possibly with some restrictions
on the signatures to get rid of unwanted actions.  The product
operation itself is in a sense a combination of a Cartesian product
(pick traces $t_{i}$ and smash them together) filtered by a
consistency rule (the smashed trace must be consistent); it acts much
like intersection (and indeed can be made identical to intersection if
we treat a trace property with a given signature as a way of
describing the set of all $T$ such that $T|\IOAsig(P_{i})$ is in
$P_{i})$.

\subsubsection{Example}

Consider two spambots $A_1$ and $A_2$ where we identify the
$\IOAspamAction(m)$ operation of $A_1$ with the
$\IOAsetMessageAction(m)$ operation of $A_2$; we'll call this combined
action $\IOAspamAction_1(m)$ to distinguish it from the output actions
of $A_2$.  We'd like to argue that the composite automaton $A_1+A_2$
satisfies the safety property (call it $P_{m}$) that any occurrence of
$\IOAspamAction(m)$ is preceded by an occurrence of
$\IOAsetMessageAction(m)$, where the signature of $P_{m}$ includes
$\IOAsetMessageAction(m)$ and $\IOAspamAction(m)$ for some specific
$m$ but no other operations.  (This is an example of where trace property signatures can be useful without being limited to actions of any specific component automaton.)

To do so, we'll prove a stronger property $P'_{m}$, which is
$P_{m}$ modified to include the $\IOAspamAction_1(m)$ action in its
signature.  Observe that $P'_m$ is the product of the safety
properties for $A_1$ and $A_2$ restricted to $\IOAsig(P'_m)$, since
the later says that any trace that includes $\IOAspamAction(m)$ has a
previous $\IOAspamAction_1(m)$ and the former says that any trace that
includes $\IOAspamAction_1(m)$ has a previous
$\IOAsetMessageAction(m)$.  Since these properties hold for the
individual $A_1$ and $A_2$, their product, and thus the restriction
$P'_m$, holds for $A_1+A_2$, and so $P_{m}$ (as a further restriction) holds for $A_1+A_2$ as well.

Now let's prove the liveness property for $A_1+A_2$, that at least one
occurrence of $\IOAsetMessageAction$ yields infinitely many
\IOAspamAction actions.  Here we let 
$L_{1} = \{ \text{at least one \IOAsetMessageAction action} ⇒
\text{infinitely many $\IOAspamAction_1$ actions} \}$ and 
$L_{2} = \{ 
\text{at least one $\IOAspamAction_1$ action}
⇒
\text{infinitely many \IOAspamAction actions} \}$.
The product of these properties is all sequences with (a) no
\IOAsetMessageAction actions or (b) infinitely many \IOAspamAction
actions, which is what we want.  This product holds if the individual
properties $L_{1}$ and $L_{2}$ hold for $A_1+A_2$, which will be the
case if we set $\IOAtask(A_1)$ and $\IOAtask(A_2)$ correctly.

\subsection{Simulation arguments}
Show that $\IOAtraces(A)$ is a subset of $\IOAtraces(B)$ (possibly
after hiding some actions of $A$) by showing a 
\index{relation!simulation}
\concept{simulation relation} 
$f:\IOAstates(A) \rightarrow \IOAstates(B)$ between states of $A$ and
states of $B$.  Requirements on $f$ are

\begin{enumerate}
 \item If $s$ is in $\IOAstart(A)$, then $f(s)$ includes some element
 of $\IOAstart(B)$.
 \item If $(s,a,s')$ is in $\IOAtrans(A)$ and $s$ is reachable, then
 for any reachable $u$ in $f(s)$, there is a sequence of actions $x$
 that takes $u$ to some $v$ in $f(s')$ with $\IOAtrace(x) =
 \IOAtrace(a)$.
\end{enumerate}

Using these we construct an execution of $B$ matching (in trace) an
execution of $A$ by starting in $f(s_{0})$ and applying the second
part of the definition to each action in the $A$ execution (including the hidden ones!)

\subsubsection{Example}

A single spambot $A$ can simulate the conjoined spambots $A_1+A_2$.
Proof: Let $f(s) = (s,s)$.  Then $f(⊥) = (⊥, ⊥)$ is a start
state of $A_1+A_2$.  Now consider a transition $(s,a,s')$ of $A$; the
action a is either (a) $\IOAsetMessageAction(m)$, giving $s' = m$;
here we let $x = \IOAsetMessageAction(m) \IOAspamAction_1(m)$ with
$\IOAtrace(x) = \IOAtrace(a)$ since $\IOAspamAction_1(m)$ is internal
and $f(s') = (m,m)$ the result of applying $x$; or (b) $a =
\IOAspamAction(m)$, which does not change $s$ or $f(s)$; the matching
$x$ is $\IOAspamAction(m)$, which also does not change $f(s)$ and has the same trace.

A different proof could take advantage of $f$ being a relation by
defining $f(s) = \{ (s,s') | s' \in \IOAstates(A_2) \}$.  Now we don't
care about the state of $A_2$, and treat a $\IOAsetMessageAction(m)$
action of $A$ as the sequence $\IOAsetMessageAction(m)$ in $A_1+A_2$
(which updates the first component of the state correctly) and treat a
$\IOAspamAction(m)$ action as $\IOAspamAction_1(m) \IOAspamAction(m)$
(which updates the second component—which we don't care about—and
has the correct trace.)  In some cases an approach of this sort is
necessary because we don't know which simulated state we are heading
for until we get an action from $A$.

Note that the converse doesn't work: $A_1+A_2$ don't simulate $A$,
since there are traces of $A_1+A_2$ (e.g. $\IOAsetMessageAction(0)
\IOAspamAction_1(0) \IOAsetMessageAction(1) \IOAspamAction(0)$) that
don't restrict to traces of $A$.  See \cite[§{}8.5.5]{Lynch1996} for a more
complicated example of how one FIFO queue can simulate two FIFO queues
and vice versa (a situation called \concept{bisimulation}).

Since we are looking at traces rather than fair traces, this kind of
simulation doesn't help much with liveness properties, but sometimes
the connection between states plus a liveness proof for $B$ can be used
to get a liveness proof for $A$ (essentially we have to argue that $A$
can't do infinitely many action without triggering a $B$-action in an
appropriate task class).  Again see \cite[§{}8.5.5]{Lynch1996}.

\backmatter

\phantomsection
\addcontentsline{toc}{part}{Bibliography}
\bibliography{notes}

\clearpage
\phantomsection
\addcontentsline{toc}{part}{Index}
\printindex

\end{document}